UNIVERSITY OF LIVERPOOL

# Model Theoretic Characterisations of Description Logics

Thesis submitted in accordance with the requirements of the University of Liverpool for the degree of Doctor in Philosophy by

Robert Edgar Felix Piro

From Celle, Germany

August 2012

Degree of Doctor in Philosophy

| | |
|---|---|
| PRIMARY SUPERVISOR | Prof. Dr. Frank Wolter<br>Department of Computer Science<br>University of Liverpool |
| SECONDARY SUPERVISOR | Dr. Boris Konev<br>Department of Computer Science<br>University of Liverpool |
| ADVISER | Dr. Clare Dixon<br>Department of Computer Science<br>University of Liverpool |
| INTERNAL EXAMINER | Dr. Clare Dixon<br>Department of Computer Science<br>University of Liverpool |
| EXTERNAL EXAMINER | Prof. Dr. Ulrike Sattler<br>School of Computer Science<br>University of Manchester |

## Abstract

The growing need for computer aided processing of knowledge has led to an increasing interest in description logics (DLs), which are applied to encode knowledge in order to make it explicit and accessible to logical reasoning.

DLs and in particular the family around the DL $\mathcal{ALC}$ have therefore been thoroughly investigated w.r.t. their complexity theory and proof theory. The question arises which expressiveness these logics actually have.

The expressiveness of a logic can be inferred by a model theoretic characterisation. On concept level, these DLs are akin to modal logics whose model theoretic properties have been investigated. Yet the model theoretic investigation of the DLs with their TBoxes, which are an original part of DLs usually not considered in context of modal logics, have remained unstudied.

This thesis studies the model theoretic properties of $\mathcal{ALC}$, $\mathcal{ALCI}$, $\mathcal{ALCQ}$, as well as $\mathcal{ALCO}$, $\mathcal{ALCQO}$, $\mathcal{ALCQIO}$ and $\mathcal{EL}$. It presents model theoretic properties, which characterise these logics as fragments of the first order logic (FO).

The characterisations are not only carried out on concept level and on concept level extended by the *universal role*, but focus in particular on TBoxes. The properties used to characterise the logics are 'natural' notions w.r.t. the logic under investigation: On the concept-level, each of the logics is characterised by an adapted form of *bisimulation* and *simulation*, respectively.

TBoxes of $\mathcal{ALC}$, $\mathcal{ALCI}$ and $\mathcal{ALCQ}$ are characterised as fragments of FO which are invariant under *global bisimulation* and *disjoint unions*. The logics $\mathcal{ALCO}$, $\mathcal{ALCQO}$ and $\mathcal{ALCQIO}$, which incorporate individuals, are characterised w.r.t. to the class $\mathbb{K}$ of all interpretations which interpret individuals as singleton sets.

The characterisations for TBoxes of $\mathcal{ALCO}$ and $\mathcal{ALCQO}$ both require, additionally to being invariant under the appropriate notion of global bisimulation and an adapted version of disjoint unions, that an FO-sentence is, under certain circumstances, preserved under *forward generated subinterpretations*.

FO-sentences equivalent to $\mathcal{ALCQIO}$-TBoxes, are—due to $\mathcal{ALCQIO}$'s inverse roles—characterised similarly to $\mathcal{ALCO}$ and $\mathcal{ALCQO}$ but have as third additional requirement that they are preserved under *generated subinterpretations*.

$\mathcal{EL}$ as sub-boolean DL is characterised on concept level as the FO-fragment which is preserved under *simulation* and preserved under *direct products*. Equally valid is the characterisation by being preserved under simulation and having *minimal models*. For $\mathcal{EL}$-TBoxes, a global version of simulation was not sufficient but FO-sentences of $\mathcal{EL}$-TBoxes are invariant under *global equi-simulation*, disjoint unions and direct products.

For each of these description logics, the *characteristic concepts* are explicated and the characterisation is accompanied by an investigation under which notion of *saturation* the logic in hand enjoys the *Hennessy-and-Milner-Property*.

As application of the results we determine the *minimal globally bisimilar* companion w.r.t. $\mathcal{ALCQO}$-bisimulation and introduce the *$\mathcal{L}_1$-to-$\mathcal{L}_2$-rewritability problem* for TBoxes, where $\mathcal{L}_1$ and $\mathcal{L}_2$ are (description) logics. The latter is the problem to decide whether or not an $\mathcal{L}_1$-TBox can be equivalently expressed as $\mathcal{L}_2$-TBox. We give algorithms which decide $\mathcal{ALCI}$-to-$\mathcal{ALC}$-rewritability and $\mathcal{ALC}$-to-$\mathcal{EL}$-rewritability.

## Acknowledgements

At first, I would like express my gratitude to British society for its openness and fairness towards strangers which have made this thesis possible. In particular, I was warmly welcomed by the Renaissance Music Group and the Southport Bee-keepers which gave me close contact and insight into the British culture and accepted me as one of their own.

Just as important is the University of Liverpool and its Department of Computer Science, who created the opportunity for non-UK students to participate in a remunerated PhD-programme which allowed myself an adequate life in Liverpool for three years. I have greatly profited from the opportunities (ESSLLI09, in-house skills programmes, trips) and facilities (e.g. library and office) that the University and the Department provided.

Special thanks go to my supervisors Boris Konev and in particular Frank Wolter with his keen and unmistakable sense for scientifically interesting problems. He was a very knowledgeable and patient contact. I would like to thank Clare Dixon, who as my adviser gave me valuable feedback and cared for my progress towards the goal.

I am also very grateful for all the opportunities to present many of the results of my thesis at international workshops and conferences.

The results of Chapter 5 were published in [91] and presented at the European Conference on Artificial Intelligence 2010 (ECAI2010).

The results of Chapter 3, Section 4.4 and Chapter 6 were published in [92] and presented at the International Joint Conferences on Artificial Intelligence (IJCAI2011).

Results of this thesis were also presented at the DL-Workshop 2010 in Waterloo, Canada and at the British Colloquium for Theoretical Computer Science in Manchester in 2012.

Thanks go also to my fellow PhD-colleagues and friends at the Department of Computer Science who created an enjoyable atmosphere and who entrusted me with the appointment as representative for the PhD-students in the Department.

I would like thank Michel Ludwig, my flatmate and dear colleague for the good start he gave me when I arrived in Liverpool, his patience and support over the two years of flatmateship. I also would like to thank Helen Parkinson who refined my qualities as flatmate during the 8 month we shared the house together.

Last but not least, I would like to give big thanks to my partner James Furlong for all his love and support especially during the last gruelling months of my doctorate.

To my mother
Gisela Elisabeth Johanna Piro

# Contents

# 1. Introduction

## 1.1 Description Logics in Context

This thesis will give a model theoretic characterisation of the description logics family around $\mathcal{ALC}$ [50, 15, 21] as well as of the description logic (DL) $\mathcal{EL}$ [10] with a special emphasis on characterising TBoxes.

A formal definition of what a logic, a characterisation etc. is, shall be given in Section 1.2 but essentially a logic is a set of syntactical entities, the formulae, of which each has a precisely defined meaning. The latter is the reason, why this mathematical approach in modern logic has been consequently pursued, despite of all the weaknesses and restrictions that come with it [89, section 1.5].

This syntactical nature and the fact that reasoning over the encoded information in logic is to a certain extent possible by mechanical syntactic manipulation has led to fruitful applications of logics in computer science.

The ever growing knowledge of our time has created the desire to not only store this knowledge as mere data, but to be able to extract, reason and manipulate it with the aid of computers. Logics seem to lend itself to accomplish this task.

Solving the problems and questions arising from expressing and managing knowledge with logics are the objectives studied in the discipline of Knowledge Representation (KR) which is considered to form a part of Artificial Intelligence.

### 1.1.1 Description Logics: A First Glimpse

DLs are considered to be logics which are predominantly derived from the $\mathcal{ALC}$-language family, which are used in KR to encode the knowledge that needs to be represented. Traditionally this knowledge is divided into several parts, mostly into ABox, TBox and occasionally RBox.

Whilst the ABox (Assertion Box) contains assertions about properties of objects in the domain which is represented, the TBox (Terminology Box) contains definitions and specialisations of the vocabulary that is used. ABoxes and RBoxes (Role Boxes) are not of interest in this thesis. They are mentioned for the sake of completeness and we shall use ABoxes as a lead-up to TBoxes.

We shall give the following examples, using informally the DL $\mathcal{EL}$. Syntax and semantics will be precisely introduced for each logic in its proper chapter. In the description logic $\mathcal{EL}$, an ABox $\mathcal{A}$ could have the following form:

$$\mathcal{A} = \{\mathsf{isParentOf}(\mathit{Anna}, \mathit{Bob}), \mathsf{Female}(\mathit{Anna}), \mathsf{Male}(\mathit{Bob})\},$$

expressing that *Anna* is a parent of *Bob* that *Anna* is female and *Bob* is male. *Bob* and *Anna* are objects of the domain about which we want to represent knowledge. isParentOf, Female and Male are three symbols from our signature which we have chosen appropriately to represented knowledge about family relations and information of a set of people.

Male and Female are unary predicate symbols, which are called *concept names* in DLs whilst isParentOf is a binary predicate symbol, called *role name*. The syntactic entities of DLs (formulae) are called *concepts*. Throughout this thesis we shall only consider DLs over unary and binary predicate symbols.

In comparison to ABoxes which merely state facts about objects, the TBox contains what one could call the binding essence or the knowledge behind these facts. We illustrate this with an $\mathcal{EL}$-TBox $\mathcal{T}$

$$\mathcal{T} = \{\mathsf{Mother} \equiv \mathsf{Female} \sqcap \exists \mathsf{isParentOf}.\top,\ \exists \mathsf{isParentOf}.\top \sqsubseteq \exists \mathsf{isAncestorOf}.\top\}$$

The *concept definition* $\mathsf{Mother} \equiv \mathsf{Female} \sqcap \exists \mathsf{isParentOf}.\top$ can be read as 'Something is a mother iff it is female and parent of something.' It gives a definition for the vocabulary item Mother. Similarly the *concept inclusion* $\exists \mathsf{isParentOf}.\top \sqsubseteq \exists \mathsf{isAncestorOf}.\top$ specialises the concept 'being parent of': 'If someone is a parent of something then (s)he is also an ancestor of something (else).'

We have to add the word 'else', as it is a weakness of $\mathcal{EL}$-TBoxes not being able to express that a parent is necessarily an ancestor of whom this person is parent of.

An $\mathcal{EL}$-RBox contains syntactical entities, containing the symbol $\sqsubseteq$ or $\equiv$ with expressions involving role names on its left-hand and right-hand side. An RBox $\mathcal{R}$ could look like, e.g., $\mathcal{R} = \{\mathsf{isParentOf} \sqsubseteq \mathsf{isAncestorOf}\}$ stating 'If someone is a parent of something then (s)he is its ancestor.' RBoxes apparently remedy this weakness of $\mathcal{EL}$-TBoxes discovered above. Many more constructs involving role names are allowed in RBoxes, but neither RBoxes nor ABoxes alike will concern us henceforth. Instead we want to concentrate on TBoxes.

With the knowledge encoded in the TBox $\mathcal{T}$ we can now deduce that *Anna* is a mother or rather that $\mathsf{Mother}(\mathit{Anna})$ is true for every family in which $\mathcal{T}$ and $\mathcal{A}$ holds. We can, from the TBox $\mathcal{T}$ alone, infer that every mother is an ancestor of something or rather $\mathsf{Mother} \sqsubseteq \exists \mathsf{isAncestorOf}.\top$.

Unlike ABox atoms which correspond to FO-sentences[1] which state properties for specific individuals, TBoxes correspond to FO-sentences which express properties of global nature that affect all elements in the domain of discourse.

### 1.1.2 Applications of Description Logic

In their capacity of encoding knowledge and defining vocabularies, DLs have enjoyed considerable success during the last decades: One direction of application was to build ontologies from scratch and thus unifying efforts and results of earlier projects and carrying on the line of research:

The ontology SNOMED CT [101], the Systematised NOMenclature of MEDicine—Clinical Terms, encodes knowledge about the human body, its parts, down to the molecular level, their known illnesses and remedies for the latter. Its purpose is to provide a consistent patient record which is transferable across the different stages of medical care.

Its roots go back to the 1950s when SNOP, the Systematized NOmenclature of Pathology, was conceived and later published in the 1960s [39] (as book). After SNOP was extended to other fields of medicine becoming SNOMED in 1974, it was finally converted into an the ontology SNOMED RT [125] in the late 1990s. From the merger with the British National Healthcare System—Clinical Terms, a coded thesaurus of clinical terms, finally SNOMED CT emerged [132]. The transition to SNOMED RT was important as this knowledge and terminology was for the first time captured as ontology using description logic with its rigorous syntax and precisely defined meaning. SNOMED CT has grown to one of the largest and most popular examples of ontologies: It encodes more than 311 000 concepts, and is, in terms of what has been seen as real-life application so far, huge in size.

Another approach was to interlink and unify vocabularies of different research projects via ontologies: A noticeable example is the Gene Ontology (GO) [40]: Deciphering the genome of living beings like humans, flies and mice was an objective of several projects. Unfortunately different structures and terms were used which make these efforts complementary but not joinable. GO applies DL in order to define a standardised vocabulary and in this way to abstract from the underlying databases. The vocabulary is enriched with annotations in the sense that it contains references which point back to the appropriate entity in the database. This also allows projects which take part in GO to model their database in a GO compatible way by using their vocabulary, which in turn grants biologists unified access to data across different areas of research.

[1] A sentence is a formula in which every variable appears in the scope of a quantification.

A recent further development is the third approach known as ontology based data access (OBDA)[32]. Similarly as in the case of GO, the ontology provides a standardised and well defined vocabulary abstracting from the database layer. But the ontology also provides knowledge about the interrelations between the database entities: OBDA uses DLs to endow the data collections with additional meaning which can later be used to make up for incomplete data. The user then accesses the database on the level of the DL, formulating the queries in the language of the ontology, which also explains the name OBDA. In this setting, the database is generally considered to be an ABox.

This recently sparked interest has been met by several papers and projects. The papers either stress a more database oriented approach, where information of the ontology is incorporated into the database and queries are rewritten [84, 93] or a more ontology oriented investigation, where reasoning techniques are adapted to cope with very large ABoxes [45, 70].

OBDA is currently under active investigation e.g. by the ExODA project which looks into 'Integrating Description Logics and Database Technologies for Expressive Ontology-Based Data Access' [49]. As another example for OBDA which applies description logic ontologies outside of life sciences we name the the OPTIQUE project [100]:

OPTIQUE aims to facilitate uniform and intuitive access for engineers to heterogeneous data sources: in oil industry, experts want to extrapolate geological data of past explorations to new exploration missions. The ontology which is to be developed is not only meant to unify and abstract from the underlying different databases, in which the geophysical data is stored, but is also aimed at facilitating query optimisation, which allows for efficient data processing on the database side. Thus a repository of 1000 Terabytes of data, stored in different databases with more then 2000 tables is to be queried.

Similarly, engineers in power plants want to relate sensor data to events, i.e. technically specified phenomena, occurring during operation. The sensor data amounts to several Terabytes and is again spread over several different databases.

Further possible applications of DLs are under investigation like concept learning and automated ontolgy building, but the three examples given have or will have very soon impact on practice. The question arises whether everything can and should be encoded in DLs and if not where the boundaries lie.

### 1.1.3 The Trade-Off between Expressivity and Complexity

Making inferences from knowledge encoded in ontologies, i.e. reasoning, allows us to discover knowledge which might or might not be obvious in the real world. The more precisely knowledge is encoded, the better it fits with the real world and the more knowledge we can discover. One can therefore think of a lot more constructs than e.g. RBoxes shortly discussed earlier, which one could introduce to a DL like $\mathcal{EL}$ in order to remedy its weaknesses in terms of what can or cannot be expressed in this language.

Though, increasing the expressivity of a language comes with a price: The reasoning complexity rises. With reasoning complexity we mean here a worst case measure for how long or how much space the expected computation may take w.r.t. the length of the input[87, 107]. As an example: One has to expect that for any given $\mathcal{ALC}$-TBox $\mathcal{T}$ and an arbitrary $\mathcal{ALC}$-concept inclusion $C \sqsubseteq D$ it might take exponential time [15] in the length of $\mathcal{T}$ and $C \sqsubseteq D$ to determine whether $C \sqsubseteq D$ follows from $\mathcal{T}$.

The reasoning complexity can quickly grow into undecidability, which is not desirable for DLs, as automated reasoning becomes severely restricted. KL-ONE [24], for example, one of the prominent efforts to formalise knowledge representation in the late 1970s [25], turned out to be undecidable.[13]( [30, 119, introduction sections] As a consequence, a host of DLs have since been developed in order to achieve different trade-offs between expressivity and reasoning complexity, always striving for decidability.

The prospect of embedding knowledge also in websites has led the World Wide Web Committee (W3C) to draw up a standard for knowledge representation languages, the OWL-specification [78, 36]. OWL is the mnemonic abbreviation for Web Ontology Language. In OWL [79], three sublanguages were specified: OWL Lite, OWL DL and OWL Full.

While OWL Lite as a restriction of OWL DL was deemed to be closer to applications, it did not yield significantly better complexity than OWL DL; The sat-isfiability problem is for OWL Lite ExpTime complete [128][2] and for OWL DL NExpTime complete [128]. OWL DL, in turn, contains various additional constructs with side conditions in order to be decidable[78, p.372]. OWL Full, finally, contains all the constructs of OWL DL, yet without the side conditions, which renders OWL Full undecidable.

The W3C issued the OWL 2 recommendation [35] containing additionally 3 profiles of tractable sublanguages: OWL 2 EL, OWL 2 QL and OWL 2 RL. The OWL 2 EL profile has the description logic $\mathcal{EL}$ as its logical underpinning. $\mathcal{EL}$ is thus a DL which is used in significant practical applications so that a characterisation of its expressivity is a creditable objective in its own right.

### 1.1.4 The Language Family around $\mathcal{ALC}$

The $\mathcal{A}$ttributive $\mathcal{L}$anguage with $\mathcal{C}$omplements, $\mathcal{ALC}$, was one of the new decidable languages proposed [122] as reaction to the discovery that formerly used methods of knowledge representation turned out to be undecidable.

$\mathcal{ALC}$ is an extension of $\mathcal{EL}$ in the sense that $\mathcal{ALC}$ is $\mathcal{EL}$ enriched by logical negation ($\neg$). As noticed in [121], it therefore corresponds on the concept level to the multi-modal logic K[18, 21, 50]. The modal logic K is also known as standard modal logic ML. The latter was already under thorough investigation and therefore provided a good theoretical foundation for $\mathcal{ALC}$.

Since $\mathcal{ALC}$-concepts are closed under boolean operations it is considered to be a kind of core logic for expressive DLs. Many extensions have been proposed [31], like inverse role names ($\mathcal{I}$) or counting quantification ($\mathcal{Q}$) and individual names ($\mathcal{O}$), which shall be explained in detail in their appropriate chapters.

Indeed the combination of all the aforementioned extensions and therefore $\mathcal{ALC}$ itself, form the underpinning of OWL Full. Since the concept level of $\mathcal{ALC}$ and its extensions can be considered as extensions of modal logics, these extensions of modal logics have been investigated from the model theoretic point of view [21]. A lot of $\mathcal{ALC}$-extensions were investigated in [85], but just on the concept level. In this sense, they were treated like modal logics in disguise.

TBoxes, an original part of knowledge representation, are not of particular interest in modal logics and therefore this thesis will give a model-theoretic characterisation of the expressiveness of these fragments of first order logic (FO) with a particular focus on TBoxes over these fragments.

[2]Tobies showed that the reasoning problem for $\mathcal{SHIF}$, the DL underpinning OWL Lite, is ExpTime complete.

### 1.1.5 Contribution

Clearly, one would want to know the expressivity of these description logics, which have received so much scientific attention and have become a focal point in applications. This thesis gives a model-theoretic characterisation for several description logics from the $\mathcal{ALC}$-family as well as for $\mathcal{EL}$ and shows how these results can be applied.

The motivation for 'characterising' the expressivity of a logic $\mathcal{L}$ is to gain knowledge about its expressiveness in different terms: Of course, one could give a trivial answer to the question what the expressivity of a language is, by saying 'everything that can be encoded by its syntax'.

The aim of characterising the expressivity of a logic $\mathcal{L}$ is therefore to give an answer to this question in different terms. The answer can be given with respect to another logic or in general:

If the answer is given with respect to another logic, it usually involves logics whose expressivity is well known and of which the logic $\mathcal{L}$ is a subset of formulae. Theorems that establish such a characterisation are sometimes called preservation theorems. A typical representative for such a characterisation is e.g. van Benthem's model-theoretic characterisation [61] of ML w.r.t. FO or the model-theoretic characterisations of several logics of Kourtonina and De Rijke [85].

A general answer, on the other hand, refers to any (abstract) logic that could possibly comprise the logic $\mathcal{L}$ that is to be characterised. These theorems give properties under which a surrounding logic satisfying these properties cannot be more expressive than $\mathcal{L}$ itself. This approach has been taken by Lindström [47, 16](originally [90]) whose theorem characterises the expressiveness of FO. In recent times logics like $\mathcal{ALC}$[112, 126] and $\mathcal{ALC}u$ [105] have been characterised on the concept level by Lindström-like theorems.

In this thesis we shall concentrate on characterisations which are given w.r.t. to FO, i.e. on the former type of answer. We are hence looking for properties which are satisfied by the logic $\mathcal{L}$, which is to be characterised, so that for every signature $\tau$ and for every formula in FO($\tau$), which satisfies the properties, there exists some logically equivalent concept of $\mathcal{L}(\tau)$.

Similarly we try to find properties which are satisfied by all TBoxes over $\mathcal{L}(\tau)$, so that for every signature $\tau$ and for every sentence in FO($\tau$), which satisfies these properties, there exists some logically equivalent TBox over $\mathcal{L}(\tau)$.

Note that we only consider characterisations in which the formulae and concepts (or sentences and TBoxes respectively) share the same symbols.

The question arises, how these properties are selected. Why certain properties and not some other properties? One could suggest the trivial property 'must be logically equivalently expressible in $\mathcal{L}$'. Indeed, there is no ultimate property but it is up to personal consideration whether or not the given properties are deemed to be canonical and useful.

In the characterisations given in this thesis, those properties for a logic $\mathcal{L}$ were chosen, for which $\mathcal{L}$ is broadly known for in the academic community: in the case of $\mathcal{ALC}$-TBoxes, say, the appropriate notion of bisimulation invariance and invariance under disjoint unions were used.

In cases where this was not enough, we decided e.g. for direct products because they are considered to be proper[3] model theoretic notions. In the case where individual names were involved, we decided for a self-evident notion, which complements invariance under disjoint unions.

It turns out that our decisions as to which properties to choose, were fruitful w.r.t. to some application: In the last chapter we show that it is possible to algorithmically decided whether or not characterising properties of $\mathcal{ALC}$ and $\mathcal{EL}$, respectively, are satisfied by TBoxes of a certain, stronger logic. Thus, we can algorithmically decide whether these TBoxes of this stronger logic can be expressed by $\mathcal{ALC}$-TBoxes or $\mathcal{EL}$-TBoxes respectively.

### 1.1.6 Structure of the Thesis

Throughout the thesis, the characterisations follow a pick and mix scheme of corresponding properties as Table 1.1 shows: each feature $\mathcal{I}$, $\mathcal{Q}$ and $\mathcal{O}$ corresponds to a certain extension of the syntax of $\mathcal{ALC}$ which is accounted for with an additional property in the model-theoretic game. In contrast, $\mathcal{EL}$ is w.r.t. simulation a restriction of $\mathcal{EL}^{\sqcup}$ and therefore needs a special property, namely being preserved under direct products, to be captured. Syntax and notions will be introduced in their appropriate chapters.

Similarly, Table 1.2 shows which properties were used in order to characterise TBoxes of the logic in the left column. In all cases the appropriate model-theoretic game is used in its global variant. Except for DLs with nominals ($\mathcal{O}$), all description logics use invariance under disjoint union. Description logics with nominals use invariance of a special notion of disjoint unions (e.g. $\biguplus^{\mathbb{K}}$ the adaption of disjoint union for $\mathcal{ALCQO}$) plus being preserved under disjoint unions of generated subinterpretations. This is also true for $\mathcal{ALCQIO}$ but for $\mathcal{ALCQIO}$ every nominal disjoint union is a disjoint union over generated subinterpretations; hence we can state invariance under disjoint unions without explicitly mentioning the preservation under disjoint unions of generated subinterpretations.

[3]Proper means here that these notions are established in (modal) model theory, e.g. [62] for disjoint unoins.

| Logic | Operators | model-theoretic game | special property |
|---|---|---|---|
| $\mathcal{ALC}$ | $\neg, \sqcap, \exists r$ | bisimulation | |
| $\mathcal{ALCI}$ | $\mathcal{ALC} + \exists r^-$ | bisimulation with inverse moves | |
| $\mathcal{ALCQ}$ | $\mathcal{ALC} + \exists^{\geq \kappa} r$ | bisimulation with successor sets | |
| $\mathcal{ALCO}$ | $\mathcal{ALC}$ + nominals | bisimulation | model-class $\mathbb{K}$ |
| $\mathcal{ALCQO}$ | $\mathcal{ALCQ}$ + nominals | bisimulation with successor sets | model-class $\mathbb{K}$ |
| $\mathcal{ALCQIO}$ | $\mathcal{ALCQO} + \exists^{\geq \kappa} r^-$ | bisimulation with inverse moves and successor sets | model-class $\mathbb{K}$ |
| $\mathcal{EL}$ | $\sqcap, \exists r$ | simulation | preserved under direct products |
| $\mathcal{EL}^{\sqcup}$ | $\mathcal{EL} + \sqcup$ | simulation | |
| $\mathcal{EL}^{\neg}$ | $\mathcal{EL}$ closed under $\sqcap$ and $\neg$ | equi-simulation | |

Table 1.1: This table shows for the DL named in the left column which syntactic constructs are allowed and which corresponding properties were used in order to characterise the logic on concept level.

We have that $\mathcal{EL}^{\neg}$-TBoxes coincide with $\mathcal{EL}^{\sqcup}$-TBoxes. $\mathcal{EL}$-TBoxes, however, relate to $\mathcal{EL}^{\sqcup}$ in an analogous fashion as $\mathcal{EL}$-concepts relate to $\mathcal{EL}^{\sqcup}$-concepts: $\mathcal{EL}$-TBoxes are those $\mathcal{EL}^{\sqcup}$-TBoxes which are preserved under direct products.

Since the schema of extension of the syntax on the one hand and extension of the model theoretic game on the other hand is always the same, the arguments in the appropriate proofs may seem somewhat repetitive and might be skipped by the reader, unless she or he needs reassurance. We therefore decided to spell out the proofs in Chapter 2 in detail, even though they can be found in textbooks and other literature.

| TBox | Disjoint union | model-theoretic game | special property |
|---|---|---|---|
| $\mathcal{ALC}$ | invariant under disjoint unions | global bisimulation | |
| $\mathcal{ALCI}$ | invariant under disjoint unions | global bisimulation with inverse moves | |
| $\mathcal{ALCQ}$ | invariant under disjoint unions | global bisimulation with successor sets | |
| $\mathcal{ALCO}$ | inv. u. coherent $\biguplus$ pres. u. coherent $\biguplus$ of gen. subinterpr. | global bisimulation | model-class $\mathbb{K}$ |
| $\mathcal{ALCQO}$ | inv. under $\biguplus^{\mathbb{K}}$ preserved under $\biguplus^{\mathbb{K}}$ of gen. subinterpr. | global bisimulation with successor sets | model-class $\mathbb{K}$ |
| $\mathcal{ALCQIO}$ | invariant under nominal disjoint unions | global bisimulation with inverse moves and successor sets | model-class $\mathbb{K}$ |
| $\mathcal{EL}$ | invariant under disjoint unions | global equi-simulation | preserved under direct products |
| $\mathcal{EL}^{\sqcup}/\mathcal{EL}^{\neg}$ | invariant under disjoint unions | global equi-simulation | |

Table 1.2: This table shows the properties which were used in order to characterise TBoxes of the logic named in the left column.

The second section of this chapter concerns itself in the first half with preliminaries: $\mathcal{ALC}$ with its syntax and semantics are introduced along with the necessary notion of interpretation and translations into other logics. The second half discusses the notion of logic, expressivity as well as compactness and defines certain abbreviations.

The second chapter then recalls van Benthem's model-theoretic characterisation of $\mathcal{ALC}$ on concept level and introduces important notions which will be used in adapted form throughout the rest of the thesis. $\mathcal{ALC}u$ will then be characterised in an equally detailed fashion, and it becomes fairly obvious how the syntax, the game and notions like type and saturation are adapted in order to accommodate the global perspective introduced by the universal role $u$. This leads up to TBoxes over $\mathcal{ALC}$ which form a fragment of $\mathcal{ALC}u$. The chapter closes with the characterisation of TBoxes over $\mathcal{ALC}$. In the third chapter $\mathcal{ALCI}$, $\mathcal{ALCI}u$ and TBoxes over $\mathcal{ALCI}$ will be characterised. The steps follow the same scheme as in Chapter 2, yet making necessary adaptions of notions introduced in Chapter 2. Since no completely new notions are introduced, things are kept shorter. Similarly $\mathcal{ALCQ}$ will be introduced and $\mathcal{ALCQ}$, $\mathcal{ALCQ}u$ and TBoxes over $\mathcal{ALCQ}$ are characterised in the second half of the chapter.

In Chapter four $\mathcal{ALCO}$, $\mathcal{ALCQO}$ and $\mathcal{ALCQIO}$ with their extension by $u$, together with TBoxes over these three DLs are characterised, following the schema of Chapter 2. We choose to characterise the logics with respect to a special class of interpretations named $\mathbb{K}$. Additionally the section treating $\mathcal{ALCQO}$ contains an extension explaining how minimal $\mathcal{ALCQO}$-bisimilar companions of an interpretation can be obtained, thus making use of the machinery introduced for the $\mathcal{ALCQO}$-characterisation.

Chapter five is devoted to the characterisation of $\mathcal{EL}$. In the course of $\mathcal{EL}$'s investigation several interesting observations were made, leading to a characterisation of not only $\mathcal{EL}$ but also of logics $\mathcal{EL}^{\sqcup}$, $\mathcal{EL}^{\neg}$ and $\mathcal{EL}u^{\neg}$, which can be considered to be newly introduced.

Chapter six gives algorithms to determine whether TBoxes can be rewritten in weaker fragments. For specific details on these rewritings and the algorithms itself, the reader is referred to the chapter itself. These algorithms therefore yield an application for the characterisation results. Finally Chapter seven concludes.

The characterisations of $\mathcal{ALCO}$, $\mathcal{ALCQIO}$ as well as $\mathcal{ALCQ}$ and $\mathcal{ALCI}$ have been published in [92] together with an investigation of its applicability presented in Chapter 7. The characterisation for $\mathcal{EL}$ was published in [91].

## 1.2 Preliminaries

Since the model-theory of $\mathcal{ALC}$-concepts or rather its modal logic counterpart ML is well established, we shall make $\mathcal{ALC}$-concepts the point of origin for our investigations. In this section we shall introduce the notion of interpretation and we shall then define the syntax and semantics of $\mathcal{ALC}$-concepts. Additionally, we give transcriptions into first order logic and modal logic.

It shall be remarked that we denote the first limit ordinal by $\omega$, and write $n < \omega$ or similar, to indicate that $n$ is a natural number. We always include the natural number naught.

### 1.2.1 Signature and Interpretation

A *signature* $\tau$ is a set of symbols in connexion with a function $\mathrm{ar} : \tau \longrightarrow \omega$ for this signature that assigns to each symbol $S \in \tau$ a natural number $\mathrm{ar}(S)$, which is called the arity of the symbol $S$.

An interpretation $\mathfrak{I}$ is a pair consisting of a symbol $\Delta$ and an interpretation function $\cdot^{\mathfrak{I}}$. The function maps $\Delta$ to some non-empty set $\Delta^{\mathfrak{I}}$, called the carrier-set of $\mathfrak{I}$. We call $\mathfrak{I}$ a *$\tau$-interpretation*, if $\cdot^{\mathfrak{I}}$ maps each symbol $S \in \tau$ to the cartesian product $(\Delta^{\mathfrak{I}})^{\mathrm{ar}(S)}$ or any possibly empty subset of it. In case $\mathrm{ar}(S) = 0$, $S^{\mathfrak{I}}$ is an element of $\Delta^{\mathfrak{I}}$. Note that we do not allow for nullary predicate symbols, which would simply be true ($\top$) or false ($\bot$), but rather for nullary function symbols, which yield a constant, which we interpret as singleton set, i.e. as set containing only one element.

However, in this thesis, we shall not treat functional symbols in general, i.e. the signatures are purely relational (except for nullary function symbols), and we shall only consider those signatures $\tau$ where each symbol $S \in \tau$ has $\mathrm{ar}(S) \leq 2$. We partition these signatures into three (disjoint) subsets namely $\mathsf{N_R}$, $\mathsf{N_C}$ and $\mathsf{N_I}$.

$\mathsf{N_R}$ contains all binary symbols of $\tau$, i.e. all $S \in \tau$ with $\mathrm{ar}(S) = 2$. They are called *role names* (therefore $\mathsf{N_R}$) or accessibility relations and we usually denote them with the letters $r$ and $s$ etc.

$\mathsf{N_C}$ contains all unary symbols, i.e. all $S \in \tau$ with $\mathrm{ar}(S) = 1$. They are called *concept names*, hence $\mathsf{N_C}$, and we usually denote them with letters $A$ and $B$.

$\mathsf{N_I}$, the set of all nullary symbols, contains all $S \in \tau$ with $\mathrm{ar}(S) = 0$. They are called *individual names* ($\mathsf{N_I}$) or objects or constants. We represent them with small letters $a$ and $b$.

In the example given earlier, the signature $\mathsf{N_C} = \{\mathsf{Male}, \mathsf{Female}, \mathsf{Mother}\}$, $\mathsf{N_R} = \{\mathsf{isPartenOf}, \mathsf{isAncestorOf}\}$ and $\mathsf{N_I} = \{\mathsf{Anna}, \mathsf{Bob}\}$ was used. If we set $\tau := \mathsf{N_C} \cup \mathsf{N_R} \cup \mathsf{N_I}$ a $\tau$-interpretation is any function $\cdot^{\mathfrak{I}}$ such that the carrier-set $\Delta^{\mathfrak{I}}$ contains at least one element and all other symbols are interpreted according to their arity. In particular, Anna and Bob must be each assigned to some element, possibly even the same element whilst all other symbols could be the empty set.

### 1.2.2 Syntax and Semantics of $\mathcal{ALC}$-concepts

The way, semantics is defined in description logics is somewhat different from the classical approach. In order to make it accessible to readers with either background, we shall give, in addition to the notation in DL, the translation to modal logic and first order logic respectively.

Let $\tau$ be a signature, i.e. relational with symbols of arity 1 or 2, thus omitting constants for now. Recall $\mathsf{N_C} \cup \mathsf{N_R} = \tau$. Instead being called *formula*, the syntactic entities of description logics are called *concepts* in description logic. The reason for this lies in the semantics: while formulae are assigned truth values for either being satisfied or not satisfied in a given interpretation, a concept is assigned the set of all elements which satisfy the concept for a given interpretation.

Depending on the signature $\tau$ the set $\mathcal{ALC}(\tau)$ of *$\mathcal{ALC}$-concepts over $\tau$* is recursively defined as follows:

1. $\top$ is a concept in $\mathcal{ALC}$ over $\tau$.
2. if $A \in \mathsf{N_C}$ then $A$ is a concept in $\mathcal{ALC}$ over $\tau$ (i.e. $A \in \mathcal{ALC}(\tau)$).
3. if $C, D \in \mathcal{ALC}(\tau)$ then $C \sqcap D \in \mathcal{ALC}(\tau)$ and $\neg C \in \mathcal{ALC}(\tau)$.
4. if $r \in \mathsf{N_R}$ and $C \in \mathcal{ALC}(\tau)$ then $\exists r.C \in \mathcal{ALC}(\tau)$.

Obviously, we avoid going into the intricacies of bracketing but we shall use brackets when needed. We define the common abbreviations for all $r \in \mathsf{N_R}$ and $C, D \in \mathcal{ALC}(\tau)$:

$$C \sqcup D := \neg(\neg C \sqcap \neg D) \quad \text{and} \quad \forall r.C := \neg\exists r.\neg C \quad \text{and} \quad C \rightarrow D := \neg C \sqcup D$$

The length $|C|$ of a concept is recursively defined as

$$|C| := \begin{cases} 1 & \text{if } C \in \mathsf{N_C}, & 1 + |D| + |E| & \text{if } C = D \sqcap E \\ 1 + |D| & \text{if } C = \neg D, & 1 + |D| & \text{if } C = \exists r.D \end{cases}$$

The *semantics of a concept* is given by the recursive definition below. Whilst, e.g., first order sentences are either true or false w.r.t. a given interpretation, the meaning of a concept w.r.t. a given interpretation is defined by the set of all elements of this interpretation at which this concept is satisfied. The interpretation function $\cdot^{\mathfrak{I}}$ is hence recursively extended for concepts in $C \in \mathcal{ALC}(\tau)$ as follows:

$$C^{\mathfrak{I}} := \begin{cases} \Delta^{\mathfrak{I}} & \text{if } C = \top \\ A^{\mathfrak{I}} & \text{if } C = A \text{ and } A \in \mathsf{N_C} \\ D^{\mathfrak{I}} \cap E^{\mathfrak{I}} & \text{if } C = D \sqcap E \\ \Delta^{\mathfrak{I}} \setminus D & \text{if } C = \neg D \\ \{d \in \Delta^{\mathfrak{I}} \mid \exists d' \in D^{\mathfrak{I}} : (d, d') \in r^{\mathfrak{I}}\} & \text{if } C = \exists r.D \end{cases}$$

### 1.2.3 Translation into Standard Modal Logic

We shall swiftly recall the one-to-one correspondence between $\mathcal{ALC}$-concepts and modal logic formulae thus making the subject matter accessible to readers with background in modal logics. Indeed, it turns out that modal logics preceded DLs by many decades: [18, Historical Overview] traces its roots back as far as 1918 and the field of modal logic was fully developed by the 1980s, but only in 1991 after [121] was published this intimate relation became clearer and was subsequently studied in more detail [88]. We shall first introduce the standard modal logic ML [18], in modal logics also known as (System) K [21]. Instead of giving a lengthy definition for the syntax, we shall give a Backus-Naur form, again implicitly assuming proper bracketing: $\varphi$ is a formula of the *standard modal logic* ML over the signature $\tau$ if

$$\varphi ::= \top \mid A \mid \chi \wedge \psi \mid \neg\chi \mid \Diamond_r \chi$$

with $A \in \mathsf{N_C}$, i.e. $A$ is unary, and $r \in \mathsf{N_R}$ and $\chi$, $\psi$ formulae of ML over $\tau$.

The semantics of a formula in ML is given for pairs consisting of an interpretation $\mathfrak{I}$ and an element $d \in \Delta^{\mathfrak{I}}$. The pair $(\mathfrak{I}, d)$ is called *pointed interpretation*, as $d$ is considered to be a distinguished point. For every formula in ML($\tau$) we recursively define $(\mathfrak{I}, d) \models \varphi$ by

$$
\begin{array}{ll}
(\mathfrak{I}, d) \models \top & \\
(\mathfrak{I}, d) \models A & \text{if } d \in A^{\mathfrak{I}} \\
(\mathfrak{I}, d) \models \chi \wedge \psi & \text{if } (\mathfrak{I}, d) \models \chi \text{ and } (\mathfrak{I}, d) \models \psi \\
(\mathfrak{I}, d) \models \neg\chi & \text{if not } (\mathfrak{I}, d) \models \chi \\
(\mathfrak{I}, d) \models \Diamond_r \chi & \text{if there is } d' \in \Delta^{\mathfrak{I}} \text{ s.t. } (d, d') \in r^{\mathfrak{I}} \text{ and } (\mathfrak{I}, d') \models \chi
\end{array}
$$

If $(\mathfrak{I}, d) \models \varphi$ we read $(\mathfrak{I}, d)$ *satisfies* $\varphi$ or equivalently $(\mathfrak{I}, d)$ *is a model of* $\varphi$.

Replacing $\wedge$ by $\sqcap$ and $\Diamond_r$ by $\exists r.$ for each $r \in \mathsf{N_R}$ yields a translation of ML-formulae into $\mathcal{ALC}$-concepts. By reversing this process, one obtains a translation from $\mathcal{ALC}$-concepts into ML formulae. If $\varphi_C$ is the translation of $C \in \mathcal{ALC}(\tau)$ into ML we have

$$C^{\mathfrak{I}} = \{d \in \Delta^{\mathfrak{I}} \mid (\mathfrak{I}, d) \models \varphi_C\}$$

i.e $C^{\mathfrak{I}}$ is the set of all elements in $\Delta^{\mathfrak{I}}$ that satisfy $\varphi_C$; analogously for $C_\varphi$ being the translation of $\varphi \in \mathrm{ML}(\tau)$ into $\mathcal{ALC}$ we have

$$(\mathfrak{I}, d) \models \varphi \iff d \in C_\varphi^{\mathfrak{I}}.$$

Indeed, $\mathcal{ALC}$-concepts have the same logical expressivity as ML. This shall be explained a bit further. A logic $\mathcal{L}_1$ is *at least as expressive as* another logic $\mathcal{L}_0$ ($\mathcal{L}_0 \leq \mathcal{L}_1$) if for all signatures $\tau$ and all $\varphi \in \mathcal{L}_0(\tau)$ there is $\psi \in \mathcal{L}_1(\tau)$ such that $(\mathfrak{I}, d) \models \varphi$ iff $(\mathfrak{I}, d) \models \psi$. So practically, for each $\tau$, $\mathcal{L}_0(\tau)$ can be regarded as subset of $\mathcal{L}_1(\tau)$.

Although in description logics, the semantics of a concept is expressed as sets of satisfying elements, rather than the model theoretic notion of a pointed interpretation being a model of a formula, it makes sense to extend the satisfaction relation $\models$ to $\mathcal{ALC}$-concepts, by setting

$$(\mathfrak{I}, d) \models C \iff d \in C^{\mathfrak{I}}.$$

This definition is consistent with the translation function, as $(\mathfrak{I}, d) \models C$ iff $(\mathfrak{I}, d) \models \varphi_C$. But this means that $\mathcal{ALC}$ is at least as expressive as ML. As the converse is true as well, both are equally expressive.

### 1.2.4 Translation into First Order Logic

Readers familiar with first order logic (FO) might find it helpful to see a translation of $\mathcal{ALC}$-concepts into FO. From the outset of DLs it was known that DLs and in particular $\mathcal{ALC}$ form fragments of FO. The relation ship of DLs and FO was after the revelations due to [121] and [88] further investigated in [22]. The translation of $\mathcal{ALC}$-concepts into FO-formulae we are about to present is in fact the adaption of the well known *standard translation* from ML into FO, which was presented in [130].

Let $\mathrm{FO}(\tau)$ denote all FO-formulae over the signature $\tau$ and let $x_0, x_1$ be two variables. For $i \in \{0, 1\}$ the *translation function* $[\cdot\,;\,\cdot] : \mathcal{ALC}(\tau) \times \mathrm{Var} \longrightarrow \mathrm{FO}(\tau)$ is recursively defined as

$$[C; x_i] := \begin{cases} \bot & C = \bot \\ A(x_i) & A = C \text{ and } A \in \tau \\ \neg[D; x_i] & C = \neg D \\ [D; x_i] \wedge [E; x_i] & C = D \sqcap E \\ \exists x_{1-i}.r(x_i, x_{1-i}) \wedge [D; x_{1-i}] & C = \exists r.D \end{cases}$$

We obtain $(\mathfrak{I}, d) \vDash C \iff \mathfrak{I}\frac{d}{x_0} \vDash [C; x_0]$, where $\frac{d}{x_0}$ assigns $d$ to the variable $x_0$. The translation function practically gives away the syntax of the FO-fragment which is equally expressive to $\mathcal{ALC}$-concepts. It also shows that $\mathcal{ALC}$-concepts are a fragment of the two-variable fragment of FO, as at most two variables, here $x_0$ and $x_1$, are needed. This property was pointed out in [53] who at the time expressed this circumstance as ‘having 2 dimensions’ and further investigated in e.g. [64].

Example 1.2.1 . In order to see, how the variable-index alternates between 0 and 1, the first order translation of the $\mathcal{ALC}$-concept $\forall r.(A \longrightarrow \exists r.B)$ shall be given, where, by our naming convention, $\mathsf{N_C} = \{A, B\}$ and $\mathsf{N_R} = \{r\}$. The concept states for every interpretation $\mathfrak{I}$: If $d \in (\forall r.(A \longrightarrow \exists r.B))^{\mathfrak{I}}$ then every $r$-successor of $d$ that satisfies $A$ has an $r$-successor that satisfies $B$.

$$
\begin{aligned}
& [\forall r.(A \longrightarrow \exists r.B); x_0] \\
= \;& \neg\exists x_1.r(x_0, x_1) \land \neg[A \to \exists r.B; x_1] \\
= \;& \forall x_1.r(x_0, x_1) \to [A \to \exists r.B; x_1] \\
= \;& \forall x_1.r(x_0, x_1) \to (A(x_1) \to [\exists r.B; x_1]) \\
= \;& \forall x_1.r(x_0, x_1) \to (A(x_1) \to \exists x_0.r(x_1, x_0) \land [B; x_0]) \\
= \;& \forall x_1.r(x_0, x_1) \to (A(x_1) \to \exists x_0.r(x_1, x_0) \land B(x_0))
\end{aligned}
$$

### 1.2.5 Abstract Logic

Before presenting a characterisation of a logic, the notion of logic itself, known from abstract model theory [16][47] shall be discussed.

A *mathematical logic* $\mathcal{L}$ is considered to be a pair $(L, \models_{\mathcal{L}})$, where $L$ is a function that assigns to every signature $\tau$ a set $L(\tau)$ whose elements are called *formulae over* $\tau$. The *satisfaction relation* $\models_{\mathcal{L}}$ is a relation between interpretations and formulae. Additionally the following compatibility conditions for all signatures $\tau_0$, $\tau_1$ and $\tau$ are required:

| | | |
|---|---|---|
| monotonicity of $L$ | $\tau_0 \subseteq \tau_1$ | $\Longrightarrow L(\tau_0) \subseteq L(\tau_1)$ |
| isomorphy invariance | $\mathfrak{I} \cong_\tau \mathfrak{H}$ | $\Longrightarrow \forall \varphi \in L(\tau) : \mathfrak{I} \models \varphi \iff \mathfrak{H} \models \varphi$ |

where $\mathfrak{I} \cong_\tau \mathfrak{H}$ means, there is a bijective function $\iota : \Delta^{\mathfrak{I}} \longrightarrow \Delta^{\mathfrak{H}}$ s.t. for all $S \in \tau$ and $d_1, \dots, d_{\mathrm{ar}(S)}$ we have

$$(d_1, \dots, d_{\mathrm{ar}(S)}) \in S^{\mathfrak{I}} \iff (\iota(d_1), \dots, \iota(d_{\mathrm{ar}(S)})) \in S^{\mathfrak{H}}$$

Indeed FO as well as $\mathcal{ALC}$ with their definition of satisfaction relation are both logics in this sense. Since all DLs considered in this thesis are fragments of FO, the satisfaction relations of these DLs are subsets of the satisfaction relation of FO.

Concerning the notation, we obviously use the same symbol for the logic itself and for the set of formulae over some signature $\tau$. E.g. $\mathcal{ALC}$ and $\mathcal{ALC}(\tau)$ instead of $\mathcal{ALC}$ and ALC$(\tau)$ or similarly. As well, we do not distinguish the satisfaction relations of different logics and hence omit the index.

A logic $\mathcal{L}_0$ is a *fragment of* another logic $\mathcal{L}_1$ if for all $\tau$ we have $\mathcal{L}_0(\tau) \subseteq \mathcal{L}_1(\tau)$ and for every interpretation $\mathfrak{I}$ and each $\varphi \in \mathcal{L}_0(\tau)$ we have $\mathfrak{I} \models_{\mathcal{L}_0} \varphi \iff \mathfrak{I} \models_{\mathcal{L}_1} \varphi$.

A formula $\varphi \in \mathcal{L}_0(\tau)$ is called *logically equivalent* to some formula $\psi \in \mathcal{L}_1(\tau)$ if for all $\tau$-interpretations $\mathfrak{I}$ we have $\mathfrak{I} \models_{\mathcal{L}_0} \varphi$ iff $\mathfrak{I} \models_{\mathcal{L}_1} \models \psi$.

A logic $\mathcal{L}_0$ is *at least as expressive as* $\mathcal{L}_1$, ($\mathcal{L}_0 \leq \mathcal{L}_1$) if for every signature $\tau$ and every formula $\varphi \in \mathcal{L}_0(\tau)$ there is a logically equivalent formula $\psi \in \mathcal{L}_1(\tau)$.

Clearly, if $\mathcal{L}_0 \leq \mathcal{L}_1$ then there is a fragment $\mathcal{L}'_1$ of $\mathcal{L}_1$ such that for each signature $\tau$, $\mathcal{L}'_1(\tau)$ contains exactly the formulae of $\mathcal{L}_0(\tau)$ which are logically equivalent. Therefore, we shall henceforth in our context call a logic $\mathcal{L}_0$ a fragment of $\mathcal{L}_1$ whenever $\mathcal{L}_0 \leq \mathcal{L}_1$; in particular, we call $\mathcal{ALC}$ and all the other DLs fragments of FO, as we have effective translation procedures to translate their concepts into FO-formulae.

It is also possible to compare the expressivity of two logics w.r.t. different signatures: In [8], the expressivity of certain knowledge representation languages were compared and successfully distinguished admitting, in particular, signature extensions. Under signature extensions a logic $\mathcal{L}_1$ is considered to be more expressive than some logic $\mathcal{L}_0$ if there is a signature $\tau_1$ and a formula $\varphi \in \mathcal{L}_1(\tau_1)$ such that for all signatures $\tau_0 \supseteq \tau_1$ there is no 'equivalent' formula $\psi \in \mathcal{L}_0(\tau_0)$.

The problem here is the notion equivalence as it has to refer to sets of interpretations in different signatures: If $\tau_0 \supseteq \tau_1$ and $\varphi \in \mathcal{L}_0(\tau_0)$ equivalent to $\psi \in \mathcal{L}(\tau_1)$ then the class $\mathrm{Mod}_{\tau_0}(\psi)$ of all $\tau_0$-models of $\psi$ comprises the subclass of all models of $\psi$ in which signature symbols in $\tau_0 \setminus \tau_1$ are interpreted as empty. This subclass has a one-to-one correspondence to the class $\mathrm{Mod}_{\tau_1}(\psi)$ of all $\tau_1$-models of $\psi$. But $\mathrm{Mod}_{\tau_0}(\psi)$ also comprises all $\psi$-models with arbitrary interpretation of symbols in $\tau_0 \setminus \tau_1$ including $\mathrm{Mod}_{\tau_0}(\varphi)$ which most likely forms a proper subclass of $\mathrm{Mod}_{\tau_0}(\psi)$. The relationship between $\mathrm{Mod}_{\tau_0}(\varphi)$ and the subclass of $\mathrm{Mod}_{\tau_0}(\psi)$ corresponding to $\mathrm{Mod}_{\tau_1}(\psi)$ is highly dependent on the logics $\mathcal{L}_0$ and $\mathcal{L}_1$. The more in generality w.r.t. signatures hence has to be paid off with a more specific notion of logical equivalence.

Where characterisations of logics are treated, e.g. in [129, 85] but also [90, 112, 126, 105] and so forth, logics are exceptionally characterised as fragments of other logics w.r.t. to the same signature. Characterisations with respect to one and the same signature will be our approach in this thesis as well.

Let $\mathcal{L}$ be a logic and $\tau$ a signature. Then for every set $\Gamma \subseteq \mathcal{L}(\tau)$ and every formula $\varphi \in \mathcal{L}(\tau)$ we define $\Gamma \models \varphi$ if for all $\tau$-interpretations $\mathfrak{I}$ we have if for all $\psi \in \Gamma$ $\mathfrak{I} \models \psi$ then $\mathfrak{I} \models \varphi$. Similarly we shall write $\varphi \models \Gamma$ as abbreviation for $\varphi \models \psi$ for all $\psi \in \Gamma$.

A logic $\mathcal{L}$ is *compact* if for every signature $\tau$ and every set of $\Gamma \subseteq \mathcal{L}(\tau)$ and every formula $\varphi \in \mathrm{FO}(\tau)$ we have $\Gamma \models \varphi$ iff there is some finite subset $\Gamma_0 \subseteq \Gamma$ such that

$\Gamma_0 \models \varphi$.

It is well known that first order logic is compact [34, 47]. By the definition of compactness, it follows that every fragment of FO must be compact, too.

Let $\xleftrightarrow{\mathcal{L}}$ be a relation on (pointed) interpretations. We call an FO-formula $\varphi(x)$ (with at most one free variable) invariant under $\xleftrightarrow{\mathcal{L}}$ if

$$(\mathfrak{I}, d) \xleftrightarrow{\mathcal{L}} (\mathfrak{H}, e) \quad \Longrightarrow \quad \left( \mathfrak{I}\frac{d}{x} \models \varphi(x) \Longleftrightarrow \mathfrak{H}\frac{e}{x} \models \varphi(x) \right).$$

Let $\mathcal{L}$ be a description logic and $\tau$ some signature. For concepts $C, D \in \mathcal{L}(\tau)$ we call $C \sqsubseteq D$ a *concept inclusion* and read $C \sqsubseteq D$ as $C$ is subsumed by $D$. We say $\mathfrak{I}$ *satisfies* or *is a model of* $C \sqsubseteq D$, in symbols $\mathfrak{I} \models C \sqsubseteq D$ if $C^{\mathfrak{I}} \subseteq D^{\mathfrak{I}}$, i.e. if for all $d \in \Delta^{\mathfrak{I}}$ we have $(\mathfrak{I}, d) \models C$ implies $(\mathfrak{I}, d) \models D$. A TBox $\mathcal{T}$ over $\mathcal{L}$ is a finite set of concept inclusions. We say $\mathfrak{I}$ *satisfies* or *is a model of* $\mathcal{T}$ if $\mathfrak{I} \models C \sqsubseteq D$ for all $C \sqsubseteq D \in \mathcal{T}$.

Let $\tau_0, \tau_1$ be signatures such that $\tau_0 \subseteq \tau_1$. Let $\mathfrak{I}$ be a $\tau_1$-interpretation. Then $\mathfrak{I}{\restriction}\tau_0$ is the $\tau_0$-interpretation $\mathfrak{K}$, where $\Delta^{\mathfrak{K}} := \Delta^{\mathfrak{I}}$ and $S^{\mathfrak{K}} := S^{\mathfrak{I}}$ for all symbols $S \in \tau_0$. We call $\mathfrak{I}{\restriction}\tau_0$ the $\tau_0$-reduct of $\mathfrak{I}$.

# 2. The characterisation of $\mathcal{ALC}$

## 2.1 The Characterisation of $\mathcal{ALC}$-Concepts

In this section we shall give a characterisation of $\mathcal{ALC}$ as FO-fragment. In essence, we repeat the characterisation given by van Benthem [129]. We shall first recall the model-theoretic invariance notion called bisimulation for $\mathcal{ALC}$ concepts. Via characteristic $\mathcal{ALC}$-concepts [61] (Characteristic formulae [47] for $\mathcal{ALC}$) we shall show that we can describe in an interpretation the neighbourhood of some element, i.e. all those elements accessible from this element up to a certain distance, sufficiently precisely in order to show that every FO-formula which is invariant under bisimulation can be expressed by a disjunction of those characteristic $\mathcal{ALC}$-concepts.

Almost all of the following has been discovered before and is succinctly presented in [61], but we shall wander on the trodden paths of modal logics in order to introduce notions and motivate our search for certain properties in later chapters.

### 2.1.1 Bisimulation and Tree-Unravelling

The notion of bisimulation was introduced to modal logics by van Benthem in [129] called p-relation and later zigzag-relation [118]. Bisimulation and its variants will be central throughout all chapters of this thesis. It will be adapted for every logic that is to be characterised. It is a model-theoretic notion, in that it relates interpretations that satisfy the same $\mathcal{ALC}$-concepts. Hence all these interpretations are models of the same $\mathcal{ALC}$-theory on concept-level.

DEFINITION 2.1.1 . Let $\mathfrak{I}$ and $\mathfrak{H}$ be interpretations of the same signature $\tau$. Then the relation $Z \subseteq \Delta^{\mathfrak{I}} \times \Delta^{\mathfrak{H}}$ is called *bisimulation* between $\mathfrak{I}$ and $\mathfrak{H}$ if for all $(d, e) \in Z$ the following holds:

1. for all $A \in \mathsf{N_C}$ we have $d \in A^{\mathfrak{I}}$ iff $e \in A^{\mathfrak{H}}$. (ATOM)

2. for all $r \in \mathsf{N_R}$ we have if there is $d_0 \in \Delta^{\mathfrak{I}}$ with $(d, d_0) \in r^{\mathfrak{I}}$ then there is $e_0 \in \Delta^{\mathfrak{H}}$ with $(e, e_0) \in r^{\mathfrak{H}}$ and $(d_0, e_0) \in Z$. (FORTH)

3. for all $r \in \mathsf{N_R}$ we have if there is $e_0 \in \Delta^{\mathfrak{H}}$ with $(e, e_0) \in r^{\mathfrak{H}}$ then there is $d_0 \in \Delta^{\mathfrak{I}}$ with $(d, d_0) \in r^{\mathfrak{I}}$ and $(d_0, e_0) \in Z$. (BACK)

We write $(\mathfrak{I}, d) \Longleftrightarrow (\mathfrak{H}, e)$ and say $(\mathfrak{I}, d)$ is *bisimilar* to $(\mathfrak{H}, e)$ if there is a bisimulation $Z$ between $\mathfrak{I}$ and $\mathfrak{H}$ such that $(d, e) \in Z$. ◇

It is sometimes more convenient to show that two pointed structures are bisimilar, if the bisimulation is described by a pebble game between two players **I** and **II**, where **II** has a winning strategy. Bisimulation games are derived from Ehrenfeucht-Fraïssé games [48]. The latter correspond to potential isomorphisms which are the analogous notion [42] of bisimulation on FO level and can be found in standard textbooks like [47]. Bisimulation games are of particular interest in characterisations of model logics over finite models [115, 104]. Although we shall not endeavour into finite model theory, we shall use them because their descriptive and illustrative nature facilitates our arguments in proofs to come.

In the *bisimulation game* $G(\mathfrak{I}, d; \mathfrak{H}, e)$, the interpretations $(\mathfrak{I}, d)$ and $(\mathfrak{H}, e)$ are considered as separate graphs. For each interpretation there is one pebble which can be placed on an element of the interpretation, and whenever a pebble is on some element $d_0 \in \Delta^{\mathfrak{I}}$ and there is $d_1 \in \Delta^{\mathfrak{I}}$ with $(d_0, d_1) \in r^{\mathfrak{I}}$ then the pebble can be *moved along an r-edge* to $d_1$. The same applies for $\mathfrak{H}$.

The position of both pebbles in each interpretation is summarised as *configuration* $(\mathfrak{I}, d_0, \mathfrak{H}, e_0)$ whenever the pebble in $\mathfrak{I}$ is placed on $d_0$ and the pebble in $\mathfrak{H}$ is placed on $e_0$.

The game $G(\mathfrak{I}, d; \mathfrak{H}, e)$ is played according to the following rules: The 0-th round starts in configuration $(\mathfrak{I}, d; \mathfrak{H}, e)$, where for all $A \in \mathsf{N_C}$: $d \in A^{\mathfrak{I}}$ iff $e \in A^{\mathfrak{H}}$ must hold, or **II** has lost the game in the 0-th round. For any configuration $(\mathfrak{I}, d_0; \mathfrak{H}, e_1)$ the following rules apply:

1. In each round **I** picks one of the two interpretations, $\mathfrak{I}$ say, and moves the pebble along some edge, say an $r$-edge emerging from the pebble's position to some element $d_1$. If there is no such edge along which player **I** could move, **II** has won the game.

2. **II** has to respond to player **I**'s move in the other interpretation, here $\mathfrak{H}$, by moving the pebble from $e_0$ along some edge with the same label, in our case an $r$-edge, to some element $e_1$ such that for all $A \in \mathsf{N_C}$: $d_1 \in A^{\mathfrak{I}}$ iff $e_1 \in A^{\mathfrak{H}}$.

In case player **II** cannot move accordingly, **II** has lost the game; otherwise a new round begins.

Clearly, these rules mirror the ATOM-property as well as the FORTH- and BACK-property, respectively, depending on which interpretation **I** chooses.

Definition 2.1.2. **II** has a *winning strategy* in the game $G(\mathfrak{I}, d; \mathfrak{H}, e)$ if she can respond to every move of **I** such that she either wins the game or can play forever. ◇

Proposition 2.1.3. **II** *has a winning strategy in the game* $G(\mathfrak{I}, d; \mathfrak{H}, e)$ *iff* $(\mathfrak{I}, d) \Longleftrightarrow (\mathfrak{H}, e)$.

Proof. For the if-direction, we define the following relation, which is obviously a bisimulation: $Z := \{(d, e) \in \Delta^{\mathfrak{I}} \times \Delta^{\mathfrak{H}} \mid \textbf{II} \text{ has a winning strategy for } G(\mathfrak{I}, d; \mathfrak{H}, e)\}$. For the only-if direction, let **II** play such that each new configuration $(\mathfrak{I}, d_0; \mathfrak{H}, e_0)$ has a pair $(d_0, e_0)$ in $Z$. To see that this is possible, let the game have reached $(\mathfrak{I}, d_0; \mathfrak{H}, e_0)$ such that $(d_0, e_0) \in Z$. Let **I** move in $\mathfrak{I}$, say, from $d_0$ along some $r$-edge to $d_1$. Then, by definition of $Z$, there must be a pair $(d_1, e_1) \in Z$ such that $e_1$ is an $r$-successor of $e_0$ in $\mathfrak{H}$ and $d_1$ is atomically equivalent to $e_1$. Hence **II** moves to $e_1$. Thus, **II** either wins the game or can play forever. □

For each signature $\tau$ the bisimulation relation is an equivalence relation on the pointed $\tau$-interpretations. We therefore talk, w.r.t. to a signature $\tau$, about bisimulation types, which are the classes formed by this equivalence relation.

The bisimulation type links interesting structures together: Tree-unravellings are interpretations which, viewed as a graph, form a tree whose elements correspond to all finite path sequences. According to [18, Notes to Chapter 2] and [20] the method of tree unravelling first appeared 1959 in [46] but became popular due to [117]. In particular, every pointed interpretation is bisimilar to its tree-unravelling:

Definition 2.1.4. Let $(\mathfrak{I}, d)$ be a pointed $\tau$ interpretation. We call $\mathfrak{I}_d$ its *tree-unravelling in d* where $\Delta^{\mathfrak{I}_d}$ is recursively defined by

1. $d \in \Delta^{\mathfrak{I}_d}$
2. Let $\bar{d} := d r_1 d_1 r_2 \cdots r_n d_n \in \Delta^{\mathfrak{I}_d}$. For all $r \in \mathsf{N_R}$: if $(d_n, d_{n+1}) \in r^{\mathfrak{I}}$ then $\bar{d} \cdot r \cdot d_{n+1} \in \Delta^{\mathfrak{I}_d}$, where $\bar{d} \cdot r \cdot d_{n+1}$ is the word we obtain by concatenation.

For all $A \in \mathsf{N_C}$ and $r \in \mathsf{N_R}$ we define

$$\begin{aligned} r^{\mathfrak{I}_d} &:= \{(\bar{d}, \bar{d}rd') \mid \bar{d}, \bar{d}rd' \in \Delta^{\mathfrak{I}_d}\} \\ A^{\mathfrak{I}_d} &:= \{\bar{d} \in \Delta^{\mathfrak{I}_d} \mid \bar{d} = d\, r_1 d_1 r_2 \cdots r_n d_n \text{ and } d_n \in A^{\mathfrak{I}}\} \end{aligned}$$

◇

Hence each element $\bar{d} = d\, r_1 d_1 r_2 \cdots r_n d_n$ in $\Delta^{\mathfrak{I}_d}$ represents a path of finite length beginning in $d$ and noting all traversed nodes and the edges by which they were accessed. $\mathfrak{I}_d$ is a proper tree-interpretation: $d$ cannot be reached via any edge and every other element in $\mathfrak{I}_d$ has exactly one predecessor and exactly one edge with which it is connected to its predecessor.

By the recursive definition of a tree-unravelling, the depth of such a tree-structure is at most countably infinite.

Observation 2.1.5 . $(\mathfrak{I}, d) \Longleftrightarrow (\mathfrak{I}_d, d)$

Proof. By maintaining configurations $(\mathfrak{I}, d'; \mathfrak{I}, d\, r_1 d_1 r_2 \cdots r_n d_n r_{n+1} d')$ **II** has a winning strategy in the game $G(\mathfrak{I}, d; \mathfrak{I}_d, d)$. □

### 2.1.2 Finite Bisimulation and Locality of $\mathcal{ALC}$-Concepts

In this section the the notion of $n$-bisimulation [55] and the appropriate $n$-round bisimulation game will be introduced. We shall show that for every $\mathcal{ALC}$-concept there is an appropriate $n < \omega$ such that this concept is invariant under $n$-bisimulation and conversely that $n$-bisimulation can, if the signature is finite, be captured by characteristic $\mathcal{ALC}$-concepts, in the sense that if two interpretations satisfy the same characteristic concept, they are $n$-bisimilar:

For every natural number $n$, the *$n$-round bisimulation game* $G_n(\mathfrak{I}, d; \mathfrak{H}, n)$ is the bisimulation game $G(\mathfrak{I}, d; \mathfrak{H}, n)$ which breaks off after $n$-rounds.

Definition 2.1.6 . **II** has a winning strategy in $G_n(\mathfrak{I}, d; \mathfrak{H}, e)$ if **II** either wins the game within $n$-rounds or can play for $n$-rounds. ◇

Equivalently, one can define $n$-bisimulation between $\mathfrak{I}$ and $\mathfrak{H}$ as a system (or family) $(Z_k)_{0 \le k \le n}$ of relations $Z_k \subseteq \Delta^{\mathfrak{I}} \times \Delta^{\mathfrak{H}}$ such that all of the following requirements are met:

1. for all $k \le n$ and all $(d_0, e_0) \in Z_k$ and all $A \in \mathsf{N_C}$ we have $d_0 \in A^{\mathfrak{I}}$ iff $e_0 \in A^{\mathfrak{H}}$ (ATOM)

2. for all $k < n$, $(d_0, e_0) \in Z_{k+1}$ and $r \in \mathsf{N_R}$ we have: if there is an $r$-successor $d_1$ of $d_0$ in $\mathfrak{I}$ then there is an $r$-successor $e_1$ of $e_0$ in $\mathfrak{H}$ such that $(d_1, e_1) \in Z_k$ (FORTH)

3. for all $k < n$, $(d_0, e_0) \in Z_{k+1}$ and $r \in \mathsf{N_R}$ we have: if there is an $r$-successor $e_1$ of $e_0$ in $\mathfrak{H}$ then there is an $r$-successor $d_1$ of $d_0$ in $\mathfrak{I}$ such that $(d_1, e_1) \in Z_k$ (BACK)

Proposition 2.1.3 equally holds for $n$-bisimulations and $n$-round bisimulation games: Every pair in $Z_k$ gives **II** to a winning strategy for the $k$-round bisimulation game, hence **II** can play $n$ rounds and conversely, if **II** has a winning strategy for $G_n(\mathfrak{I}, d; \mathfrak{H}, e)$ then for each $k \leq n$ we obtain a relation

$$Z_k := \{(d_0, e_0) \mid \textbf{II} \text{ has a winning strategy for } G_k(\mathfrak{I}, d_0; \mathfrak{H}, e_0)\}$$

such that $(Z_k)_{k \leq n}$ forms an $n$-bisimulation between $\mathfrak{I}$ and $\mathfrak{H}$ with $(d, e) \in Z_n$.

Analogously to bisimulation, we write $(\mathfrak{I}, d) \Longleftrightarrow_n (\mathfrak{H}, e)$ if there is an $n$-bisimulation $(Z_k)_{k \leq n}$ between $\mathfrak{I}$ and $\mathfrak{H}$ with $(d, e) \in Z_n$.

Observation 2.1.7 . *For all $\tau$ and all $n < \omega$ we have*

1. $\Longleftrightarrow_n$ *is an equivalence relation on the pointed interpretations*

2. *If $(\mathfrak{I}, d) \Longleftrightarrow_{n+1} (\mathfrak{H}, e)$ then $(\mathfrak{I}, d) \Longleftrightarrow_n (\mathfrak{H}, e)$*

3. *If $(\mathfrak{I}, d) \Longleftrightarrow (\mathfrak{H}, e)$ then $(\mathfrak{I}, d) \Longleftrightarrow_n (\mathfrak{H}, e)$*

Interestingly, the concepts of $\mathcal{ALC}$ can be stratified according to the nesting-depth of their quantifications, which we shall call rank. The fragments of $\mathcal{ALC}$-concepts, which are induced by this stratification, are preserved by the appropriate notion of $n$-bisimulation, where $n$ is the rank of the concepts.

For all $\tau$ the *rank* of an $C \in \mathcal{ALC}(\tau)$ is defined as follows:

$$\operatorname{rank} C := \begin{cases} 0 & C \in \mathsf{N_C} \text{ or } C = \bot \\ \max\{\operatorname{rank} D, \operatorname{rank} E\} & C = D \sqcap E \\ \operatorname{rank} D & C = \neg D \\ 1 + \operatorname{rank} D & C = \exists r.D, \text{ for all } r \in \mathsf{N_R} \end{cases}$$

Proposition 2.1.8 . *Let $C$ be an $\mathcal{ALC}$-concept of rank $\leq n$ as well as $(\mathfrak{I}, d)$ and $(\mathfrak{H}, e)$ pointed interpretations. If $(\mathfrak{I}, d) \Longleftrightarrow_n (\mathfrak{H}, e)$ then $d \in C^{\mathfrak{I}} \iff e \in C^{\mathfrak{H}}$.*

Proof. The proof is carried out via induction upon $n$. For $n = 0$, $C$ is w.l.o.g. a concept name. As **II** has a winning strategy for $G_0(\mathfrak{I}, d; \mathfrak{H}, e)$, she does not lose the 0-th round. Hence $d$ and $e$ are atomically equivalent and therefore $d \in C^{\mathfrak{I}}$ iff $e \in C^{\mathfrak{H}}$.

For $n + 1$ the cases $C = D \sqcap E$ and $C = \neg D$ can be reduced to showing the equivalence for $D$ and $E$ respectively. We can therefore w.o.l.g. assume that $C = \exists r.D$ for some $r \in \mathsf{N_R}$.

Let $d \in (\exists r.D)^{\mathfrak{I}}$, thus there is an $r$-successor $d_0$ of $d$ in $\mathfrak{I}$ such that $d_0 \in D^{\mathfrak{I}}$. As **II** has a winning strategy for $G_{n+1}(\mathfrak{I}, d; \mathfrak{H}, e)$, there is an $r$-successor $e_0$ of $e$ in $\mathfrak{H}$ such that **II** has a winning strategy for $G_n(\mathfrak{I}, d_0; \mathfrak{H}, e_0)$. The induction hypothesis yields now $e_0 \in D^{\mathfrak{H}}$ and hence $e \in (\exists r.D)^{\mathfrak{H}}$. The only-if direction follows the same rationale. □

Corollary 2.1.9. *If $(\mathfrak{I}, d) \Longleftrightarrow (\mathfrak{H}, e)$ then for every $\mathcal{ALC}$-concept $C$ we have $d \in C^{\mathfrak{I}}$ iff $e \in C^{\mathfrak{H}}$.*

Although the corollary is obvious, it reveals the interesting fact that no property is expressible by $\mathcal{ALC}$ which could distinguish the tree-unravelling from its original structure. Furthermore each concept $C$ is local:

Let $\mathfrak{I}^{\ell}_d$ be defined as the restriction of the tree-unravelling $\mathfrak{I}_d$ to path-elements up to length $\ell$:

$$\begin{aligned} \Delta^{\mathfrak{I}^{\ell}_d} &:= \{ d r_0 d_1 r_1 \cdots r_k d_k \mid k < \ell \} \\ A^{\mathfrak{I}^{\ell}_d} &:= A^{\mathfrak{I}_d} \cap \Delta^{\mathfrak{I}^{\ell}_d} \\ r^{\mathfrak{I}^{\ell}_d} &:= r^{\mathfrak{I}_d} \cap \left( \Delta^{\mathfrak{I}^{\ell}_d} \times \Delta^{\mathfrak{I}^{\ell}_d} \right) \end{aligned}$$

Lemma 2.1.10. *For all $\ell < \omega$ we have $(\mathfrak{I}, d) \Longleftrightarrow_{\ell} (\mathfrak{I}^{\ell}_d, d)$.*

Proof. Consider the game $G_{\ell}(\mathfrak{I}_d, d; \mathfrak{I}^{\ell}_d, d)$. As the game starts uniformly on $d$ in both interpretations, the path-elements in each configuration get successively longer with each round, but do not exceed length $\ell$. By simply copying every move of **I**, **II** has a winning strategy in $G_{\ell}(\mathfrak{I}_d, d; \mathfrak{I}^{\ell}_d, d)$.

As $(\mathfrak{I}, d)$ is in particular $\ell$-bisimilar to $(\mathfrak{I}_d, d)$ and $\Longleftrightarrow_{\ell}$ is transitive we have $(\mathfrak{I}, d) \Longleftrightarrow_{\ell} (\mathfrak{I}^{\ell}_d, d)$. □

Definition 2.1.11. An $\mathcal{ALC}$-concept $C$ is *$\ell$-local* if for every interpretation $\mathfrak{I}$ and every $d \in \Delta^{\mathfrak{I}}$ we have $d \in C^{\mathfrak{I}}$ iff $d \in C^{\mathfrak{I}^{\ell}_d}$. ◇

We remark that the notion of locality stems from classical model theory where it appears in the context of Gaifman-Graphs [54] for which the notion of being $\ell$-local is defined in much greater generality. It is not unusual though to restrict locality in the context of standard modal logics and therefore also in our context to forward reachable elements [103].

Corollary 2.1.12. *Let $C$ be an $\mathcal{ALC}$-concept of rank $\ell$. Then $C$ is $\ell$-local.*

This result shows that $\mathcal{ALC}$-concepts can only express properties whose truth can be checked within finite depth from distinguished point. $\mathcal{ALC}$-concepts are also oblivious to copies of successors, i.e. the number of successors, which are bisimilar to each other, nor can $\mathcal{ALC}$ distinguish elements that have predecessors from those that do not.

$\mathcal{ALC}$-concepts are not affected by disjoint unions with other structures either: Let $\mathfrak{I}$ and $\mathfrak{H}$ be two structures such that $\Delta^{\mathfrak{I}}$ and $\Delta^{\mathfrak{H}}$ are disjoint. Then $\mathfrak{I} \uplus \mathfrak{H}$ is called the *disjoint union* of $\mathfrak{I}$ and $\mathfrak{H}$, has the carrier set $\Delta^{\mathfrak{I}\uplus\mathfrak{H}} := \Delta^{\mathfrak{I}} \uplus \Delta^{\mathfrak{H}}$ and $A^{\mathfrak{I}\uplus\mathfrak{H}} := A^{\mathfrak{I}} \uplus A^{\mathfrak{H}}$ for all $A \in \mathsf{N_C}$ and $r^{\mathfrak{I}\uplus\mathfrak{H}} := r^{\mathfrak{I}} \uplus r^{\mathfrak{H}}$ for all $r \in \mathsf{N_R}$.

Since the tree-structure retains only those elements accessible on the connected component of its root, we have $\mathfrak{I}_d = (\mathfrak{I} \uplus \mathfrak{H})_d$, so locality of $\mathcal{ALC}$-concepts entails for all $d \in \mathfrak{I}$ and every interpretation $\mathfrak{H}$ that $(\mathfrak{I}, d) \models C \iff (\mathfrak{I} \uplus \mathfrak{H}, d) \models C$.

The notion of being $\ell$-local for $\mathcal{ALC}$-concepts or modal logic formulae however plays a central role in the characterisation of standard modal logics: de Rijke [113] used locality as characteristic property to define *abstract modal logics* (abstract in the sense of generic) which he named *finite rank*, alluding to the statement Corollary 2.1.12. Clearly, locality is also the reason why we can decide for an $\mathcal{ALC}$-concept, whether or not it is satisfiable. For an $\ell$-local concept $C$, we only need to build every tree of depth $\ell$ involving only the (finitely many) signature symbols of $C$. The result goes back to [124] where the decidability of the 2-variable fragment of FO without equality is proven. Later, the result is extended in [97] allowing equality, though using completely different techniques. A complexity theoretical investigation is conducted in [86, 68] showing PSPACE-completeness for the satisfiability problem of standard modal logic formulae. Fourteen years after [86] an $\mathcal{ALC}$-specific version showing PSPACE-completeness for concept satisfiability is presented along with the logic $\mathcal{ALC}$ itself in [122]. How a naïve search for a model could be accomplished will be clearer when looking at characteristic concepts.

### 2.1.3 Characteristic Concepts

We shall now introduce characteristic concepts. If the signature $\tau$ is finite we can assemble a concept that captures the $n$-bisimulation type of an interpretation w.r.t. $\tau$. This means that every interpretation which satisfies this concept is $n$-bisimilar to our original interpretation and vice versa.

Characteristic formulae in FO are also known as Hintikka-formulae [71] where they are intimately connected to finite Ehrenfeucht-Fraïssé games. Characteristic formulae [61] or here, characteristic concepts, are the straight forward adaption of Hintikka-formulae to finite bisimulation.

Before defining these concepts we shall introduce a short-hand notation: For an $n$-element set $\Gamma$ of concepts whose elements are enumerated $C_0, ...C_{n-1}$, we set $\bigsqcap \Gamma := C_0 \sqcap C_1 \sqcap \ldots \sqcap C_{n-1}$. Since there are only finitely many enumerations for the elements in $\Gamma$, there are only finitely many different outcomes for $\bigsqcap \Gamma$. Analogously we define $\bigsqcup \Gamma$.

Definition 2.1.13 . Let $\tau$ be a finite signature and $(\mathfrak{I}, d)$ an interpretation. Then the characteristic concept $X^n_{\mathfrak{I},d}$ for $(\mathfrak{I}, d)$ on level $n$ w.r.t. $\tau$ is recursively defined as follows:

$$X^0_{\mathfrak{I},d} := \bigsqcap\{A \in \mathsf{N_C} \mid d \in A^{\mathfrak{I}}\} \sqcap \neg \bigsqcup\{A \in \mathsf{N_C} \mid d \notin A^{\mathfrak{I}}\}$$

$$X^{n+1}_{\mathfrak{I},d} := X^0_{\mathfrak{I},d} \sqcap \bigsqcap_{r \in \mathsf{N_R}} \left[\bigsqcap\{\exists r.X^n_{\mathfrak{I},d_0} \mid (d, d_0) \in r^{\mathfrak{I}}\} \sqcap \forall r.\bigsqcup\{X^n_{\mathfrak{I},d_0} \mid (d, d_0) \in r^{\mathfrak{I}}\}\right]$$

◇

We shall give an intuition for the characteristic concepts after Observation 2.1.15. In order to claim that $X^n_{\mathfrak{I},d}$ is an $\mathcal{ALC}$-concept we have to show that for each level $k < n$ there are at most finitely many characteristic concept, as we do not allow infinite disjunctions or conjunctions.

Lemma 2.1.14 . *Let $\tau$ be a finite signature. There are at most finitely many different characteristic concepts on level n*

Proof. The proof is carried out by induction upon $n$. For $n = 0$ the characteristic concept is already determined by the non-negated concept names occurring in it, which could be any subset of $\mathsf{N_C}$. Since $\mathsf{N_C}$ is finite, there are only finitely many concepts on level 0. On level $n + 1$, the bracketed conjunction is already determined by those concepts with existential operator, for each $r \in \mathsf{N_R}$. The induction hypothesis yields a finite cardinality for the number of characteristic concepts on level $n$. The concepts with existential operator form a subset of the characteristic concepts on level $n$ and hence there are only finitely many different of these bracketed conjunctions for each $r \in \mathsf{N_R}$. But also $\mathsf{N_R}$ was finite and hence, combined with the finitely many characteristic concepts on level 0, there are solely finitely many characteristic concepts on level $n+1$. □

Observation 2.1.15. *Let $\tau$ be a finite signature and let $(\mathfrak{I}, d)$ be a pointed $\tau$-interpretation. For every natural number n we have*

1. $d \in (X^n_{\mathfrak{I},d})^{\mathfrak{I}}$
2. $\operatorname{rank} X^n_{\mathfrak{I},d} = n$

The observation is immediate. Before presenting the next proposition we shall give an intuitive explanation what a characteristic $\mathcal{ALC}$-concept encodes. Assume therefore we have two pointed $\tau$-interpretations $(\mathfrak{I}, d)$ and $(\mathfrak{H}, e)$. For level 0, $X^n_{\mathfrak{I},d}$ states exactly which concept names are satisfied by $d$ and which are not satisfied. If $e$ satisfies $X^{n+1}_{\mathfrak{I},d}$, it must be atomically equivalent to $d$ and it must provide for each $r$-successor of $d$ an 'appropriate' $r$-successor on level $n$ and all of $e$'s $r$-successors must be 'appropriate' for at least one $r$-successor of $d$. If appropriate means $n$-bisimilar, and this is what will be shown now, then the left-hand side of the bracketed conjunction ensures the FORTH-condition and the right-hand side the BACK-condition for an $n+1$-bisimulation between $d$ and $e$.

Proposition 2.1.16. *Let $\tau$ be a finite signature and $X^n_{\mathfrak{I},d}$ the characteristic formula w.r.t. to this signature. Then for any interpretation $\mathfrak{H}$ we have*

$$e \in (X^n_{\mathfrak{I},d})^{\mathfrak{H}} \iff (\mathfrak{I}, d) \Longleftrightarrow_n (\mathfrak{H}, e).$$

Proof. '$\Longrightarrow$': The proof is carried out via induction upon $n$. We show the induction step, whose proof rests on the intuition given for characteristic concepts. As $e \in (X^{n+1}_{\mathfrak{I},d})^{\mathfrak{I}}$ we have $e \in X^0_{\mathfrak{I},d}$ and so **II** does not lose the 0-th round.

For any challenge of **I** with some $r$-successor $d'$ of $d$ in $\mathfrak{I}$, there is $\exists r.X^n_{\mathfrak{I},d'}$ amongst the left-hand side of the bracketed conjunction. Hence $e \in (\exists r.X^n_{\mathfrak{I},d'})^{\mathfrak{H}}$ and so there is an $r$-successor $e'$ of $e$ with $e' \in (X^n_{\mathfrak{I},d'})^{\mathfrak{H}}$ for which, by induction hypothesis, **II** has a winning strategy in $G_n(\mathfrak{I}, d'; \mathfrak{H}, e')$.

In case **I** challenges **II** with some $r$-successor $e'$ of $e$ in $\mathfrak{H}$, we know that $e$ satisfies the right-hand side of the bracketed conjunction, hence $e \in (\forall r. \bigsqcup\{X^n_{\mathfrak{I},d'} \mid$

$(d, d') \in r^{\mathfrak{I}}\})^{\mathfrak{H}}$, i.e. $e' \in (X^n_{\mathfrak{I},d'})^{\mathfrak{H}}$ for some $r$-successor $d'$ of $d$. The induction hypothesis yields again that **II** has a winning strategy for $G_n(\mathfrak{I}, d'; \mathfrak{H}, e')$.

Thus **II** can ward off any challenge from **I** in $n+1$ rounds, and hence **II** has a winning strategy for $G_{n+1}(\mathfrak{I}, d; \mathfrak{H}, e)$.

'$\Longleftarrow$': For the only-if direction we use from Observation 2.1.15 that $\operatorname{rank} X^n_{\mathfrak{I},d} = n$ as well as $d \in (X^n_{\mathfrak{I},d})^{\mathfrak{I}}$ and from Proposition 2.1.8 that, since $(\mathfrak{I}, d) \Longleftrightarrow_n (\mathfrak{H}, e)$ by assumption, we have $d \in (X^n_{\mathfrak{I},d})^{\mathfrak{I}}$ iff $e \in (X^n_{\mathfrak{I},d})^{\mathfrak{H}}$ □

Let $\mathrm{Th}_n(\mathfrak{I}, d) := \{C \in \mathcal{ALC}(\tau) \mid d \in C^{\mathfrak{I}} \text{ and } \operatorname{rank} C \leq n\}$. We shall summarise the results from Propositions 2.1.16 and 2.1.8 as well as Observation 2.1.15 in the following

Theorem 2.1.17 . *For finite $\tau$ and $\tau$-interpretations $(\mathfrak{I}, d)$, $(\mathfrak{H}, e)$ the following statements are equivalent for all $n < \omega$*

1. $e \in (X^n_{\mathfrak{I},d})^{\mathfrak{H}}$
2. $(\mathfrak{I}, d) \Longleftrightarrow_n (\mathfrak{H}, e)$
3. $\mathit{Th}_n(\mathfrak{I}, d) = \mathit{Th}_n(\mathfrak{H}, e)$

The question arises if—although we cannot construct a single formula for infinite signatures—$\mathrm{Th}_n(\mathfrak{I}, d) = \mathrm{Th}_n(\mathfrak{H}, e)$ iff $(\mathfrak{I}, d) \Longleftrightarrow_n (\mathfrak{H}, e)$ holds. Clearly the only-if direction is true, but for the if-direction a counter-example can be constructed:

Example 2.1.18 . Let $\mathsf{N_C} = \mathbb{N}$ and $\mathsf{N_R} = \{r\}$. Let further $\mathfrak{I}$ be the interpretation, where $d \in \Delta^{\mathfrak{I}}$ has for each finite subset $M \subseteq \mathbb{N}$ an $r$-successor $d_M$ with $(\mathfrak{I}, d) \models n$ iff $n \in M$. Let now $(\mathfrak{H}, e)$ be a copy of $(\mathfrak{I}, d)$, where we delete an arbitrary $r$-successor $e_M$ from $\mathfrak{H}$.

It is clear that **II** cannot find an appropriate answer when **I** challenges her by moving in $\mathfrak{I}$ from $d$ to $d_M$: No response in $\mathfrak{H}$ would be atomically equivalent to $d_M$ and hence *not* $(\mathfrak{I}, d) \Longleftrightarrow (\mathfrak{H}, e)$. We shall show that nevertheless for every $C \in \mathcal{ALC}(\tau)$ we have

$$d \in C^{\mathfrak{I}} \iff e \in C^{\mathfrak{I}}$$

in the root element $d$ and $e$.

The proof is carried out by induction: Atomic equivalence is clear by construction. We can reduce the step-case to $C = \exists r.D$ where $\operatorname{rank} D = 0$. By assuming $D$ is in disjunctive normal form, we can further reduce this case to $D$ being a conjunction of negated and non-negated concept names.

Let $M^-$ be the set of negated concept names and $M^+$ be the set of non-negated concept names in $D$. Clearly $(\mathfrak{I}, d_{M^+}) \models D$ and hence $(\mathfrak{I}, d) \models C$; we only need to consider the if-direction. If $e_{M^+}$ is still present in $\mathfrak{H}$ we are done. Otherwise take $n := \max M^+ \cup M^-$ and set $M := M^+ \cup \{n+1\}$. Then $M \subseteq \mathbb{N}$ is still finite and so $e_M \in \Delta^{\mathfrak{H}}$. We obtain $(\mathfrak{H}, e_M) \models D$, which shows that $(\mathfrak{H}, e) \models C$.

### 2.1.4 Saturation and the Theorem of Hennessy and Milner

What saturation means differs and is dependent on the notion of type. A type of an element is nothing else but its theory, in our case the set of all $\mathcal{ALC}$-concepts that are satisfied by this element. It turns out that there are interpretations which do not realise all the types which are possible and we shall shortly see, what this means.

Such an interpretation can be extended to an interpretation which realises all possible types and this interpretation is called saturated (with types). We shall give an adapted notion of type and saturation for each logic that we treat.

After the setback exhibited in Example 2.1.18 that even 1-bisimilarity cannot be captured when $\tau$ is infinite, we shall discover that saturated interpretations are in this respect a remarkable class of interpretations. The Hennessy-Milner-Theorem will summarise the special property of saturated interpretations.

As mentioned earlier, the equivalence relation $\Longleftrightarrow_{n+1}$ is a refinement of $\Longleftrightarrow_n$ and $\Longleftrightarrow$ is a refinement of $\Longleftrightarrow_n$ for all natural numbers $n$. The question arises whether the limit for the chain of refinements $(\Longleftrightarrow_n)_{n<\omega}$ finally coincides with $\Longleftrightarrow$.

Definition 2.1.19. $(\mathfrak{I}, d) \Longleftrightarrow_\omega (\mathfrak{H}, e)$ if for all $n < \omega$ we have $(\mathfrak{I}, d) \Longleftrightarrow_n (\mathfrak{H}, e)$. ◊

Analogously an $\omega$-bisimulation $Z_\omega$ is a system of bisimulations $(Z_n)_{n<\omega}$ between $\mathfrak{I}$ and $\mathfrak{H}$. **II** has a winning strategy in $G_\omega(\mathfrak{I}, d; \mathfrak{H}, e)$ if she has a winning strategy in $G_n(\mathfrak{I}, d; \mathfrak{H}, e)$ for all $n < \omega$, or equivalently $(d, e) \in Z_n$ for all $n < \omega$.

As we have counter-examples for $\mathrm{Th}_n(\mathfrak{I}, d) = \mathrm{Th}_n(\mathfrak{H}, e)$ implies $(\mathfrak{I}, d) \Longleftrightarrow_n (\mathfrak{H}, e)$ on each level $n$, we cannot hope to have $\mathrm{Th}(\mathfrak{I}, d) = \mathrm{Th}(\mathfrak{H}, e)$ iff $(\mathfrak{I}, d) \Longleftrightarrow_\omega (\mathfrak{H}, e)$ in general.

Even for finite signatures $(\mathfrak{I}, d) \Longleftrightarrow_\omega (\mathfrak{H}, e)$ iff $(\mathfrak{I}, d) \Longleftrightarrow (\mathfrak{H}, e)$ fails as the following counter-example shows:

Example 2.1.20. Let $\Delta^{\mathfrak{I}} := \{(k, n) \in \omega \times \omega \mid 1 \leq k \leq n\} \cup \{(0, 0)\}$, $\mathsf{N_C} := \emptyset$ and

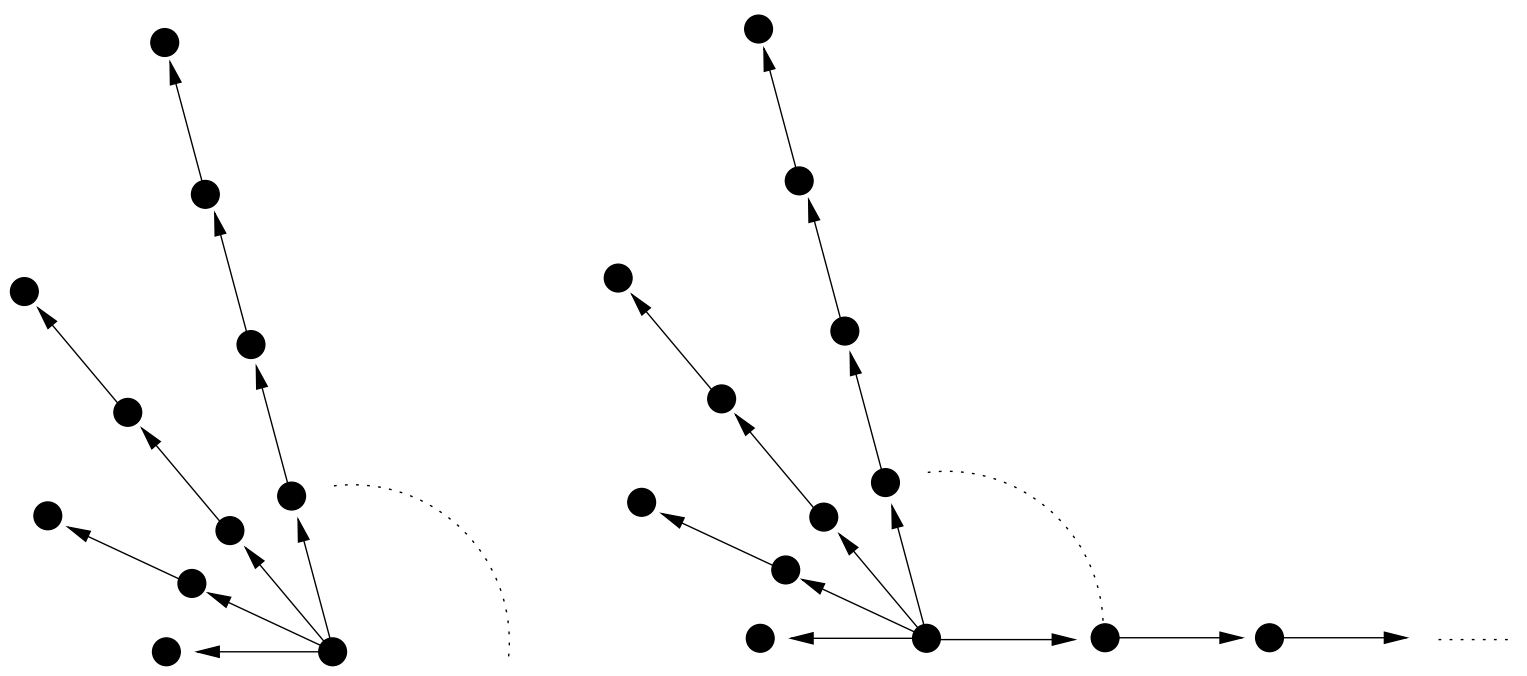

Figure 2.1: On the left the interpretation $\mathfrak{I}$ with point of origin $(0, 0)$ and on the right the interpretation $\mathfrak{H}$ with point of origin $(0, 0)$ and its infinite path.

$\mathsf{N_R} := \{r\}$ where $r^{\mathfrak{I}} \subseteq \Delta^{\mathfrak{I}} \times \Delta^{\mathfrak{I}}$ is defined as

$$r^{\mathfrak{I}} := \{((k, n), (k + 1, n)) \mid k < n \text{ and } 1 \leq n < \omega\} \cup \{((0, 0), (1, n)) \mid n < \omega\}.$$

Hence $(0, 0)$ is the root of a tree so that for every $n < \omega$ a non-branching, non-cyclic chain $(1, n), \ldots, (n, n)$ of $r$-successors emerges from $(0, 0)$, which has length $n$. Therefore $\mathfrak{I}$ looks like bristles sprouting from the end of a broom, where each bristle has finite length. Let now $(\mathfrak{H}, (0, 0))$ be a copy of $(\mathfrak{I}, (0, 0))$ where $(0, 0)$ has additionally an infinite $r$-successor path. More concrete: $\Delta^{\mathfrak{H}} := \Delta^{\mathfrak{I}} \cup \{(0, n) \mid n < \omega\}$ and $r^{\mathfrak{H}} := r^{\mathfrak{I}} \cup \{((0, n), (0, n + 1)) \mid n < \omega\}$.

We show that $(\mathfrak{I}, (0, 0)) \Longleftrightarrow_{\omega} (\mathfrak{H}, (0, 0))$. Let $n < \omega$ be arbitrary. We have to show that **II** has a winning strategy in $G_n(\mathfrak{I}, (0, 0); \mathfrak{H}, (0, 0))$: In which ever branch **I** might move in his first challenge, **II** will move into the branch with length $\min\{\mathit{length}, n\}$ where *length* is the length of the branch **I** has moved to. **II** thus makes sure that both parties can move at least $\min\{\mathit{length}, n\}$ further rounds. Either both will reach the end in their chains after $\min\{\mathit{length}, n\} - 1$ rounds and **II** wins, or **II** can play $n$ rounds and therefore wins.

We show that **I** has a winning strategy in the game $G(\mathfrak{I}, (0, 0); \mathfrak{H}, (0, 0))$, i.e. **II** loses the game: Let **I** move first in $\mathfrak{H}$ from $(0, 0)$ into the infinite chain. According to the rules **II** must move in $\mathfrak{I}$ from $(0, 0)$ inevitably into some finite chain of depth $n$, say. Let **I** move in the infinite path of $\mathfrak{H}$ for further $n$-times. **II** cannot ward off the last challenge as she is at the end of the chain in $\mathfrak{I}$ and loses.

But there is a class of structures for which not only $\Longleftrightarrow_{\omega}$ immediately upgrades to $\Longleftrightarrow$ but where even equality in $\mathcal{ALC}$-theories is enough to entail bisimilarity, and despite of Example 2.1.18 even for infinite signatures $\tau$! We therefore lift the restriction to finite signatures and allow signatures of arbitrary cardinality.

In order to define this class of structures we have to introduce the notion of $\mathcal{ALC}$-types and saturated interpretations.

Definition 2.1.21 . Let $\tau$ be a signature and $(\mathfrak{I}, d)$ a pointed $\tau$-interpretation. For each $r \in \mathsf{N_R}$ we call

1. $\Gamma$ an *r-type* of $d$, if for every finite subset $\Gamma_0 \subseteq \Gamma$ we have $d \in (\exists r. \bigsqcap \Gamma_0)^{\mathfrak{I}}$.
2. an $r$-type $\Gamma$ of $d$ realised at $d$, if there is an $r$-successor $d_0$ of $d$ in $\mathfrak{I}$ such that $d_0 \in \bigcap_{C \in \Gamma} C^{\mathfrak{I}}$.

◇

As we predominantly need this notion in its adaption for $\mathcal{ALC}$, we shall refer to $\mathcal{ALC}$-$r$-types simply as $r$-types or even only types if $r$ is clear. Similarly we refer to the notion of $\mathcal{ALC}$-saturation, which will be introduced next, simply as saturation.

Definition 2.1.22 . Let $\tau$ be a signature and $\mathfrak{I}$ a $\tau$-interpretation. Then $\mathfrak{I}$ is *saturated* if for every $d \in \Delta^{\mathfrak{I}}$ and every $r \in \mathsf{N_R}$ every $r$-type of $d$ is realised at $d$.

◇

Proposition 2.1.23 (Hennessy-Milner). *For saturated pointed interpretations* $(\mathfrak{I}, d)$, $(\mathfrak{H}, e)$ *we have*

$$\mathit{Th}(\mathfrak{I}, d) = \mathit{Th}(\mathfrak{H}, e) \iff (\mathfrak{I}, d) \Longleftrightarrow (\mathfrak{H}, e)$$

Proof. The only-if direction is immediate. Assume $\mathrm{Th}(\mathfrak{I}, d) = \mathrm{Th}(\mathfrak{H}, e)$. Then $d$ and $e$ are atomically equivalent and so **II** has not lost the 0-th round. We show that **II** has a winning strategy by maintaining configurations $(\mathfrak{I}, d'; \mathfrak{H}, e')$ such that $\mathrm{Th}(\mathfrak{I}, d') = \mathrm{Th}(\mathfrak{H}, e')$:

Assume the game has reached the configuration $(\mathfrak{I}, d'; \mathfrak{H}, e')$ with $\mathrm{Th}(\mathfrak{I}, d') = \mathrm{Th}(\mathfrak{H}, e')$ and **I** moves from $d'$ along an $r$-edge to $d''$. Then for every finite subset $\Gamma_0 \subseteq \mathrm{Th}(\mathfrak{I}, d'')$ we have $d' \in \exists r. \bigsqcap \Gamma_0$, so $\exists r. \bigsqcap \Gamma_0 \in \mathrm{Th}(\mathfrak{I}, d')$. As $\mathrm{Th}(\mathfrak{I}, d') = \mathrm{Th}(\mathfrak{H}, e')$ we have $\exists r. \bigsqcap \Gamma_0 \in \mathrm{Th}(\mathfrak{H}, e')$ for every such $\Gamma_0 \subseteq \mathrm{Th}(\mathfrak{I}, d'')$.

Hence $\mathrm{Th}(\mathfrak{I}, d'')$ is an $r$-type of $e'$ and since $\mathfrak{H}$ is saturated, $\mathrm{Th}(\mathfrak{I}, d'')$ is realised at $e'$, meaning, there is some $r$-successor $e''$ of $e'$ with $e \in \bigcap_{C \in \mathrm{Th}(\mathfrak{I}, d'')} C^{\mathfrak{H}}$. Since $\mathcal{ALC}$ is closed under negation we obtain $\mathrm{Th}(\mathfrak{I}, d'') = \mathrm{Th}(\mathfrak{H}, e'')$.

By moving to $e''$, **II** can again obtain a configuration that satisfies the requirement. The same rationale can be applied for any other $r \in \mathsf{N_R}$ and in case **I** would have challenged **II** in $\mathfrak{H}$. Hence **II** has a winning strategy in $G(\mathfrak{I}, d; \mathfrak{H}, e)$. $\square$

In Example 2.1.18, we saw that under infinite signatures $\mathcal{ALC}$-concepts cannot even capture 1-bisimilarity let alone bisimilarity. Example 2.1.20 showed that even $\omega$-bisimilarity does not necessarily coincide with full bisimilarity. Thus, the surprising thing about the Hennessy-Milner-Theorem is not only that $\Longleftrightarrow_\omega$ would upgrade to $\Longleftrightarrow$ but that even for infinite signatures merely having the same $\mathcal{ALC}$-theory is enough for saturated interpretations to be bisimilar!

Note that the Hennessy-Milner-Theorem in its original form in [69] was restricted to the class of finitely branching interpretations. Its generalisation to saturated interpretations is due to [73].

### A Remark on Saturation and Types

Saturation and in particular $\omega$-saturation are long standing notions in classical model theory [109, 34], which are closely connected to the notion of type. Poizat describes in [109] the type of an element as being nothing else but its theory, i.e. the set of all FO-formulae with at most one free variable which are satisfied by this element.

An $n$-type $t_A(d)$ of an element $d$ in a $\tau$-interpretation $\mathfrak{I}$ is, for a fixed $n$-element set $A \subseteq \Delta^{\mathfrak{I}}$ (parameter set), the type of $d$ where the signature $\tau$ is extended by a set of fresh constant symbols $\{ c_a \mid a \in A\}$ and each $c_a$ is considered to be interpreted as $c_a^{\mathfrak{I}} = a$. If $A$ is clear, we may refer to $t_A(d)$ just as type, omitting the $n$.

So $t_A(d)$ describes $d$ relative to the elements in $A$. Note that $t_A(d)$ contains $t_\varnothing(d)$ i.e. the theory of $d$ without the signature extensions, as well as for each $a \in A$ all sentences obtained by replacing the free variable in formulae from $t_\varnothing(a)$ with the constant symbol $c_a$.

A non-realised type of $\mathfrak{I}$ over the parameter set $A$, is a set of FO-formulae over $\tau \cup \{ c_a \mid a \in A\}$ with at most one free variable such that each of its finite subsets is satisfied by some element of the interpretation but no element of $\mathfrak{I}$ satisfies all formulae at once.

As we have seen in Example 2.1.20, there are interpretations which have non-realised types. This is where saturation comes into play: An interpretation is *$\omega$-saturated* if it realises all types over finite parameter sets or in other words if this interpretation has no non-realised types over finite parameter sets. From classical model theory it is known that every interpretation $\mathfrak{I}$ has an $\omega$-saturated interpretation, which is an elementary extension of $\mathfrak{I}$ (see [109, 34]). In particular the interpretations and its elementary extension are elementary equivalent, i.e. they satisfy the same FO-sentences.

Potential isomorphisms [47] play in the context of FO the same role as bisimulation in the context of $\mathcal{ALC}$-concepts: If a potential isomorphism exists between two interpretations they are elementary equivalent. The converse is in general not true, just as little as satisfying the same $\mathcal{ALC}$-concepts implies bisimilarity (cf. Examples 2.1.18 and 2.1.20).

In this respect, $\omega$-saturated interpretations are of interest as they form a special class of interpretations in which elementary equivalence between two interpretations implies that there is a potential isomorphism between these two interpretations.

We thus have explored analogue notions of type and saturation in the context of $\mathcal{ALC}$ on concept level for which the Theorem of Hennessy and Milner yields a result analogue to FO namely that $\mathcal{ALC}$-saturated interpretations $(\mathfrak{I}, d)$ and $(\mathfrak{H}, e)$ which satisfy the same $\mathcal{ALC}$-concepts are bisimilar.

So what is the relationship between an FO-type and its adaption to the $\mathcal{ALC}$-setting? In principle, an $\mathcal{ALC}$-type $\Gamma$ of an element $d$ is the set of all $\mathcal{ALC}$-concepts which are satisfied by this element, i.e. its $\mathcal{ALC}$-theory. Since we can always switch to the tree-unravelling of an interpretation, we describe an element by its $\mathcal{ALC}$-theory and its position in this tree-unravelling, i.e. its predecessor and the label on the edge with which it is connected to this predecessor. Let this unique predecessor be $e \in \Delta^{\mathfrak{I}}$ such that $(e, d) \in r^{\mathfrak{I}}$. The predecessor $e$ takes the role of a single parameter relative to which $d$ is described; hence $\Gamma$ is actually a sort of 1-type. Yet $\mathcal{ALC}$ does not incorporate constant symbols and so we cannot incorporate the predecessor in $\Gamma$ itself, as the FO-formula $r(c_e, x)$ would have done. We therefore have to explicitly refer to $\Gamma$ as *$\exists r$-type of $e$*.

As we have seen, this notion of $\mathcal{ALC}$-type is 'sufficient' in order to obtain the Theorem of Hennessy and Milner. It turns out that every $\omega$-saturated interpretation is already $\mathcal{ALC}$-saturated. So, instead of explaining for each DL $\mathcal{L}$ we are about to investigate, how we could obtain an $\mathcal{L}$-saturated extension of our interpretation (cf. [105]), we shall simply use the result from classical model theory which states that every interpretation has an $\omega$-saturated extension.

### 2.1.5 $\mathcal{ALC}$-Concepts as Bisimulation Invariant Fragment of FO

Equipped with this knowledge, we shall now characterise $\mathcal{ALC}$-concepts as fragment of FO, restating and reproving a variant of van Benthem's characterisation theorem [129] for standard modal logic. In what follows, we shall first state the theorem and then prove several lemmas that we need before proving the theorem itself. In particular we shall show that every FO-formula which is invariant under bisimulation is already invariant under $n$-bisimulation for some $n < \omega$, a view particularly stressed in [112]. Using characteristic $\mathcal{ALC}$-concepts we can then construct for every bisimulation invariant FO-formula an $\mathcal{ALC}$-concept over the same signature which is equivalent to this FO-formula.

Theorem 2.1.24 . *If $\varphi(x)$ is a bisimulation invariant FO-formula over $\tau$ then there is an $\mathcal{ALC}$-concept $C$ over $\tau$ such that $\mathfrak{I}\frac{d}{x} \models \varphi(x)$ iff $(\mathfrak{I}, d) \models C$.*

We occasionally write $(\mathfrak{I}, d) \models \varphi(x)$ instead of $\mathfrak{I}\frac{d}{x} \models \varphi(x)$. As FO has the finite occurrence property, i.e. every formula uses only a finite number of different symbols, we may for this section assume that $\tau$ contains exactly the same role and concept names of $\varphi(x)$ and therefore that $\tau$ is finite.

Lemma 2.1.25 . *If $\varphi$ is invariant under bisimulation then $\varphi$ is invariant under $n$-bisimulation.*

The proof of Lemma 2.1.25 will make use the following lemma in which we shall use compactness of FO. We use an equivalent variant to the way compactness was defined in Section 1.2.5: If for a set $\Gamma \subseteq \mathrm{FO}(\tau)$ all finite subsets $\Gamma_0 \subseteq \Gamma$ are satisfiable, then $\Gamma$ is satisfiable.

Lemma 2.1.26 . *If $\varphi$ is an FO-formula and invariant under $\Longleftrightarrow$ but not invariant under $\Longleftrightarrow_n$ for any $n < \omega$ then there are two interpretations $(\mathfrak{I}, d)$ $(\mathfrak{H}, e)$ such that $Th(\mathfrak{I}, d) = Th(\mathfrak{H}, e)$ but $\mathfrak{I} \models \varphi$ and $\mathfrak{H} \not\models \varphi$*

Proof. Let the premise be true. Then for each $n < \omega$ there are pointed interpretations $(\mathfrak{I}, d)$ and $(\mathfrak{H}, e)$ such that $(\mathfrak{I}, d) \Longleftrightarrow_n (\mathfrak{H}, e)$ yet $(\mathfrak{I}, d) \models \varphi$ and $(\mathfrak{H}, e) \models \neg\varphi(x)$. So none of the

$$\mathfrak{X}_n := \{X^n_{\mathfrak{I},d} \in \mathcal{ALC}(\tau) \mid \exists (\mathfrak{I}, d),\ (\mathfrak{H}, e) \,:\, \varphi(x) \mathrel{=\!\!|} (\mathfrak{I}, d) \Longleftrightarrow_n (\mathfrak{H}, e) \not\models \varphi(x)\}$$

is empty. As there are only finitely many characteristic concepts on each level $n$, $\bigsqcup \mathfrak{X}_n$ is an $\mathcal{ALC}$-concept. Let $\mathfrak{X} := \{\bigsqcup \mathfrak{X}_n \mid n < \omega\}$. Then, by compactness of FO, there is an interpretation $(\mathfrak{I}, d) \models \mathfrak{X}$ with $(\mathfrak{I}, d) \models \varphi$. By the definition of $\mathfrak{X}_n$, for every $n < \omega$ there is $(\mathfrak{H}_n, e_n) \not\models \varphi(x)$ with $(\mathfrak{I}, d) \Longleftrightarrow_n (\mathfrak{H}_n, e_n)$. Hence again by compactness, $\{X^n_{\mathfrak{I},d} \mid n < \omega\} \cup \{\neg\varphi(x)\}$ has a model $(\mathfrak{H}, e)$ with $(\mathfrak{H}, e) \not\models \varphi(x)$.

This implies $(\mathfrak{I}, d) \Longleftrightarrow_\omega (\mathfrak{H}, e)$ and finally $\mathrm{Th}(\mathfrak{I}, d) = \mathrm{Th}(\mathfrak{H}, e)$, so that $(\mathfrak{I}, d)$ and $(\mathfrak{H}, e)$ are the interpretations we are looking for.

□

Proof of Lemma 2.1.25. Let $(\mathfrak{I}, d)$ and $(\mathfrak{H}, e)$ be as in Lemma 2.1.26. As mentioned above, we can find $\omega$-saturated extensions $(\mathfrak{I}^*, d)$ for $(\mathfrak{I}, d)$ and $(\mathfrak{H}^*, e)$ for $(\mathfrak{H}, e)$ such that they satisfy the same FO formulae as their original models. This implies in particular that $(\mathfrak{I}^*, d) \models \varphi(x)$ and $(\mathfrak{H}^*, e) \not\models \varphi(x)$, and since $\mathcal{ALC}$-concepts correspond to FO-formulae that $\mathrm{Th}(\mathfrak{I}, d) = \mathrm{Th}(\mathfrak{I}^*, d)$ and $\mathrm{Th}(\mathfrak{H}, e) = \mathrm{Th}(\mathfrak{H}^*, e)$.

These $\omega$-saturated extensions are also saturated in the sense of $\mathcal{ALC}$. By the theorem of Hennessy and Milner we therefore infer $(\mathfrak{I}^*, d) \Longleftrightarrow (\mathfrak{H}^*, e)$. But this is a contradiction to $\varphi(x)$ being invariant under bisimulation. Hence there is an $n$ such that $\varphi(x)$ is invariant under $n$-bisimulation. □

Proof of Theorem 2.1.24. As Lemma 2.1.25 shows, $\varphi$ is $n$-bisimulation invariant for some $n < \omega$. Let $C_\varphi := \{X^n_{\mathfrak{I},d} \mid (\mathfrak{I}, d) \models \varphi(x)\}$. $C_\varphi$ is finite as on every level $n$ only finitely many characteristic concepts exist. Hence $\bigsqcup C_\varphi$ is an $\mathcal{ALC}$-concept. We have $\varphi(x) \models \bigsqcup C_\varphi$. But we also have $\bigsqcup C_\varphi \models \varphi(x)$, so $\varphi(x)$ is logically equivalent to $C_\varphi$. □

## 2.2 The Characterisation of $\mathcal{ALC}u$-Concepts

$\mathcal{ALC}u$ is $\mathcal{ALC}$ extended with the universal role $u$, i.e. a role which is for every interpretation $\mathfrak{I}$ interpreted as $u^{\mathfrak{I}} = \Delta^{\mathfrak{I}} \times \Delta^{\mathfrak{I}}$. Thus it allows each element to be reachable by a forward edge labelled $u$ from each element in the carrier set.

In the field of modal logics, the corresponding extension of the standard modal logic ML is known as standard modal logic with global modality ML[∀] [21]. The global modality is therefore sometimes also called universal modality [18, Notes to Chapter 7]. In description logics it is called value restriction on the universal role (existential quantification on the universal role), expressed by $\forall u$ ($\exists u$), where $u$ is a role name, representing the universal role [13]. Whilst the global modality ∀ is in a modal context considered to be a logical symbol, value restrictions on the universal role are considered to be a common value restriction over a special role. Since we do not want to consider $u$ to be part of the signature, we shall follow the modal perspective and consider $u$ to be logical symbol.

Several reasons led to this decision: We would need to restrict signatures $\tau$ to those with $u \in \tau$ as otherwise we would look at plain $\mathcal{ALC}$-concepts. Additionally the model class for $\mathcal{ALC}u$-concepts would need to be restricted to the class of interpretations with $u^{\mathfrak{I}} = \Delta^{\mathfrak{I}} \times \Delta^{\mathfrak{I}}$. This, in turn, would mean that also the FO-fragment to which the set of $\mathcal{ALC}u$-concepts corresponds needs to be restricted to this class. Meanwhile the actual feature, namely that global quantification corresponds to the unrestricted FO-quantifying over the free variable in the translated $\mathcal{ALC}$-part of the concept, would be obfuscated. We shall therefore consider $u$ as logical symbol which does not belong to the signature.

It will turn out that the extension of $\mathcal{ALC}$ with universal quantification has a straight forward correspondence in the notion of global bisimulation, the notion of forest unravelling and global $\mathcal{ALC}$-saturation. It is therefore difficult to find references to whom all these notions are attributed [18, Notes for Chapter 7]. A first thorough investigation has been performed in [62].

Note that our notion of universal quantification using the universal role $u$ should not be confused with the notion of a universal role mentioned in [76, Section 9.2.4]: the latter is interpreted as the reflexive and transitive closure of the union over all relations in an interpretation and simply encodes reachability. In modal logics the universal role with this definition has been established as master modality [18, cf. Chapter 6.5 and Notes for Chapter 7]. Our universal quantification is strictly more than the reflexive, transitive closure of the union over all relations in a given interpretation!

EXAMPLE 2.2.1. Let $\Delta^{\mathfrak{I}} = \{d, e\}$ and the signature $\tau = \{R\}$ where $R^{\mathfrak{I}} = \varnothing$. The reflexive, transitive closure over $R$ is $\{(d, d), (e, e)\}$, the universal role in our sense, however, is equal to the set $\{(d, d), (d, e), (e, d), (e, e)\}$ and thus different.

As we have seen in our little example at the beginning, TBoxes seem to express general interrelations between concepts which must hold for all individuals. It turns out that TBoxes form a fragment of this logic $\mathcal{ALC}u$. We shall use the characterisation result presented here to characterise TBoxes later on.

Obviously, in an interpretation $\mathfrak{I}$ which interprets $u$ as $u^{\mathfrak{I}} = \Delta^{\mathfrak{I}} \times \Delta^{\mathfrak{I}}$, all notions for $\mathcal{ALC}$ like bisimulation, type and saturation would implicitly turn out to coincide with the notions we shall introduce further down. Yet point is that signatures $\tau$ over which we shall formulate TBoxes are not required to incorporate a universal rule. Furthermore, even if $\tau$ would contain a symbol $u$ one would have to restrict the class of interpretations to those which interpret $u$ as a universal role.

We therefore treat $u$ as a logical symbol and adapt the notions introduced in $\mathcal{ALC}$ in such a way that the extension of $\mathcal{ALC}$ with universal quantification is accommodated. In this sense, the reader should not expect great revelations, nor is it necessary to follow each proof in detail: The proof are only meant as reassurance and convenience for the reader.

### 2.2.1 Syntax, Semantics and Normal Form of $\mathcal{ALC}u$-Concepts

The description logics $\mathcal{ALC}u$ allows, additionally to all $\mathcal{ALC}$-operators, for the universal role $u$. The syntax is then as follows

$$C ::= \top \mid A \mid D \sqcap E \mid \neg D \mid \exists r.D \mid \exists u.D$$

where $A \in \mathsf{N}_\mathsf{C}$, $r \in \mathsf{N}_\mathsf{R}$, $D, E \in \mathcal{ALC}u(\tau)$ and $u$ is the universal role. The semantics is exactly the same as for $\mathcal{ALC}$ where $u^{\mathfrak{I}} := \Delta^{\mathfrak{I}} \times \Delta^{\mathfrak{I}}$ for every interpretation $\mathfrak{I}$. Though, we do not consider $u$ to be symbol in our signature, but a logical symbol, like $\exists$ or $\sqcap$ which comes with the logic. Again, we use the abbreviation $\forall u.D := \neg\exists u.\neg D$.

As the universal role connects all elements with each other, the interpretation of a globally quantified concept, i.e. a concept of the form $\exists u.D$ or $\forall u.D$, is independent of the distinguished point. Instead of $(\mathfrak{I}, d) \models \exists u.D$ we may also write

$\mathfrak{I} \models \exists u.D$ and similarly for $\forall u$.

The interpretation of a globally quantified concept under any interpretation $\mathfrak{I}$ can either be $\emptyset$ or $\Delta^{\mathfrak{I}}$; we have $(\exists u.D)^{\mathfrak{I}} = \emptyset$ iff $D^{\mathfrak{I}} = \emptyset$ and $(\exists u.D)^{\mathfrak{I}} = \Delta^{\mathfrak{I}}$ otherwise. In the case of $(\forall u.D)^{\mathfrak{I}}$ we have $(\forall u.D)^{\mathfrak{I}} = \Delta^{\mathfrak{I}}$ iff $D^{\mathfrak{I}} = \Delta^{\mathfrak{I}}$ and $(\forall u.D)^{\mathfrak{I}} = \emptyset$ otherwise.

Of course, for concepts that have non-globally and globally quantified parts, like $A \sqcap \exists u.E$, we still need the distinguished point to evaluate whether or not $(\mathfrak{I}, d) \models A \sqcap \exists u.E$.

The translation into first order logic is given by the continuation of the translation function for $\mathcal{ALC}$ (Section 1.2.4) on $\mathcal{ALC}u$-concepts with $[\exists u.C; x_i] := \exists x_i.[C; x_i]$, where $i \in \{0, 1\}$.

Interestingly, concepts in $\mathcal{ALC}u$ split into a global and a local part where the global part is local in nature: The local part of an $\mathcal{ALC}u$-concept $C$ needs to be evaluated at the distinguished element and is therefore comparable to an $\mathcal{ALC}$-concept. The global part, however, must be satisfied at every element of an interpretation, but at each element, it behaves like an $\mathcal{ALC}$-concept, i.e. it only talks about properties of elements which are reachable from this specific element via edges.

One way to highlight this property is to define a normal-form for $\mathcal{ALC}u$-concepts which represents the original concept by a boolean connexion of $\mathcal{ALC}$-concepts (the local component) and those $\mathcal{ALC}u$-concepts which are composed of a global quantifier followed by an $\mathcal{ALC}$-concept.

Definition 2.2.2. An $\mathcal{ALC}u$-concept is in *disjunctive normal form*, if it has the following structure where $D_{i,j}$ is an $\mathcal{ALC}$-concept and $Q_{i,j} \in \{\exists, \forall\}$:

$$\bigsqcup_{i<m} \left( D_{i,0} \sqcap \bigsqcap_{j<n_i} Q_{i,j} u.D_{i,j+1} \right)$$

◊

Lemma 2.2.3. *Every $\mathcal{ALC}u$-concept $C$ can be rewritten as logically equivalent concept in disjunctive normal form.*

Proof. The proof is carried out by induction over $\mathcal{ALC}u$ concepts: Atomic concepts are already in normal form. In case $C = D \sqcap E$ we my apply the induction hypothesis to $D$ and $E$ and obtain

$$\begin{aligned} C = & \left(\bigsqcup_{i<m} D_{i,0} \sqcap \bigsqcap_{j<n_i} Q_{i,j}u.D_{i,j+1}\right) \sqcap \left(\bigsqcup_{r<k} E_{r,0} \sqcap \bigsqcap_{s<\ell_r} Q_{r,s}u.E_{r,s+1}\right) \\ = & \bigsqcup_{i<m} \left(D_{i,0} \sqcap \bigsqcap_{j<n_i} Q_{i,j}u.D_{i,j+1} \sqcap \left(\bigsqcup_{r<k} E_{r,0} \sqcap \bigsqcap_{s<\ell_r} Q_{r,s}u.E_{r,s+1}\right)\right) \\ = & \bigsqcup_{i<m} \bigsqcup_{r<k} \left(D_{i,0} \sqcap \bigsqcap_{j<n_i} Q_{i,j}u.D_{i,j+1} \sqcap E_{r,0} \sqcap \bigsqcap_{s<\ell_r} Q_{r,s}u.E_{r,s+1}\right) \end{aligned}$$

where the last line is a concept in disjunctive normal form. In a similar way, the induction hypothesis can be applied to $D$ in the case $C = \neg D$.

$$C = \neg \bigsqcup_{i<m} D_{i,0} \sqcap \bigsqcap_{j<n_i} Q_{i,j}u.D_{i,j+1} = \bigsqcap_{i<m} D'_{i,0} \sqcup \bigsqcup_{j<n_i} Q'_{i,j}u.D'_{i,j+1}$$

where $D'_{i,j}$ is the negation of $D_{i,j}$ and $Q'_{i,j}$ is the dual of $Q_{i,j}$. This can be considered as a conjunction of disjunctive normal forms, which can be rendered into a disjunctive normal form by applying the distributivity laws like in the case $C = D \sqcap E$.

To treat the case $C = \exists t.D$ where $t \in \mathsf{N_R} \cup \{u\}$, we first show that $\exists t. \bigsqcup_{i<m} D_{i,0} \sqcap \bigsqcap_{j<n_i} Q_{i,j}u.D_{i,j+1}$ is logically equivalent to $\bigsqcup_{i<m}(\exists t.D_{i,0}) \sqcap \bigsqcap_{j<n_i} Q_{i,j}u.D_{i,j+1}$. For then $\exists t.D_{i,0}$ is either a globally quantified or still an $\mathcal{ALC}$-concept, depending on $t$; in any case, $\bigsqcup_{i<m}(\exists t.D_{i,0}) \sqcap \bigsqcap_{j<n_i} Q_{i,j}u.D_{i,j+1}$ is a disjunctive normal form.

Clearly, $\exists t. \bigsqcup_{i<m} D_{i,0} \sqcap \bigsqcap_{j<n_i} Q_{i,j}u.D_{i,j+1} \equiv \bigsqcup_{i<m} \exists t.(D_{i,0} \sqcap \bigsqcap_{j<n_i} Q_{i,j}u.D_{i,j+1})$. For arbitrary $\mathfrak{I}$ and $d \in \Delta^{\mathfrak{I}}$, we have $d \in \exists t.(D_{i,0} \sqcap \bigsqcap_{j<n_i} Q_{i,j}u.D_{i,j+1})^{\mathfrak{I}}$ for some $i < m$, iff there is some $t$-successor $d'$ of $d$, with $d' \in D^{\mathfrak{I}}_{i,0} \cap \bigcap_{j<n_i}(Q_{i,j}u.D_{i,j+1})^{\mathfrak{I}}$. As remarked above $\bigcap_{j<n_i}(Q_{i,j}u.D_{i,j+1})^{\mathfrak{I}} \neq \emptyset$ iff $\bigcap_{j<n_i}(Q_{i,j}u.D_{i,j+1})^{\mathfrak{I}} = \Delta^{\mathfrak{I}}$, hence $d' \in D^{\mathfrak{I}}_{i,0} \cap \bigcap_{j<n_i}(Q_{i,j}u.D_{i,j+1})^{\mathfrak{I}}$ iff $d' \in D^{\mathfrak{I}}_{i,0}$ and $\bigcap_{j<n_i}(Q_{i,j}u.D_{i,j+1})^{\mathfrak{I}} = \Delta^{\mathfrak{I}}$.

The latter is the case iff $d \in \left(\left(\exists t.D_{i,0}\right) \sqcap \bigsqcap_{j<n_i} Q_{i,j}u.D_{i,j+1}\right)^{\mathfrak{I}}$, which proves our claim

$$\exists t. \bigsqcup_{i<m} D_{i,0} \sqcap \bigsqcap_{j<n_i} Q_{i,j}u.D_{i,j+1} \equiv \bigsqcup_{i<m} (\exists t.D_{i,0}) \sqcap \bigsqcap_{j<n_i} Q_{i,j}u.D_{i,j+1}$$

and conludes the induction step for $C = \exists t.D$. □

### 2.2.2 Global Bisimulation, Finite Global Bisimulation and Characteristic $\mathcal{ALC}u$-Concepts

The notion of bisimulation for $\mathcal{ALC}$ can now be lifted onto the global level for $\mathcal{ALC}u$:

Definition 2.2.4. Let $\mathfrak{I}_0$ and $\mathfrak{I}_1$ be two structures. Then $\mathfrak{I}_0 \xleftrightarrow{g} \mathfrak{I}_1$ if both of the following conditions hold

1. for all $d \in \Delta^{\mathfrak{I}_0}$ there is $e \in \Delta^{\mathfrak{I}_1}$ with $(\mathfrak{I}_0, d) \Longleftrightarrow (\mathfrak{I}_1, e)$
2. for all $e \in \Delta^{\mathfrak{I}_1}$ there is $d \in \Delta^{\mathfrak{I}_0}$ with $(\mathfrak{I}_0, d) \Longleftrightarrow (\mathfrak{I}_1, e)$

We write $(\mathfrak{I}_0, d) \overset{g}{\Longleftrightarrow} (\mathfrak{I}_1, e)$ if $\mathfrak{I}_0 \overset{g}{\Longleftrightarrow} \mathfrak{I}_1$ and $(\mathfrak{I}_0, d) \Longleftrightarrow (\mathfrak{I}_1, e)$. ◇

In particular Otto, e.g. [102], defines global bisimulation via games with reset rounds: at the beginning of every round, **I** can call a reset-round, where **I** chooses one of the interpretations and places its pebble on any element in this interpretation. **II** can place the pebble of the other structure on some arbitrary element which must be atomically equivalent. If she cannot find such an element, she loses the game. Otherwise a new $\mathcal{ALC}$-bisimulation game starts in this configuration.

Our definition of global bisimulation, although equivalent to games with reset rounds, are very close to the intuition of the local nature of globally quantified $\mathcal{ALC}u$ concepts, in that we require for every element in one interpretation a locally bisimilar partner in the other interpretation.

In analogy to $\mathcal{ALC}$-bisimulation, we can introduce a bounded bisimulation $\overset{g}{\Longleftrightarrow}_n$ by replacing $\Longleftrightarrow$ with $\Longleftrightarrow_n$ in the definition above. To make use of this stratification, we continue the rank-function defined in Section 2.1.2 on $\mathcal{ALC}u$-concepts, by setting $\operatorname{rank} \exists u.C := \operatorname{rank} C$, therefore ignoring global quantification.

We want to show that the disjunctive normal form does not increase the rank of a concept:

LEMMA 2.2.5 . *Let $C$ be an $\mathcal{ALC}u$-concept then there is a concept $C'$ in disjunctive normal form such that $C \equiv C'$ and* $\operatorname{rank} C' \leq \operatorname{rank} C$.

The proof is readily carried out by induction over the structure of $\mathcal{ALC}u$-concepts. The inequality between the ranks is necessary as the following example shows: Let $r \in \mathsf{N_R}$ and $C := \exists r.(\top \sqcap \forall u.\exists s.\top)$ then $C' := (\exists r.\top) \sqcap \forall u.\exists s.\top$ is its normal form with $\operatorname{rank} C' < \operatorname{rank} C$.

OBSERVATION 2.2.6 . *$\mathcal{ALC}u$ is invariant under $\overset{g}{\Longleftrightarrow}$.*

With the following lemma and the remark that for all $n < \omega$

$$(\mathfrak{I}_0, d) \overset{g}{\Longleftrightarrow} (\mathfrak{I}_1, e) \quad \Longrightarrow \quad (\mathfrak{I}_0, d) \overset{g}{\Longleftrightarrow}_{n+1} (\mathfrak{I}_1, e) \quad \Longrightarrow \quad (\mathfrak{I}_0, d) \overset{g}{\Longleftrightarrow}_n (\mathfrak{I}_1, e)$$

the observation is clear.

LEMMA 2.2.7 . *Let $C \in \mathcal{ALC}u$ and* $\operatorname{rank} C \leq n$ *then for all pointed interpretations*

$(\mathfrak{I}_0, d), (\mathfrak{I}_1, e)$ *with* $(\mathfrak{I}_0, d) \overset{g}{\Longleftrightarrow}_n (\mathfrak{I}_1, e)$ *we have* $d \in C^{\mathfrak{I}_0}$ *iff* $e \in C^{\mathfrak{I}_1}$.

Proof. Under the given preconditions let $d \in C^{\mathfrak{I}_0}$. Then $C$ has a disjunctive normal form $C' := \bigsqcup_{i<m} D_{i,0} \sqcap \bigsqcap_{j<n_i} Q_{i,j}u.D_{i,j+1}$ such that rank $C' \leq$ rank $C$. Let $i < m$ such that $d \in (D_{i,0} \sqcap \bigsqcap_{j<n_i} Q_{i,j}u.D_{i,j+1})^{\mathfrak{I}_0}$. By definition $C'$ all $D_{i,j}$ are $\mathcal{ALC}$-concepts of rank $\leq n$. In particular $(\mathfrak{I}_0, d) \overset{g}{\Longleftrightarrow}_n (\mathfrak{I}_1, e)$ entails $(\mathfrak{I}_0, d) \Longleftrightarrow_n (\mathfrak{I}_1, e)$ so, by Theorem 2.1.17, we have $e \in D^{\mathfrak{I}_1}_{i,0}$. If $Q_{i,j} = \exists$ then there is $d' \in D^{\mathfrak{I}_0}_{i,j+1}$ which finds some $e' \in \Delta^{\mathfrak{I}_1}$ such that $(\mathfrak{I}_0, d') \Longleftrightarrow_n (\mathfrak{I}_1, e')$, so $e \in (\exists u.D_{i,j+1})^{\mathfrak{I}_1}$. In the other case, where $Q_{i,j} = \forall$, we have $\Delta^{\mathfrak{I}_0} = D^{\mathfrak{I}_0}_{i,j+1}$. But every $e' \in \Delta^{\mathfrak{I}_1}$ finds a $d' \in \Delta^{\mathfrak{I}_0}$ such that $(\mathfrak{I}_0, d') \Longleftrightarrow_n (\mathfrak{I}_1, e')$. Again with Theorem 2.1.17 one obtains that all $e' \in \Delta^{\mathfrak{I}_1}$ must be in in $D^{MI_1}_{i,j+1}$ which entails that $e \in \forall u.D_{i,j+1}$. Hence $e \in (D_{i,0} \sqcap \bigsqcap_{j<n_i} Q_{i,j}u.D_{i,j+1})^{\mathfrak{I}_1}$. The symmetry of $\overset{g}{\Longleftrightarrow}$ clinches the argument for the converse direction. □

Given, the signature $\tau$ is finite, characteristic concepts for $\mathcal{ALC}$ were introduced in Definition 2.1.13. Similarly we can define characteristic concepts for $\mathcal{ALC}u$; indeed their construction is based on $\mathcal{ALC}$-concepts and is straight forward: In order to ensure that a $\tau$-interpretation $\mathfrak{I}_1$ is globally $n$-bisimilar to some interpretation $\mathfrak{I}_0$,

1. for every element in $d \in \Delta^{\mathfrak{I}_0}$ there must be some $n$-bisimilar element in $\mathfrak{I}_1$, hence $\mathfrak{I}_1$ has to satisfy $\bigsqcap\{\exists u.X^n_{\mathfrak{I}_0,d} \mid d \in \Delta^{\mathfrak{I}_0}\}$.
2. every element in $\mathfrak{I}_1$ must be $n$-bisimilar to some element $\mathfrak{I}_0$, i.e. $\mathfrak{I}_1$ has to satisfy $\forall u. \bigsqcup\{X^n_{\mathfrak{I}_0,d} \mid d \in \Delta^{\mathfrak{I}_0}\}$.

In both cases $X^n_{\mathfrak{I}_0,d}$ are the characteristic $\mathcal{ALC}$-concepts from Definition 2.1.13, with respect to the given signature. Since there are only finitely many characteristic concepts on each level, the sets given in item 1. and 2. are finite. So we obtain the characteristic $\mathcal{ALC}u$-concept for $\mathfrak{I}_0$

$$X^n_{\mathfrak{I}_0} := \bigsqcap\{\exists u.X^n_{\mathfrak{I}_0,d} \mid d \in \Delta^{\mathfrak{I}_0}\} \sqcap \forall u. \bigsqcup\{X^n_{\mathfrak{I}_0,d} \mid d \in \Delta^{\mathfrak{I}_0}\}$$

The characteristic $\mathcal{ALC}u$ concept for a pointed interpretation $(\mathfrak{I}_0, d)$ is then given by $X^n_{\mathfrak{I}_0,d} := X^n_{\mathfrak{I}_0} \sqcap Y^n_{\mathfrak{I}_0,d}$ where $X^n_{\mathfrak{I}_0}$ is the characteristic $\mathcal{ALC}u$-concept defined above and $Y^n_{\mathfrak{I}_0,d}$ is the characteristic $\mathcal{ALC}$-concept.

### 2.2.3 $\mathcal{ALC}u$-Concepts as Global Bisimulation Invariant Fragment of FO

We can extend the notion of $r$-types given in Definition 2.1.21 immediately on $\mathcal{ALC}u$ by allowing $r \in \mathsf{N_R} \cup \{u\}$. We shall give an explicit definition anyway. Note that types are still subsets of $\mathcal{ALC}$ rather than of $\mathcal{ALC}u$:

Definition 2.2.8 . Let $\tau$ be a signature and $(\mathfrak{I}, d)$ a pointed $\tau$-interpretation.

1. For each $r \in \mathsf{N_R} \cup \{u\}$ we call $\Gamma \subseteq \mathcal{ALC}(\tau)$ an *r-type* of $d$, if for every finite subset $\Gamma_0 \subseteq \Gamma$ we have $d \in \exists r. \bigsqcap \Gamma_0$. In case $r = u$ we also say $\Gamma$ is a $u$-type of $\mathfrak{I}$.

2. For each $r \in \mathsf{N_R} \cup \{u\}$, an $r$-type $\Gamma$ of $d$ is realised at $d$, if there is an $r$-successor $d_0$ of $d$ in $\mathfrak{I}$ such that $d_0 \in \bigcap_{C \in \Gamma} C^{\mathfrak{I}}$. If $r = u$ we also say that $\Gamma$ is realised in $\mathfrak{I}$.

3. $\mathfrak{I}$ is saturated if for all $r \in \mathsf{N_R} \cup \{u\}$ and all $d \in \Delta^{\mathfrak{I}}$ every $r$-type $\Gamma$ of $d$ is realised at $d$.

◇

As the theorem of Hennessy-Milner was mainly aimed at the standard modal logic ML, we shall say that an arbitrary logic $\mathcal{L}$ has the Hennessy-Milner Property, if it satisfies an analogous theorem. More formally: Let $\mathcal{L}$ be a logic and $\overset{\mathcal{L}}{\Longleftrightarrow}$ an appropriate variant of bisimulation for $\mathcal{L}$. Similarly let $\mathcal{L}$-saturation be an appropriate notion of saturation w.r.t. $\mathcal{L}$.

Definition 2.2.9 . $\mathcal{L}$ has the *Hennessy-Milner-Property* if for $\mathcal{L}$-saturated $\tau$-interpretations $(\mathfrak{I}, d)$ and $(\mathfrak{H}, e)$ the following is true

$$\mathrm{Th}_{\mathcal{L}}(\mathfrak{I}, d) = \mathrm{Th}_{\mathcal{L}}(\mathfrak{H}, e) \quad \Longleftrightarrow \quad (\mathfrak{I}, d) \overset{\mathcal{L}}{\Longleftrightarrow} (\mathfrak{H}, e)$$

where $\mathrm{Th}_{\mathcal{L}}(\mathfrak{I}, d) = \{C \in \mathcal{L}(\tau) \mid d \in C^{\mathfrak{I}}\}$. ◇

The Hennessy-Milner-Property obviously not only depends on the logic itself but also on the chosen model-theoretic relation $\overset{\mathcal{L}}{\Longleftrightarrow}$ on the interpretation and the notion of saturation. Clearly, in every instance where we prove the Hennessy-Milner-Property it will be clear what $\mathcal{L}$, the concrete variant of bisimulation $\overset{\mathcal{L}}{\Longleftrightarrow}$ and the notion of $\mathcal{L}$-saturation is.
or our concrete case of $\mathcal{ALC}u$ we now (re)define $\mathrm{Th}(\mathfrak{I}, d)$ as follows: For every $\tau$-interpretation $\mathfrak{I}$, we define $\mathrm{Th}(\mathfrak{I}, d) := \{C \in \mathcal{ALC}u(\tau) \mid d \in C^{\mathfrak{I}}\}$.

Theorem 2.2.10. *$\mathcal{ALC}u$ has the Hennessy and Milner-Property, i.e. for any saturated interpretation $(\mathfrak{I}, d)$ and $(\mathfrak{H}, e)$ we have*

$$Th(\mathfrak{I}, d)Th(\mathfrak{H}, e) \iff (\mathfrak{I}, d) \overset{g}{\longleftrightarrow} (\mathfrak{H}, e)$$

Proof. '$\Longrightarrow$': Let $d' \in \Delta^{\mathfrak{I}}$ be arbitrary and $\mathrm{Th}_{\mathcal{ALC}}(\mathfrak{I}, d') := \{C \in \mathcal{ALC} \mid d' \in C\}$ of $d \in \Delta^{\mathfrak{I}}$ its $\mathcal{ALC}$-Theory. $\mathrm{Th}_{\mathcal{ALC}}(\mathfrak{I}, d')$ is a $u$-type in $\mathfrak{H}$. Since $\mathfrak{H}$ is saturated, there is an element $e' \in \Delta^{\mathfrak{H}}$ such that $e' \in \bigcap\{C^{\mathfrak{H}} \mid C \in \mathrm{Th}_{\mathcal{ALC}}(\mathfrak{I}, d')\}$. Applying the result from Hennessy and Milner (Proposition 2.1.23) we obtain that $(\mathfrak{I}, d') \longleftrightarrow (\mathfrak{H}, e')$. Hence there for all $d'$ there is an element $e' \in \Delta^{\mathfrak{H}}$ such that $(\mathfrak{I}, d') \longleftrightarrow (\mathfrak{H}, e')$. With the same argument we can find for every $e' \in \Delta^{\mathfrak{H}}$ a locally bisimilar element $d' \in \Delta^{\mathfrak{I}}$. Hence $\mathfrak{I} \overset{g}{\longleftrightarrow} \mathfrak{H}$. As by assumption $e \in \bigcap\{C^{\mathfrak{H}} \mid C \in \mathrm{Th}_{\mathcal{ALC}}(\mathfrak{I}, d)\}$, we also have $(\mathfrak{I}, d) \overset{g}{\longleftrightarrow} (\mathfrak{H}, e)$ and so $(\mathfrak{I}, d) \overset{g}{\longleftrightarrow} (\mathfrak{H}, e)$. The converse is entailed by Observation 2.2.6. □

Merely requiring that $\exists u$-types are realised in a in interpretation is not enough!

Example 2.2.11. Consider Example 2.1.20, with its interpretations $\mathfrak{I}$ and $\mathfrak{H}$. We consider two new interpretations, one being $\mathfrak{K}_0$ containing two disjoint copies of $\mathfrak{H}$ and one being $\mathfrak{K}_1$ containing a copy of $\mathfrak{I}$ and a copy $\mathfrak{H}$, who are both disjoint.

Both $\mathfrak{K}_0$ and $\mathfrak{K}_1$ realise all $\exists u$-types and although $\mathfrak{K}_0 \overset{g}{\longleftrightarrow}_n \mathfrak{K}_1$ for every $n < \omega$, $(\mathfrak{K}_1, d)$ where $d$ is the root element of the copy of $\mathfrak{I}$ is not fully bisimilar to any element in $\mathfrak{K}_0$: The only bisimilar candidates to $d$ in $\mathfrak{K}_0$ are the root elements of the two copies of $\mathfrak{H}$. But Example 2.1.20 showed that $d$ and the root element of $\mathfrak{H}$ is not bisimilar.

Theorem 2.2.12. *Let $\varphi(x)$ be an FO formula over $\tau$ which is invariant under $\overset{g}{\longleftrightarrow}$ then there is an $\mathcal{ALC}u$-concept over $\tau$ such that for all $\tau$-interpretations $\mathfrak{I}$ we have $\mathfrak{I}\frac{d}{x} \models \varphi(x)$ iff $(\mathfrak{I}, d) \models C$.*

Using the same rational as in Lemma 2.1.25 and 2.1.26 we can show the equivalent of Lemma 2.1.25:

Lemma 2.2.13. *If $\varphi$ is invariant under $\overset{g}{\longleftrightarrow}$ then there is some $n < \omega$ s.t. $\varphi$ is invariant under $\overset{g}{\longleftrightarrow}_n$.*

Proof of Theorem 2.2.12. The concept $C$ in Theorem 2.2.12 is then given by $C := \bigsqcup\{X^n_{\mathfrak{I},d} \mid \mathfrak{I}\frac{d}{x} \models \varphi(x)\}$ where $n$ is given by Lemma 2.2.13 and $X^n_{\mathfrak{I},d}$ is the characteristic $\mathcal{ALC}u$-concept. Again, $C$ is well defined as there are only finitely many different characteristic concepts on each level $n$.

It is clear that $\mathfrak{I}\frac{d}{x} \models \varphi(x)$ implies $(\mathfrak{I}, d) \models C$. If, for the converse, $(\mathfrak{I}, d) \models C$ then there is $\mathfrak{H}\frac{e}{x} \models \varphi(x)$ and $(\mathfrak{I}, d) \overset{g}{\Longleftrightarrow}_n (\mathfrak{H}, e)$. As $\varphi$ is invariant under global $n$-bisimulation we have $\mathfrak{I}\frac{d}{x} \models \varphi(x)$. □

We call $\mathfrak{I}^F := \bigcup_{d \in \Delta^{\mathfrak{I}}} \mathfrak{I}_d$ the *forest unravelling* of $\mathfrak{I}$, where $\mathfrak{I}_d$ is the tree-unravelling of $\mathfrak{I}$ in $d$ (cf. Definition 2.1.4).

Observation 2.2.14 . $\mathfrak{I}^F \overset{g}{\Longleftrightarrow} \mathfrak{I}$

Proof. Clearly, for every element $d \in \Delta^{\mathfrak{I}}$ there is the root element of $\mathfrak{I}_d$ which is bisimilar to $d$. Hence $\mathfrak{I}^F \overset{g}{\Longleftrightarrow} \mathfrak{I}$. We have to show the opposite direction:

From the winning-strategy given in Observation 2.1.5 we infer immediately that for every element $\bar{d} \in \Delta^{\mathfrak{I}_d}$ there is some element $d' \in \Delta^{\mathfrak{I}}$ such that $(\mathfrak{I}_d, \bar{d}) \Longleftrightarrow (\mathfrak{I}, d')$. The element $d'$ is the last element in the path $\bar{d}$. This shows that for all $\bar{d} \in \Delta^{\mathfrak{I}^F}$ there is an element $d \in \Delta^{\mathfrak{I}}$ such that $(\mathfrak{I}^F, \bar{d}) \Longleftrightarrow (\mathfrak{I}, d)$ and □

This Observation and the characterisation shows that $\mathcal{ALC}u$-concepts cannot distinguish between an interpretation and its forest unravelling. It also shows that $\mathcal{ALC}u$-concepts are local in the sense that these global quantification merely encapsulate ordinary local $\mathcal{ALC}$-concepts.

Though, they are sensitive to disjoint unions: $\forall u.A \sqcup \forall u.B$ can be satisfied by one interpretation containing exactly one element satisfying $A$ and by another interpretation containing one element satisfying only $B$. The disjoint union of both interpretations however does not satisfy $\forall u.A \sqcup \forall u.B$!

## 2.3 The Characterisation of $\mathcal{ALC}$-TBoxes

TBoxes play a great role in knowledge representation as pointed out in the introduction. We shall give a characterisation for these TBoxes and later in Chapter 6 an application of these characterisation results.

A characterisation result has been given in the context of global definability in modal logic [43]. It does not concern TBoxes directly but characterises elementary[1] classes which are closed under surjective bisimulations and disjoint unions as those whose interpretations are models of universally quantified translations of modal formulae.

Our equivalent characterisation result differs in that we characterise TBoxes as being equivalent to exactly the fragment of FO formulae which is invariant under TBoxes and invariant under global bisimulation. The more specialised notion of global bisimulation (every global bisimulation is surjective) is countered by the broader notion of invariance under disjoint unions which requires a formula also to be preserved under subinterpretations; We therefore characterise TBoxes not only as fragment of FO but also as fragment of $\mathcal{ALC}u$ and present our characterisation result as extension of the $\mathcal{ALC}u$-characterisation.

[1]Defined by a set of first order formulae.

$\mathcal{ALC}$*-TBoxes* are finite sets of *concept inclusion* $C \sqsubseteq D$ where $C, D$ are $\mathcal{ALC}$-concepts. An $\mathcal{ALC}$-concept is satisfied by an interpretation $\mathfrak{I}$, $\mathfrak{I} \models C \sqsubseteq D$, iff $C^{\mathfrak{I}} \subseteq D^{\mathfrak{I}}$. An $\mathcal{ALC}$-TBox $\mathcal{T}$ is true or *satisfied* by an interpretation $\mathfrak{I}$, $\mathfrak{I} \models \mathcal{T}$ iff for all $C \sqsubseteq D \in \mathcal{T}$ we have $\mathfrak{I} \models C \sqsubseteq D$.

We define $\operatorname{rank}(C \sqsubseteq D) := \max\{\operatorname{rank} C, \operatorname{rank} D\}$ and

$$\operatorname{rank} \mathcal{T} := \max\{\operatorname{rank}(C \sqsubseteq D) \mid C \sqsubseteq D \in \mathcal{T}\}.$$

The length $|\mathcal{T}|$ of a TBox $\mathcal{T}$ is defined as $|\mathcal{T}| := \sum_{C \sqsubseteq D \in \mathcal{T}} |C| + |D|$

We have $\mathfrak{I} \models C \sqsubseteq D$ iff $C^{\mathfrak{I}} \subseteq D^{\mathfrak{I}}$ iff for all $d \in \mathfrak{I}$ we have $(\mathfrak{I}, d) \models C \to D$. Hence an $\mathcal{ALC}$-concept inclusion can be rewritten as $\mathcal{ALC}u$-concept: $\forall u.C \to D$. This means in particular that the set of all TBoxes is a fragment of $\mathcal{ALC}u$ and therefore invariant under global bisimulation.

The question, which property singles out those FO-sentences that are equivalent to an $\mathcal{ALC}$-TBox, reduces to the question which property singles out those global $\mathcal{ALC}u$-concepts that are equivalent to $\mathcal{ALC}$-TBoxes.

As we shall see in the next proposition, the closure of $\mathcal{ALC}$-TBoxes under $\neg$ and $\sqcup$ has the same expressivity as global $\mathcal{ALC}u$-concepts. We shall specify what global $\mathcal{ALC}u$-concepts are and how we interpret $\mathcal{ALC}$-TBoxes connected by $\sqcup$ and $\neg$:

A *global* $\mathcal{ALC}u$*-concept* is an $\mathcal{ALC}u$-concept which is logically equivalent to a disjunctive normal form $\bigsqcup_{i<m} D_{i,0} \sqcap \bigsqcap_{j<n_i} Q_{i,j} u.D_{i,j+1}$ as introduced in Definition 2.2.2 such that for all $i < m$ we have $D_{i,0} = \top$ and $Q_{i,j} \in \{\exists, \forall\}$. Note that all $D_{i,j}$ are $\mathcal{ALC}$-concepts, and thus are local in nature but each of them is globally quantified. These global concepts are equivalent to global bisimulation invariant FO-sentences.

We recursively define elements of the set which contains $\mathcal{ALC}$-TBoxes and is closed under $\sqcup$ and $\neg$ as follows: $\mathcal{C} ::= \mathcal{T} \mid \mathcal{C}_0 \sqcup \mathcal{C}_1 \mid \neg \mathcal{C}_0$ where $\mathcal{T}$ is an $\mathcal{ALC}$-TBox and $\mathcal{C}_0, \mathcal{C}_1$ are both elements of this set. We extend the usual satisfaction relation for TBoxes: $\mathfrak{I} \models C_0 \sqcup C_1$ if $\mathfrak{I} \models C_0$ or $\mathfrak{I} \models C_1$ and $\mathfrak{I} \models \neg C_0$ if not $\mathfrak{I} \models C_0$.

Proposition 2.3.1 . *Let $C$ be a global $\mathcal{ALC}u$-concept. Then there is boolean combination $\mathcal{C}$ of $\mathcal{ALC}$-TBoxes such that $\mathfrak{I} \models C$ iff $\mathfrak{I} \models \mathcal{C}$.*

Proof. Assume $\bigsqcup_{i<m} \bigsqcap_{j<n_i} Q_{i,j} u.D_{i,j+1}$ would be a disjunctive normal form of $C$. In case $Q_{i,j} = \exists$, one can rewrite $Q_{i,j} u.D_{i,j+1}$ equivalently as the negated TBox $\neg\{D_{i,j+1} \sqsubseteq \bot\}$. If $Q_{i,j} = \forall$ then $Q_{i,j} u.D_{i,j+1}$ is equivalent to the TBox $\{\top \sqsubseteq D_{i,j+1}\}$. Replacing all conjuncts by their rewritings renders the normal form into a boolean connexion of $\mathcal{ALC}$-TBoxes which is logically equivalent to $C$. □

### 2.3.1 Invariance under Disjoint Unions

The missing property which highlights the universal character of $\mathcal{ALC}$-TBoxes in the sense that $\mathcal{ALC}$-TBoxes are $\forall u$-quantified implications of $\mathcal{ALC}$-concepts is being invariant under disjoint unions.

Let $I$ be an index set and $(\mathfrak{I}_i)_{i \in I}$ a family of interpretations. We can enforce pairwise disjointness of the carrier-sets by setting $\Delta^{\mathfrak{I}'_i} := \Delta^{\mathfrak{I}_i} \times \{i\}$ and adapting the interpretation of symbols accordingly. We may therefore assume w.l.o.g. that all carrier-sets are disjoint. We then define the *disjoint union* $\biguplus_{i \in I} \mathfrak{I}_i$ as follows

$$
\begin{array}{lll}
\Delta^{\biguplus_i \mathfrak{I}_i} & := \bigcup_{i \in I} \Delta^{\mathfrak{I}_i} & \text{recall } \Delta^{\mathfrak{I}_i} \cap \Delta^{\mathfrak{I}_j} = \emptyset \\
A^{\biguplus_i \mathfrak{I}_i} & := \bigcup_{i \in I} A^{\mathfrak{I}_i} & \text{for all } A \in \mathsf{N_C} \\
r^{\biguplus_i \mathfrak{I}_i} & := \bigcup_{i \in I} r^{\mathfrak{I}_i} & \text{for all } r \in \mathsf{N_R}
\end{array}
$$

Complementary to disjoint unions we have generated subinterpretations: Let $\mathfrak{I}$ be an interpretation and $G \subseteq \Delta^{\mathfrak{I}}$. The *generated subinterpretation* $\mathfrak{U}$ of $\mathfrak{I}$ over $G$ is essentially all connected components containing elements of $G$. Formally this is defined as follows:

$$
\begin{array}{lll}
\Delta^{\mathfrak{U}} & := \bigcap \mathfrak{G} & \\
A^{\mathfrak{U}} & := A^{\mathfrak{I}} \cap \Delta^{\mathfrak{U}} & \text{for all } A \in \mathsf{N_C} \\
r^{\mathfrak{U}} & := r^{\mathfrak{I}} \cap \Delta^{\mathfrak{U}} \times \Delta^{\mathfrak{U}} & \text{for all } r \in \mathsf{N_R}
\end{array}
$$

where $\mathfrak{G}$ contains all subsets $S \subseteq \Delta^{\mathfrak{I}}$ such that $G \subseteq S$ and for all $r \in \mathsf{N_R}$ we have if $d \in S$ and $(d, e) \in r^{\mathfrak{I}}$ or $(e, d) \in r^{\mathfrak{I}}$ then $e \in S$.

The last requirement reminds somewhat of a functional closure. As result, elements in the generated subinterpretation are extracted together with their connected component. Therefore a generated subinterpretation $\mathfrak{U}$ is not merely a restriction of the underlying structure $\mathfrak{I}$ to a subset $G$ of its carrier-set. This fact highlights the following lemma:

LEMMA 2.3.2. *Let $\mathfrak{U}$ be a generated subinterpretation of $\mathfrak{I}$. Then for all $d \in \Delta^{\mathfrak{U}}$ we have* $(\mathfrak{U}, d) \Longleftrightarrow (\mathfrak{I}, d)$

PROOF. We show that player **II** can maintain configurations $(\mathfrak{U}, d'; \mathfrak{I}, d')$ throughout the game. The requirement is true for the start configuration. Assume the game has reached configuration $(\mathfrak{U}, d'; \mathfrak{I}, d')$ As $\mathfrak{U}$ is a subinterpretation of $\mathfrak{I}$ it is clear that **II** can always copy challenges from **I**, if he chooses to play in $\mathfrak{U}$.

In case **I** challenges **II** with an $r$-successor $d''$ of $d'$ in $\mathfrak{I}$. As $d' \in \Delta^{\mathfrak{U}}$, $d'$ finds itself in all $S \in \mathfrak{G}$, used to construct $\mathfrak{U}$. By definition for all $S \in \mathfrak{G}$, also $d'' \in S$. Hence $d'' \in \Delta^{\mathfrak{U}}$ and as $\mathfrak{U}$ is a subinterpretation of $\mathfrak{I}$, $d''$ is an $r$-successor of $d'$ and a valid configuration $(\mathfrak{U}, d''; \mathfrak{I}, d'')$ is reached. □

DEFINITION 2.3.3. A concept is *invariant under disjoint union*, if for every family $(\mathfrak{I}_i, d_i)_{i \in I}$ of pointed interpretations we have

$$\forall i \in I : (\mathfrak{I}_i, d_i) \models C \quad \Longleftrightarrow \quad \forall j \in I : (\biguplus_{i \in I} \mathfrak{I}_i, d_j) \models C$$

a TBox is invariant under disjoint unions if

$$\forall i \in I : \mathfrak{I}_i \models \mathcal{T} \quad \Longleftrightarrow \quad \forall j \in I : \biguplus_{i \in I} \mathfrak{I}_i \models \mathcal{T}.$$

◇

OBSERVATION 2.3.4. *$\mathcal{ALC}$-TBoxes are invariant under disjoint unions.*

Before we give a proof we furnish two lemmas stating that $\mathcal{ALC}$-TBoxes are preserved under disjoint unions and preserved under generated subinterpretations:

LEMMA 2.3.5. *$\mathcal{ALC}$-TBoxes are preserved under subinterpretations, i.e. if $\mathfrak{I} \models \mathcal{T}$ and $\mathfrak{U}$ is a generated substructre of $\mathfrak{I}$ then $\mathfrak{U} \models \mathcal{T}$.*

PROOF. Assume $C \sqsubseteq D \in \mathcal{T}$ and $d \in C^{\mathfrak{U}}$. With Lemma 2.3.2 we infer $d \in C^{\mathfrak{I}}$. As $\mathfrak{I} \models \mathcal{T}$, we have $C^{\mathfrak{I}} \subseteq D^{\mathfrak{I}}$ and therefore $d \in D^{\mathfrak{I}}$, which, again via Lemma 2.3.2, entails $d \in D^{\mathfrak{U}}$. □

Lemma 2.3.6 . *$\mathcal{ALC}$-TBoxes are preserved under disjoint unions, i.e. for every $\mathcal{ALC}$-TBox $\mathcal{T}$ holds*

$$\forall i \in I : \mathfrak{I}_i \models \mathcal{T} \implies \biguplus_{i \in I} \mathfrak{I}_i \models \mathcal{T}$$

Proof. For every $d \in C^{\biguplus_i \mathfrak{I}_i}$ there is $i \in I$ such that $d \in \Delta^{\mathfrak{I}_i}$. Since $\mathfrak{I}_i$ can be recovered as generated subinterpretation of $\biguplus_{i \in I} \mathfrak{I}_i$ by setting $G := \Delta^{\mathfrak{I}_i}$, Lemma 2.3.2 yields $d \in C^{\mathfrak{I}_i}$. As $\mathfrak{I}_i \models \mathcal{T}$ by assumption, $d \in D^{\mathfrak{I}_i}$, whence again with 2.3.2 $d \in D^{\biguplus_i \mathfrak{I}_i}$ is inferred, which proves the claim. □

Proof of Observation 2.3.4. The if direction is given by Lemma 2.3.6. It remains to show that $\biguplus_{i \in I} \mathfrak{I}_i \models \mathcal{T}$ implies $\mathfrak{I}_i \models \mathcal{T}$ for all $i \in I$. Let $G := \Delta^{\mathfrak{I}_i}$. Then the generated subinterpretation of $\mathfrak{U}$ from $G$ in $\biguplus_{i \in I} \mathfrak{I}_i$ satisfies $\mathcal{T}$. But $\mathfrak{U}$ is identical to $\mathfrak{I}_i$, too. So $\mathfrak{I}_i \models \mathcal{T}$. □

### 2.3.2 $\mathcal{ALC}$-TBoxes as FO-Fragment

We can now give a characterisation for $\mathcal{ALC}$-TBoxes.

Theorem 2.3.7 . *Every sentence $\varphi \in \mathrm{FO}(\tau)$ which is invariant under global bisimulation and invariant under disjoint unions is equivalent to some $\mathcal{ALC}$-TBox $\mathcal{T}$ over $\tau$.*

Proof. It is sufficient to so show that there is a possibly infinite set cons $\varphi$ of $\mathcal{ALC}$-concept inclusions such that $\varphi$ logically equivalent to cons $\varphi$. As $\mathcal{ALC}u$ is a fragment of FO, it has compactness and therefore there is a finite subset $\mathcal{T} \subseteq$ cons $\varphi$ such that $\mathcal{T}$ is logically equivalent to $\varphi$.

Let cons $\varphi := \{C \sqsubseteq D \mid C, D \in \mathcal{ALC}(\tau) \text{ and } \forall \mathfrak{I} : \mathfrak{I} \models \varphi \implies \mathfrak{I} \models C \sqsubseteq D\}$. Clearly, $\varphi \models$ cons $\varphi$. Assume $\varphi$ is not logically equivalent to cons $\varphi$, i.e. there is $\mathfrak{I} \models$ cons $\varphi$ and $\mathfrak{I} \models \neg\varphi$. Let

$$T := \{p \subseteq \mathcal{ALC}(\tau) \mid \text{ there is } (\mathfrak{I}, d) \models p \text{ with } \mathfrak{I} \models \varphi\}$$

and for each $p \in T$ let $\mathfrak{I}_p$ be a model of $\varphi$ such that there is $d_p \in \Delta^{\mathfrak{I}_p}$ for which $(\mathfrak{I}_p, d_p) \models p$.

Let $\mathfrak{I} := \biguplus_{p \in T} \mathfrak{I}_p$. As $\varphi$ is invariant under disjoint unions $\mathfrak{I}$ is a model of $\varphi$. As $\varphi$ is not logically equivalent to cons $\varphi$ there is an interpretation $\mathfrak{H} \models$ cons $\varphi \cup \{\neg\varphi\}$. For $\mathfrak{K} := \mathfrak{I} \uplus \mathfrak{H}$, invariance under disjoint unions entails $\mathfrak{K} \models$ cons $\varphi \cup \{\neg\varphi\}$. Still for every $p \in T$ there is $d_p \in \Delta^{\mathfrak{K}}$ such that $(\mathfrak{K}, d_p) \models p$. As $\mathfrak{I}$ is a subinterpretation of $\mathfrak{H}$, $\mathfrak{I}$ is a generated subinterpretation of $\mathfrak{K}$ and therefore every $d \in \Delta^{\mathfrak{I}}$ is locally bisimilar to $d \in \Delta^{\mathfrak{K}}$.

Let now $d \in \Delta^{\mathfrak{K}} \cap \Delta^{\mathfrak{H}}$ and assume there is no element in $\mathfrak{I}$ that satisfies the type $p := \mathrm{Th}_{\mathcal{ALC}}(\mathfrak{K}, d)$. Then $p$ is not satisfiable with $\varphi$. Let $[p; x_0] := \{[C; x_0] \mid C \in p\}$ be the set of FO-formulae that contains all FO-rewritings of concepts in $p$. Since $\varphi$ cannot be satisfied together with $p$, the set $\{\varphi\} \cup [p; x_0]$ is unsatisfiable. Compactness of *FO* yields a finite subset $P(x_0) \subseteq [p; x_0]$ such that $P(x_0) \cup \{\varphi\}$ is unsatisfiable. All formulae in $P(x_0)$ can be re-translated into $\mathcal{ALC}$-concepts, whence we obtain $P \subseteq \mathcal{ALC}(\tau)$. In particular for every element $d$ in any model $\mathfrak{H}$ of $\varphi$ we have $d \notin (\bigsqcap P)^{\mathfrak{H}}$ or rather $(\bigsqcap P)^{\mathfrak{H}} \subseteq \emptyset$, which entails $\bigsqcap P \sqsubseteq \bot \in \mathrm{cons}\, \varphi$. Since $\mathfrak{K} \models \mathrm{cons}\, \varphi$, this contradicts the assumption there would be some element $d \in \Delta^{\mathfrak{K}}$ such that $d$ satisfies $p$.

Hence for every element in $\mathfrak{K}$ there is a counterpart in $\mathfrak{I}$ satisfying the same type. Let $\mathfrak{K}^*$ be a saturated extension of $\mathfrak{K}$ and $\mathfrak{I}^*$ be the saturated extension of $\mathfrak{I}$ such that still $\mathfrak{I}^* \models \varphi$ and $\mathfrak{K}^* \models \neg\varphi$. From the theorem by Hennessy and Milner 2.2.10. We obtain $\mathfrak{I}^* \overset{g}{\Longleftrightarrow} \mathfrak{K}^*$ which is absurd as $\mathfrak{I}^* \models \varphi$ and $\mathfrak{K}^* \models \neg\varphi$ yet $\varphi$ is supposed to be invariant under global bisimulation at the same time. This contradicts the assumption $\varphi$ is not logically equivalent to cons $\varphi$ and concludes the proof. □

This concludes the chapter for $\mathcal{ALC}$. We have seen characterisations of the expressiveness of $\mathcal{ALC}$-concepts, the $\mathcal{ALC}u$-concepts and finally a characterisation for $\mathcal{ALC}$-TBoxes. The expressiveness of $\mathcal{ALC}$ and $\mathcal{ALC}u$ is closely related to the capability of $\mathcal{ALC}$-bisimulation and $\mathcal{ALC}u$-bisimulation (global $\mathcal{ALC}$-bisimulation) to distinguish interpretations: Properties that get lost under tree-unravellings and forest-unravellings respectively cannot be expressed in $\mathcal{ALC}$ and $\mathcal{ALC}u$, respectively, which is the reason for the decidability [131, 63].

Examples for what cannot be expressed is most notably transitivity amongst others like the multiplicity of elements of a certain type or properties of predecessors of elements, as they all get lost under tree-unravellings.

The characterisation of TBoxes revealed that additionally no properties can be expressed which are not invariant under disjoint unions. Examples are disjunctions of TBoxes, like $\{\top \sqsubseteq A\} \sqcup \{\top \sqsubseteq B\}$, as the disjoint union of two interpretations $\mathfrak{I}_A$ where all elements satisfy $A$ but not $B$ and $\mathfrak{I}_B$ where all elements satisfy $B$ but not $A$ does not satisfy the TBox.

Similarly every property which requires the presence of an element of a specific type, e.g. $\exists A$, cannot be expressed: Let $\mathfrak{I}$ be an interpretation of two separate elements $d_A$ satisfying $A$ and $d$ not satisfying $A$. The generated sub-interpretation of $d$ will only contain $d$ and hence will not satisfy $A$ at any element.

# 3. The Characterisation of $\mathcal{ALCI}$ and $\mathcal{ALCQ}$

## 3.1 $\mathcal{ALCI}$

### 3.1.1 The Characterisation of $\mathcal{ALCI}$-Concepts

In this section we shall investigate an extension of $\mathcal{ALC}$, namely $\mathcal{ALCI}$. $\mathcal{ALCI}$ possess, unlike $\mathcal{ALC}$, the capability to not only express properties concerning successors but also properties of predecessors. This is achieved by introducing an additional quantification operator. In particular, from a given distinguished element, all elements in this element's connected component are now 'accessible' to the language. Still, we cannot directly express numbers of successors or the number of elements of a certain type.

To accommodate the increase in expressivity, a new model theoretic notion, *inverse bisimulation* is introduced. In analogy but without detailed explanation, we reassure ourselves that the new notion of bisimulation fits into the framework:

1. inverse bisimulation can be stratified and characteristic concepts exist
2. $\mathcal{ALCI}$-concepts have, together with inverse bisimulation, the Hennessy-and-Milner-Property
3. inverse bisimulation characterises $\mathcal{ALCI}$-concepts as fragment of FO.

We then carry on to extend $\mathcal{ALCI}$ to $\mathcal{ALCI}u$ for which we prove that the three items just given apply as well. The characterisation of $\mathcal{ALCI}$-TBoxes is then analogous to $\mathcal{ALC}$-TBoxes.

Being an extension and very similar in its model-theoretic behaviour to $\mathcal{ALC}$, $\mathcal{ALCI}$ is seldom treated separately in literature on its model-theory. As examples for $\mathcal{ALCI}$ on concept level not being treated separately, see [18, Chapter 7] but also [104]. This is clearly in contrast to literature on reasoning over $\mathcal{ALCI}$ where rewritings have been investigated which reduce in particular $\mathcal{ALCI}$-TBoxes to $\mathcal{ALC}$-TBoxes, whose size is polynomial in the size of the $\mathcal{ALCI}$-TBox. As Literature related to this approach can be named [41, 58] and later for $\mathcal{ALCI}$ [33] as well as [44] with another approach. The reason for this discrepancy is that reasoning over $\mathcal{ALCI}$-TBoxes in practice does not behave as well as $\mathcal{ALC}$-TBoxes, contrarily to what their model theoretic analogies suggest.

### Syntax and Semantic of $\mathcal{ALCI}$-Concepts

Each $\mathcal{ALCI}$-concept $C$ over $\tau$ can be constructed as follows:

$$C ::= \top \mid A \mid D \sqcap E \mid \neg D \mid \exists r.D \mid \exists r^{-}.D$$

where $A \in \mathsf{N_C}$, $r \in \mathsf{N_R}$ and $D, E$ are $\mathcal{ALCI}$-concepts. The set of $\mathcal{ALCI}$-concepts over $\tau$ is denoted by $\mathcal{ALCI}(\tau)$. The interpretation of these concepts follows exactly $\mathcal{ALC}$ (cf. the definition on page 25) and is simply extended for interpretations $\mathfrak{I}$ and elements $d \in \Delta^{\mathfrak{I}}$ by

$$d \in (\exists r^{-}.D)^{\mathfrak{I}} \text{ if there is } d' \in \Delta^{\mathfrak{I}} \text{ such that } (d', d) \in r^{\mathfrak{I}} \text{ and } d' \in D^{\mathfrak{I}}$$

Note that we do not treat $r^{-}$ as a separate symbol of our signature. Although the minus is annotated as superscript of $r$, we understand $\exists \cdot^{-}$ rather as a circumscript around the symbol $r$, which indicates that the exists quantification ranges over predecessors. We shall however allow $r^{-}$ to be treated like a separate signature symbol when explicitly stated.

Example 3.1.1 . Whilst $\mathcal{ALC}$ could only speak about $r$-successors, $\mathcal{ALCI}$ can now also talk about predecessors: 'Every $r$-predecessor satisfies $B$' can be expressed by $\forall r^{-}.B$ in $\mathcal{ALCI}$. But clearly, $\mathcal{ALCI}$ can still express every concept that could be expressed by $\mathcal{ALC}$.

For any two pointed structures $(\mathfrak{I}, d)$ and $(\mathfrak{H}, e)$ the *$\mathcal{ALCI}$-bisimulation* game $G(\mathfrak{I}, d; \mathfrak{H}, e)$ is played by two players **I** and **II**. The game starts in the configuration $(\mathfrak{I}, d; \mathfrak{H}, e)$, i.e. one of the two pebbles is placed on $d$ and the other one on $e$. For every configuration $(\mathfrak{I}, d'; \mathfrak{H}, e')$ during the game, the following rules apply:

1. For every $A \in \mathsf{N_C}$ we have $d' \in A^{\mathfrak{I}}$ iff $e' \in A^{\mathfrak{H}}$ or **II** has lost the game.

2. In every turn of the game, **I** can choose one of the structures, $\mathfrak{I}$ say, and some $r \in \mathsf{N_R}$ to challenge **II** by moving the pebble along an outgoing or incoming $r$-edge to some adjacent element $d''$. If player **I** cannot move then **II** has won the game.

3. **II** must move the pebble in the other structure, here $\mathfrak{H}$, from $e'$ along some $r$-edge in the same direction to some element $e''$. The same direction means, if **I** took an incoming(outgoing) edge labelled with $r \in \mathsf{N_R}$ from $d''$ to $d'$ then **II** must move along an incoming(outgoing) $r$-edge from $e'$ to some element $e''$. Thus the next configuration $(\mathfrak{I}, d''; \mathfrak{H}, e'')$ is reached and a new round begins. If **II** cannot move accordingly, she has lost the game.

In comparison to the $\mathcal{ALC}$ game, this game is extended by the possibility to move in the reverse direction of the arrows: if the game has reached the configuration $(\mathfrak{I}, d'; \mathfrak{H}, e')$ and $(d'', d') \in r^{\mathfrak{I}}$ then **I** can challenge **II** by moving from $d'$ to $d''$.

Definition 3.1.2. Two pointed structures $(\mathfrak{I}, d)$ and $(\mathfrak{H}, e)$ are *$\mathcal{ALCI}$-bisimilar*, $(\mathfrak{I}, d) \xleftrightarrow{i} (\mathfrak{H}, e)$, if **II** has a winning strategy in the $\mathcal{ALCI}$-bisimulation game $G(\mathfrak{I}, d; \mathfrak{H}, e)$, i.e. if **II** can respond to any challenge of **I** such that she does not lose any round. ◊

Observation 3.1.3.

1. *$\xleftrightarrow{i}$ is an equivalence relation on the set of pointed interpretations*

2. *Being $\mathcal{ALCI}$-bisimilar implies being $\mathcal{ALC}$-bisimilar.*

Again the notion of $\mathcal{ALCI}$-bisimilarity can be graded into $\mathcal{ALCI}$-$n$-bisimilarity where two pointed structures $(\mathfrak{I}, d)$, $(\mathfrak{H}, e)$ are $\mathcal{ALCI}$-$n$-bisimilar, $(\mathfrak{I}, d) \xleftrightarrow{i}_n (\mathfrak{H}, e)$, if **II** can play in the bisimulation game $G(\mathfrak{I}, d; \mathfrak{H}, e)$ for at least $n$-rounds.

Observation 3.1.4. *For all $n < \omega$ we have*

1. *$\xleftrightarrow{i}_n$ is an equivalence relation on the set of pointed interpretations*

2. *If $(\mathfrak{I}, d) \xleftrightarrow{i}_{n+1} (\mathfrak{H}, e)$ then $(\mathfrak{I}, d) \xleftrightarrow{i}_n (\mathfrak{H}, e)$*

3. *If $(\mathfrak{I}, d) \xleftrightarrow{i} (\mathfrak{H}, e)$ then $(\mathfrak{I}, d) \xleftrightarrow{i}_n (\mathfrak{H}, e)$*

4. *If $(\mathfrak{I}, d) \xleftrightarrow{i}_n (\mathfrak{H}, e)$ then $(\mathfrak{I}, d) \Longleftrightarrow_n (\mathfrak{H}, e)$*

Like for $\mathcal{ALC}$-bisimulation, $(\mathfrak{I}, d) \xleftrightarrow{i}_n (\mathfrak{H}, e)$ for all $n < \omega$ does not imply $(\mathfrak{I}, d) \xleftrightarrow{i} (\mathfrak{H}, e)$ nor does it imply $(\mathfrak{I}, d) \Longleftrightarrow (\mathfrak{H}, e)$. As counterexample for such an implication can serve the Counterexample 2.1.20 given for $\mathcal{ALC}$.

Speaking of tree-unravellings, we can immediately apply the Definition 2.1.4 for tree-unravellings to $\mathcal{ALCI}$, by treating $r^{-}$ as separate role names in the signature. Except for isolated elements, this leads always to infinite tree-unravellings. The analog of Observation 2.1.5 stating $(\mathfrak{I}, d) \xleftrightarrow{i} (\mathfrak{I}_d, d)$ is readily shown.

The rank -function can be continued from $\mathcal{ALC}$-concepts onto $\mathcal{ALCI}$-concepts by setting $\operatorname{rank} \exists r^{-}.D := 1 + \operatorname{rank} D$ for all $r \in \mathsf{N_R}$ and every $\mathcal{ALCI}$-concept $D$. A proof by induction upon the rank of $\mathcal{ALCI}$-concepts then shows for all $\mathcal{ALCI}$-concepts $D$:

$$\text{if } (\mathfrak{I}, d) \xleftrightarrow{i}_n (\mathfrak{H}, e) \text{ and } \operatorname{rank} D = n \text{ then } d \in D^{\mathfrak{I}} \text{ iff } e \in D^{\mathfrak{H}}.$$

In case the signature is finite, characteristic concepts can be constructed that allow to describe $\mathcal{ALCI}$ $n$-bisimilarity types: Let $(\mathfrak{I}, d)$ be an interpretation and $\tau := \mathsf{N_R} \cup \mathsf{N_C}$ a finite signature. The *characteristic $\mathcal{ALCI}$-concept* for each level $n < \omega$ are recursively defined as

$$X^0_{\mathfrak{I},d} := \bigsqcap\{A \in \mathsf{N_C} \mid d \in A^{\mathfrak{I}}\} \sqcap \bigsqcap\{\neg A \mid A \in \mathsf{N_C} \text{ and } d \notin A^{\mathfrak{I}}\}$$

$$\begin{aligned} X^{n+1}_{\mathfrak{I},d} := X^0_{\mathfrak{I},d} \sqcap \textstyle\bigsqcap_{r \in \mathsf{N_R}} \bigsqcap\{\exists r.X^n_{\mathfrak{I},d'} \mid (d,d') \in r^{\mathfrak{I}}\} \ \sqcap \forall r. \bigsqcup\{X^n_{\mathfrak{I},d'} \mid (d,d') \in r^{\mathfrak{I}}\} \\ \sqcap \bigsqcap\{\exists r^{-}.X^n_{\mathfrak{I},d'} \mid (d',d) \in r^{\mathfrak{I}}\} \sqcap \forall r^{-}. \bigsqcup\{X^n_{\mathfrak{I},d'} \mid (d',d) \in r^{\mathfrak{I}}\} \end{aligned}$$

Again, these concepts exist in $\mathcal{ALCI}$ because $\tau$ is finite, and therefore on every level there are only finitely many different characteristic concepts:

Observation 3.1.5. *Let the signature $\tau := \mathsf{N_R} \cup \mathsf{N_C}$ be finite and $(\mathfrak{I}, d)$ and $(\mathfrak{H}, e)$ be $\tau$-interpretations. The following statements are equivalent for all characteristic $\mathcal{ALCI}$-concepts*

1. $(\mathfrak{H}, e) \models X^n_{\mathfrak{I},d}$
2. $(\mathfrak{I}, d) \xleftrightarrow{i}_n (\mathfrak{H}, e)$
3. $\mathit{Th}_n(\mathfrak{I}, d) = \mathit{Th}_n(\mathfrak{H}, e)$

*where $\mathit{Th}_n(\mathfrak{I}, d) := \{C \in \mathcal{ALCI}(\tau) \mid d \in C^{\mathfrak{I}}$ and $\operatorname{rank} C \leq n\}$ is the $\mathcal{ALCI}$-theory of $(\mathfrak{I}, d)$.*

Proof. 1. $\Longrightarrow$ 2. The proof is similar to the proof of Proposition 2.1.16 which is the analogous statement for $\mathcal{ALC}$. It can be carried out via induction upon $n$ and differs to the induction step of Proposition 2.1.16 in that **I** can also challenge **II** by moving in reverse direction which is analog to the cases where **I** moves in forward direction. Nevertheless, we shall shortly discuss the case where **I** challenges **II** in $\mathfrak{I}$ by moving in reverse direction from $d$ along some $r$-edge to $d'$. Then $(\mathfrak{H}, e)$ satisfies the subconcept $\exists r^{-}.X^{n}_{\mathfrak{I},d'}$ of $X^{n+1}_{\mathfrak{I},d}$ and so there is some $e' \in \Delta^{\mathfrak{H}}$ with $(e', e) \in r^{\mathfrak{H}}$ such that $(\mathfrak{H}, e') \models X^{n}_{\mathfrak{I},d'}$. The induction hypothesis holds which yields that **II** has a winning strategy for $n$-further rounds, i.e. in total $n+1$ rounds.

2. $\Longrightarrow$ 3. has been informally stated earlier on. The proof is analogous to Proposition 2.1.8 which states the same claim for $\mathcal{ALC}$.

For 3. $\Longrightarrow$ 1. it is easy to see that $X^{n}_{\mathfrak{I},d} \in \mathrm{Th}_n(\mathfrak{I}, d)$. Hence $(\mathfrak{H}, e) \models X^{n}_{\mathfrak{I},d}$.

□

### Saturation and the Hennessy-Milner-Property for $\mathcal{ALCI}$

As in earlier sections, after exploring properties and model-theoretic notions directly connected with the logic in question, we are now gathering the necessary ingredients for a characterisation of $\mathcal{ALCI}$. An adapted notion of saturation shall be established and the Hennessy-Milner-Property with respect to this saturation will be proven.

Definition 3.1.6 . $(\mathfrak{I}, d) \xleftrightarrow{i}_{\omega} (\mathfrak{H}, e)$ if for all $n < \omega$ we have $(\mathfrak{I}, d) \xleftrightarrow{i}_{n} (\mathfrak{H}, e)$. ◇

We have to adapt and extend the definition of types that has been made in definition 2.1.21.

Definition 3.1.7 . Let $\tau$ be a signature and $(\mathfrak{I}, d)$ a pointed interpretation. Then for each $r \in \mathsf{N_R}$ we call

1. $\Gamma \subseteq \mathcal{ALCI}(\tau)$ an $r$-type of $d$ if for every finite subset $\Gamma_0 \subseteq \Gamma$ we have $d \in \exists r. \bigsqcap \Gamma_0$.
2. $\Gamma \subseteq \mathcal{ALCI}(\tau)$ an $r^{-}$-type of $d$ if for every finite subset $\Gamma_0 \subseteq \Gamma$ we have $d \in \exists r^{-}. \bigsqcap \Gamma_0$.
3. an $r$-type of $d$ realised at $d$ if there is an $r$-successor $d'$ in $\mathfrak{I}$ such that $d' \in \bigcap_{C \in \Gamma} C^{\mathfrak{I}}$
4. an $r^{-}$-type of $d$ realised at $d$ if there is an $r$-predecessor $d'$ in $\mathfrak{I}$ such that $d' \in \bigcap_{C \in \Gamma} C^{\mathfrak{I}}$

◇

Definition 3.1.8 . An interpretation $\mathfrak{I}$ is saturated if for all $d \in \Delta^{\mathfrak{I}}$ and all $r \in \mathsf{N_R}$ every $r$-type and every $r^{-}$-type of $d$ is realised at $d$. ◇

Let $\mathrm{Th}(\mathfrak{I}, d) := \{C \in \mathcal{ALCI}(\tau) \mid (\mathfrak{I}, d) \models C\}$ for every pointed $\tau$-interpretation $(\mathfrak{I}, d)$.

Proposition 3.1.9 . *$\mathcal{ALCI}$ has the Hennessy-Milner-Property, i.e. for saturated interpretations $(\mathfrak{I}, d)$ and $(\mathfrak{H}, e)$ we have $\mathrm{Th}(\mathfrak{I}, d) = \mathrm{Th}(\mathfrak{H}, e)$ iff $(\mathfrak{I}, d) \xleftrightarrow{i} (\mathfrak{H}, e)$.*

Proof. The proof follows the scheme we used to show the analogue result for $\mathcal{ALC}$ in Proposition 2.1.23. We show that for every round in the $\mathcal{ALCI}$-bisimulation game $G(\mathfrak{I}, d; \mathfrak{H}, e)$ **II** can a achieve configuration $(\mathfrak{I}, d'; \mathfrak{H}, e')$ such that $\mathrm{Th}(\mathfrak{I}, d') = \mathrm{Th}(\mathfrak{H}, e')$.

We thus show that the relation $\mathrm{Th}(\mathfrak{I}, d') = \mathrm{Th}(\mathfrak{H}, e')$ satisfies atomic equivalence and is closed under the forth- and back-property. As we can use the argument symmetrically for all cases we shall show exactly one case: Assume the game has reached the configuration $(\mathfrak{I}, d'; \mathfrak{H}, e')$ and **I** challenges **II** in $\mathfrak{I}$ by moving from $d'$ via an incoming edge labelled with $r \in \mathsf{N_R}$ to $d''$. As $e'$ and $d'$ has the same theory, for every finite subset $\Gamma_0 \subseteq \mathrm{Th}(\mathfrak{I}, d'')$ we have $d' \in (\exists r^-. \bigsqcap \Gamma_0)^{\mathfrak{I}}$ and therefore $e \in (\exists r^-. \bigsqcap \Gamma_0)^{\mathfrak{H}}$. Hence $\mathrm{Th}(\mathfrak{I}, d'')$ forms an $r^-$-type at $e'$ and since $\mathfrak{H}$ is saturated, there is an $r$-predecessor $e''$ of $e'$ that satisfies $\mathrm{Th}(\mathfrak{I}, d'')$. **II** can choose this $e''$ and thus reaches a new configuration $(\mathfrak{I}, d''; \mathfrak{H}, e'')$ that satisfies $\mathrm{Th}(\mathfrak{I}, d'') = \mathrm{Th}(\mathfrak{H}, e'')$. □

## $\mathcal{ALCI}$ as $\mathcal{ALCI}$-Bisimulation Invariant Fragment of FO

Proposition 3.1.10 . *Every* FO *formula that is invariant under $\xleftrightarrow{i}$ is invariant under $\xleftrightarrow{i}_n$ for some $n < \omega$.*

Proof. The proof uses the same construction as in Lemma 2.1.25; The only difference is that we now refer to characteristic $\mathcal{ALCI}$-concepts instead of characteristic $\mathcal{ALC}$-concepts:

The construction yields, if the claim of this proposition would not be true, an FO-formula $\varphi(x)$ and two interpretations $(\mathfrak{I}, d)$ and $(\mathfrak{H}, e)$ would exist, which share the same $\mathcal{ALCI}$-theory, yet $(\mathfrak{I}, d) \models \varphi(x)$ and $(\mathfrak{H}, e) \not\models \varphi(x)$.

For each of them we can find a saturated extension $(\mathfrak{I}^*, d) \models \varphi(x)$, and $(\mathfrak{H}^*, e) \not\models \varphi(x)$ respectively, which has the same FO-theory and hence the same $\mathcal{ALCI}$-theory as $(\mathfrak{I}, d)$, and $(\mathfrak{H}, e)$ respectively. We can use the Hennessy-Milner-Property of $\mathcal{ALCI}$ stated in Proposition 3.1.9 to argue that $(\mathfrak{I}^*, d) \xleftrightarrow{i} (\mathfrak{H}^*, e)$. This is a contradiction of $\varphi(x)$ being invariant under $\xleftrightarrow{i}$ which proves the proposition. □

With this lemma we can characterise $\mathcal{ALCI}$ as fragment of FO in the following way:

THEOREM 3.1.11 . *$\mathcal{ALCI}$ is the $\overset{i}{\longleftrightarrow}$-invariant fragment of* FO

PROOF. Let $\varphi(x)$ be an FO-formula over some signature $\tau := \mathsf{N_C} \cup \mathsf{N_R}$ such that $\varphi(x)$ is invariant under $\overset{i}{\longleftrightarrow}$. We have to show that there is a logically equivalent $\mathcal{ALCI}$-concept $C_\varphi$.

W.l.o.g. we assume that $\tau$ is finite and Proposition 3.1.10 yields some $n < \omega$ such that $\varphi(x)$ is invariant under $\overset{i}{\longleftrightarrow}_n$. Let $X^n_{\mathfrak{I},d}$ denote the characteristic $\mathcal{ALCI}$-concept for $(\mathfrak{I}, d)$ on level $n$ for $\tau$. Then $C_\varphi := \bigsqcup\{X^n_{\mathfrak{I},d} \mid (\mathfrak{I}, d) \models \varphi(x)\}$. As mentioned below, the definition of characteristic $\mathcal{ALCI}$-concepts, there are only finitely many such concepts on each level $n$ for $\tau$. Hence the disjunction $C_\varphi$ is an $\mathcal{ALCI}$-concept and $\varphi(x) \models C_\varphi$. Invariance under $\overset{i}{\longleftrightarrow}_n$ also yields $\bigsqcup C_\varphi \models \varphi(x)$ and thus $\varphi(x)$ is logically equivalent to $C_\varphi$. □

### 3.1.2 The Characterisation of $\mathcal{ALCI}u$-Concepts

Similar to $\mathcal{ALC}$ we extend $\mathcal{ALCI}$ by the universal role $u$ with $u^\mathfrak{I} = \Delta^\mathfrak{I} \times \Delta^\mathfrak{I}$ for all interpretations $\mathfrak{I}$. Although $u$ is a logical constant in $\mathcal{ALCI}u$ and not a symbol of the signature, it can be treated like any other role-name. Note that it would not make sense to introduce the inverse role $u^-$ for $u$, as all elements are connected by 'forward' $u$-arrows anyway. Hence any concept $\exists u^-.C$ is logically equivalent to $\exists u.C$ and so no expressivity would be added. We thus define $\mathcal{ALCI}u$-concepts as follows:

$$C ::= \top \mid A \mid D \sqcap E \mid \neg D \mid \exists r.D \mid \exists r^-.D \mid \exists u.D$$

where $A \in \mathsf{N_C}$, $D, E$ are $\mathcal{ALCI}u$-concepts and $r \in \mathsf{N_R}$. The usual abbreviations $\forall, \sqcup, \rightarrow$ etc. apply. We can extend the rank-function from $\mathcal{ALCI}$-concepts to $\mathcal{ALCI}u$-concepts by simply ignoring universal roles, thus setting $\operatorname{rank} \exists u.C := \operatorname{rank} C$.

DEFINITION 3.1.12 . Interpretations $\mathfrak{I}$ and $\mathfrak{H}$ are globally $\mathcal{ALCI}$-bisimilar, $\mathfrak{I} \overset{gi}{\longleftrightarrow} \mathfrak{H}$ if both of the following conditions are met

1. for all $d \in \Delta^\mathfrak{I}$ there is $e \in \Delta^\mathfrak{H}$ with $(\mathfrak{I}, d) \overset{i}{\longleftrightarrow} (\mathfrak{H}, e)$
2. for all $e \in \Delta^\mathfrak{H}$ there is $d \in \Delta^\mathfrak{I}$ with $(\mathfrak{I}, d) \overset{i}{\longleftrightarrow} (\mathfrak{H}, e)$

We write $(\mathfrak{I}, d) \overset{gi}{\longleftrightarrow} (\mathfrak{H}, e)$ if $\mathfrak{I} \overset{gi}{\longleftrightarrow} \mathfrak{H}$ and $(\mathfrak{I}, d) \overset{i}{\longleftrightarrow} (\mathfrak{H}, e)$. ◇

This notion for global $\mathcal{ALCI}$-bisimulation is analogous to the global $\mathcal{ALC}$-bisimulation given in Definition 2.2.4. Again, the global notion can be stratified with respect to $n < \omega$ into $\xleftrightarrow{gi}_n$ with the obvious definition. The following chain of implications hold for all $n < \omega$:

$$(\mathfrak{I}, d) \xleftrightarrow{gi} (\mathfrak{H}, e) \implies (\mathfrak{I}, d) \xleftrightarrow{gi}_{n+1} (\mathfrak{H}, e) \implies (\mathfrak{I}, d) \xleftrightarrow{gi}_n (\mathfrak{H}, e)$$

Definition 3.1.13. Characteristic $\mathcal{ALCI}u$-concepts are defined for a finite signature $\tau := \mathsf{N_C} \cup \mathsf{N_R}$ and an interpretation $\mathfrak{I}$ as follows:

$$X^n_{\mathfrak{I}} := \bigsqcap\{\exists u.X^n_{\mathfrak{I},d} \mid d \in \Delta^{\mathfrak{I}}\} \sqcap \forall u.\bigsqcup\{X^n_{\mathfrak{I},d} \mid d \in \Delta^{\mathfrak{I}}\}$$

where $X^n_{\mathfrak{I},d}$ is the characteristic $\mathcal{ALCI}$-concept for $\tau$. ◇

We shall give a short explanation. Characteristic $\mathcal{ALCI}u$-concepts are composed of two parts: One requires that the interpretation which satisfies $X^n_{\mathfrak{I}}$ provides elements for each $n$-$\mathcal{ALCI}$-bisimulation type realised in $\mathfrak{I}$. So if $\mathfrak{H} \models X^n_{\mathfrak{I}}$, then for all $d \in \Delta^{\mathfrak{I}}$ there must be $e \in \Delta^{\mathfrak{H}}$ such that $(\mathfrak{I}, d) \xleftrightarrow{i}_n (\mathfrak{H}, e)$. The second part restricts the occurrence of $n$-$\mathcal{ALCI}$-bisimulation types. Every interpretation $\mathfrak{H} \models X^n_{\mathfrak{I}}$ can only realise $n$-$\mathcal{ALCI}$-bisimulation types which are present in $\mathfrak{I}$. Hence for all $e \in \Delta^{\mathfrak{H}}$ there is $d \in \Delta^{\mathfrak{I}}$ such that $(\mathfrak{I}, d) \xleftrightarrow{i}_n (\mathfrak{H}, e)$. So we have that $\mathfrak{I} \xleftrightarrow{gi}_n \mathfrak{H}$.

Observation 3.1.14. *The following is equivalent for all $n < \omega$ and finite signature $\tau$:*

1. $(\mathfrak{H}, e) \models X^n_{\mathfrak{I}} \sqcap X^n_{\mathfrak{I},d}$
2. $(\mathfrak{I}, d) \xleftrightarrow{gi}_n (\mathfrak{H}, e)$
3. $\mathit{Th}_n(\mathfrak{I}, d) = \mathit{Th}_n(\mathfrak{H}, e)$

*where $\mathfrak{I}, \mathfrak{H}$ are $\tau$-interpretations, $X^n_{\mathfrak{I}}$ is the characteristic $\mathcal{ALCI}u$-concept for $\tau$, $X^n_{\mathfrak{I},d}$ is the characteristic $\mathcal{ALCI}$-concept for $\tau$ and $\mathit{Th}_n(\mathfrak{I}) := \{C \in \mathcal{ALCI}u(\tau) \mid (\mathfrak{I}, d) \models C$ and $\operatorname{rank} C \leq n\}$*

This Observation is a global version of the anlogous Observation 3.1.5. Since we omitted this statement for $\mathcal{ALC}u$ we shall give a proof in full length, yet the reader might skip it as it is analog to the proof in Observation 3.1.5 and simply deals with the extra arguments for global $\mathcal{ALCI}$-bismilarity.

Proof. For 1. $\Longrightarrow$ 2. From Observation 3.1.5 we obtain $(\mathfrak{I}, d) \overset{gi}{\longleftrightarrow}_n (\mathfrak{H}, e)$ immediately. We now show $\mathfrak{I} \overset{gi}{\longleftrightarrow}_n \mathfrak{H}$ Let $d_0 \in \Delta^{\mathfrak{I}}$ be arbitrary. Then $\exists u.X^n_{\mathfrak{I},d_0}$ is amongst the conjuncts of $X^n_{\mathfrak{I}}$. As $\mathfrak{H} \models X^n_{\mathfrak{I}}$ there is an element $e_0 \in \Delta^{\mathfrak{H}}$ such that $(\mathfrak{H}, e_0) \models X^n_{\mathfrak{I},d_0}$. From Observation 3.1.5 we infer that $(\mathfrak{I}, d_0) \overset{i}{\longleftrightarrow}_n (\mathfrak{H}, e_0)$.

Conversely, let $e_0 \in \Delta^{\mathfrak{H}}$ arbitrary. As $\mathfrak{H} \models X^n_{\mathfrak{I}}$, there must be a disjunct in $\forall u.\{X^n_{\mathfrak{I},d_0} \mid d_0 \in \Delta^{\mathfrak{I}}\}$ such that $(\mathfrak{H}, e_0) \models X^n_{\mathfrak{I},d_0}$. Hence Observation 3.1.5 implies $(\mathfrak{I}, d_0) \overset{i}{\longleftrightarrow}_n (\mathfrak{H}, e_0)$. Both arguments together show $\mathfrak{I} \overset{gi}{\longleftrightarrow}_n \mathfrak{H}$ and with $(\mathfrak{I}, d) \overset{i}{\longleftrightarrow}_n (\mathfrak{H}, e)$ we obtain $(\mathfrak{I}, d) \overset{gi}{\longleftrightarrow}_n (\mathfrak{H}, e)$.

For 2. $\Longrightarrow$ 3. let $(\mathfrak{I}, d) \overset{gi}{\longleftrightarrow}_n (\mathfrak{H}, e)$. We show by induction upon the construction of concepts $C \in \mathcal{ALCI}u(\tau)$ with $\operatorname{rank} C \leq n$ that $(\mathfrak{I}, d) \models C$ iff $(\mathfrak{H}, e) \models C$. The claim is true for atomic concepts as in particular $(\mathfrak{I}, d) \overset{i}{\longleftrightarrow}_n (\mathfrak{H}, e)$. Conjunctions and negations are immediately yield by the induction hypothesis.

In case $C = \exists u.D$ and $\mathfrak{I} \models C$, there is $d_0 \in \Delta^{\mathfrak{I}}$ such that $(\mathfrak{I}, d_0) \models D$. Since $\mathfrak{I} \overset{gi}{\longleftrightarrow}_n \mathfrak{H}$ there is $e_0 \in \Delta^{\mathfrak{H}}$ such that $(\mathfrak{I}, d_0) \overset{i}{\longleftrightarrow}_n (\mathfrak{H}, e_0)$. Still $\mathfrak{I} \overset{gi}{\longleftrightarrow}_n \mathfrak{H}$ and thus $(\mathfrak{I}, d_0) \overset{gi}{\longleftrightarrow}_n (\mathfrak{H}, e_0)$. The induction hypothesis yields that $(\mathfrak{H}, e_0) \models D$ and so $\mathfrak{H} \models \exists u.D$. The converse can be shown with the same rationale.

In case $C = \exists r^-.D$ and $(\mathfrak{I}, d) \models C$, we have $1 \leq \operatorname{rank} C \leq n$ and there is an $r$-predecessor $d_0$ of $d$ such that $(\mathfrak{I}, d_0) \models D$. As $(\mathfrak{I}, d) \overset{i}{\longleftrightarrow}_n (\mathfrak{H}, e)$ there is an $r$-predecessor of $e_0$ such that $(\mathfrak{I}, d_0) \overset{i}{\longleftrightarrow}_{n-1} (\mathfrak{H}, e_0)$. In particular, $\mathfrak{I} \overset{gi}{\longleftrightarrow}_n \mathfrak{H}$ implies $\mathfrak{I} \overset{gi}{\longleftrightarrow}_{n-1} \mathfrak{H}$, hence $(\mathfrak{I}, d_0) \overset{gi}{\longleftrightarrow}_{n-1} (\mathfrak{H}, e_0)$. Since $\operatorname{rank} D \leq n-1$ the induction hypothesis applies and $(\mathfrak{H}, e_0) \models D$. Hence $(\mathfrak{H}, e) \models \exists r^-.D$. The converse can be argued in the same way. The case where $C = \exists r.D$ is explained analogously.

3. $\Longrightarrow$ 1. is immediate, as $\operatorname{rank} X^n_{\mathfrak{I}} \leq n$ and $\operatorname{rank} X^n_{\mathfrak{I},d} \leq n$. □

Since we can treat $u$ like a role name, we extend the notion of $\mathcal{ALCI}$-saturation (cf. Definition 3.1.7) to $\mathcal{ALCI}u$-saturation by simply additionally requiring that all $\mathcal{ALCI}$-$u$-types must be realised as well. Types, though, are still subsets of $\mathcal{ALCI}$. This is analogous to the definition of $\mathcal{ALC}$-$u$-types in Definition 2.2.8.

Let $\operatorname{Th}(\mathfrak{I}, d) := \{C \in \mathcal{ALCI}u(\tau) \mid (\mathfrak{I}, d) \models C\}$ for every pointed $\tau$-interpretation $(\mathfrak{I}, d)$.

Proposition 3.1.15 . *$\mathcal{ALCI}u$ has the Hennessy-Milner-Property, i.e. for two saturated $\mathcal{ALCI}u$-interpretations $(\mathfrak{I}, d)$ and $(\mathfrak{H}, e)$ we have*

$$\mathit{Th}(\mathfrak{I}, d) = \mathit{Th}(\mathfrak{H}, e) \iff (\mathfrak{I}, d) \overset{gi}{\longleftrightarrow} (\mathfrak{H}, e).$$

Proof. As $\mathfrak{I}$ and $\mathfrak{H}$ are $\mathcal{ALCI}$-saturated, the Hennessy-Milner-Property of $\mathcal{ALCI}$ yields $(\mathfrak{I}, d) \xleftrightarrow{i} (\mathfrak{H}, e)$. For each $d_0 \in \Delta^{\mathfrak{I}}$ and every finite subset $\Gamma_0 \subseteq \mathrm{Th}(\mathfrak{I}, d_0)$ of its $\mathcal{ALCI}$-theory we have $\mathfrak{I} \models \exists u. \bigsqcap \Gamma_0$. As $\mathfrak{I}$ satisfies exactly the same $\mathcal{ALCI}u$-concepts as $\mathfrak{H}$, we have $\mathfrak{H} \models \exists u. \bigsqcap \Gamma_0$ and thus the $\mathcal{ALCI}$-theory of $d_0$ forms a $u$-type in $\mathfrak{H}$. As $\mathfrak{H}$ is $\mathcal{ALCI}u$-saturated there is an $e_0$ that satisfies this type. The Hennessy-Milner-Property of $\mathcal{ALCI}$ yields $(\mathfrak{I}, d_0) \xleftrightarrow{i} (\mathfrak{H}, e_0)$. With the same arguments it is shown that for every $e_0 \in \Delta^{\mathfrak{H}}$ there is some $d_0 \in \Delta^{\mathfrak{I}}$ such that $(\mathfrak{I}, d_0) \xleftrightarrow{i} (\mathfrak{H}, e_0)$ which proves $\mathfrak{I} \xleftrightarrow{gi} \mathfrak{H}$ and thus $(\mathfrak{I}, d) \xleftrightarrow{gi} (\mathfrak{H}, e)$. □

Proposition 3.1.16. *Let $\varphi(x)$ be an* FO*-formula that is invariant under $\xleftrightarrow{gi}$ then $\varphi(x)$ is invariant under $\xleftrightarrow{gi}_n$ for some $n < \omega$.*

Proof. Like in the proof of Proposition 3.1.10 the proof rests on an adaption of Lemma 2.1.25: Assume this would not be the case then for every $n < \omega$ we have $(\mathfrak{I}, d) \Longleftrightarrow_n (\mathfrak{H}, e)$ yet $\mathfrak{I}\frac{d}{x} \models \varphi(x)$ and $\mathfrak{H}\frac{e}{x} \not\models \varphi(x)$.

$$\mathfrak{X}^n := \{X^n_{\mathfrak{I}} \sqcap X^n_{\mathfrak{I},d} \mid \exists(\mathfrak{H}, e) : \varphi(x) \dashv (\mathfrak{I}, d) \xleftrightarrow{gi}_n (\mathfrak{H}, e) \not\models \varphi(x)\}$$

where $X^n_{\mathfrak{I}} \sqcap X^n_{\mathfrak{I},d}$ is the conjunction of the characteristic $\mathcal{ALCI}u$-concept and characteristic $\mathcal{ALCI}$-concept over the signature $\tau$ which contains exactly the symbols used in $\varphi$ and thus is finite. As mentioned before, on every level $n < \omega$ there are only finitely many characteristic $\mathcal{ALCI}u$- and $\mathcal{ALCI}$-concepts, so $\mathfrak{X}^n$ is finite.

Every finite subset of $\mathfrak{X} := \{\bigsqcup \mathfrak{X}^n \mid n < \omega\}$ is satisfiable with $\varphi(x)$ so that the compactness of FO yields the existence of some model $(\mathfrak{K}, d)$ for $\mathfrak{X} \cup \{\varphi(x)\}$. In turn $\{X^n_{\mathfrak{K}} \sqcap X^n_{\mathfrak{K},d} \mid n < \omega\}$ is satisfiable with $\neg\varphi(x)$ as $(\mathfrak{K}, d) \models \bigsqcup \mathfrak{X}^n$ for all $n < \omega$.

Hence there are two pointed interpretations with the same $\mathcal{ALCI}u$-theory distinguished by $\varphi(x)$. W.l.o.g. we may assume that these interpretations are $\mathcal{ALCI}u$-saturated since for both of them $\omega$-saturated extensions exist which share their $\mathcal{ALCI}u$-theory yet are still distinguished by $\varphi(x)$. The Hennessy-Milner-Property of $\mathcal{ALCI}u$ yields that these $\omega$-saturated extensions are globally $\mathcal{ALCI}$-bisimilar. But this contradicts that they are distinguished by some globally $\mathcal{ALCI}$-bisimulation invariant FO-formula and thus proves the claim. □

Theorem 3.1.17. *Every* FO*-formula over $\tau$ which is invariant under global $\mathcal{ALCI}$-bisimulation is equivalent to some $\mathcal{ALCI}u$-concept.*

Proof. Let $\varphi(x) \in \mathrm{FO}(\tau)$. As FO has the finite occurrence property, i.e. every FO-formula contains only finitely many signature symbols, we may assume that $\tau$ is finite. Let $n < \omega$ be the natural number for which $\varphi(x)$ is invariant under $\xleftrightarrow{gi}_n$. Then $\varphi(x)$ is logically equivalent to the disjunction of $C_\varphi := \{X^n_{\mathfrak{I}} \sqcap X^n_{\mathfrak{I},d} \mid (\mathfrak{I}, d) \models \varphi(x)\}$. Since there are only finitely many different characteristic formulae on each level $n$, $\bigsqcup C_\varphi$ is a concept in $\mathcal{ALCI}u$. □

We say $C \in \mathcal{ALCI}u(\tau)$ is a *global* $\mathcal{ALCI}u$*-concept* if $C \equiv \bigsqcup\{X^n_{\mathfrak{I}} \mid (\mathfrak{I}, d) \models C\}$. A global concept is therefore analogous to the notion of a sentence in FO.

Corollary 3.1.18. *Every* FO*-sentence that is invariant under global* $\mathcal{ALCI}$*-bisimulation is logically equivalent to some global* $\mathcal{ALCI}u$*-concept.*

### 3.1.3 The Characterisation of $\mathcal{ALCI}$-TBoxes

In this section, the notion of $\mathcal{ALCI}$-TBoxes is to be introduced and then the important property, being invariant under disjoint unions, shall be shown. A characterisation is then immediate.

Proposition 3.1.19. *Every global* $\mathcal{ALCI}u$*-concept is equivalent to some boolean combination of* $\mathcal{ALCI}$*-TBoxes.*

Proof. Every characteristic $\mathcal{ALCI}u$-concept $X^n_{\mathfrak{I}} = \bigsqcap\{\exists u.X^n_{\mathfrak{I},d} \mid d \in \Delta^{\mathfrak{I}}\} \sqcap \forall u. \bigsqcup\{X^n_{\mathfrak{I},d} \mid d \in \Delta^{\mathfrak{I}}\}$ can be rewritten as the following boolean combination of $\mathcal{ALCI}u$-TBoxes:

$$\bigsqcap_{d \in \Delta^{\mathfrak{I}}} \neg\{X^n_{\mathfrak{I},d} \sqsubseteq \bot\} \sqcap \{\top \sqsubseteq \bigsqcup_{d \in \Delta^{\mathfrak{I}}} X^n_{\mathfrak{I},d}\}$$

As global $\mathcal{ALCI}u$-concepts are equivalent to disjunctions of characteristic $\mathcal{ALCI}u$-concepts, this proves the claim. □

We may use the exact same notion of disjoint union and generated substructure introduced after in Section 2.3.1 on page 58. Note that we defined a substructure to be the union over the (full) connected component of each element in the generator set $G$. Although, e.g., the union of all tree-unravellings in elements of $G$ would have sufficed for $\mathcal{ALC}$, we decided for one notion which carries through all the description logics we treat with mild adaptions when nominals are involved.

We formulate analogously to Lemma 2.3.2

Lemma 3.1.20. *Let* $\mathfrak{U}$ *be a generated substructure of* $\mathfrak{I}$*. For all* $d \in \Delta^{\mathfrak{U}}$ *we have* $(\mathfrak{U}, d) \xleftrightarrow{i} (\mathfrak{I}, d)$

Proof. II has a winning-strategy in the game $G(\mathfrak{U}, d; \mathfrak{I}, d)$ if she maintains configurations $(\mathfrak{U}, d', \mathfrak{I}, d')$ throughout the game. Clearly, if she can maintain these kinds of configurations, atomic equivalence is always given. It hence suffices to show that every path $dr_0d_1r_1d_1 \cdots r_{n-1}d_n$ with $n < \omega$ in $\mathfrak{I}$, where $r_n \in \mathsf{N_R} \cup \{r^- \mid r \in \mathsf{N_R}\}$ and $d_i \in \Delta^{\mathfrak{I}}$, is also a path in $\mathfrak{U}$ and vice versa.

The proof is carried out via the length of the path. As $d \in \Delta^{\mathfrak{U}}$ the claim is true for paths of length 0. Let $d \cdots d_{n-1}r_{n-1}d_n$ is a path in $\mathfrak{I}$ of length $n+1$ then $d \ldots d_{n-1}$ is a path in $\mathfrak{U}$ by induction hypothesis. In particular $d_{n-1}$ is in every subset $S \subseteq \Delta^{\mathfrak{I}}$ which comprises the generator set $G$ and which contains for all $r \in \mathsf{N_R}$ every predecessor and successor of elements in $S$. We shall just tackle the case where $r_{n-1} = r^-$, where $r \in \mathsf{N_R}$. Then $d_n$ is an $r$-predecessor of $d_{n-1}$. This element is contained in $\bigcap \mathfrak{G}$ and therefore contained in $\Delta^{\mathfrak{U}}$. As for all $r \in \mathsf{N_R}$, $r^{\mathfrak{U}}$ is merely the restriction of $r^{\mathfrak{I}}$ to the carrier-set of $\mathfrak{U}$, the given path is a path in $\mathfrak{U}$. The latter argument also explains, why every path in $\mathfrak{U}$ is a path in $\mathfrak{I}$. □

Proposition 3.1.21 . *$\mathcal{ALCI}$-TBoxes are invariant under disjoint unions.*

Proof. Let $\mathcal{T}$ be an $\mathcal{ALCI}$-TBox and $(\mathfrak{I}_i)_{i \in I}$ a family of interpretations. Assume that $\biguplus \mathfrak{I}_i \models \mathcal{T}$. Then $\mathfrak{I}_i$ is a generated substructure of $\biguplus \mathfrak{I}_i$, where the generator-set $G := \Delta^{\mathfrak{I}_i}$. Assume now $C \sqsubseteq D \in \mathcal{T}$ and $d \in C^{\mathfrak{I}_i}$ then $(\biguplus \mathfrak{I}_i, d) \xleftrightarrow{i} (\mathfrak{I}_i, d)$ and hence $d \in C^{\biguplus \mathfrak{I}_i}$. As $\biguplus_{i \in I} \mathfrak{I}_i \models \mathcal{T}$, we have $d \in D^{\biguplus \mathfrak{I}_i}$ and since $(\biguplus \mathfrak{I}, d) \xleftrightarrow{i} (\mathfrak{I}_i, d)$ the bisimulation invariance of $\mathcal{ALCI}$ ensures $d \in D^{\mathfrak{I}_i}$. This shows that $\mathfrak{I}_i \models C \sqsubseteq D$ for all $C \sqsubseteq D \in \mathcal{T}$ which entails $\mathfrak{I}_i \models \mathcal{T}$.

Assume $\mathfrak{I}_i \models \mathcal{T}$ for all $i \in I$. As each $\mathfrak{I}_i$ is a generated substructure of $\biguplus_{i \in I} \mathfrak{I}_i$ we have $(\mathfrak{I}_i, d) \xleftrightarrow{i} (\biguplus_{i \in I} \mathfrak{I}_i, d)$ for all $d \in \Delta^{\mathfrak{I}_i}$ and the invariance of $\mathcal{ALCI}$ under $\mathcal{ALCI}$-bisimulation yields $(\mathfrak{I}_i, d) \models C \iff (\biguplus_{i \in I} \mathfrak{I}_i, d) \models C$ for every $\mathcal{ALCI}$-concept $C$. Assume now $C \sqsubseteq D \in \mathcal{T}$ and $d \in C^{\biguplus \mathfrak{I}_i}$ then $d \in C^{\mathfrak{I}_i}$ for some $i \in I$ and as $\mathfrak{I}_i \models \mathcal{T}$ we have $d \in D^{\mathfrak{I}_i}$ which entails $d \in D^{\biguplus \mathfrak{I}_i}$ as $(\mathfrak{I}_i, d) \models D \iff (\biguplus_{i \in I} \mathfrak{I}_i, d) \models D$. Hence $\biguplus_{i \in I} \mathfrak{I}_i \models C \sqsubseteq D$ for all $C \sqsubseteq D \in \mathcal{T}$ which had to be shown. □

Theorem 3.1.22 . *Every sentence $\varphi \in \mathrm{FO}(\tau)$ which is invariant under global $\mathcal{ALCI}$-bisimulation and invariant under disjoint unions is equivalent to some $\mathcal{ALCI}$-TBox over $\tau$.*

Proof. Let $\mathrm{cons}\,\varphi := \{C \sqsubseteq D \mid C, D \in \mathcal{ALCI}(\tau) \mid \varphi \models C \sqsubseteq D\}$, so $\varphi \models \mathrm{cons}\,\varphi$. If we can prove that $\mathrm{cons}\,\varphi \models \varphi$ then, by compactness of FO, there is a finite subset $\mathcal{T} \subseteq \operatorname{cons} \varphi$ such that $\mathcal{T} \models \varphi$. $\mathcal{T}$ would be a logically equivalent $\mathcal{ALCI}$-TBox to $\varphi$.

The proof for $\operatorname{cons} \varphi \models \varphi$ is carried out via proof by contradiction. Assume $\operatorname{cons} \varphi \not\models \varphi$ then there is an interpretation $\mathfrak{K} \models \operatorname{cons} \varphi \cup \{\neg\varphi\}$. Let

$$T := \{p \in \mathcal{ALCI}(\tau) \mid \text{there is } (\mathfrak{I}, d) \models p \text{ and } \mathfrak{I} \models \varphi\}$$

and let for every $p \in T$ be $\mathfrak{I}_p$ some $\tau$-interpretation s.t. there is $d_p \in \Delta^{\mathfrak{I}_p}$ with $(\mathfrak{I}_p, d_p) \models p$ and $\mathfrak{I}_p \models \varphi$. We set $\mathfrak{H} := \biguplus\{\mathfrak{I}_p \mid p \in T\}$. As $\varphi$ is invariant under disjoint unions, $\mathfrak{H} \models \varphi$. Let $\mathfrak{I} := \mathfrak{K} \uplus \mathfrak{H}$. Then $\mathfrak{I} \models \operatorname{cons} \varphi$ as every concept-inclusion of $\operatorname{cons} \varphi$, considered as TBox, is invariant under disjoint unions. Yet $\mathfrak{I} \not\models \varphi$ for the same reason.

We now show $\mathfrak{I} \overset{gi}{\longleftrightarrow} \mathfrak{H}$. W.l.o.g. we assume that $\mathfrak{I}$ and $\mathfrak{H}$ are $\omega$-saturated. Using the Hennessy-Milner-Property, it suffices to show for every element in one structure that there is an element with the same $\mathcal{ALCI}$-theory in the other structure. As $\mathfrak{H}$ is a generated substructure of $\mathfrak{I}$ we settle on proving for all $d \in \Delta^{\mathfrak{I}}$ there is $e \in \Delta^{\mathfrak{H}}$ such that $\operatorname{Th}(\mathfrak{I}, d) = \operatorname{Th}(\mathfrak{H}, e)$.

Assume this element $e$ would not exist in $\mathfrak{H}$. Then $\operatorname{Th}(\mathfrak{I}, d)$ must be unsatisfiable with $\varphi$. For otherwise $p := \operatorname{Th}(\mathfrak{I}, d)$ is in $T$ and $(\mathfrak{H}, d_p) \models p \cup \{\varphi\}$. By the compactness of FO there is a finite subset $\Gamma_0 \subseteq \operatorname{Th}(\mathfrak{I}, d)$ such that $\Gamma_0 \cup \{\varphi\}$ is unsatisfiable and hence $\varphi \models \bigsqcap \Gamma_0 \sqsubseteq \bot$. But then $\Gamma_0 \sqsubseteq \bot \in \operatorname{cons} \varphi$ and so the generated substructure of $d$ from $\mathfrak{I}$ does not satisfy $\operatorname{cons} \varphi$; a contradiction to $\operatorname{cons} \varphi$ being invariant under disjoint union.

This proved $\mathfrak{I} \overset{gi}{\longleftrightarrow} \mathfrak{H}$, which is, in turn, a contradiction to $\varphi$ being invariant under global $\mathcal{ALCI}$-bisimulation. So $\operatorname{cons} \varphi \models \varphi$ which had to be shown. □

## 3.2 $\mathcal{ALCQ}$

In this chapter, the description logic $\mathcal{ALCQ}$ shall be introduced, which allows for counting quantifications, also known as graded modal operators in modal logic [60, 52, 51] or qualified number restriction in description logics [74]. The latter explains the $Q$ in $\mathcal{ALCQ}$.

These operators allow to express not only all properties $\mathcal{ALC}$ can express, but additionally they allow to specify at how many $r$-successors for $r \in \mathsf{N}_\mathsf{R}$ of a distinguished element, a certain concept is satisfied. But only a finite number of successors can be specified.

There is another description logic $\mathcal{ALCN}$ [75, 14] whose model-theoretic char-acterisation on concept-level was treated in [85]. $\mathcal{N}$ stands here for number restriction: In comparison to $\mathcal{ALCQ}$, it can specify a finite minimum number of $r$-successors, but it cannot specify which concepts these $r$-successors satisfy. The latter is the reason why number restrictions in $\mathcal{ALCN}$ are called unqualified, whilst number restrictions in $\mathcal{ALCQ}$ are qualified. Its expressivity (for fixed signature) is therefore strictly less than $\mathcal{ALCQ}$.

Example 3.2.1 . Although we do not have introduced the syntax of $\mathcal{ALCQ}$, we shall give an example of how the two logics differ. Whilst $\mathcal{ALCQ}$ allows to express $\exists^{\geq 3}.A$ which is 'there are at least three $r$-successors that satisfy $A$' concepts from $\mathcal{ALCN}$ can only express $\exists^{\geq 3}.\top$ 'there are at least three $r$-successors', yet they cannot *qualify* what properties these three $r$-successors have.

It shall be remarked that the model theoretic characterisation in [85] for $\mathcal{ALCN}$ was only carried out on concept level. TBoxes over $\mathcal{ALCN}$ or the extension with the universal role were not considered.

In comparison to $\mathcal{ALCI}$, we are not able to address predecessors of a distinguished element.

We shall investigate $\mathcal{ALCQ}$-concepts in the usual way, defining a model theoretic notion of $\mathcal{ALCQ}$ which is used to characterise $\mathcal{ALCQ}$ as FO-fragment. The model-theoretic notion of counting bisimulation could have also be defined using bijective functions on successor sets but further down we shall give good reasons for why we chose this notion in the specific way. We present an explicit construction for characteristic concepts and show that they capture the appropriate bisimulation level.

We then lift $\mathcal{ALCQ}$-bisimulation onto the global level and give a characterisation for $\mathcal{ALCQ}u$, where we allow us to present a variant with a counting global quantifier. The usual extension with a non-counting global quantifier $\mathcal{ALCQ}u_1$ is a special case of $\mathcal{ALCQ}u$. $\mathcal{ALCQ}$-TBoxes are then a fragment of $\mathcal{ALCQ}u_1$ which is characterised as FO-fragment.

### 3.2.1 The Characterisation of $\mathcal{ALCQ}$-Concepts

$\mathcal{ALCQ}$ itself was first proposed in [74] as generalisation of ordinary quantification and unqualified number restrictions. This paper concentrated on the satisfiability problem and investigated its complexity. Since then, reasoning with $\mathcal{ALCQ}$ has had some attention (e.g. [74, 127, 65]) as $\mathcal{ALCQ}$ is an important fragment of $\mathcal{ALCQIO}$ and hence $\mathcal{SROIQ}$ [77] which underpins OWL 2. Though, the model-theoretic properties of $\mathcal{ALCQ}$ have not been investigated so far. There is one explicit model-theoretic investigation of graded modalities [114] where in other cases this extension of standard modal logic is merely mentioned [61] or completely omitted [18]. The results of this section have been published in [92] and were novel.

### Syntax and Semantics

For every signature $\tau$, the syntax of an $\mathcal{ALCQ}$-concept $C$ over $\tau$ is defined as follows:

$$C ::= \top \mid A \mid D \sqcap E \mid \neg D \mid \exists^{\geq\kappa} r.D$$

where $A \in \mathsf{N_C}$, $D, E$ are $\mathcal{ALCQ}$-concepts over $\tau$ and $r \in \mathsf{N_R}$ with $\kappa < \omega$. The set of all $\mathcal{ALCQ}$-concepts over $\tau$ is denoted as $\mathcal{ALCQ}(\tau)$. We use the common abbreviations $\sqcup$, $\rightarrow$ etc.

The notation $\exists^{\geq\kappa}$, more common in context of FO with counting quantifiers (e.g. [123]), is not the usual one used in Description Logics. Instead, $\exists^{\geq\kappa} r.C$ is in the DL literature consistently expressed as $\geq \kappa\ r\ C$. We, however, want to appeal to the analogy of the existential quantification $\exists r.C$, and $\exists r^{-}.C$ respectively, which we have introduced so far. Our particular notation naturally begs the question whether $\forall^{\geq\kappa}$-quantifications are possible, and its answer will explain why qualifying number restrictions $\geq \kappa\ r\ C$ do not have a separate symbol equivalent to a value restriction in DLs.

The semantic is explained by recursively continuing the interpretation function $\cdot^{\mathfrak{I}}$ on concepts $C \in \mathcal{ALCQ}(\tau)$ as follows:

$$C^{\mathfrak{I}} := \begin{cases} C^{\mathfrak{I}} & \text{if } C \in \mathsf{N_C} \\ D^{\mathfrak{I}} \cap E^{\mathfrak{I}} & \text{if } C = D \sqcap E \\ \Delta^{\mathfrak{I}} \setminus D & \text{if } C = \neg D \\ \{d \in \Delta^{\mathfrak{I}} \mid |r^{\mathfrak{I}}(d) \cap D^{\mathfrak{I}}| \geq \kappa\} & \text{if } C = \exists^{\geq\kappa} r.D \end{cases}$$

where $|M|$ yields the cardinality of a set $M$ and $r^{\mathfrak{I}}(d) := \{e \in \Delta^{\mathfrak{I}} \mid (d, e) \in r^{\mathfrak{I}}\}$. Hence $|r^{\mathfrak{I}}(d) \cap D^{\mathfrak{I}}|$ is the number of $r$-successors of $d$ in $\mathfrak{I}$ that satisfy $D$. The interpretation function coincides with the interpretation function for $\mathcal{ALC}$ from page 25 in all but one case: the existential quantification. The interpretation function for $\mathcal{ALCQ}$ is simply a generalisation of the interpretation function for $\mathcal{ALC}$ in that it interprets $\exists^{\geq 1} r.C$ like the concept $\exists r.C$. In particular, $\mathcal{ALCQ}$ subsumes $\mathcal{ALC}$.

We can define several abbreviations:

$$\begin{aligned}
\exists^{>n}r.D &:= \exists^{\geq n+1}r.D \\
\exists^{\leq n}r.D &:= \neg\exists^{\geq n+1}r.D \\
\exists^{<n}r.D &:= \neg\exists^{\geq n}r.D \\
\exists^{=n}r.D &:= (\exists^{\geq n}r.D) \sqcap (\neg\exists^{\geq n+1}r.D)
\end{aligned}$$

We shall merely check that $\exists^{<n}r.D$ meets our intuition as all other abbreviations are then straight forward:

$$\begin{aligned}
(\exists^{<n}r.D)^{\mathfrak{I}} &= \Delta^{\mathfrak{I}} \setminus \{d \in \Delta^{\mathfrak{I}} \mid |r^{\mathfrak{I}}(d) \cap D^{\mathfrak{I}}| \geq n\} \\
&= \{d \in \Delta^{\mathfrak{I}} \mid \text{ not } |r^{\mathfrak{I}}(d) \cap D^{\mathfrak{I}}| \geq n\} \quad = \{d \in \Delta^{\mathfrak{I}} \mid |r^{\mathfrak{I}}(d) \cap D^{\mathfrak{I}}| < n\}
\end{aligned}$$

Clearly $\exists^{<0}r.D$ always yields the empty set and is therefore logically equivalent to $\bot$.

$\mathcal{ALCQ}$ is at least as expressive as $\mathcal{ALC}$ as every $\exists r$-quantification simply needs to be replaced by $\exists^{\geq 1}r$. We can translate $\forall^{\geq n}r.D := \neg\exists^{\geq n}r.\neg D$ to 'all but less than $n$ $r$-successors satisfy $D$' or 'from the $n$-th $r$-successor onwards all satisfy $D$'. It is analogue to the $\forall r$ quantification in the sense that one could read them out as 'from the first $r$-successor onwards all satisfy $D$'.

But in context of $\mathcal{ALCQ}$, it is difficult to make sense of $\forall^{\leq n}r.D$ which is logically equivalent to $\exists^{\geq n+1}r.\neg D$. It cannot be satisfied by an element with less than $n+1$-successors and even an element that satisfies $\forall^{\leq n}r.D$ and has at least $n+1$ $r$-successors needs not to have any $r$-successors that satisfy $D$. We shall therefore not define these kinds of abbreviations.

Note that $\mathcal{ALCQ}$ only allows for finite cardinality restrictions. For otherwise, an $\mathcal{ALCQ}$-concept could state that the distinguished element has infinitely many successors. But this would catapult $\mathcal{ALCQ}$ outside of FO: In case $\mathcal{ALCQ}$ could express 'there are uncountably many $r$-successors', the Löwenheim-Skolem-Property [47, 72, 109] of FO would be violated: This property states that every FO-formula can be satisfied by some countable (finite or infinite) interpretation.

If $\mathcal{ALCQ}$ could express 'there are only countably infinite many $r$-successors' then $\{\exists^{\geq\kappa}r.\top \mid \kappa < \omega\} \models \exists^{\geq\omega}r.\top$. Compactness would then yield a finite subset of $\{\exists^{\geq\kappa}r.\top \mid \kappa < \omega\}$ and therefore a finite ordinal $\lambda$ such that $\{\exists^{\geq\kappa}r.\top \mid \kappa \leq \lambda\} \models \exists^{\geq\omega}r.\top$, which is obviously false. Hence $\mathcal{ALCQ}$ would not be compact, and hence could not be a fragment of FO.

In the next step, we shall define the rank-function for $\mathcal{ALCQ}$-concepts and we shall give a grade-function which yields the greatest $\kappa$ that occurs as grade in a concept:

$$\operatorname{rank} C := \begin{cases} 0 & \text{if } C \in \mathsf{N_C} \\ \max\{\operatorname{rank} D, \operatorname{rank} E\} & \text{if } C = D \sqcap E \\ \operatorname{rank} D & \text{if } C = \neg D \\ 1 + \operatorname{rank} D & \text{if } C = \exists^{\geq \kappa} r.D \text{ for all } \kappa < \omega \end{cases}$$

$$\operatorname{grade} C := \begin{cases} 1 & \text{if } C \in \mathsf{N_C} \\ \max\{\operatorname{grade} D, \operatorname{grade} E\} & \text{if } C = D \sqcap E \\ \operatorname{grade} D & \text{if } C = \neg D \\ \max\{\kappa, \operatorname{grade} D\} & \text{if } C = \exists^{\geq \kappa} r.D \end{cases}$$

We always have $\operatorname{grade} C < \omega$: only natural numbers are admitted as grades and every $\mathcal{ALCQ}$-concept $C$ is of finite length.

We shall define a translation into FO recursively on the structure of $\mathcal{ALCQ}$-concepts. In contrast to $\mathcal{ALC}$-concepts which indeed translate into 'plain' FO-formulae, $\mathcal{ALCQ}$-concepts are translated into $\mathrm{FO}^{\equiv}$-formulae, i.e. into formulae of First Order Logic with equality ($\equiv$). Contrary to previous chapters, we need $\kappa + 1$-many variables to translate an $\mathcal{ALCQ}$-concept of grade $\kappa$ equivalently into FO:

$$[C; x_i] := \begin{cases} C(x_i) & \text{if } C \in \mathsf{N_C} \\ [D; x_i] \land [E; x_i] & \text{if } C = D \sqcap E \\ \neg[D; x_i] & \text{if } C = \neg D \end{cases}$$

as usual, and in case $C = \exists^{\geq \kappa} r.D$, we have

$$[C; x_i] := \exists (x_\ell)_{\ell \in \kappa \setminus \{i\}} \bigwedge_{k, \ell \in \kappa \setminus \{i\}} \{\neg x_k \equiv x_\ell \mid k \neq \ell\} \land \bigwedge_{k \in \kappa \setminus \{i\}} r(x_i, x_k) \land [D; x_k]$$

A simple inductive argument shows that the translation is correct; in particular $\mathfrak{I}\frac{d}{x_i} \models [\exists^{\geq \kappa} r.D; x_i]$ iff there are at least $\kappa$ many pairwise distinct $r$-successors of $d$ which all satisfy the FO-translation of $D$ hence iff $(\mathfrak{I}, d) \models \exists^{\geq \kappa} r.D$.

Indeed, this shows that $\mathcal{ALCQ}$ is a fragment of FO and hence $\mathcal{ALCQ}$ is compact.

$\mathcal{ALCQ}$ Bisimulation and Properties

The notion of $\mathcal{ALCQ}$-bisimulation can be found in [114] and is here expressed as game, similar to $\mathcal{ALCI}$-bisimulation and $\mathcal{ALC}$-bisimulation.

Let $\mathfrak{I}, \mathfrak{H}$ be two $\tau$-interpretations. The *$\mathcal{ALCQ}$-bisimulation game* $G(\mathfrak{I}, d; \mathfrak{H}, e)$ is a round-based game played by two players **I** and **II** as follows: The state of the game is captured in a bijection $\beta : D \longrightarrow E$ where $D \subseteq \Delta^{\mathfrak{I}}$ and $E \subseteq \Delta^{\mathfrak{H}}$.

The game $G(\mathfrak{I}, d; \mathfrak{H}, e)$ starts with the bijection $d \longmapsto e$. If not $d \in A^{\mathfrak{I}} \iff e \in A^{\mathfrak{H}}$ for all $A \in \mathsf{N_C}$ then **II** has lost the game in the 0-th round.

Each round is played as follows: It starts with the bijection resulting from the previous round; let the bijection be $\beta : D \longrightarrow E$. **I** chooses one element from $D$ or $E$ and some $r \in \mathsf{N_R}$. Say **I** chooses $e \in E$ and $r \in \mathsf{N_R}$. He then plays a non-empty, finite subset $E' \subseteq r^{\mathfrak{H}}(e)$. If **I** cannot deliver such a subset, **II** wins the game.

**II** then has to play a subset from the $r$-successors of the element which is related to $e$ by $\beta$. In our case this is some $D' \subseteq r^{\mathfrak{I}}(d)$ with $d = \beta^{-1}(e)$. $D'$ must be chosen in way such that there is a bijection $\beta : D' \longrightarrow E'$ that satisfies $d' \in A^{\mathfrak{I}}$ iff $\beta(d') \in A^{\mathfrak{H}}$ for all $A \in \mathsf{N_C}$ and all $d' \in D'$. If **II** cannot deliver such a set $D'$ then **II** has lost the game. The next round starts with the bijection $\beta'$.

Definition 3.2.2. **II** has a *winning strategy* in $G(\mathfrak{I}, d; \mathfrak{H}, e)$, if she wins the game or can ward off any challenge of **I** and thus never loses a round. We then say, $(\mathfrak{I}, d)$ is $\mathcal{ALCQ}$-bisimilar to $(\mathfrak{H}, e)$ and denote this with $(\mathfrak{I}, d) \overset{<\omega}{\Longleftrightarrow} (\mathfrak{H}, e)$. ◇

We introduce a restricted version of the game which is bound in rounds and in the cardinality of sets that **I** and **II** may play: $G^{\kappa}_{n}(\mathfrak{I}, d; \mathfrak{H}, e)$ is the *n-round game* in which only sets of positive cardinality of at most $\kappa < \omega$ may be played. **II** wins the game, if she does not lose within $n$ rounds.

Definition 3.2.3. **II** has a *winning strategy* in $G^{\kappa}_{n}(\mathfrak{I}, d; \mathfrak{H}, e)$ if she can ward off challenges of **I** up to positive cardinalities $\kappa < \omega$ during the first $n$ rounds. We then say, $(\mathfrak{I}, d)$ is $(\kappa, n)$-bisimilar to $(\mathfrak{H}, e)$ and denote this with $(\mathfrak{I}, d) \overset{\leq\kappa}{\Longleftrightarrow}_n (\mathfrak{H}, e)$. ◇

Before we move on to characteristic $\mathcal{ALCQ}$-concepts, a remark should be made. Assume, $\mathcal{ALCQ}$-bisimulation would be defined as relation $Z$ which requires for an element $(d, e) \in Z$ a full bijection $\beta : r^{\mathfrak{I}}(d) \longrightarrow r^{\mathfrak{H}}(e)$ between all $r$-successors such that $(d_0, \beta(d_0)) \in Z$ for all $d_0 \in r^{\mathfrak{I}}$. Then this coincides with the notion we have given above, only if we restrict ourselves finitely branching interpretations, i.e. every node has but finitely many successors.

But having the finite branching property is a severe restriction imposed on the interpretations we allow and clearly it is not the right notion when characterising $\mathcal{ALCQ}$ over arbitrary interpretations. We shall argue more about this after we have proved that $\mathcal{ALCQ}$ satisfies the Hennessy-Milner-Property (cf. Proposition 3.2.12).

Proposition 3.2.4. *For all* $C \in \mathcal{ALCQ}(\tau)$ *with* $\operatorname{rank} C \leq n$ *and* $\operatorname{grade} C \leq \kappa$ *we have*

$$(\mathfrak{I}, d) \xleftrightarrow{\leq\kappa}_n (\mathfrak{H}, e) \quad\Longrightarrow\quad (\mathfrak{I}, d) \models C \iff (\mathfrak{H}, e) \models C.$$

Proof. The proof is carried out by induction upon $n < \omega$. $n = 0$ is immediate. Assume $(\mathfrak{I}, d) \xleftrightarrow{\leq\kappa}_{n+1} (\mathfrak{H}, e)$. Let $C \in \mathrm{Th}^{\kappa}_{n+1}(\mathfrak{I}, d)$ arbitrary. We may w.l.o.g. assume $C = \exists^{\geq\lambda} r.D$ where $\lambda \leq \kappa$, $D \in \mathcal{ALCQ}$ with $\operatorname{rank} D \leq n$ and $\operatorname{grade} D \leq \kappa$. Then $(\mathfrak{I}, d) \models C$ iff there is a set $D_0 \subseteq D^{\mathfrak{I}} \cap r^{\mathfrak{I}}(d)$ with $|D_0| = \lambda$. Since $(\mathfrak{I}, d) \xleftrightarrow{\leq\kappa}_{n+1} (\mathfrak{H}, e)$ there is a set $E_0 \subseteq r^{\mathfrak{H}}(e)$ with $|E_0| = \lambda$ and a bijection $\beta : D_0 \longrightarrow E_0$ such that $(\mathfrak{I}, d_0) \xleftrightarrow{\leq\kappa}_n (\mathfrak{H}, \beta(d_0))$ for all $d_0 \in D_0$. The induction hypothesis yields $(\mathfrak{H}, \beta(d_0)) \models D$ for all $d_0 \in D_0$ and so $(\mathfrak{H}, e) \models \exists^{\geq\lambda} r.D$. The only-if direction is derived in the same manner. □

### Characteristic $\mathcal{ALCQ}$-Concepts

It does not make much sense to define characteristic $\mathcal{ALCQ}$-concepts for $\kappa = 0$, as domain and image of each bijection during the game would be empty. So even the start bijection would not fit this scheme. Hence for finite signature $\tau$ we define characteristic $\mathcal{ALCQ}$-concepts for each positive $\kappa < \omega$.

Due to the restriction to sets of cardinality $\kappa$, our players can only count $r$-successors of a certain $(\kappa, n)$-bisimulation-type up to $\kappa$. The following construct will accommodate this handicap.

Let therefore $\tau$ be finite and $\mathfrak{I}$ a $\tau$-interpretation with $d \in \Delta^{\mathfrak{I}}$. $\#\langle n, r(d, e)\rangle :=$ $|\{d_1 \in r^{\mathfrak{I}}(d) \mid (\mathfrak{I}, d_1) \models X^{\kappa,n}_{\mathfrak{I},e}\}|$ i.e. the number of all $r$-successors of $d$ that satisfy the same characteristic $\mathcal{ALCQ}$-concept as $e$ on level $(\kappa, n)$. Clearly, the expression $\#\langle n, r(d, e)\rangle$ is underspecified as it does not mention $\mathfrak{I}$ and $\kappa$. But we shall always have this expression in context of the characteristic concept, where it belongs to and for which we shall know $\kappa$. We now create a function which yields a suitable number-restriction for quantifiers:

$$?\langle n, r(d,e)\rangle := \begin{cases} = \#\langle n, r(d,e)\rangle & \text{if } \#\langle n, r(d,e)\rangle < \kappa \\ \geq \kappa & \text{else} \end{cases}$$

Note that $?\langle n, r(d,e)\rangle$ yields a string: if $\kappa = 4$ and $\#\langle n, r(d,e)\rangle = 3$ then $?\langle n, r(d,e)\rangle$ yields the string '$= 3$'. If $\#\langle n, r(d,e)\rangle = 5$ then $?\langle n, r(d,e)\rangle$ yields the string '$\geq 4$'.

The *characteristic $\mathcal{ALCQ}$-concept* is recursively defined upon $n < \omega$ defined as follows:

$$\begin{aligned} X^{\kappa,0}_{\mathfrak{I},d} &:= \bigsqcap\{A \in \mathsf{N_C} \mid d \in A^{\mathfrak{I}}\} \sqcap \bigsqcap\{\neg A \mid d \notin A^{\mathfrak{I}}\} \\ X^{\kappa,n+1}_{\mathfrak{I},d} &:= X^{\kappa,0}_{\mathfrak{I},d} \sqcap \bigsqcap\nolimits_{r \in \mathsf{N_R}} \{\exists^{?\langle n, r(d,e)\rangle} r.X^{\kappa,n}_{\mathfrak{I},e} \mid e \in r^{\mathfrak{I}}(d)\} \\ &\qquad \sqcap \bigsqcap\nolimits_{r \in \mathsf{N_R}} \forall^{\geq 1} r. \bigsqcup\{X^{\kappa,n}_{\mathfrak{I},e} \mid e \in r^{\mathfrak{I}}(d)\} \end{aligned}$$

$\forall^{\geq 1} r. \bigsqcup\{X^{\kappa,n}_{\mathfrak{I},e} \mid e \in r(d)\}$ merely assures that no $(\kappa, n)$-bisimulation-types are allowed, which are not satisfied by some $e \in r(d)$.

Lemma 3.2.5. *For every positive $\kappa < \omega$ and all $n < \omega$ the characteristic $\mathcal{ALCQ}$-concept $X^{\kappa,n}_{\mathfrak{I},d}$ is well defined.*

Proof. This Lemma is the analogue of Lemma 2.1.14 which stated that characteristic $\mathcal{ALC}$-concepts are well defined. We merely have to check, whether all sets mentioned in such a concept are finite. For fixed positive $\kappa < \omega$, we show by induction upon $n < \omega$ that there are only finitely many characteristic concepts on each level $n$. In contrast to Lemma 2.1.14, we now have to consider quantifications with different cardinalities. Since they only occur in the step case we omit the base case.

In the step-case, we assume that there are only finitely many characteristic $\mathcal{ALCQ}$-concepts on level $n$. For each $r \in \mathsf{N_R}$, a concept can be paired up with a quantifier from $\{\exists^{=1}, \ldots, \exists^{=\kappa-1}, \exists^{\geq\kappa}\}$. There are only $\kappa$ many quantifiers and $\mathsf{N_R}$ is finite, hence the number of subsets containing concepts of the form $\exists^{?\langle n, r(d,e)\rangle} r.X^{\kappa,n}_{\mathfrak{I},e}$ is finite. Therefore there are also just finitely many combinations of

$$\bigsqcap_{r \in \mathsf{N_R}} \{\exists^{?\langle n, r(d,e)\rangle} r.X^{\kappa,n}_{\mathfrak{I},e} \mid e \in r^{\mathfrak{I}}(d)\} \quad \text{and} \quad \bigsqcap_{r \in \mathsf{N_R}} \forall^{\geq 1} r. \bigsqcup\{X^{\kappa,n}_{\mathfrak{I},e} \mid e \in r(d)\}$$

which shows that there are only finitely many formulae $X^{\kappa,n+1}_{\mathfrak{I},d}$ on level $n + 1$. From this follows that all characteristic $\mathcal{ALCQ}$-concepts are well defined. □

Proposition 3.2.6. *If $(\mathfrak{I}, d) \models X^{\kappa,n}_{\mathfrak{H},e}$ then $(\mathfrak{I}, d) \overset{\leq\kappa}{\Longleftrightarrow}_n (\mathfrak{H}, e)$*

We shall need the following lemma during the proof the proposition. For given $\kappa$, $n$ and $(\mathfrak{I}, d)$ and $(\mathfrak{H}, e)$ we state

LEMMA 3.2.7. *Assume* $(\mathfrak{M}, w) \models X^{\kappa,n}_{\mathfrak{N},v} \Longrightarrow (\mathfrak{M}, w) \xleftrightarrow{\leq\kappa}_n (\mathfrak{N}, v)$ *for all* $\mathfrak{M}, \mathfrak{N}$ *with* $w \in \Delta^{\mathfrak{M}}$ *and* $v \in \Delta^{\mathfrak{N}}$*. Then for all* $(\mathfrak{K}, g)$

$$\Big((\mathfrak{K}, g) \models X^{\kappa,n}_{\mathfrak{I},d} \text{ and } (\mathfrak{K}, g) \models X^{\kappa,n}_{\mathfrak{H},e}\Big) \Longrightarrow X^{\kappa,n}_{\mathfrak{I},d} \equiv X^{\kappa,n}_{\mathfrak{H},e}$$

PROOF. From the premise we obtain $(\mathfrak{I}, d) \xleftrightarrow{\leq\kappa}_n (\mathfrak{K}, g) \xleftrightarrow{\leq\kappa}_n (\mathfrak{H}, e)$. Assume that for some $(\mathfrak{M}, w)$, we have $(\mathfrak{M}, w) \models X^{\kappa,n}_{\mathfrak{I},d}$. Then $(\mathfrak{M}, w) \xleftrightarrow{\leq\kappa}_n (\mathfrak{I}, d)$ and since concepts of grade $\leq \kappa$ and rank $\leq n$ are invariant under $\xleftrightarrow{\leq\kappa}_n$ and $(\mathfrak{H}, e) \models X^{\kappa,n}_{\mathfrak{H},e}$ we have $(\mathfrak{M}, w) \models X^{\kappa,n}_{\mathfrak{H},e}$, which shows $X^{\kappa,n}_{\mathfrak{I},d} \models X^{\kappa,n}_{\mathfrak{H},e}$. In the same way, one shows $X^{\kappa,n}_{\mathfrak{H},e} \models X^{\kappa,n}_{\mathfrak{I},d}$ whence $X^{\kappa,n}_{\mathfrak{I},d} \equiv X^{\kappa,n}_{\mathfrak{H},e}$ follows. □

PROOF OF PROPOSITION 3.2.6. The claim is immediate for $n = 0$. For $n+1$ assume that **I** challenges **II** in the start configuration with some $r$-successor set $D$ with $|D| \leq \kappa$ of $d$ in $\mathfrak{I}$.

We shall partition $D$ and $r^{\mathfrak{H}}(e)$ into equivalence classes and match them accordingly. For $D$ we set $d_0 \sim_{\mathfrak{I}} d_1$ iff $(\mathfrak{I}, d_0) \models X^{\kappa,n}_{\mathfrak{H},e_0}$ and $(\mathfrak{I}, d_1) \models X^{\kappa,n}_{\mathfrak{H},e_0}$ for any $e_0 \in r^{\mathfrak{H}}(e)$. This relation is reflexive because $(\mathfrak{I}, d) \models \forall^{\geq 1} r. \bigsqcup\{X^{\kappa,n}_{\mathfrak{H},e_0} \mid e_0 \in r^{\mathfrak{H}}(e)\}$ and so every $d_0 \in D$ satisfies some $X^{\kappa,n}_{\mathfrak{H},e_0}$. The relation is trivially symmetric and transitivity follows from the induction hypothesis with Lemma 3.2.7. For $r^{\mathfrak{H}}(e)$ we set $e_0 \sim_{\mathfrak{H}} e_1$ iff $X^{\kappa,n}_{\mathfrak{H},e_0} \equiv X^{\kappa,n}_{\mathfrak{H},e_1}$.

Because $|D| \leq \kappa$, every class in $D/\sim_{\mathfrak{I}}$ is $\leq \kappa$. Let for $[d_0] \in D/\sim_{\mathfrak{I}}$ be $e_0 \in r^{\mathfrak{H}}(e)$ such that $(\mathfrak{I}, d_0) \models X^{\kappa,n}_{\mathfrak{H},e_0}$.

We observe that all $d' \in [d_0]$ satisfy $X^{\kappa,n}_{\mathfrak{H},e_0}$ because each pair $(d'_0, d'_1) \in \sim_{\mathfrak{I}}$ satisfies a common $X^{\kappa,n}_{\mathfrak{H},e'}$ and the induction hypothesis yields $(\mathfrak{I}, d'_0) \xleftrightarrow{\leq\kappa} (\mathfrak{H}, e') \xleftrightarrow{\leq\kappa}_n (\mathfrak{I}, d'_1)$. In particular, $(\mathfrak{I}, d') \xleftrightarrow{\leq\kappa}_n (\mathfrak{I}, d_0)$ and so $(\mathfrak{I}, d') \models X^{\kappa,n}_{\mathfrak{H},e_0}$.

Together with $(\mathfrak{I}, d) \models \exists^{?\langle n, r(e,e_0)\rangle} r.X^{\kappa,n}_{\mathfrak{H},e_0}$ our observation shows $|[d_0]| \leq \#\langle n, r(e, e_0)\rangle$. Hence for each class in $D/\sim_{\mathfrak{I}}$ we can fix an injective function $\beta_{[d_0]} : [d_0] \longrightarrow [e_0]$, where $(\mathfrak{I}, d_0) \models \exists^{?\langle n, r(e,e_0)\rangle} r.X^{\kappa,n}_{\mathfrak{H},e_0}$.

We set $\beta : d_1 \longmapsto \beta_{[d_1]}(d_1)$ and have to show that $\beta$ is injective. If $[\beta(d_0)] = [\beta(d_1)]$ then $X^{\kappa,n}_{\mathfrak{H},\beta(d_0)} \equiv X^{\kappa,n}_{\mathfrak{H},\beta(d_1)}$ by definition of $\sim_{\mathfrak{H}}$. By definition of $\sim_{\mathfrak{I}}$, in turn, $[d_0] = [d_1]$. Since $\beta_{[d_0]}$ was injective, $d_0 = d_1$.

Setting $E := \beta(D)$ yields the necessary response for **II**: $\beta : D \longrightarrow E$ is injective and for each $d' \in D$ with where $\beta_{[d']} : [d'] \longrightarrow [e_0]$ we have $(\mathfrak{I}, d') \models X^{\kappa,n}_{\mathfrak{H},e_0}$. Since $\beta(d_0) \in [e_0]$ we have by definition of $\sim_{\mathfrak{H}}$ that $X^{\kappa,n}_{\mathfrak{H},e_0} \equiv X^{\kappa,n}_{\mathfrak{H},\beta(d')}$ from which we infer $(\mathfrak{I}, d') \models X^{\kappa,n}_{\mathfrak{H},\beta(d')}$ and so the induction hypothesis yields $(\mathfrak{I}, d') \overset{\leq\kappa}{\Longleftrightarrow}_n (\mathfrak{H}, \beta(d'))$. So **II** can play for $n$ more rounds, giving her a winning strategy in this game.

Assume **I** challenges **II** with some $r$-successor set $|E| \leq \kappa$ of $e$ in $\mathfrak{H}$. Let now $\sim_{\mathfrak{I}}$ be defined as above, yet continued on $r^{\mathfrak{I}}(d)$ and let $\sim_{\mathfrak{H}}$ be defined as above yet restricted to $E$.

For each class $[e_0]$ we can fix an injection $\beta_{[e_0]} : [e_0] \longrightarrow [d_0]$ such that $(\mathfrak{I}, d_0) \models X^{\kappa,n}_{\mathfrak{H},e_0}$: The class $[d_0]$ exists because $\#\langle n, r(e, e_0)\rangle > 0$ and $(\mathfrak{I}, d) \models \exists^{?\langle n,r(e,e_0)\rangle} r.X^{\kappa,n}_{\mathfrak{H},e_0}$. Since $|E| \leq \kappa$ we have also $|[e_0]| \leq \#\langle n, r(e, e_0)\rangle \leq |[d_0]|$. The latter is true as all elements in $r^{\mathfrak{I}}(d)$ that satisfy $X^{\kappa,n}_{\mathfrak{H},e_0}$ are in $[d_0]$ by the definition of $\sim_{\mathfrak{I}}$ and $(\mathfrak{I}, d) \models \exists^{?\langle n,r(e,e_0)\rangle} r.X^{\kappa,n}_{\mathfrak{H},e_0}$.

We set $\beta : e_0 \longmapsto \beta_{[e_0]}(e_0)$ for all $e_0 \in E$. We have to show that $\beta$ is injective. Again for elements from the same class this is trivial. Assume $e_0 \neq e_1$ and $[e_0] \neq [e_1]$. Then $X^{\kappa,n}_{\mathfrak{H},e_0} \not\equiv X^{\kappa,n}_{\mathfrak{H},e_1}$ by definition of $\sim_{\mathfrak{H}}$. From the definition of $\sim_{\mathfrak{H}}$ follows that all elements in $\beta_{[e_0]}([e_0])$ satisfy $X^{\kappa,n}_{\mathfrak{H},e_0}$. Therefore no element in $\beta_{[e_0]}([e_0])$ may satisfy $X^{\kappa,n}_{\mathfrak{H},e_1}$ for otherwise Lemma 3.2.7 shows $X^{\kappa,n}_{\mathfrak{H},e_0} \equiv X^{\kappa,n}_{\mathfrak{H},e_1}$. Hence $\beta_{[e_0]}([e_0]) \neq \beta_{[e_1]}([e_1])$ and so both classes are disjoint.

Setting $D := \beta(E)$ yields the appropriate response for **II**: $\beta^{-1} : D \longrightarrow E$ is bijective. By definition of $\beta^{-1}$, each $d' \in D$ satisfies $X^{\kappa,n}_{\mathfrak{H},e_0}$ for some $e_0 \in [\beta^{-1}(d')]$. By definition of $\sim_{\mathfrak{H}}$ we have $X^{\kappa,n}_{\mathfrak{H},e_0} \equiv X^{\kappa,n}_{\mathfrak{H},\beta^{-1}(d)}$. Therefore $(\mathfrak{I}, d') \models X^{\kappa,n}_{\mathfrak{H},\beta^{-1}(d)}$ and the induction hypothesis shows that **II** has a winning strategy for the next $n$-rounds, which proves her winning strategy for this game. □

We define $\mathrm{Th}^{\kappa}_n(\mathfrak{I}, d) := \{C \in \mathcal{ALCQ}(\tau) \mid d \in C^{\mathfrak{I}},\ \text{grade}\, C \leq \kappa \text{ and rank}\, C \leq n\}$ and deliver the following important theorem:

Theorem 3.2.8. *Let $\tau$ be finite and let $\mathfrak{I}, \mathfrak{H}$ be $\tau$-interpretations. The following three statements are equivalent for all positive $\kappa < \omega$ and all $n < \omega$:*

1. $(\mathfrak{H}, e) \models X^{\kappa,n}_{\mathfrak{I},d}$
2. $(\mathfrak{I}, d) \overset{\leq\kappa}{\Longleftrightarrow}_n (\mathfrak{H}, e)$
3. $\mathit{Th}^{\kappa}_n(\mathfrak{I}, d) = \mathit{Th}^{\kappa}_n(\mathfrak{H}, e)$

Proof.

'1. $\Longrightarrow$ 2.' was given by Proposition 3.2.6

'2. $\Longrightarrow$ 3.' was proven by Proposition 3.2.4.

'3. $\Longrightarrow$ 1.' is immediate.

□

Saturation and Hennessy-Milner-Property

We lift the restriction of signatures $\tau$ being finite.

Definition 3.2.9 . Let $\tau$ be a signature and $(\mathfrak{I}, d)$ a pointed $\tau$-interpretation.

1. Let $\kappa < \omega$ be positive, $r \in \mathsf{N_R}$. Then $\Gamma \subseteq \mathcal{ALCQ}(\tau)$ is called $\exists^{\geq\kappa} r$*-type* of $(\mathfrak{I}, d)$ if for all finite $\Gamma_0 \subseteq \Gamma$ we have $(\mathfrak{I}, d) \models \exists^{\geq\kappa} r. \bigsqcap \Gamma_0$.

2. An $\exists^{\geq\kappa} r$-type $\Gamma$ of $(\mathfrak{I}, d)$ is *realised at d* if there is a set $D \subseteq r^{\mathfrak{I}}(d)$ with $|D| \geq \kappa$ such that for all $d_0 \in D$ we have $(\mathfrak{I}, d) \models \Gamma$.

◇

Definition 3.2.10 . An interpretation $\mathfrak{I}$ is $\mathcal{ALCQ}$*-saturated*, if for all positive $\kappa < \omega$, all $r \in \mathsf{N_R}$ and every $d \in \Delta^{\mathfrak{I}}$, every $\exists^{\geq\kappa} r$-type of $(\mathfrak{I}, d)$ is realised at $d$. ◇

Regarding the subsection Remark on Saturation and Types on page 44, one notices that also for $\mathcal{ALCQ}$-types the unique predecessor in the tree-unravelling forms the only parameter and hence $\omega$-saturation is enough to obtain $\mathcal{ALCQ}$-saturation. We state this fact in the following observation:

Observation 3.2.11 . *Every $\omega$-saturated interpretation is $\mathcal{ALCQ}$-saturated*

Proposition 3.2.12 . *$\mathcal{ALCQ}$-concepts have the Hennessy-Milner-property, i.e. if $\mathfrak{I}$ and $\mathfrak{H}$ are both $\mathcal{ALCQ}$-saturated then*

$$Th(\mathfrak{I}, d) = Th(\mathfrak{H}, e) \quad \textit{iff} \quad (\mathfrak{I}, d) \overset{<\omega}{\Longleftrightarrow} (\mathfrak{H}, e)$$

*where* $Th(\mathfrak{I}, d) := \{C \in \mathcal{ALCQ}(\tau) \mid (\mathfrak{I}, d) \models C\}$.

Like for $\mathcal{ALC}$ (Proposition 2.1.23, **II** can win by maintaining configurations such that corresponding pebbles are always on elements with equivalent of theories. It might be interesting for the reader to see how we deal with successor-sets $D$, played as challenges by **I** which contain elements with different theories.

Proof. **II** has a winning strategy for this game, if she can maintain configurations, i.e. bijections, $\beta : D \longrightarrow E$ such that $\mathrm{Th}(\mathfrak{I}, d_0) = \mathrm{Th}(\mathfrak{H}, \beta(d_0))$ for all $d_0 \in D$. The start-configuration is clearly of this form. Let the game have reached a configuration of this kind and let **I** choose $d_0$ whose image is $e_0$ under $\beta$ then $\mathrm{Th}(\mathfrak{I}, d_0) = \mathrm{Th}(\mathfrak{H}, e_0)$.

Assume **I** challenges **II** by playing some finite set $D \subseteq r^{\mathfrak{I}}(d_0)$ with $\kappa := |D|$. We partition $D$ by the equivalence relation $d_0 \equiv d_1$ iff $\mathrm{Th}(\mathfrak{I}, d_0) = \mathrm{Th}(\mathfrak{I}, d_1)$. Then

each class $[d_1] \in D/\sim_{\mathfrak{I}}$ has at most $\kappa$ elements.

Let $[d_1] \in D/\sim_{\mathfrak{I}}$ be arbitrary and $\mathrm{Th}^{\omega}([d_1]) := \{C \in \mathcal{ALCQ}(\tau) \mid (\mathfrak{I}, d_1) \models C\}$. As $(\mathfrak{I}, d_0) \models \exists^{\geq |[d_1]|} r. \bigsqcap \Gamma_0$ and $\mathrm{Th}(\mathfrak{I}, d_0) = \mathrm{Th}(\mathfrak{H}, e_0)$, we have $(\mathfrak{H}, e_0) \models \exists^{\geq |[d_1]|} r. \bigsqcap \Gamma_0$ for all finite $\Gamma_0 \subseteq \mathrm{Th}^{\omega}([d_0])$ and so $\mathrm{Th}^{\omega}([d_1])$ is an $\exists^{\geq |[d_1]|} r$-type of $e_0$. As $(\mathfrak{H}, e_0)$ is saturated, $\mathrm{Th}^{\omega}([d_1])$ is realised at $e_0$ and hence there is a set $E_{d_1} \subseteq r^{\mathfrak{H}}(e_0)$ with at least $|[d_1]|$ many elements that satisfy $\mathrm{Th}^{\omega}([d_1])$. Thus there is an injection $\beta_{[d_1]} : [d_1] \longrightarrow E_{d_1}$. Note that this injection is chosen independently of the representative.

Clearly, for any two different classes $[d_1], [d_2] \in D/\sim_{\mathfrak{I}}$ we have $\mathrm{Th}^{\omega}([d_1]) \neq \mathrm{Th}^{\omega}([d_2])$ and therefore $E_{d_1} \cap E_{d_2} = \varnothing$, i.e. $\beta_{[d_1]}$ and $\beta_{[d_2]}$ have disjoint ranges. Hence $\beta : d_1 \mapsto \beta_{[d_1]}(d_1)$ is a bijection between $D$ and $\beta(D)$ such that $\mathrm{Th}(\mathfrak{I}, d_1) = \mathrm{Th}(\mathfrak{H}, \beta(d_1))$ for all $d_1 \in D$.

Hence **II** can ward off any challenge $D$ with $|D| < \omega$ from **I** in $\mathfrak{I}$. A similar argument shows that **II** can respond to any such challenge in $\mathfrak{H}$, too. This proves, she has a winning strategy in the game and so $(\mathfrak{I}, d) \overset{<\omega}{\Longleftrightarrow} (\mathfrak{H}, e)$. □

Observing the results from $\mathcal{ALC}$, $\mathcal{ALCI}$ and $\mathcal{ALCQ}$, it seems that the triplet of notions logic, bisimulation and saturation form a unit. If these notions are chosen in a manner such that they complement each other, one obtains the Hennessy-Milner-Property and eventually a characterisation.

This phenomenon is not restricted to description logics but is known in classical model theory as well: elementary equivalent $\omega$-saturated interpretations are partially isomorphic [109, 34].

This unit would have been broken for $\mathcal{ALCQ}$ if we had required for $\mathcal{ALCQ}$-bisimulations $Z$ that for each element $(d, e) \in Z$ there is a bijection $\beta : r^{\mathfrak{I}}(d) \longrightarrow r^{\mathfrak{H}}(e)$.

Example 3.2.13 . We denote the smallest infinite limit ordinal with $\omega$ as usual and use $\omega_1$ to denote the smallest uncountable limit ordinal which contains all countable ordinals. In particular $\omega_1$ is uncountable. Consider $\mathfrak{I}$ with $\Delta^{\mathfrak{I}} = \{*\} \cup \omega$ and $r^{\mathfrak{I}} := \{*\} \times \omega$ and $\mathfrak{H}$ with $\Delta^{\mathfrak{H}} = \{*\} \cup \omega_1$ and $r^{\mathfrak{I}} := \{*\} \times \omega_1$. Then $\mathfrak{I}$ and $\mathfrak{H}$ are both $\omega$-saturated and elementary equivalent and hence they are partially isomorphic. Clearly, they would be distinguished by the $\mathcal{ALCQ}$-bisimulation which requires bijections between successor sets and the Hennessy-Milner-Property would not be given for $\mathcal{ALCQ}$-concepts, not even under the strong notion of $\omega$-saturation.

For exactly this reason, we opted for a model-theoretic notion which requires only bijections between finite subsets of the successor sets.

Proposition 3.2.14. *If $\varphi \in \mathrm{FO}(\tau)$ is invariant under $\overset{<\omega}{\Longleftrightarrow}$ then there is a positive $\kappa < \omega$ and $n < \omega$ such that $\varphi$ is invariant under $\overset{\leq\kappa}{\Longleftrightarrow}_n$.*

Lemma 3.2.15. *If $\varphi \in \mathrm{FO}(\tau)$ and $\varphi$ is not invariant under $\overset{\leq\kappa}{\Longleftrightarrow}_n$ for any $n < \omega$ and any positive $\kappa < \omega$ then there are pointed $\tau$-interpretations $(\mathfrak{I}, d), (\mathfrak{H}, e)$ such that*

$$\varphi(x) \dashv (\mathfrak{I}, d) \equiv_{\mathcal{ALCQ}} (\mathfrak{H}, e) \models \neg\varphi(x)$$

W.l.o.g. we may assume for the rest this section that $\tau$ contains exactly the symbols of $\varphi$ and is hence finite. We define $I := (\omega \setminus \{0\}) \times \omega$.

The proof is basically the proof of Lemma 2.1.25. The only difference is, that we deal with two indices and therefore have to use the supremum instead of the maximum.

Proof. Let $\mathfrak{X}^{\kappa,n} := \{X^{\kappa,n}_{\mathfrak{I},d} \mid \exists(\mathfrak{I}, d), (\mathfrak{H}, e) : \varphi(x) \dashv (\mathfrak{I}, d) \overset{\leq\kappa}{\Longleftrightarrow}_n (\mathfrak{H}, e) \models \neg\varphi(x)\}$, where $X^{\kappa,n}_{\mathfrak{I},d}$ is the characteristic $\mathcal{ALCQ}$-concept for $(\mathfrak{I}, d)$ on level $(\kappa, n) \in I$ over $\tau$.

As required by the premise of this lemma, for each $(\kappa, n) \in I$ no $\mathfrak{X}^{\kappa,n}$ is empty. There are only finitely many characteristic $\mathcal{ALCQ}$-concepts on level $(\kappa, n)$ and so $\mathfrak{X}^{\kappa,n}$ is finite. Therefore $\bigsqcup \mathfrak{X}^{\kappa,n}$ is an $\mathcal{ALCQ}$-concept for all $(\kappa, n) \in I$.

$\mathfrak{X} := \{\bigsqcup \mathfrak{X}^{\kappa,n} \mid (\kappa, n) \in I\}$ is satisfiable with $\varphi$: For any finite set $\mathfrak{X}_0 \subseteq \mathfrak{X}$ let $(\mu, m)$ be the supremum in $I$ of all pairs $(\kappa, n)$ with $\bigsqcup \mathfrak{X}^{\kappa,n}$ occurring in $\mathfrak{X}_0$. Since $\mathfrak{X}^{\mu,m} \neq \emptyset$ there are $\tau$-interpretations $(\mathfrak{I}, d), (\mathfrak{H}, e)$ such that $\varphi \dashv (\mathfrak{I}, d) \overset{\leq\mu}{\Longleftrightarrow}_m (\mathfrak{H}, e) \not\models \varphi$. This implies $(\mathfrak{I}, d) \overset{\leq\kappa}{\Longleftrightarrow}_n (\mathfrak{H}, e)$ for all $\kappa \leq \mu$ and $n \leq m$ and so $X^{\kappa,n}_{\mathfrak{I},d} \in \mathfrak{X}^{\kappa,n}$ for all $\kappa \leq \mu$ and $n \leq m$. Hence $(\mathfrak{I}, d) \models \mathfrak{X}_0 \cup \{\varphi\}$.

As every finite subset of $\mathfrak{X}$ is satisfiable with $\varphi$, the compactness of FO yields a $\tau$-model $(\mathfrak{K}, c)$ of $\mathfrak{X} \cup \{\varphi\}$.

We shall show that $\mathfrak{X}_{\mathfrak{K},c} := \{X^{\kappa,n}_{\mathfrak{K},c} \mid (\kappa, n) \in I\}$ is satisfiable with $\neg\varphi$. Let $\mathfrak{X}_0 \subseteq \mathfrak{X}_{\mathfrak{K},c}$ be finite, and let $(\mu, m)$ be the supremum in $I$ of all $(\kappa, n)$ with $X^{\kappa,n}_{\mathfrak{K},c}$ occurring in $\mathfrak{X}_0$. As $(\mathfrak{K}, c) \models \mathfrak{X}$, there is $X^{\mu,m}_{\mathfrak{I},d} \in \mathfrak{X}^{\mu,m}$ satisfied by $(\mathfrak{K}, c)$. For $(\mathfrak{I}, d)$ there is a $(\mu, m)$-bisimilar partner $(\mathfrak{H}, e)$, which is a model of $\neg\varphi$. It follows that

$$(\mathfrak{K}, c) \overset{\leq\mu}{\Longleftrightarrow}_m (\mathfrak{I}, d) \overset{\leq\mu}{\Longleftrightarrow}_m (\mathfrak{H}, e)$$

and in particular that $(\mathfrak{H}, e) \models X^{\kappa,n}_{\mathfrak{K},c}$ for all $\kappa \leq \mu$ and $n \leq m$. Hence $(\mathfrak{H}, e) \models \mathfrak{X}_0 \cup \{\neg\varphi\}$. As every finite subset of $\mathfrak{X}_{\mathfrak{K},c}$ is satisfiable with $\neg\varphi$, compactness of FO yields a $\tau$-model $(\mathfrak{M}, g)$ of $\mathfrak{X}_{\mathfrak{K},c} \cup \{\neg\varphi\}$. With Theorem 3.2.8 it follows that

$(\mathfrak{M}, g) \equiv_{\mathcal{ALCQ}} (\mathfrak{K}, c)$. Which proves the claim. □

Proof of Proposition 3.2.14. Assume there would not be $(\kappa, n) \in I$ such that $\varphi$ is invariant under $\xleftrightarrow{\leq\kappa}_n$. Lemma 3.2.15 furnishes two pointed interpretations $(\mathfrak{I}, d)$ and $(\mathfrak{H}, e)$ such that

$$\varphi \dashv (\mathfrak{I}, d) \equiv_{\mathcal{ALCQ}} (\mathfrak{H}, e) \models \neg\varphi.$$

Their $\omega$-saturated extensions $(\mathfrak{I}^*, d)$ and $(\mathfrak{H}^*, e)$ satisfy the same FO-theory as the originals and in particular they realise the FO-type of $d$ and $e$ so that

$$\varphi \dashv (\mathfrak{I}^*, d) \equiv_{\mathcal{ALCQ}} (\mathfrak{H}^*, e) \models \neg\varphi.$$

But since $\mathcal{ALCQ}$ has the Hennessy-Milner-Property, we have $(\mathfrak{I}^*, d) \xleftrightarrow{<\omega} (\mathfrak{H}^*, e)$. This is absurd, as $\varphi$ is $\xleftrightarrow{<\omega}$ invariant. It shows that there must be $(\kappa, n) \in I$ such that $\varphi$ is $\xleftrightarrow{\leq\kappa}_n$ invariant. □

Theorem 3.2.16 . *Every* FO*-formula over* $\tau$ *which is invariant under* $\xleftrightarrow{<\omega}$ *is logically equivalent to some* $\mathcal{ALCQ}$*-concept* $C$ *over* $\tau$.

The proof uses the exact same rational as the proof for 2.1.24.

Proof. Let $\varphi(x)$ be an FO-formula over $\tau$ which is invariant under $\xleftrightarrow{<\omega}$. From Proposition 3.2.14, we obtain $(\kappa, n) \in I$ so that $\varphi$ is $\xleftrightarrow{\leq\kappa}_n$ invariant.

Let $C_\varphi$ contain all characteristic $\mathcal{ALCQ}$-concepts $X^{\kappa,n}_{\mathfrak{I},d}$ over symbols occurring in $\varphi$ such that $(\mathfrak{I}, d) \models \varphi(x)$. $C_\varphi$ is finite because there are only finitely many characteristic $\mathcal{ALCQ}$-concepts on level $(\kappa, n)$. Clearly $\varphi(x) \models \bigsqcup C_\varphi$.

But also $\bigsqcup C_\varphi \models \varphi(x)$: Let $(\mathfrak{H}, e) \models \bigsqcup C_\varphi$ then there is $(\mathfrak{I}, d) \models \varphi(x)$ such that $X^{\kappa,n}_{\mathfrak{I},d} \in C_\varphi$ and $(\mathfrak{H}, e) \models X^{\kappa,n}_{\mathfrak{I},d}$. Now $(\mathfrak{H}, e) \xleftrightarrow{\leq\kappa}_n (\mathfrak{I}, d)$ and since $\varphi$ is invariant under $(\kappa, n)$-bisimulation we have $(\mathfrak{H}, e) \models \varphi$. □

### 3.2.2 The Characterisation of $\mathcal{ALCQ}u$-Concepts

$\mathcal{ALCQ}u$ has been introduced in [11] and is the extension of $\mathcal{ALCQ}$ with a universal role. Unlike the description logics $\{\mathcal{ALC}\} \times \{\mathcal{Q}, \mathcal{I}, \varepsilon\}$ that have been treated so far, we have here the opportunity to introduce a graded universal role. The syntax for $\mathcal{ALCQ}u$-concepts $C$ over a signature $\tau$ would look as follows

$$C ::= \top \mid A \mid D \sqcap E \mid \neg D \mid \exists^{\geq\kappa} r.D \mid \exists^{\geq\kappa} u.D$$

where $A \in \mathsf{N_C}$, $E, D$ are $\mathcal{ALCQ}u$-concepts, $r \in \mathsf{N_R}$, $\kappa < \omega$ an $u$ is a logical symbol which does not belong to $\tau$. The set of all $\mathcal{ALCQ}u$-concepts over $\tau$ is denoted as $\mathcal{ALCQ}u(\tau)$. The semantics is simply the extension of the $\mathcal{ALCQ}$-interpretation onto $\mathcal{ALCQ}u$ by setting

$$(\mathfrak{I}, d) \models \exists^{\geq\kappa} u.D \iff |D^{\mathfrak{I}}| \geq \kappa$$

This is independent of the distinguished element $d$ and we hence also write $\mathfrak{I} \models \exists^{\geq\kappa} u.D$.

The rank-function ignores $\exists^{\geq\kappa} u$: $\operatorname{rank}(\exists^{\geq\kappa} u.D) := \operatorname{rank} D$ whilst the grade-function is defined as $\operatorname{grade} \exists^{\geq\kappa} u.D = \max\{\kappa, \operatorname{grade} D\}$ and hence treats $\exists^{\geq\kappa} u$ like any other quantification $\exists^{\geq\kappa} r$:

Definition 3.2.17. Two interpretations $\mathfrak{I}, \mathfrak{H}$ are globally $\mathcal{ALCQ}$-bisimilar, $\mathfrak{I} \xleftrightarrow{<\omega}{}^{\forall} \mathfrak{H}$, if both of the following is true:

1. for all finite $D \subseteq \Delta^{\mathfrak{I}}$ there is an injective function $\beta : D \longrightarrow \Delta^{\mathfrak{H}}$ with $(\mathfrak{I}, d) \xleftrightarrow{<\omega} (\mathfrak{H}, \beta(d))$ for all $d \in D$.
2. for all finite $E \subseteq \Delta^{\mathfrak{H}}$ there is an injective function $\beta : E \longrightarrow \Delta^{\mathfrak{I}}$ with $(\mathfrak{I}, \beta^{-1}(e)) \xleftrightarrow{<\omega} (\mathfrak{H}, e)$ for all $e \in E$.

Analogously we say $\mathfrak{I}$ and $\mathfrak{H}$ are globally $\mathcal{ALCQ}$-$(\kappa, n)$-bisimilar, $\mathfrak{I} \xleftrightarrow{\leq\kappa}{}^{\forall}_{n} \mathfrak{H}$, if both of the following holds:

1. for all $D \subseteq \Delta^{\mathfrak{I}}$ with $|D| \leq \kappa$ there is an injective function $\beta : D \longrightarrow \Delta^{\mathfrak{H}}$ with $(\mathfrak{I}, d) \xleftrightarrow{\leq\kappa}_{n} (\mathfrak{H}, \beta(d))$ for all $d \in D$.
2. for all $E \subseteq \Delta^{\mathfrak{H}}$ with $|E| \leq \kappa$ there is an injective function $\beta : E \longrightarrow \Delta^{\mathfrak{H}}$ with $(\mathfrak{I}, \beta^{-1}(e)) \xleftrightarrow{\leq\kappa}_{n} (\mathfrak{H}, e)$ for all $e \in E$.

As usual we define $(\mathfrak{I}, d) \xleftrightarrow{\leq\kappa}{}^{\forall}_{n} (\mathfrak{I}, e)$ if $\mathfrak{I} \xleftrightarrow{\leq\kappa}{}^{\forall}_{n} \mathfrak{H}$ and $(\mathfrak{I}, d) \xleftrightarrow{\leq\kappa}_{n} (\mathfrak{H}, e)$ and similarly for $< \omega$ etc. ◇

### Characteristic Concepts for $\mathcal{ALCQ}u$

Throughout this section let $\tau$ be finite and $I := (\omega \setminus \{0\}) \times \omega$. Let a $\tau$-interpretation $\mathfrak{I}$ and $(\kappa, n) \in I$ be given. We define for all $d \in \Delta^{\mathfrak{I}}$

$$?\kappa(d) := \begin{cases} = |\{d_0 \in \Delta^{\mathfrak{I}} \mid (\mathfrak{I}, d_0) \models X^{\kappa,n}_{\mathfrak{I},d}\}| & \text{if } |\{d_0 \in \Delta^{\mathfrak{I}} \mid (\mathfrak{I}, d_0) \models X^{\kappa,n}_{\mathfrak{I},d}\}| < \kappa \\ \geq \kappa & \text{else,} \end{cases}$$

where $X^{\kappa,n}_{\mathfrak{I},d}$ is the characteristic $\mathcal{ALCQ}$-concept over $\tau$ for $(\mathfrak{I}, d)$ on level $(\kappa, n)$. $?\kappa(d)$ evaluates to a string, e.g. '$= 2$' or '$\geq \kappa$' for the given cardinal $\kappa$. It might be read out as 'the number of type $d$ up to $\kappa$'.

In connexion with a characteristic concept $X^{\kappa,n}_{\mathfrak{I},d}$ the necessary parameters $\mathfrak{I}$, $n$ and $\kappa$ to evaluate $?\kappa(d)$ are given; hence for $\exists^{?\kappa(d)}u.X^{\kappa,n}_{\mathfrak{I},d}$ we shall not specify them explicitly and we may use $?\kappa(d)$ in other contexts, when the corresponding characteristic concept is deducible.

The *global characteristic $\mathcal{ALCQ}u$-concept over $\tau$ on level* $(\kappa, n)$ is defined as

$$X^{\kappa,n}_{\mathfrak{I}} := \bigsqcap\{\exists^{?\kappa(d)}u.X^{\kappa,n}_{\mathfrak{I},d} \mid d \in \Delta^{\mathfrak{I}}\} \sqcap \forall^{\geq 1}u.\bigsqcup\{X^{\kappa,n}_{\mathfrak{I},d} \mid d \in \Delta^{\mathfrak{I}}\}$$

again, where $X^{\kappa,n}_{\mathfrak{I},d}$ is the characteristic $\mathcal{ALCQ}$-concept over $\tau$ for $(\mathfrak{I}, d)$ on level $(\kappa, n)$. $X^{\kappa,n}_{\mathfrak{I}}$ is well defined as there are only up to $\kappa$ many different global quantifiers $\{\exists^{=1}, \ldots \exists^{=\kappa-1}, \exists^{\geq\kappa}\}$ and on each level $(\kappa, n)$ there are only finitely many characteristic $\mathcal{ALCQ}$-concepts (cf. proof of Lemma 3.2.5).

If $C$ is constructed by the following scheme then $C$ is called a *global $\mathcal{ALCQ}u$-concept over $\tau$*:

$$C ::= \exists^{\geq\kappa}u.D \mid E \sqcap F \mid \neg E$$

where $D \in \mathcal{ALCQ}(\tau)$ and $\kappa < \omega$ and $E, F$ are global $\mathcal{ALCQ}u$-concepts over $\tau$; clearly then $C \in \mathcal{ALCQ}u(\tau)$. Apparently, global $\mathcal{ALCQ}u$-concepts do not depend on a distinguished element. Let

$$\mathrm{Th}^{\kappa}_{n}(\mathfrak{I}) := \{C \in \mathcal{ALCQ}u(\tau) \mid \mathfrak{I} \models C, C \text{ is global, rank } C \leq n, \text{ grade } C \leq \kappa\}.$$

Note that not all $\mathcal{ALCQ}u$-concepts which do not depend on a distinguished elements are captured with our definition of a global concept: we exclude all concepts with nested global operators, e.g. $\exists^{\geq 1}u.\exists^{\geq 1}u.\top$. However, Corollary 3.2.20 will show that all $\mathcal{ALCQ}u$-concepts that do not depend on distinguished elements are logically equivalent to some global $\mathcal{ALCQ}u$-concept. In this sense, global $\mathcal{ALCQ}u$-concepts can be considered as a normal form of those $\mathcal{ALCQ}u$-concepts independent of distinguished elements.

PROPOSITION 3.2.18 . *Let $(\kappa, n) \in I$ and $\tau$ finite. The following three statements are equivalent:*

1. $\mathfrak{H} \models X^{\kappa,n}_{\mathfrak{I}}$

2. $\mathfrak{I} \overset{\leq\kappa\,\forall}{\Longleftrightarrow}_{n} \mathfrak{H}$

3. $Th^{\kappa}_n(\mathfrak{I}) = Th^{\kappa}_n(\mathfrak{H})$

The proof uses the same technique as the proof of Proposition 3.2.6: We partition successor sets by factorising them over equivalent theories.

Proof. '1. $\Longrightarrow$ 2.' Let $D \subseteq \Delta^{\mathfrak{I}}$ with $|D| \leq \kappa$. We define the following equivalence relation on $D$: $d_0 \sim_{\mathfrak{I}} d_1$ iff $X^{\kappa,n}_{\mathfrak{I},d_0} \equiv X^{\kappa,n}_{\mathfrak{I},d_1}$.

We introduce the following relation on $\Delta^{\mathfrak{H}}$: $e_0 \sim_{\mathfrak{H}} e_1$ iff there is $d_1 \in \Delta^{\mathfrak{I}}$ with $(\mathfrak{H}, e_0) \models X^{\kappa,n}_{\mathfrak{I},d_1}$ and $(\mathfrak{H}, e_1) \models X^{\kappa,n}_{\mathfrak{I},d_1}$ for some $d_1 \in \Delta^{\mathfrak{I}}$. Since $\mathfrak{H} \models \forall^{\geq 1} u. \bigsqcup\{X^{\kappa,n}_{\mathfrak{I},d} \mid d \in \Delta^{\mathfrak{I}}\}$ this relation is reflexive. It is symmetric by definition.

For transitivity assume $e_0 \sim_{\mathfrak{H}} e_1 \sim_{\mathfrak{H}} e_2$. Then there is $d_0, d_2 \in \Delta^{\mathfrak{I}}$ such that $(\mathfrak{H}, e_1) \models X^{\kappa,n}_{\mathfrak{I},d_0}$ and $(\mathfrak{H}, e_0) \models X^{\kappa,n}_{\mathfrak{I},d_0}$ as well as $(\mathfrak{H}, e_1) \models X^{\kappa,n}_{\mathfrak{I},d_2}$ and $(\mathfrak{H}, e_2) \models X^{\kappa,n}_{\mathfrak{I},d_2}$. Lemma 3.2.7 yields $X^{\kappa,n}_{\mathfrak{I},d_0} \equiv X^{\kappa,n}_{\mathfrak{H},e_1} \equiv X^{\kappa,n}_{\mathfrak{I},d_2}$, which shows that $e_0 \sim_{\mathfrak{H}} e_2$.

Let $[d] \in D/\sim_{\mathfrak{I}}$ be arbitrary. $\mathfrak{H} \models \exists^{?\kappa(d)} u.X^{\kappa,n}_{\mathfrak{I},d}$ so there is an equivalence class $[e] \in \Delta^{\mathfrak{H}}/\sim_{\mathfrak{H}}$ such that $(\mathfrak{H}, e) \models X^{\kappa,n}_{\mathfrak{I},d}$ with $|[e]| ?\kappa(d)$ respectively. We have $[d] \subseteq \{d_0 \in \Delta^{\mathfrak{I}} \mid (\mathfrak{I}, d_0) \models X^{\kappa,n}_{\mathfrak{I},d}\}$ and since $|\{d_0 \in \Delta^{\mathfrak{I}} \mid (\mathfrak{I}, d_0) \models X^{\kappa,n}_{\mathfrak{I},d}\}| ?\kappa(d)$ we have $|[d]| \leq |[e]|$. Hence there is an injection $\beta_{[d]} : [d] \longrightarrow [e]$, which we choose for the class $[d]$ and independently from the representative.

We define $\beta : d \longmapsto \beta_{[d]}(d)$ for all $d \in D$. This function is well-defined as the domains of $\beta_{[d_0]}$ and $\beta_{[d_1]}$ are disjoint if $[d_0] \neq [d_1]$ and still injective because $[\beta(d_0)] = [\beta(d_1)]$ implies $X^{\kappa,n}_{\mathfrak{I},d_0} \equiv X^{\kappa,n}_{\mathfrak{I},d_1}$ (see transitivity of $\sim_{\mathfrak{H}}$) and therefore $[d_0] = [d_1]$; as $\beta_{[d_0]}$ was injective this implies $d_0 = d_1$. For all $d \in D$ Theorem 3.2.8 yields $(\mathfrak{I}, d) \overset{\leq\kappa}{\Longleftrightarrow}_n (\mathfrak{H}, \beta(d))$ since $(\mathfrak{H}, \beta(d)) \models X^{\kappa,n}_{\mathfrak{I},d}$. This shows 1. in Definition 3.2.17

For 2. in Definition 3.2.17 let $E \subseteq \Delta^{\mathfrak{H}}$ with $|E| \leq \kappa$ and let $[e] \in E/\sim_{\mathfrak{H}}$ be arbitrary. For some $d \in \Delta^{\mathfrak{I}}$ we have $\mathfrak{H} \models \exists^{?\kappa(d)} u.X^{\kappa,n}_{\mathfrak{I},d}$ and $(\mathfrak{H}, e) \models X^{\kappa,n}_{\mathfrak{I},d}$. For this $[d] \in \Delta^{\mathfrak{I}}/\sim_{\mathfrak{I}}$ there is an injective function $\beta_{[e]} : [e] \longrightarrow [d]$. Again $\beta : e \longmapsto \beta_{[e]}(e)$ is an injective function for all $e \in E$ such that $(\mathfrak{H}, e) \overset{\leq\kappa}{\Longleftrightarrow}_n (\mathfrak{I}, \beta(e))$. This shows 2. and so $\mathfrak{I} \overset{\leq\kappa\,\forall}{\Longleftrightarrow}_n \mathfrak{H}$.

'2. $\Longrightarrow$ 3.' We show the claim by induction over the structure of $C$. For the base case assume $\exists^{\geq\lambda} u.D \in \mathrm{Th}^{\kappa}_n(\mathfrak{I})$ where $\lambda \leq \kappa$. Hence there is $D_0 \subseteq D^{\mathfrak{I}}$ with $|D_0| \leq \kappa$. Since $\mathfrak{I} \overset{\leq\kappa\,\forall}{\Longleftrightarrow}_n \mathfrak{H}$ there is an injection $\beta : D_0 \longrightarrow \Delta^{\mathfrak{H}}$ such that $(\mathfrak{I}, d) \overset{\leq\kappa}{\Longleftrightarrow}_n (\mathfrak{H}, \beta(d))$ for all $d \in D_0$. Theorem 3.2.8 yields $\beta(D_0) \subseteq D^{\mathfrak{H}}$ and so $\mathfrak{H} \models \exists^{\geq\lambda} u.D$. With the same arguments the only-if direction is shown. Conjunctions and negations are immediate through the induction hypothesis.

'3. $\Longrightarrow$ 1.' Every global characteristic $\mathcal{ALCQ}u$-concept is a global concept. □

What has been shown in Proposition 3.2.18 for interpretations without distinguished element, can now be extended to interpretations which feature a dis-tinguished element. In this case, characteristic $\mathcal{ALCQ}u$-concepts and bounded $\mathcal{ALCQ}u$-bisimulation are assembled from global characteristic $\mathcal{ALCQ}u$-concepts and $\mathcal{ALCQ}$-concepts and similarly from $\mathcal{ALCQ}u$-bisimulation and $\mathcal{ALCQ}$-bisimulation. The only new notion is the $\mathcal{ALCQ}u$-theory of an interpretation with distinguished element:

$$\mathrm{Th}_n^\kappa(\mathfrak{I}, d) := \{C \in \mathcal{ALCQ}u(\tau) \mid (\mathfrak{I}, d) \models C, \text{rank } C \leq n \text{ and grade } C \leq \kappa\}$$

Proposition 3.2.19. *Let* $(\kappa, n) \in I$ *and* $\tau$ *finite. The following three statements are equivalent:*

1. $(\mathfrak{H}, e) \models X_{\mathfrak{I}}^{\kappa,n} \sqcap X_{\mathfrak{I},d}^{\kappa,n}$
2. $(\mathfrak{I}, d) \overset{\leq\kappa\,\forall}{\Longleftrightarrow}_n (\mathfrak{H}, e)$
3. $Th_n^\kappa(\mathfrak{I}, d) = Th_n^\kappa(\mathfrak{H}, e)$

Proof. '1. $\Longrightarrow$ 2.' is immediate from Proposition 3.2.18 and Theorem 3.2.8. '2. $\Longrightarrow$ 3.' Let $\kappa < \omega$ be fix but arbitrary. The proof is carried out by induction over $n$ and the structure of concepts, using the componentwise order: We show that for all $n < \omega$ and all $C \in \mathcal{ALCQ}u$

$$\text{if grade } C \leq \kappa \text{ and rank } C \leq n \text{ and } (\mathfrak{I}, d) \overset{\leq\kappa\,\forall}{\Longleftrightarrow}_n (\mathfrak{H}, e) \text{ then}$$
$$(\mathfrak{I}, d) \models C \iff (\mathfrak{H}, e) \models C.$$

If $C$ is an atomic concept then $C \in \mathcal{ALCQ}$; $(\mathfrak{I}, d) \overset{\leq\kappa\,\forall}{\Longleftrightarrow}_0 (\mathfrak{H}, e)$ implies $(\mathfrak{I}, d) \overset{\leq\kappa}{\Longleftrightarrow}_0 (\mathfrak{H}, e)$, and so $(\mathfrak{I}, d) \models C \iff (\mathfrak{H}, e) \models C$. If $C = \neg D$ our induction hypothesis yields $(\mathfrak{I}, d) \models D \iff (\mathfrak{H}, e) \models D$ but the latter is true iff $(\mathfrak{I}, d) \models C \iff (\mathfrak{H}, e) \models C$. Again using the induction hypothesis, the case $C = D \sqcap E$ can be treated.

Let $C = \exists^{\geq\lambda} u.D$ where $\lambda \leq \kappa$ and assume $\mathfrak{I} \models C$. Then there is a subset of $D_0 \subseteq D^{\mathfrak{I}}$ with $|D_0| = \lambda$. Since $\mathfrak{I} \overset{\leq\kappa\,\forall}{\Longleftrightarrow}_n \mathfrak{H}$ there is an injection $\beta : D_0 \longrightarrow \Delta^{\mathfrak{H}}$ such that $(\mathfrak{I}, d_0) \overset{\leq\kappa}{\Longleftrightarrow}_n (\mathfrak{H}, \beta(d_0))$ for all $d_0 \in D_0$. In particular $(\mathfrak{I}, d_0) \overset{\leq\kappa\,\forall}{\Longleftrightarrow}_n (\mathfrak{H}, \beta(d_0))$ for all $d_0 \in D_0$. The induction hypothesis yields $\beta(D_0) \subseteq D^{\mathfrak{H}}$. Hence $(\mathfrak{H}, \beta(d_0)) \models \exists^{\geq\lambda} u.D$. The only-if direction is inferred similarly.

Finally assume $C = \exists^{\geq\lambda} r.D$ with $\lambda \leq \kappa$. In this case rank $C = n + 1$. $(\mathfrak{I}, d) \models C$ iff there is $D_0 \subseteq r^{\mathfrak{I}}(d) \cap D^{\mathfrak{I}}$ such that $|D_0| = \lambda$. Since $(\mathfrak{I}, d) \overset{\leq\kappa}{\Longleftrightarrow}_{n+1} (\mathfrak{H}, e)$ there is an injective function $\beta : D_0 \longrightarrow r^{\mathfrak{H}}(e)$ such that $(\mathfrak{I}, d_0) \overset{\leq\kappa}{\Longleftrightarrow}_n (\mathfrak{H}, \beta(d_0))$ for all $d_0 \in D_0$. Since $\mathfrak{I} \overset{\leq\kappa\,\forall}{\Longleftrightarrow}_{n+1} \mathfrak{H}$ we have also $\mathfrak{I} \overset{\leq\kappa\,\forall}{\Longleftrightarrow}_n \mathfrak{H}$ and so $(\mathfrak{I}, d_0) \overset{\leq\kappa\,\forall}{\Longleftrightarrow}_n (\mathfrak{H}, \beta(d_0))$ for each $d_0 \in D_0$. The induction hypothesis for $D$ with rank $D \leq n$ and grade $D \leq \lambda$ yields $\beta(D_0) \subseteq r^{\mathfrak{H}}(e) \cap D^{\mathfrak{H}}$. Hence $(\mathfrak{H}, e) \models \exists^{\geq\lambda} r.D$. The same argument apply for the only-if direction. This proves the claim.

'3. $\Longrightarrow$ 1.' is immediate. □

Corollary 3.2.20. *Every concept $C \in \mathcal{ALCQ}u(\tau)$ such that $\mathfrak{I} \overset{<\omega\forall}{\Longleftrightarrow} \mathfrak{H}$ implies $\mathfrak{I} \models C \iff \mathfrak{H} \models C$ is equivalent to a global concept $D \in \mathcal{ALCQ}u(\tau)$.*

Proof. Let $\kappa :=$ grade $C$ and $n :=$ rank $C$ as well as $X_C := \{X_{\mathfrak{I}}^{\kappa,n} \mid \mathfrak{I} \models C\}$. We show that $\bigsqcup X_C$ is logically equivalent to $C$: Clearly $C \models \bigsqcup X_C$. If, on the other hand, $\mathfrak{H} \models \bigsqcup X_C$ then Proposition 3.2.18 yields a model $\mathfrak{I}$ of $C$ with $\mathfrak{I} \overset{\leq\kappa\forall}{\Longleftrightarrow}_n \mathfrak{H}$. Form Proposition 3.2.18 we gather $\mathfrak{H} \models C$. Hence $X_C \models C$. □

## Saturation and Hennessy-Milner-Property

We lift the restriction on $\tau$ being finite and allow signatures of arbitrary size.

Definition 3.2.21. Let $\tau$ be a signature, $\mathfrak{I}$ a $\tau$-interpretation and let $\Gamma \subseteq \mathcal{ALCQ}(\tau)$.

1. $\Gamma$ is an $\exists^{\geq\kappa}u$*-type* of $\mathfrak{I}$ if for every finite subset $\Gamma_0 \subseteq \Gamma$ we have $\mathfrak{I} \models \exists^{\geq\kappa}u.\bigsqcap\Gamma_0$
2. An $\exists^{\geq\kappa}u$-type of $\mathfrak{I}$ is *realised* in $\mathfrak{I}$ if there is a set $D \subseteq \Delta^{\mathfrak{I}}$ with $|D| \geq \kappa$ such that for all $d \in D$ we have $(\mathfrak{I}, d) \models \Gamma$.

A $\tau$-interpretation $\mathfrak{I}$ is $\mathcal{ALCQ}u$*-saturated*, if all $\exists^{\geq\kappa}u$-types are realised in $\mathfrak{I}$ and $\mathfrak{I}$ is $\mathcal{ALCQ}$-saturated, i.e. if for all $d \in \Delta^{\mathfrak{I}}$, $r \in \mathsf{N_R}$ and all $\kappa < \omega$ all $\exists^{\geq\kappa}r$-types of $(\mathfrak{I}, d)$ are realised at $d$. ◇

Proposition 3.2.22. *$\mathcal{ALCQ}u$ has the Hennessy-Milner property, i.e. for any two $\mathcal{ALCQ}u$-saturated $\tau$-interpretations $\mathfrak{I}, \mathfrak{H}$ the following holds:*

$$Th(\mathfrak{I}) = Th(\mathfrak{H}) \iff \mathfrak{I} \overset{<\omega}{\Longleftrightarrow} \mathfrak{H}.$$

*where $Th(\mathfrak{I}) := \{C \in \mathcal{ALCQ}u(\tau) \mid C$ is global and $\mathfrak{I} \models C\}$.*

Proof. Let $D \subseteq \Delta^{\mathfrak{I}}$ be a finite set. We factorise this set by $\equiv_{\mathcal{ALCQ}}$. Every class $\mathfrak{d}$ in this partition has some finite cardinality $\kappa$. Let $\Gamma$ be the $\mathcal{ALCQ}$-theory of the elements in $\mathfrak{d}$ and $\kappa := |\mathfrak{d}|$. Then $\mathfrak{I} \models \exists^{\geq\kappa}u.\bigsqcap\Gamma_0$ for all finite $\Gamma_0 \subseteq \Gamma$. Since $\mathfrak{I} \equiv_{\mathcal{ALCQ}u} \mathfrak{H}$ we have $\mathfrak{H} \models \exists^{\geq\kappa}u.\bigsqcap\Gamma_0$ and so $\Gamma$ forms an $\exists^{\geq\kappa}u$-type in $\mathfrak{H}$. As $\mathfrak{H}$ is saturated, this type is realised and a subset $E_{\mathfrak{d}}$ of cardinality $\kappa$ can be extracted.

Hence for every class $\mathfrak{d}$ in the partition of $D$ there is an injection $\beta_\mathfrak{d} : \mathfrak{d} \longrightarrow E_\mathfrak{d}$. We set $\beta : d \longmapsto \beta_\mathfrak{d}(d)$ for all $d \in D$ where $\mathfrak{d} \in D/{\equiv_{\mathcal{ALCQ}}}$ is the class of $d$. $\beta : D \longrightarrow \Delta^\mathfrak{H}$ satisfies $(\mathfrak{I}, d) \equiv_{\mathcal{ALCQ}} (\mathfrak{H}, \beta(d))$ for all $d \in D$ and is well defined and injective: It is well defined as intersections between classes are empty and it is injective since corresponding sets $E_{\mathfrak{d}_0}$ and $E_{\mathfrak{d}_1}$ are disjoint if $\mathfrak{d}_0$ and $\mathfrak{d}_1$ are disjoint, as the elements of $E_{\mathfrak{d}_0}$ and $E_{\mathfrak{d}_0}$ satisfy different $\mathcal{ALCQ}$-theories.

As $\mathfrak{I}$ and $\mathfrak{H}$ are in particular $\mathcal{ALCQ}$-saturated, the Hennessy-Milner-Property of $\mathcal{ALCQ}$ yields $(\mathfrak{I}, d) \overset{<\omega}{\Longleftrightarrow} (\mathfrak{H}, \beta(d))$ and so 1. in Definition 3.2.17 is satisfied. The same arguments apply in order to show 2. Hence $\mathfrak{I} \overset{<\omega}{\Longleftrightarrow} \mathfrak{H}$.

The only-if direction follows from the invariance of $\mathcal{ALCQ}u$-concepts under $\overset{<\omega}{\Longleftrightarrow}$. □

## The Characterisation Theorem for $\mathcal{ALCQ}u$-Concepts

For this section, let $I := (\omega \setminus \{0\}) \times \omega$. W.l.o.g. we may assume that the signature $\tau$ is finite.

Theorem 3.2.23. *Let $\varphi(x) \in \mathrm{FO}(\tau)$. Then $\varphi$ is equivalent to an $\mathcal{ALCQ}u$-concept over $\tau$ iff $\varphi$ is invariant under $\overset{<\omega}{\Longleftrightarrow}^\forall$.*

The following proposition is analogous to Proposition 3.2.14. Instead of providing an additional Lemma like Lemma 3.2.15 we have just one proof, which first proves the analogue statement of Lemma 3.2.15 and then clinches the final argument. Since the proof is only there to point out the details but uses the exact same rationale as Lemma 3.2.15 and Proposition 3.2.14 respectively, the reader may simply skip the proof.

Proposition 3.2.24. *Every FO-formula that is invariant under $\overset{<\omega}{\Longleftrightarrow}^\forall$ is invariant under $\overset{<\kappa}{\Longleftrightarrow}^\forall_n$ for some $(\kappa, n) \in I$.*

Proof. We first show that if the claim does not hold, pointed interpretations $(\mathfrak{I}^*, d)$ and $(\mathfrak{H}^*, e)$ exists such that

$$\varphi(x) \dashv (\mathfrak{I}^*, d) \overset{<\omega}{\Longleftrightarrow}^\forall (\mathfrak{H}^*, e) \models \neg\varphi(x).$$

Assume that the statement of our proposition is not true. Then for every $(\kappa, n) \in I$ there are pointed $\tau$-interpretations $(\mathfrak{I}, d), (\mathfrak{H}, e)$ such that $\varphi(x) \dashv (\mathfrak{I}, d) \overset{\leq\kappa}{\Longleftrightarrow}^\forall_n$

$(\mathfrak{H}, e) \models \neg\varphi(x)$. The set

$$\mathfrak{X}^{\kappa,n} := \{X_{\mathfrak{I}}^{\kappa,n} \sqcap X_{\mathfrak{I},d}^{\kappa,n} \mid \exists(\mathfrak{I}, d), (\mathfrak{H}, e) : \varphi(x) \dashv (\mathfrak{I}, d) \overset{\leq\kappa\,\forall}{\Longleftrightarrow}_n (\mathfrak{H}, e) \models \neg\varphi\},$$

where $X_{\mathfrak{I}}^{\kappa,n}$ is the characteristic $\mathcal{ALCQ}u$-concept for $\mathfrak{I}$ over $\tau$ on level $(\kappa, n)$ and $X_{\mathfrak{I},d}^{\kappa,n}$ is the characteristic $\mathcal{ALCQ}$-concept, is finite and not empty. Compactness of FO yields that $\mathfrak{X} := \{\bigsqcup \mathfrak{X}^{\kappa,n} \mid (\kappa, n) \in I\}$ is satisfiable with $\varphi$ and so we obtain some model $(\mathfrak{I}, d)$ of $\mathfrak{X} \cup \{\varphi(x)\}$.

For this $(\mathfrak{I}, d)$, let $X_{\mathfrak{I},d} := \{X_{\mathfrak{I}}^{\kappa,n} \sqcap X_{\mathfrak{I},d}^{\kappa,n} \mid (\kappa, n) \in I\}$. Since $(\mathfrak{I}, d) \models \mathfrak{X}$, every finite subset of $\mathfrak{X}_{\mathfrak{I},d}$ is satisfiable with $\neg\varphi$. Compactness of FO yields: there is $(\mathfrak{H}, e) \models \mathfrak{X}_{\mathfrak{I},d} \cup \{\neg\varphi(x)\}$.

It follows $\varphi(x) \dashv (\mathfrak{I}, d) \equiv_{\mathcal{ALCQ}u} (\mathfrak{H}, e) \models \neg\varphi(x)$. Both interpretations have $\omega$-saturated extensions $(\mathfrak{I}^*, d)$ and $(\mathfrak{H}^*, e)$ such that $\varphi(x) \dashv (\mathfrak{I}^*, d) \equiv_{\mathcal{ALCQ}u} (\mathfrak{H}^*, e) \models \neg\varphi(x)$. The Hennessy-Milner-Property of $\mathcal{ALCQ}u$ (cf. Proposition 3.2.22) yields

$$\varphi(x) \dashv (\mathfrak{I}^*, d) \overset{<\omega\,\forall}{\Longleftrightarrow} (\mathfrak{H}^*, e) \models \neg\varphi(x).$$

This however contradicts the assumption $\varphi$ would be invariant under $\overset{<\omega\,\forall}{\Longleftrightarrow}$. Hence there must be $(\kappa, n) \in I$ such that $\varphi$ is invariant under $\overset{\leq\kappa\,\forall}{\Longleftrightarrow}_n$. □

Proof of Theorem 3.2.23. Assume $\varphi \in \mathrm{FO}(\tau)$ is invariant under $\mathcal{ALCQ}u$-bisimulation. Then $\varphi$ is invariant under $\overset{\leq\kappa\,\forall}{\Longleftrightarrow}_n$ for some $(\kappa, n) \in I$. Let $C_\varphi := \{X_{\mathfrak{I}}^{\kappa,n} \sqcap X_{\mathfrak{I},d}^{\kappa,n} \mid (\mathfrak{I}, d) \models \varphi(x)\}$. Then $C_\varphi$ is finite and $\varphi \models \bigsqcup C_\varphi$.

Assume now $(\mathfrak{H}, e) \models \bigsqcup C_\varphi$. Then $(\mathfrak{H}, e) \models X_{\mathfrak{I}}^{\kappa,n} \sqcap X_{\mathfrak{I},d}^{\kappa,n}$ for some $X_{\mathfrak{I}}^{\kappa,n} \sqcap X_{\mathfrak{I},d}^{\kappa,n} \in C_\varphi$. So $(\mathfrak{H}, e) \overset{\leq\kappa\,\forall}{\Longleftrightarrow}_n (\mathfrak{I}, d)$ and since $\varphi$ is invariant under $\overset{\leq\kappa\,\forall}{\Longleftrightarrow}_n$ we have $(\mathfrak{H}, e) \models \varphi(x)$. This shows $\bigsqcup C_\varphi \models \varphi(x)$ and hence proves the theorem. □

Corollary 3.2.25 . *Every sentence $\varphi \in \mathrm{FO}(\tau)$ which is invariant under $\overset{<\omega\,\forall}{\Longleftrightarrow}$ is equivalent to some global concept $C \in \mathcal{ALCQ}u(\tau)$.*

Proof. $\varphi$ is invariant under $\overset{\leq\kappa\,\forall}{\Longleftrightarrow}_n$ for some $\kappa, n < \omega$. Let $C_\varphi := \{X_{\mathfrak{I}}^{\kappa,n} \mid \mathfrak{I} \models \varphi\}$. Then $C_\varphi$ is finite and $\varphi \models \bigsqcup C_\varphi$. If $\mathfrak{H} \models \bigsqcup C_\varphi$ then $\mathfrak{H} \models X_{\mathfrak{I}}^{\kappa,n}$ for some $\mathfrak{I} \models \varphi$ and from the invariance under $\overset{\leq\kappa\,\forall}{\Longleftrightarrow}_n$ of $\varphi$ we obtain $\mathfrak{H} \models \varphi$. □

### $\mathcal{ALCQ}u_1$

We give a characterisation for $\mathcal{ALCQ}u_1$, i.e. the logic which only allows $\exists^{\geq 1}u$-quantifications.

Definition 3.2.26 . $\mathfrak{I} \xleftrightarrow{<\omega}{}^{Q} \mathfrak{H}$ if both of the following holds:

1. for all $d \in \Delta^{\mathfrak{I}}$ there is some $e \in \Delta^{\mathfrak{H}}$ such that $(\mathfrak{I}, d) \xleftrightarrow{<\omega} (\mathfrak{H}, e)$
2. for all $e \in \Delta^{\mathfrak{H}}$ there is some $d \in \Delta^{\mathfrak{I}}$ such that $(\mathfrak{I}, d) \xleftrightarrow{<\omega} (\mathfrak{H}, e)$

Analogously we define $\mathfrak{I} \xleftrightarrow{\leq\kappa}{}^{Q}_{n} \mathfrak{H}$ if both of the following holds:

1. for all $d \in \Delta^{\mathfrak{I}}$ there is some $e \in \Delta^{\mathfrak{H}}$ such that $(\mathfrak{I}, d) \xleftrightarrow{\leq\kappa}_{n} (\mathfrak{H}, e)$
2. for all $e \in \Delta^{\mathfrak{H}}$ there is some $d \in \Delta^{\mathfrak{I}}$ such that $(\mathfrak{I}, d) \xleftrightarrow{\leq\kappa}_{n} (\mathfrak{H}, e)$

We also write $(\mathfrak{I}, d) \xleftrightarrow{<\omega}{}^{Q} (\mathfrak{H}, e)$ if $\mathfrak{I} \xleftrightarrow{<\omega}{}^{Q} \mathfrak{H}$ and $(\mathfrak{I}, d) \xleftrightarrow{<\omega} (\mathfrak{H}, e)$. We define $(\mathfrak{I}, d) \xleftrightarrow{\leq\kappa}{}^{Q}_{n} (\mathfrak{H}, e)$ similarly. ◇

Let $\mathfrak{I}$ be an interpretation. Then its forest-unravelling $\mathfrak{I}^F$ is defined as $\biguplus_{d\in\Delta^{\mathfrak{I}}} \mathfrak{I}_d$, where $\mathfrak{I}_d$ is the tree-unravelling of $\mathfrak{I}$ in $d$. We denote the last letter in $\bar{d}$ by $\lambda\bar{d}$

Lemma 3.2.27 . *For arbitrary* $d \in \Delta^{\mathfrak{I}}$ *we have* $(\mathfrak{I}, d) \xleftrightarrow{<\omega} (\mathfrak{I}^F, \bar{d})$ *with* $\lambda\bar{d} = d$

Proof. **II** has a winning strategy if she maintains configurations $\beta : D \longrightarrow E$ such that $\lambda \circ \beta(d) = d$ for all $d \in D$. In particular, $\beta(d)$ and $d$ are then atomically equivalent.

The start-configuration meets this requirement and by definition of the tree-unravelling $\bar{d}$ is atomically equivalent to $\lambda\bar{d}$, which is $d$; **II** does not lose the 0-th round.

Assume now the game has reached $\beta : D \longrightarrow E$ which meets the requirement and **I** chooses $d \in D$ and some finite set of $r$-successors $D'$ of $d$ for some $r \in \mathsf{N_R}$. Since $\lambda\bar{d} = d$ the definition of the tree-unravelling yields $\bar{d} \cdot r \cdot d' \in \Delta^{\mathfrak{I}^F}$ for all $d' \in D'$. Hence $\beta' : d' \longmapsto \bar{d} \cdot r \cdot d'$ is a bijection between $D'$ and $\{ d \cdot r \cdot d' \mid d' \in D' \}$ which meets the requirement. Again by definition, $\bar{d} \cdot r \cdot d'$ is atomically equivalent to $d'$ which shows that **II** can respond with an admissible move.

Assume **I** challenges **II** with some $r$-successor set $E'$ for some $r \in \mathsf{N_R}$ and $\bar{d} \in E$. Every element $\bar{d}' \in E'$ has the form $\bar{d} \cdot r \cdot \lambda\bar{d}'$. Hence $\lambda\bar{d}_0 \neq \lambda\bar{d}_1$ if $\bar{d}_0 = \bar{d}_1$ and $\lambda\bar{d}'$ is an $r$-successor of $\lambda\bar{d}$ for all $\bar{d}' \in E$. Hence $\beta : \bar{d}' \mapsto \lambda\bar{d}'$ is injective and $\beta^{-1}$ is injective and satisfies the requirement.

This shows that **II** has a winning strategy. □

Corollary 3.2.28 . $\mathfrak{I} \xleftrightarrow{<\omega}{}^{Q} \mathfrak{I}^F$

$\mathfrak{I} \xleftrightarrow{<\omega}{}^{\forall} \mathfrak{I}^F$ does not hold in general as every element $d$ in $\mathfrak{I}$ becomes a root of its own tree-unravelling in $\mathfrak{I}^F$ and several copies might be produced during the forest unravelling, if it is successor of other elements. Hence the number of elements which are $\mathcal{ALCQ}$-bisimilar to $d$ rises $\mathfrak{I}^F$ which destroys global $\mathcal{ALCQ}$-bisimilarity between $\mathfrak{I}$ and $\mathfrak{I}^F$.

Definition 3.2.29 . For finite $\tau$ the characteristic $\mathcal{ALCQ}u_1$-concept for $(\mathfrak{I}, d)$ on level $(n, \kappa)$ is defined as follows:

$$Y_{\mathfrak{I}}^{\kappa,n} := \bigsqcap\{\exists^{\geq 1}u.X_{\mathfrak{I},d}^{\kappa,n} \mid d \in \Delta^{\mathfrak{I}}\} \sqcap \forall^{\geq 1}u.\bigsqcup\{X_{\mathfrak{I},d}^{\kappa,n} \mid d \in \Delta^{\mathfrak{I}}\}$$

where $X_{\mathfrak{I},d}^{\kappa,n}$ is the characteristic $\mathcal{ALCQ}$-concept on level $(\kappa, n)$. ◇

$$\mathrm{Th}_{Q,n}^{\kappa}(\mathfrak{I}) := \{C \in \mathcal{ALCQ}u_1(\tau) \mid \mathfrak{I} \models C, C \text{ is global }, \text{rank } C \leq n, \text{grade } C \leq \kappa\}$$

Proposition 3.2.30 . *Let $\tau$ be finite and let $\mathfrak{I}, \mathfrak{H}$ be two $\tau$-interpretations The following three statements are equivalent:*

1. $\mathfrak{H} \models Y_{\mathfrak{I}}^{\kappa,n}$
2. $\mathfrak{I} \overset{\leq\kappa\, Q}{\Longleftrightarrow}_n \mathfrak{H}$
3. $Th_{Q,n}^{\kappa}(\mathfrak{I}) = Th_{Q,n}^{\kappa}(\mathfrak{H})$

Proof. '1. $\Longrightarrow$ 2.' For every $d \in \Delta^{\mathfrak{I}}$, there is $e \in \Delta^{\mathfrak{H}}$ such that $(\mathfrak{H}, e) \models X_{\mathfrak{I},d}^{\kappa,n}$. Theorem 3.2.8 yields $(\mathfrak{I}, d) \overset{\leq\kappa}{\Longleftrightarrow}_n (\mathfrak{H}, e)$. Conversely, for every $e \in \Delta^{\mathfrak{H}}$ there is a subformula $X_{\mathfrak{I},d}^{\kappa,n}$ of $\forall^{\geq 1}u.\bigsqcup\{X_{\mathfrak{I},d}^{\kappa,n} \mid d \in \Delta^{\mathfrak{I}}\}$ such that $(\mathfrak{H}, e) \models X_{\mathfrak{I},d}^{\kappa,n}$. Again Theorem 3.2.8 yields $(\mathfrak{I}, d) \overset{\leq\kappa}{\Longleftrightarrow}_n (\mathfrak{H}, e)$ and so $\mathfrak{I} \overset{\leq\kappa\, Q}{\Longleftrightarrow}_n \mathfrak{H}$.

'2. $\Longrightarrow$ 3.' Let $n, \kappa < \omega$ be arbitrary and assume $\mathfrak{I} \overset{\leq\kappa\, Q}{\Longleftrightarrow}_n \mathfrak{H}$. We show by induction upon the structure of $C$ that $C \in \mathrm{Th}_{Q,n}^{\kappa}(\mathfrak{I})$ iff $C \in \mathrm{Th}_{Q,n}^{\kappa}(\mathfrak{H})$. The induction hypothesis readily yields negation and conjunction. Let $C = \exists^{\geq 1}u.D$; recall that $D \in \mathcal{ALCQ}(\tau)$ since is $C$ is global. $\mathfrak{I} \models C$ iff there is $d \in \Delta^{\mathfrak{I}}$ such that $(\mathfrak{I}, d) \models D$. For $d \in \Delta^{\mathfrak{I}}$ there is $e \in \Delta^{\mathfrak{H}}$ such that $(\mathfrak{I}, d) \overset{\leq\kappa}{\Longleftrightarrow}_n (\mathfrak{H}, e)$ and with Theorem 3.2.8 we obtain $(\mathfrak{H}, e) \models D$. Hence $\mathfrak{H} \models \exists^{\geq 1}u.D$ which is $C \in \mathrm{Th}_{Q,n}^{\kappa}(\mathfrak{H})$. One can show with the same arguments that $\mathrm{Th}_{Q,n}^{\kappa}(\mathfrak{H}) \subseteq \mathrm{Th}_{Q,n}^{\kappa}(\mathfrak{I})$.

'3. $\Longrightarrow$ 1.' is immediate. □

Let $\tau$ be of arbitrary size. An interpretation is $\mathcal{ALCQ}u_1$*-saturated*, if it is $\mathcal{ALCQ}$-saturated and every $\exists^{\geq 1}u$-type is realised. Clearly, every $\mathcal{ALCQ}u$-saturated interpretation is $\mathcal{ALCQ}u_1$-saturated. The proof of the following proposition is straight forward.

Proposition 3.2.31 . *$\mathcal{ALCQ}u_1$ has the Hennessy-Milner property, i.e. for any two $\mathcal{ALCQ}u_1$-saturated $\tau$-interpretations $\mathfrak{I}$, $\mathfrak{H}$ the following holds*

$$Th(\mathfrak{I}) = Th(\mathfrak{H}) \iff \mathfrak{I} \overset{\leq\kappa}{\underset{}{\longleftrightarrow}}{}^{Q} \mathfrak{H}$$

Proof. Let $d \in \Delta^{\mathfrak{I}}$ be fix but arbitrary. We have $\mathfrak{H} \models \exists^{\geq 1} u.\Gamma_0$ for all finite $\Gamma_0 \subseteq \mathrm{Th}_n^{\kappa}(\mathfrak{I}, d)$ since $\mathfrak{I} \models \exists^{\geq 1} u.\Gamma_0$ and $\mathfrak{I} \equiv_{\mathcal{ALCQ}u_1} \mathfrak{H}$. Hence $\mathrm{Th}_n^{\kappa}(\mathfrak{I}, d)$ is an $\exists^{\geq 1} u$-type in $\mathfrak{H}$ and so there is $e \in \Delta^{\mathfrak{H}}$ such that $(\mathfrak{H}, e) \models \mathrm{Th}_n^{\kappa}(\mathfrak{I}, d)$. In particular, $\mathfrak{I}$ and $\mathfrak{H}$ are $\mathcal{ALCQ}$-saturated and the Hennessy-Milner-Property for $\mathcal{ALCQ}$ yields $(\mathfrak{I}, d) \overset{<\omega}{\longleftrightarrow} (\mathfrak{H}, e)$. This shows 1. in Definition 3.2.26. We obtain 2. in the same way and hence $\mathfrak{I} \overset{\leq\kappa}{\longleftrightarrow}{}^{Q} \mathfrak{H}$. The converse is immediate. □

For the rest of this section, we assume w.l.o.g. that $\tau$ contains only symbols occurring in $\varphi$. Recall that $I = \omega \setminus \{0\} \times \omega$ (page 87).

Proposition 3.2.32 . *Every $\varphi \in$ FO($\tau$) formula which is invariant under $\overset{<\omega}{\longleftrightarrow}{}^{Q}$ is invariant under $\overset{\leq\kappa}{\longleftrightarrow}{}^{Q}_{n}$ for some $(\kappa, n) \in I$.*

Proof. Assume for every $(\kappa, n) \in I$ there are pointed $\tau$-interpretations $(\mathfrak{I}, d), (\mathfrak{H}, e)$ such that $\varphi(x) \dashv (\mathfrak{I}, d) \overset{\leq\kappa}{\longleftrightarrow}{}^{Q}_{n} (\mathfrak{H}, e) \models \neg\varphi(x)$. Let

$$\mathfrak{Y}^{\kappa,n} := \{ Y_{\mathfrak{I}}^{\kappa,n} \sqcap X_{\mathfrak{I},d}^{\kappa,n} \mid \exists (\mathfrak{I}, d), (\mathfrak{H}, e) : \varphi(x) \dashv (\mathfrak{I}, d) \overset{\leq\kappa}{\longleftrightarrow}{}^{Q}_{n} (\mathfrak{H}, e) \models \neg\varphi(x) \},$$

where $Y_{\mathfrak{I}}^{\kappa,n}$ is the characteristic $\mathcal{ALCQ}u_1$-concept for $\mathfrak{I}$ over $\tau$ on level $(\kappa, n)$ and $X_{\mathfrak{I},d}^{\kappa,n}$ is the characteristic $\mathcal{ALCQ}$-concept on the level $(\kappa, n)$. $\mathfrak{Y}^{\kappa,n}$ is finite and not empty. Compactness of FO yields that $\mathfrak{Y} := \{\bigsqcup \mathfrak{Y}^{\kappa,n} \mid (\kappa, n) \in I\}$ is satisfiable with $\varphi$ and so we obtain some $(\mathfrak{I}, d) \models \mathfrak{Y} \cup \{\varphi(x)\}$.

Let $\mathfrak{Y}_{\mathfrak{I},d} := \{ Y_{\mathfrak{I}}^{\kappa,n} \sqcap X_{\mathfrak{I},d}^{\kappa,n} \mid (\kappa, n) \in I\}$. Since $(\mathfrak{I}, d) \models \mathfrak{Y}$, every finite subset of $\mathfrak{Y}_{\mathfrak{I},d}$ is satisfiable with $\neg\varphi$. Compactness of FO yields, there is $(\mathfrak{H}, e) \models \mathfrak{Y}_{\mathfrak{I},d} \cup \{\neg\varphi(x)\}$.

It follows $\varphi(x) \dashv (\mathfrak{I}, d) \equiv_{\mathcal{ALCQ}u} (\mathfrak{H}, e) \models \neg\varphi(x)$. Both interpretations have $\omega$-saturated extensions $(\mathfrak{I}^*, d)$ and $(\mathfrak{H}^*, e)$ such that $\varphi(x) \dashv (\mathfrak{I}^*, d) \equiv_{\mathcal{ALCQ}u} (\mathfrak{H}^*, e) \models \neg\varphi(x)$. The Hennessy-Milner-Property of $\mathcal{ALCQ}u_1$ (cf. Proposition 3.2.22) yields

$$\varphi(x) \dashv (\mathfrak{I}^*, d) \overset{<\omega}{\longleftrightarrow}{}^{Q} (\mathfrak{H}^*, e) \models \neg\varphi(x).$$

This contradicts the assumption $\varphi$ would be invariant under $\overset{<\omega}{\longleftrightarrow}{}^{Q}$. Hence there must be $(\kappa, n) \in I$ such that $\varphi$ is invariant under $\overset{\leq\kappa}{\longleftrightarrow}{}^{Q}_{n}$. □

Theorem 3.2.33 . *Every formula $\varphi(x) \in$ FO($\tau$) that is invariant under $\overset{<\omega}{\longleftrightarrow}{}^{Q}$ is equi-* *valent to some* $C \in \mathcal{ALCQ}u_1(\tau)$.

Proof. Proposition 3.2.32 yields that $\varphi$ is invariant under $\overset{\geq\kappa,Q}{\Longleftrightarrow}_n$ for some $(\kappa, n) \in I$. Let $C_\varphi := \{Y^{\kappa,n}_{\mathfrak{I}} \sqcap X^{\kappa,n}_{\mathfrak{I},d} \mid (\mathfrak{I}, d) \models \varphi(x)\}$. Then $\varphi(x) \models \bigsqcup C_\varphi$. We also have $\bigsqcup C_\varphi \models \varphi(x)$: every $C_\varphi$-satisfying pointed interpretation is $\overset{\leq\kappa,Q}{\Longleftrightarrow}_n$ bisimilar to some model $(\mathfrak{I}, d)$ of $\varphi(x)$. Since $\varphi(x)$ is invariant $\overset{\leq\kappa,Q}{\Longleftrightarrow}_n$, this interpretation satisfies $\varphi(x)$. □

Corollary 3.2.34. *Every sentence* $\varphi \in \mathrm{FO}(\tau)$ *which is invariant under* $\overset{<\omega,Q}{\Longleftrightarrow}$ *is equivalent to some global concept* $C \in \mathcal{ALCQ}u_1(\tau)$.

### 3.2.3 The Characterisation of $\mathcal{ALCQ}$-TBoxes

We call $C \sqsubseteq D$ an $\mathcal{ALCQ}$*-concept inclusion* if $C$ and $D$ are $\mathcal{ALCQ}$-concepts. An interpretation $\mathfrak{I}$ *is a model of* $C \sqsubseteq D$ or *satisfies* $C \sqsubseteq D$, in symbols $\mathfrak{I} \models C \sqsubseteq D$ if $C^{\mathfrak{I}} \subseteq D^{\mathfrak{I}}$. A finite set $\mathcal{T}$ of $\mathcal{ALCQ}$-concept inclusions is called $\mathcal{ALCQ}$*-TBox* and $\mathfrak{I} \models \mathcal{T}$ if $\mathfrak{I}$ is a model of every concept-inclusion $C \sqsubseteq D \in \mathcal{T}$.

Proposition 3.2.35. *Every global* $\mathcal{ALCQ}u_1$*-concept over* $\tau$ *is equivalent to a boolean combination of* $\mathcal{ALCQ}$*-TBoxes.*

Proof. First consider the global concept $\exists^{\geq 1} u.D$, where $D$ is an $\mathcal{ALCQ}$-concept. Then $\exists^{\geq 1} u.D$ is logically equivalent to $\neg\forall^{\geq 1} u.\neg D$, which is logically equivalent to $\neg\{D \sqsubseteq \bot\}$. The inductive definition of global concepts allows to recursively replace global concepts $C$ yielding a boolean combination of $\mathcal{ALCQ}$-TBoxes. □

Proposition 3.2.36. $\mathcal{ALCQ}$*-TBoxes are invariant under disjoint unions, i.e. for any family* $(\mathfrak{I}_i)_{i\in I}$ *of* $\tau$*-interpretations and any* $\mathcal{ALCQ}$*-TBox over* $\tau$ *we have*

$$(\forall i \in I : \mathfrak{I} \models \mathcal{T}) \iff \biguplus_{i\in I} \mathfrak{I} \models \mathcal{T}$$

Proof. '$\Longrightarrow$' Let for all $i \in I$ be $\mathfrak{I}_i \models \mathcal{T}$ and assume $C \sqsubseteq D \in \mathcal{T}$. Assume $d \in C^{\biguplus \mathfrak{I}_i}$. Then there is $i \in I$ such that $d \in \Delta^{\mathfrak{I}_i}$ and $(\mathfrak{I}, d) \overset{<\omega}{\Longleftrightarrow} (\biguplus_{i\in I}\mathfrak{I}_i, d)$. Hence $d \in C^{\mathfrak{I}_i}$ and since $\mathfrak{I}_i \models \mathcal{T}$ also $d \in D^{\mathfrak{I}_i}$. But $(\mathfrak{I}, d) \overset{<\omega}{\Longleftrightarrow} (\biguplus_{i\in I}\mathfrak{I}_i, d)$ and so $d \in D^{\biguplus \mathfrak{I}_i}$. Hence $\biguplus_{i\in I}\mathfrak{I}_i \models C \sqsubseteq D$. This shows $\biguplus_{i\in I}\mathfrak{I}_i \models \mathcal{T}$.

'$\Longleftarrow$' Let $\biguplus_{i\in I}\mathfrak{I} \models \mathcal{T}$ and assume $C \sqsubseteq D \in \mathcal{T}$ and let $i \in I$ be arbitrary. For any $d \in C^{\mathfrak{I}_i}$ we have $(\mathfrak{I}_i, d) \overset{<\omega}{\Longleftrightarrow} (\biguplus_{i\in I}\mathfrak{I}_i, d)$. Therefore $d \in C^{\biguplus \mathfrak{I}_i}$. Since $\biguplus_{i\in I}\mathfrak{I}_i \models \mathcal{T}$, we also have $d \in D^{\biguplus \mathfrak{I}_i}$. Again $(\mathfrak{I}_i, d) \overset{<\omega}{\Longleftrightarrow} (\biguplus_{i\in I}\mathfrak{I}_i, d)$ and so $d \in D^{\mathfrak{I}_i}$. Hence $\mathfrak{I}_i \models$ $C \sqsubseteq D$ and this shows $\mathfrak{I}_i \models \mathcal{T}$. Since $i$ was arbitrary in $I$ we have $\forall i \in I : \mathfrak{I} \models \mathcal{T}$. □

THEOREM 3.2.37. *Every* FO*-sentence which is invariant under* $\xleftrightarrow{<\omega}_Q$ *and invariant under disjoint unions is equivalent to an* $\mathcal{ALCQ}$*-TBox.*

PROOF. Let $\varphi \in \mathrm{FO}(\tau)$ which is invariant under $\xleftrightarrow{<\omega}$ and disjoint unions and let $\mathrm{cons}\,\varphi = \{C \sqsubseteq D \mid C, D \in \mathcal{ALCQ}(\tau) \text{ and } \varphi \models C \sqsubseteq D\}$. Clearly $\varphi \models \mathrm{cons}\,\varphi$.

Assume $\varphi$ would not be equivalent to any $\mathcal{ALCQ}$-TBox; then every finite $\mathcal{T} \subseteq \mathrm{cons}\,\varphi$, which is an $\mathcal{ALCQ}$-TBox, must be satisfiable with $\neg\varphi$. Compactness of FO yields that $\mathrm{cons}\,\varphi \cup \{\neg\varphi\}$ is satisfiable by some interpretation $\mathfrak{I}_{\neg\varphi}$.

Let $T := \{p \subseteq \mathcal{ALCQ}(\tau) \mid p \cup \{\varphi\} \text{ is satisfiable }\}$. For each $p \in T$ there is a model $(\mathfrak{I}_p, d_p)$ of $p \cup \{\varphi\}$. Since $\varphi$ is invariant under disjoint unions $\mathfrak{H} := \biguplus_{p \in T} \mathfrak{I}_p$ is a model of $\varphi$. Let $\mathfrak{K} := \mathfrak{H} \uplus \mathfrak{I}_{\neg\varphi}$. Both, $\mathfrak{H}$ and $\mathfrak{I}_{\neg\varphi}$ are models of $\mathrm{cons}\,\varphi$ and therefore $\mathfrak{K} \models \mathrm{cons}\,\varphi$. But $\varphi$ is invariant under disjoint unions, which implies that $\mathfrak{K} \not\models \varphi$.

Let $\mathfrak{H}^*$ and $\mathfrak{K}^*$ be the $\omega$-saturated extension of $\mathfrak{H}$ and $\mathfrak{K}$ respectively. We show that $\mathfrak{H}^* \xleftrightarrow{<\omega}_Q \mathfrak{K}^*$; for then we derive a contradiction to $\varphi$ being invariant under $\xleftrightarrow{<\omega}_Q$ which, in turn, proves that $\varphi$ must be equivalent to some $\mathcal{ALCQ}$-TBox.

Clearly, for all $d \in \Delta^{\mathfrak{H}^*}$ we have $(\mathfrak{H}^*, d) \xleftrightarrow{<\omega} (\mathfrak{K}^*, d)$. Let now $e \in \Delta^{\mathfrak{K}^*}$. If any finite subset $p \subseteq \mathrm{Th}_{\mathcal{ALCQ}}(\mathfrak{K}^*, e) := \{C \in \mathcal{ALCQ}(\tau) \mid e \in C^{\mathfrak{K}^*}\}$ is not satisfiable with $\varphi$ then $\varphi \models \left(\bigsqcap p\right) \sqsubseteq \bot$ and $\mathfrak{K}^*$ would not be a model of $\mathrm{cons}\,\varphi$. So $p \in T$ and $(\mathfrak{H}, d_p) \models p$ for all finite subsets $p \subseteq \mathrm{Th}_{\mathcal{ALCQ}}(\mathfrak{K}^*, e)$. Therefore, $\mathrm{Th}_{\mathcal{ALCQ}}(\mathfrak{K}^*, e)$ is an $\exists^{\geq 1}u$-type in $\mathfrak{H}^*$ and must be realised by some element $d \in \Delta^{\mathfrak{H}^*}$ because $\mathfrak{H}^*$ is saturated. The Hennessy-Milner-Property of $\mathcal{ALCQ}$ yields $(\mathfrak{H}^*, d) \xleftrightarrow{<\omega} (\mathfrak{K}^*, e)$. Hence for all $e \in \Delta^{\mathfrak{K}^*}$ there is $d \in \Delta^{\mathfrak{H}^*}$ with $(\mathfrak{H}^*, d) \xleftrightarrow{<\omega} (\mathfrak{K}^*, e)$. This shows $\mathfrak{H}^* \xleftrightarrow{\leq\kappa}_Q \mathfrak{K}^*$ and derives the contradiction. □

Using adapted notions, we have proved characterisation theorems of $\mathcal{ALCQ}$ on different levels. The notions used and in particular $\mathcal{ALCQ}$-bisimulation are in good coherence with those used to characterise the different fragments of $\mathcal{ALC}$ and $\mathcal{ALCI}$: The Hennessy-Milner-Property could be proved for $\mathcal{ALCQ}$-concepts, $\mathcal{ALCQ}u$-concepts and $\mathcal{ALCQ}u_1$-concepts and we found the counterparts of characteristic $\mathcal{ALC}$-concepts for all three mentioned logics. All this reassures us in the choice of our conceptualisation.

Additionally we obtain insights into the expressiveness of $\mathcal{ALCQ}$: it only allows to talk about finite successor sets and co-finite successor sets (i.e. their complement w.r.t. all successors of the element is finite) in a precise way, as we can only state that something is true for more than a finite cardinality, $\exists^{\geq\kappa}.C$, or at most a finite cardinality $\exists^{\leq\kappa}.C$. This only changes for theories, e.g. $\{\exists^{\geq\kappa}.C \mid \kappa < \omega\}$. Furthermore, even though we gained control over the multiplicity of successor elements, $\mathcal{ALCQ}$ and $\mathcal{ALCQ}u_1$ only allow to encode FO-properties which are invariant under tree-unravellings. This is different for $\mathcal{ALCQ}u$ which allows for the control of multiplicity on the global level.

# 4. The Characterisation of $\mathcal{ALCO}$, $\mathcal{ALCQO}$ and $\mathcal{ALCQIO}$

$\mathcal{ALCO}$ [4, 3] (and its extensions $\mathcal{ALCQO}$ and $\mathcal{ALCQIO}$) is an extension of $\mathcal{ALC}$ by individuals, also called nominals, whose letter 'o' explains the $\mathcal{O}$ in $\mathcal{ALCO}$. The classical counterpart of individuals are constants in FO, i.e. those symbols, which are interpreted as a nullary function, yielding exactly one element. Nominals were first investigated within modal logics, called Hybrid Logics, already in the 1950s by Prior (cf. [18, Notes to Chapter 7]). In different settings, nominals were re-investigated over the years in [28, 108, 17] until they eventually became an integral part of DLs [88].

In the spirit of description logics, individual names are rather interpreted as singleton subsets of the carrier-set of an interpretation, which brings certain technical difficulties along as we shall see.

Traditionally [23], individual names occur in ABoxes and are not considered to be a feature of the logic: An *$\mathcal{ALC}$-ABox assertion* is an expression $C(a)$ or $r(a, b)$ where $a$ and $b$ are individuals, $C$ is a concept and $r$ is a role name. A finite set of $\mathcal{ALC}$-ABox assertions is called an *$\mathcal{ALC}$-ABox* and an interpretation $\mathfrak{I}$ satisfies $C(a)$ if $a^{\mathfrak{I}} \in C^{\mathfrak{I}}$ or rather $a^{\mathfrak{I}} \subseteq C^{\mathfrak{I}}$ as individual names are interpreted as singleton sets. Similarly $\mathfrak{I}$ satisfies $r(a, b)$ if $(a^{\mathfrak{I}}, b^{\mathfrak{I}}) \in r^{\mathfrak{I}}$.

So in an ABox, we can express properties for a certain individual $a$, say, but we cannot express that e.g. a certain property holds for all elements which are not $a$. Description logics which are extended by individual names can now refer to these specific objects in a concept and therefore allow to refer to them in TBoxes as well:

Consider $\mathsf{N_I} := \{\mathsf{Anne}, \mathsf{John}\}$ and $\mathsf{N_R} := \{\mathsf{isSisterOf}, \mathsf{isBrotherOf}, \mathsf{isChildOf}\}$

and $\mathsf{N_C} := \{\mathsf{Boy}\}$. With individuals it is now possible to express the TBox

$$\mathcal{T} := \{(\exists \mathsf{isChildOf}.\mathsf{Anne}) \sqcap \neg \mathsf{John} \sqsubseteq (\exists \mathsf{isSisterOf}.\mathsf{John}) \sqcup (\exists \mathsf{isBrotherOf}.\mathsf{John})\}$$

which states that every child of Anne which is not John himself, must be a brother or a sister of John. This extends the expressivity of the original language $\mathcal{ALC}$ and generalises TBoxes as now all ABox-facts, provided the constants are interpreted as the appropriate elements, are expressible as TBox Axioms:

$$\begin{aligned} \mathcal{A} &:= \{\mathsf{isChildOf}(\mathsf{John}, \mathsf{Anne}), \mathsf{Boy}(\mathsf{John})\} \\ \mathcal{T} &:= \{\mathsf{John} \sqsubseteq \exists \mathsf{isChildOf}.\mathsf{Anne},\ \mathsf{John} \sqsubseteq \mathsf{Boy}\} \end{aligned}$$

$\mathcal{T}$ is logically equivalent to $\mathcal{A}$; Hence reasoning with TBoxes in $\mathcal{ALCO}$ subsumes reasoning with knowledge bases consisting of TBoxes and ABoxes.

In this chapter we shall investigate in sequence $\mathcal{ALCO}$, $\mathcal{ALCQO}$ and $\mathcal{ALCQIO}$. In the development of $\mathcal{ALCO}$ we shall justify and explain issues arising with the extension of individuals. Whilst there is little new concerning the behaviour of $\mathcal{ALCO}$-concepts and $\mathcal{ALCO}u$-concepts, we shall require a modified notion of substructure and disjoint union. These results were published in [92].

The altered notion of disjoint unions made the investigation of $\mathcal{ALCQO}$ which underpins $\mathcal{SHOQ}$ [80] particularly interesting. The model-theoretic investigation of $\mathcal{ALCQO}$ is therefore novel.

Finally we present the results for $\mathcal{ALCQIO}$, where the incorporation of inverse quantification eases the technical difficulties which arise in $\mathcal{ALCQO}$. The results for $\mathcal{ALCQIO}$ were published in [92] as well.

## 4.1 The Model Class $\mathbb{K}$

With $\mathsf{N_I}$ we denote the set of *individual names* in our signature. Most of the time, we shall use letters $a$ and $b$ for individual names. We continue the recursive definition of the interpretation function for $\mathcal{ALC}$-concepts by defining $a^{\mathfrak{I}}$ to be a singleton subset of $\Delta^{\mathfrak{I}}$. In particular no interpretation of an individual name is empty, though we do not require different individual names to have different interpretations. Hence we do not have the *unique name assumption.*

Although we are inclined to think the interpretation of $a$ as single element, the individual, we define $a^{\mathfrak{I}}$ to be a singleton set since $(C \sqcap a)^{\mathfrak{I}} = C^{\mathfrak{I}} \cap a^{\mathfrak{I}}$ is only a well formed set-theoretic term, if $a$ is interpreted as set. This is the typical description-logic flavour in contrast to the more classical approach e.g. in FO where such symbols would be interpreted as constants.

This comes along with the technical difficulty mentioned above: As individual symbols are not interpreted as constants but as predicates with one element, we are not looking at a special type of signatures containing constants but at a special class $\mathbb{K}$ of interpretations, namely those interpretations that assign a singleton set to every symbol in $\mathsf{N_I}$.

Note that similar to expressions like $\mathrm{Mod}_\tau$ or $\mathrm{Mod}_\tau\varphi$ denoting the class of all $\tau$-interpretations and the class of all $\tau$-models of $\varphi$, $\mathbb{K}$ is not a set in the sense of the von Neumann Universe as it contains elements of non-bounded cardinality. Hence constructions like $\biguplus \mathbb{K}$ or $\bigtimes \mathbb{K}$ are not well defined and we shall abstain from expressions like $\mathfrak{I} \in \mathbb{K}$. Instead we shall refer to '$\mathfrak{I}$ in $\mathbb{K}$' whenever we want to express that the interpretation $\mathfrak{I}$ assigns a singleton set to every $a \in \mathsf{N_I}$.

Since we explicitly want to refer to interpretations in $\mathbb{K}$ we need to adapt the notion of satisfiability and we need a restricted satisfaction relation:

Definition 4.1.1 .

1. $(\mathfrak{I}, d) \models_{\mathbb{K}} C$ if $\mathfrak{I}$ in $\mathbb{K}$ and $(\mathfrak{I}, d) \models_{\mathcal{ALC}} C$ where symbols in $\mathsf{N_I}$ are treated like symbols in $\mathsf{N_C}$.

2. $(\mathfrak{I}, d) \not\models_{\mathbb{K}} C$ if $\mathfrak{I}$ in $\mathbb{K}$ and $(\mathfrak{I}, d) \not\models_{\mathcal{ALC}} C$ where symbols in $\mathsf{N_I}$ are treated like symbols in $\mathsf{N_C}$. Hence $\mathfrak{I} \models_{\mathbb{K}} \neg C$ iff $\mathfrak{I} \not\models_{\mathbb{K}} C$.

3. A $\mathcal{ALCO}$-concept $C$ is satisfiable if there is $\mathfrak{I}$ in $\mathbb{K}$ such that $\mathfrak{I} \models_{\mathbb{K}} C$.

4. $\varphi \models_{\mathbb{K}} \psi$ if for all $\mathfrak{I}$ in $\mathbb{K}$ we have $\mathfrak{I} \models_{\mathbb{K}} \varphi$ implies $\mathfrak{I} \models_{\mathbb{K}} \psi$.

◇

For every FO-formula $\varphi(x)$, we define $\mathfrak{I}\frac{d}{x} \models_{\mathbb{K}} \varphi(x)$ if $\mathfrak{I}$ in $\mathbb{K}$ and $\mathfrak{I}\frac{d}{x} \models_{\mathrm{FO}} \varphi(x)$ and analogously to the above $\mathfrak{I}\frac{d}{x} \not\models_{\mathbb{K}} \varphi(x)$.

This might raise the question, whether $\mathcal{ALCO}$ is compact with respect to $\mathbb{K}$. Especially when later an arbitrary bisimulation invariant FO-formula is used.

Definition 4.1.2 . We define '$a \in \mathsf{N_I}$' $:= [\exists x_0.a(x_0) \wedge \forall x_0 x_1.(a(x_0) \wedge a(x_1)) \longrightarrow x_0 \equiv x_1]$ for every $a \in \mathsf{N_I}$. ◇

So '$a \in \mathsf{N_I}$' is simply the FO-sentence which states that $a$ is interpreted by exactly one element.

Proposition 4.1.3. *If every finite subset of* $\Gamma \subseteq \mathrm{FO}(\tau)$ *has a* $\tau$*-model in* $\mathbb{K}$ *then* $\Gamma$ *has a* $\tau$*-model in* $\mathbb{K}$.

Proof. Let $\varphi_a$ :=‘$a \in \mathsf{N_I}$’ and $\Theta := \{\varphi_a \mid a \in \mathsf{N_I}\}$. Every $\tau$-interpretation in $\mathbb{K}$ is a model of $\Theta$ and vice versa. So, since every finite subset $\Gamma_0$ has a model in $\mathbb{K}$, every finite subset of $\Gamma \cup \Theta$ is satisfiable. By the compactness of FO, $\Gamma \cup \Theta$ is satisfiable and hence $\Gamma$ has a $\tau$-model in $\mathbb{K}$. □

We use the standard translation of $\mathcal{ALC}$ to FO for $\mathcal{ALCO}$-concepts, where we treat individual names like concept names. This readily shows that also $\mathcal{ALCO}$ is compact over $\mathbb{K}$.

Observation 4.1.4. *Every interpretation in* $\mathfrak{I}$ *in* $\mathbb{K}$ *has an* $\omega$*-saturated extension* $\mathfrak{I}^*$ *in* $\mathbb{K}$ *that satisfies the same* FO*-theory.*

Proof. if $\mathfrak{I}$ in $\mathbb{K}$ then $\mathfrak{I}$ is a model of the FO-sentence $\varphi$ :=‘$a \in \mathsf{N_I}$’ for all $a \in \mathsf{N_I}$. As $\mathfrak{I}$ and $\mathfrak{I}^*$ have the same FO-theory, i.e. they satisfy the same sentences, $\mathfrak{I}^*$ is a model of every $\varphi$ :=‘$a \in \mathsf{N_I}$’, too, and so $\mathfrak{I}^*$ in $\mathbb{K}$. □

## 4.2 $\mathcal{ALCO}$

We start with the usual introduction of syntax and semantics and we shall then elaborate on the special circumstances that arise with the interpretation of individual names as singleton predicates.

From there on we follow the scheme we previously applied: we shall use $\mathcal{ALC}$-bisimulation as model-theoretic relation between pointed interpretations, shortly discuss saturation and the Hennessy-Milner-Property and finally give a characterisation for $\mathcal{ALCO}$-concepts within FO. The same scheme is applied for $\mathcal{ALCO}u$-concepts.

For $\mathcal{ALCO}$-TBoxes we have to introduce an adapted notion of disjoint union and an additional property that explains how, under the given constraints, FO-formulae can be preserved in tree-unravellings. We finally give a characterisation for $\mathcal{ALCO}$-TBoxes.

### 4.2.1 The Characterisation of $\mathcal{ALCO}$-Concepts

Syntax and Semantics

The syntax of an $\mathcal{ALCO}$-concept $C$ is then given by

$$C ::= \top \mid a \mid A \mid D \sqcap E \mid \neg D \mid \exists r.D$$

where $a \in \mathsf{N_I}$, $A \in \mathsf{N_C}$, $r \in \mathsf{N_R}$ and $D, E$ are $\mathcal{ALCO}$-concepts. We use the common abbreviations $\bot$, $\sqcup$, $\rightarrow$, $\forall r$ etc.

The increase in expressivity for $\mathcal{ALCO}$ on concept-level is rather moderate, but it is now possible to express the whole theory of an individual which is reachable from the distinguished element in finitely many steps by an infinite set of concepts. As an example, take an individual which is reachable within two steps via an $s$ and an $r$ labelled edge:

$$\{\forall s.\forall r.(a \rightarrow C) \mid C \in \mathcal{ALCO}(\tau) \text{ with } (\mathfrak{I}, a^{\mathfrak{I}}) \models C\}$$

Pinning down the whole $\mathcal{ALC}$-theory for a single element in $\mathcal{ALC}$ was previously only possible in the distinguished element itself. We saw Example 2.1.18 in which the inability to express the whole $\mathcal{ALC}$-theory for a specific successor-element led to the failure of capturing even 1-bisimilarity. We shall make use of this new ability when characterising $\mathcal{ALCO}$-TBoxes.

As shortcoming of $\mathcal{ALCO}$, we can still only express properties of individuals which are reachable from the distinguished element. This problem will be overcome with the introduction of the universal role.

As the interpretation as singleton set is given as feature of the interpretation and not as property of the logic, individual names can be treated as concept names. We are hence looking at a special class of interpretations $\mathbb{K}$. Thus the definition for $\mathcal{ALC}$-bisimulation can be immediately used as $\mathcal{ALCO}$-bisimulation. This is distinctively different from the notion of hybrid bisimulation introduced in [2], where elements are 'stored' in the tuples of the bisimulation. For the following statements the proofs of the analogous statements for $\mathcal{ALC}$ can be used:

Observation 4.2.1 . *$\Longleftrightarrow$ is an equivalence relation on the set of pointed interpretations in $\mathbb{K}$. For all pointed interpretations in $\mathbb{K}$ and every $n < \omega$ we have*

1. *$\Longleftrightarrow_n$ is an equivalence relation on pointed interpretations in $\mathbb{K}$*

2. *If $(\mathfrak{I}, d) \Longleftrightarrow_{n+1} (\mathfrak{H}, e)$ then $(\mathfrak{I}, d) \Longleftrightarrow_n (\mathfrak{H}, e)$*

3. *If* $(\mathfrak{I}, d) \Longleftrightarrow (\mathfrak{H}, e)$ *then* $(\mathfrak{I}, d) \Longleftrightarrow_n (\mathfrak{H}, e)$

Observation 4.2.2. *For pointed interpretations* $(\mathfrak{I}, d), (\mathfrak{H}, e)$ *in* $\mathbb{K}$*, all* $n < \omega$ *and every* $\mathcal{ALCO}$*-concept* $D$ *we have*

$$\text{If } (\mathfrak{I}, d) \Longleftrightarrow_n (\mathfrak{H}, e) \text{ and } \operatorname{rank} D = n \text{ then } d \in D^{\mathfrak{I}} \text{ iff } e \in D^{\mathfrak{H}}.$$

Let $\tau = \mathsf{N_C} \cup \mathsf{N_R} \cup \mathsf{N_I}$ be finite. *Characteristic* $\mathcal{ALCO}$*-concepts* $X^n_{\mathfrak{I},d}$ for a $\tau$-interpretation $\mathfrak{I}$ on level $n$ are defined as characteristic concepts for $\mathcal{ALC}$ (cf. Definition 2.1.13) where again elements in $\mathsf{N_I}$ are treated as concept names.

Due to treating individual names as concept names, characteristic $\mathcal{ALCO}$-concepts incorporate individual names on level 0 and the definition for characteristic $\mathcal{ALCO}$-concepts on level $n \geq 1$ can be immediately transferred from $\mathcal{ALC}$. One has to be careful, though: For $\mathcal{ALC}$, the structure of characteristic concepts suggested that every concept $C$ that had the syntactical make up of a characteristic formula was indeed a characteristic formula, i.e. there was a model $(\mathfrak{I}, d)$ and $n < \omega$ such that $C \equiv X^n_{\mathfrak{I},d}$. This is not the case anymore:

$$X^1 = (a \sqcap b) \sqcap \exists r.(a \sqcap \neg b) \sqcap \forall r.(a \sqcap \neg b)$$

does not have a model in $\mathbb{K}$ yet would be equivalent to some characteristic concept for some model on level 1 if $\mathsf{N_C} := \{a, b\}, \mathsf{N_R} := \{r\}$ and $\mathsf{N_I} = \emptyset$. Caveats of this type can be avoided if we always refer to an interpretation $(\mathfrak{I}, d)$ in the index of $X^n_{\mathfrak{I},d}$ which is an interpretation in the class $\mathbb{K}$. For then every element that is not identical (we stay with the idea of individuals) with the interpretation of an individual name, $a$ say, satisfies $\neg a$.

Proposition 4.2.3. *For finite* $\tau$ *and pointed* $\tau$*-interpretations* $(\mathfrak{I}, d)$*,* $(\mathfrak{H}, e)$ *in* $\mathbb{K}$ *the following statement is equivalent for all* $n < \omega$

1. $(\mathfrak{H}, e) \models_{\mathbb{K}} X^n_{\mathfrak{I},d}$

2. $(\mathfrak{I}, d) \Longleftrightarrow_n (\mathfrak{H}, e)$

3. $\mathit{Th}_n(\mathfrak{I}, d) = \mathit{Th}_n(\mathfrak{H}, e)$

*where* $\mathit{Th}_n := \{C \in \mathcal{ALCO} \mid (\mathfrak{I}, d) \models_{\mathbb{K}} C \text{ and } \operatorname{rank} C = n\}$

We lift the restriction to finite signatures. Definition 2.1.21 for $\mathcal{ALC}$-types, realisation of types and the notion of saturation (Definition 2.1.22) are the same for $\mathcal{ALCO}$. Of course now an $\exists r.$-type of an element $d$ in a $\tau$-interpretation $\mathfrak{I}$ is a set $\Gamma \subseteq \mathcal{ALCO}(\tau)$ such that for every finite subset $\Gamma_0 \subseteq \Gamma$ we have $(\mathfrak{I}, d) \models \exists r. \bigsqcap \Gamma_0$.

Proposition 4.2.4 . *$\mathcal{ALCO}$ has the Hennessy-Milner-Property.*

$\mathcal{ALCO}$-Concepts as Bisimulation Invariant FO-Fragment over $\mathbb{K}$

We need to be careful not to neglect signature symbols as we usually do, in order to assure that we really show the claim for interpretations in $\mathbb{K}$. Recall (page 30) that $\mathfrak{I}\restriction\tau$ is the $\tau$-reduct of $\mathfrak{I}$.

Lemma 4.2.5 . *Let $\tau$ be a signature, $\varphi \in \mathrm{FO}(\tau)$ be invariant under $\Longleftrightarrow_\dagger$ in $\mathbb{K}$ with $\dagger < \omega$ or omitted, and $\tau_\varphi \subseteq \tau$ exactly the set of symbols of $\tau$ occurring in $\varphi$. Then $(\mathfrak{I}\restriction\tau_\varphi, d) \Longleftrightarrow_\dagger (\mathfrak{H}\restriction\tau_\varphi, e)$ implies $\mathfrak{I}\frac{d}{x} \models \varphi(x)$ iff $\mathfrak{H}\frac{e}{x} \models \varphi(x)$*

Proof. We add an element of $\mathsf{N_I}$ to $\tau_\varphi$. Assume $(\mathfrak{I}, d)$ and $(\mathfrak{H}, e)$ are $\tau$-interpretations in $\mathbb{K}$ and $(\mathfrak{I}, d) \Longleftrightarrow_\dagger (\mathfrak{H}, e)$. Let $a \in \mathsf{N_I} \cap \tau_\varphi$. We obtain $\mathfrak{I}_0$ from $\mathfrak{I}$ by adding a fresh element $d'$. In an analogous manner we obtain $\mathfrak{H}_0$ by adding $e'$. We obtain $\mathfrak{I}_1$ by redefining $b^{\mathfrak{I}_1} := \{d'\}$ for all $b \in \mathsf{N_I} \setminus \tau_\varphi$ and interpreting all other symbols $S \in (\mathsf{N_C} \cup \mathsf{N_R}) \setminus \tau_\varphi$ as $S^{\mathfrak{I}_1} := \emptyset$. In an analogously fashion we obtain $\mathfrak{H}_1$. We have

$$(\mathfrak{I}, d) \Longleftrightarrow (\mathfrak{I}_0, d) \cong_{\tau_\varphi} (\mathfrak{I}_1, d) \Longleftrightarrow_\dagger (\mathfrak{H}_1, e) \cong_{\tau_\varphi} (\mathfrak{H}_0, e) \Longleftrightarrow (\mathfrak{H}, e).$$

The chain holds for the following reasons: Adding disconnected elements goes unnoticed by $\Longleftrightarrow$, so $(\mathfrak{I}, d) \Longleftrightarrow (\mathfrak{I}_0, d)$. Moving or removing labels or edges outside of $\tau_\varphi$ cannot be detected by isomorphisms restricted to $\tau_\varphi$, so $(\mathfrak{I}_0, d) \cong_{\tau_\varphi} (\mathfrak{I}_1, d)$. Since the connected component of $d$ in $\mathfrak{I}$ is unaltered w.r.t. $\tau_\varphi$ in $\mathfrak{I}_1$ we have $(\mathfrak{I}\restriction\tau_\varphi, d) \Longleftrightarrow (\mathfrak{I}_1, d)$. Analogously we obtain $(\mathfrak{H}\restriction\tau_\varphi, e) \Longleftrightarrow (\mathfrak{H}_1, e)$ and so $(\mathfrak{I}\restriction\tau_\varphi, d) \Longleftrightarrow_\dagger (\mathfrak{H}\restriction\tau_\varphi, e)$ yields $(\mathfrak{I}_1, d) \Longleftrightarrow_\dagger (\mathfrak{H}_1, e)$. Both $\mathfrak{I}_1$ and $\mathfrak{H}_1$ are $\tau$-interpretations in $\mathbb{K}$.

We show that $\mathfrak{I}\frac{d}{x} \models \varphi(x)$ iff $\mathfrak{H}\frac{e}{x} \models \varphi(x)$: Both $\mathfrak{I}$ and $\mathfrak{I}_0$ are in $\mathbb{K}$ and $(\mathfrak{I}, d) \Longleftrightarrow (\mathfrak{I}_0, d)$ always implies $(\mathfrak{I}, d) \Longleftrightarrow_\dagger (\mathfrak{I}_0, d)$, we have $\mathfrak{I}\frac{d}{x} \models_{\mathbb{K}} \varphi(x)$ iff $\mathfrak{I}_0\frac{d}{x} \models_{\mathbb{K}} \varphi(x)$. Every FO-formula is invariant under isomorphisms of their signatures, hence $\mathfrak{I}_0\frac{d}{x} \models_{\mathbb{K}} \varphi(x)$ iff $\mathfrak{I}_1\frac{d}{x} \models_{\mathbb{K}} \varphi(x)$. Again, invariance under $\Longleftrightarrow_\dagger$ in $\mathbb{K}$ yields $\mathfrak{I}_1\frac{d}{x} \models_{\mathbb{K}} \varphi(x)$ iff $\mathfrak{H}_1\frac{e}{x} \models_{\mathbb{K}} \varphi(x)$ and using the reverse order of our arguments we obtain $\mathfrak{H}_1\frac{e}{x} \models_{\mathbb{K}} \varphi(x)$ iff $\mathfrak{H}\frac{e}{x} \models_{\mathbb{K}} \varphi(x)$. □

Lemma 4.2.6 . *If $\varphi$ is bisimulation invariant in $\mathbb{K}$ then there is $n < \omega$ such that $\varphi$ is $n$-bisimulation invariant in $\mathbb{K}$.*

This lemma is not a corollary of Lemma 2.1.25! We admit FO-formulae which are invariant under bisimulation in $\mathbb{K}$, and hence we admit more formulae than those which are bisimulation invariant within the class of all $\tau$-interpretations. In particular $\varphi$ :=‘$a \in \mathsf{N_I}$’ which is equivalent to $\top$ in $\mathbb{K}$ is such an invariant formula. Nevertheless, the arguments in the proof of Lemma 2.1.25 are analogous to those that would be used in a proof for Lemma 4.2.6, which we thus omit. We just have to recall that FO is compact w.r.t. $\mathbb{K}$ (cf. Proposition 4.1.3).

THEOREM 4.2.7. *If $\varphi(x) \in \mathrm{FO}(\tau)$ is bisimulation invariant in $\mathbb{K}$ then there is an $\mathcal{ALC}$-concept $C$ over $\tau$ such that $\mathfrak{I}\frac{d}{x} \models_{\mathbb{K}} \varphi(x)$ iff $(\mathfrak{I}, d) \models_{\mathbb{K}} C$.*

PROOF. Let $\varphi(x)$ be a bisimulation invariant FO-formula over $\tau$. According to Lemma 4.2.6, $\varphi(x)$ is $n$-bisimulation invariant for some $n < \omega$. We define $\tau_\varphi$ as all the symbols in $\tau$ that appear in $\varphi$, hence $\tau_\varphi$ is finite.

Let cons $\varphi := \{X^n_{\mathfrak{I}\restriction\tau_\varphi,d} \in \mathcal{ALCO}(\tau_\varphi) \mid \mathfrak{I}\frac{d}{x} \models_{\mathbb{K}} \varphi(x)\}$, where $X^n_{\mathfrak{I}\restriction\tau_\varphi,d}$ is the characteristic $\mathcal{ALCO}$-concept for $(\mathfrak{I}\restriction\tau_\varphi, d)$ on level $n < \omega$. As $\tau_\varphi$ is finite, cons $\varphi$ is well defined and finite, and so $\bigsqcup$ cons $\varphi$ is a concept in $\mathcal{ALCO}(\tau_\varphi)$. We shall show that $\bigsqcup$ cons $\varphi$ is the $\mathcal{ALCO}$-concept that is logically equivalent to $\varphi(x)$. We have $\mathfrak{I}\frac{d}{x} \models_{\mathbb{K}} \varphi(x)$, whenever $X^n_{\mathfrak{I}\restriction\tau_\varphi,d} \in$ cons $\varphi$ and so $(\mathfrak{I}, d) \models_{\mathbb{K}} \bigsqcup$ cons $\varphi$ entailing $\varphi(x) \models_{\mathbb{K}} \bigsqcup$ cons $\varphi$.

To show the converse, assume $\mathfrak{I}$ is a $\tau$-interpretation in $\mathbb{K}$ and $(\mathfrak{I}, d) \models_{\mathbb{K}} \bigsqcup$ cons $\varphi$. Then there is $X^n_{\mathfrak{H}\restriction\tau_\varphi,e} \in$ cons $\varphi$ such that $(\mathfrak{I}, d) \models_{\mathbb{K}} X^n_{\mathfrak{H}\restriction\tau_\varphi,e}$ and $\mathfrak{H}\frac{e}{x} \models_{\mathbb{K}} \varphi(x)$; now $(\mathfrak{I}\restriction\tau_\varphi, d) \Longleftrightarrow_n (\mathfrak{H}\restriction\tau_\varphi, e)$, and so Lemma 4.2.5 yields $\mathfrak{I}\frac{d}{x} \models_{\mathbb{K}} \varphi(x)$. □

### 4.2.2 The Characterisation of $\mathcal{ALCO}u$-Concepts

The syntax of $\mathcal{ALCO}u$ is recursively defined as follows

$$C ::= \top \mid a \mid A \mid D \sqcap E \mid \neg D \mid \exists r.D \mid \exists u.D$$

where $a \in \mathsf{N_I}$, $A \in \mathsf{N_C}$, $D, E \in \mathcal{ALCO}$ as well as $r \in \mathsf{N_R}$ and $u$ being the universal role. The definitions for the rank-function and global bisimulation stay the same. We shall use the usual abbreviations $\bot$, $\sqcup$, $\rightarrow$, $\forall r$, $\forall u$ etc.

In comparison to $\mathcal{ALCO}$-concepts which can address individuals only if they are in the scope of the distinguished element, $\mathcal{ALCO}u$ can now address arbitrary individuals elsewhere through expressions like $\forall u.(a \rightarrow C)$ which is equivalent to the @-operator of hybrid logic [21]: For every individual name $a \in \mathsf{N_I}$, hybrid logic has an operator $@_a$ where $@_a\,\varphi$, for some hybrid logic formula $\varphi$, is satisfied by an interpretation $\mathfrak{I}$ iff $(\mathfrak{I}, a^{\mathfrak{I}})$ satisfies $\varphi$. In $\mathcal{ALCO}$ the latter is true iff $\mathfrak{I} \models$ $\forall u.(a \to C_\varphi)$ for a suitable translation $C_\varphi$ of $\varphi$.

Observation 4.2.8 . *For all $n < \omega$ the following is true*

$$(\mathfrak{I}, d) \overset{g}{\Longleftrightarrow} (\mathfrak{H}, e) \Longrightarrow (\mathfrak{I}, d) \overset{g}{\Longleftrightarrow}_{n+1} (\mathfrak{H}, e) \Longrightarrow (\mathfrak{I}, d) \overset{g}{\Longleftrightarrow}_{n} (\mathfrak{H}, e)$$

For finite signature $\tau := \mathsf{N_R} \cup \mathsf{N_C} \cup \mathsf{N_I}$ the global characteristic $\mathcal{ALCO}u$-concepts are defined as we have done before:

$$X^n_{\mathfrak{I}} := \bigsqcap \{\exists u.X^n_{\mathfrak{I},d} \mid d \in \Delta^{\mathfrak{I}}\} \sqcap \forall u. \bigsqcup \{X^n_{\mathfrak{I},d} \mid d \in \Delta^{\mathfrak{I}}\}$$

where $X^n_{\mathfrak{I},d}$ is the characteristic $\mathcal{ALCO}$-concept with respect to $\tau$.

Proposition 4.2.9 . *The following is equivalent for each $n < \omega$ and finite signature $\tau$ and all $\tau$-interpretations $\mathfrak{I}, \mathfrak{H}$ in $\mathbb{K}$*

1. $(\mathfrak{H}, e) \models_{\mathbb{K}} X^n_{\mathfrak{I}} \sqcap X^n_{\mathfrak{I},d}$
2. $(\mathfrak{I}, d) \overset{g}{\Longleftrightarrow}_{n} (\mathfrak{H}, e)$
3. $\mathit{Th}_n(\mathfrak{I}, d) = \mathit{Th}_n(\mathfrak{H}, e)$

*where $X^n_{\mathfrak{I}}$ is the global characteristic $\mathcal{ALCO}u$-concept for $\tau$, $X^n_{\mathfrak{I},d}$ is the characteristic $\mathcal{ALCO}$-concept for $\tau$ and $\mathit{Th}_n(\mathfrak{I}, d) := \{C \in \mathcal{ALCO}u(\tau) \mid (\mathfrak{I}, d) \models C$ and $\operatorname{rank} C \leq n\}$.*

As this claim is true for all $\tau$-interpretations, it is in particular true for $\tau$-interpretations in $\mathbb{K}$. We lift the restriction on $\tau$ being finite. The next definition for types is the same as for $\mathcal{ALC}u$ (cf. Definition 2.2.8) and will yield appropriate notions for $\mathcal{ALCO}u$-saturation.

Definition 4.2.10 . Let $\tau$ be a signature and $\mathfrak{I}$ a $\tau$-interpretation. Let $r \in \mathsf{N_R}$ then

1. $\Gamma \subseteq \mathcal{ALCO}(\tau)$ is an *$r$-$\mathcal{ALCO}$-type* of $(\mathfrak{I}, d)$ with $d \in \Delta^{\mathfrak{I}}$ if for every finite subset $\Gamma_0 \subseteq \Gamma$ we have $d \in (\exists r. \bigsqcap \Gamma_0)^{\mathfrak{I}}$.
2. an $r$-type $\Gamma$ of $(\mathfrak{I}, d)$ is *realised at $d$* if there is some $r$-successor $d' \in \Delta^{\mathfrak{I}}$ such that $d' \in \bigcap_{C \in \Gamma} C^{\mathfrak{I}}$.
3. $\Gamma \subseteq \mathcal{ALCO}(\tau)$ is a *$u$-$\mathcal{ALCO}$-type* in $\Delta^{\mathfrak{I}}$ if for every finite subset $\Gamma_0 \subseteq \Gamma$ we have $\mathfrak{I} \models \exists u. \bigsqcap \Gamma_0$.
4. a $u$-type of $\mathfrak{I}$ is *realised in $\mathfrak{I}$* if there is $d \in \Delta^{\mathfrak{I}}$ such that $d \in \bigcap_{C \in \Gamma} C^{\mathfrak{I}}$.

◇

Definition 4.2.11 . A $\tau$-interpretation $\mathfrak{I}$ in $\mathbb{K}$ is $\mathcal{ALCO}u$-saturated if

1. for all $d \in \Delta^{\mathfrak{I}}$ and every $r \in \mathsf{N_R}$ every $r$-$\mathcal{ALCO}$-type of $d$ is realised at $d$
2. every $u$-$\mathcal{ALCO}$-type of $\mathfrak{I}$ is realised in $\mathfrak{I}$.

◇

Proposition 4.2.12 . *$\mathcal{ALCO}u$ has the Hennessy-Milner-Property, i.e. for two $\mathcal{ALCO}u$-saturated $\tau$-interpretations $\mathfrak{I}, \mathfrak{H}$ in $\mathbb{K}$ we have*

$$\mathit{Th}(\mathfrak{I}, d) = \mathit{Th}(\mathfrak{H}, e) \iff (\mathfrak{I}, d) \overset{g}{\longleftrightarrow} (\mathfrak{H}, e).$$

*where $\mathit{Th}(\mathfrak{I}, d) := \{C \in \mathcal{ALCO}u(\tau) \mid (\mathfrak{I}, d) \models C\}$ and analogously for $\mathit{Th}(\mathfrak{H}, e)$.*

Proposition 2.2.10 carries over to interpretations in $\mathbb{K}$ and thus proves this proposition. It shall be remarked again that every interpretation $\mathfrak{I}$ in $\mathbb{K}$ has an $\omega$-saturated interpretation $\mathfrak{I}^*$. As this interpretation shares the same FO-theory with $\mathfrak{I}$, it satisfies the FO-sentences $\varphi :=$'$a \in \mathsf{N_I}$' for all $a \in \mathsf{N_I}$. hence $\mathfrak{I}^*$ in $\mathbb{K}$. In particular for every $d \in \Delta^{\mathfrak{I}^*}$ every $r$-type is realised at $d$ and every $u$-$\mathcal{ALCO}$-type in $\mathfrak{I}^*$ is realised in $\mathfrak{I}^*$, thus $\mathfrak{I}^*$ is $\mathcal{ALCO}u$-saturated.

Lemma 4.2.13 . *Let $\tau$ be a signature, $\varphi \in \mathrm{FO}(\tau)$ be invariant under $\overset{g}{\longleftrightarrow}_\dagger$ in $\mathbb{K}$ with $\dagger < \omega$ or omitted, and $\tau_\varphi \subseteq \tau$ exactly the set of symbols of $\tau$ occurring in $\varphi$. Then $(\mathfrak{I}{\restriction}\tau_\varphi, d) \overset{g}{\longleftrightarrow}_\dagger (\mathfrak{H}{\restriction}\tau_\varphi, e)$ implies $\mathfrak{I}\frac{d}{x} \models \varphi(x)$ iff $\mathfrak{H}\frac{e}{x} \models \varphi(x)$*

In comparison to Lemma 4.2.5 we cannot simply add a new element to the interpretation as this could destroy $\overset{g}{\longleftrightarrow}$. Instead we have to choose one of the individuals on which we shall 'park' all non-used individual names.

Proof. We assume w.l.o.g. that $\mathsf{N_I}$ is not empty and we add an element of $\mathsf{N_I}$ to $\tau_\varphi$. Assume $(\mathfrak{I}, d)$ and $(\mathfrak{H}, e)$ are $\tau$-interpretations in $\mathbb{K}$ and $(\mathfrak{I}, d) \longleftrightarrow_\dagger (\mathfrak{H}, e)$. Let $a \in \mathsf{N_I} \cap \tau_\varphi$. We obtain $\mathfrak{I}_0$ from $\mathfrak{I}$ by redefining $b^{\mathfrak{I}_0} := a^{\mathfrak{I}}$ for all $b \in \mathsf{N_I} \setminus \tau_\varphi$ and interpreting all other symbols $S \in (\mathsf{N_C} \cup \mathsf{N_R}) \setminus \tau_\varphi$ as $S^{\mathfrak{I}_0} := \emptyset$. In an analogously fashion we obtain $\mathfrak{H}_1$. We have

$$(\mathfrak{I}, d) \cong_{\tau_\varphi} (\mathfrak{I}_0, d) \longleftrightarrow_\dagger (\mathfrak{H}_0, e) \cong_{\tau_\varphi} (\mathfrak{H}, e).$$

The chain holds for the following reasons: Moving or removing labels or edges outside of $\tau_\varphi$ cannot be detected by isomorphisms restricted to $\tau_\varphi$, so $(\mathfrak{I}, d) \cong_{\tau_\varphi}$ $(\mathfrak{I}_0, d)$. Since the connected component of $d$ in $\mathfrak{I}$ is unaltered w.r.t. $\tau_\varphi$ in $\mathfrak{I}_0$ we have $(\mathfrak{I}\restriction\tau_\varphi, d) \Longleftrightarrow (\mathfrak{I}_0, d)$. Analogously we obtain $(\mathfrak{H}\restriction\tau_\varphi, e) \Longleftrightarrow (\mathfrak{H}_0, e)$ and so $(\mathfrak{I}\restriction\tau_\varphi, d) \Longleftrightarrow_\dagger (\mathfrak{H}\restriction\tau_\varphi, e)$ yields $(\mathfrak{I}_0, d) \Longleftrightarrow_\dagger (\mathfrak{H}_0, e)$. Both $\mathfrak{I}_0$ and $\mathfrak{H}_0$ are $\tau$-interpretations in $\mathbb{K}$.

We show that $\mathfrak{I}\frac{d}{x} \models \varphi(x)$ iff $\mathfrak{H}\frac{e}{x} \models \varphi(x)$: Every FO-formula is invariant under isomorphisms of their signatures, hence $\mathfrak{I}\frac{d}{x} \models_\mathbb{K} \varphi(x)$ iff $\mathfrak{I}_0\frac{d}{x} \models_\mathbb{K} \varphi(x)$. Invariance under $\Longleftrightarrow_\dagger$ in $\mathbb{K}$ yields $\mathfrak{I}_0\frac{d}{x} \models_\mathbb{K} \varphi(x)$ iff $\mathfrak{H}_0\frac{e}{x} \models_\mathbb{K} \varphi(x)$ and using argument from above we obtain $\mathfrak{H}_0\frac{e}{x} \models_\mathbb{K} \varphi(x)$ iff $\mathfrak{H}\frac{e}{x} \models_\mathbb{K} \varphi(x)$. □

Proposition 4.2.14. *Let $\varphi(x)$ be an* FO*-formula that is invariant under $\overset{g}{\Longleftrightarrow}$ then $\varphi(x)$ is invariant under $\overset{g}{\Longleftrightarrow}_n$ for some $n < \omega$.*

Similarly to Lemma 4.2.6 the proof of the analogue Lemma 2.2.13 for $\mathcal{ALCu}$ carries over to this proposition.

Theorem 4.2.15. *Every* FO*-formula over $\tau$ which is satisfiable in $\mathbb{K}$ and invariant under $\overset{g}{\Longleftrightarrow}$ is equivalent to some $\mathcal{ALCO}u$-concept in $\mathbb{K}$*

Proof. Let $\tau_\varphi$ be the part of $\tau$ occurring in $\varphi$, and let $\tau_\varphi$ w.l.o.g. contain at least one element, say $b \in \mathsf{N_I}$. Furthermore let $\mathrm{cons}\,\varphi := \{X^n_{\mathfrak{I}\restriction\tau_\varphi} \sqcap X^n_{\mathfrak{I}\restriction\tau_\varphi, d} \mid \mathfrak{I}$ in $\mathbb{K}$ and $(\mathfrak{I}, d) \models \varphi(x)\}$, where $X^n_{\mathfrak{I}\restriction\tau_\varphi}$ is the characteristic $\mathcal{ALCO}u$-concept for $\tau$ and $X^n_{\mathfrak{I}\restriction\tau_\varphi, d}$ is the characteristic $\mathcal{ALCO}$-concept for $\tau$. This set is non-empty as $\varphi(x)$ is satisfiable over $\mathbb{K}$ and $\tau_\varphi$ is finite, so that the characteristic concepts exist. Moreover $\mathrm{cons}\,\varphi$ itself is finite and hence $\bigsqcup \mathrm{cons}\,\varphi \in \mathcal{ALCO}u(\tau_\varphi)$. We show that $\bigsqcup \mathrm{cons}\,\varphi$ is the concept we are looking for.

If $(\mathfrak{I}, d) \models \varphi(x)$ then $X^n_{\mathfrak{I}\restriction\tau_\varphi} \sqcap X^n_{\mathfrak{I}\restriction\tau_\varphi, d} \in \mathrm{cons}\,\varphi$. $(\mathfrak{I}\restriction\tau_\varphi, d) \models X^n_{\mathfrak{I}\restriction\tau_\varphi} \sqcap X^n_{\mathfrak{I}\restriction\tau_\varphi, d}$ and since every formula is invariant under signature extensions, $\mathfrak{I} \cong_{\tau_\varphi} \mathfrak{I}\restriction\tau_\varphi$ yields $(\mathfrak{I}, d) \models_\mathbb{K} \bigsqcup \mathrm{cons}\,\varphi$.

Assume, on the other hand, $(\mathfrak{I}, d) \models_\mathbb{K} \bigsqcup \mathrm{cons}\,\varphi$. Then $(\mathfrak{I}, d)$ is a model of $X^n_{\mathfrak{H}\restriction\tau_\varphi} \sqcap X^n_{\mathfrak{H}\restriction\tau_\varphi, e} \in \mathrm{cons}\,\varphi$ for some $(\mathfrak{H}, e) \models \varphi(x)$ and so is $(\mathfrak{I}\restriction\tau_\varphi, d)$. This entails $(\mathfrak{I}\restriction\tau_\varphi, d) \overset{g}{\Longleftrightarrow}_n (\mathfrak{H}\restriction\tau_\varphi, e)$ and so Lemma 4.2.13 yields $(\mathfrak{I}, d) \models \varphi(x)$. □

### 4.2.3 The Characterisation of $\mathcal{ALCO}$-TBoxes

The structure of this section is dictated by the necessary notions which are needed to characterise $\mathcal{ALCO}$-TBoxes. Merely considered as $\mathcal{ALC}$-TBoxes, they are indeed invariant under disjoint unions. Yet, the class $\mathbb{K}$ is not closed under disjoint unions, as for example, the disjoint union of $\mathfrak{I}, \mathfrak{H}$ in $\mathbb{K}$ would have to assign $a^\mathfrak{I}$ and $a^{\mathfrak{K}}$ to the individual name $a \in \mathsf{N_I}$ and hence would not be a member of $\mathbb{K}$ anymore.

The solution to this problem is to merge the individuals into one element. But another problem arises: The $\mathcal{ALCO}$-TBox $\{a \sqcap \exists r.\top \sqsubseteq \neg A\}$, where $a \in \mathsf{N_I}$, $A \in \mathsf{N_C}$ and $r \in \mathsf{N_R}$, allows—amongst others—for the following two interpretations $\mathfrak{I}, \mathfrak{K}$ in $\mathbb{K}$, where $a^{\mathfrak{I}} \subseteq A^{\mathfrak{I}}$ and $(a^{\mathfrak{K}}, a^{\mathfrak{K}}) \in r^{\mathfrak{K}}$. What should the individual $a^{\mathfrak{I} \uplus \mathfrak{K}}$ as outcome of the merging between $a^{\mathfrak{I}}$ and $a^{\mathfrak{K}}$ satisfy in order to make $\mathcal{T}$ true?

A definite answer cannot be given. Hence we do not allow disjoint unions of arbitrary interpretations in $\mathbb{K}$, but we require them to be coherent in the sense that two individuals of the same name can only be merged, if they are bisimilar.

Our road-map is therefore as follows: First, we have to formally define the process of merging and show properties we can infer from the resulting interpretations. After this, we shall give our adapted notions of disjoint unions and, in this case, also the adapted disjoint union of generated substructures before we finally prove the theorem.

### Factorising Interpretations

The merging of elements will be performed by factorising with a suitable equivalence relation. We hence define what the factor-interpretation of an interpretation is and prove a proposition on such interpretations for when they are induced by an equivalence relation which is respected by bisimulation.

Let $\mathfrak{I}$ be a $\tau$-interpretation and $E$ an equivalence relation on $\Delta^{\mathfrak{I}} \times \Delta^{\mathfrak{I}}$. For every $d \in \Delta^{\mathfrak{I}}$ we define $[d]_E := \{e \in \Delta^{\mathfrak{I}} \mid (d, e) \in E\}$ to be the class of $d$ under $E$. The *factor-interpretation* $\mathfrak{I}/E$ of $\mathfrak{I}$ induced by $E$ is defined as follows:

$$\begin{array}{lll} \Delta^{\mathfrak{I}/E} & := \{[d]_E \mid d \in \Delta^{\mathfrak{I}}\} & \\ a^{\mathfrak{I}/E} & := \{\mathfrak{d} \in \Delta^{\mathfrak{I}/E} \mid \exists e \in \mathfrak{d} : e \in a^{\mathfrak{I}}\} & \text{where } a \in \mathsf{N_I} \\ A^{\mathfrak{I}/E} & := \{\mathfrak{d} \in \Delta^{\mathfrak{I}/E} \mid \exists e \in \mathfrak{d} : e \in A^{\mathfrak{I}}\} & \text{where } A \in \mathsf{N_C} \\ r^{\mathfrak{I}/E} & := \{(\mathfrak{d}, \mathfrak{e}) \in \Delta^{\mathfrak{I}/E} \times \Delta^{\mathfrak{I}/E} \mid \exists d \in \mathfrak{d}.\exists e \in \mathfrak{e} : (d, e) \in r^{\mathfrak{I}}\} & \text{where } r \in \mathsf{N_R} \end{array}$$

Proposition 4.2.16 . *Let $\mathfrak{I}$ be an interpretation and let $E$ be an equivalence relation on $\Delta^{\mathfrak{I}}$ such that $(\mathfrak{I}, d) \leftrightarrow (\mathfrak{I}, e)$ for all $(d, e) \in E$. Then*

$$(\mathfrak{I}, d) \leftrightarrow (\mathfrak{I}, e) \iff (\mathfrak{I}, d) \leftrightarrow (\mathfrak{I}/E, [e]_E)$$

Proof. '$\Longrightarrow$': We assume $(\mathfrak{I}, d) \leftrightarrow (\mathfrak{I}, e)$ and have to show that **II** has a winning strategy for the game $G(\mathfrak{I}, d; \mathfrak{I}/E, [e]_E)$.

As a first intermediate step we show that under the condition $(\mathfrak{I}, d) \Longleftrightarrow (\mathfrak{I}, e)$, $d$ is atomically equivalent to $[e]_E$: if $d \in A^{\mathfrak{I}}$ then, since $e$ is bisimilar to $e$ also $e \in A^{\mathfrak{I}}$ and according to the definition of $A^{\mathfrak{I}/E}$ we have $[e]_E \in A^{\mathfrak{I}/E}$. If $[e]_E \in A^{\mathfrak{I}/E}$ by definition of $A^{\mathfrak{I}/E}$ there is some $e_0 \in [e]_E$ such that $e_0 \in A^{\mathfrak{I}}$. Now $e_0 \in [e]_E$ iff $(e_0, e) \in E$ and by requirement for $E$ we have $e_0$ is bisimilar to $e$. Hence $e \in A^{\mathfrak{I}}$ and since $d$ is bisimilar to $e$ by assumption, we have $d \in A^{\mathfrak{I}}$. This proves the intermediate claim.

We now show that **II** has a winning strategy if she maintains configurations $(\mathfrak{I}, d_0; \mathfrak{I}/E, [e_0]_E)$ such that $(\mathfrak{I}, d_0) \Longleftrightarrow (\mathfrak{I}, e_0)$. Clearly the start-configuration of this type. As $d$ and $[e]_E$ are atomically equivalent, **II** does not lose the 0-th round. Assume the game has reached configuration $(\mathfrak{I}, d_0; \mathfrak{I}/E, [e_0]_E)$ and $(\mathfrak{I}, d_0) \Longleftrightarrow (\mathfrak{I}, e_0)$.

Let **I** challenge **II** by moving in $\mathfrak{I}$ from $d_0$ via an edge labelled with $r \in \mathsf{N_R}$ to $d_1$. Since $d_0$ is bisimilar to $e_0$ by assumption, there is some $r$-successor $e_1$ of $e_0$ in $\mathfrak{I}$ such that $(\mathfrak{I}, d_1) \Longleftrightarrow (\mathfrak{I}, e_1)$. By the definition of $r^{\mathfrak{I}/E}$ we have $([e_0], [e_1]) \in r^{\mathfrak{I}/E}$. Hence **II** can move via an $r$-edge from $[e_0]$ to $[e_1]$. As shown at the beginning, $d_1$ and $[e_1]$ are atomically equivalent because $(\mathfrak{I}, d_1) \Longleftrightarrow (\mathfrak{I}, e_1)$. In this sense the new configuration $(\mathfrak{I}, d_1; \mathfrak{I}/E[e_1]_E)$ is admissible and satisfies the requirement that $d_1$ is bisimilar to $e_1$.

Let **I** challenge **II** by moving in $\mathfrak{I}/E$ from $[e_0]_E$ via some $r$-edge to $[e_1]_E$. By definition of $r^{\mathfrak{I}/E}$ there are $e_2 \in [e_0]_E$ and $e_3 \in [e_1]_E$ such that $(e_2, e_3) \in r^{\mathfrak{I}/E}$. But $e_2 \in [e_0]_E$ also implies that $(\mathfrak{I}, e_0) \Longleftrightarrow (\mathfrak{I}, e_2)$ by the requirement for $E$. Hence there must be an $r$-successor $e_1'$ of $e_0$ such that $(\mathfrak{I}, e_1') \Longleftrightarrow (\mathfrak{I}, e_3)$. As $(\mathfrak{I}, d_0) \Longleftrightarrow (\mathfrak{I}, e_0)$ there must be an $r$-successor $d_1$ of $d_0$ in $\mathfrak{I}$ such that $(\mathfrak{I}, d_1) \Longleftrightarrow (\mathfrak{I}, e_1') \Longleftrightarrow (\mathfrak{I}, e_3)$. **II** moves from $d_0$ via the $r$-edge to $d_1$. For then $(\mathfrak{I}, d_1) \Longleftrightarrow (\mathfrak{I}, e_3)$ and since $[e_1]_E = [e_3]_E$ be obtain a configuration $(\mathfrak{I}, d_1; \mathfrak{I}/E, [e_3]_E)$. As shown at the beginning $d_1$ and $[e_3]_E$ are atomically equivalent and hence this configuration is admissible. This proves

$$(\mathfrak{I}, d) \Longleftrightarrow (\mathfrak{I}, e) \Longrightarrow (\mathfrak{I}, d) \Longleftrightarrow (\mathfrak{I}/E, [e]_E).$$

'$\Longleftarrow$': Assume $(\mathfrak{I}, d) \Longleftrightarrow (\mathfrak{I}/E, [e]_E)$. We have to show that **II** has a winning strategy in the Game $G(\mathfrak{I}, d; \mathfrak{I}, e)$.

Again as an intermediary step we show that under the condition $(\mathfrak{I}, d) \Longleftrightarrow (\mathfrak{I}/E, [e]_E)$, $d$ is atomically equivalent to $e$: Assume $d \in A^{\mathfrak{I}}$. Then $[e]_E \in A^{\mathfrak{I}/E}$ and by definition of $A^{\mathfrak{I}/E}$ there is some element $e_0 \in [e]_E$ such that $e_0 \in A^{\mathfrak{I}}$. But then also, $(e_0, e) \in E$ which entails that $(\mathfrak{I}, e_0) \Longleftrightarrow (\mathfrak{I}, e)$, whence $e \in A^{\mathfrak{I}}$ is inferred. If, conversely, $e \in A^{\mathfrak{I}}$ then by the definition of $A^{\mathfrak{I}/E}$ also $[e] \in A^{\mathfrak{I}/E}$ and since $(\mathfrak{I}, d) \Longleftrightarrow (\mathfrak{I}/E, [e]_E)$, we have $d \in A^{\mathfrak{I}}$, which proves our claim.

We now show that **II** has a winning strategy if she maintains configurations $(\mathfrak{I}, d_0; \mathfrak{I}, e_0)$ such that $(\mathfrak{I}, d_0) \Longleftrightarrow (\mathfrak{I}/E, [e_0]_E)$. The start configuration is of the desired kind, and as just shown $d$ is atomically equivalent to $e$; so **II** does not lose the 0-th round. Assume that the game has reached configuration $(\mathfrak{I}, d_0; \mathfrak{I}, e_0)$.

Let first **I** challenge **II** by moving from $d_0$ via some $r$-edge to $d_1$ in $\mathfrak{I}$. As $(\mathfrak{I}, d_0) \Longleftrightarrow (\mathfrak{I}/E, [e_0]_E)$, there is some $r$-successor $[e_1]_E$ of $[e_0]_E$ such that $(\mathfrak{I}, d_1) \Longleftrightarrow (\mathfrak{I}/E, [e_1]_E)$. By the definition of $r^{\mathfrak{I}/E}$ there must be $e_2 \in [e_0]_E$ and $e_3 \in [e_1]_E$ such that $(e_2, e_3) \in r^{\mathfrak{I}}$. We also have $(e_2, e_0) \in E$ and by requirement for $E$ this implies $(\mathfrak{I}, e_0) \Longleftrightarrow (\mathfrak{I}, e_2)$. Hence there is an $r$-successor $e_1'$ of $e_0$ such that $(\mathfrak{I}, e_1') \Longleftrightarrow (\mathfrak{I}, e_3)$. We now have to show that $(\mathfrak{I}, d_1) \Longleftrightarrow (\mathfrak{I}/E, [e_1']_E)$. For then **II** is able to move from $e_0$ via an $r$-edge to $e_1'$ and obtaining an admissible configuration such that $(\mathfrak{I}, d_1) \Longleftrightarrow (\mathfrak{I}/E, [e_1']_E)$.

We know $(\mathfrak{I}, e_1') \Longleftrightarrow (\mathfrak{I}, e_3)$, and since $e_3 \in [e_1]_E$, we also know $(\mathfrak{I}, e_3) \Longleftrightarrow (\mathfrak{I}, e_1)$. We infer from '$\Longrightarrow$' that $(\mathfrak{I}/E, [e_1']_E) \Longleftrightarrow (\mathfrak{I}, e_3)$ and $(\mathfrak{I}, e_3) \Longleftrightarrow (\mathfrak{I}/E, [e_1]_E)$. Together with the fact that $(\mathfrak{I}, d_1) \Longleftrightarrow (\mathfrak{I}/E, [e_1]_E)$, transitivity of $\Longleftrightarrow$ yields $(\mathfrak{I}, d_1) \Longleftrightarrow (\mathfrak{I}/E, [e_1']_E)$

Conversely, let **I** challenge **II** by moving from $e_0$ via some $r$-edge to $e_1$. Then, by definition of $r^{\mathfrak{I}/E}$ we obtain $[e_1]_E$ being an $r$-successor of $[e_0]_E$. Since $(\mathfrak{I}, d_0) \Longleftrightarrow (\mathfrak{I}/E, [e_0]_E)$, there is an $r$-successor $d_1$ of $d_0$ such that $(\mathfrak{I}, d_1) \Longleftrightarrow (\mathfrak{I}/E, [e_1]_E)$. As shown above, under this condition $d_1$ is atomically equivalent to $e_1$ and hence **II** reaches, by moving from $d_0$ via this $r$-edge to $d_1$, an admissible configuration $(\mathfrak{I}, d_1; \mathfrak{I}, e_1)$ such that $(\mathfrak{I}, d_1) \Longleftrightarrow (\mathfrak{I}/E, [e_1]_E)$. This proves

$$(\mathfrak{I}, d) \Longleftrightarrow (\mathfrak{I}/E, [e]_E) \implies (\mathfrak{I}, d) \Longleftrightarrow (\mathfrak{I}, e).$$

□

Factor-interpretations of this kind are also known as *bisimulation-quotients* [61]. If $E := \{(d, d') \in \Delta^{\mathfrak{I}} \times \Delta^{\mathfrak{H}} \mid (\mathfrak{I}, d) \Longleftrightarrow (\mathfrak{I}, d')\}$, i.e. $E$ is the largest auto-bisimulation then $\mathfrak{I}/E$ is a minimal bisimilar companion for a given interpretation. Proposition 4.2.16, however, does not require $E$ to be the largest bisimulation.

It is difficult to trace the origins of bisimulation-quotients. Factor-interpretations as such are fairly common in (universal) algebra [29] and the origins of bisimu-lation in process algebras [118] suggest that bisimulation quotients have been explored early on. [118] thus quotes [1] where bisimulation-quotients already appear as exercise in lecture notes.

### Coherent Interpretations and Coherent Disjoint Unions

In the following, we shall introduce coherent interpretations and in turn coherent disjoint unions, which will provide an adequate replacement for arbitrary disjoint unions in $\mathcal{ALC}$. The construction to be given for coherent disjoint unions will ensure that they are objects in $\mathbb{K}$. Note that in the definitions to come, we do not assume that interpretations are members of $\mathbb{K}$ unless explicitly stated; individual names in $\mathsf{N_I}$ are thus treated like concept names.

Definition 4.2.17. A $\tau$-interpretation $\mathfrak{I}$ is *coherent* if for all $a \in \mathsf{N_I}$ and all $d, e \in a^{\mathfrak{I}}$ we have $(\mathfrak{I}, d) \iff (\mathfrak{I}, e)$. ◇

For all coherent $\tau$-interpretations $\mathfrak{I}$ we define the following relation:

$$E_{\mathfrak{I}} := \{(d, d) \mid d \in \Delta^{\mathfrak{I}}\} \cup \bigcup_{a \in \mathsf{N_I}} a^{\mathfrak{I}} \times a^{\mathfrak{I}}.$$

Lemma 4.2.18. *$E_{\mathfrak{I}}$ is an equivalence relation on $\Delta^{\mathfrak{I}}$ such that for all $(d, e) \in E_{\mathfrak{I}}$ : $(\mathfrak{I}, d) \iff (\mathfrak{I}, e)$.*

Proof. We first show that for all $(d, e) \in E_{\mathfrak{I}}$ : $(\mathfrak{I}, d) \iff (\mathfrak{I}, e)$: Assume $d \neq e$ then $d, e \in a^{\mathfrak{I}}$ for some $a \in \mathsf{N_I}$. As $\mathfrak{I}$ is coherent, all elements in $a$ are bisimilar to each other, so are $d$ and $e$.

Reflexivity holds by definition. If $(d, e) \in E_{\mathfrak{I}}$ and $d \neq e$ then $(d, e) \in a^{\mathfrak{I}} \times a^{\mathfrak{I}}$ and so is $(e, d)$. This shows symmetry. For transitivity assume $(d_0, d_1), (d_1, d_2) \in E_{\mathfrak{I}}$. The claim is trivial if either $d_0 = d_1$ or $d_1 = d_2$. Assume therefore $d_0 \neq d_1$ and $d_1 \neq d_2$. Then there is $a \in \mathsf{N_I}$ such that $(d_0, d_1) \in a^{\mathfrak{I}} \times a^{\mathfrak{I}}$ and there is $b \in \mathsf{N_I}$ such that $(d_1, d_2) \in b^{\mathfrak{I}} \times b^{\mathfrak{I}}$. From the latter follows by requirement for $E$ that $(\mathfrak{I}, d_1) \iff (\mathfrak{I}, d_2)$ and so $d_2 \in a^{\mathfrak{I}}$, hence $(d_0, d_2) \in E$. □

Corollary 4.2.19. *$\mathfrak{I}/E_{\mathfrak{I}}$ in $\mathbb{K}$ if $\mathfrak{I}$ is a coherent interpretation with $a^{\mathfrak{I}} \neq \emptyset$ for all $a \in \mathsf{N_I}$.*

Definition 4.2.20. A family $(\mathfrak{I}_i)_{i \in I}$ of $\tau$-interpretations is called *coherent*, if $a^{\mathfrak{I}_i} \neq \emptyset$ for all $i \in I$ and for all $i, j \in I$, all $a \in \mathsf{N_I}$ and all $d \in a^{\mathfrak{I}_i}$ and $e \in a^{\mathfrak{I}_j}$ we have $(\mathfrak{I}_i, d) \iff (\mathfrak{I}_j, e)$. ◇

Observation 4.2.21. *The disjoint union* $\biguplus_{i\in I}\mathfrak{I}_i$ *of a coherent family* $(\mathfrak{I}_i)_{i\in I}$ *of* $\tau$-*interpretations is a coherent interpretation. Therefore*

1. $(\biguplus_{i\in I}\mathfrak{I}_i, d) \Longleftrightarrow (\biguplus_{i\in I}\mathfrak{I}_i/E_{\biguplus\mathfrak{I}_i}, [d]_{E_{\biguplus\mathfrak{I}_i}})$, *for all* $d \in \Delta^{\biguplus\mathfrak{I}_i}$,
2. $\biguplus_{i\in I}\mathfrak{I}_i \overset{g}{\Longleftrightarrow} \biguplus_{i\in I}\mathfrak{I}_i/E_{\biguplus\mathfrak{I}_i}$,
3. $\biguplus_{i\in I}\mathfrak{I}_i/E_{\biguplus\mathfrak{I}_i}$ *in* $\mathbb{K}$.

We omit the proof as 1. follows from Proposition 4.2.16, 2. follows from 1. and 3. is an instance of Corollary 4.2.19.

Definition 4.2.22. An FO-sentence $\varphi$ is invariant under coherent disjoint union, if for a coherent family of $\tau$-interpretations we have

$$\forall i \in I : \mathfrak{I}_i \models \varphi \quad \Longleftrightarrow \quad (\biguplus_{i\in I}\mathfrak{I}_i)/E_{\biguplus_{i\in I}\mathfrak{I}_i} \models \varphi.$$

◇

We also need to provide an additional notion, namely being preserved under coherent forward generated subinterpretations: In the case of $\mathcal{ALC}$, invariance under disjoint unions not only entailed that an FO-sentence was preserved by disjoint unions but also these sentences were preserved under generated subinterpretations. Together with the invariance of global bisimulation which allowed for forest unravellings, these FO-sentences were preserved under arbitrary tree-unravellings.

The definition for invariance under coherent disjoint unions however does not allow for arbitrary tree-unravellings or the disassembling of interpretations: interpretations for which an FO-sentence is invariant under coherent disjoint unions are all required to be in $\mathbb{K}$ which forbids disassembling of interpretations in general and tree-unravellings in particular as otherwise individual names would remain being interpreted as empty.

But for a characterisation analogue to $\mathcal{ALC}$ (cf. Theorem 2.3.7) we need a replacement for this property which allowed us to use parts of an interpretation. We shall therefore introduce forward generated subinterpretations, which do not cut out complete connected components of an element, but only those parts of a connected component which contain reachable elements from a distinguished element; therefore *forward generated.* FO-sentences will then be required to be preserved in a coherent disjoint union of these forward generated subinterpreta-tions, This seems to be an ad hoc solution at first, but it is not, as we shall discuss after having given the definitions.

Definition 4.2.23. A *forward generated subinterpretation* $\mathfrak{K}$ *of* $\mathfrak{I}$ *in* $d \in \Delta^{\mathfrak{I}}$ is defined as $\mathfrak{K} := \mathfrak{I}{\restriction}\Delta^{\mathfrak{K}}$ where $\Delta^{\mathfrak{K}} := \{d' \in \Delta^{\mathfrak{I}} \mid \exists d \cdots d' \in \Delta^{\mathfrak{I}_d}\}$ with $\Delta^{\mathfrak{I}_d}$ the carrier-set of the tree-unravelling of $\mathfrak{I}$ in $d$. ◇

In essence, $\Delta^{\mathfrak{K}}$ contains all elements reachable in finitely many steps from $d$ as $\Delta^{\mathfrak{K}}$ contains all last letters of path-elements in $\Delta^{\mathfrak{I}_d}$. Note that a forward generated subinterpretation need not to be in $\mathbb{K}$ as individuals not reachable from the distinguished element $d$ will be lost and hence interpreted as empty.

Observation 4.2.24. *Let* $\mathfrak{I}$ *be a* $\tau$*-interpretation and* $\mathfrak{K}$ *a forward generated subinterpretation of* $\mathfrak{I}$ *in some element* $d_0 \in \Delta^{\mathfrak{I}}$*. Then* $(\mathfrak{K}, d) \Longleftrightarrow (\mathfrak{I}, d)$ *for all* $d \in \Delta^{\mathfrak{K}}$*.*

It is clear that $\mathfrak{K}$ contains all elements reachable from $d$. Hence **II** can copy every challenge set out by **I** and thus has a winning-strategy.

Definition 4.2.25. A sentence $\varphi$ is preserved under coherent disjoint unions of forward generated subinterpretations, if for a family $(\mathfrak{I}_i)_{i\in I}$ of $\tau$-interpretations

$$\forall i \in I : \mathfrak{I}_i \models \varphi \quad \Longrightarrow \quad \biguplus_{i\in I} \mathfrak{K}_i / E_{\biguplus \mathfrak{K}_i} \models \varphi$$

where $\mathfrak{K}_i$ is the forward generated subinterpretation of $\mathfrak{I}_i$ in some element $d \in \Delta^{\mathfrak{I}}$ and $\biguplus_{i\in I} \mathfrak{K}_i$ is a coherent $\tau$-interpretation. ◇

Observation 4.2.26. *$\mathcal{ALCO}$-TBoxes are*

1. *invariant under global bisimulation*
2. *invariant under coherent disjoint unions*
3. *preserved under coherent disjoint unions of forward generated subinterpretations.*

Proof. Let $\mathcal{T}$ be an $\mathcal{ALCO}$-TBox and $(\mathfrak{I}_i)_{i\in I}$ a coherent family of $\tau$-interpretations.

1. As every $\mathcal{ALCO}$-TBox is merely an $\mathcal{ALC}$-TBox, the claim is immediate.
2. As individual names are treated like concept names, we may consider $\mathcal{T}$ as $\mathcal{ALC}$-TBox. It then follows that $\mathcal{T}$ is invariant under disjoint union and hence $\forall i \in I : \mathfrak{I} \models \mathcal{T}$ iff $\biguplus_{i\in I} \mathfrak{I}_i \models \mathcal{T}$. As Observation 4.2.21 assures, $\biguplus_{i\in I}\mathfrak{I}_i \overset{g}{\Longleftrightarrow} \biguplus_{i\in I}/E_{\biguplus\mathfrak{I}_i}$ and as $\mathcal{ALCO}$-TBoxes are invariant under global $\mathcal{ALC}$-bisimulation we obtain $\biguplus_{i\in I}\mathfrak{I}_i \models \mathcal{T}$ iff $\biguplus_{i\in I}\mathfrak{I}_i/E_{\biguplus\mathfrak{I}_i} \models \mathcal{T}$.

3. $\mathcal{T}$, considered as $\mathcal{ALC}$-TBox, is preserved under forward generated subinterpretations: If $\mathfrak{I} \models \mathcal{T}$ and $C \sqsubseteq D \in \mathcal{T}$ then every element in $\mathfrak{I}$ satisfies the $\mathcal{ALC}$-concept of $C \to D$. Since $(\mathfrak{K}, d) \Longleftrightarrow (\mathfrak{I}, d)$ for all $d \in \Delta^{\mathfrak{K}}$ also every element in $\mathfrak{K}$ satisfies $C \to D$ and so $\mathfrak{K} \models C \sqsubseteq D$.

   As $\mathcal{T}$ is also preserved under disjoint union, $\biguplus_{i\in I}\mathfrak{K}_i \models \mathcal{T}$. Assume additionally that $\biguplus_{i\in I}\mathfrak{K}_i$ is coherent. Then $\biguplus_{i\in I}\mathfrak{K}_i \overset{g}{\Longleftrightarrow} \biguplus_{i\in I}\mathfrak{K}_i/E_{\biguplus\mathfrak{K}_i}$ and invariance of $\mathcal{T}$ under global bisimulation yields $\biguplus_{i\in I}\mathfrak{K}_i/E_{\biguplus\mathfrak{K}_i} \models \mathcal{T}$.

□

At this point we have not made the requirement that structures must interpret every $a \in \mathsf{N_I}$ with a non-empty set, short of interpreting this set as singleton. Observation 4.2.26 is strengthened if we require additionally that every member of the family $(\mathfrak{I}_i)_{i\in I}$ is in $\mathbb{K}$. We then obtain that every $\mathcal{ALCO}$-TBox $\mathcal{T}$ is *invariant under coherent disjoint unions in* $\mathbb{K}$, i.e. $\forall i \in I : \mathfrak{I}_i \models_{\mathbb{K}} \mathcal{T} \iff (\biguplus \mathfrak{I}_i)/E_{\biguplus\mathfrak{I}_i} \models_{\mathbb{K}} \mathcal{T}$. Similarly every $\mathcal{ALCO}$-TBox $\mathcal{T}$ is preserved under coherent disjoint unions of tree-unravellings in $\mathbb{K}$.

Concerning whether or not being preserved under the coherent disjoint union of forward generated subinterpretation is ad hoc, we have witnessed in Observation 4.2.26, TBoxes are preserved from $\mathfrak{K}$ to $\mathfrak{I}$ if every element in $\mathfrak{K}$ finds a bisimilar partner in $\mathfrak{I}$. This is in particular true for interpretations $\mathfrak{K}, \mathfrak{I}$ in $\mathbb{K}$. Thus FO-sentences which can be expressed as TBoxes must have this property. But instead of subjecting FO-sentences to an abstract and non-constructive condition like 'is preserved under total bisimulations (i.e. the domain of the bisimulation relation $S$ coincides with $\Delta^{\mathfrak{K}}$) in $\mathbb{K}$' we give a constructive procedure how to obtain such interpretations.

### The Characterisation of $\mathcal{ALCO}$-TBoxes as FO-Fragment over $\mathbb{K}$

We shall now state the theorem and furnish another lemma towards the proof.

Theorem 4.2.27 . *Let $\tau$ be a signature and $\varphi \in \mathrm{FO}(\tau)$ a sentence which is*

1. *invariant under bisimulation in* $\mathbb{K}$

2. *invariant under coherent disjoint unions in* $\mathbb{K}$

3. *preserved under coherent disjoint unions of forward generated subinterpretations in $\mathbb{K}$*

*then $\varphi$ is equivalent to an $\mathcal{ALCO}$-TBox over $\tau$.*

Before giving the lemma, we define ${\mathsf{N_R}}^* := \bigcup_{n<\omega} {\mathsf{N_R}}^n$ where ${\mathsf{N_R}}^0 := \{\varepsilon\}$ and ${\mathsf{N_R}}^{n+1} = \mathsf{N_R} \times {\mathsf{N_R}}^n$ for all $n < \omega$. Intuitively, ${\mathsf{N_R}}^*$ contains all words (of finite length) over the alphabet $\mathsf{N_R}$. We speak of an *$r^*$-path*, if we mean a sequence of edges $(d_i, d_{i+1}) \in r_i$ for $i \in \{0, \dots, n\}$ such that $r^*$ corresponds to the word $r_0 \cdots r_n$. For every $\mathcal{ALCO}$-concept $C$ and every $r^* \in {\mathsf{N_R}}^*$ we recursively define

$$\exists r^* C := \begin{cases} C & \text{if } r^* = \varepsilon \\ \exists s.\exists s^* C & \text{if } r^* = s \cdot s^*. \end{cases}$$

We use $\forall r^* C$ as abbreviation for $\neg(\exists r^* \neg C)$. Every unfolding of $\exists r^*$ leads to a finite or empty chain of quantifications and for every expression $\exists r^* C$ we obtain a proper $\mathcal{ALCO}$-concept if $C$ is a proper $\mathcal{ALCO}$-concept.

E.g. the semantics in $\mathbb{K}$ of $\forall r^*.(a \to C)$ is, 'whenever the individual labelled $a$ is reachable via some $r^*$-path then it satisfies $C$.' $\{\forall r^*(a \to C) \mid r^* \in {\mathsf{N_R}}^*\}$ amounts to: 'if the individual is reachable it satisfies $C$.'

Definition 4.2.28. An element $e$ is *in the scope of* an element $d$ or *$d$ sees $e$*, if for some $r^* \in {\mathsf{N_R}}^*$ there is an $r^*$-path beginning with $d$ and ending with $e$. ◇

Let for some $\tau$, $\varphi$ be as required by Theorem 4.2.27. Assume there is $\mathfrak{H} \models \text{cons } \varphi \cup \{\neg\varphi\}$. Let $\mathrm{Th}(\mathfrak{H}, e) := \{C \in \mathcal{ALCO}(\tau) \mid e \in C^{\mathfrak{H}}\}$ and $\mathrm{Th}(\mathfrak{H}, \mathsf{N_I}) := \{a \sqsubseteq C \mid C \in \mathrm{Th}(\mathfrak{H}, a^{\mathfrak{H}}) \text{ and } a \in \mathsf{N_I}\}$.

Lemma 4.2.29. *For each $e \in \Delta^{\mathfrak{H}}$ there is a model $\mathfrak{I}_e$ in $\mathbb{K}$ of $\varphi$ with an element $d \in \Delta^{\mathfrak{I}}$ such that $(\mathfrak{I}_e, d) \models \mathrm{Th}(\mathfrak{H}, e)$ and if $a \in \mathsf{N_I}$ is in the scope of $d$ then $(\mathfrak{I}_e, a^{\mathfrak{I}_e}) \models \mathrm{Th}(\mathfrak{H}, a^{\mathfrak{H}})$.*

Proof. Assume the claim would not hold. Then there is some $e \in \Delta^{\mathfrak{H}}$ such that $\{\varphi\} \cup \mathrm{Th}(\mathfrak{H}, e)$ together with the set

$$\Gamma := \{\forall r^*(a \to C) \mid a \in \mathsf{N_I}, r^* \in {\mathsf{N_R}}^* \text{ and } C \in \mathrm{Th}(\mathfrak{H}, a^{\mathfrak{H}})\}$$

is unsatisfiable in $\mathbb{K}$. As FO is compact w.r.t. to $\mathbb{K}$ there is a finite subset $\Gamma_0 \subseteq \Gamma$ and some finite set $T \subseteq \mathrm{Th}(\mathfrak{H}, e)$ such that for every model $\mathfrak{M}$ of $\varphi$ in $\mathbb{K}$ and all $d \in \Delta^{\mathfrak{M}}$ we have $(\mathfrak{M}, d) \models \neg(\bigsqcap T \sqcap \bigsqcap \Gamma_0)$. Hence for all $d \in \Delta^{\mathfrak{M}}$ we obtain

$$(\mathfrak{M}, d) \models \left(\neg \bigsqcap T\right) \sqcup \bigsqcup \{\exists r^*(a \sqcap \neg C) \mid \forall r^*(a \to C) \in \Gamma_0\}$$

So $\mathfrak{M} \models \bigsqcap T \sqsubseteq \bigsqcup\{\exists r^*(a \sqcap \neg C) \mid \forall r^*(a \to C) \in \Gamma_0\}$; This is true for all models of $\varphi$ in $\mathbb{K}$ and it follows

$$\bigsqcap T \sqsubseteq \bigsqcup \{\exists r^*(a \sqcap \neg C) \mid \forall r^*(a \to C) \in \Gamma_0\} \in \operatorname{cons} \varphi$$

As $\mathfrak{H} \models \operatorname{cons} \varphi$ and $(\mathfrak{H}, e) \models \bigsqcap T$, there must be a path from $e$ to $a^{\mathfrak{H}}$ for some $a \in \mathsf{N_I}$ such that this $a^{\mathfrak{H}}$ does not satisfy the type of $\operatorname{Th}(\mathfrak{H}, a^{\mathfrak{H}})$. But this is absurd and hence $\mathfrak{I}_e$ must exist. □

Proof of Theorem 4.2.27. We shall prove further down that $\neg\varphi$ is unsatisfiable in $\mathbb{K}$ with the following set

$$\operatorname{cons} \varphi := \{C \sqsubseteq D \mid C, D \in \mathcal{ALCO}(\tau) \text{ and } \varphi \models_{\mathbb{K}} C \sqsubseteq D\}.$$

For then, as FO is compact over $\mathbb{K}$, the unsatisfiability of $\operatorname{cons} \varphi \cup \{\neg\varphi\}$ implies that there is a finite subset $\mathcal{T} \subseteq \operatorname{cons} \varphi$ which is unsatisfiable with $\neg\varphi$ in $\mathbb{K}$ and hence $\mathcal{T} \models_{\mathbb{K}} \varphi$. As $\varphi \models_{\mathbb{K}} \mathcal{T}$, the $\mathcal{ALCO}$-TBox is logically equivalent to $\varphi$ in $\mathbb{K}$ and so $\mathcal{T}$ is the $\mathcal{ALCO}$-TBox we were looking for; this proves the theorem.

For the sake of contradiction let $\mathfrak{H}$ be an $\omega$-saturated model in $\mathbb{K}$ of $\operatorname{cons} \varphi \cup \{\neg\varphi\}$. Let $T$ be the set of all $p \subseteq \mathcal{ALCO}(\tau)$ such that

$$\{\varphi\} \cup p \cup \{\forall r^*(a \to C) \mid a \in \mathsf{N_I}, r^* \in \mathsf{N_R}^* \text{ and } C \in \operatorname{Th}(\mathfrak{H}, a^{\mathfrak{H}})\} \text{ is satisfiable in } \mathbb{K}.$$

For each $p \in T$, we set $\mathfrak{I}_p$ to be the forward generated subinterpretation in $d_p$ of the $\omega$-saturated interpretation $\mathfrak{I}$ in $\mathbb{K}$ with

$$(\mathfrak{I}, d_p) \models_{\mathbb{K}} \{\varphi\} \cup p \cup \{\forall r^*(a \to C) \mid a \in \mathsf{N_I}, r^* \in \mathsf{N_R}^* \text{ and } C \in \operatorname{Th}(\mathfrak{H}, a^{\mathfrak{H}})\}.$$

As $\mathfrak{I}_p$ and $\mathfrak{H}$ are saturated, $(\mathfrak{H}, a^{\mathfrak{H}}) \iff (\mathfrak{I}_p, a^{\mathfrak{I}})$ for all $a$ in the scope of $d_p$. Hence $\biguplus_{p \in T} \mathfrak{I}_p$ is a coherent interpretation and $\mathfrak{J} := \biguplus_{p \in T} \mathfrak{I}_p / E_{\biguplus \mathfrak{I}_p}$ is a coherent disjoint union of forward generated subinterpretations. Lemma 4.2.29 applied to the case $e = a^{\mathfrak{H}}$ yields $\operatorname{Th}(\mathfrak{H}, a^{\mathfrak{H}}) \in T$ for all $a \in \mathsf{N_I}$, and so $a^{\mathfrak{J}} \neq \emptyset$. Since $\mathfrak{J}$ emerges from a factorisation by $E_{\biguplus \mathfrak{I}_p}$ there is at most one element in $a^{\mathfrak{J}}$ for all $a \in \mathsf{N_I}$; hence $\mathfrak{J}$ in $\mathbb{K}$. In particular we have $(\mathfrak{J}, a^{\mathfrak{J}}) \iff (\mathfrak{H}, a^{\mathfrak{H}})$ for all $a \in \mathsf{N_I}$.

$\mathcal{ALCO}$-theories and $\varphi$ itself are preserved under coherent disjoint unions of forward generated subinterpretations in $\mathbb{K}$, so $\mathfrak{J} \models_{\mathbb{K}} \varphi$ and for all $p \in T$ there is $[d_p]_E \in \Delta^{\mathfrak{J}}$ such that $(\mathfrak{J}, [d_p]) \models p$.

We set $\mathfrak{K}$ to be the $\omega$-saturated extension of $(\mathfrak{J} \uplus \mathfrak{H})/E_{\mathfrak{J}\uplus\mathfrak{H}}$. Again $\mathfrak{K}$ is in $\mathbb{K}$ and $(\mathfrak{K}, a^{\mathfrak{K}}) \Longleftrightarrow (\mathfrak{H}, a^{\mathfrak{H}})$ for all $a \in \mathsf{N}_\mathsf{I}$. Also $\mathfrak{K} \models \text{cons}\,\varphi$ as every finite subset $\mathcal{T} \subseteq \text{cons}\,\varphi$ is an $\mathcal{ALCO}$-TBox satisfied by both interpretations and therefore preserved by coherent disjoint unions. But $\mathfrak{K} \models_{\mathbb{K}} \neg\varphi$ since $\varphi$ is *invariant* under coherent disjoint unions and $\mathfrak{H} \not\models_{\mathbb{K}} \varphi$.

We have to show $\mathfrak{J} \overset{g}{\Longleftrightarrow} \mathfrak{K}$. Lemma 4.2.18 tells us that $E_0 := E_{\mathfrak{J}\uplus\mathfrak{H}}$ is suitable for Proposition 4.2.16, whence $(\mathfrak{J}, d) \Longleftrightarrow (\mathfrak{K}, [d]_{E_0})$ for all $d \in \Delta^{\mathfrak{J}}$ is inferred. Let now $[e]_{E_0} \in \Delta^{\mathfrak{K}}$ such that $e \in \Delta^{\mathfrak{H}}$. Then $[d_{\text{Th}(\mathfrak{H},e)}]_{E_1}$ with $E_1 = E_{\biguplus \mathfrak{J}_p}$ is in $\mathfrak{J}$ and satisfies $\text{Th}(\mathfrak{H}, e)$. Its class $[e]_{E_0} \in \Delta^{\mathfrak{K}}$ satisfies $\text{Th}(\mathfrak{H}, e)$ as well and since $\mathfrak{K}$ and $\mathfrak{J}$ are both $\mathcal{ALCO}$-saturated, the Hennessy-Milner-Property of $\mathcal{ALCO}$ yields $(\mathfrak{K}, [e]_{E_0}) \Longleftrightarrow (\mathfrak{J}, [d_{\text{Th}(\mathfrak{H},e)}]_{E_1})$.

So $\mathfrak{K}, \mathfrak{J}$ in $\mathbb{K}$ and $\mathfrak{J} \overset{g}{\Longleftrightarrow} \mathfrak{K}$. Since $\varphi$ is invariant under bisimulation between two interpretations in $\mathbb{K}$, $\mathfrak{K} \models \varphi$. But this is absurd and so there cannot be model of $\text{cons}\,\varphi \cup \{\neg\varphi\}$ in $\mathbb{K}$. □

Concluding it shall be remarked that individuals introduce some kind of orientation in the sense that we need specifically forward generated subinterpretations. In Lemma 4.2.29 coherent models of $\varphi$ are created for every $\mathcal{ALCO}$-$u$-type realised in $\mathfrak{H}$. In order to prove this lemma we had to derive a contradiction. This contradiction however had to be in terms of $\text{cons}\,\varphi$ i.e. in terms of positive concept inclusions because we only new $\mathfrak{H} \models_{\mathbb{K}} \text{cons}\,\varphi$ and $\mathfrak{H} \models_{\mathbb{K}} \neg\varphi$. It is therefore not possible to create interpretations which are per se coherent with $\mathfrak{H}$:

Assume for a moment we wanted to prove that for some element $e \in \Delta^{\mathfrak{H}}$ there is some model of $\varphi$ and $\text{Th}(\mathfrak{H}, e)$ which is coherent with $\mathfrak{H}$. In this instance we had to show that either

$$\text{Th}(\mathfrak{H}, e) \cup \{\varphi\} \cup \bigcup_{a \in \mathsf{N}_\mathsf{I}} \{a \sqsubseteq C \mid C \in \text{Th}(\mathfrak{H}, a^{\mathfrak{H}})\}$$

or

$$\text{Th}(\mathfrak{H}, e) \cup \{\varphi\} \cup \bigcup_{a \in \mathsf{N}_\mathsf{I}} \{\exists u.a \sqcap C \mid C \in \text{Th}(\mathfrak{H}, a^{\mathfrak{H}})\}$$

is satisfiable in $\mathbb{K}$. If this is not the case, compactness of FO over $\mathbb{K}$ yields for the former set a finite subset $T_0 \subseteq \text{Th}(\mathfrak{H}, e)$ and a finite subset $\Gamma \subseteq \bigcup_{a \in \mathsf{N}_\mathsf{I}} \{a \sqsubseteq C \mid$

$C \in \mathrm{Th}(\mathfrak{H}, a^{\mathfrak{H}})\}$ and so we obtain

$$\varphi \models \left(\bigsqcap T_0\right) \to \bigsqcup\{\exists u.a \sqcap \neg C \mid a \sqsubseteq C \in \Gamma\}$$

contradicts the satisfiability of the former set, without leading to a contradiction in $\mathfrak{H}$. Similarly, compactness yields for some finite sets $T_0 \subseteq \mathrm{Th}(\mathfrak{H}, e)$ and $\Gamma \subseteq \bigcup_{a \in \mathsf{N}_\mathsf{I}} \{\exists u.a \sqcap C \mid C \in \mathrm{Th}(\mathfrak{H}, a^{\mathfrak{H}})\}$ to

$$\varphi \models \left(\bigsqcap T_0\right) \to \bigsqcup\{a \sqsubseteq \neg C \mid a \sqsubseteq C \in \Gamma\}$$

without leading to a contradiction in $\mathfrak{H}$—unless we require that those individuals $a$ are mentioned in $\Gamma$ are in the scope of $e$. Hence we need forward generated subinterpretations to cut away individuals outside the scope of $e$, which may not be bisimilar to individuals in $\mathfrak{H}$.

The stronger requirement, used in the $\mathcal{ALC}$-TBox characterisation, that an FO-sentence $\varphi$ need not only to be invariant under bisimulation and disjoint unions w.r.t. to a certain restricted class of interpretations but w.r.t. all interpretations, implicitly included the preservation under forward generated subinterpretations.

## 4.3 $\mathcal{ALCQO}$

### 4.3.1 The Characterisation of $\mathcal{ALCQO}$-Concepts

The syntax for an $\mathcal{ALCQO}$-concept $C$ over $\tau = \mathsf{N_I} \cup \mathsf{N_C} \cup \mathsf{N_R}$ is given by

$$C ::= \top \mid a \mid A \mid D \sqcap E \mid \neg D \mid \exists^{\geq \kappa} r.D$$

where $a \in \mathsf{N_I}$, $A \in \mathsf{N_C}$, $r \in \mathsf{N_R}$, $\kappa < \omega$ and $D, E$ are $\mathcal{ALCQO}$-concepts over $\tau$. The set of all $\mathcal{ALCQO}$-concepts over $\tau$ is denoted by $\mathcal{ALCQO}(\tau)$. As usual $\bot$, $\sqcup$, $\rightarrow$, $\forall^{\geq \kappa}$ etc. are considered to be abbreviations.

We use the semantics defined in the chapter for $\mathcal{ALCO}$ where concepts are interpreted over a special class $\mathbb{K}$ of interpretations, which assign singleton subsets of their carrier-sets to individual names. Yet in principle we treat individual names like concept names and allow $\mathcal{ALCQO}$-concepts to be interpreted outside of $\mathbb{K}$ like $\mathcal{ALCQ}$-concepts.

Note that concepts of the form $\exists^{\geq 2} r.a$ do not have a model in $\mathbb{K}$ and $(\mathfrak{I}, d) \models \exists^{\geq 1} r.a$ iff $(\mathfrak{I}, d) \models \exists^{=1} r.a$ for all $a \in \mathsf{N_I}$ and all pointed interpretations $(\mathfrak{I}, d)$ in $\mathbb{K}$.

$\mathcal{ALCQO}$-bisimulation is merely $\mathcal{ALCQ}$-bisimulation as defined on page 80. Characteristic $\mathcal{ALCQO}$-concepts are obtained like their $\mathcal{ALCQ}$-counterpart (cf. page 82). Theorem 3.2.8 and therefore invariance of $\mathcal{ALCQO}$-concepts under $\mathcal{ALCQ}$-bisimulation is immediate. $\mathcal{ALCQO}$-saturation is the same as $\mathcal{ALCQ}$-saturation as stipulated by Definition 3.2.10 using $\mathcal{ALCQ}$-types (cf. Definition 3.2.9) and Proposition 3.2.12 stating the Hennessy-Milner-Property applies also for $\mathcal{ALCQ}$. As mentioned in Observation 4.1.4, every interpretation in $\mathbb{K}$ has an $\omega$-saturated extension in $\mathbb{K}$ and therefore an $\mathcal{ALCQO}$-saturated extension in $\mathbb{K}$.

#### The Characterisation Theorem for $\mathcal{ALCQO}$-Concepts

Let $\tau$ be arbitrary and $I := (\omega \setminus \{0\}) \times \omega$. We may assume w.l.o.g. that $\tau$ contains at least one element in $\mathsf{N_I}$, for otherwise this case would reduce to $\mathcal{ALCQ}$.

Lemma 4.3.1 . *Let $\varphi(x) \in \mathrm{FO}(\tau)$ be invariant under $\xleftrightarrow{*}_{\dagger}$ in $\mathbb{K}$, where $* \in \{\text{`}< \omega\text{'}\} \cup \{\text{`}\leq \kappa\text{'} \mid \kappa < \omega\}$ and $\dagger < \omega$ or simply omitted. Let $\tau_\varphi \subseteq \tau$ contain all signature symbols occurring in $\varphi$ and $\mathfrak{I}, \mathfrak{H}$ in $\mathbb{K}$.*

$$(\mathfrak{I} \restriction \tau_\varphi, d) \xleftrightarrow{*}_{\dagger} (\mathfrak{H} \restriction \tau_\varphi, e) \quad \Longrightarrow \quad (\mathfrak{I}, d) \models \varphi(x) \iff (\mathfrak{H}, e) \models \varphi(x)$$

Proof. We add a fresh element $d_0$ to $\mathfrak{I}$ obtaining $\mathfrak{I}_0$ and a fresh element $e_0$ to $\mathfrak{H}$ obtaining $\mathfrak{H}_0$. We have

$$(\mathfrak{I}, d) \overset{<\omega}{\Longleftrightarrow} (\mathfrak{I}_0, d) \cong_{\tau_\varphi} (\mathfrak{I}_0 \restriction \tau_\varphi, d) \overset{*}{\Longleftrightarrow}_\dagger (\mathfrak{H}_0 \restriction \tau_\varphi, e) \cong_{\tau_\varphi} (\mathfrak{H}_0, e) \overset{<\omega}{\Longleftrightarrow} (\mathfrak{H}, e)$$

We set $\mathfrak{I}_1$ to be $\mathfrak{I}_0 \restriction \tau_\varphi$ where $a^{\mathfrak{I}_1} := \{d_0\}$ for all $a \in \mathsf{N_I} \setminus \tau_\varphi$ and all other symbols in $\tau \setminus (\tau_\varphi \cup \mathsf{N_I})$ are interpreted as $\emptyset$. Analogously we set $\mathfrak{H}_1$ to be $\mathfrak{H}_0 \restriction \tau_\varphi$ where $a^{\mathfrak{H}_1} := \{e_0\}$ and again all other symbols in $\tau \setminus (\tau_\varphi \cup \mathsf{N_I})$ are interpreted as $\emptyset$. We have $\mathfrak{I}_1, \mathfrak{H}_1$ in $\mathbb{K}$ and $(\mathfrak{I}_1, d) \overset{*}{\Longleftrightarrow}_\dagger (\mathfrak{H}_1, e)$.

Let now $(\mathfrak{I}, d) \models \varphi$; $(\mathfrak{I}, d) \overset{<\omega}{\Longleftrightarrow} (\mathfrak{I}_0, d)$ always implies $(\mathfrak{I}, d) \overset{*}{\Longleftrightarrow}_\dagger (\mathfrak{I}_0, d)$, so invariance of $\varphi(x)$ under $\overset{*}{\Longleftrightarrow}_\dagger$ implies $(\mathfrak{I}_0, d) \models \varphi(x)$. Since all first order formulae are invariant under isomorphims of their signature we have $(\mathfrak{I}_0 \restriction \tau_\varphi, d) \models \varphi(x)$. But we also have $\mathfrak{I}_0 \restriction \tau_\varphi \cong_{\tau_\varphi} \mathfrak{I}_1$ and so $(\mathfrak{I}_1, d) \models \varphi(x)$. Invariance under $\overset{*}{\Longleftrightarrow}_\dagger$ yields $(\mathfrak{H}_1, e) \models \varphi(x)$ and using the chain of arguments backwards we can infer that $(\mathfrak{H}, e) \models \varphi(x)$. The only-if direction is obtained similarly. □

Proposition 4.3.2 . *Every formula $\varphi \in \mathrm{FO}(\tau)$ which is invariant under $\overset{<\omega}{\Longleftrightarrow}$ in $\mathbb{K}$ is invariant under $\overset{\leq\kappa}{\Longleftrightarrow}_n$ in $\mathbb{K}$ for some $(\kappa, n) \in I$.*

On one hand, the proof would use the exact same rationale as Proposition 3.2.24 for $\mathcal{ALCQ}$ only that we now would have to be careful with the signature $\tau$ and the class $\mathbb{K}$. On the other hand Proposition 4.2.14 deals with $\tau$ and $\mathbb{K}$ accordingly but does not need to take care about different $\kappa$. Due to the fact that we are right in the intersection of both other proves, we shall skip the proof entirely and rather refer the reader to Proposition 4.3.7

Theorem 4.3.3 . *Let $\tau$ be a signature. For all $\varphi(x) \in \mathrm{FO}(\tau)$, which are invariant under $\overset{<\omega}{\Longleftrightarrow}$ in $\mathbb{K}$ exists a concept $C \in \mathcal{ALCQO}(\tau)$ such that $\varphi(x)$ is logically equivalent to $C$.*

Proof. According to Proposition 4.3.2 $\varphi$ is invariant under $\overset{\leq\kappa}{\Longleftrightarrow}_n$ for some $(\kappa, n) \in I$. Let $\tau_\varphi \subseteq \tau$ contain all signature symbols occuring in $\varphi$ and let $C := \bigsqcup\{X^{\kappa,n}_{\mathfrak{I}\restriction\tau_\varphi,d} \mid (\mathfrak{I}, d) \models \varphi(x)\}$, where $X^{\kappa,n}_{\mathfrak{I}\restriction\tau_\varphi,d}$ is the characteristic $\mathcal{ALCQO}$-concept of $(\mathfrak{I}, d)$ over the signature $\tau_\varphi$ on level $(\kappa, n)$.

Since $\varphi(x) \models C$ it remains to show $C \models \varphi(x)$. Assume $(\mathfrak{H}, e) \models X^{\kappa,n}_{\mathfrak{K}\restriction\tau_\varphi,c}$ with $X^{\kappa,n}_{\mathfrak{K}\restriction\tau_\varphi,c} \in \{X^{\kappa,n}_{\mathfrak{I}\restriction\tau_\varphi,d} \mid (\mathfrak{I}, d) \models \varphi(x)\}$. Then $(\mathfrak{H} \restriction \tau_\varphi, e) \overset{\leq\kappa}{\Longleftrightarrow}_n (\mathfrak{K} \restriction \tau_\varphi, c)$ and Lemma 4.3.1 yields $(\mathfrak{H}, e) \models \varphi(x)$. Hence $C \models \varphi(x)$ which shows that $C$ is logically equivalent to $\varphi(x)$. □

### 4.3.2 The Characterisation of $\mathcal{ALCQO}u$-Concepts

The syntax for $\mathcal{ALCQO}u$-concepts $C$ is recursively given by

$$C ::= \top \mid a \mid A \mid D \sqcap E \mid \neg D \mid \exists^{\geq \kappa} r.D \mid \exists^{\geq \kappa} u.D$$

where $a \in \mathsf{N_I}$, $A \in \mathsf{N_C}$, $r \in \mathsf{N_R}$, $\kappa < \omega$ and $D, E$ are $\mathcal{ALCQO}$-concepts over $\tau$. $u$ is a logical symbol, i.e. it is not part of the signature. The set of all $\mathcal{ALCQO}$-concepts over $\tau$ is denoted by $\mathcal{ALCQO}u(\tau)$. As usual $\bot$, $\rightarrow$, $\sqcup$, $\forall^{\geq \kappa}$ etc. are considered to be abbreviations.

Modulo whether a symbol $S$ belongs to $\mathsf{N_C}$ or $\mathsf{N_I}$, $\mathcal{ALCQ}u$-concepts already capture $\mathcal{ALCQO}u$-concepts: an $\mathcal{ALCQO}u$-concept $C$ can simply be translated into $D := C \sqcap \bigsqcap_{a \in \mathsf{N_I}(C)} \exists^{=1} u.a$ where $\mathsf{N_I}(C) := \{a \in \mathsf{N_I} \mid a \text{ occurs in } C\}$. $D$ is now an $\mathcal{ALCQ}u$-concept in which all elements of $\mathsf{N_I}(C)$ belong to $\mathsf{N_C}$.

However, our characterisations so far focused on interpretations from the class $\mathbb{K}$ especially with its ramifications on tree-unravellings and forest-unravellings respectively. This one-to-one correspondence between $\mathcal{ALCQO}u$-concepts and $\mathcal{ALCQ}u$-concepts makes the fact explicit that also $\mathcal{ALCQ}u$ is not invariant under forest-unravellings as elements multiply.

The question arises, whether a characterisation w.r.t. $\mathbb{K}$ is sensible, as $\mathcal{ALCQO}u$-concepts like $\exists^{\geq 2} r.a$ and $\exists^{\geq 2} u.a$ with $a \in \mathsf{N_I}$ do not have a model in $\mathbb{K}$ and $\exists^{\geq 1} t.a \models \exists^{=1} t.a$ with $a \in \mathsf{N_I}$ and $t \in \mathsf{N_R} \cup \{u\}$ is always true in $\mathbb{K}$. We do not gain much insight:

The notion of $\mathcal{ALCQO}u$-bisimulation is given by $\mathcal{ALCQ}u$-bisimulation and for finite signatures $\tau$ the characteristic $\mathcal{ALCQO}u$-concepts $X_{\mathfrak{J}}^{\kappa,n}$ for $\mathfrak{J}$ on level $(\kappa, n)$ is given by the characteristic $\mathcal{ALCQ}u$-concept $X_{\mathfrak{J}}^{\kappa,n}$ (cf. page 90) where individual names in $\mathsf{N_I}$ are treated like concept names.

Proposition 3.2.18 holds for characteristic $\mathcal{ALCQO}u$-concepts and $\mathcal{ALCQO}u$-theories and thus shows the invariance of $\mathcal{ALCQO}u$-concept under $\xleftrightarrow{\leq\kappa}_n$ and $\xleftrightarrow{<\omega}$ for appropriate $(\kappa, n) \in I$. As demonstrated in the proof of Proposition 3.2.22 the Hennessy-Milner-Property can be inferred for $\mathcal{ALCQO}u$, where we use the notion of $\mathcal{ALCQ}u$-saturation and $\mathcal{ALCQ}u$-types as given in Definition 3.2.21 and below.

One needs to prove a lemma analogously to Lemma 4.3.6 further down and can then proceed with the following proposition and theorem:

Proposition 4.3.4. *Let $\tau$ be an arbitrary signature. If $\varphi(x) \in \mathrm{FO}(\tau)$ and $\varphi$ is invari-*

*ant under $\xleftrightarrow{<\omega}{}^{\forall}$ in $\mathbb{K}$ then $\varphi$ is invariant under $\xleftrightarrow{\leq\kappa}{}^{\forall}_{n}$ for some $(\kappa, n) \in I$.*

THEOREM 4.3.5 . *Every formula $\varphi(x) \in \mathrm{FO}(\tau)$ which is invariant under $\xleftrightarrow{<\omega}{}^{\forall}$ in $\mathbb{K}$ is logically equivalent to some concept $C \in \mathcal{ALCQO}u(\tau)$.*

### The Characterisation of $\mathcal{ALCQO}u_1$-Concepts as FO-Fragment over $\mathbb{K}$

$\mathcal{ALCQO}u_1$ is the extension of $\mathcal{ALCQO}$ with $\exists^{\geq 1} u$. We use $\xleftrightarrow{<\omega}{}^{Q}$, the $\mathcal{ALCQ}u_1$-bisimulation, as defined in 3.2.26 and for finite signatures we declare characteristic $\mathcal{ALCQO}u_1$-concepts to be the corresponding characteristic $\mathcal{ALCQ}u_1$-concept defined in 3.2.29 where individual names have been treated like concept names. Proposition 3.2.30 translates immediately to interpretations in $\mathbb{K}$. $\mathcal{ALCQ}u_1$-saturation is sufficient to show the Hennessy-Milner-Property of $\mathcal{ALCQO}u_1$ (cf. analogue Proposition 3.2.31).

LEMMA 4.3.6 . *Let $\varphi(x) \in \mathrm{FO}(\tau)$ be invariant under $\xleftrightarrow{*}{}^{Q}_{\dagger}$ in $\mathbb{K}$, where $* \in \{`< \omega'\} \cup \{`\leq \kappa' \mid \kappa < \omega\}$ and $\dagger < \omega$ or simply omitted. Let $\tau_\varphi \subseteq \tau$ contain the signature symbols occurring in $\varphi$, as well as at least one $a \in \mathsf{N_I}$ and let $\mathfrak{I}, \mathfrak{H}$ in $\mathbb{K}$.*

$$(\mathfrak{I}\restriction\tau_\varphi, d) \xleftrightarrow{*}{}^{Q}_{\dagger} (\mathfrak{H}\restriction\tau_\varphi, e) \quad \Longrightarrow \quad (\mathfrak{I}, d) \models \varphi(x) \iff (\mathfrak{H}, e) \models \varphi(x)$$

The proof follows the exact same rationale as Lemma 4.2.13 and is therefore omitted.

PROPOSITION 4.3.7 . *Every formula $\varphi \in \mathrm{FO}(\tau)$ which is invariant under $\xleftrightarrow{<\omega}{}^{Q}$ in $\mathbb{K}$ is invariant under $\xleftrightarrow{\leq\kappa}{}^{Q}_{n}$ in $\mathbb{K}$ for some $(\kappa, n) \in I$.*

PROOF. Assume the claim would be false. Then for every $(\kappa, n) \in I$ there are pointed interpretations $(\mathfrak{I}, d), (\mathfrak{H}, e)$ in $\mathbb{K}$ such that $\varphi(x) \dashv (\mathfrak{I}, d) \xleftrightarrow{\leq\kappa}{}^{Q}_{n} (\mathfrak{H}, e) \models \neg\varphi(x)$. Let $\tau_\varphi$ contain all signature symbols occurring in $\varphi$, at least one symbol from $\mathsf{N_I}$ and be finite; in case $\varphi$ does not contain a symbol from $\mathsf{N_I}$ we simply add a symbol $a \in \mathsf{N_I}$ to $\tau_\varphi$.

$$\mathfrak{Y}^{\kappa,n} := \{ Y^{\kappa,n}_{\mathfrak{I}\restriction\tau_\varphi} \sqcap X^{\kappa,n}_{\mathfrak{I}\restriction\tau_\varphi,d} \mid \exists(\mathfrak{I}, d), (\mathfrak{H}, e) \text{ in } \mathbb{K} : \varphi(x) \dashv (\mathfrak{I}, d) \xleftrightarrow{\leq\kappa}{}^{Q}_{n} (\mathfrak{H}, e) \models \neg\varphi(x)\}$$

where $Y^{\kappa,n}_{\mathfrak{I}\restriction\tau_\varphi}$ is the characteristic $\mathcal{ALCQO}u_1$-concept and $X^{\kappa,n}_{\mathfrak{I}\restriction\tau_\varphi,d}$ is the characteristic $\mathcal{ALCQO}$-concept for $(\mathfrak{I}, d)$ over $\tau_\varphi$ on level $(\kappa, n)$. Each $\mathfrak{Y}^{\kappa,n}$ is finite and not empty. We set $\mathfrak{Y} := \{\bigsqcup \mathfrak{Y}^{\kappa,n} \mid (\kappa, n) \in I\}$.

Similar to $\mathcal{ALCQO}u$, $Y^{\mu,m}_{\mathfrak{I}\restriction\tau_\varphi} \sqcap X^{\mu,m}_{\mathfrak{I}\restriction\tau_\varphi,d} \models \mathfrak{Y}_0$ for every finite $\mathfrak{Y}_0 \subseteq \mathfrak{Y}$, where $(\mu, m)$ is the supremum in $I$ of all $(\kappa, n)$ with $\bigsqcup \mathfrak{Y}^{\kappa,n} \in \mathfrak{Y}_0$ and $Y^{\mu,m}_{\mathfrak{I}\restriction\tau_\varphi} \sqcap X^{\mu,m}_{\mathfrak{I}\restriction\tau_\varphi,d} \in \mathfrak{Y}^{\mu,m}$. This shows that every finite $\mathfrak{Y}_0 \subseteq \mathfrak{Y}$ has a model for $\mathfrak{Y}_0 \cup \{\varphi(x)\}$ in $\mathbb{K}$. Compactness of FO over $\mathbb{K}$ yields a pointed model $(\mathfrak{I}, d)$ for $\mathfrak{Y} \cup \{\varphi(x)\}$ in $\mathbb{K}$.

Let $\mathfrak{Y}_{\mathfrak{I},d} := \{Y^{\mu,m}_{\mathfrak{I}\restriction\tau_\varphi} \sqcap X^{\mu,m}_{\mathfrak{I}\restriction\tau_\varphi,d} \mid (\kappa, n) \in I\}$ where $Y^{\kappa,n}_{\mathfrak{I},d}$ is the characteristic $\mathcal{ALCQO}u_1$-concept for $(\mathfrak{I}, d)$ over $\tau_\varphi$ on level $(\kappa, n)$. Every finite $\mathfrak{Y}_0 \subseteq \mathfrak{Y}_{\mathfrak{I},d}$ is satisfiable with $\neg\varphi$: Let $(\mu, m)$ be the supremum in $I$ for all $(\kappa, n)$ with $Y^{\kappa,n}_{\mathfrak{I}\restriction\tau_\varphi} \sqcap X^{\kappa,n}_{\mathfrak{I}\restriction\tau_\varphi,d} \in \mathfrak{Y}_0$. Then $Y^{\mu,m}_{\mathfrak{I}\restriction\tau_\varphi} \sqcap X^{\mu,m}_{\mathfrak{I}\restriction\tau_\varphi,d} \models \mathfrak{Y}_0$ and since $(\mathfrak{I}, d) \models \mathfrak{Y}$, there is $(\mathfrak{H}, e)$ in $\mathbb{K}$ with $(\mathfrak{I}\restriction\tau_\varphi, d) \xleftrightarrow{\leq\kappa\, Q}_n (\mathfrak{H}\restriction\tau_\varphi, e)$, where $(\mathfrak{H}, e) \models \neg\varphi(x)$. In particular, $(\mathfrak{H}, e) \models \{Y^{\mu,m}_{\mathfrak{I}\restriction\tau_\varphi} \sqcap X^{\mu,m}_{\mathfrak{I}\restriction\tau_\varphi,d}, \neg\varphi(x)\}$ and so $\mathfrak{Y}_0 \cup \{\neg\varphi(x)\}$ is satisfiable in $\mathbb{K}$. Compactness of FO over $\mathbb{K}$ yields a model $(\mathfrak{H}, e)$ in $\mathbb{K}$ for $\mathfrak{Y}_{\mathfrak{I},d} \cup \{\neg\varphi(x)\}$.

The Hennessy-Milner-Property yields $\varphi(x) \dashv (\mathfrak{I}^*\restriction\tau_\varphi, d) \xleftrightarrow{<\omega\, Q} (\mathfrak{H}^*\restriction\tau_\varphi, e) \models \neg\varphi(x)$ for the $\omega$-saturated extensions $\mathfrak{I}^*$, $\mathfrak{H}^*$ in $\mathbb{K}$ of $\mathfrak{I}$ and $\mathfrak{H}$ respectively. In contradiction to this, Lemma 4.3.6 yields $(\mathfrak{H}^*, e) \models \varphi(x)$. This shows that there must be $(\kappa, n) \in I$ such that $\varphi$ is invariant under $\xleftrightarrow{\leq\kappa\, Q}_n$. □

Theorem 4.3.8. *Every formula $\varphi(x) \in \mathrm{FO}(\tau)$ which is invariant under $\xleftrightarrow{<\omega\, Q}$ in $\mathbb{K}$ is equivalent to some concept $C \in \mathcal{ALCQO}u_1(\tau)$*

Proof. Let $\tau_\varphi$ comprise the symbols occurring in $\varphi$. If $\varphi$ does not contain an element of $\mathsf{N_I}$, we simply choose an element from $\mathsf{N_I}$ and add this to $\tau_\varphi$.

According to Proposition 4.3.7 $\varphi$ is invariant under $\xleftrightarrow{\leq\kappa\, Q}_n$ for some $(\kappa, n) \in I$. Let $C := \{Y^{\kappa,n}_{\mathfrak{I}\restriction\tau_\varphi} \sqcap X^{\kappa,n}_{\mathfrak{I}\restriction\tau_\varphi,d} \mid (\mathfrak{I}, d) \text{ in } \mathbb{K} \text{ and } (\mathfrak{I}, d) \models \varphi(x)\}$ where $Y^{\kappa,n}_{\mathfrak{I}\restriction\tau_\varphi}$ is the characteristic $\mathcal{ALCQO}u_1$-concept and $X^{\kappa,n}_{\mathfrak{I}\restriction\tau_\varphi,d}$ is the characteristic $\mathcal{ALCQO}$-concept for $(\mathfrak{I}\restriction\tau_\varphi, d)$ over $\tau_\varphi$ on level $(\kappa, n)$.

$\varphi(x)$ is logically equivalent to $\bigsqcup C$: We show $\bigsqcup C \models \varphi(x)$. Let $(\mathfrak{H}, e) \models \bigsqcup C$. Then there is $(\mathfrak{I}, d)$ in $\mathbb{K}$ such that $Y^{\kappa,n}_{\mathfrak{I}\restriction\tau_\varphi} \sqcap X^{\kappa,n}_{\mathfrak{I}\restriction\tau_\varphi,d} \in C$ and $(\mathfrak{H}, e) \models Y^{\kappa,n}_{\mathfrak{I}\restriction\tau_\varphi} \sqcap X^{\kappa,n}_{\mathfrak{I}\restriction\tau_\varphi,d}$. Hence $(\mathfrak{H}\restriction\tau_\varphi, e) \xleftrightarrow{\leq\kappa\, Q}_n (\mathfrak{I}\restriction\tau_\varphi, d)$ and Lemma 4.3.6 yields $(\mathfrak{H}, e) \models \varphi(x)$. □

Corollary 4.3.9. *If the sentence $\varphi \in \mathrm{FO}(\tau)$ is invariant under $\xleftrightarrow{<\omega\, Q}$ in $\mathbb{K}$ then $\varphi$ is logically equivalent to some global concept $C \in \mathcal{ALCQO}u_1(\tau)$.*

### 4.3.3 The Characterisation of $\mathcal{ALCQO}$-TBoxes

$C \sqsubseteq D$ is called a *$\mathcal{ALCQO}$-concept inclusion* over $\tau$ if $C, D \in \mathcal{ALCQO}(\tau)$. An *$\mathcal{ALCQO}$-TBox* is a finite set of $\mathcal{ALCQO}$-concept inclusions. An interpretation $\mathfrak{I}$ *satisfies* $C \sqsubseteq D$ or *is a model of* $C \sqsubseteq D$, $\mathfrak{I} \models C \sqsubseteq D$, if $C^{\mathfrak{I}} \subseteq D^{\mathfrak{I}}$. An $\mathcal{ALCQO}$-TBox $\mathcal{T}$ *is satisfied by* $\mathfrak{I}$, $\mathfrak{I} \models \mathcal{T}$, iff $\mathfrak{I} \models C \sqsubseteq D$ for all $C \sqsubseteq D \in \mathcal{T}$.

Proposition 3.2.35 transfers to $\mathcal{ALCQO}u_1$-concepts and so we can deduce that every global $\mathcal{ALCQO}u_1$-concept is equivalent to a boolean $\mathcal{ALCQO}$-TBox of the same signature.

Considering $\mathcal{ALCQO}$-TBoxes simply as $\mathcal{ALCQ}$-TBoxes without referring to the class $\mathbb{K}$, the following is true for interpretations, in particular, of course, for those contained in $\mathbb{K}$.

Observation 4.3.10 . *Let $(\mathfrak{I}_i)_{i\in I}$ be a family of $\tau$-interpretations. $\mathcal{ALCQO}$-TBoxes are*

1. *invariant under global $\mathcal{ALCQ}$-bisimulation $\overset{<\omega}{\Longleftrightarrow}_{\mathcal{Q}}$*
2. *invariant under disjoint unions*

The the former is immediate because every TBox is equivalent to some $\mathcal{ALCQO}u_1$-concept. The latter follows directly from the proof of the corresponding Proposition 3.2.36 for $\mathcal{ALCQ}$-TBoxes.

But like in $\mathcal{ALCO}$ the disjoint union of elements in $\mathbb{K}$ is not contained in $\mathbb{K}$ as soon as $\mathsf{N_I} \neq \emptyset$ since individual names are no longer interpreted as singleton sets but become predicates with more than one element. Similar to $\mathcal{ALCO}$ we need interpretations to be coherent, i.e. for each $a \in \mathsf{N_I}$ and all $i, j \in I$ we have $(\mathfrak{I}_i, a^{\mathfrak{I}_i}) \overset{<\omega}{\Longleftrightarrow} (\mathfrak{I}_j, a^{\mathfrak{I}_j})$. The solution for $\mathcal{ALCO}$ was simple: By factorisation, individuals with the same name were merged in an equivalence class. A simple example shows now that this is no longer possible for $\mathcal{ALCQO}$ as the number of successors is violated:

Example 4.3.11 . Let $\mathfrak{I}$ contain two elements $d_0$ and $d_1$ with $a^{\mathfrak{I}} := d_0$ and $(d_0, d_1) \in r^{\mathfrak{I}}$ and $\mathfrak{H}$ contain three elements $e_0$, $e_1$ and $e_2$ where $a^{\mathfrak{H}} := e_0$ and $(e_0, e_1), (e_0, e_2) \in r^{\mathfrak{H}}$. Both, $\mathfrak{I}$ and $\mathfrak{H}$ satisfy $a \sqsubseteq \exists^{\leq 2} r.\top$. By simple merging $a^{\mathfrak{I}}$ and $a^{\mathfrak{H}}$ we would obtain 3 $r$-successors of $a^{\mathfrak{I} \uplus \mathfrak{H}}$.

One could try to fix this problem by also merging successor nodes. But how to merge these successor nodes? A simple factorisation by $\overset{<\omega}{\Longleftrightarrow}$ would merge $\mathcal{ALCQ}$-bisimilar successors, again not preserving the number of successors: Our example would end up with only one successor node. Even for tree-interpretations, the examples can be arbitrarily complicated because the problem of surplus successor propagates with each merged node.

But what we actually need is merely altering the structures so that they are still globally $\mathcal{ALCQ}$-bisimilar to their original and stay in $\mathbb{K}$. The solution given here is the following:

Let $(\mathfrak{I}_i)_{i\in I}$ be a family of coherent interpretations. For the sake of simplicity we fix just one $i \in I$ at the moment. We perform a forest unravelling on each structure $\mathfrak{I}_j$ where $j \in I$ and are now possibly outside of $\mathbb{K}$. We replace every subtree whose root is labelled with an individual name $a$ by the tree with its root labelled $a$ in $\mathfrak{I}_i$ obtaining for each $\mathfrak{I}_j^F$ an $\mathfrak{I}_j^{F'}$. We then have for all $j \in I$

1. every subtree in $\mathfrak{I}_j^{F'}$ with a root labelled $a \in \mathsf{N_I}$ is isomorphic to the subtree starting with $a^{\mathfrak{I}_i}$ in $\mathfrak{I}_i$.

2. $\mathfrak{I}_j^{F'}$ is still globally $\mathcal{ALCQ}$-bisimilar to $\mathfrak{I}_j^F$ since we had coherent interpretations and so the subtrees starting with the same individual name are bisimilar.

In order to get back into $\mathbb{K}$ we now factorise each interpretation by equality of the last letter in each element. Recall: forest unravellings (cf. page 56) have the collection of all finite paths $d_0 \cdot r_0 \cdot d_1 \cdot r_1 \dot{d}_2 \cdots r_{n-1} \cdot d_n$ as carrier set. Hence different elements with the same last letter are in fact one and the same element reached via different paths. So factorising by this relation is the inverse of tree-unravelling.

Since we have altered our interpretations by replacing subtrees by $\mathcal{ALCQ}$-bisimilar subtrees, each quotient structures $\mathfrak{I}_j^{F'}/E$ is bisimilar to $\mathfrak{I}_j$. But $\mathfrak{I}_j^{F'}/E$ is also in $\mathbb{K}$ as every element labelled with an individual name $a$ had as last letter $a^{\mathfrak{I}_i}$.

One has therefore the possibility to assume that a coherent family has w.l.o.g. identical tree-unravellings in their individuals. A factorisation of the disjoint union of $\mathfrak{I}_i^{F'}$ over all $i \in I$ by the equality of the last letter, yields an interpretation in $\mathbb{K}$. In particular this equivalence relation respects numbers of $r$-successors for all $r \in \mathsf{N_R}$ up to $\omega$ and the resulting interpretation is globally $\mathcal{ALCQ}$-bisimilar to the disjoint union of the original family $(\mathfrak{I}_i)_{i\in I}$.

So: the aim is to uniformly replace $\overset{<\omega}{\Longleftrightarrow}$-subtrees in tree-unravellings of interpretations and later on to factorise by the equivalence relation. We shall perform this in some generality as we later on have to deal with cases where we cannot simply choose one single $i \in I$.

Construction of Coherent Sub-Interpretations.

Let $I$ be an index set and $(\mathfrak{I}_i)_{i\in I}$ a family of pairwise disjoint $\tau$-interpretations, which are not necessarily coherent. A system of substitutions $(\sigma_i)_{i\in I}$ is a family

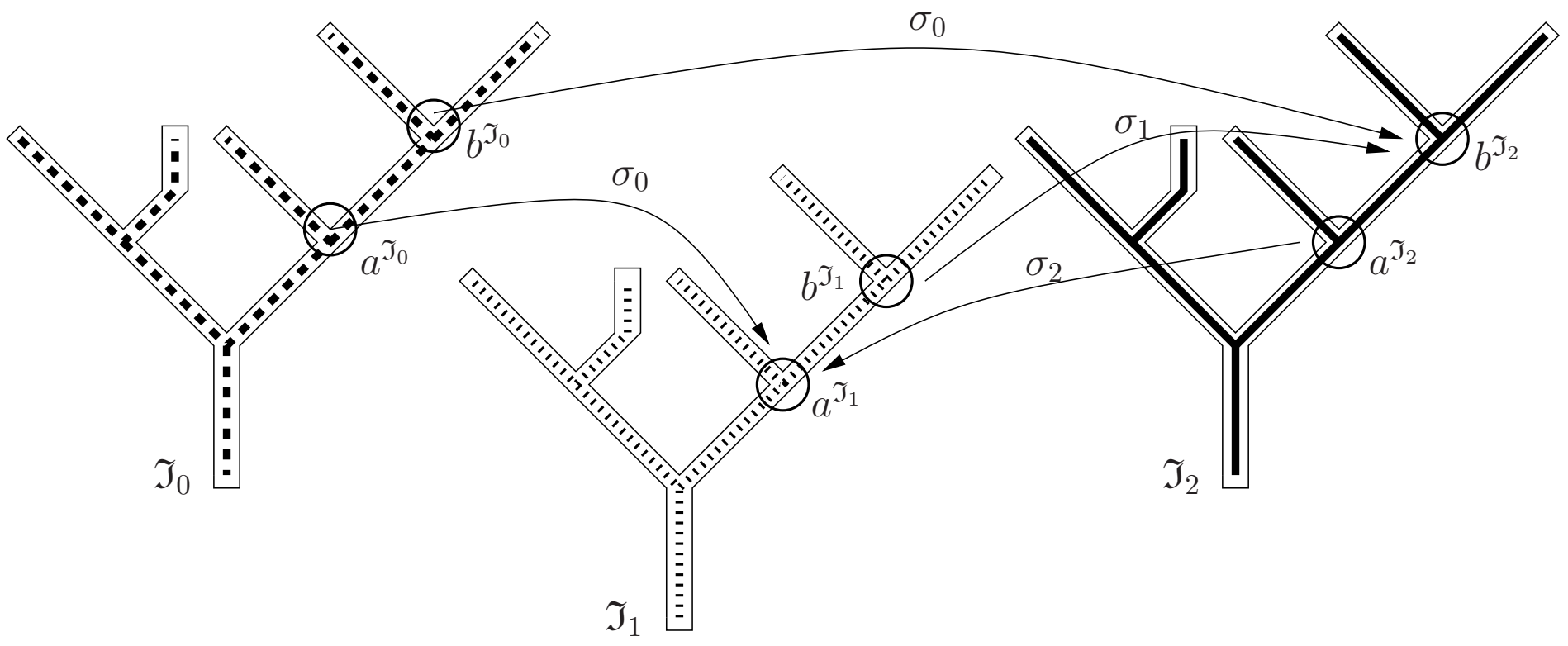

Figure 4.1: This picture exemplifies a system of substitutions $(\sigma_i)_{i\in I}$: Depicted is a family of pairwise disjoint $\tau$-interpretations $\mathfrak{I}_0, \mathfrak{I}_1, \mathfrak{I}_2$ where, as the picture suggests, $(\mathfrak{I}_i, a^{\mathfrak{I}_i}) \overset{\leq\omega}{\Longleftrightarrow} (\mathfrak{I}_j, a^{\mathfrak{I}_j})$ and similarly for $b$ for all $i, j \in \{0, 1, 2\}$. We have chosen one index for each individual $a$ and $b$ in $\mathsf{N}_\mathsf{I}$, $i_a = 1$ and $i_b = 2$, where the individuals of the other interpretations are mapped to.

of mappings such that for each $i \in I$

$$\sigma_i : \Delta^{\mathfrak{I}_i} \supseteq \mathbb{D}_i \longrightarrow (\bigcup_{i\in I} \Delta^{\mathfrak{I}_i}) \setminus (\bigcup_{i\in I} \mathbb{D}_i)$$

where $\mathbb{D}_i$ denotes the domain of $\sigma_i$.

We have excluded the domains from the range in order to exclude compositions of substitutions which would allow some sort of 'forwarding' and eventually infinite or circular chains of substitutions whose result would be undefined.

We define $[\cdot] : \bigcup_{i\in I} \mathbb{D}_i \longrightarrow I$ where $d_m \longmapsto j$ if $d_m \in \mathbb{D}_i \Longrightarrow \sigma_i(d_m) \in \Delta^{\mathfrak{I}_j}$. This function is well defined as all interpretations were pairwise disjoint.

We continue each $\sigma_i$ to a mapping on the forest unravelling $\mathfrak{I}_i^F$ of $\mathfrak{I}_i$. The mapping $\bar{\sigma}_i$ is recursively defined over the length of the path-elements and we shall consider at first only those path-elements that contain some $d \in \mathbb{D}_i$. Let $\bar{d} = d_0 \cdots d_n$ have length $n + 1$ and let $m$ be the smallest index such that $d_m \in \mathbb{D}_i$. Then

$$\begin{aligned}\bar{\sigma}_i(\bar{d}) := \; & \{d_0 \cdots d_{m-1} \cdot r_{m-1}\} \times \{\sigma_i(d_m)\} \times \\ & \Big(\{\varepsilon\} \cup \bigcup\nolimits_{r\in \mathsf{N}_\mathsf{R}} \{r \cdot \bar{\sigma}_{[d_m]}(\bar{e}) \mid \sigma_i(d_m) \cdot r \cdot \bar{e} \text{ in } \mathfrak{I}^F_{[d_m]} \text{ and } |\bar{e}| = n - m\}\Big)\end{aligned}$$

Note that $\bar{e}$ is properly shorter than $\bar{d}$ which has a length of $n + 1$. Additionally we added the empty-word $\varepsilon$ to avoid $\sigma_i(\bar{d})$ evaluating to the empty set in case $d_m$ is a leaf in $\mathfrak{I}^F_{[d_m]}$. In this case $\sigma_i(\bar d) = \{d_0 \cdots d_{m-1} \cdot r_{m-1} \cdot \sigma_i(d_m)\}$.

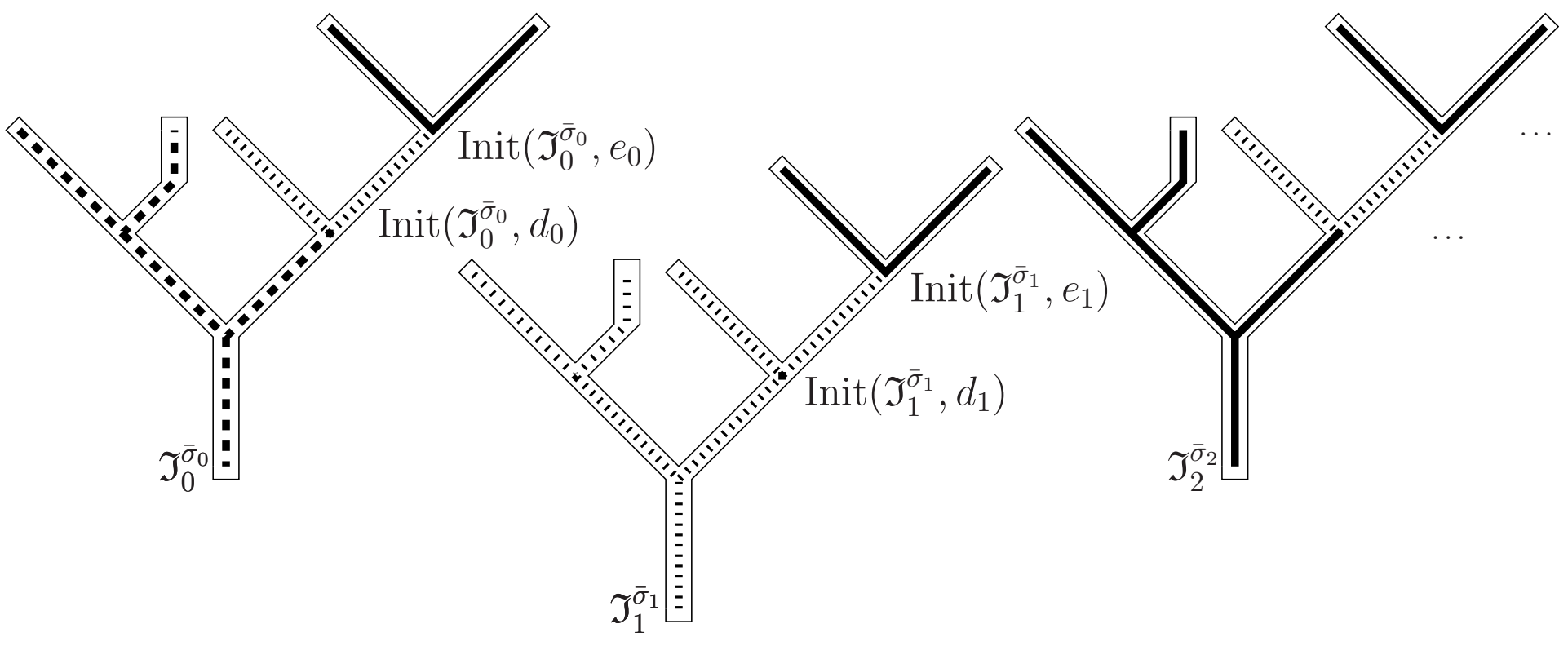

Figure 4.2: We refer to the initial situation in Figure 4.1. This figure merely gives an intuition of what happens. It does not take into account that $\mathfrak{I}_i^{\bar\sigma_i}$ are in fact forest-unravellings, etc. However, after the substitution $\bar\sigma_1$ is performed on $\mathfrak{I}_1$, the subtree $\mathrm{Init}(\mathfrak{I}_1, b^{\mathfrak{I}_1})$ is 'replaced' with the subtree $\mathrm{Init}(\mathfrak{I}_2, b^{\mathfrak{I}_2})$, yielding $\mathrm{Init}(\mathfrak{I}_1^{\bar\sigma_1}, d_1)$, where $d_1$ corresponds to the former $a^{\mathfrak{I}_1}$. In $\mathfrak{I}_0$, $\mathrm{Init}(\mathfrak{I}_0, a^{\mathfrak{I}_0})$ is replaced by $\mathrm{Init}(\mathfrak{I}_1^{\bar\sigma_1}, d_1)$ which results in the subtree $\mathrm{Init}(\mathfrak{I}_0^{\bar\sigma_0}, d_0)$ where, similarly, $d_0$ corresponds to $a^{\mathfrak{I}_0}$. In $\mathfrak{I}_2$, $\mathrm{Init}(\mathfrak{I}_2, a^{\mathfrak{I}_2})$ is replaced with $\mathrm{Init}(\mathfrak{I}_1^{\bar\sigma_1}, d_1)$.

What $\bar\sigma_i$ essentially does, is to delete from the tree-unravelling in $d_0$ contained in $\mathfrak{I}_i$ the largest subtree that starts with an element $d_m$ from $\mathbb{D}_i$ and then glues the fully substituted subtree (recursion) emerging from $\sigma_i(d_m)$ in its place. It accomplishes this by replacing each element $\bar d$ with prefix $d_0 \cdots d_{m-1} \cdot r_{m-1} \cdot d_m$ with *all* paths of the from the fully substituted tree (recursion) of $d_0$ which have the same length $|\bar d|$ and prefix $d_0 \cdots d_{m-1} \cdot r_{m-1} \cdot d_m$. It uses the fact that the tree-unravelling of $\mathfrak{I}_{[d_m]}$ in $\sigma_i(d_m) \cdot r \cdot e_0$ is isomorphic to the tree-unravelling of $\mathfrak{I}_{[d_m]}$ in $e_0$. This essentially shortens the path-element which is to be recursively substituted and yields the well-foundedness of the recursion. Clearly, the elements of this fully substituted tree might be 'imported' multiple times, but this not of importance.

If $\bar d$ does not contain any element from $\mathbb{D}_i$ then $\bar\sigma_i(\bar d) = \{\bar d\}$. With $\lambda\bar d$ we denote the last element of the tuple $\bar d$. For every $i \in I$ we define $\mathfrak{I}_i^{\bar\sigma_i}$ as

$$
\begin{aligned}
\Delta^{\mathfrak{I}_i^{\bar\sigma_i}} &:= \bigcup\{\bar\sigma_i(\bar d) \mid \bar d \in \Delta^{\mathfrak{I}_i^F}\} & \\
S^{\mathfrak{I}_i^{\bar\sigma_i}} &:= \{\bar d \in \Delta^{\mathfrak{I}_i^{\bar\sigma_i}} \mid \lambda\bar d \in \textstyle\bigcup_{i\in I} S^{\mathfrak{I}_i}\} & S \in \mathsf{N_I} \cup \mathsf{N_C} \\
r^{\mathfrak{I}_i^{\bar\sigma_i}} &:= \{(\bar d, \bar d') \in \Delta^{\mathfrak{I}_i^{\bar\sigma_i}} \times \Delta^{\mathfrak{I}_i^{\bar\sigma_i}} \mid \bar d' = \bar d \cdot r \cdot \lambda\bar d'\} & r \in \mathsf{N_R}
\end{aligned}
$$

It is obvious from the definition of roles $r^{\mathfrak{I}_i^{\bar{\sigma}_i}}$ that $\mathfrak{I}_i^{\bar{\sigma}_i}$ is a forest interpretation. Yet $\mathfrak{I}_i^{\bar{\sigma}_i}$ is in general not the tree-unravelling of an interpretation, i.e. not every last letter $\lambda\bar{d}$ of some element $\bar{d}$ in $\mathfrak{I}_i^{\bar{\sigma}_i}$ occurs as path-element in $\mathfrak{I}_i^{\bar{\sigma}_i}$

Let $\Delta^{\mathrm{Init}(\mathfrak{I},\bar{d})} := \{\bar{d}' \in \Delta^{\mathfrak{I}} \mid \bar{d}$ is initial piece of $\bar{d}'\}$ and $\mathrm{Init}(\mathfrak{I},\bar{d}) := \mathfrak{I}\restriction\Delta^{\mathrm{Init}(\mathfrak{I},\bar{d})}$ It is not obvious that $\mathrm{Init}(\mathfrak{I}_i^{\bar{\sigma}_i},\bar{d})$ is a tree. By definition of $\mathfrak{I}_i^{\bar{\sigma}_i}$, it is a forest interpretation, but there could be 'gaps' between the elements, causing them to be in different connected components. In what follows we shall therefore show that the intuition given by the figures above is true: We obtain forest-like interpretations, whose subtrees emerging from some element in the domain of $\sigma_i$ is isomorphic to the subtree emerging from its image under $\sigma_i$. In order to not get lost we shall describe the structure: Proposition 4.3.12 rests on Lemma 4.3.13 whilst Proposition 4.3.18 uses Lemmas 4.3.15 as well as 4.3.16 and 4.3.17.

Proposition 4.3.12 . *Every element $\bar{d}$ in $\mathfrak{I}_i^{\bar{\sigma}_i}$ with $|\bar{d}| > 1$ has a predecessor in $\mathfrak{I}_i^{\bar{\sigma}_i}$.*

We shall give the proof after the following lemma which states: the predecessor of $e \in \bar{\sigma}_i(\bar{d})$ is in $\bar{\sigma}_i(\text{predecessor}(\bar{d}))$.

Lemma 4.3.13 . *For all $\bar{d}$ with $|\bar{d}| > 1$ we have for all $i \in I$: if $\bar{d} \in \Delta^{\mathfrak{I}_i^F}$ and $\tilde{d}$ is its predecessor, i.e. there is $r \in \mathsf{N_R}$ with $\tilde{d} \cdot r \cdot \lambda\bar{d}$, then*

$$\textit{for all } \bar{e} \in \bar{\sigma}_i(\bar{d}) \textit{ exists a predecessor } \tilde{e} \in \bar{\sigma}_i(\tilde{d}) \textit{ in } \mathfrak{I}_i^{\bar{\sigma}_i}$$

Proof. We prove the claim by induction upon the length of $\bar{d}$. It is easy to show the claim directly if the predecessor does not contain an element from $\mathbb{D}_i$. Let therefore $\bar{d} \in \Delta^{\mathfrak{I}_i^F}$ such that $\tilde{d}$ contains an element in $\mathbb{D}_i$. Let $m$ be the smallest index of $\bar{d}$ such that $d_m \in \mathbb{D}_i$ and let $j$ such that $\sigma_i(d_m) \in \Delta^{\mathfrak{I}_j}$.

Assume $\bar{e} \in \bar{\sigma}_i(\bar{d})$ then there is $r \in \mathsf{N_R}$ and $\bar{g} \in \Delta^{\mathfrak{I}_j^F}$ with $\sigma_i(d_m) \cdot r \cdot \bar{g} \in \mathfrak{I}_j^F$ such that

$$\bar{e} \in \{d_0 \cdots d_{m-1} \cdot r_{m-1}\} \times \{\sigma_i(d_m)\} \times r \cdot \bar{\sigma}_j(\bar{g})$$

If $\bar{g}$ is a singleton then $|\tilde{d}| = m + 1$ and so $\bar{\sigma}_i(\tilde{d}) = \{d_0 \cdots d_{m-1} \cdot r_{m-1} \cdot \sigma_i(d_m)\}$ which contains the predecessor of $\bar{e}$ and hence proves the claim. Otherwise, $\bar{g}$ has a predecessor $\tilde{g}$ in $\mathfrak{I}_j^{\bar{\sigma}_j}$ and by induction hypothesis $\bar{\sigma}(\tilde{g})$ contains the predecessor $e_{m+1} \cdots e_{|\bar{e}|-2}$ of $e_{m+1} \cdots e_{|\bar{e}|-1}$. Since $\sigma_i(d_m) \cdot r \cdot \bar{g}$ is an element in $\mathfrak{I}_j^F$, so is $\sigma_i(d_m) \cdot r \cdot \tilde{g}$ and hence

$$\{d_0 \cdots d_{m-1} \cdot r_{m-1}\} \times \{\sigma_i(d_m)\} \times r \cdot \bar{\sigma}_j(\tilde{g}) \subseteq \bar{\sigma}_i(\tilde{d})$$

which proves the claim. □

Proof of Proposition 4.3.12. As $\bar{\sigma}_i$ preserves lengths, we have for all $\bar{d}$ in $\mathfrak{I}_i^{\bar{\sigma}_i}$ with $|\bar{d}| > 1$: there is $\bar{d}'$ in $\mathfrak{I}_i^F$ with $|\bar{d}'| = |\bar{d}|$ such that $\bar{d} \in \sigma_i(\bar{d}')$ and hence for the predecessor $\bar{d}''$ of $\bar{d}'$ there is a predecessor contained in $\bar{\sigma}_i(\bar{d}'')$. □

Corollary 4.3.14. *Init*$(\mathfrak{I}_i^{\bar{\sigma}_i}, \bar{d})$ *is a tree.*

Lemma 4.3.15. *For all* $\bar{e}$ *and* $i \in I$*: if* $\bar{e} \in \Delta^{\mathfrak{I}_i^{\bar{\sigma}_i}}$ *and* $\lambda\bar{e} \in \Delta^{\mathfrak{I}_j}$ *then* $\lambda\bar{e} \in \Delta^{\mathfrak{I}_j^{\bar{\sigma}_j}}$.

Proof. We lay the induction base for a singleton $e$. The claim is immediate, if $i = j$. If $i \neq j$, there is $d \in \Delta^{\mathfrak{I}_i^F}$ such that $\sigma_i(d) = e$. By the definition of $\sigma_i$ this can only be the case if $e \notin \mathbb{D}_j$. Hence $\bar{\sigma}_j(e) = \{e\}$ and so $e \in \Delta^{\mathfrak{I}_j^{\bar{\sigma}_j}}$.

For the step case, let $|\bar{e}| > 1$ and $\bar{d} \in \mathfrak{I}_i^F$ such that $\bar{e} \in \bar{\sigma}_i(\bar{d})$. If $\bar{d}$ does not contain any elements for $\mathbb{D}_i$. Then $\bar{e} = \bar{d}$ and in particular $\lambda\bar{d} \notin \mathbb{D}_i$. Hence $\lambda\bar{d} \in \Delta^{\mathfrak{I}_i^{\sigma_i}}$.

Otherwise there is a minimal $m < |\bar{d}|$ with $d_m \in \mathbb{D}_i$ and some $\sigma_i(d_m) \cdot r \cdot \bar{g} \in \Delta^{\mathfrak{I}_j^F}$ such that either

$$\bar{e} = d_0 \cdots d_{m-1} \cdot r_{m-1} \cdot \sigma_i(d_m) \quad \text{or} \quad \bar{e} \in \{d_0 \cdots d_{m-1} \cdot r_{m-1}\} \times \{\sigma_i(d_m)\} \times r \cdot \bar{\sigma}_j(\bar{g})$$

In both cases $\sigma_i(d_m) \notin \mathbb{D}_j$ by definition of $\sigma_i$ and hence, in the former case $\sigma_i(d_m) \in \Delta^{\mathfrak{I}_j^{\sigma_j}}$. In the latter case $\tilde{e} := e_{m+1} \cdots e_{|\bar{e}|-1}$ is in $\bar{\sigma}_j(\bar{g})$ and hence $\tilde{e} \in \Delta^{\mathfrak{I}_j^{\bar{\sigma}_j}}$. Since $|\tilde{e}| < |\bar{e}|$ the induction hypothesis yields for $\tilde{e}$ that if $\lambda\tilde{e} \in \Delta^{\mathfrak{I}_k^F}$ then $\lambda\tilde{e} \in \Delta^{\mathfrak{I}_k^{\bar{\sigma}_k}}$. □

For the following proposition, it is important to keep in mind that up to now, we have not made any requirements that $\sigma_i$ has to map elements to elements with a special property, e.g. being bisimilar to its pre-image. Proposition 4.3.18 imposes a weak restriction, namely that Init$(\mathfrak{I}_i^F, d)$ and Init$(\mathfrak{I}_{[d]}^F, \sigma_i(d))$ have the same depth for all $i \in I$ and $d \in \mathbb{D}_i$, i.e. for all $\bar{d} \in \Delta^{\text{Init}(\mathfrak{I}_i^F, d)}$ exists $\bar{e} \in \Delta^{\text{Init}(\mathfrak{I}_{[d]}^F, \sigma_i(d))}$ with $|\bar{d}| = |\bar{e}|$ and vice versa.

Lemma 4.3.16. *If* $\bar{d} \in \Delta^{\mathfrak{I}_i^{\sigma_i}}$ *with* $\bar{d} = d_0 \cdots d_n$, $|\bar{d}| > 1$ *and* $d_0 \in \Delta^{\mathfrak{I}_i}$ *then* $d_1 \cdots d_{|\bar{d}|-1} \in \Delta^{\mathfrak{I}_i^{\sigma_i}}$

Proof. Let $\bar{g} \in \mathfrak{I}_i^F$ such that $\bar{e} \in \bar{\sigma}_i(\bar{g})$. Since $\bar{\sigma}_i(g_0) = \{d_0\} \subseteq \Delta^{\mathfrak{I}_i^{\bar{\sigma}_i}}$ we may w.l.o.g. assume that $g_0 = d_0$. If $\bar{g}$ does not contain any elements in $\mathbb{D}_i$ the claim is trivial. Otherwise let $m < |\bar{g}| - 1$ minimal such that $g_m \in \mathbb{D}_i$. Then there is $r \in \mathsf{N_R}$ and $\tilde{e} \in \mathfrak{I}_j^F$ such that $\sigma_i(g_m) \cdot r \cdot \tilde{e} \in \mathfrak{I}_j^F$ and $\bar{d} \in \{g_0 \cdots g_{m-1} \cdot r_{m-1} \cdot \sigma(g_m)\} \times r \cdot \bar{\sigma}_j(\tilde{e})$.

Now

$$\begin{aligned} & \{g_0 \cdots g_{m-1} \cdot r_{m-1} \cdot \sigma(g_m)\} \times r \cdot \bar{\sigma}_j(\tilde{e}) \\ = & \{g_0\} \times r_0 \cdot \{g_1 \cdots g_{m-1} \cdot r_{m-1} \cdot \sigma(g_m)\} \times r \cdot \bar{\sigma}_j(\tilde{e}) \\ \subseteq & \{g_0\} \times r_0 \cdot \bar{\sigma}_j(\tilde{g}) \end{aligned}$$

where $\tilde{g} := g_1 \cdots g_{|\tilde{g}|-1}$. Since $\tilde{g} \in \Delta^{\mathfrak{I}_i^F}$ we have $d_1 \cdots d_{|\bar{d}|-1} \in \Delta^{\mathfrak{I}_i^{\bar{\sigma}_i}}$. □

LEMMA 4.3.17 . *Assume the trees Init*($\mathfrak{I}_i^F, d$) *and Init*($\mathfrak{I}_{[d]}^F, \sigma_i(d)$) *have the same depth for all* $d \in \mathbb{D}_i$ *and* $i \in I$. *For* $d_0 \cdot r_0 \cdot d_1 \in \mathfrak{I}_i^{\bar{\sigma}_i}$ *with* $d_0 \in \mathfrak{I}_i^F$ *we have*

$$\bar{d} \in \mathit{Init}(\mathfrak{I}_i^{\bar{\sigma}_i}, d_1) \Longrightarrow (d_0 \cdot r \cdot \bar{d}) \in \mathit{Init}(\mathfrak{I}_i^{\bar{\sigma}_i}, d_0)$$

PROOF. We distinguish the case $d_1 \in \Delta^{\mathfrak{I}^i}$: Then $d_0 \cdot r_0 \cdot d_1 \in \mathfrak{I}_i^F$. For $\bar{d} \in \mathrm{Init}(\mathfrak{I}_i^{\bar{\sigma}_i}, d_1)$ exists $\bar{g} \in \mathrm{Init}(\mathfrak{I}_i^F, d_1)$ such that $\bar{d} \in \bar{\sigma}_i(\bar{g})$. Since $d_0 \cdot r_0 \cdot d_1 \in \mathfrak{I}_i^F$ and $g_1 = d_1$ also $d_0 \cdot r_0 \cdot \bar{g} \in \mathfrak{I}_i^F$. It is $\bar{\sigma}_i(d_0 \cdot r_0 \cdot \bar{g}) = \{d_0\} \times r_0 \cdot \bar{\sigma}_i(\bar{g})$ which proves $\bar{d} \in \Delta^{\mathrm{Init}(\mathfrak{I}_i^{\bar{\sigma}_i}, d_0)}$.

If $d_1 \in \Delta^{\mathfrak{I}_j}$ with $j \neq i$ then there is $\tilde{g} \in \mathrm{Init}(\mathfrak{I}_i^F, d_0)$ such that $\bar{\sigma}_i(\tilde{g}) = \{d_0 \cdot r_0 \cdot d_1\}$, where $\sigma_i(\tilde{g}_1) = d_1$. Let $\bar{d} \in \mathrm{Init}(\mathfrak{I}_i^{\bar{\sigma}_i}, d_1)$ arbitrary, then there is $\bar{q} \in \mathfrak{I}_i^F$ such that $\bar{d} \in \bar{\sigma}_i(\bar{q})$. From the premise of this lemma we infer $\mathrm{Init}(\mathfrak{I}_i^F, \tilde{g}_1)$ and $\mathrm{Init}(\mathfrak{I}_i^F, q_0)$ have the same depth. Hence there is $\bar{g}$ in $\mathrm{Init}(\mathfrak{I}_i^F, \tilde{g}_1)$ with $|\bar{g}| = |\bar{q}|$ such that $\bar{\sigma}_i(\bar{g}) = \bar{\sigma}_i(\bar{q})$. Since $\tilde{g}_1 = g_0$ we also know that $d_0 \cdot r_0 \cdot \bar{g} \in \mathfrak{I}_i^F$. Hence $d_0 \cdot r_0 \cdot \bar{\sigma}_i(\bar{g}) = \bar{\sigma}_i(d_0 \cdot r_0 \cdot \bar{g}) \subseteq \Delta^{\mathrm{Init}(\mathfrak{I}_i^{\bar{\sigma}_i}, d_0)}$ which proves that $\bar{d} \in \Delta^{\mathrm{Init}(\mathfrak{I}_i^{\bar{\sigma}_i}, d_0)}$. □

PROPOSITION 4.3.18 . *If the trees Init*($\mathfrak{I}_i^F, d$) *and Init*($\mathfrak{I}_{[d]}^F, \sigma_i(d)$) *have the same depth for all* $d \in \mathbb{D}_i$ *and* $i \in I$ *then* $\iota$ *exists for every* $\bar{e} \in \Delta^{\mathfrak{I}_i^{\bar{\sigma}_i}}$ *and is an isomorphism where*

$$\iota : \mathit{Init}(\mathfrak{I}_i^{\bar{\sigma}_i}, \bar{e}) \longrightarrow \mathit{Init}(\mathfrak{I}_j^{\bar{\sigma}_j}, \lambda\bar{e}) : \bar{e} \cdot r \cdot \bar{e}' \longmapsto \lambda\bar{e} \cdot r \cdot \bar{e}' \text{ with } \lambda\bar{e} \in \Delta^{\mathfrak{I}_j}$$

PROOF. The proof is carried out by induction upon the length of $\bar{e}$. Let for the base case $e$ be a single lettered path-element. It suffices to show $\mathrm{id} : \mathrm{Init}(\mathfrak{I}_i^{\bar{\sigma}_i}, e) \longrightarrow \mathrm{Init}(\mathfrak{I}_j^{\bar{\sigma}_j}, e)$ exists and is surjective. For then, due to the definition of $\mathfrak{I}_i^{\bar{\sigma}_i}$ and $\mathfrak{I}_j^{\bar{\sigma}_j}$ respectively, this function is an isomorphism. If $e \in \Delta^{\mathfrak{I}_i}$, i.e. $i = j$ the claim is trivial. Assume $e \in \Delta^{\mathfrak{I}_j}$ with $j \neq i$.

Totality: Let $\bar{e} \in \mathrm{Init}(\mathfrak{I}_i^{\bar{\sigma}_i}, e)$. Then there is $\bar{g} \in \Delta^{\mathfrak{I}_i^F}$ such that $\bar{e} \in \bar{\sigma}_i(\bar{g})$. Since $\bar{e}$ starts with $e \notin \Delta^{\mathfrak{I}_i}$, for any such $\bar{g}$ we have $g_0 \in \sigma_i^{-1}(e)$, i.e. $g_0 \in \mathbb{D}_i$ and hence the evaluation of $\bar{\sigma}_i(\bar{g})$ implies that there must be $\bar{d} \in \mathfrak{I}_j^F$ and $r \in \mathsf{N_R}$ with $e \cdot r \cdot \bar{d} \in \mathfrak{I}_j^F$ so that $\bar{e} \in \{e\} \times r \cdot \bar{\sigma}_j(\bar{d})$.

With $e \in \sigma_i(\mathbb{D}_i)$ the definition of $\sigma_i$ yields $e \notin \mathbb{D}_j$ and so $\{e\} \times r \cdot \bar{\sigma}_j(\bar{d}) = \bar{\sigma}_j(e \cdot r \cdot \bar{d})$ which proves that $\bar{e} \in \Delta^{\mathfrak{I}_j^{\bar{\sigma}_j}}$. It follows immediately that $\bar{e} \in \mathrm{Init}(\mathfrak{I}_j^{\bar{\sigma}_j}, e)$.

Surjectivity: Assume $\bar{e} \in \Delta^{\mathrm{Init}(\mathfrak{I}_j^{\bar{\sigma}_j}, e)}$. There is $\bar{g} \in \Delta^{\mathfrak{I}_j^F}$ with $\bar{e} \in \bar{\sigma}_j(\bar{g})$. We know that $\bar{g} \in \mathrm{Init}(\mathfrak{I}_j^F, e)$ because $\bar{\sigma}_j(g_0) = \{e\}$ and $e \notin \mathbb{D}_j$.

Let $d \in \sigma^{-1}(e)$ be arbitrary. Since $\mathrm{Init}(\mathfrak{I}_i^F, d)$ and $\mathrm{Init}(\mathfrak{I}_j^F, e)$ have the same depth, there is $\bar{d} \in \Delta^{\mathrm{Init}(\mathfrak{I}_i^F, d)}$ such that $|\bar{d}| = |\bar{g}|$. For appropriate $r_0 \in \mathsf{N_R}$

$$\begin{aligned}\bar{\sigma}_i(\bar{d}) &= \{\sigma_i(d)\} \times \bigcup_{r \in \mathsf{N_R}} \{r \cdot \bar{\sigma}_j(\tilde{g}) \mid e \cdot r \cdot \tilde{g} \in \Delta^{\mathfrak{I}_j^F} \text{ and } |\tilde{g}| = |\bar{g}| - 1\} \\ &\supseteq \{e\} \times r_0 \cdot \bar{\sigma}_j(g_1 \cdots g_{|\bar{g}|-1}) = \bar{\sigma}_j(\bar{g}) \ni \bar{e}\end{aligned}$$

This shows $\bar{e} \in \Delta^{\mathrm{Init}(\mathfrak{I}_i^{\bar{\sigma}_i}, e)}$ and hence id is surjective.

In the step case we have to show that $\iota : \mathrm{Init}(\mathfrak{I}_i^{\bar{\sigma}_i}, \bar{e}) \longrightarrow \mathrm{Init}(\mathfrak{I}_j^{\bar{\sigma}_j}, \lambda\bar{e})$ of the form given above exists for $|\bar{e}| > 1$.

We distinguish the case $e_0 \in \Delta^{\mathfrak{I}_i}$. Since $|\bar{e}| > 1$ and all elements have predecessors, $e_0 \cdot r_0 \cdot e_1 \in \Delta^{\mathfrak{I}_i^{\bar{\sigma}_i}}$. Lemma 4.3.16 yields that for all $\bar{d}$ in $\mathrm{Init}(\mathfrak{I}_i^{\bar{\sigma}_i}, e_0 \cdot r_0 \cdot e_1)$ we have $d_1 \cdots d_{|\bar{d}|-1} \in \Delta^{\mathfrak{I}_i^{\bar{\sigma}_i}}$ giving rise to an injection

$$\iota_1 : \mathrm{Init}(\mathfrak{I}_i^{\bar{\sigma}_i}, e_0 \cdot r \cdot e_1) \longrightarrow \mathrm{Init}(\mathfrak{I}_i^{\bar{\sigma}_i}, e_1) : \bar{d} \longmapsto d_1 \cdots d_{|\bar{d}|-1}.$$

$\iota_0$ is also surjective: Lemma 4.3.17 entails that whenever an element starts with $\bar{d} \in \Delta^{\mathrm{Init}(\mathfrak{I}_i^{\bar{\sigma}_i}, e_1)}$ we have $e_0 \cdot r_0 \cdot \bar{d} \in \Delta^{\mathrm{Init}(\mathfrak{I}_i^{\bar{\sigma}_i}, e_0 \cdot r_0 \cdot e_1)}$.

Hence $\iota_1$ is an isomorphism between $\mathrm{Init}(\mathfrak{I}_i^{\bar{\sigma}_i}, e_0 \cdot r_0 \cdot e_1)$ and $\mathrm{Init}(\mathfrak{I}_i^{\bar{\sigma}_i}, e_1)$. For the isomorphic image $\mathrm{Init}(\mathfrak{I}_i^{\bar{\sigma}_i}, e_1 \cdots e_{|\bar{e}|-1})$ of $\mathrm{Init}(\mathfrak{I}_i^{\bar{\sigma}_i}, \bar{e})$ we can apply our induction hypothesis and obtain an isomorphism

$$\iota_2 : \mathrm{Init}(\mathfrak{I}_i^{\bar{\sigma}_i}, e_1 \cdots e_{|\bar{e}|-1}) \longrightarrow \mathrm{Init}(\mathfrak{I}_j^{\bar{\sigma}_j}, e_{|\bar{e}|-1}) : \bar{d} \longmapsto d_1 \cdots d_{|\bar{d}|-1}$$

The composition $\iota_2 \circ \iota_1 : \mathrm{Init}(\mathfrak{I}_i^{\bar{\sigma}_i}, \bar{e}) \longrightarrow \mathrm{Init}(\mathfrak{I}_i^{\bar{\sigma}_i}, \lambda\bar{e})$ is the isomorphism we are looking for.

In the other case, where $e_0 \notin \Delta^{\mathfrak{I}_i}$, we have $e_0 \in \Delta^{\mathfrak{I}_i^{\bar{\sigma}_i}}$ because all elements have predecessors according to Proposition 4.3.12. Let $e_0 \in \Delta^{\mathfrak{I}_j}$. Since $e_0 \in \sigma_i(\mathbb{D}_i)$ it follows that $e_0 \notin \mathbb{D}_j$ and hence $e_0 \in \Delta^{\mathfrak{I}_j^{\bar{\sigma}_j}}$. Like in the induction base we can construct an isomorphism $\mathrm{id} : \mathrm{Init}(\mathfrak{I}_i^{\bar{\sigma}_i}, e_0) \longrightarrow \mathrm{Init}(\mathfrak{I}_j^{\bar{\sigma}_j}, e_0)$.

Since then $e_0 \in \Delta^{\mathfrak{I}_j}$ we can construct in the same manner as above an isomorphism $\iota_2 \circ \iota_1 : \mathrm{Init}(\mathfrak{I}_j^{\bar{\sigma}_j}, \bar{e}) \longrightarrow \mathrm{Init}(\mathfrak{I}_k^{\bar{\sigma}_k}, \lambda\bar{e})$ and hence obtain the required isomorphism $\iota_2 \circ \iota_1 \circ \mathrm{id} : \mathrm{Init}(\mathfrak{I}_i^{\bar{\sigma}_i}, \bar{e}) \longrightarrow \mathrm{Init}(\mathfrak{I}_k^{\bar{\sigma}_k}, \lambda\bar{e})$. □

Observation 4.3.19. *If $\iota : Init(\mathfrak{I}_0, d_0) \longrightarrow Init(\mathfrak{I}_1, d_1)$ is an isomorphism then $(\mathfrak{I}_0, d_0) \overset{<\omega}{\Longleftrightarrow} (\mathfrak{I}_1, d_1)$.*

Henceforth we shall apply $\bar{\sigma}_i$ to single letters, pretending they were path elements (for which $\bar{\sigma}_i$ is defined), whenever we want to show a certain property for all elements in the range of $\sigma_i$ or those which are not in the domain $\mathbb{D}_{\sigma_i}$ of $\sigma_i$.

We call $\bar{\sigma}_i$ *injective on successors-sets*, if for all $r \in \mathsf{N_R}$ and for every $d \in \Delta^{\mathfrak{I}_i}$ the restriction $\bar{\sigma}_i : r^{\mathfrak{I}_i}(d) \longrightarrow \bar{\sigma}_i(r^{\mathfrak{I}_i}(d))$ is injective.

Proposition 4.3.20. *If for all $i \in I$ every $\bar{\sigma}_i$ is injective on successor-sets and for all $i \in I$ and all $d \in \mathbb{D}_i$ we have $(\mathfrak{I}_i, d) \overset{<\omega}{\Longleftrightarrow} (\mathfrak{I}_j, \sigma_i(d))$, we have $\mathfrak{I}_i \overset{<\omega\mathcal{Q}}{\Longleftrightarrow} \mathfrak{I}_i^{\bar{\sigma}_i}$*

Lemma 4.3.21. *For all $d \in \Delta^{\mathfrak{I}_i}$: $(\mathfrak{I}_i, d) \overset{<\omega}{\Longleftrightarrow} (\mathfrak{I}_i^{\bar{\sigma}_i}, \bar{\sigma}_i(d))$*

Proof. For every $\bar{d} \in \Delta^{\mathfrak{I}_i^{\sigma_i}}$ we call $d_{m+1}$ a *glue-point*, if for any $j, k \in I$ with $j \neq k$ we have $d_m \in \Delta^{\mathfrak{I}_j}$ and $d_{m+1} \in \Delta^{\mathfrak{I}_k}$. We always call $d_0$ a glue-point! The last glue-point of an element is the glue-point with the highest index.

Let **II** maintain configurations $(\mathfrak{I}_i, d'; \mathfrak{I}_i^{\sigma_i}, \bar{d}')$ such that $(\mathfrak{I}_i, d') \overset{<\omega}{\Longleftrightarrow} (\mathfrak{I}_k, \lambda\bar{d}')$, where the last glue-point of $\bar{d}'$ is in $\Delta^{\mathfrak{I}_k}$. $(*)$

This requirement is true for the start-configuration, in particular because we assume $(\mathfrak{I}_i, d) \overset{<\omega}{\Longleftrightarrow} (\mathfrak{I}_j, \sigma_i(d))$ for all $d \in \mathbb{D}_i$. By definition of $\mathfrak{I}_i^{\bar{\sigma}_i}$ both $(\mathfrak{I}_i, d)$ and $(\mathfrak{I}_i^{\bar{\sigma}_i}, \bar{\sigma}_i(d))$ are atomically equivalent.

Assume we have reached configuration $(\mathfrak{I}_i, d'; \mathfrak{I}_i^{\bar{\sigma}_i}, \bar{d}')$ with $\lambda\bar{d}' \in \Delta^{\mathfrak{I}_j}$ which satisfies the requirement $(*)$ and **I** challenges **II** with a finite set of $r$-successor $D$ of $d'$ in $\mathfrak{I}_i$. Then there is $\beta : D \longrightarrow r^{\mathfrak{I}_j}(d')$ such that $(\mathfrak{I}_i, d'') \overset{<\omega}{\Longleftrightarrow} (\mathfrak{I}_j, \beta(d''))$ for all $d'' \in D$.

For each $d'' \in D$ we have $\lambda\bar{d}' \cdot r \cdot \beta(d'') \in \mathfrak{I}_j^F$ and since $\bar{\sigma}_j(\lambda\bar{d}') = \{\lambda\bar{d}'\}$ (cf. Lemma 4.3.15) we have $\bar{\sigma}_j(\lambda\bar{d}' \cdot r \cdot \beta(d'')) = \lambda\bar{d}' \cdot r \cdot \bar{\sigma}_j \circ \beta(d'')$ and hence $\lambda\bar{d}' \cdot r \cdot \bar{\sigma}_j \circ \beta(d'')$ is an $r$-successor of $\lambda\bar{d}'$ in $\mathfrak{I}_j^{\bar{\sigma}_j}$.

Let $\iota : \mathrm{Init}(\mathfrak{I}_i^{\bar{\sigma}_i}, \bar{d}') \longrightarrow \mathrm{Init}(\mathfrak{I}_j^{\bar{\sigma}_j}, \lambda\bar{d})$ be the isomorphism constructed in Proposition 4.3.18. Using $\iota^{-1}(\lambda\bar{d}' \cdot r \cdot \bar{\sigma}_j \circ \beta(d'')) = \bar{d}' \cdot r \cdot \bar{\sigma}_j \circ \beta(d'')$ we find a set $E = \{\bar{d}' \cdot r \cdot \bar{\sigma}_j \circ \beta(d'')) \mid d'' \in D\}$ of $r$-successors of $\bar{d}'$ in $\mathfrak{I}_i^{\sigma_i}$. Since $\beta$ is injective and $\bar{\sigma}_j$ is injective on successor-sets, $|E| = |D|$.

We set $\beta' : D \longrightarrow E : d'' \longmapsto \bar{d}' \cdot r \cdot \bar{\sigma}_j \circ \beta(d'')$ as new configuration and show that it meets the requirement $(*)$: For all $d'' \in D$, we have if $\beta(d'') \notin \mathbb{D}_j$ then $\bar{\sigma}_j \circ \beta(d'') = \{\beta(d'')\}$ and by definition of $\beta$ we get $(\mathfrak{I}_i, d'') \overset{<\omega}{\Longleftrightarrow} (\mathfrak{I}_j, \beta(d''))$. If otherwise $\beta(d'') \in \mathbb{D}_j$ then $\bar{\sigma}_j \circ \beta(d'') = \{\sigma_j \circ \beta(d'')\}$ and the requirement for $\sigma_j$ entails $(\mathfrak{I}_j, \beta(d'')) \overset{<\omega}{\Longleftrightarrow} (\mathfrak{I}_k, \sigma_j \circ \beta(d''))$. Transitivity of $\overset{<\omega}{\Longleftrightarrow}$ yields $(\mathfrak{I}_i, d'') \overset{<\omega}{\Longleftrightarrow} (\mathfrak{I}_k, \sigma_j \circ \beta(d''))$.

In both cases the definition of $\mathfrak{I}_i^{\bar{\sigma}_i}$ yields that each $\bar{d}' \cdot r \cdot \bar{\sigma}_j \circ \beta(d'')$ is atomically equivalent to $\bar{\sigma}_j \circ \beta(d'')$. Hence $\beta'$ is a valid configuration that satisfies the requirement $(*)$.

Assume now **I** challenges **II** by playing a set of $r$-successors $E$ of $\bar{d}'$ in $\mathfrak{I}_i^{\bar{\sigma}_i}$. Let $\iota : \mathrm{Init}(\mathfrak{I}_i^{\bar{\sigma}_i}, \bar{d}') \longrightarrow \mathrm{Init}(\mathfrak{I}_j^{\bar{\sigma}_j}, \lambda\bar{d}')$ be the isomorphism constructed in Proposition 4.3.18.

We have $\lambda\bar{d}' \notin \mathbb{D}_j$: For the requirement $\lambda\bar{d} \in \Delta^{\mathfrak{I}_j}$ Lemma 4.3.15 ensures $\lambda\bar{d}' \in \Delta^{\mathfrak{I}_j^{\bar{\sigma}_j}}$. The latter and $\lambda\bar{d}' \in \mathbb{D}_j$ would result in $\sigma_j(\lambda\bar{d}') = \lambda\bar{d}'$ which violates the type of $\sigma_j$.

Since $\bar{\sigma}_j$ is injective for successor-sets we can determine a unique $g_{\bar{e}} \in \Delta^{\mathfrak{I}_j}$ for each $\bar{e} \in E$ such that $\iota(\bar{e}) = \lambda\bar{d}' \cdot r \cdot \bar{\sigma}_j(g_{\bar{e}}) = \bar{\sigma}_j(\lambda\bar{d}' \cdot r \cdot g_{\bar{e}})$. Hence we have an $r$-successor-set $D' := \{g_{\bar{e}} \in r^{\mathfrak{I}_j}(\lambda\bar{d}') \mid \bar{e} \in E\}$ of $\lambda\bar{d}'$ in $\mathfrak{I}_j$ for which there is an injection $\beta' : D' \longrightarrow r^{\mathfrak{I}_i}(d')$ such that $(\mathfrak{I}_i, \beta'(g_{\bar{e}})) \overset{<\omega}{\Longleftrightarrow} (\mathfrak{I}_j, g_{\bar{e}})$ for all $g_{\bar{e}} \in D'$.

We show that the configuration $\beta : \bar{e} \longmapsto \beta'(g_{\bar{e}})$ meets requirement $(*)$ and is valid: If $g_{\bar{e}} \notin \mathbb{D}_j$ then $\lambda\bar{e} = \lambda \circ \iota(\bar{e}) = \bar{\sigma}_j(g_{\bar{e}}) = g_{\bar{e}}$ and as just shown $(\mathfrak{I}_i, \beta'(g_{\bar{e}})) \overset{<\omega}{\Longleftrightarrow} (\mathfrak{I}_j, g_{\bar{e}})$, so $(\mathfrak{I}_i, \beta(\bar{e})) \overset{<\omega}{\Longleftrightarrow} (\mathfrak{I}_j, \lambda\bar{e})$.

Or $g_{\bar{e}} \in \mathbb{D}_j$. Then $(\mathfrak{I}_j, g_{\bar{e}}) \overset{<\omega}{\Longleftrightarrow} (\mathfrak{I}_k, \sigma_j(g_{\bar{e}}))$ as stipulated for all elements in $\mathbb{D}_j$. Transitivity yields again $(\mathfrak{I}_i, \beta'(g_{\bar{e}})) \overset{<\omega}{\Longleftrightarrow} (\mathfrak{I}_k, \sigma_j(g_{\bar{e}}))$, i.e. $(\mathfrak{I}_i, \beta(\bar{e})) \overset{<\omega}{\Longleftrightarrow} (\mathfrak{I}_k, \lambda\bar{e})$. So $\beta$ satisfies the requirement.

The construction of $\mathfrak{I}_i^{\sigma_i}$ yields $\bar{e}$ is atomically equivalent to $\lambda\bar{e}$ which is atomically equivalent to $\beta(\bar{e})$. Hence $\beta$ is a valid configuration. □

Proof of Proposition 4.3.20. Lemma 4.3.21 shows one direction. We only have to show: For every $\bar{d} \in \Delta^{\mathfrak{I}_i^{\sigma_i}}$ there is $d \in \Delta^{\mathfrak{I}_i}$ such that $(\mathfrak{I}, d) \overset{<\omega}{\Longleftrightarrow} (\mathfrak{I}_i^{\sigma_i}, \bar{d})$.

Let $\bar{d} \in \Delta^{\mathfrak{I}_i^{\sigma_i}}$ be arbitrary and let $d_0$ be the first element of $\bar{d}$. $\bar{d} \in \Delta^{\mathrm{Init}(\mathfrak{I}_i^{\sigma_i}, d_0)}$. Corollary 4.3.14 tells us that $\mathrm{Init}(\mathfrak{I}_i^{\sigma_i}, d_0)$ is a tree and hence there is a path from $d_0$ to $\bar{d}$.

Let $d' \in \bar{\sigma}_i^{-1}(d_0)$. Lemma 4.3.21 now yields that $(\mathfrak{I}_i, d') \overset{<\omega}{\Longleftrightarrow} (\mathfrak{I}_i^{\sigma_i}, d_0)$. In particular **II** could pretend being challenged by **I** along the path from $d_0$ to $\bar{d}$ coming up with some $d \in \Delta^{\mathfrak{I}_i}$ such that $(\mathfrak{I}, d) \overset{<\omega}{\Longleftrightarrow} (\mathfrak{I}_i^{\sigma_i}, \bar{d})$. This proves the claim. □

We set

$$E := \{(\bar{d}, \bar{d}') \in \Delta^{\biguplus_{i\in I} \mathfrak{I}_i^{\sigma_i}} \times \Delta^{\biguplus_{i\in I} \mathfrak{I}_i^{\sigma_i}} \mid \lambda\bar{d} = \lambda\bar{d}'\}$$

Observation 4.3.22. $(\mathfrak{I}_i^{\sigma_i}, \bar{d}) \overset{<\omega}{\Longleftrightarrow} (\biguplus_{i\in I} \mathfrak{I}_i^{\sigma_i}/E, [\bar{d}])$

Proof. **II** has a winning strategy if she maintains configurations $\beta : d' \longmapsto [\bar{d}]$. Every $\bar{d}'$ is atomically equivalent to $[\bar{d}']$. The start-configuration is therefore valid.

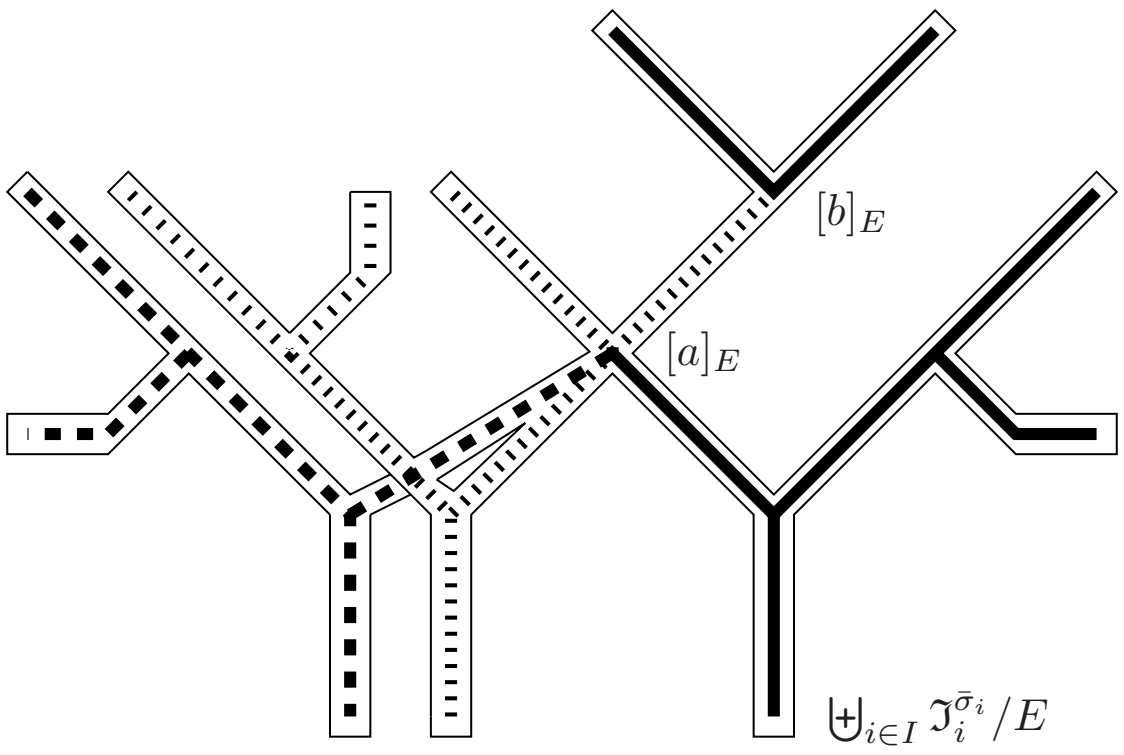

Figure 4.3: Depicted is the outcome of the factorisation performed on the interpretations from Figure 4.2: Since the elements in $\mathrm{Init}(\mathfrak{I}_i^{\bar{\sigma}_i}, d_i)$ are identical modulo some initial segment, we can identify them accross the different $\mathfrak{I}_i^{\bar{\sigma}_i}$ and merge them into equivalence classes. This ensures that in particular the individuals with the same name all fall into the same class; hence $\biguplus_{i\in I} \mathfrak{I}_i^{\bar{\sigma}_i}/E \in \mathbb{K}$.

Assume we have reached a configuration $\beta : D' \longrightarrow E'$ of the desired kind and for some $\bar{d}' \in D'$ **I** plays a finite $r$-successor set $D''$ of $\bar{d}'$ in $\mathfrak{I}_i^{\sigma_i}$. Let $E'' := \{[\bar{d}''] \mid \bar{d}'' \in D''\}$.

Each $\bar{d}'' \in D''$ is an $r$-successor of $\bar{d}'$. By definition of $r^{\biguplus_{i\in I}\mathfrak{I}_i^{\sigma_i}/E}$ each $[\bar{d}'']$ is an $r$-successor of $[\bar{d}']$ in $\biguplus_{i\in I} \mathfrak{I}_i^{\sigma_i}/E$. We also have $|D''| = |E''|$: Since $\mathfrak{I}_i^{\sigma_i}$ is a forest interpretation, $\bar{d}''$ is an $r$-successor of $\bar{d}'$ iff $\bar{d}'' = \bar{d}' \cdot r \cdot \lambda\bar{d}''$. Hence two elements in $D''$ are distinguished iff their last letter differs, i.e. for $d_0'', d_1'' \in D$ we have $\bar{d}_0'' \neq \bar{d}_1''$ iff $[\bar{d}_0''] \neq [\bar{d}_1'']$ and so $\beta' : D'' \longrightarrow E'' : \bar{d}'' \longmapsto [\bar{d}'']$ is a valid configuration.

Assume on the other hand that **II** is challenged by **I** with an $r$-successor set $E''$ of some $[\bar{d}'] \in E'$. By definition of $r^{\biguplus_{i\in I}\mathfrak{I}_i^{\bar{\sigma}_i}/E}$, each element $\mathfrak{e} \in E''$ contains an $r$-successor $\bar{d}''$ of some element $\bar{d}_{\mathfrak{e}}' \in [\bar{d}']$. Assume $\bar{d}_{\mathfrak{e}}' \in \Delta^{\mathfrak{I}_k^{\bar{\sigma}_k}}$. Proposition 4.3.18 yields isomorphisms $\iota_0 : \mathrm{Init}(\mathfrak{I}_i^{\sigma_i}, \bar{d}') \longrightarrow \mathrm{Init}(\mathfrak{I}_j^{\sigma_j}, \lambda\bar{d}')$ and $\iota_1 : \mathrm{Init}(\mathfrak{I}_k^{\bar{\sigma}_k}, \bar{d}_{\mathfrak{e}}') \longrightarrow \mathrm{Init}(\mathfrak{I}_j^{\sigma_j}, \lambda\bar{d}')$. Define $\bar{d}_{\mathfrak{e}}'' := \iota_0^{-1} \circ \iota_1(d'')$. Then $\bar{d}_{\mathfrak{e}}''$ is an $r$-successor of $\bar{d}'$ in $\mathfrak{I}_i^{\bar{\sigma}_i}$ with $\lambda\bar{d}_{\mathfrak{e}}'' = \lambda\bar{d}''$. Hence $\bar{d}_{\mathfrak{e}}''$ is atomically equivalent to $\mathfrak{e}$.

Set $\beta' : E'' \longrightarrow D'' : \mathfrak{e} \longmapsto \bar{d}_{\mathfrak{e}}''$ where $D'' := \{\bar{d}_{\mathfrak{e}}'' \mid \mathfrak{e} \in E''\}$. $D''$ is an $r$-successor set of $\bar{d}'$. We also have if $\bar{d}_{\mathfrak{e}}'' = \bar{d}_{\mathfrak{e}'}''$ iff $\lambda\bar{d}_{\mathfrak{e}}'' = \lambda\bar{d}_{\mathfrak{e}'}''$ iff $\mathfrak{e} = [\lambda\bar{d}_{\mathfrak{e}}''] = [\lambda\bar{d}_{\mathfrak{e}'}''] = \mathfrak{e}'$. Hence $\beta'$ satisfies the requirement $(*)$, is bijective and therefore valid. □

Corollary 4.3.23. $\biguplus_{i\in I} \mathfrak{I}_i \overset{<\omega Q}{\Longleftrightarrow} \biguplus_{i\in I} \mathfrak{I}_i^{\bar{\sigma}_i}/E$

Proof. If $d \in \Delta^{\biguplus_{i\in I}\mathfrak{I}_i}$ then there is $i \in I$ with $d \in \Delta^{\mathfrak{I}_i}$. Lemma 4.3.21 and Observation 4.3.22 yield that $(\mathfrak{I}_i, d) \overset{\leq\omega}{\Longleftrightarrow} (\biguplus_{i\in I}\mathfrak{I}_i^{\bar{\sigma}_i}/E, [\bar{\sigma}_i(d)])$. Hence $(\biguplus \mathfrak{I}_i, d) \overset{\leq\omega}{\Longleftrightarrow} (\biguplus_{i\in I}\mathfrak{I}_i^{\bar{\sigma}_i}/E, [\bar{\sigma}_i(d)])$.

For $\mathfrak{d} \in \Delta^{\biguplus_{i\in I}\mathfrak{I}_i^{\bar{\sigma}_i}/E}$ there is $\bar{d} \in \mathfrak{d}$ and $i \in I$ such that $\bar{d} \in \Delta^{\mathfrak{I}_i^{\bar{\sigma}_i}}$. Observation 4.3.22 yields $(\mathfrak{I}_i^{\bar{\sigma}_i}, \bar{d}) \overset{\leq\omega}{\Longleftrightarrow} (\biguplus_{i\in I}\mathfrak{I}_i^{\bar{\sigma}_i}/E, \mathfrak{d})$. Proposition 4.3.20 yields that there is $d \in \Delta^{\mathfrak{I}_i}$ with $(\mathfrak{I}_i, d) \overset{\leq\omega}{\Longleftrightarrow} (\mathfrak{I}_i^{\bar{\sigma}_i}, \bar{d})$. Hence $(\biguplus_{i\in I}\mathfrak{I}_i, d) \overset{\leq\omega}{\Longleftrightarrow} (\biguplus_{i\in I}\mathfrak{I}_i^{\bar{\sigma}_i}/E, \mathfrak{d})$. □

### The Characterisation of $\mathcal{ALCQO}$-TBoxes as FO-Fragment

We shall now introduce the notions necessary for the characterisation. Throughout this section we shall assume that families of $\tau$-interpretations $(\mathfrak{I}_i)_{i\in I}$ have pairwise disjoint carrier-sets!

It is now that we need our construction for coherent disjoint unions in all its generality: The generated subinterpretations with which we shall deal further down will not interpret all individual names and hence we need for each subinterpretation $\mathfrak{I}_i$ its own substitution $\sigma_i$.

Let $\mathfrak{K}_i$ be a family $\tau$-interpretations with

1. for all $a \in \mathsf{N_I}$ there is $i \in I$ with $a^{\mathfrak{K}_i} \neq \emptyset$
2. if $a^{\mathfrak{K}_i} \neq \emptyset$ then $|a^{\mathfrak{K}_i}| = 1$ for all $a \in \mathsf{N_I}$ and $i \in I$.
3. if $a^{\mathfrak{K}_i} \neq \emptyset \neq a^{\mathfrak{K}_j}$ then $(\mathfrak{I}_i, a^{\mathfrak{I}_i}) \overset{\leq\omega}{\Longleftrightarrow} (\mathfrak{I}_j, a^{\mathfrak{I}_j})$ for any $i, j \in I$ and all $a \in \mathsf{N_I}$.

Every family of *coherent* $\tau$-interpretations (cf. Definition 4.2.20) satisfies the requirements 1. 2. and 3. from above.

Note that if 2. and 3. is true, from $a, b \in \mathsf{N_I}$ with $a^{\mathfrak{K}_i} = b^{\mathfrak{K}_i} \neq \emptyset$ follows for all $i \in I$ that $a^{\mathfrak{K}_i} = b^{\mathfrak{K}_i}$. This is important for the existence of the function $[\cdot]$ which we shall define now.

For this family of $\tau$-interpretations, there is a function $[\cdot] : \mathsf{N_I} \longrightarrow I$ such that $a^{\mathfrak{K}_{[a]}} \neq \emptyset$ and if $a^{\mathfrak{K}_i} = b^{\mathfrak{K}_i} \neq \emptyset$ for any $i \in I$ then $[a] = [b]$. We define $\mathsf{N_I}(\mathfrak{K}_i) := \{a \in \mathsf{N_I} \mid a^{\mathfrak{K}_i} \neq \emptyset\}$ for every $i \in I$. For every $i \in I$ we define $\sigma_i : a^{\mathfrak{K}_i} \longmapsto a^{\mathfrak{K}_{[a]}}$ with $\mathbb{D}_i := \{a^{\mathfrak{K}_i} \mid a \in \mathsf{N_I}(\mathfrak{K}_i) \text{ with } [a] \neq i\}$.

The requirement $[a] \neq i$ avoids that $a^{\mathfrak{K}_i}$ in $\mathbb{D}_i$ and allows for every $j \in I$ that $\sigma_j$ is of the type $\mathbb{D}_j \longrightarrow \bigcup_{i\in I}\mathfrak{I}_i \setminus \bigcup_{i\in I}\mathbb{D}_i$. In particular, we have that for every $a \in \mathsf{N_I}$ all but one element of $\bigcup_{i\in I} a_i^{\mathfrak{I}}$ is in $\bigcup_{i\in I}\mathbb{D}_i$. All elements in $\bigcup_{i\in I} a^{\mathfrak{I}_i}$ are mapped to this single element in $\bigcup_{i\in I} a_i^{\mathfrak{I}} \setminus \bigcup_{i\in I}\mathbb{D}_i$

The next lemma proves that this system $(\sigma_i)_{i\in I}$ is admissible in the sense of Proposition 4.3.20.

LEMMA 4.3.24 . *For all $i \in I$, $\bar{\sigma}_i$ is injective on successor sets.*

PROOF. Let $d \in \Delta^{\mathfrak{K}_i}$ and $r \in \mathsf{N_R}$ be arbitrary. Consider $d_0', d_1' \in r^{\mathfrak{K}_i}(d)$ with $d_0' \neq d_1'$. We distinguish two cases: if none of them satisfies any individual name then $\bar{\sigma}_i(d_0') = \{d_0'\} \neq \{d_1'\} = \bar{\sigma}_i(d_1')$. If only one of them or both satisfy an individual name then $(\mathfrak{I}_i, d_0) \overset{\omega}{\not\Longleftrightarrow} (\mathfrak{I}_i, d_1)$. Hence also $(\mathfrak{I}_j, \bar{\sigma}_i(d_0)) \overset{\omega}{\not\Longleftrightarrow} (\mathfrak{I}_k, \bar{\sigma}_i(d_1))$ and so $\bar{\sigma}_i(d_0) \neq \bar{\sigma}_i(d_1)$. □

LEMMA 4.3.25 . *$\biguplus_{i \in I} \mathfrak{K}_i^{\bar{\sigma}_i}$ is in $\mathbb{K}$.*

PROOF. Let $\mathfrak{d}_0, \mathfrak{d}_1 \in a^{\biguplus_{i \in I} \mathfrak{K}_i^{\bar{\sigma}_i}/E}$. Then there is $\bar{d}_0 \in \mathfrak{d}_0$ and $\bar{d}_1 \in \mathfrak{d}_1$ such that $\bar{d}_0 \in a^{\mathfrak{I}_k^{\bar{\sigma}_k}}$ and $\bar{d}_1 \in a^{\mathfrak{I}_\ell^{\bar{\sigma}_\ell}}$.

Proposition 4.3.18 yields an isomorphism $\iota_0 : \mathrm{Init}(\mathfrak{K}_k^{\bar{\sigma}_k}, \bar{d}_0) \longrightarrow \mathrm{Init}(\mathfrak{I}_m^{\bar{\sigma}_m}, \lambda\bar{d}_0)$ where $\lambda\bar{d}_0 \in \Delta^{\mathfrak{I}_m}$. Since all members of the family were pairwise disjoint, $m$ is uniquely determined. The only possibility that $\lambda\bar{d}_0 \in \Delta^{\mathfrak{K}_m^{\bar{\sigma}_m}}$ and $\lambda\bar{d}_0 \in a^{\mathfrak{K}_m}$ at the same time is for $m = [a]$. For otherwise $\lambda\bar{d}_0$ would have been replaced by $a^{\mathfrak{K}_{[a]}}$ which is, due to the disjointness, not in $\Delta^{\mathfrak{I}_m}$.

The same applies for $\bar{d}_1 \in \mathfrak{d}_1$ and so $\lambda\bar{d}_0 = \lambda\bar{d}_1$. But this shows that $\mathfrak{d}_0 = \mathfrak{d}_1$ and hence that $a^{\biguplus_{i \in I} \mathfrak{K}_i^{\bar{\sigma}_i}/E}$ is interpreted by at most one element. Since $\bar{\sigma}_{[a]}(a^{\mathfrak{K}_{[a]}}) = \{a^{\mathfrak{K}_{[a]}}\}$, we have $a^{\mathfrak{K}_{[a]}} \subseteq \Delta^{\biguplus_{i \in I} \mathfrak{I}_i^{\bar{\sigma}_i}}$ and therefore each $a \in \mathsf{N_I}$ is interpreted by exactly one element in $\biguplus_{i \in I} \mathfrak{K}_i^{\bar{\sigma}_i}/E$. □

Corollary 4.3.23 shows that it does not matter which function $[\cdot] : \mathsf{N_I} \longrightarrow I$ such that $a^{\mathfrak{K}_{[a]}} \neq \emptyset$ we choose, we shall always obtain $\biguplus_{i \in I} \mathfrak{K}_i \overset{<\omega Q}{\Longleftrightarrow} \biguplus_{i \in I} \mathfrak{K}_i^{\bar{\sigma}_i}/E$. Since all these disjoint unions are globally $\mathcal{ALCQ}$-bisimilar we simply give

DEFINITION 4.3.26 . For a family of $\tau$-interpretations $(\mathfrak{I}_i)_{i \in I}$ that satisfy requirement 1. 2. 3. the *coherent disjoint union* $\biguplus_{i \in I}^{\mathbb{K}} \mathfrak{I}_i$ is defined as $\biguplus_{i \in I} \mathfrak{I}_i^{\bar{\sigma}_i}/E$, where $(\sigma_i)_{i \in I}$ is a fix but arbitrary system such that $\sigma_i : a^{\mathfrak{I}_i} \longmapsto a^{\mathfrak{I}_{[a]}}$ with $\mathbb{D}_i := \{a^{\mathfrak{I}_i} \mid a \in \mathsf{N_I}(\mathfrak{I}_i) \text{ with } [a] \neq i\}$ for all $i \in I$. ◇

The invariance of $\mathcal{ALCQO}$-TBoxes under global bisimulation yields the following result.

$$(\forall i \in I : \mathfrak{K}_i \models \mathcal{T}) \Longleftrightarrow \biguplus_{i \in I}{}^{\mathbb{K}} \mathfrak{K}_i \models \mathcal{T}$$

$\mathfrak{K}$ is called a *forward generated subinterpretation* of $\mathfrak{I}$ in $d$ if it is equal to the restriction of $\mathfrak{I}$ to all the elements that are reachable from the element $d \in \Delta^{\mathfrak{I}_i}$ (cf. Definition 4.2.23).

Definition 4.3.27 . For a family of $\tau$-interpretations $(\mathfrak{I}_i)_{i\in I}$ in $\mathbb{K}$ a family of $\tau$-interpretations is called family of *coherent generated subinterpretation* if it satisfies 1. 2. and 3. of the requirements from above. ◇

Definition 4.3.28 . Let $\tau$ be a signature and $\varphi \in \mathrm{FO}(\tau)$ a sentence. Then $\varphi$ is

1. invariant under $\overset{<\omega}{\Longleftrightarrow}^{Q}$ in $\mathbb{K}$ iff for any two interpretations $\mathfrak{I}, \mathfrak{H}$ in $\mathbb{K}$ we have
$$\mathfrak{I} \overset{<\omega}{\Longleftrightarrow}^{Q} \mathfrak{H} \Longrightarrow \mathfrak{I} \models \varphi \text{ iff } \mathfrak{H} \models \varphi$$
2. invariant under coherent disjoint unions in $\mathbb{K}$ iff for every family $(\mathfrak{I}_i)_{i\in I}$ of coherent $\tau$-interpretations in $\mathbb{K}$, i.e. $\mathfrak{I}_i$ in $\mathbb{K}$ for all $i \in I$, we have
$$(\forall i \in I : \mathfrak{I}_i \models \varphi) \iff \biguplus\nolimits_{i\in I}^{\mathbb{K}} \mathfrak{I}_i \models \varphi$$
3. preserved under coherent disjoint unions of generated subinterpretations in $\mathbb{K}$ iff for any family of coherent subinterpretations $(\mathfrak{K}_i)_{i\in I}$ of some family of $\tau$-interpretations $(\mathfrak{I}_i)_{i\in I}$ in $\mathbb{K}$ we have
$$(\forall i \in I : \mathfrak{I}_i \models \varphi) \Longrightarrow \biguplus\nolimits_{i\in I}^{\mathbb{K}} \mathfrak{K}_i \models \varphi$$

◇

Since $\mathcal{ALCQ}$-TBoxes and therefore also $\mathcal{ALCQO}$-TBoxes are invariant under the normal disjoint union, i.e. in particular they are preserved under generated subinterpretations, $\mathcal{ALCQO}$-TBoxes are preserved under coherent disjoint unions of generated subinterpretations.

Theorem 4.3.29 . *Let $\tau$ be a signature and $\varphi \in \mathrm{FO}(\tau)$ a sentence which is*

1. *invariant under $\overset{<\omega}{\Longleftrightarrow}^{Q}$ in $\mathbb{K}$*
2. *invariant under coherent disjoint unions in $\mathbb{K}$*
3. *preserved under coherent disjoint unions of forward generated subinterpretations in $\mathbb{K}$*

*then $\varphi$ is equivalent to an $\mathcal{ALCQO}$-TBox over $\tau$.*

Let $\varphi \in \mathrm{FO}(\tau)$ be as required by Theorem 4.3.29 and let

$$\operatorname{cons}\varphi := \{C \sqsubseteq D \mid C, D \in \mathcal{ALCQO}(\tau) \text{ and } \varphi \models_{\mathbb{K}} C \sqsubseteq D\}.$$

Similarly to $\mathcal{ALCO}$, we define $\mathsf{N_R}^*$ to be the set of all words over the alphabet $\mathsf{N_R}$, including the empty word $\varepsilon$ and we define for all $r^* \in \mathsf{N_R}^*$ and every $\mathcal{ALCQO}$-

concept $C$

$$\exists r^*.C := \begin{cases} C & \text{if } r = \varepsilon \\ \exists s_0^{\geq 1}.\exists s_1^* C & \text{if } r = s_0 \cdot s_1^* \text{ with } s_0 \in \mathsf{N_R} \text{ and } s_1^* \in \mathsf{N_R}^* \end{cases}$$

where we use $\forall r^* C$ as abbreviation for $\neg\exists r^*\neg C$. For an interpretation $\mathfrak{H}$ and $e \in \Delta^{\mathfrak{H}}$ we say that an individual $a^{\mathfrak{H}}$ is *in the scope of* $e$ if there is $r^* \in \mathsf{N_R}^*$ such that $e \in (\exists r^*.a)^{\mathfrak{H}}$. We define

$$\mathrm{Th}(\mathfrak{H}, e) := \{C \in \mathcal{ALCQO}(\tau) \mid (\mathfrak{H}, e) \models C\}$$

Lemma 4.3.30. *If there is a model $\mathfrak{H}$ in $\mathbb{K}$ of* cons $\varphi \cup \{\neg\varphi\}$ *then for every $e \in \Delta^{\mathfrak{H}}$ there is a model $\mathfrak{I}_e$ in $\mathbb{K}$ of $\varphi$ with an element $d \in \Delta^{\mathfrak{I}}$ such that $(\mathfrak{I}_e, d) \models \mathit{Th}(\mathfrak{H}, e)$, and if $a \in \mathsf{N_I}$ is in the scope of $d$ then $(\mathfrak{I}_e, a^{\mathfrak{I}_e}) \models \mathit{Th}(\mathfrak{H}, a^{\mathfrak{H}})$.*

Proof. Assume the claim to be false. Then there is $e \in \Delta^{\mathfrak{H}}$ such that $\{\varphi\} \cup \mathrm{Th}(\mathfrak{H}, e)$ together with the set

$$\Gamma := \{\forall r^*(a \to C) \mid a \in \mathsf{N_I}, r^* \in \mathsf{N_R}^*,\ C \in \mathrm{Th}(\mathfrak{H}, a^{\mathfrak{H}})\}$$

is unsatisfiable in $\mathbb{K}$.

Since FO is compact over $\mathbb{K}$ there is a finite set $T \subseteq \mathrm{Th}(\mathfrak{H}, e)$ and $\Gamma_0 \subseteq \Gamma$ such that $\{\varphi\} \cup T \cup \Gamma_0$ is unsatisfiable, i.e. for every model $\mathfrak{M}$ in $\mathbb{K}$ of $\varphi$ and every $d \in \Delta^{\mathfrak{M}}$ we have $(\mathfrak{M}, d) \models \neg(\bigsqcap T \sqcap \bigsqcap \Gamma_0)$. Hence for every $d \in \Delta^{\mathfrak{M}}$ we obtain

$$(\mathfrak{M}, d) \models \neg \bigsqcap T \sqcup \bigsqcup \{\exists r^*(a \sqcap \neg C) \mid \forall r^*(a \to C) \in \Gamma_0\}.$$

So $\mathfrak{M} \models \bigsqcap T \sqsubseteq \bigsqcup \{\exists r^*(a \sqcap \neg C) \mid \forall r^*(a \to C) \in \Gamma_0\}$ and since this is true for all models $\varphi$ in $\mathbb{K}$ it follows

$$\bigsqcap T \sqsubseteq \bigsqcup \{\exists r^*(a \sqcap \neg C) \mid \forall r^*(a \to C) \in \Gamma_0\} \in \mathrm{cons}\,\varphi.$$

Since $\mathfrak{H} \models \mathrm{cons}\,\varphi$ there is a path from $e$ to some individual $a^{\mathfrak{H}}$ such that this individual does not satisfy the type $\mathrm{Th}(\mathfrak{H}, a^{\mathfrak{H}})$. A contradiction! □

Proof of Theorem 4.3.29. Since $\varphi \models_{\mathbb{K}} \mathrm{cons}\,\varphi$ it remains to show that $\mathrm{cons}\,\varphi \models_{\mathbb{K}} \varphi$. For the sake of contradiction, assume $\mathrm{cons}\,\varphi \not\models_{\mathbb{K}} \varphi$. Then there must be model $\mathfrak{H}$ in $\mathbb{K}$ of $\mathrm{cons}\,\varphi \cup \{\neg\varphi\}$ for which we can w.l.o.g. assume that it is $\omega$-saturated.

Let $\Gamma$ be defined as in the proof of Lemma 4.3.30 and let

$$P := \{p \subseteq \mathcal{ALCQO}(\tau) \mid p \cup \Gamma \cup \{\varphi\} \text{ satisfiable in } \mathbb{K}\}$$

Let further for each $p \in P$ be $(\mathfrak{I}_p, d_p)$ a pointed, $\omega$-saturated interpretation in $\mathbb{K}$ such that $(\mathfrak{I}_p, d_p) \models_{\mathbb{K}} p \cup \Gamma \cup \{\varphi\}$ for each $p \in P$. We define $\mathfrak{K}_p$ be the forward generated subinterpretation of $\mathfrak{I}_p$ in $d_p$. Each $\mathfrak{K}_p$ is still $\mathcal{ALCQO}$-saturated.

For this family $(\mathfrak{I}_p)_{p\in P}$, the forward generated subinterpretations $(\mathfrak{K}_p)_{p\in P}$ satisfy the following properties:

1. for all $a \in \mathsf{N_I}$ there is $\mathfrak{K}_p$ with $a^{\mathfrak{K}_p} \neq \emptyset$
2. if $a^{\mathfrak{K}_p} \neq \emptyset$ then $|a^{\mathfrak{K}_p}| = 1$ for all $a \in \mathsf{N_I}$ and $p \in P$
3. if $a^{\mathfrak{K}_p} \neq \emptyset \neq a^{\mathfrak{K}_q}$ then $(\mathfrak{I}_p, a^{\mathfrak{K}_p}) \xleftrightarrow{<\omega} (\mathfrak{K}_q, a^{\mathfrak{I}_q})$ for any $p, q \in P$ and all $a \in \mathsf{N_I}$.

The first item is explained by the fact that $\{a\} \in P$ for all $a \in \mathsf{N_I}$ as Lemma 4.3.30 for $e \in a^{\mathfrak{H}}$ shows. The second item explains itself by the fact that each $\mathfrak{K}_p$ is generated subinterpretation of $\mathfrak{I}_p$ in $\mathbb{K}$ and the third item is true since $\mathfrak{K}_p \models \Gamma$ and every element in $\mathfrak{K}_p$ is in the scope of the root, which entails $(\mathfrak{K}_p, a^{\mathfrak{K}_p}) \models \mathrm{Th}(\mathfrak{H}, a^{\mathfrak{H}})$ for all $a \in \mathsf{N_I}$ with $a^{\mathfrak{K}_p} \neq \emptyset$. The Hennessy-Milner-Property for $\mathcal{ALCQO}$ yields the full $\mathcal{ALCQO}$-bisimilarity. Therefore $(\mathfrak{K}_p)_{p\in P}$ is a family coherent generated subinterpretation of the family $(\mathfrak{I}_i)_{p\in P}$.

Hence we set $\mathfrak{K} := \biguplus^{\mathbb{K}}_{p\in P} \mathfrak{K}_p$ to be the coherent disjoint unions generated subinterpretations and Lemma 4.3.25 confirms that $\mathfrak{K}$ in $\mathbb{K}$. Note that $\mathfrak{K}_p \models \mathcal{T}$ with $\mathcal{T} := \{a \sqsubseteq C \mid a \in \mathsf{N_I} \text{ and } C \in \mathrm{Th}(\mathfrak{H}, a^{\mathfrak{H}})\}$. Hence the coherent disjoint union $\mathfrak{K}$ of the $\mathfrak{K}_p$ will satisfy $\mathcal{T}$.

Furthermore we have by the $\mathcal{ALCQO}$ saturatedness of $\mathfrak{K}_p$ for every $a \in \mathsf{N_I}$ that if $a^{\mathfrak{K}_p} \neq \emptyset$ then $(\mathfrak{K}_p, a^{\mathfrak{K}_p}) \xleftrightarrow{<\omega} (\mathfrak{H}, a^{\mathfrak{H}})$. Proposition 4.3.20 and Observation 4.3.22 show that there is $\mathfrak{d} \in \mathfrak{K}$ for which $(\mathfrak{K}_p, a^{\mathfrak{K}_p}) \xleftrightarrow{<\omega} (\mathfrak{K}, \mathfrak{d})$. It immediately follows that $\{\mathfrak{d}\} = a^{\mathfrak{K}}$ which proves that $\mathfrak{H}$ and $\mathfrak{K}$ are coherent.

Since $\varphi$ is preserved by the coherent disjoint unions of generated subinterpretations, we have $\mathfrak{K} \models_{\mathbb{K}} \varphi$. Let $\mathfrak{I} := \mathfrak{K} \uplus^{\mathbb{K}} \mathfrak{H}$ be the coherent disjoint union. Since $\varphi$ is invariant under coherent disjoint unions and $\mathfrak{H} \not\models_{\mathbb{K}} \varphi$ we have $\mathfrak{I} \models_{\mathbb{K}} \neg\varphi$.

For each $e \in \Delta^{\mathfrak{K}}$ there is $d \in \Delta^{\mathfrak{I}}$ with $(\mathfrak{K}, e) \xleftrightarrow{<\omega} (\mathfrak{I}, d)$ as Proposition 4.3.20 and Observation 4.3.22 shows. Let $\mathfrak{d} \in \Delta^{\mathfrak{I}}$ be arbitrary. If $\mathfrak{d}$ contains an element $\bar{d}$ from $\mathfrak{K}^{\sigma_{\mathfrak{K}}}$ then Observation 4.3.22 yields $(\mathfrak{K}^{\sigma_{\mathfrak{K}}}, \bar{d}) \xleftrightarrow{<\omega} (\mathfrak{I}, \mathfrak{d})$. Proposition 4.3.20 provides a $d \in \Delta^{\mathfrak{K}}$ such that $(\mathfrak{K}, d) \xleftrightarrow{<\omega} (\mathfrak{K}^{\sigma_{\mathfrak{K}}}, \bar{d})$. Transitivity of $\xleftrightarrow{<\omega}$ then yields $(\mathfrak{K}, d) \xleftrightarrow{<\omega} (\mathfrak{I}, \mathfrak{d})$.

Assume $\bar{d}$ is in $\mathfrak{H}^{\sigma_{\mathfrak{H}}}$. Again Observation 4.3.22 and Proposition 4.3.20 provide a $d \in \Delta^{\mathfrak{H}}$ such that $(\mathfrak{H}, d) \overset{<\omega}{\Longleftrightarrow} (\mathfrak{I}, \mathfrak{d})$. Lemma 4.3.30 shows that there is $p \in P$ with $p = \mathrm{Th}(\mathfrak{H}, d)$. Since $\mathfrak{K}_p$ and $\mathfrak{H}$ are $\mathcal{ALCQO}$-saturated, we obtain $(\mathfrak{K}_p, d_p) \overset{<\omega}{\Longleftrightarrow} (\mathfrak{H}, d)$. Yet again, Proposition 4.3.20 and Observation 4.3.22 show that there is $e \in \mathfrak{K}$ such that $(\mathfrak{K}_p, d_p) \overset{<\omega}{\Longleftrightarrow} (\mathfrak{K}, e)$ and transitivity yields $(\mathfrak{K}, e) \overset{<\omega}{\Longleftrightarrow} (\mathfrak{I}, \mathfrak{d})$

Assembling the whole chain of arguments, we obtain $\mathfrak{K} \overset{<\omega}{\Longleftrightarrow}{}^{g} \mathfrak{I}$. A contradiction to the assumption that $\varphi$ would be invariant under global $\mathcal{ALCQ}$-bisimulation over $\mathbb{K}$. Hence we infer that cons $\varphi \vDash_{\mathbb{K}} \varphi$. □

### 4.3.4 Minimal $\mathcal{ALCQ}u_1$-Bisimilar and $\mathcal{ALCQ}u$-Bisimilar Companions

In what follows, we shall use the substitutions $\sigma$ introduced in Section 4.3.3 on page 131 ff. We shall use $\bar{\sigma}(d)$ very frequently as abbreviation for '$\sigma(d)$ if $d$ in $\mathbb{D}_\sigma$, i.e. the domain of $\sigma$, or $d$'. This is, indeed, the definition of $\bar{\sigma}(d)$. Hence we flip back and forth between stating properties for $\sigma$ and putting them with $\bar{\sigma}$ into formulae, where the property will trivially hold for elements $d \notin \mathbb{D}_\sigma$. The important bit is that $d$ will be always a letter in the context of $\bar{\sigma}$, i.e. a single element of a path in the tree-unravelling.

The Construction of Coherent Substructures (Section 4.3.3) was kept as general as possible and, as a side note, we shall now show how this can be used to obtain the smallest bisimilar companion of an interpretation.

In the context of $\mathcal{ALC}u$, the smallest $\overset{g}{\Longleftrightarrow}$-bisimilar companion for an interpretation $\mathfrak{I}$ can be obtained through factorisation by the largest auto-bisimulation

$$\Longleftrightarrow := \{(d_0, d_1) \in \Delta^{\mathfrak{I}} \times \Delta^{\mathfrak{I}} \mid (\mathfrak{I}, d_0) \Longleftrightarrow (\mathfrak{I}, d_1)\}.$$

But this does not carry over to $\mathcal{ALCQ}$:

Example 4.3.31 . Consider a 3-clique of nodes $\mathfrak{I}$ or more concrete $\Delta^{\mathfrak{I}} := \{0, 1, 2\}$ with $r^{\mathfrak{I}} := \{(d, d') \mid d < d' \text{ or } d' < d\}$. Clearly, for all $d, d' \in \Delta^{\mathfrak{I}}$ we have $(\mathfrak{I}, d) \overset{<\omega}{\Longleftrightarrow} (\mathfrak{I}, d')$ and the factorisation by the largest $\mathcal{ALCQ}u_1$-bisimulation relation for $\mathfrak{I}$, which is $\Delta^{\mathfrak{I}} \times \Delta^{\mathfrak{I}}$, would yield a singleton set with a reflexive $r$-edge. Yet, this element has only one successor whilst every element in $\mathfrak{I}$ has two. Hence, this factor-interpretation is not $\mathcal{ALCQ}u_1$-bisimilar to $\mathfrak{I}$.

Essentially, in the case of $\mathcal{ALC}u$, the smallest bisimilar companion reuses elements of the same bisimulation type. For $\mathcal{ALCQ}u_1$ this is not as straight forward, as one has to additionally preserve the number of successors up to $\omega$. In order to construct such a smallest $\mathcal{ALCQ}u_1$-bisimilar companion, we shall construct the substitution $\sigma : \mathbb{D}_\sigma \longrightarrow \Delta^{\mathfrak{I}^F} \setminus \mathbb{D}_\sigma$ such that $\sigma$ is injective on successor-sets up to $\omega$ by distinguishing several cases: Let $\mathfrak{I}$ be an interpretation and $\mathfrak{I}^F$ its forest unravelling. Let $\mathfrak{d} \in \mathfrak{I}^F / \overset{<\omega}{\Longleftrightarrow}$ and

$$E_{\mathfrak{d}} := \left\{ (\bar{d}_0, \bar{d}_1) \in \mathfrak{d}^2 \mid \bar{d}_0 = \bar{d}_1 \text{ or } \exists\, \bar{d} \in \Delta^{\mathfrak{I}^F}. \exists r \in \mathsf{N_R} : \bar{d}_0 = \bar{d} \cdot r \cdot \lambda \bar{d}_0 \text{ and } \bar{d}_1 = \bar{d} \cdot r \cdot \lambda \bar{d}_1 \right\}$$

i.e. $\bar{d}_0$ and $\bar{d}_1$ are both $r$-successors of a common predecessor. In particular root elements are in relation to themselves. $E_{\mathfrak{d}}$ is an equivalence relation which groups those elements in classes which are, for the same $r \in \mathsf{N_R}$, $r$-successors of the same predecessor. Root elements are the only elements in their class. For each $\mathfrak{d} \in \mathfrak{I}^F / \overset{<\omega}{\Longleftrightarrow}$, we define $\sigma$ piecewise via restrictions $\sigma{\restriction}\mathfrak{c}$ of $\sigma$ to classes $\mathfrak{c} \in \mathfrak{d}/E_{\mathfrak{d}}$:

1. In case $\mathfrak{d}/E_{\mathfrak{d}}$ contains a largest class $\mathfrak{c}$ which is finite, we set $\sigma{\restriction}\mathfrak{c}' : \mathfrak{c}' \longrightarrow \mathfrak{c}$ to be some injective function for all $\mathfrak{c}' \in \mathfrak{d}/E_{\mathfrak{d}}$ with $\mathfrak{c}' \neq \mathfrak{c}$.

2. In case $\mathfrak{d}/E_{\mathfrak{d}}$ contains an infinite class $\mathfrak{c}'$ we set, by using the axiom of choice, $\mathfrak{c}$ to be an countably infinite subset of $\mathfrak{c}'$ and define for each class $\mathfrak{c}_0 \in \mathfrak{d}/E_{\mathfrak{d}}$:

   If $\mathfrak{c}_0 = \mathfrak{c}'$ we set $\sigma{\restriction}(\mathfrak{c}_0 \setminus \mathfrak{c}) : (\mathfrak{c}_0 \setminus \mathfrak{c}) \longrightarrow \mathfrak{c}$ to be an arbitrary function.

   If $\mathfrak{c}_0 \neq \mathfrak{c}'$ is infinite, let, by the means of the axiom of choice, be $\mathfrak{c}_1$ be a countably infinite subset of $\mathfrak{c}_0$ and set $\sigma{\restriction}\mathfrak{c}_1 : \mathfrak{c}_1 \longrightarrow \mathfrak{c}$ to be an injective function and for $\sigma{\restriction}(\mathfrak{c}_0 \setminus \mathfrak{c}_1) : (\mathfrak{c}_0 \setminus \mathfrak{c}_1) \longrightarrow \mathfrak{c}$ an arbitrary function.

   If $\mathfrak{c}_0$ is finite, let $\sigma{\restriction}\mathfrak{c}_0 : \mathfrak{c}_0 \longrightarrow \mathfrak{c}$ be some injective function.

3. If $\mathfrak{d}/E_{\mathfrak{d}}$ contains infinitely many finite classes forming an infinite ascending chain of cardinalities, we fix for each cardinality $\kappa < \omega$ a class $\mathfrak{c}_\kappa$ with $|\mathfrak{c}_\kappa| = \kappa$, if it exists, and set $\sigma{\restriction}\mathfrak{c}' : \mathfrak{c}' \longrightarrow \mathfrak{c}_\kappa$ for all classes $\mathfrak{c}' \in \mathfrak{d}/E_{\mathfrak{d}}$ with $\mathfrak{c}' \neq \mathfrak{c}_\kappa$ and $|\mathfrak{c}'| = \kappa$.

With this definition of $\sigma$ we can obtain a globally $\mathcal{ALCQO}$-bisimilar companion of $\mathfrak{I}^F$ on which $\sigma$ is bijective on $r$-successors for all $r \in \mathsf{N_R}$:

Consider item 2. Let $M$ be the set of all elements in an infinite class $\mathfrak{c}' \in \mathfrak{d}/E_{\mathfrak{d}}$ which do not belong to the range of $\sigma$ and which are not in a countably infinite subset $\mathfrak{c}_1$ of a class $\mathfrak{c}_0 \in \mathfrak{d}/E_{\mathfrak{d}}$ which had been chosen to be injectively mapped by $\sigma$. We set $\mathfrak{H} := \mathfrak{I}^F{\restriction}(\Delta^{\mathfrak{I}} \setminus M)$.

Proposition 4.3.32. *$\mathfrak{I}^F \overset{<\omega}{\Longleftrightarrow}{}^{\varrho}\, \mathfrak{H}$ and $\sigma$ is injective on successor sets in $\mathfrak{H}$.*

Lemma 4.3.33 . *For all* $\bar{d} \in \Delta^{\mathfrak{H}}$ *we have:* $(\mathfrak{I}^F, \bar{d}) \overset{<\omega}{\Longleftrightarrow} (\mathfrak{H}, \bar{d})$.

Proof. **II** has a winning strategy if she maintains configurations $\beta : D \longrightarrow E$ such that $(\mathfrak{I}^F, \bar{d}_0) \overset{<\omega}{\Longleftrightarrow} (\mathfrak{I}^F, \beta(\bar{d}_0))$. The atomic equivalence of $\bar{d}_0$ and $\beta(\bar{d}_0)$ is immediately clear for all these configurations. Since the start-configuration satisfies the requirement, **II** does not lose the 0-th round.

Assume, the game has reached a configuration $\beta_0 : D_0 \longrightarrow E_0$ which meets the requirement and assume, **I** challenges **II** by playing for some $\bar{d}_0 \in D_0$ some finite $r$-successor set $D_1$.

We partition $r^{\mathfrak{I}} \circ \beta(\bar{d}_0)$ into classes of the same $\overset{<\omega}{\Longleftrightarrow}$-type. We also partition $D_1$ into $D_1 / \overset{<\omega}{\Longleftrightarrow}$. Let $\mathfrak{d} \in D_1 / \overset{<\omega}{\Longleftrightarrow}$. Since $(\mathfrak{I}^F, \bar{d}_0) \overset{<\omega}{\Longleftrightarrow} (\mathfrak{I}^F, \beta(\bar{d}_0))$ every element in $D_1$ has a class $\mathfrak{e} \in r^{\mathfrak{I}} \circ \beta(\bar{d}_0) / \overset{<\omega}{\Longleftrightarrow}$ which contains $\mathcal{ALCQ}$-bisimilar elements.

Either $\mathfrak{e} \cap M = \emptyset$ then all elements in $\mathfrak{e}$ are still present in $\mathfrak{H}$ and so, using her winning strategy for $G(\mathfrak{I}^F, \bar{d}_0; \mathfrak{I}^F, \beta(\bar{d}_0))$, **II** can set $\beta_1 \restriction \mathfrak{d} : \mathfrak{d} \longrightarrow \mathfrak{e}$, injectively mapping $\mathfrak{d}$ into $\mathfrak{e}$ such that $(\mathfrak{I}^F, \bar{d}_1) \overset{<\omega}{\Longleftrightarrow} (\mathfrak{I}^F, \beta_1(\bar{d}_1))$ for all $d_1 \in \mathfrak{d}$.

Or $\mathfrak{e} \cap M \neq \emptyset$. We know that there is, according to 2. a countably infinite $\mathfrak{e}_0 \subseteq \mathfrak{e}$ for which $\sigma$ is injective and hence its elements are not contained in $M$. **II** can set $\beta_1 \restriction \mathfrak{d} : \mathfrak{d} \longmapsto \mathfrak{e}_0$ injectively mapping $\mathfrak{d}$ into $\mathfrak{e}_0$ such that $(\mathfrak{I}^F, \bar{d}_1) \overset{<\omega}{\Longleftrightarrow} (\mathfrak{I}^F, \beta_1(\bar{d}_1))$ for all $d_1 \in \mathfrak{d}$.

In any case, **II** can find a new configuration $\beta_1 : D_1 \longrightarrow \beta(D_1)$ that yields the requirement. If conversely **I** challenges **II** by playing some finite $r$-successor set $E_1$, **II** can find, by virtue of $(\mathfrak{I}^F, \bar{d}_0) \overset{<\omega}{\Longleftrightarrow} (\mathfrak{I}^F, \beta(\bar{d}_0))$, an $r$-successor set $D_1$ of $\bar{d}_0$ such that $\beta_1 : D_1 \longrightarrow E_1$ is an injection and $(\mathfrak{I}^F, \bar{d}_1) \overset{<\omega}{\Longleftrightarrow} (\mathfrak{I}^F, \beta(\bar{d}_1))$ for all $d_1 \in \mathfrak{d}$. □

Proof of Proposition 4.3.32. Since we have Lemma 4.3.33, we only have to show the for elements in $\mathfrak{I}$ and in particular for all $\bar{d} \in M$ that there is $\bar{e} \in \Delta^{\mathfrak{H}}$ such that $(\mathfrak{I}^F, \bar{d}) \overset{<\omega}{\Longleftrightarrow} (\mathfrak{H}, \bar{e})$.

Assume $\bar{d}$ is in $M$. Then $\bar{d}$ is an $r$-successor of some element and hence located on some tree in $\mathfrak{I}$ with root $\bar{d}_0$, say. This root is, as mentioned above, the only element in its class and hence cannot be an element of $M$. So $\bar{d}_0$ is present in $\mathfrak{H}$. Lemma 4.3.33 yields a winning strategy for **II** in the game $G(\mathfrak{I}^F, \bar{d}_0; \mathfrak{H}, \bar{d}_0)$ and **II** can pretend to be challenged in $\mathfrak{I}^F$ along the path from $\bar{d}_0$ to $\bar{d}$ playing according to her winning strategy, so that she can finally come up with some $e \in \Delta^{\mathfrak{H}}$ such that $(\mathfrak{I}^F, \bar{d}) \overset{<\omega}{\Longleftrightarrow} (\mathfrak{H}, \bar{e})$. □

Proposition 4.3.32 shows that we can w.l.o.g. switch to some globally $\mathcal{ALCQ}$-bisimilar companion for which $\bar{\sigma}$ is injective on $r$-successor sets.

Observation 4.3.34 . $\mathfrak{I} \underset{\longleftrightarrow}{\ll\omega}_{\mathcal{Q}} \mathfrak{I}^F \underset{\longleftrightarrow}{\ll\omega}_{\mathcal{Q}} (\mathfrak{I}^F)^{\bar{\sigma}}$.

This observation is a direct consequence of Proposition 4.3.20. Note that $\cdot^{\sigma}$ will create a forest unravelling of the forest unravelling. We set

$$E := \{(\bar{d}, \bar{d}') \in \Delta^{(\mathfrak{I}^F)^{\bar{\sigma}}} \times \Delta^{(\mathfrak{I}^F)^{\bar{\sigma}}} \mid \lambda\bar{d} = \lambda\bar{d}'\}$$

Proposition 4.3.35 . *$(\mathfrak{I}^F)^{\bar{\sigma}}/E$ is a minimal $\mathcal{ALCQ}u_1$ bisimilar companion of $\mathfrak{I}$.*

Proof. Assume there is a $\mathfrak{H}$ with $\mathfrak{H} \underset{\longleftrightarrow}{\ll\omega}_{\mathcal{Q}} \mathfrak{I}$ such that $|\Delta^{\mathfrak{H}}| < |\Delta^{(\mathfrak{I}^F)^{\bar{\sigma}}/E}|$. Then for some bisimilarity type $\mathfrak{d} \in \Delta^{(\mathfrak{I}^F)^{\bar{\sigma}}/E}/ \underset{\longleftrightarrow}{\ll\omega}$ there is $\mathfrak{e} \in \Delta^{\mathfrak{H}}/ \underset{\longleftrightarrow}{\ll\omega}$ such that $|\mathfrak{e}| < |\mathfrak{d}|$. Since every class $\mathfrak{d}$ is at most countable, $\mathfrak{e}$ is finite. We shall go through the definition of $\sigma$ to reveal a the contradiction. Let $\mathfrak{d}_0 \in \Delta^{\mathfrak{I}^F}/ \underset{\longleftrightarrow}{\ll\omega}$ be of the same $\underset{\longleftrightarrow}{\ll\omega}$-type as $\mathfrak{d}$ i.e. for all $d_0 \in \mathfrak{d}_0$ and $d \in \mathfrak{d}$ we have $(\mathfrak{I}^F, d_0) \underset{\longleftrightarrow}{\ll\omega} ((\mathfrak{I}^F)^{\bar{\sigma}}, d)$.

If case 2. applies to $\mathfrak{d}_0$ then there was some $r \in \mathsf{N_R}$ and an infinite set of $r$-successors of some element $d'$ in $\mathfrak{I}$ containing elements of the same $\underset{\longleftrightarrow}{\ll\omega}$-type as $\mathfrak{e}$. Since each element in $\mathfrak{H}$ has at most $|\mathfrak{e}|$ many successors, for any $e \in \Delta$, I could challenge **II** in the $\mathcal{ALCQ}u_1$-bisimulation game $G(\mathfrak{I}, d'; \mathfrak{H}, e)$, with a subset $D$ of size $|\mathfrak{e}|+1$ of this $r$-successor set of $d'$. **II** must respond with some $r$-successor set $E$ of size $|\mathfrak{e}|+1$, yielding a configuration $\beta : D \longrightarrow E$ such that $(\mathfrak{I}, d_0) \underset{\longleftrightarrow}{\ll\omega} (\mathfrak{H}, \beta(d_0))$ is false for at least one element $d_0 \in D$. Clearly **II** loses the game $G(\mathfrak{I}, d'; \mathfrak{H}, e)$ for all $e \in \Delta^{\mathfrak{H}}$ and hence $\mathfrak{H} \underset{\longleftrightarrow}{\ll\omega}_{\mathcal{Q}} \mathfrak{I}$ is false. A contradiction to our assumption.

In case 1. applies to $\mathfrak{d}_0$ then there is $r \in \mathsf{N_R}$ and some $r$-successor set of some element in $\mathfrak{I}^F$ and therefore some $r$-successor set of some element $d'$ in $\mathfrak{I}$ which contains $|\mathfrak{d}|$ many elements of the same $\underset{\longleftrightarrow}{\ll\omega}$-type. Since $|\mathfrak{e}| < |\mathfrak{d}|$ no element in $\mathfrak{H}$ has as many $r$-successors as $d'$ and hence $\mathfrak{H} \underset{\longleftrightarrow}{\ll\omega}_{\mathcal{Q}} \mathfrak{I}$ is false. A contradiction to our assumption.

In case 3. applies to $\mathfrak{d}_0$, there is for every finite cardinality $\kappa$ some $r$ and a finite set of $r$-successors $D$ of some element $d' \in \Delta^{\mathfrak{I}}$ such that $|D| > \kappa$ of the same $\underset{\longleftrightarrow}{\ll\omega}$-type as elements in $\mathfrak{e}$. Let in particular $d' \in \Delta^{\mathfrak{I}}$ be such that there is some $r \in \mathsf{N_R}$ and some $r$-successor set $D$ of $d'$ with $|D| > |\mathfrak{e}|$ of the same $\underset{\longleftrightarrow}{\ll\omega}$-type as elements in $\mathfrak{e}$. Clearly, no element in $\mathfrak{H}$ can be $\mathcal{ALCQ}$-bisimilar to $d'$ and hence $\mathfrak{H} \underset{\longleftrightarrow}{\ll\omega}_{\mathcal{Q}} \mathfrak{I}$ is false. A contradiction to our assumption. □

It should be remarked that if root elements of $\mathfrak{I}^F$ are treated as group by $E_{\mathfrak{d}}$, i.e.

for each $\mathfrak{d} \in \mathfrak{I}^F / \overset{<\omega}{\Longleftrightarrow}$

$$E_{\mathfrak{d}} := \left\{ \begin{array}{l} (\bar{d}_0, \bar{d}_1) \in \mathfrak{d} \mid \bar{d}_0 \text{ and } \bar{d}_1 \text{ are root elements or} \\ \exists\, \bar{d} \in \Delta^{\mathfrak{I}^F}. \exists r \in \mathsf{N_R} : \bar{d} \cdot r \cdot \lambda \bar{d}_0 \text{ and } \bar{d} \cdot r \cdot \lambda \bar{d}_1 \end{array} \right\}$$

we can not only preserve the number (up to $\omega$) of successors, but also the global number of elements of a certain $\overset{<\omega}{\Longleftrightarrow}$-type.

Example 4.3.36 . Take the interpretation $\mathfrak{I}$ with $\Delta^{\mathfrak{I}} := \{0, 1\}$ where all predicates are empty. The construction before would yield $(\mathfrak{I}^F)^\sigma / E$ with exactly on element, whilst the new definition of $E_{\mathfrak{d}}$ yields interpretation isomorphic to $\mathfrak{I}$, i.e. with two elements.

We hence make, without proof, for the new definition of $E_{\mathfrak{d}}$ the following

Observation 4.3.37 . $\mathfrak{I} \overset{<\omega}{\Longleftrightarrow}{}^{\forall} (\mathfrak{I}^F)^{\bar{\sigma}}$.

Note that in case one has a finite interpretation $\mathfrak{I}$ and needs to stay finite, but wants to find the smallest $\mathcal{ALCQ}u_1$- or $\mathcal{ALCQ}u$-bisimilar companion, it is not necessary to perform a complete tree-unravelling. One can stop as soon as one detects the same element twice in a path-element: We define this *partial tree-unravelling* $\mathfrak{I}^P$ of $\mathfrak{I}$ by setting $\mathfrak{I}^P := \mathfrak{I}^F \restriction \Delta^{\mathfrak{I}^P}$ where

$$\Delta^{\mathfrak{I}^P} := \{\bar{d} \in \Delta^{\mathfrak{I}^F} \mid \text{ for all } d \in \Delta^{\mathfrak{I}} \text{ which occur twice in } \bar{d} \text{ we have } \lambda \bar{d} = d\}.$$

The length of the elements in $\Delta^{\mathfrak{I}^P}$ are bounded by $|\Delta^{\mathfrak{I}}| + 1$ as exactly one element can occur twice and hence $\mathfrak{I}^P$ is finite iff $\mathfrak{I}$ is finite.

These partial tree-unravellings are known from blocking techniques [50] applied in tableau based reasoning [50, 122] which were for DLs first applied in [7] and have been since then refined and improved [11, 27, 98]. In contrast to blocking techniques we do not build interpretations from scratch but we block the tree-unravelling of a given interpretation. This makes our blocking condition, namely occurring twice in a path, rather simple in comparison to the strategies aimed at detecting as soon as possible when the generation of new elements can be stopped [59].

The substitution $\sigma$ is constructed piecewise as follows: For each $\mathfrak{d}' \in \mathfrak{I}^F / \overset{<\omega}{\Longleftrightarrow}$ let $\mathfrak{d} := \mathfrak{d}' \cap \Delta^{\mathfrak{I}^P}$, i.e. $\mathfrak{d}$ contains all elements of $\Delta^{\mathfrak{I}^P}$ that would have been $\mathcal{ALCQ}$-bisimilar in $\mathfrak{I}^F$. Again, we set

$$E_{\mathfrak{d}} := \left\{ (\bar{d}_0, \bar{d}_1) \in \mathfrak{d}^2 \mid \bar{d}_0 = \bar{d}_1 \text{ or } \exists\, \bar{d} \in \Delta^{\mathfrak{I}^F}.\exists r \in \mathsf{N}_{\mathsf{R}} : \bar{d}_0 = \bar{d} \cdot r \cdot \lambda\bar{d}_0 \text{ and } \bar{d}_1 = \bar{d} \cdot r \cdot \lambda\bar{d}_1 \right\}$$

and choose a maximal class in $\mathfrak{c}_0 \in \mathfrak{d}/E_{\mathfrak{d}}$. We define $\mathfrak{c}$ to be the class $\mathfrak{c}_0$ where all elements $\bar{d} = d_0 \cdots d_i \cdots d_n$ in which some letter occurs twice, concretely where $d_i = d_n$ with $i \neq n$, is replaced by $d_0 \cdots d_i$. With this replacement we do not 'lose' any elements: All elements in $\mathfrak{c}_0$ are distinguished by their last letter. The new element's last letter is equal to the old element's last letter and hence is still distinguished from all other elements in $\mathfrak{c}_0$. This argument can iteratively applied for each replacement, whence $|\mathfrak{c}_0| = |\mathfrak{c}|$ is inferred.

For each class $\mathfrak{c}' \in \mathfrak{d}/E_{\mathfrak{d}}$ we set $\sigma\restriction\mathfrak{c}' : \mathfrak{c}' \setminus (\mathfrak{c}' \cap \mathfrak{c}) \longrightarrow \mathfrak{c} \setminus (\mathfrak{c}' \cap \mathfrak{c})$ to be some injective function. $\sigma$ is injective on successor-sets and due to the replacements maps elements only on elements in which no letter occurs twice.

We define the following equivalence relation, where, due to the fact that $\bar{\sigma}$ operates on the forest-unravelling of $\mathfrak{I}^P$ we have to apply $\lambda$ twice:

$$\Theta := \{(\bar{d}_0, \bar{d}_1) \in (\mathfrak{I}^P)^{\bar{\sigma}} \mid \lambda \circ \lambda\bar{d}_0 = \lambda \circ \lambda\bar{d}_1\}$$

For all $\mathfrak{d}$ in $(\mathfrak{I}^P)^{\bar{\sigma}}/E$ we define $d_{\mathfrak{d}} := \lambda \circ \lambda\bar{d}$ for some arbitrary $\bar{d} \in \mathfrak{d}$. $d_{\mathfrak{d}}$ is uniquely determined by the definition of $\Theta$.

Proposition 4.3.38 . *For all $d_0 \in \Delta^{\mathfrak{I}}$ we have $(\mathfrak{I}, d_0) \overset{<\omega}{\Longleftrightarrow} ((\mathfrak{I}^P)^{\bar{\sigma}}/\Theta, \mathfrak{d})$ if $(\mathfrak{I}, d_0) \overset{<\omega}{\Longleftrightarrow} (\mathfrak{I}, d_{\mathfrak{d}})$.*

Proof. We show that **II** has a winning strategy if she maintains configurations $\beta : D \longrightarrow E$ such that $(\mathfrak{I}, d_0) \overset{<\omega}{\Longleftrightarrow} (\mathfrak{I}, d_{\beta(d_0)})$ for all $d_0 \in D$.

We shall first show that for configurations which satisfy this requirement, $d_0$ and $\beta(d_0)$ are atomically equivalent: Let $\bar{d} \in \beta(d_0)$ be arbitrary. Since $\lambda \circ \lambda\bar{d}$ is atomically equivalent to $d_0$, we shall merely show that $[\bar{d}]$ is atomically equivalent to $\lambda\circ\lambda\bar{d}$: $\lambda\circ\lambda\bar{d} \in A^{\mathfrak{I}}$ iff $\lambda\bar{d} \in A^{\mathfrak{I}^P}$ iff $\bar{d} \in A^{(\mathfrak{I}^P)^{\bar{\sigma}}}$ iff $[\bar{d}] \in A^{(\mathfrak{I}^P)^{\bar{\sigma}}/E}$. The only-if direction of the latter follows from the fact that all elements $\bar{d}' \in [\bar{d}]$ share the same element $\lambda \circ \lambda\bar{d}'$.

Assume the configuration $\beta : D \longrightarrow E$ with $(\mathfrak{I}, d_0) \overset{<\omega}{\Longleftrightarrow} (\mathfrak{I}, d_{\beta(d_0)})$ for all $d_0 \in D$ has been reached. Let **I** play for some $r \in \mathsf{N}_{\mathsf{R}}$ an $r$-successor set $D'$ of some $d \in D$.

Let $\bar{d}_0 \cdots \bar{d}_n \in \beta(d)$ be fix but arbitrary; note that $\bar{d}_i \in \mathfrak{I}^P$ for all $i \in \{0, \ldots, n\}$ as $\bar{\sigma}$ operates on the tree-unravelling of $\mathfrak{I}^P$. All letters $\bar{d}_i$ in $\bar{d}_0 \cdots \bar{d}_n$ are the outcome of the substitution by $\bar{\sigma}$, i.e. they are in the range of $\bar{\sigma}$. As the range of $\sigma$ is excluded from its domain $\mathbb{D}_\sigma$, especially $\bar{d}_n \notin \mathbb{D}_\sigma$ and so $\bar{d}_n$ is in $(\mathfrak{I}^P)^\sigma$. By the definition of $E$, we have $\bar{d}_n \in \beta(d)$, so $[\bar{d}_n] = \beta(d)$.

The definition of $\sigma$ implies that no letter appears twice in $\bar{d}_n$. Therefore $\bar{d}_n \in \mathfrak{I}^P$ and has all its (immediate) successors from $\mathfrak{I}^F$ in $\mathfrak{I}^P$. Since $(\mathfrak{I}, d) \overset{<\omega}{\Longleftrightarrow} (\mathfrak{I}, \lambda \bar{d}_n)$, meaning in particular $(\mathfrak{I}, d) \overset{<\omega}{\Longleftrightarrow} (\mathfrak{I}^F, \bar{d}_n)$, there is an $r$-successor set $E_0$ of $\bar{d}_n$ in $\mathfrak{I}^P$ and $\beta_0 : D' \longrightarrow E_0$ such that $\beta_0$ is injective and $(\mathfrak{I}, d_0') \overset{<\omega}{\Longleftrightarrow} (\mathfrak{I}^F, \beta_0(d_0'))$ for all $d_0' \in D'$. Since $\mathfrak{I}^P$ contains all successors of $\bar{d}_n$, for all $d_0' \in D'$ we have $\bar{d}_n \cdot r \cdot \beta_0(d_0') \in (\Delta^{\mathfrak{I}^P})^F$ on which $\bar{\sigma}$ is defined! In particular

$$\bar{\sigma}(\bar{d}_n \cdot r \cdot \beta_0(d_0')) = \bar{d}_n \cdot r \cdot \bar{\sigma}(\beta_0(d_0'))$$

as $\bar{d}_n \notin \mathbb{D}_\sigma$. Note that for each element $\bar{e} \in \mathfrak{I}^P$ we have $\bar{\sigma}(e) = \sigma(e)$ if $e$ in $\mathbb{D}_\sigma \subseteq \mathfrak{I}^P$ or $\bar{\sigma}(e) = e$ otherwise. As $\sigma(e)$ and $e$ is in $\mathfrak{I}^P$ it follows that $\bar{\sigma}(e) \in \mathfrak{I}^F$. By the definition of $\sigma$, for all $e \in \mathfrak{I}^P$ we have $(\mathfrak{I}^F, e) \overset{<\omega}{\Longleftrightarrow} (\mathfrak{I}^F, \bar{\sigma}(e))$ and in particular

$$(\mathfrak{I}, d_0') \overset{<\omega}{\Longleftrightarrow} (\mathfrak{I}^F, \beta(d_0')) \overset{<\omega}{\Longleftrightarrow} (\mathfrak{I}^F, \bar{\sigma} \circ \beta_0(d_0')) \overset{<\omega}{\Longleftrightarrow} (\mathfrak{I}, \lambda \circ \bar{\sigma} \circ \beta_0(d_0')),$$

the latter by definition of $\mathfrak{I}^F$, for all $d_0' \in D'$.

As $\sigma$ is injective on successors, $\bar{\sigma}$ maps $E_0$ injectively to an $r$-successor set of $\bar{d}_n$ in $(\mathfrak{I}^P)^{\bar{\sigma}}$. We shall show the following intermediate claim: $\lambda \circ \lambda(\bar{d}_n \cdot r \cdot \bar{\sigma} \circ \beta_0(d_0'))$ must be different from $\lambda \circ \lambda(\bar{d}_n \cdot r \cdot \bar{\sigma} \circ \beta_0(d_1'))$ for all $d_0', d_1' \in D'$ with $d_0' \neq d_1'$:

Assume $d_0' \neq d_1'$. Then $e_0' := \beta_0(d_1') \neq \beta_0(d_1') =: e_1'$ because $\beta_0$ is injective. Either $e_0'$ and $e_1'$ have a different $\overset{<\omega}{\Longleftrightarrow}$-type, then $\bar{\sigma}$ maps them to elements in $\mathfrak{I}^P$ with different $\overset{<\omega}{\Longleftrightarrow}$-type in $\mathfrak{I}^F$ and so $\lambda \circ \bar{\sigma}(e_0') \neq \lambda \circ \bar{\sigma}(e_1')$. Or, if they have the same $\overset{<\omega}{\Longleftrightarrow}$-type, $\bar{\sigma}$ maps them injectively to a maximal $r'$-successor set of some element for some $r'$ or a set which contains a root, containing elements of the same $\overset{<\omega}{\Longleftrightarrow}$-type. In our case, it must be a set of $r'$-successors, as sets containing roots are singletons sets. Each of these successors, since they are successors of a common predecessor in $\mathfrak{I}^P$, must have a different last element. In case these elements have been replaced, as happened from $\mathfrak{c}_0$ to $\mathfrak{c}$ above, their last elements were preserved. Hence $\lambda \circ \bar{\sigma}(e_0') \neq \lambda \circ \sigma(e_1')$, which shows the intermediate claim.

The intermediate claim shows $[\bar{d}_n \cdot r \cdot \bar{\sigma} \circ \beta_0(d_0')] \neq [\bar{d}_n \cdot r \cdot \bar{\sigma} \circ \beta_0(d_1')]$ and each $[\bar{d}_n \cdot r \cdot \bar{\sigma} \circ \beta_0(d_0')]$ is an $r$-successor of $[\bar{d}_n]$. **II** can thus find an injection

$$\beta' : D' \longrightarrow E' : d_0' \longmapsto [(\bar{d}_n \cdot r \cdot \bar{\sigma} \circ \beta_0(d_0'))]$$

where $E' := \beta'(D')$ is an $r$-successor set of $[\bar{d}_n]$ such that for all $d'_0 \in D'$ we have $(\mathfrak{I}, d'_0) \xleftrightarrow{\leq\omega} (\mathfrak{I}, \lambda \circ \lambda(\bar{d}_n \cdot r \cdot \bar{\sigma} \circ \beta_0(d'_0))$.

Let conversely **I** challenge **II** by playing an $r$-successor set $E'$ for some $\beta(d_0) \in E$. Let $\mathfrak{e} \in E'$ be fix but arbitrary. By the definition of $(\mathfrak{I}^P)^{\bar{\sigma}}$ this is only the case, if there is an element $\bar{d}_0 \cdots \bar{d}_n \in \beta(d_0)$ and an element $\bar{d}_0 \cdots \bar{d}_n \cdot r \cdot \bar{d}_{n+1} \in \mathfrak{e}$. As $\bar{d}_n \notin \mathbb{D}_\sigma$, $\bar{d}_n$ and $\bar{d}_n \cdot r \cdot \bar{d}_{n+1}$ are in $(\mathfrak{I}^P)^\sigma$ and we can, since $\sigma$ is injective on $r$-successors, uniquely determine an $r$-successor $\bar{d}'_{n+1}$ in $\mathfrak{I}^P$ such that $\bar{d}_n \cdot r \cdot \bar{d}_{n+1} = \bar{d}_n \cdot r \cdot \bar{\sigma}(\bar{d}'_{n+1})$ and $\bar{d}_n \cdot r \cdot \bar{d}'_{n+1} \in (\mathfrak{I}^P)^F$. From the latter follows that $\lambda\bar{d}_n = d_{\beta(d_0)}$ had an $r$-successor $\lambda\bar{d}'_{n+1} \in \mathfrak{I}$ and by the definition of $\sigma$ it follows that $(\mathfrak{I}^F, \bar{d}'_{n+1}) \xleftrightarrow{\leq\omega} (\mathfrak{I}^F, \bar{\sigma}(\bar{d}'_{n+1}))$ which entails that $(\mathfrak{I}, \lambda\bar{d}'_{n+1}) \xleftrightarrow{\leq\omega} (\mathfrak{I}, \lambda \circ \bar{\sigma}(\bar{d}'_{n+1}))$

For each class $\mathfrak{e} \in E'$ we set $d_\mathfrak{e}$ to be the appropriate $\lambda\bar{d}'_{n+1} \in \mathfrak{I}$ as just determined in the previous paragraph. Each $d_\mathfrak{e}$ is an $r$-successor of $d_{\beta(d_0)}$. Since $d_\mathfrak{e} \neq d_{\mathfrak{e}'}$ whenever $\mathfrak{e} \neq \mathfrak{e}'$ for all $\mathfrak{e}, \mathfrak{e}' \in E'$, there is a bijection $\beta_0 : \mathfrak{e} \longmapsto d_\mathfrak{e}$ between $E'$ and $D_0 := \{d_\mathfrak{e} \mid \mathfrak{e} \in E'\}$.

Since $(\mathfrak{I}, d_0) \xleftrightarrow{\leq\omega} (\mathfrak{I}, d_{\beta(d_0)})$ **II** can find an $r$-successor set $D'$ such that $\beta' : D' \longrightarrow D_0$ is injective and $(\mathfrak{I}, d'_0) \xleftrightarrow{\leq\omega} (\mathfrak{I}, \beta'(d'_0))$ for all $d'_0 \in D'$. Hence $\beta_0 \circ \beta' : D' \longrightarrow E'$ is injective and satisfies the requirement! This shows that **II** has a winning strategy. □

Proposition 4.3.39 . $(\mathfrak{I}^P)^{\bar{\sigma}}/\Theta$ *is the minimal* $\mathcal{ALCQ}u$*-bisimilar companion of* $\mathfrak{I}$.

Proof. We first show that $(\mathfrak{I}^P)^{\bar{\sigma}}/\Theta \xleftrightarrow{\leq\omega}{}^{\mathcal{Q}} \mathfrak{I}$. Every element $d \in \Delta^\mathfrak{I}$ is present in $\mathfrak{I}^P$ and mapped by $\bar{\sigma}$ to some element in $\mathfrak{I}^P$, such that $(\mathfrak{I}, d) \xleftrightarrow{\leq\omega} (\mathfrak{I}^P, \bar{\sigma}(d))$. In particular $(\mathfrak{I}, d) \xleftrightarrow{\leq\omega} (\mathfrak{I}, \lambda \circ \bar{\sigma}(d))$. Since $\bar{\sigma}(d) \notin \mathbb{D}_\sigma$, $\bar{\sigma}(d) \in (\mathfrak{I}^P)^{\bar{\sigma}}$ and so with Proposition 4.3.38 we obtain that $(\mathfrak{I}, d) \xleftrightarrow{\leq\omega} ((\mathfrak{I}^P)^{\bar{\sigma}}/E, [\bar{\sigma}(d)])$.

For any $\mathfrak{d}$ we have with Proposition 4.3.38 $(\mathfrak{I}, d_\mathfrak{d}) \xleftrightarrow{\leq\omega} ((\mathfrak{I}^P)^{\bar{\sigma}}/\Theta, \mathfrak{d})$ where $d_\mathfrak{d}$ is defined as before Proposition 4.3.38.

Finally $(\mathfrak{I}^P)^{\bar{\sigma}}/\Theta$ is minimal, as for every $\xleftrightarrow{\leq\omega}$-type we used one $r$-successor set of one common predecessor $\bar{d} \in \mathfrak{I}^F$ for some $r \in \mathsf{N_R}$. Any interpretation $\mathfrak{H}$ with less elements has at least one $\xleftrightarrow{\leq\omega}$-type with lesser elements. Hence there is no element $e \in \Delta^\mathfrak{H}$ such that $(\mathfrak{I}, \lambda\bar{d}) \xleftrightarrow{\leq\omega} (\mathfrak{H}, e)$. □

It is not difficult to see that with the same redefinition of $E_\mathfrak{d}$ as before, we can obtain the $\mathcal{ALCQ}u$-bisimilar companion of $\mathfrak{I}$. Similarly, using classes $\mathfrak{d} \in \mathfrak{I}^F / \xleftrightarrow{\leq\kappa}_n$ will yield appropriate companions $\mathfrak{H}$ such that $\mathfrak{I} \xleftrightarrow{\leq\kappa\,\forall}_n \mathfrak{H}$.

## 4.4 $\mathcal{ALCQIO}$

This section will introduce $\mathcal{ALCQIO}$ and give again characterisation theorems for the standard logic, the extension with a universal role and for TBoxes. As an extension of $\mathcal{ALCQI}$ with nominals, the reasoning complexity of $\mathcal{ALCQIO}$ has been investigated in [128]. It is noted in [128] that [57] investigates $\mathcal{ALCQO}$ and $\mathcal{ALCIO}$ yet not the combination $\mathcal{ALCQIO}$. The DL $\mathcal{ALCQIO}$ is of particular interest since $\mathcal{SHOIQ}$ [81], the description logic underpinning OWL DL, extends $\mathcal{ALCQIO}$ by allowing also RBoxes which contain role hierarchies and explicitly transitive roles. Similarly $\mathcal{ALCQIO}u_1$ is akin to $\mathcal{SROIQ}$ [77] which extends $\mathcal{ALCQIO}u$ and is the underpinning description logics of OWL 2 [38].

### 4.4.1 The Characterisation of $\mathcal{ALCQIO}$-Concepts

Syntax and Semantics

As for $\mathcal{ALCO}$ and $\mathcal{ALCQO}$ the signature $\tau$ comprises role names $\mathsf{N_R}$, concept names $\mathsf{N_C}$ and individual names $\mathsf{N_I}$. The syntax for an $\mathcal{ALCQIO}$-concept $C$ over $\tau$ is given by

$$C ::= \top \mid a \mid A \mid D \sqcap E \mid \neg D \mid \exists^{\geq\kappa} r.D \mid \exists^{\geq\kappa} r^{-}.D$$

where $a \in \mathsf{N_I}$, $A \in \mathsf{N_C}$, $r \in \mathsf{N_R}$, $\kappa < \omega$ and $D, E$ are $\mathcal{ALCQIO}$-concepts over $\tau$. The set of all $\mathcal{ALCQIO}$-concepts over $\tau$ is denoted by $\mathcal{ALCQIO}(\tau)$. As usual $\bot$, $\rightarrow$, $\sqcup$, $\forall^{\geq\kappa}$ etc. are considered to be abbreviations.

The semantics for $\mathcal{ALCQIO}$ is explained by recursively continuing the interpretation function $\cdot^{\mathfrak{I}}$ on concepts $C \in \mathcal{ALCO}(\tau)$ as follows:

$$C^{\mathfrak{I}} := \begin{cases} \{d \in \Delta^{\mathfrak{I}} \mid |r^{\mathfrak{I}}(d) \cap D^{\mathfrak{I}}| \geq n\} & \text{if } C = \exists^{\geq n} r.D \\ \{d \in \Delta^{\mathfrak{I}} \mid |r^{-\mathfrak{I}}(d) \cap D^{\mathfrak{I}}| \geq n\} & \text{if } C = \exists^{\geq n} r^{-}.D \end{cases}$$

where $|M|$ yields the cardinality of a set $M$, $r^{\mathfrak{I}}(d) := \{e \in \Delta^{\mathfrak{I}} \mid (d, e) \in r^{\mathfrak{I}}\}$ and $r^{-\mathfrak{I}}(d) := \{e \in \Delta^{\mathfrak{I}} \mid (e, d) \in r^{\mathfrak{I}}\}$, i.e. the set of predecessors of $d$ in $\mathfrak{I}$. Hence $|r^{-\mathfrak{I}}(d) \cap D^{\mathfrak{I}}|$ is the number of $r$-predecessors of $d$ in $\mathfrak{I}$ that satisfy $D$.

Rank-function and grade-function for $\mathcal{ALCQIO}$ are straight forward extensions of the rank- and grade-function for $\mathcal{ALCQO}$ (page 78) to $\mathcal{ALCQIO}$. The translation into $\mathrm{FO}^{\equiv}$ is the translation for $\mathcal{ALCQ}$ into $\mathrm{FO}^{\equiv}$, which is extended for

$C = \exists^{\geq\kappa} r^{-}.D$ as follows

$$[C; x_i] := \exists (x_\ell)_{\ell \in \kappa \setminus \{i\}}. \bigwedge_{k,\ell \in \kappa \setminus \{i\}} \{\neg x_k \equiv x_\ell \mid k \neq \ell\} \wedge \bigwedge_{k \in \kappa \setminus \{i\}} r(x_k, x_i) \wedge [D; x_k]$$

Note again that we treat individuals like ordinary concepts and interpret $\mathcal{ALCQIO}$-concepts over the class $\mathbb{K}$ (cf. Section 4.1). Hence $[a; x_i]$ does not translate to a constant, but as all other concept names to an atomic predicate formula $a(x_i)$.

Although we have gained a lot more expressivity by inverse roles and graded modal operators, we are still not as expressive as e.g. the two-variable fragment [53] of FO: reflexivity $\forall x.r(x, x)$ or role inclusion $\forall x, y.r(x, y) \rightarrow s(x, y)$ etc. cannot be expressed by an $\mathcal{ALCQIO}$-concept.

However, $\mathcal{ALCQIO}$ is, since it is a fragment of FO, compact. Since $\mathbb{K}$ is an elementary class, i.e. it can be defined by a set of FO-formulae $\Gamma$, also FO is compact over $\mathbb{K}$For a proof see Proposition 4.1.3.

### $\mathcal{ALCQIO}$-Bisimulation and Properties

The *$\mathcal{ALCQIO}$-bisimulation game* $G(\mathfrak{I}, d; \mathfrak{H}, e)$ is formalised as the bisimulation game for $\mathcal{ALCQ}$ (cf. Section 3.2.1), again treating individual names like concept names, yet now **I** can challenge **II** additionally by playing sets of predecessors:

The state of the game is captured in a bijection $\beta : D \longrightarrow E$ where $D \subseteq \Delta^{\mathfrak{I}}$ and $E \subseteq \Delta^{\mathfrak{H}}$. **I** chooses one element from $D$ or $E$ and some $r \in \mathsf{N}_\mathsf{R}$, say $d' \in D$, and can challenge **II** either with a finite subset $D' \subseteq r^{\mathfrak{I}}(d')$ or a finite subset $D' \subseteq r^{-\mathfrak{I}}(d')$.

**II** then has to respond accordingly with a set $E' \subseteq r^{\mathfrak{H}}(\beta(d'))$ and $E' \subseteq r^{-\mathfrak{H}}(\beta(d'))$ respectively, such that $\beta' : D' \longrightarrow E'$ is a bijection such that $d''$ is atomically equivalent to $\beta(d'')$ for all $d'' \in D$. If **II** cannot deliver $E'$ accordingly **II** has lost the game.

In every other aspect the game proceeds as for $\mathcal{ALCQ}$.

Definition 4.4.1. **II** has a *winning strategy* in $G(\mathfrak{I}, d; \mathfrak{H}, e)$, if she can ward off any challenge of **I** and thus never loses a round. We then say, $(\mathfrak{I}, d)$ is $\mathcal{ALCQIO}$-bisimilar to $(\mathfrak{H}, e)$ and denote this with $(\mathfrak{I}, d) \xleftrightarrow{<\omega} (\mathfrak{H}, e)$. ◇

We introduce a restricted version of the game which is bound in the number rounds and in the cardinality of sets that **I** may play: $G_n^\kappa(\mathfrak{I}, d; \mathfrak{H}, e)$ is the $n$-round game in which only sets of positive cardinality of at most $\kappa < \omega$ may be played. **II** wins the game, if she does not lose for within $n$ rounds.

Definition 4.4.2. **II** has a winning strategy in $G_n^\kappa(\mathfrak{I}, d; \mathfrak{H}, e)$ if she can ward off challenges of **I** up to positive cardinalities of $\kappa < \omega$ during the first $n$ rounds. We then say, $(\mathfrak{I}, d)$ is $(\kappa, n)$-bisimilar to $(\mathfrak{H}, e)$ and denote this with $(\mathfrak{I}, d) \xleftrightarrow{\le\kappa}_n (\mathfrak{H}, e)$. ◇

We make the following observations for our notion of $\mathcal{ALCQIO}$-bisimulation (and $\mathcal{ALCQI}$-bisimulation respectively) which are also true for $\mathcal{ALCQ}$-bisimulation:

$\xleftrightarrow{<\omega}$ and $\xleftrightarrow{\le\kappa}_n$ for all $n < \omega$ and all positive $\kappa < \omega$ form equivalence relations on the set of pointed $\tau$-interpretations. We have $\xleftrightarrow{<\omega} \subseteq \xleftrightarrow{\le\kappa}_n$ and $\xleftrightarrow{\le\kappa}_n \subseteq \xleftrightarrow{\le\kappa'}_{n'}$ for all $n, n' < \omega$ and all positive $\kappa, \kappa' < \omega$, where $n \le n'$ *and* $\kappa \le \kappa'$. For all $n < \omega$ the sequence $(\xleftrightarrow{\le\kappa}_n)_{0<\kappa<\omega}$ is an approximation for $\xleftrightarrow{<\omega}_n$ in the sense that $\bigcap_{0<\kappa<\omega} \xleftrightarrow{\le\kappa}_n = \xleftrightarrow{<\omega}_n$. Similarly $\bigcap_{0<\kappa<\omega} \xleftrightarrow{\le\kappa} = \xleftrightarrow{<\omega}$.

Since individual names are interpreted like concept names the following proposition can be stated without referring to the class $\mathbb{K}$; they all hold in particular in $\mathbb{K}$; we are giving properties of $\mathcal{ALCQI}$, as it were.

Proposition 4.4.3. *Every $\mathcal{ALCQIO}$-concept $C$ with* grade $C \le \kappa$ *and* rank $C \le n$ *is invariant under $\xleftrightarrow{\le\kappa}_n$.*

The proof is, barring predecessor sets, analogous the proof for Proposition 3.2.4 and is therefore omitted. For finite signature $\tau$ we define *characteristic $\mathcal{ALCQIO}$-concepts* for each positive $\kappa < \omega$. Let therefore $\tau$ be finite and $\mathfrak{I}$ a $\tau$-interpretation with $d \in \Delta^{\mathfrak{I}}$.

$$\#\langle n, r(d, e)\rangle := |\{d_1 \in r^{\mathfrak{I}}(d) \mid (\mathfrak{I}, d_1) \models X^{\kappa,n}_{\mathfrak{I},e}\}|$$

i.e. the number of all $r$-successors of $d$ that satisfy the same characteristic $\mathcal{ALCQIO}$-concept as $e$ on level $n$ and let similarly

$$\#\langle n, r^-(d, e)\rangle := |\{d_1 \in r^{-\mathfrak{I}}(d) \mid (\mathfrak{I}, d_1) \models X^{\kappa,n}_{\mathfrak{I},e}\}|$$

be the number of all $r$-predecessors of $d$ that satisfy $X^{\kappa,n}_{\mathfrak{I},e}$. We set

$$?\langle n, r(d, e)\rangle := \begin{cases} = \#\langle n, r(d, e)\rangle & \text{if } \#\langle n, r(d, e)\rangle < \kappa \\ \ge \kappa & \text{else} \end{cases}$$

and analogously we set the same for $?\langle n, r^-(d, e)\rangle$. Note that $?\langle n, r(d, e)\rangle$ delivers a string: If $\kappa = 5$ and $\#\langle n, r(d, e)\rangle = 6$ then $?\langle n, r^-(d, e)\rangle$ evaluates to '$\ge 5$'. For fixed $\kappa$ the characteristic $\mathcal{ALCQIO}$-concept is recursively upon $n < \omega$ defined as:

$$
\begin{aligned}
X^{\kappa,0}_{\mathfrak{I},d} &:= \textstyle\bigsqcap\{A \in \mathsf{N_C} \mid d \in A^{\mathfrak{I}}\} \sqcap \textstyle\bigsqcap\{\neg A \mid d \notin A^{\mathfrak{I}}\} \\
X^{\kappa,n+1}_{\mathfrak{I},d} &:= X^{\kappa,0}_{\mathfrak{I},d} \sqcap \textstyle\bigsqcap_{r \in \mathsf{N_R}} \{\exists^{?\langle n, r(d,e)\rangle} r.X^{\kappa,n}_{\mathfrak{I},e} \mid e \in r^{\mathfrak{I}}(d)\} \\
&\quad \sqcap \textstyle\bigsqcap_{r \in \mathsf{N_R}} \{\exists^{?\langle n, r^-(d,e)\rangle} r.X^{\kappa,n}_{\mathfrak{I},e} \mid e \in r^{-\mathfrak{I}}(d)\} \\
&\quad \sqcap \textstyle\bigsqcap_{r \in \mathsf{N_R}} \forall^{\geq 1} r. \textstyle\bigsqcup\{X^{\kappa,n}_{\mathfrak{I},e} \mid e \in r^{\mathfrak{I}}(d)\} \sqcap \forall^{\geq 1} r^-. \textstyle\bigsqcup\{X^{\kappa,n}_{\mathfrak{I},e} \mid e \in r^{-\mathfrak{I}}(d)\}
\end{aligned}
$$

An inductive argument shows that each set cannot contain infinitely many concepts. Hence, for all positive $\kappa < \omega$ and all $n$ the characteristic $\mathcal{ALCQIO}$-concept is well defined. Clearly, $(\mathfrak{I}, d) \models X^{\kappa,n}_{\mathfrak{I},d}$ for all positive $\kappa < \omega$ and all $n < \omega$.

Proposition 4.4.4 . *If* $(\mathfrak{I}, d) \models X^{\kappa,n}_{\mathfrak{H},e}$ *then* $(\mathfrak{I}, d) \overset{\leq\kappa}{\Longleftrightarrow}_n (\mathfrak{H}, e)$

Again the proof can be copied almost word for word from the proof of Proposition 3.2.6, which makes the correspondent claim for $\mathcal{ALCQ}$, and is therefore omitted. Analogously to Theorem 3.2.8, we can make the following summarising statement:

Theorem 4.4.5 . *Let* $\tau$ *be finite and let* $\mathfrak{I}, \mathfrak{H}$ *be* $\tau$*-interpretations. The following three statements are equivalent:*

1. $(\mathfrak{H}, e) \models X^{\kappa,n}_{\mathfrak{I},d}$
2. $(\mathfrak{I}, d) \overset{\leq\kappa}{\Longleftrightarrow}_n (\mathfrak{H}, e)$
3. $Th^{\kappa}_n(\mathfrak{I}, d) = Th^{\kappa}_n(\mathfrak{H}, e)$

Proof. '1.$\Longrightarrow$ 2.' follows from Proposition 4.4.4, '2.$\Longrightarrow$ 3.' follows from Proposition 4.4.3 and '3.$\Longrightarrow$ 1.' follows from the fact that rank $X^{\kappa,n}_{\mathfrak{I},d} = n$ and grade $X^{\kappa,n}_{\mathfrak{I},d} \leq \kappa$. □

### Saturation and Hennessy-Milner-Property

We lift the restriction of signatures $\tau$ being finite and admit signatures of arbitrary size. We still need not to refer to the class $\mathbb{K}$. The definition of type is the one for $\mathcal{ALCQ}$ extended by a definition of $\exists^{\geq\kappa} r^-$-types.

Definition 4.4.6 . Let $\tau$ be a signature and $(\mathfrak{I}, d)$ a pointed $\tau$-interpretation.

1. Let $\kappa < \omega$ be positive, $r \in \mathsf{N_R}$. Then $\Gamma \subseteq \mathcal{ALCO}(\tau)$ is called $\exists^{\geq\kappa} r$*-type* of $(\mathfrak{I}, d)$ if for all finite $\Gamma_0 \subseteq \Gamma$ we have $(\mathfrak{I}, d) \models \exists^{\geq\kappa} r. \bigsqcap \Gamma_0$. Similarly we define $\exists^{\geq\kappa} r^-$-types.

2. An $\exists^{\geq\kappa}r$-type $\Gamma$ of $(\mathfrak{I}, d)$ is *realised at d* if there is a set $D \subseteq r^{\mathfrak{I}}(d)$ with $|D| \geq \kappa$ such that for all $d_0 \in D$ we have $(\mathfrak{I}, d_0) \models \Gamma$.

3. An $\exists^{\geq\kappa}r^{-}$-type $\Gamma$ of $(\mathfrak{I}, d)$ is *realised at d* if there is a set $D \subseteq r^{-\mathfrak{I}}(d)$ with $|D| \geq \kappa$ such that for all $d_0 \in D$ we have $(\mathfrak{I}, d_0) \models \Gamma$.

$\diamond$

Definition 4.4.7. An interpretation $\mathfrak{I}$ is *$\mathcal{ALCQIO}$-saturated*, if for all positive $\kappa < \omega$, all $r \in \mathsf{N_R}$ and every $d \in \Delta^{\mathfrak{I}}$, every $\exists^{\geq\kappa}r$-type and every $\exists^{\geq\kappa}r^{-}$-type of $(\mathfrak{I}, d)$ is realised at $d$. $\diamond$

Observation 4.4.8. *Every $\omega$-saturated interpretation is $\mathcal{ALCQIO}$-saturated*

Proposition 4.4.9. *$\mathcal{ALCQIO}$ has the Hennessy-Milner-property, i.e. if $\mathfrak{I}$ and $\mathfrak{H}$ are both $\mathcal{ALCQIO}$-saturated then*

$$(\mathfrak{I}, d) \equiv_{\mathcal{ALCQIO}} (\mathfrak{H}, e) \quad \textit{iff} \quad (\mathfrak{I}, d) \xleftrightarrow{<\omega} (\mathfrak{H}, e),$$

*where* $Th(\mathfrak{I}, d) := \{C \in \mathcal{ALCQIO}(\tau) \mid (\mathfrak{I}, d) \models C\}$ *and analogously for* $(\mathfrak{H}, e)$.

The proof is omitted as it follows exactly the rationale in the proof of Proposition 3.2.12: **II** has maintain $(\mathfrak{I}, d') \equiv_{\mathcal{ALCQIO}} (\mathfrak{H}, \beta(d'))$ for every configuration $\beta : D \longrightarrow E$ and $d' \in D$ during the game.
It is also possible to restrict the proposition to the fragment which allows for graded operators up to $\kappa$. We then obtain the analogue result

Definition 4.4.10. An interpretation $\mathfrak{I}$ is *$\mathcal{ALCQIO}$-$\kappa$-saturated*, if for all positive $\lambda < \kappa$, all $r \in \mathsf{N_R}$ and every $d \in \Delta^{\mathfrak{I}}$, every $\exists^{\geq\lambda}r$-type and every $\exists^{\geq\lambda}r^{-}$-type of $(\mathfrak{I}, d)$ is realised at $d$. $\diamond$

Proposition 4.4.11. *$\mathcal{ALCQIO}^{\kappa}$ has the Hennessy-Milner-property, i.e. if $\mathfrak{I}$ and $\mathfrak{H}$ are both $\mathcal{ALCQIO}$-$\kappa$-saturated then $Th^{\kappa}(\mathfrak{I}, d) = Th^{\kappa}(\mathfrak{H}, e)$ iff $(\mathfrak{I}, d) \xleftrightarrow{\leq\kappa} (\mathfrak{H}, e)$, where $Th^{\kappa}(\mathfrak{I}, d) = \{C \in \mathcal{ALCQIO}(\tau) \mid (\mathfrak{I}, d) \models C$ and* $\mathrm{rank}\, C \leq \kappa\}$ *and similarly for $(\mathfrak{H}, e)$*

The adaption of the proof given for Proposition 3.2.12 is not difficult.

### The Characterisation Theorem for $\mathcal{ALCQIO}$-Concepts over $\mathbb{K}$

We shall now present the characterisation theorem for $\mathcal{ALCQIO}$-concepts w.r.t. to $\mathbb{K}$. Yet, with the notions for $\mathcal{ALCQIO}$-saturatedness and $\mathcal{ALCQIO}$-character-istic concepts, which were presented without referring to $\mathbb{K}$, the steps taken for the characterisation of $\mathcal{ALCQIO}$-concepts can be readily extended to $\mathcal{ALCQI}$.

Let $\tau$ be arbitrary and $I := (\omega \setminus \{0\}) \times \omega$. We may assume w.l.o.g. that $\tau$ contains at least one element in $\mathsf{N}_\mathsf{I}$, for otherwise this case would reduce to $\mathcal{ALCQI}$.

Lemma 4.4.12 . *Let $\varphi(x) \in \mathrm{FO}(\tau)$ be invariant under $\xleftrightarrow{*}_\dagger$ in $\mathbb{K}$, where $* \in \{`< \omega'\} \cup \{`\leq \kappa' \mid \kappa < \omega\}$ and $\dagger < \omega$ or simply omitted. Let $\tau_\varphi \subseteq \tau$ contain all signature symbols occurring in $\varphi$ and $\mathfrak{I}, \mathfrak{H}$ in $\mathbb{K}$.*

$$(\mathfrak{I} \restriction \tau_\varphi, d) \xleftrightarrow{*}_\dagger (\mathfrak{H} \restriction \tau_\varphi, e) \implies (\mathfrak{I}, d) \models \varphi(x) \iff (\mathfrak{H}, e) \models \varphi(x)$$

We can directly use the proof for Lemma 4.3.1

Proposition 4.4.13 . *Every formula $\varphi \in \mathrm{FO}(\tau)$ which is invariant under $\xleftrightarrow{<\omega}$ in $\mathbb{K}$ is invariant under $\xleftrightarrow{\leq\kappa}_n$ in $\mathbb{K}$ for some $(\kappa, n) \in I$.*

A proof would verbatim repeat the proof of Proposition 4.3.2 and is thus omitted.

Theorem 4.4.14 . *Let $\tau$ be a signature. For all $\varphi(x) \in \mathrm{FO}(\tau)$, which are invariant under $\xleftrightarrow{<\omega}$ in $\mathbb{K}$ exists a concept $C \in \mathcal{ALCQIO}(\tau)$ such that $\varphi(x)$ is logically equivalent to $C$.*

Proof. According to Proposition 4.4.13 $\varphi$ is invariant under $\xleftrightarrow{\leq\kappa}_n$ for some $(\kappa, n) \in I$. Let $\tau_\varphi \subseteq \tau$ contain all signature symbols occuring in $\varphi$ and let $C := \bigsqcup\{X^{\kappa,n}_{\mathfrak{I}\restriction\tau_\varphi,d} \mid (\mathfrak{I}, d) \models \varphi(x)\}$, where $X^{\kappa,n}_{\mathfrak{I}\restriction\tau_\varphi,d}$ is the characteristic $\mathcal{ALCQIO}$-concept of $(\mathfrak{I}, d)$ over the signature $\tau_\varphi$ on level $(\kappa, n)$.

Since $\varphi(x) \models C$ it remains to show $C \models \varphi(x)$. Assume $(\mathfrak{H}, e) \models X^{\kappa,n}_{\mathfrak{K}\restriction\tau_\varphi,c}$ with $X^{\kappa,n}_{\mathfrak{K}\restriction\tau_\varphi,c} \in \{X^{\kappa,n}_{\mathfrak{I}\restriction\tau_\varphi,d} \mid (\mathfrak{I}, d) \models \varphi(x)\}$. Then $(\mathfrak{H}\restriction\tau_\varphi, e) \xleftrightarrow{\leq\kappa}_n (\mathfrak{K}\restriction\tau_\varphi, c) \models \varphi$ and Lemma 4.4.12 yields $(\mathfrak{H}, e) \models \varphi(x)$. Hence $C \models \varphi(x)$ which shows that $C$ is logically equivalent to $\varphi(x)$. □

### 4.4.2 The Characterisation of $\mathcal{ALCQIO}u$-Concepts

Similarly to $\mathcal{ALCQO}u$, the characterisation of $\mathcal{ALCQIO}u$ over $\mathbb{K}$ does not make much sense as such since we can already control the cardinality via the universal graded role: $\exists^{=1} u.a$ can be stated in $\mathcal{ALCQIO}u$ and can be used to force concept names being interpreted as singleton. However, we have only indirectly investigated the description logic $\mathcal{ALCQI}$ so far. So results stated for $\mathcal{ALCQIO}u$ are, without problems, reducible to $\mathcal{ALCQI}$.

As for $\mathcal{ALCQ}u$ and $\mathcal{ALCQO}u$, we first give a characterisation for graded universal roles and we shall later, in preparation for TBoxes, restrict the universal role to $\exists^{\geq 1}$ quantifications. Each $\mathcal{ALCQIO}u$-concept $C$ over the signature $\tau$ is recursively defined by

$$C ::= \top \mid a \mid A \mid D \sqcap E \mid \neg D \mid \exists^{\geq\kappa} r.D \mid \exists^{\geq\kappa} r^{-}.D \mid \exists^{\geq\kappa} u.D$$

where $a \in \mathsf{N_I}$, $A \in \mathsf{N_C}$, $r \in \mathsf{N_R}$, $\kappa < \omega$ and $D, E$ are $\mathcal{ALCQIO}u$-concepts over $\tau$. $u$ is a logical symbol, i.e. it is not part of the signature. The set of all $\mathcal{ALCQIO}u$-concepts over $\tau$ is denoted by $\mathcal{ALCQIO}u(\tau)$. As usual $\bot$, $\rightarrow$, $\sqcup$, $\forall^{\geq\kappa}$ etc. are considered to be abbreviations.

The interpretation function for $\mathcal{ALCQIO}$-concepts is extended to $\mathcal{ALCQIO}u$-concepts by setting $\mathfrak{I} \models \exists^{\geq\kappa} u.D$ iff $|D^{\mathfrak{I}}| \geq \kappa$. Similar to $\mathcal{ALCQO}$, the grade function is extended to $\mathcal{ALCQIO}u$-concepts by setting grade $\exists^{\geq\kappa} u.D := \max\{\kappa, D\}$, i.e. actually considering it and rank $\exists^{\geq\kappa} u.D :=$ rank $D$, i.e. ignoring the universal role.

Definition 4.4.15 . Two interpretations $\mathfrak{I}, \mathfrak{H}$ are *$\mathcal{ALCQIO}u$-bisimilar*, $\mathfrak{I} \xleftrightarrow{<\omega}{}^{\forall} \mathfrak{H}$, if both of the following is true:

1. for all finite $D \subseteq \Delta^{\mathfrak{I}}$ there is an injective function $\beta : D \longrightarrow \Delta^{\mathfrak{H}}$ with $(\mathfrak{I}, d) \xleftrightarrow{<\omega} (\mathfrak{H}, \beta(d))$ for all $d \in D$.
2. for all finite $E \subseteq \Delta^{\mathfrak{H}}$ there is an injective function $\beta : E \longrightarrow \Delta^{\mathfrak{I}}$ with $(\mathfrak{I}, \beta(e)) \xleftrightarrow{<\omega} (\mathfrak{H}, e)$ for all $e \in E$.

Analogously we say $\mathfrak{I}$ and $\mathfrak{H}$ are *$\mathcal{ALCQIO}u$-$(\kappa, n)$-bisimilar*, $\mathfrak{I} \xleftrightarrow{\leq\kappa}{}^{\forall}_{n} \mathfrak{H}$, iff

1. for all $D \subseteq \Delta^{\mathfrak{I}}$ with $|D| \leq \kappa$ there is an injective function $\beta : D \longrightarrow \Delta^{\mathfrak{H}}$ with $(\mathfrak{I}, d) \xleftrightarrow{\leq\kappa}_{n} (\mathfrak{H}, \beta(d))$ for all $d \in D$.
2. for all $E \subseteq \Delta^{\mathfrak{H}}$ with $|E| \leq \kappa$ there is an injective function $\beta : E \longrightarrow \Delta^{\mathfrak{H}}$ with $(\mathfrak{I}, \beta(e)) \xleftrightarrow{\leq\kappa}_{n} (\mathfrak{H}, e)$ for all $e \in E$.

As usual, we define $(\mathfrak{I}, d) \xleftrightarrow{\leq\kappa}{}^{\forall}_{n} (\mathfrak{I}, e)$ if $\mathfrak{I} \xleftrightarrow{\leq\kappa}{}^{\forall}_{n} \mathfrak{H}$ and $(\mathfrak{I}, d) \xleftrightarrow{\leq\kappa}_{n} (\mathfrak{H}, e)$ and similarly for $< \omega$ etc. ◇

It is straight forward to see that all these relations are equivalence relations on the set of $\tau$-interpretations for each signature $\tau$ and that every $\mathcal{ALCQIO}u$-concept $C$ of rank $C \leq n$ and grade $C \leq \kappa$ is invariant under $\xleftrightarrow{\leq\kappa}{}^{\forall}_{n}$. Again, this is true for all $\tau$-interpretations, also those outside of $\mathbb{K}$.

Characteristic Concepts for $\mathcal{ALCQIO}u$

Since the universal role does not occur in its inverse form $u^{-}$ as it would mean the same as $u$ on its own, this section is almost identical to $\mathcal{ALCQIO}$. Throughout this section let $\tau$ be finite and $I := (\omega \setminus \{0\}) \times \omega$

Let a $\tau$-interpretation $\mathfrak{I}$ and $(\kappa, n) \in I$ be given. We define for all $d \in \Delta^{\mathfrak{I}}$

$$?\kappa(d) := \begin{cases} = |\{d_0 \in \Delta^{\mathfrak{I}} \mid (\mathfrak{I}, d_0) \models X^{\kappa,n}_{\mathfrak{I},d}\}| & \text{if } |\{d_0 \in \Delta^{\mathfrak{I}} \mid (\mathfrak{I}, d_0) \models X^{\kappa,n}_{\mathfrak{I},d}\}| < \kappa \\ \geq \kappa & \text{else,} \end{cases}$$

where $X^{\kappa,n}_{\mathfrak{I},d}$ is the characteristic $\mathcal{ALCQIO}$-concept over $\tau$ for $(\mathfrak{I}, d)$ on level $(\kappa, n)$

The *global characteristic* $\mathcal{ALCQIO}u$*-concept over* $\tau$ *on level* $(\kappa, n)$ is defined as

$$X^{\kappa,n}_{\mathfrak{I}} := \bigsqcap\{\exists^{?\kappa(d)} u.X^{\kappa,n}_{\mathfrak{I},d} \mid d \in \Delta^{\mathfrak{I}}\} \sqcap \forall^{\geq 1} u. \bigsqcup\{X^{\kappa,n}_{\mathfrak{I},d} \mid d \in \Delta^{\mathfrak{I}}\}$$

again, where $X^{\kappa,n}_{\mathfrak{I},d}$ is the characteristic $\mathcal{ALCQIO}$-concept over $\tau$ for $(\mathfrak{I}, d)$ on level $(\kappa, n)$.

If $C$ is constructed by the following scheme, $C$ is called a *global* $\mathcal{ALCQIO}u$*-concept* over $\tau$: $C ::= \exists^{\geq\kappa} u.D \mid E \sqcap F \mid \neg E$ where $D \in \mathcal{ALCQIO}(\tau)$ and $\kappa < \omega$ and $E, F$ are global $\mathcal{ALCQIO}u$-concepts over $\tau$; clearly then $C \in \mathcal{ALCQIO}u(\tau)$. Apparently, global $\mathcal{ALCQIO}u$-concepts do not depend on a distinguished element. Let

$$\mathrm{Th}^{\kappa}_{n}(\mathfrak{I}) := \{C \in \mathcal{ALCQIO}u(\tau) \mid \mathfrak{I} \models C, C \text{ is global, rank } C \leq n, \text{ grade } C \leq \kappa\}.$$

Proposition 4.4.16 . *Let* $(\kappa, n) \in I$ *and* $\tau$ *finite. The following three statements are equivalent:*

1. $\mathfrak{H} \models X^{\kappa,n}_{\mathfrak{I}}$
2. $\mathfrak{I} \overset{\leq\kappa\,\forall}{\Longleftrightarrow}_{n} \mathfrak{H}$
3. $Th^{\kappa}_{n}(\mathfrak{I}) = Th^{\kappa}_{n}(\mathfrak{H})$

The arguments used in the proof of Proposition 3.2.18 for $\mathcal{ALCQ}u$ can be immediately transferred to show Proposition 4.4.25.

Proposition 4.4.17 . *Let* $(\kappa, n) \in I$ *and* $\tau$ *finite. The following three statements are equivalent:*

1. $(\mathfrak{H}, e) \models X^{\kappa,n}_{\mathfrak{I}} \sqcap X^{\kappa,n}_{\mathfrak{I},d}$

2. $(\mathfrak{I}, d) \xleftrightarrow{\leq\kappa\,\forall}_n (\mathfrak{H}, e)$

3. $\mathit{Th}_n^\kappa(\mathfrak{I}, d) = \mathit{Th}_n^\kappa(\mathfrak{H}, e)$

*where* $\mathit{Th}_n^\kappa(\mathfrak{I}, d) := \{C \in \mathcal{ALCQIO}u(\tau) \mid (\mathfrak{I}, d) \models C,\ \operatorname{rank} C \leq n, \operatorname{grade} C \leq \kappa\}$ *and similarly for* $(\mathfrak{H}, e)$.

The proposition is the combination of Theorem 4.4.5 and Proposition 4.4.25.

### Saturation and Hennessy-Milner-Property

The signature $\tau$ needs not to be finite anymore but can have arbitrary size.

Definition 4.4.18 . Let $\tau$ be a signature, $\mathfrak{I}$ a $\tau$-interpretation and let $\Gamma \subseteq \mathcal{ALCQIO}(\tau)$.

1. $\Gamma$ is an $\exists^{\geq\kappa}u$*-type* of $\mathfrak{I}$ if for every finite subset $\Gamma_0 \subseteq \Gamma$ we have $\mathfrak{I} \models \exists^{\geq\kappa}u. \bigsqcap \Gamma_0$
2. An $\exists^{\geq\kappa}u$-type of $\mathfrak{I}$ is *realised* in $\mathfrak{I}$ if there is a set $D \subseteq \Delta^{\mathfrak{I}}$ with $|D| \geq \kappa$ such that for all $d \in D$ we have $(\mathfrak{I}, d) \models \Gamma$.

◇

A $\tau$-interpretation $\mathfrak{I}$ is $\mathcal{ALCQIO}u$*-saturated*, if it is $\mathcal{ALCQIO}$-saturated and for all positive $\kappa < \omega$ all $\exists^{\geq\kappa}u$-types are realised.

Proposition 4.4.19 . *$\mathcal{ALCQIO}u$ has the Hennessy-Milner property, i.e. for any two $\mathcal{ALCQIO}u$-saturated $\tau$-interpretations $\mathfrak{I}, \mathfrak{H}$ the following holds:*

$$\mathfrak{I} \equiv_{\mathcal{ALCQIO}u} \mathfrak{H} \iff \mathfrak{I} \xleftrightarrow{<\omega} \mathfrak{H}.$$

Again, one can transfer the arguments from Proposition 3.2.22, which shows the Hennessy-Milner-Property for $\mathcal{ALCQ}u$, in order to show Proposition 4.4.19.

Similar to the Hennessy-Milner-Property for $\mathcal{ALCQIO}$, one can call a $\tau$-interpretation $\mathcal{ALCQIO}u$*-$\kappa$-saturated*, whenever it is $\mathcal{ALCQIO}$-$\kappa$-saturated and all $\exists^{\geq\lambda}u$-types for $\lambda < \kappa$ are realised. Then the following variant of the Hennessy-and-Milner property holds:

Proposition 4.4.20 . *$\mathcal{ALCQIO}u$ has the Hennessy-Milner property, i.e. for any two $\mathcal{ALCQIO}u$-$\kappa$-saturated $\tau$-interpretations $\mathfrak{I}, \mathfrak{H}$ the following holds:*

$$\mathit{Th}^\kappa(\mathfrak{I}) = \mathit{Th}^\kappa(\mathfrak{H}) \iff \mathfrak{I} \xleftrightarrow{\leq\kappa} \mathfrak{H}$$

*where* $\mathit{Th}^\kappa := \{C \in \mathcal{ALCQIO}u(\tau) \mid \mathfrak{I} \models C$ *and* $\operatorname{grade} C \leq \kappa\}$ *and similarly for* $\mathfrak{H}$.

### The Characterisation Theorem for $\mathcal{ALCQIO}u$-Concepts

We shall give a characterisation theorem for $\mathcal{ALCQI}u$-concepts instead of $\mathcal{ALCQIO}u$-concepts.

Proposition 4.4.21. *Let $\tau$ be an arbitrary signature. If $\varphi(x) \in \mathrm{FO}(\tau)$ and $\varphi$ is invariant under $\overset{<\omega}{\longleftrightarrow}{}^{\forall}$ then $\varphi$ is invariant under $\overset{\leq\kappa}{\longleftrightarrow}{}^{\forall}_{n}$ for some $(\kappa, n) \in I$.*

The proof can simply be copied from Proposition 3.2.14, provided one refers to the adapted notions of characteristic $\mathcal{ALCQIO}u$-concepts and $\mathcal{ALCQIO}u$-bisimulation.

Theorem 4.4.22. *Every formula $\varphi(x) \in \mathrm{FO}(\tau)$ which is invariant under $\overset{<\omega}{\longleftrightarrow}{}^{\forall}$ is logically equivalent to some concept $C \in \mathcal{ALCQI}u(\tau)$.*

With an analogous lemma to Lemma 4.4.27, and Proposition 4.4.21 one can proof the following theorem, whose proof has been omitted as it is analogous to the proof of Theorem 4.4.29:

Theorem 4.4.23. *Every formula $\varphi(x) \in \mathrm{FO}(\tau)$ which is invariant under $\overset{<\omega}{\longleftrightarrow}{}^{\forall}$ in $\mathbb{K}$ is logically equivalent to some concept $C \in \mathcal{ALCQIO}u(\tau)$.*

### $\mathcal{ALCQIO}u_1$

$\mathcal{ALCQIO}u_1$ is the extension of $\mathcal{ALCQIO}$ with $\exists^{\geq 1}u$.

Definition 4.4.24. Two $\tau$-interpretation $\mathfrak{I}$, $\mathfrak{H}$ are globally $\mathcal{ALCQIO}$-bisimilar, $\mathfrak{I} \overset{\leq\kappa}{\longleftrightarrow}{}^{\varrho} \mathfrak{H}$, if both of the following holds:

1. for every $d \in \Delta^{\mathfrak{I}}$ there is $e \in \Delta^{\mathfrak{H}}$ with $(\mathfrak{I}, d) \overset{<\omega}{\longleftrightarrow} (\mathfrak{H}, e)$
2. for every $e \in \Delta^{\mathfrak{H}}$ there is $d \in \Delta^{\mathfrak{I}}$ with $(\mathfrak{I}, d) \overset{<\omega}{\longleftrightarrow} (\mathfrak{H}, e)$

◇

We define $\mathfrak{I} \overset{\leq\kappa}{\longleftrightarrow}{}^{\varrho}_{n} \mathfrak{H}$ analogously. Again we abbreviate $(\mathfrak{I}, d) \overset{<\omega}{\longleftrightarrow}{}^{\varrho} (\mathfrak{H}, e)$ if $\mathfrak{I} \overset{<\omega}{\longleftrightarrow}{}^{\varrho} \mathfrak{H}$ and $(\mathfrak{I}, d) \overset{<\omega}{\longleftrightarrow} (\mathfrak{H}, e)$ and similarly we abbreviate $(\mathfrak{I}, d) \overset{\leq\kappa}{\longleftrightarrow}{}^{\varrho}_{n} (\mathfrak{H}, e)$.

For finite signature $\tau$, and $(\kappa, n) \in I$ characteristic $\mathcal{ALCQIO}u_1$-concepts for some $\tau$-interpretation $\mathfrak{I}$ are defined as

$$Y^{\kappa,n}_{\mathfrak{I}} := \bigsqcap\{\exists^{\geq 1}u.X^{\kappa,n}_{\mathfrak{I},d} \mid d \in \Delta^{\mathfrak{I}}\} \sqcap \forall^{\geq 1}u.\bigsqcup\{X^{\kappa,n}_{\mathfrak{I},d} \mid d \in \Delta^{\mathfrak{I}}\}$$

where $X^{\kappa,n}_{\mathfrak{I},d}$ is the characteristic $\mathcal{ALCQIO}$-concept over $\tau$ on level $(\kappa, n)$.

Proposition 4.4.25 . *Let* $(\kappa, n) \in I$ *and* $\tau$ *finite. The following three statements are equivalent:*

1. $\mathfrak{H} \models Y^{\kappa,n}_{\mathfrak{I}}$
2. $\mathfrak{I} \xleftrightarrow{\leq\kappa\, Q}_n \mathfrak{H}$
3. $Th^{\kappa}_n(\mathfrak{I}) = Th^{\kappa}_n(\mathfrak{H})$

*with* $Th^{\kappa}_n(\mathfrak{I}) := \{C \in \mathcal{ALCQIO}u_1(\tau) \mid \mathfrak{I} \models C, C \text{ global}, \operatorname{rank} C \leq n, \operatorname{grade} C \leq \kappa\}$.

Proof. '1.$\Longrightarrow$ 2.' If $\mathfrak{H} \models Y^{\kappa,n}_{\mathfrak{I}}$ then for every $d \in \Delta^{\mathfrak{I}}$ $\mathfrak{H} \models \exists^{\geq 1} u.X^{\kappa,n}_{\mathfrak{I},d}$ and so there is $e \in \Delta^{\mathfrak{H}}$ with $(\mathfrak{I}, d) \xleftrightarrow{\leq\kappa}_n (\mathfrak{H}, e)$, and conversely, since $\mathfrak{H} \models \forall^{\geq 1} u. \bigsqcup\{X^{\kappa,n}_{\mathfrak{I},d} \mid d \in \Delta^{\mathfrak{I}}\}$, for every $e \in \Delta^{\mathfrak{H}}$ there is some $d \in \Delta^{\mathfrak{I}}$ with $\mathfrak{H} \models X^{\kappa,n}_{\mathfrak{I},d}$, and so $(\mathfrak{I}, d) \xleftrightarrow{\leq\kappa}_n (\mathfrak{H}, e)$.

'2.$\Longrightarrow$ 3.' The proof is carried out by induction upon the structure of global concepts $C \in \mathcal{ALCQIO}u_1(\tau)$ with $\operatorname{grade} C \leq \kappa$ and $\operatorname{rank} C \leq n$: We show $\mathfrak{I} \models \exists^{\geq 1} u.C \Longrightarrow \mathfrak{I} \models \exists^{\geq 1} u.C$: Assume the premise is true. Then there is $d \in \Delta^{\mathfrak{I}}$ with $(\mathfrak{I}, d) \models C$. Since $\mathfrak{I} \xleftrightarrow{\leq\kappa\, Q}_n \mathfrak{H}$ there is $e \in \Delta^{\mathfrak{H}}$ with $(\mathfrak{I}, d) \xleftrightarrow{\leq\kappa}_n (\mathfrak{H}, e)$ and since $C \in \mathcal{ALCQIO}u_1(\tau)$ we have $(\mathfrak{H}, e) \models C$. The converse is obtained the same way. The step cases are readily yielded by the induction hypothesis.

'3.$\Longrightarrow$ 1.' Is immediate. □

Let the signature $\tau$ be arbitrary. We call a $\tau$-interpretation $\mathfrak{I}$ $\mathcal{ALCQIO}u_1$-*saturated* if it is $\mathcal{ALCQIO}$-saturated (cf. Definition 4.4.6) and all $\exists^{\geq 1} u$-types (cf. Definition 4.4.18) are realised.

Proposition 4.4.26 . $\mathcal{ALCQIO}u_1$ *has the Hennessy-Milner property, i.e. for any two* $\mathcal{ALCQIO}u_1$*-saturated* $\tau$*-interpretations* $\mathfrak{I}, \mathfrak{H}$ *the following holds:*

$$Th(\mathfrak{I}) = Th(\mathfrak{H}) \iff \mathfrak{I} \xleftrightarrow{<\omega\, Q} \mathfrak{H}.$$

*where* $Th(\mathfrak{I}) := \{C \in \mathcal{ALCQIO}u_1(\tau) \mid \mathfrak{I} \models C, C \text{ global}\}$.

Proof. Let $d \in \Delta^{\mathfrak{I}}$ arbitrary. Then $\mathfrak{I} \models \exists^{\geq 1} u. \bigsqcap \Gamma_0$ for each finite subset of its $\mathcal{ALCQIO}$-theory $\mathrm{Th}(\mathfrak{I}, d) := \{C \in \mathcal{ALCQIO}(\tau) \mid d \in C^{\mathfrak{I}}\}$. Since $\mathrm{Th}(\mathfrak{I}) = \mathrm{Th}(\mathfrak{H})$ we have $\mathfrak{H} \models \exists^{\geq 1} u. \bigsqcap \Gamma_0$ and so $\mathrm{Th}(\mathfrak{I}, d)$ is a type of $\mathfrak{H}$ which is realised by some $e \in \Delta^{\mathfrak{H}}$. The Hennessy-Milner-Property of $\mathcal{ALCQIO}$ stated in Proposition 4.4.9 shows $(\mathfrak{I}, d) \xleftrightarrow{<\omega} (\mathfrak{H}, e)$.

The same arguments can be used to show that for every $e \in \Delta^{\mathfrak{H}}$ there is $d \in \Delta^{\mathfrak{I}}$ such that $(\mathfrak{I}, d) \overset{<\omega}{\Longleftrightarrow} (\mathfrak{H}, e)$, which proves the if-direction. The only-if direction is readily inferred from Proposition 4.4.25 □

This proposition can easily be extended to pointed interpretations. Again, the Hennessy-Milner-Property holds for every positive cardinality $\kappa < \omega$:

$$\mathrm{Th}^{\kappa}(\mathfrak{I}, d) = \mathrm{Th}^{\kappa}(\mathfrak{H}, d) \iff (\mathfrak{I}, d) \overset{<\kappa}{\Longleftrightarrow}^{\mathcal{Q}} (\mathfrak{H}, d).$$

where $\mathrm{Th}^{\kappa}(\mathfrak{I}, d) := \{C \in \mathcal{ALCQIO}u_1(\tau) \mid (\mathfrak{I}, d) \models C \text{ with rank } C \leq \kappa\}$. Notice that the universal role still has only grade 1, whilst arbitrary positive grades $\kappa < \omega$ are allowed on the level of $\mathcal{ALCQIO}$-concepts.

Lemma 4.4.27 . *Let $\varphi(x) \in \mathrm{FO}(\tau)$ be invariant under $\overset{<\omega}{\Longleftrightarrow}^{\mathcal{Q}}$ in $\mathbb{K}$. Let $\tau_\varphi \subseteq \tau$ contain the signature symbols occurring in $\varphi$, as well as at least one $a \in \mathsf{N}_\mathsf{I}$ and let $\mathfrak{I}, \mathfrak{H}$ in $\mathbb{K}$.*

$$(\mathfrak{I} \upharpoonright \tau_\varphi, d) \overset{<\omega}{\Longleftrightarrow}^{\mathcal{Q}} (\mathfrak{H} \upharpoonright \tau_\varphi, e) \quad \Longrightarrow \quad (\mathfrak{I}, d) \models \varphi(x) \iff (\mathfrak{H}, e) \models \varphi(x)$$

The proof for this lemma can be copied from the proof of Lemma 4.4.27 in a straight forward way.

Proposition 4.4.28 . *Every formula $\varphi \in \mathrm{FO}(\tau)$ which is invariant under $\overset{<\omega}{\Longleftrightarrow}^{\mathcal{Q}}$ in $\mathbb{K}$ is invariant under $\overset{\leq\kappa}{\Longleftrightarrow}^{\mathcal{Q}}_n$ in $\mathbb{K}$ for some $(\kappa, n) \in I$.*

We skip the proof as it uses the same rational as Proposition 4.3.7.

Theorem 4.4.29 . *Every formula $\varphi(x) \in \mathrm{FO}(\tau)$ which is invariant under $\overset{<\omega}{\Longleftrightarrow}^{\mathcal{Q}}$ in $\mathbb{K}$ is equivalent to some concept $C \in \mathcal{ALCQIO}u_1(\tau)$*

Proof. Let $\tau_\varphi$ comprise the symbols occurring in $\varphi$. If $\varphi$ does not contain an element of $\mathsf{N}_\mathsf{I}$, we simply choose an element from $\mathsf{N}_\mathsf{I}$ and add this to $\tau_\varphi$.

According to Proposition 4.4.28 $\varphi$ is invariant under $\overset{\leq\kappa}{\Longleftrightarrow}^{\mathcal{Q}}_n$ for some $(\kappa, n) \in I$. Let $C := \{ Y^{\kappa,n}_{\mathfrak{I}\upharpoonright\tau_\varphi} \sqcap X^{\kappa,n}_{\mathfrak{I}\upharpoonright\tau_\varphi,d} \mid (\mathfrak{I}, d) \text{ in } \mathbb{K} \text{ and } (\mathfrak{I}, d) \models \varphi(x)\}$ where $Y^{\kappa,n}_{\mathfrak{I}\upharpoonright\tau_\varphi}$ is the characteristic $\mathcal{ALCQIO}u_1$-concept and $X^{\kappa,n}_{\mathfrak{I}\upharpoonright\tau_\varphi,d}$ is the characteristic $\mathcal{ALCQIO}$-concept for $(\mathfrak{I}, d)$ over $\tau_\varphi$ on level $(\kappa, n)$.

We show $\bigsqcup C \models \varphi(x)$. Let $(\mathfrak{H}, e) \models \bigsqcup C$. Then there is $(\mathfrak{I}, d)$ in $\mathbb{K}$ such that $X^{\kappa,n}_{\mathfrak{I},d} \in C$ and $(\mathfrak{H}, e) \models X^{\kappa,n}_{\mathfrak{I},d}$. Hence $(\mathfrak{H} \upharpoonright \tau_\varphi, e) \overset{\leq\kappa}{\Longleftrightarrow}^{\mathcal{Q}}_n (\mathfrak{I} \upharpoonright \tau_\varphi, d)$ and Lemma 4.4.27 yields $(\mathfrak{H}, e) \models \varphi(x)$. □

### 4.4.3 The Characterisation of $\mathcal{ALCQIO}$-TBoxes

Interestingly, the characterisation of $\mathcal{ALCQIO}$-TBoxes is easier than for $\mathcal{ALCQO}$: the additional expressivity of the inverse role makes every type of an element describe the whole connected component where the element is located.

Since one element of a connected component is enough, we need not realising more types of other elements in this connected component, which led in case of $\mathcal{ALCQO}$ to new realisations of individuals. We dealt with the resulting excess amount of individuals by cutting out their generated subinterpretations and engraft some uniformly chosen representative of their bisimulation type in their place; a rather technical endeavour, to say the least.

For $\mathcal{ALCQIO}$-TBoxes this problem is reduced to a simple pick-and-mix procedure, which makes it neat and transparent.

For the following observation we need not to refer to the class $\mathbb{K}$. In particular, this observation could be also made for $\mathcal{ALCQI}$-TBoxes.

Observation 4.4.30 . *Let $(\mathfrak{I}_i)_{i\in I}$ be a family of $\tau$-interpretations. $\mathcal{ALCQIO}$-TBoxes are*

1. *invariant under global $\mathcal{ALCQIO}$-bisimulation $\overset{<\omega}{\Longleftrightarrow}{}^{g}$*
2. *invariant under disjoint unions*

Proof.

1. Since every $\mathcal{ALCQIO}$-TBox is equivalent to some $\mathcal{ALCQIO}u_1$-concept, the claim follows directly from Theorem 4.4.29.
2. Note that connected components are transferred without change or alteration into the disjoint union. Since the game for (local) $\mathcal{ALCQIO}$-bisimulation takes place only within a connected component, elements in the disjoint union are $\mathcal{ALCQIO}$-bisimilar to their counterparts in the original structure.

   Assume $\mathcal{T}$ is an $\mathcal{ALCQIO}$-TBox and let $(\mathfrak{I}_i)_{i\in I}$ be a family of $\tau$-interpretations. We have to show

$$(\forall i \in I : \mathfrak{I} \models \mathcal{T}) \iff \biguplus_{i\in I} \mathfrak{I}_i \models \mathcal{T}.$$

   '$\Longrightarrow$': Assume $\forall i \in I : \mathfrak{I} \models \mathcal{T}$. We have to show for all $C \sqsubseteq D \in \mathcal{T}$ that $C^{\biguplus_{i\in I}\mathfrak{I}_i} \subseteq D^{\biguplus_{i\in I}\mathfrak{I}_i}$. Assume $d \in C^{\biguplus_{i\in I}\mathfrak{I}_i}$. Then there is $i \in I$ such that $d \in \Delta^{\mathfrak{I}_i}$. Since $(\mathfrak{I}_i, d) \overset{<\omega}{\Longleftrightarrow} (\biguplus_{i\in I} \mathfrak{I}_i, d)$ we have $d \in C^{\mathfrak{I}_i}$ and since $\mathfrak{I}_i \models \mathcal{T}$ by assumption, it follows that $d \in D^{\mathfrak{I}_i}$, which—because $(\mathfrak{I}_i, d) \overset{<\omega}{\Longleftrightarrow} (\biguplus_{i\in I} \mathfrak{I}_i, d)$—entails $d \in D^{\biguplus_{i\in I} \mathfrak{I}_i}$.

If on the other hand $\biguplus_{i\in I} \mathfrak{I}_i \models \mathcal{T}$ with $C \sqsubseteq D \in \mathcal{T}$ and we have $d \in C^{\mathfrak{I}_i}$ for some $i \in I$ and some $d \in \Delta^{\mathfrak{I}_i}$ then $(\mathfrak{I}_i, d) \overset{<\omega}{\Longleftrightarrow} (\biguplus_{i\in I} \mathfrak{I}_i, d)$ implies $d \in C^{\biguplus_{i\in I} \mathfrak{I}_i}$ which entails, because $\biguplus_{i\in I} \mathfrak{I}_i \models \mathcal{T}$ with $C \sqsubseteq D \in \mathcal{T}$, automatically $d \in D^{\biguplus_{i\in I} \mathfrak{I}_i}$. The fact that $(\mathfrak{I}_i, d) \overset{<\omega}{\Longleftrightarrow} (\biguplus_{i\in I} \mathfrak{I}_i, d)$ implies $d \in D^{\mathfrak{I}_i}$. So $C^{\mathfrak{I}_i} \subseteq D^{\mathfrak{I}_i}$ which shows $\mathfrak{I}_i \models C \sqsubseteq D$.

□

We define $\mathsf{N_R}^* := \bigcup_{n<\omega} \mathsf{N_R}^n$ where $\mathsf{N_R}^0 := \{\varepsilon\}$ and

$$\mathsf{N_R}^{n+1} = (\mathsf{N_R} \cup \{r^- \mid r \in \mathsf{N_R}\}) \times \mathsf{N_R}^n$$

for all $n < \omega$. Intuitively, $\mathsf{N_R}^*$ contains all words (of finite length) over the alphabet $\mathsf{N_R} \cup \{r^- \mid r \in \mathsf{N_R}\}$. We speak of an $r^*$*-path*, if we mean a finite sequence of edges $(d_i, d_{i+1}) \in r_i$ for $i \in \{0, \ldots, n\}$ such that $r^*$ corresponds to the word $r_0 \cdots r_n$. For every $\mathcal{ALCQIO}$-concept $C$ and every $r^* \in \mathsf{N_R}^*$ we recursively define

$$\exists r^* C := \begin{cases} C & \text{if } r^* = \varepsilon \\ \exists s.\exists s^* C & \text{if } r^* = s \cdot s^*. \end{cases}$$

We use $\forall r^* C$ as abbreviation for $\neg(\exists r^* \neg C)$. Every unfolding of $\exists r^*$ leads to a finite or empty chain of quantifications and for every expression $\exists r^* C$ we obtain a proper $\mathcal{ALCQIO}$-concept if $C$ is a proper $\mathcal{ALCQIO}$-concept.

The following definition of generated subinterpretation was equivalently given on page 58 but shall be repeated for convenience. A generated subinterpretations in one element $d$ is exactly the connected component of $d$.

Definition 4.4.31 . Let $\tau$ be an arbitrary signature. $\mathfrak{K}$ is called *generated subinterpretation* of $\mathfrak{I}$ in $d \in \Delta^{\mathfrak{I}}$ if $\mathfrak{K} := \mathfrak{I} \upharpoonright \Delta^{\mathfrak{K}}$, and $\mathfrak{K}$ contains exactly those elements reachable on a finite path from $d$ (including $d$) in $\mathfrak{I}$. ◇

Definition 4.4.32 . Let $\tau$ be a signature and $\varphi \in \mathrm{FO}(\tau)$ a sentence. Then $\varphi$ is

1. invariant under $\overset{<\omega}{\Longleftrightarrow}^{\mathcal{Q}}$ in $\mathbb{K}$ if for any two interpretations $\mathfrak{I}, \mathfrak{H}$ in $\mathbb{K}$ we have
$$\mathfrak{I} \overset{<\omega}{\Longleftrightarrow}^{\mathcal{Q}} \mathfrak{H} \Longrightarrow \mathfrak{I} \models \varphi \text{ iff } \mathfrak{H} \models \varphi$$
2. *invariant under nominal disjoint unions* in $\mathbb{K}$ if the following holds: Let $(\mathfrak{I}_i)_{i\in I}$ be a family of $\tau$-interpretations in $\mathbb{K}$ such that $\mathfrak{I}_i \models_{\mathbb{K}} \varphi$ for all $i \in I$.

Let $\mathsf{N_I}(\mathfrak{K}) := \{a \in \mathsf{N_I} \mid a^{\mathfrak{K}} \neq \emptyset\}$ and $(\mathfrak{K}_i)_{i \in I}$ be a family of generated subinterpretations, where $\mathfrak{K}_i$ is a generated subinterpretation of some $d_i \in \Delta^{\mathfrak{I}_i}$ in $\mathfrak{I}_i$ for each $i \in I$ such that

(a) $\bigcup_{i \in I} \mathsf{N_I}(\mathfrak{K}_i) = \mathsf{N_I}$

(b) $\mathsf{N_I}(\mathfrak{K}_i) \cap \mathsf{N_I}(\mathfrak{K}_j) = \emptyset$ for all $i, j \in I$ with $i \neq j$.

then $\forall i \in I : \mathfrak{I}_i \models \varphi$ iff $\biguplus_{i \in I} \mathfrak{K}_i \models_{\mathbb{K}} \varphi$.

◇

In particular, every sentence $\varphi$ which is invariant under nominal disjoint unions, is preserved under under all subinterpretations of $\mathfrak{I}$ which contain all individuals and which are closed under $r$-predecessors and $r$-successors for $r \in \mathsf{N_R}$.

Before the characterisation theorem for $\mathcal{ALCQIO}$-TBoxes will be given, we shall make a remark about connected components. In the proof of the characterisation theorem, we shall claim that, if a set $\Gamma := \{\forall r^*.\neg a \mid a \in \mathsf{N_I} \text{ and } r^* \in \mathsf{N_R}^*\}$ is satisfiable with a set $p \subseteq \mathcal{ALCQIO}(\tau)$ and some FO-sentence $\varphi$ then there is a model $(\mathfrak{H}, e)$ such that no element in the connected component of $e$ satisfies any $a \in \mathsf{N_I}$.

But satisfiability only yields that no *finite* path $r^* \in \mathsf{N_R}^*$ emerging from $e$ can lead to an element which satisfies some $a \in \mathsf{N_I}$. The question is, whether there could be, in infinite distance, some element $e_0$ that satisfies some $a \in \mathsf{N_I}$: Take, e.g. the signature $\mathsf{N_I} := \{a\}$, $\mathsf{N_C} = \emptyset$ and $\mathsf{N_R} := \{<\}$ and the $\tau$-interpretation with $\Delta^{\mathfrak{H}} := \mathfrak{P}(\omega)$, the power-set of all natural numbers. Set $a^{\mathfrak{H}} := \{\omega\}$ and set $M < N$ iff there is $n < \omega$ such that $M \cup \{n\} = N$. Is $\emptyset \in \Delta^{\mathfrak{H}}$ connected to $\omega \in \Delta^{\mathfrak{H}}$?

Apart from the fact that paths are defined as finite sequences of edges and connected components are defined as all elements that are reachable in finitely many steps, $\omega$ and $\emptyset$ are not connected as the following observation will show in a more general setting:

Observation 4.4.33 . *Only elements in finite distance are connected*

Proof. Let $\mathfrak{H}$ be an arbitrary interpretation over some signature $\tau$ and set

$$E := \{(e_0, e_1) \in \Delta^{\mathfrak{H}} \times \Delta^{\mathfrak{H}} \mid \exists r^* \in \mathsf{N_R}^* : e_0 \text{ is connected to } e_1 \text{ by an } r^*\text{-path}\}$$

Note that $r^*$ describes a finite path. $E$ is an equivalence relation: it is reflexive as $\varepsilon \in \mathsf{N_R}^*$, it is symmetric since two elements connected by an $r^*$-path are connected an rev($r^*$)-path. Where $\mathrm{rev}(\varepsilon) := \varepsilon$, $\mathrm{rev}(r \cdot r^*) := \mathrm{rev}(r^*) \cdot r^-$ and $\mathrm{rev}(r^- \cdot r^*) := \mathrm{rev}(r^*) \cdot r$. Finally $E$ is transitive as $e_0$ is connected by a path $r_0^* \cdot r_1^*$ to $e_2$ whenever $e_0$ is connected by a path $r_0^*$ to $e_1$ and $e_1$ is connected by a path $r_1^*$ to $e_2$.

We construct the interpretation $\mathfrak{U}$: We set $\Delta^{\mathfrak{U}} := \Delta^{\mathfrak{H}}$ and for every $r \in \mathsf{N_R}$, we set $r^{\mathfrak{U}} := \bigcup_{e_0 \in \Delta^{\mathfrak{H}}} r^{\mathfrak{U}} \cap ([e_0] \times [e_0])$ and $S^{\mathfrak{U}} := \{e_0 \in \Delta^{\mathfrak{H}} \mid e_0 \in S^{\mathfrak{H}}\}$ for all $S \in \mathsf{N_C} \cup \mathsf{N_I}$. By definition of $r^{\mathfrak{U}}$, there cannot be an $r$-edge between two different classes.

We show that $\mathrm{id} : \mathfrak{H} \longrightarrow \mathfrak{U}$ is an isomorphism. Clearly, $r^{\mathfrak{U}} \subseteq \mathfrak{I}$ so it remains to show that $(e_0, e_1) \in r^{\mathfrak{H}} \implies (e_0, e_1) \in r^{\mathfrak{U}}$ for all $r \in \mathsf{N_R}$. If the former is true then $[e_0] = [e_1]$ by definition, and so $(e_0, e_1) \in r^{\mathfrak{H}} \cap ([e_0] \times [e_0])$, hence $(e_0, e_1) \in r^{\mathfrak{U}}$. □

Observation 4.4.33 shows that $\varnothing$ and $\omega$ are not connected. Even the transitive closure of $<$ from our example above does not change the fact that $\varnothing$ and $\omega$ are not connected, showing that the subset relation $\subseteq$ must properly comprise the transitive closure of $<$. It follows that if two elements are not connected by some finite path, they are not connected at all.

Theorem 4.4.34. *Let $\tau$ be an arbitrary signature and $\varphi \in \mathrm{FO}(\tau)$ a sentence such that $\varphi$ is*

1. *invariant under $\overset{\le_\kappa \mathcal{Q}}{\Longleftrightarrow}$ in $\mathbb{K}$*
2. *invariant under nominal disjoint unions of generated subinterpretations in $\mathbb{K}$*

Proof. Let $\tau_\varphi$ be the signature symbols used in $\varphi$ containing at least one $c \in \mathsf{N_I}$. If $\tau_\varphi$ does not contain such an element, we simply choose one from $\mathsf{N_I}$ and add it to $\tau_\varphi$. Let this $c \in \mathsf{N_I}$ be fixed for the rest of the proof.

$$\mathrm{cons}_{\tau_\varphi} := \{C \sqsubseteq D \mid C, D \in \mathcal{ALCQIO}(\tau_\varphi) \text{ s.t. } \varphi \models_{\mathbb{K}} C \sqsubseteq D\}$$

Assume $\mathrm{cons}_{\tau_\varphi} \varphi \not\models_{\mathbb{K}} \varphi$. Then there must be some $\mathfrak{H}$ in $\mathbb{K}$ model for $\mathrm{cons}\, \varphi \cup \{\neg\varphi\}$.

We shall first show that every $\mathcal{ALCQIO}$-type of an element in $\mathfrak{H}$, which is not located on a component which hosts an individual, can be realised in a model of $\varphi$ on a component which again has no individual: Let $\Gamma := \{\forall^{\ge 1} r^* \neg a \mid a \in \mathsf{N_I}\}$ and

$$P := \{p \subseteq \mathcal{ALCQIO}(\tau_\varphi) \mid p \cup \mathrm{cons}_{\tau_\varphi} \varphi \cup \Gamma \text{ satisfiable in } \mathbb{K}\}$$

Every $p \in P$ is satisfiable with $\Gamma \cup \{\varphi\}$ in $\mathbb{K}$: For otherwise the compactness of FO over $\mathbb{K}$ provides a finite subset $p_0 \subseteq p$ and a finite subset $\Gamma_0$ such that $p_0 \cup \Gamma_0 \cup \{\varphi\}$ is unsatisfiable in $\mathbb{K}$, yielding that for all models $\mathfrak{K}$ in $\mathbb{K}$ of $\varphi$ and all $d \in \Delta^{\mathfrak{K}}$ we have $(\mathfrak{K}, d) \models \neg(\bigsqcap p_0 \sqcap \bigsqcap \Gamma_0)$, which is true iff $\mathfrak{K} \models \bigsqcap p_0 \sqcap \bigsqcap \Gamma_0 \sqsubseteq \bot$. Since every model $\mathfrak{K}$ in $\mathbb{K}$ of $\varphi$ satisfies this concept inclusion, we have $\bigsqcap p_0 \sqcap \bigsqcap \Gamma_0 \sqsubseteq \bot \in \operatorname{cons}_{\tau_\varphi} \varphi$, a contradiction to the satisfiability of $p \cup \operatorname{cons}_{\tau_\varphi} \varphi \cup \Gamma$ in $\mathbb{K}$.

Let $\mathfrak{C}(e_p)$ be the connected component of $e$ in $\mathfrak{U}$ where $(\mathfrak{U}, e) \models p \cup \Gamma \cup \{\varphi\}$.

In the next step, we partition $\mathsf{N_I} \cap \tau_\varphi$ in a way that all individuals which occur in the same generated subinterpretation in $\mathfrak{H}$ are in the same class:

$$E := \{(a, b) \in (\tau_\varphi \cap \mathsf{N_I}) \times (\tau_\varphi \cap \mathsf{N_I}) \mid (\mathfrak{H}, a^{\mathfrak{H}}) \models \exists^{\geq 1} r^*.b\}$$

With the same arguments as in Observation 4.4.33, one can show that $E$ is an equivalence relation. Since $\tau_\varphi$ is finite $(\tau_\varphi \cap \mathsf{N_I})/E$ has finitely many classes, we can choose for each class one representative which we collect in the set $N$. Each type $\operatorname{Th}(\mathfrak{H}, a_i^{\mathfrak{H}})$ with $a \in N$ can be be satisfied in $\mathbb{K}$ with $\Gamma_a := \{\forall^{\geq 1} r^* \neg b \mid b \in (\tau_\varphi \cap \mathsf{N_I}) \setminus [a]\}$ and $\varphi$. Again, otherwise we could infer that $\varphi \models_{\mathbb{K}} \bigsqcap \Gamma'_a \sqcap T_0 \sqsubseteq \bot$, where $\Gamma'_a \subseteq \Gamma_a$ and $T_0 \subseteq \operatorname{Th}(\mathfrak{H}, a^{\mathfrak{H}})$, a contradiction to $\mathfrak{H} \models \operatorname{cons}_{\tau_\varphi} \varphi$.

Let $(\mathfrak{K}_a, a^{\mathfrak{K}_a})$ in $\mathbb{K}$ be a model of $\operatorname{cons}_{\tau_\varphi} \cup \operatorname{Th}(\mathfrak{H}, a^{\mathfrak{H}}) \cup \Gamma_a$ for all $a \in N$. With the same model-theoretic construction as above we can ensure that no individual from any other class than $[a^{\mathfrak{H}}]$ is on the same component as $a^{\mathfrak{K}_a}$.

Additionally, we may assume that all individual names $a \in \mathsf{N_I} \setminus \tau_\varphi$ are assigned to the individual $c^{\mathfrak{K}_a}$ we fixed above: If this is not the case, we can switch to the interpretation $\mathfrak{K}'_a$ which is $\mathfrak{K}_a$ where exactly $a^{\mathfrak{K}'_a} := c^{\mathfrak{K}_a}$ for all $a \in \mathsf{N_I} \setminus \tau_\varphi$. We have then $\mathfrak{K}_a \cong_{\tau_\varphi} \mathfrak{K}'_a$ and hence $(\mathfrak{K}'_a, a^{\mathfrak{K}'}) \models \operatorname{Th}(\mathfrak{H}, a^{\mathfrak{H}}) \cup \Gamma_a \cup \{\varphi\}$.

Let $\mathfrak{C}(a)$ be the connected component of $a$ in $\mathfrak{K}_a$ for all $a \in N$ and let $\mathfrak{C}(e_p)$ be the connected component for each $p \in P$ as discussed above. We know for

$$\mathfrak{K} := \biguplus_{p \in P} \mathfrak{C}(e_p) \uplus \biguplus_{a \in N} \mathfrak{C}(a)$$

that $\mathfrak{K}$ in $\mathbb{K}$ and since $\varphi$ is invariant under nominal disjoint unions we know that $\mathfrak{K} \models \varphi$.

We set $\mathfrak{I} := \mathfrak{H} \uplus \biguplus_{p \in P} \mathfrak{C}(e_p)$ and infer that $\mathfrak{I}$ in $\mathbb{K}$. Since $\mathfrak{H} \not\models_{\mathbb{K}} \varphi$ and $\varphi$ is invariant under nominal disjoint unions we have $\mathfrak{I} \models_{\mathbb{K}} \neg\varphi$.

Let $\mathfrak{I}^*$ be the $\omega$-saturated extension of $\mathfrak{I} \restriction (\tau_\varphi \cup \mathsf{N_I})$ and likewise $\mathfrak{K}^*$ the one for $\mathfrak{K} \restriction (\tau_\varphi \cup \mathsf{N_I})$. I.e. we are looking at $\tau$-interpretations in $\mathbb{K}$ where $S^{\mathfrak{I}^*} = \emptyset$ for all symbols $S \in (\mathsf{N_C} \cup \mathsf{N_R}) \setminus \tau_\varphi$ and likewise for $\mathfrak{K}^*$. We still have $\mathfrak{I}^* \models_{\mathbb{K}} \varphi$ and $\mathfrak{K}^* \models_{\mathbb{K}} \neg\varphi$

We show $\mathfrak{K}^* \xleftrightarrow{\leq\kappa}{}^{Q} \mathfrak{J}^*$. Let $d \in \mathfrak{K}^*$. Then each finite subset $\Gamma_0 \subseteq \mathrm{Th}(\mathfrak{K}^*, d)$ was realised at some $d_{\Gamma_0} \in \Delta^{\mathfrak{K}}$. Note that $\mathrm{Th}(\mathfrak{K}^*, d) \subseteq \mathcal{ALCQIO}(\tau_\varphi \cup \mathsf{N_I})$. In case $d_{\Gamma_0}$ finds itself on one of the connected components $\mathfrak{C}(d_p)$ with $p \in P$ then $d_{\Gamma_0} \in \Delta^{\mathfrak{J}}$ and so $\mathfrak{J} \models \exists^{\geq 1} u. \bigsqcap \Gamma_0$.

Otherwise there is some $a \in N$ and some path $r^*$ such that $(\mathfrak{K}, a^{\mathfrak{K}}) \models \exists^{\geq 1} r^*. \bigsqcap \Gamma_0$. Since $(\mathfrak{K}, a^{\mathfrak{K}}) \models \mathrm{Th}(\mathfrak{H}\restriction(\tau_\varphi \cup \mathsf{N_I}), a^{\mathfrak{H}})$, we also know that $(\mathfrak{H}, a^{\mathfrak{H}}) \models \exists^{\geq 1} r^*. \bigsqcap \Gamma_0$. This shows that $\mathfrak{J} \models \exists^{\geq 1} u. \bigsqcap \Gamma_0$.

Since both interpretations $\mathfrak{K}^*$ and $\mathfrak{J}^*$ are in particular $\mathcal{ALCQIO}u_1$-saturated, there must be $d' \in \Delta^{\mathfrak{J}^*}$ with $(\mathfrak{J}^*, d') \models \mathrm{Th}(\mathfrak{K}^*, d)$. The Hennessy-Milner-Property for $\mathcal{ALCQIO}u_1$ shows that $(\mathfrak{K}^*, d) \xleftrightarrow{<\omega}{}^{Q} (\mathfrak{J}^*, d')$.

In the same way, we can show that for every $d \in \Delta^{\mathfrak{J}^*}$ there is $d' \in \Delta^{\mathfrak{K}^*}$ with $(\mathfrak{J}^*, d) \xleftrightarrow{<\omega}{}^{Q} (\mathfrak{K}^*, d')$, which proves $\mathfrak{K}^* \xleftrightarrow{\leq\kappa}{}^{Q} \mathfrak{J}^*$. But now $\mathfrak{K}^* \models_{\mathbb{K}} \varphi$ and $\mathfrak{J} \models_{\mathbb{K}} \neg\varphi$. A contradiction which proves that $\mathrm{cons}_{\tau_\varphi} \models \varphi$. □

We have seen that $\mathcal{ALCQIO}$-TBoxes could be characterised by invariance under global $\mathcal{ALCQIO}$-bisimulation and a pick-and-mix procedure of connected components so that a new interpretation in $\mathbb{K}$ is obtained. This is considerably less technical than in the case for $\mathcal{ALCQO}$.

Surely, the spy-point technique comes to mind when thinking about the question as to why the characterisation of $\mathcal{ALCQIO}$-TBoxes is easier than $\mathcal{ALCQO}$-TBoxes. The spy-point technique is used (e.g. [19, 2, 50]) in order to 'internalise' TBoxes [121]: Reasoning techniques for $\mathcal{ALCQIO}$-concepts say, can be reused for $\mathcal{ALCQIO}$-TBoxes by rewriting $\mathcal{ALCQIO}$-concept subsumptions $C \sqsubseteq D$ as $\mathcal{ALCQIO}$-concepts:

Let $\tau$ be a signature and let $a$ be a fresh individual for $\tau$ the *the spy-point translation* $\cdot^{\mathrm{spy}}$ is in [128] recursively defined on $\mathcal{ALCQIO}$-concepts as follows:

$$C^{\mathrm{spy}} := \begin{cases} C & \text{if } C \in \mathsf{N_I} \cup \mathsf{N_C} \\ D^{\mathrm{spy}} \sqcap E^{\mathrm{spy}} & \text{if } C = D \sqcap E \\ \neg(D^{\mathrm{spy}}) & \text{if } C = \neg D \\ \exists^{\geq\kappa} r.D^{\mathrm{spy}} \sqcap \exists^{\geq 1}\mathrm{spy}^-.a & \text{if } C = \exists^{\geq\kappa} r.D \end{cases}$$

If we assume w.l.o.g. that the $\mathcal{ALCQIO}$-TBox contains only one concept subsumption of the form $\top \sqsubseteq D$. The spy-point translation $\mathcal{T}^{\mathrm{spy}}$ of $\mathcal{T}$ is then the $\mathcal{ALCQIO}$-concept $a \sqcap D^{\mathrm{spy}} \sqcap \forall \mathrm{spy}.D^{\mathrm{spy}}$. It turns out that the TBox and its spy-point translation are equi-satisfiable (cf. [128, Lemma 5.24]), i.e. $\mathcal{T}$ has a model iff $\mathcal{T}^{\mathrm{spy}}$ has a model. A model $\mathfrak{H}$ for $\mathcal{T}$ can be extracted from the model $(\mathfrak{J}, a^{\mathfrak{J}}) \models \mathcal{T}^{\mathrm{spy}}$ by restricting $\mathfrak{I}$ to only those elements which are either $a^{\mathfrak{I}}$ itself or connected to $a^{\mathfrak{I}}$ via a spy-edge.

The argument therefore would be that $\mathcal{ALCQIO}$ can, via this detour, internalise $\mathcal{ALCQIO}$-TBoxes which accounts for the relatively moderate increase of expressive power which expresses itself in the transparent notion of nominal disjoint union used for the characterisation of $\mathcal{ALCQIO}$-TBoxes.

But the spy-point technique makes heavily use of signature extensions which are not in the spirit of our model theoretic characterisation where signatures are sacrosanct. In a setting of strict invariant signatures, it is the inverse property which yields the explanation: The $\mathcal{ALCQIO}$-theory of a point describes the whole connected component. In this sense we do not have forward generated subinterpretations like in $\mathcal{ALCQO}$ but always fully connected components. These fully connected components reduce disjoint unions to pick and mix procedures and make complicated unions of interpretations obsolete.

Although $\mathcal{ALCQIO}u_1$ is close to $\mathcal{SROIQ}$, which allows TBoxes to contain concepts with universal operator, it does not make sense to give a separate characterisation, as these $\mathcal{ALCQIO}u_1$-TBoxes capture the whole expressivity of the global concepts in $\mathcal{ALCQIO}u_1$.

# 5. The characterisation of $\mathcal{EL}$

## 5.1 $\mathcal{EL}$

$\mathcal{EL}$ emerged, as mentioned in the introduction Chapter 1, from the formalisation of large medical ontologies like SNOMED CT. It is considered to be a sub-boolean description logic as it is not closed under all boolean operators $\sqcap, \sqcup, \neg$ but only under $\sqcap$. Its expressiveness is compared with $\mathcal{ALC}$ and further extensions which we treated in preceding chapters relatively weak. As benefit from this sacrifice it has good complexity (PTIME-complete) [66, 9, 26, 10, 12] when it comes to reasoning. Since reasoning under $\mathcal{EL}$ is tractable, the World Wide Web Committee (W3C) has specified the OWL 2 $\mathcal{EL}$ profile [37], which makes $\mathcal{EL}$ particularly interesting.

### 5.1.1 Syntax and Semantics

As mentioned $\mathcal{EL}$ does not contain negations and disjunctions. Since we consider disjunctions to be syntactical abbreviations, its syntax is, in the style given for DLs in the previous chapters, merely $\mathcal{ALC}$ without negation: $C$ is an *$\mathcal{EL}$-concept* over the signature $\tau$ if

$$C ::= \bot \mid \top \mid A \mid D \sqcap E \mid \exists r.D$$

where $A \in \mathsf{N_C}$, $r \in \mathsf{N_R}$ and $D, E$ are $\mathcal{EL}$-concepts over $\tau$. Like for $\mathcal{ALC}$, we consider $\mathsf{N_I} = \emptyset$. We denote with $\mathcal{EL}(\tau)$ the set of all $\mathcal{EL}$-concepts over $\tau$ including the atomic concept $\bot$. Note that $\bot$ does not allow the formulation of any kind of negation but simply states inconsistency and is never true.

As $\mathcal{EL}$ is a fragment of $\mathcal{ALC}$, we simply apply the interpretation for $\mathcal{ALC}$-concepts to $\mathcal{EL}$-concepts and similarly we can restrict the rank-function for $\mathcal{ALC}$-concepts (page 35) to $\mathcal{EL}$-concepts.

### 5.1.2 $\mathcal{EL}$-Simulation

As our aim is to find a characterisation for $\mathcal{EL}$, we need to find a model theoretic notion which allows, restricted to finite signatures and finite number of rounds, to be described by characteristic concepts, as done with all description logics previously mentioned.

Since we do not have negations and in particular no $\forall$-quantifications, there is no possibility to describe a symmetrical notion like bisimulation for $\mathcal{ALC}$. The notion in hand is simply *simulation*. We shall explain the appropriate game, at first not restricted to finite rounds or finite signature. Let therefore $\tau$ be arbitrary.

Definition 5.1.1 (Simulation-Game). Let $(\mathfrak{I}, d)$ and $(\mathfrak{H}, e)$ be $\tau$ interpretations. The game $G(\mathfrak{I}, d; \mathfrak{H}, e)$ is played by two players named **I** and **II**

The start-configuration of $G(\mathfrak{I}, d; \mathfrak{H}, e)$ is $(\mathfrak{I}, d; \mathfrak{H}, e)$. For each configuration $(\mathfrak{I}, d_0; \mathfrak{H}, e_0)$, **I** chooses some $r \in \mathsf{N_R}$ and an $r$-successor $d_1$ of $d_0$ in $\mathfrak{I}$. If **I** cannot make such a move, **II** wins the game. If **II** is able to respond with an $r$-successor $e_1$ of $e_0$ such that

$$\forall A \in \mathsf{N_I} : d_1 \in A^{\mathfrak{I}} \Longrightarrow e_1 \in A^{\mathfrak{H}}$$

holds, a new configuration $(\mathfrak{I}, d_1; \mathfrak{H}, e_1)$ of $G(\mathfrak{I}, d; \mathfrak{H}, e)$ is reached; **II** looses otherwise. The start-configuration may be considered as the outcome of the 0-th round and $d \in A^{\mathfrak{I}} \Longrightarrow e \in A^{\mathfrak{H}}$ must hold for all $a \in \mathsf{N_C}$ too, or **II** has already lost the game. ◇

Definition 5.1.2 (Winning-strategy). **II** has a *winning-strategy* in $G(\mathfrak{I}, d; \mathfrak{H}, e)$ or $(\mathfrak{I}, d) \Longrightarrow (\mathfrak{H}, e)$, if she finds a response in $G(\mathfrak{I}, d; \mathfrak{H}, e)$ for any challenge of **I** and does not loose any round. ◇

Note that $(\mathfrak{I}, d) \Longrightarrow (\mathfrak{H}, e)$ and $(\mathfrak{H}, e) \Longrightarrow (\mathfrak{I}, d)$ does, in general, *not* imply $(\mathfrak{I}, d) \Longleftrightarrow (\mathfrak{H}, e)$. The reason for this is that simulation cannot capture negative information about concepts names.

Example 5.1.3. Take $\Delta^{\mathfrak{I}} := \{d, b, c\}$ with $r^{\mathfrak{I}} := \{(d, b), (d, c)\}$ and $A := \{c, b\}$ and $B := \{b\}$, whilst $\Delta^{\mathfrak{H}} := \{e, g\}$ $r^{\mathfrak{I}} := \{(e, g)\}$ and $B, A := \{g\}$. Clearly $(\mathfrak{I}, d) \Longrightarrow (\mathfrak{H}, e)$ and $(\mathfrak{H}, e) \Longrightarrow (\mathfrak{I}, d)$, but $(\mathfrak{I}, d)$ is not bisimilar to $(\mathfrak{H}, e)$: **I** challenging **II** in the bisimulation game $G(\mathfrak{I}, d; \mathfrak{H}, e)$ by moving from $d$ to $c$ cannot be responded by **II** in $\mathfrak{H}$ as $g$ does not satisfy $\neg B$.

Bisimulation follows if one requires **II** to respond to challenges of **I** throughout the game with *equisimilar* elements. I.e. for any configuration $(\mathfrak{I}, d'; \mathfrak{H}, e')$ in the game $G(\mathfrak{I}, d; \mathfrak{H}, e)$ we require $(\mathfrak{I}, d') \Longrightarrow (\mathfrak{H}, e')$ and $(\mathfrak{H}, e') \Longrightarrow (\mathfrak{I}, d')$. The notion of equisimilarity shall be elaborated further down.

The fact that $(\mathfrak{I}, d) \Longrightarrow (\mathfrak{H}, e)$ and $(\mathfrak{H}, e) \Longrightarrow (\mathfrak{I}, d)$ does not imply bisimulation, has been tried to be remedied in [85] by requiring atomic equivalence for configurations occurring during the game. This notion is clearly too strong for $\mathcal{EL}$ since, as mentioned just now, $\mathcal{EL}$ cannot express negated concept names. Hence this notion of simulation cannot be captured by characteristic $\mathcal{EL}$-concepts in the way they will be introduced further down.

With a simple proof by induction, one can show that every concept $C \in \mathcal{EL}(\tau)$ is preserved under simulation. The crucial point is here that $C$ is only preserved, i.e. if $(\mathfrak{I}, d) \Longrightarrow (\mathfrak{H}, e)$ then $(\mathfrak{I}, d) \models C \Longrightarrow (\mathfrak{H}, e) \models C$. The converse $(\mathfrak{H}, e) \models C$ implies $(\mathfrak{I}, d) \models C$ cannot be expected.

Definition 5.1.4 . **II** has a winning strategy in the $n$-round game $G_n(\mathfrak{I}, d; \mathfrak{H}, e)$ if she can ward off any challenges of **I** during the first $n$ rounds. We denote this by $(\mathfrak{I}, d) \Longrightarrow_n (\mathfrak{K}, e)$ ◇

For every signature $\tau$ the relation $\Longrightarrow_n$ is a reflexive and transitive relation on the set of all pointed $\tau$-interpretations. For all $n < \omega$ we have $(\mathfrak{I}, d) \Longrightarrow_{n+1} (\mathfrak{H}, e)$ implies $(\mathfrak{I}, d) \Longrightarrow_n (\mathfrak{H}, e)$ and $(\mathfrak{I}, d) \Longrightarrow (\mathfrak{H}, e)$ implies $(\mathfrak{I}, d) \Longrightarrow_n (\mathfrak{H}, e)$. A proof by induction shows for all $C \in \mathcal{EL}(\tau)$ with rank $C \leq n$

$$(\mathfrak{I}, d) \Longrightarrow_n (\mathfrak{H}, e) \Longrightarrow (\mathfrak{I}, d) \models C \Longrightarrow (\mathfrak{H}, e) \models C$$

### Characteristic $\mathcal{EL}$-Concepts

Let, throughout this section, $\tau$ be finite and $\mathfrak{I}$ a $\tau$-interpretation. For each level $n < \omega$ we define the *characteristic $\mathcal{EL}$-concept* for $(\mathfrak{I}, d)$ on level $n$ recursively as follows

$$\begin{aligned} X^0_{\mathfrak{I},d} &:= \textstyle\bigsqcap\{A \in \mathsf{N_C} \mid d \in A^{\mathfrak{I}}\} \\ X^{n+1}_{\mathfrak{I},d} &:= X^0_{\mathfrak{I},d} \sqcap \textstyle\bigsqcap_{r \in \mathsf{N_R}} \bigsqcap\{\exists r.X^n_{\mathfrak{I},d'} \mid (d, d') \in r^{\mathfrak{I}}\} \end{aligned}$$

Proposition 5.1.5 . *For all pointed $\tau$-interpretations $(\mathfrak{I}, d)$, $(\mathfrak{H}, e)$ the following statements are equivalent:*

1. $(\mathfrak{H}, e) \models X^n_{\mathfrak{I},d}$
2. $(\mathfrak{I}, d) \Longrightarrow_n (\mathfrak{H}, e)$
3. $(\mathfrak{H}, e) \models Th_n(\mathfrak{I}, d)$

*where* $Th_n(\mathfrak{I}, d) := \{C \in \mathcal{EL}(\tau) \mid d \in C^{\mathfrak{I}}$ *and* $\operatorname{rank} C \leq n\}$

Proof. Since '2.$\Longrightarrow$ 3.' follows from the observation made directly before this section about characteristic $\mathcal{EL}$-concepts and '3.$\Longrightarrow$1.' follows from the fact that $X^n_{\mathfrak{I},d}$ has rank $\leq n$, we settle on '1.$\Longrightarrow$2.'

If $n = 0$ and $(\mathfrak{H}, e) \models X^0_{\mathfrak{I},d}$ then $e$ satisfies all concept names satisfied by $d$ and so **II** wins the game $G_0(\mathfrak{I}, d; \mathfrak{H}, e)$. In the step case, we assume that **I** chooses $r \in \mathsf{N_R}$ and challenges **II** in the start-configuration in $\mathfrak{I}$ with some $r$-successor $d'$ of $d$. Then $\exists r.X^n_{\mathfrak{I},d'}$ is a conjunct of $X^{n+1}_{\mathfrak{I},d}$ and since $(\mathfrak{H}, e) \models X^{n+1}_{\mathfrak{I},d}$ we obtain that there must be an $r$-successor $e'$ of $e$ such that $(\mathfrak{H}, e') \models X^n_{\mathfrak{I},d'}$.

This $r$-successor $e'$ satisfies the same concept names as $d'$ and also satisfies $X^n_{\mathfrak{I},d'}$. The induction hypothesis yields a winning strategy for **II** in $G_n(\mathfrak{I}, d'; \mathfrak{H}, e')$. As the challenge of **I** was arbitrary, it shows that **II** has a winning strategy in the $n+1$-round game. □

### $\mathcal{EL}$-Saturation and the Hennessy-Milner-Property for $\mathcal{EL}$

We do not impose the requirement that $\tau$ must be finite any more. We shall give an appropriate notion of type and saturation for $\mathcal{EL}$

Definition 5.1.6 . Let $\tau$ be a signature and let $\mathfrak{I}$ be $\tau$-Interpretation with $d \in \Delta^{\mathfrak{I}}$.

1. A set of $\Gamma \subseteq \mathcal{EL}(\tau)$ is called $\exists r$*-type* of $d$ in $\mathfrak{I}$ if for all finite subsets $\Gamma_0 \subseteq \Gamma$ there is an $r$-successor $d'$ of $d$ such that $d' \in (\bigsqcap \Gamma_0)^{\mathfrak{I}}$.
2. An $\exists r$-type $\Gamma$ is *realised* at $d$ if there is an $r$-successor $d'$ of $d$ such that $(\mathfrak{I}, d') \models \Gamma$.
3. An interpretation is $\mathcal{EL}$-saturated if for all $d \in \Delta^{\mathfrak{I}}$ and all $r \in \mathsf{N_R}$ every $\exists r$-type of $d$ is realised at $d$.

◊

Observation 5.1.7 . *Every $\omega$-saturated interpretation is $\mathcal{EL}$-saturated.*

Proposition 5.1.8 . *$\mathcal{EL}$ has the Hennessy-Milner-Property, i.e. if $\mathfrak{H}$ is $\mathcal{EL}$-saturated then $Th(\mathfrak{I}, d) \subseteq Th(\mathfrak{H}, e)$ iff $(\mathfrak{I}, d) \Longrightarrow (\mathfrak{H}, e)$*

It is a question whether this property should be called Hennessy-Milner-Property as it has this asymmetric appearance. Though it will turn out that we can use this proposition to prove the Hennessy-Milner-Property for the equisimilar invariant fragment of FO. In comparison to the Hennessy-Milner-Theroem 2.1.23, we just make sure that all arguments go through in the asymmetric case.

Proof. We show that **II** has a winning strategy in the simulation game $(\mathfrak{I}, d; \mathfrak{H}, e)$ iff she maintains configurations $(\mathfrak{I}, d'; \mathfrak{H}, e')$ such that $\mathrm{Th}(\mathfrak{I}, d') \subseteq \mathrm{Th}(\mathfrak{H}, e')$. In particular all these configurations are valid as every $A \in \mathsf{N}_\mathsf{I}$ which is satisfied at $d'$ is also satisfied at $e'$. This start-configuration meets the requirement.

Assume, the game has reached the configuration $(\mathfrak{I}, d_0; \mathfrak{H}, e_0)$ with $\mathrm{Th}(\mathfrak{I}, d_0) \subseteq \mathrm{Th}(\mathfrak{H}, e_0)$ and **I** challenges **II** by choosing some $r \in \mathsf{N}_\mathsf{R}$ and moving to an $r$-successor $d_1$ of $d_0$ in $\mathfrak{I}$. For each finite set $\Gamma_0 \subseteq \mathrm{Th}(\mathfrak{I}, d_1)$ we have $(\mathfrak{I}, d_0) \models \exists r. \bigsqcap \Gamma_0$ and since $\mathrm{Th}(\mathfrak{I}, d_0) \subseteq \mathrm{Th}(\mathfrak{H}, e_0)$, $\mathrm{Th}(\mathfrak{I}, d_1)$ forms an $\exists r$-type of $(\mathfrak{H}, e_0)$. As $\mathfrak{H}$ is $\mathcal{EL}$-saturated, this type is realised at $e_0$, i.e. there is some $r$-successor of $e_1$ such that $(\mathfrak{H}, e_1) \models \mathrm{Th}(\mathfrak{I}, d_1)$. The configuration $(\mathfrak{I}, d_1; \mathfrak{H}, e_1)$ meets the requirement. This proves that **II** has a winning strategy in the simulation game. □

### 5.1.3 The $\mathcal{EL}^{\sqcup}$ Extension of $\mathcal{EL}$

The notion of simulation as consequent development from the notion of bisimulation, cannot prevent disjunctions from being preserved: Let $C, D$ be $\mathcal{EL}$-concepts over some signature $\tau$ and assume the pointed $\tau$-interpretation $(\mathfrak{I}, d) \models C$ then $(\mathfrak{I}, d) \models C$ *or* $(\mathfrak{I}, d) \models D$. For any pointed $\tau$-interpretation $(\mathfrak{H}, e)$ with $(\mathfrak{I}, d) \Longrightarrow (\mathfrak{H}, e)$ we obtain that $(\mathfrak{H}, e) \models C$ *or* $(\mathfrak{H}, e) \models D$ since $C$ is preserved under simulation.

It shows that indeed simulation is too strong for $\mathcal{EL}$. We therefore introduce the following mild extension $\mathcal{EL}^{\sqcup}$ of $\mathcal{EL}$ which is for any signature $\tau$ recursively defined by

$$C ::= \bot \mid \top \mid A \mid D \sqcap E \mid D \sqcup E \mid \exists r.D$$

where $A \in \mathsf{N}_\mathsf{C}$, $r \in \mathsf{N}_\mathsf{R}$, and $D, E$ are $\mathcal{EL}^{\sqcup}$-concepts over $\tau$. We define $\mathcal{EL}^{\sqcup}(\tau)$ as the set of all $\mathcal{EL}^{\sqcup}$ concepts over $\tau$ together with $\bot$ as atomic concept.

$\mathcal{EL}^{\sqcup}$ is simply the closure of $\mathcal{EL}$ over $\sqcup$ as disjunctions commute with existential quantifications: $\exists r.(C \sqcup D) \equiv \exists r.C \sqcup \exists r.D$. We can define the semantic for $\mathcal{EL}^{\sqcup}$-concepts by restricting the $\mathcal{ALC}$ interpretation to $\mathcal{EL}^{\sqcup}$. The $\mathcal{ALC}$ semantic will however interpret $C \sqcup D$ as abbreviation for $\neg(\neg C \sqcap \neg D)$ which does not change its semantic meaning and the fact that it is negation free. Similarly we can restrict the rank function for $\mathcal{ALC}$ to $\mathcal{EL}^{\sqcup}$.

A proof by induction shows that $\mathcal{EL}^{\sqcup}$ is preserved under simulation. Since $\mathcal{EL}(\tau) \subseteq \mathcal{EL}^{\sqcup}(\tau)$ for every signature $\tau$ we obtain immediately

1. characteristic concepts for $\mathcal{EL}^{\sqcup}$ which are simply the characteristic $\mathcal{EL}$-concepts
2. that Proposition 5.1.5 transfers immediately to $\mathcal{EL}^{\sqcup}$ with $\mathrm{Th}_n(\mathfrak{I}, d)$ redefined as $\{C \in \mathcal{EL}^{\sqcup} \mid d \in C^{\mathfrak{I}}$ and rank $C \leq n\}$
3. that $\mathcal{EL}^{\sqcup}$ has, using the Definition 5.1.6 for $\mathcal{EL}$-types and $\mathcal{EL}$-saturation, the Hennessy-Milner-Property (c.f. Proposition 5.1.8).

We can thus promptly give the characterisation theorem for $\mathcal{EL}^{\sqcup}$

Theorem 5.1.9. *For every signature $\tau$ and every* FO*-formula $\varphi(x)$ over $\tau$ which is preserved under $\Longrightarrow$ there is a logically equivalent $\mathcal{EL}^{\sqcup}$-concept over $\tau$.*

Proof. We may w.l.o.g. assume that $\tau$ contains exactly the signature symbols occurring in $\varphi$ and is therefore finite. Any FO-formula which is preserved under $\Longrightarrow$ is for some $n < \omega$ already preserved under $\Longrightarrow_n$: In particular $\varphi$ is invariant under $\Longleftrightarrow$ and hence $\varphi$ is logically equivalent to some $\mathcal{ALC}$-concept $C_{\mathcal{ALC}}$ over $\tau$. Let $n :=$ rank $C_{\mathcal{ALC}}$ and assume that $(\mathfrak{I}, d) \Longrightarrow_n (\mathfrak{H}, e)$. For the tree-unravellings $\mathfrak{I}^T_d$ of $\mathfrak{I}$ in $d$ and analogously $\mathfrak{H}^T_e$ of $\mathfrak{H}$ in $e$ we have

$$(\mathfrak{I}, d) \Longleftrightarrow (\mathfrak{I}^T_d, d) \Longleftrightarrow_n (\mathfrak{I}^T_d{\restriction}n, d) \Longrightarrow (\mathfrak{H}^T_e{\restriction}n, e) \Longleftrightarrow_n (\mathfrak{H}^T_e, e) \Longleftrightarrow (\mathfrak{H}, e),$$

where $\mathfrak{I}^T_d{\restriction}n$ is the restriction of $\mathfrak{I}^T_d$ to those elements which are reachable within $n$-steps from $d$ and likewise for $\mathfrak{H}^T_e{\restriction}n$.

We set $C_\varphi := \bigsqcup\{X^n_{\mathfrak{I},d} \mid (\mathfrak{I}, d) \models \varphi(x)\}$ where $X^n_{\mathfrak{I},d}$ is the characteristic $\mathcal{EL}$-concept over $\tau$. The disjunction $C_\varphi$ is up to logical equivalence finite and so $C_\varphi \in \mathcal{EL}^{\sqcup}(\tau)$. Since trivially $\varphi(x) \models C_\varphi$ it remains to show that $C_\varphi \models \varphi(x)$.

Assume $(\mathfrak{H}, e) \models C_\varphi$. Then there is a model $(\mathfrak{I}, d)$ of $\varphi(x)$ such that $X^n_{\mathfrak{I},d}$ is a disjunct of $C_\varphi$ and $(\mathfrak{H}, e) \models X^n_{\mathfrak{I},d}$. Proposition 5.1.5 yields $(\mathfrak{I}, d) \Longrightarrow_n (\mathfrak{H}, e)$. And since $\varphi$ was preserved by $\Longrightarrow_n$, it follows that $(\mathfrak{H}, e) \models \varphi(x)$. □

### 5.1.4 Characterisation of $\mathcal{EL}$ as FO-Fragment

As shown in Theorem 5.1.9, $\mathcal{EL}^{\sqcup}$ is the fragment preserved by simulation. The question arises, which model theoretic property distinguishes $\mathcal{EL}$ from $\mathcal{EL}^{\sqcup}$? Note that disjoint unions, which were used to single out TBoxes from the boolean closure over TBoxes, cannot be used: The validity of a non-global concept at some element completely depends on the connected neighbourhood of this element which is not affected by disjoint unions. Hence we need a notion which bars disjunctions from $\mathcal{EL}$-concepts, namely the minimal model property used in [91].

EXAMPLE 5.1.10. Disjoint unions are enough to exclude $\sqcup$ 'between' TBoxes: Assume we had a disjunction of two TBoxes $\{\top \sqsubseteq A\} \sqcup \{\top \sqsubseteq B\}$ which is logically equivalent to the $\mathcal{ALC}u$-concept $\forall u.(\top \to A) \sqcup \forall u.(\top \to B)$. Note that this is *not* an $\mathcal{EL}^{\sqcup}$-TBox but a TBox from the boolean closure over all $\mathcal{EL}$-TBoxes.

Then this disjunction of TBoxes is not preserved under disjoint unions: Let $\Delta^{\mathfrak{I}} = \{d\}$ with $A^{\mathfrak{I}} = \{d\}$ and $B^{\mathfrak{I}} = \emptyset$ and let $\Delta^{\mathfrak{H}} = \{e\}$ with $A^{\mathfrak{H}} = \emptyset$ and $B^{\mathfrak{I}} = \{e\}$. Then all elements in $\mathfrak{I}$ satisfy $A$ and so $\mathfrak{I} \models \{\top \sqsubseteq A\} \sqcup \{\top \sqsubseteq B\}$. Similarly, all elements in $\mathfrak{H}$ satisfy $B$ which entails $\mathfrak{H} \models \{\top \sqsubseteq A\} \sqcup \{\top \sqsubseteq B\}$. However the disjoint union of the two satisfies neither disjunct: $A^{\mathfrak{I} \uplus \mathfrak{H}} = \{d\}$ and $B^{\mathfrak{I} \uplus \mathfrak{H}} = \{e\}$.

However, the disjoint union is not enough to exclude disjunctions from *concepts*. Let $(\mathfrak{I}, d), (\mathfrak{H}, e)$ be models of the $\mathcal{EL}^{\sqcup}$-concept $A \sqcup B$. Following our remark on page 37, $\mathcal{ALC}$-concepts and in particular $\mathcal{EL}$-concepts are not affected by the disjoint union. Hence $(\mathfrak{I} \uplus \mathfrak{H}, d) \models A \sqcup B$ and $(\mathfrak{I} \uplus \mathfrak{H}, e) \models A \sqcup B$.

DEFINITION 5.1.11. A formula $\varphi$ over vocabulary $\tau$ has a *minimal model* w.r.t. some logic $\mathcal{L}$ if there is a $\tau$-interpretation $\mathfrak{I}_\varphi$ such that for all $\mathcal{L}$-formulae $\psi$ over $\tau$ we have $\mathfrak{I}_\varphi \models \psi$ iff $\varphi \models \psi$. ◇

If a formula has a minimal model it can only be equivalent to 'trivial' disjunctions. To illustrate this, let $\varphi \in \mathcal{L}(\tau)$ some formula that has a minimal model $\mathfrak{I}_\varphi$ w.r.t. $\mathcal{L}$ which would be equivalent to a disjunction $\bigsqcup_{i \in I} \psi_i$ of formulae $\psi_i \in \mathcal{L}(\tau)$. Then $\mathfrak{I}_\varphi \models \psi_i$ for some $i \in I$ and so $\varphi \models \psi_i$. We show $\psi_j \models \psi_i$ for all $j \in I$: Let $j \in I$ be arbitrary and assume for some $\mathfrak{H} \models \psi_j$ then of course $\mathfrak{H} \models \bigsqcup_{i \in I} \psi_i$ which is equivalent to $\mathfrak{H} \models \varphi$. Hence $\mathfrak{H} \models \psi_i$. This shows $\psi_j \models \psi_i$. In case $j \neq i$ one can remove $\psi_j$ from the disjunction and obtain a logically equivalent disjunction. Since this is true for all $j \neq i$ in $I$, we have $\varphi$ is logically equivalent to $\bigsqcup_{i \in I} \psi_i$ which is, in turn,logically equivalent to $\psi_i$.

DEFINITION 5.1.12. Let $(\mathfrak{I}_i)_{i \in I}$ be a family of $\tau$-interpretations. Then $\bigtimes_{i \in I} \mathfrak{I}_i$ is defined by

1. $\Delta^{\times \mathfrak{I}_i} := \bigtimes_{i \in I} \Delta^{\mathfrak{I}_i}$
2. $A^{\times \mathfrak{I}_i} := \{\bar{d} \in \Delta^{\times \mathfrak{I}_i} \mid \forall i \in I : \bar{d}_i \in A^{\mathfrak{I}_i}\}$
3. $r^{\times \mathfrak{I}_i} := \{(\bar{d}, \bar{d}') \in \Delta^{\times \mathfrak{I}_i} \times \Delta^{\times \mathfrak{I}_i} \mid \forall i \in I : (\bar{d}_i, \bar{d}_{i'}) \in r^{\mathfrak{I}_i}\}$

where $\bar{d}_i$ denotes the $i$-th component of $d$. ◇

We do not restrict the index set $I$ to be finite but define the product for arbitrary families $(\mathfrak{I}_i)_{i \in I}$. In case $i \in I$ is infinite, we can interpret $\Delta^{\bigtimes \mathfrak{I}_i}$ as set of all functions $\bar{d} : I \longrightarrow \biguplus \Delta^{\mathfrak{I}_i}$ such that $\bar{d}_i \in \Delta^{\mathfrak{I}_i}$. Also note that the general assumption 'any product of a non-empty family of non-empty sets is not empty' requires the axiom of choice.

Apparently, the minimal model property is what we are looking for. For $\mathcal{EL}$, the minimal model property is the outcome of $\mathcal{EL}$ being invariant under direct products: A sentence $\varphi$ of some logic $\mathcal{L}$ over the signature $\tau$ is *invariant under direct products*, if for every family of $\tau$-interpretations $(\mathfrak{I}_i)_{i \in I}$ we have

$$(\forall i \in I : \mathfrak{I}_i \models \varphi) \iff \bigtimes_{i \in I} \mathfrak{I}_i \models \varphi$$

Proposition 5.1.13. *$\mathcal{EL}$ is invariant under arbitrary direct products.*

Proof. Let $\tau$ be arbitrary. We shall show by induction upon the structure of $\mathcal{EL}(\tau)$-concepts that for any family of pointed $\tau$-interpretations $(\mathfrak{I}_i, d_i)$ we have

$$\forall i \in I : (\mathfrak{I}_i, d_i) \models C \iff (\bigtimes_{i \in I} \mathfrak{I}_i, (d_i)_{i \in I}) \models C$$

The claim follows for atomic concepts from the definition of the direct products. The claim for conjunctions can be directly obtained by the induction hypothesis. In case $C = \exists r.D$ and assume for all $i \in I$ that $(\mathfrak{I}, d_i) \models \exists r.D$. Then there is an $r$-successor $d'_i$ of $d_i$ such that $(\mathfrak{I}, d'_i) \models D$.

Let $(d'_i)_{i \in I}$ be the sequence of all those $r$-successors. By the definition of the direct product, we have $(d'_i)_{i \in I}$ is an $r$-successor of $(d_i)_{i \in I}$ in $\bigtimes_{i \in I} \mathfrak{I}_i$. The induction hypothesis yields that $(\bigtimes_{i \in I} \mathfrak{I}_i, (d'_i)_{i \in I}) \models D$. Hence $(\bigtimes_{i \in I} \mathfrak{I}_i, (d_i)_{i \in I}) \models \exists r.D$.

For the only-if direction we know that if $(\bigtimes_{i \in I} \mathfrak{I}_{i \in I}, (d_i)_{i \in I}) \models \exists r.D$ then there is an $r$-successor $(d'_i)_{i \in I}$ of $(d_i)_{i \in I}$ such that $(\bigtimes_{i \in I} \mathfrak{I}_{i \in I}, (d'_i)_{i \in I}) \models D$. By the definition of the direct product, each $d'_i$ is an $r$-successor of $d_i$ in $\mathfrak{I}_i$. The induction hypothesis yields that $(\mathfrak{I}_i, d'_i) \models D$ for all $i \in I$, and hence $(\mathfrak{I}_i, d_i) \models \exists r.D$ for all $i \in I$. □

Lemma 5.1.14. $(\bigtimes_{i \in I} \mathfrak{I}_i, (d_i)_{i \in I}) \Longrightarrow (\mathfrak{I}_i, d_i)$ *for all* $i \in I$.

Proof. Maintaining configurations $(\bigtimes_{i \in I} \mathfrak{I}_i, (d'_i)_{i \in I}; \mathfrak{I}_i, d'_i)$, i.e. $d_i$ is always in the sequence $(d_i)_{i \in I}$, exhibits a winning strategy for **II**. Any such configuration is valid in the way that for every atomic concept $A$ Proposition 5.1.13 shows $(\mathfrak{I}_i, d_i) \models A$.

The start configuration satisfies the requirement. Assume we have reached the configuration $(\bigtimes_{i\in I}\mathfrak{I}_i, (d'_i)_{i\in I}; \mathfrak{I}_i, d'_i)$. Let **I** challenge **II** by moving in $\bigtimes_{i\in I}\mathfrak{I}_i$ from $(d'_i)_{i\in I}$ to $(d''_i)_{i\in I}$. Then $d''_i$ is an $r$-successor of $d'_i$ in $\mathfrak{I}_i$ for all $i \in I$. Hence **II** can respond with $d''_i$ in $\mathfrak{I}_i$, reaching a valid configuration that satisfies the requirement. □

It is not difficult to show that the projection $\pi_i : \bigtimes_{i\in I}\mathfrak{I}_i \longrightarrow \mathfrak{I}_i : (d_i)_{i\in I} \longmapsto d_i$ is a surjective homomorphism, i.e. for all role names $r \in \mathsf{N}_\mathsf{R}$ we have

$$((d_i)_{i\in I}, (d'_i)_{i\in I}) \in r^{\bigtimes_{i\in I}\mathfrak{I}_i} \implies (\pi_i(d_i)_{i\in I}, \pi_i(d'_i)_{i\in I}) \in r^{\mathfrak{I}_i}$$

and analogously for concept names. Lyndon's theorem [34, 95]([116]) proves that the FO-fragment, which is preserved under surjective homomorphisms are positive FO-sentences. Positive FO-sentences are those, which are composed from (non-negated) atoms and closed under conjunctions, disjunctions and arbitrary quantifications. Lemma 5.1.14 reformulates the claim within our scope.

Before we show that $\mathcal{EL}$ has minimal models, we recursively define over $n < \omega$ the following $\tau$-model $\mathfrak{M}(X)$ for each characteristic $\mathcal{EL}$-concept $X$ over $\tau_\varphi$: Assume $X = X^n_{\mathfrak{H},e}$ for some $\tau$-interpretation $\mathfrak{H}$. For all $S \in \tau \setminus \tau_\varphi$ we set $S^{\mathfrak{M}(X)} = \emptyset$.

If $n = 0$ then $X = X^0_{\mathfrak{H},e}$ and we set $\Delta^{\mathfrak{M}}(X) = \{d\}$ and for all $A$ which occur as conjuncts in $X^0_{\mathfrak{H},e}$ we define $A^{\mathfrak{M}(X)} := \{d\}$, whereas all $A \in \tau_\varphi$ that do not occur in $X^0_{\mathfrak{H},e}$ are interpreted empty.

If $n = n + 1$ then $X = X^0_{\mathfrak{H},e} \sqcap \bigsqcap_{r\in\mathsf{N}_\mathsf{R}} \bigsqcap\{\exists r.X^n_{\mathfrak{H},e'} \mid (e, e') \in r^{\mathfrak{H}}\}$. We set

$$\mathfrak{M}(X) := \mathfrak{M}(X^0_{\mathfrak{H},e}) \uplus \biguplus\{\mathfrak{M}(X^n_{\mathfrak{H},e'}) \mid \exists r.X^n_{\mathfrak{H},e'} \text{ is a conjunct in } X\}$$

Let $d$ be the element in $\mathfrak{M}(X^0_{\mathfrak{H},e})$. We redefine $r^{\mathfrak{M}(X)}$ for all $r \in \mathsf{N}_\mathsf{R}$

$$r^{\mathfrak{M}(X)} := \bigcup\{(d, d') \mid (\mathfrak{M}(X), d') \models X^n_{\mathfrak{H},e'} \text{ and } \exists r.X^n_{\mathfrak{H},e'} \text{ is a conjunct in } X\}$$

$\mathfrak{M}(X)$ is up to isomorphy uniquely determined for $X$. It is also a finite tree-interpretation of depth $n$ that satisfies $X$, and in particular $(\mathfrak{M}(X), d) \Longrightarrow_n (\mathfrak{H}, e)$ if $X = X^n_{\mathfrak{H},e}$.

Proposition 5.1.15. *$\mathcal{EL}$ has the minimal model property, i.e. for all signatures $\tau$ and all concepts $C \in \tau$ there is a pointed $\tau$-model $(\mathfrak{I}, d)$ of $C$ such that for all concepts*

$D \in \mathcal{EL}(\tau)$ *we have*

$$(\mathfrak{I}, d) \models D \iff C \models D.$$

Proof. Let $\tau$ be a signature and $C \in \mathcal{EL}(\tau)$ be arbitrary with $n := \operatorname{rank} C$ and let $\tau_\varphi \subseteq \tau$ be the set of all symbols occurring in $C$. We set

$$I := \{X^n_{\mathfrak{H}\restriction\tau_\varphi, e} \mid \mathfrak{H} \text{ is a } \tau\text{-interpretation and } (\mathfrak{H}, e) \models C\},$$

where $X^n_{\mathfrak{H}\restriction\tau_\varphi, e}$ is the characteristic $\mathcal{EL}$-concept for $(\mathfrak{H}, e)$ over $\tau_\varphi$ on level $n$ (cf. page 174). As discussed, there are only finitely many different characteristic $\mathcal{EL}$-concepts over $\tau$ on each level $n$ and hence $I$ is finite.

For each $i \in I$, i.e. $i = X^n_{\mathfrak{H}\restriction\tau_\varphi, e}$ for some $(\mathfrak{H}, e) \models C$, we set $(\mathfrak{I}_i, d_i) := (\mathfrak{M}(i), d)$. Since $(\mathfrak{I}_i, d_i) \models X^n_{\mathfrak{H}, e}$ and $C$ is preserved under $n$-simulation, we have $(\mathfrak{H}, e) \cong_{\tau_\varphi} (\mathfrak{H}\restriction\tau_\varphi, e) \Longrightarrow_n (\mathfrak{I}_i, d_i)$ and so $(\mathfrak{I}_i, d_i) \models C$. By Proposition 5.1.13, $\mathcal{EL}$-concepts are invariant under direct products and so we obtain $(\bigtimes_{i\in I} \mathfrak{I}_i, \bar{d}) \models C$.

We have to show that $(\bigtimes_{i\in I} \mathfrak{I}_i, \bar{d}) \models D$ iff $C \models D$. Assume $(\bigtimes_{i\in I} \mathfrak{I}_i, \bar{d}) \models D$ and $(\mathfrak{H}, e) \models C$ for some $(\mathfrak{H}, e)$. Then $(\mathfrak{I}_i, d_i)$ with $i = X^n_{\mathfrak{H}\restriction\tau_\varphi, e}$ is a factor in $(\bigtimes_{i\in I} \mathfrak{I}_i, \bar{d})$ and since $(\bigtimes_{i\in I} \mathfrak{I}_i, \bar{d}) \Longrightarrow (\mathfrak{I}_i, d_i)$ and $(\mathfrak{I}_i, d_i) \Longrightarrow (\mathfrak{H}, e)$, we have $(\mathfrak{H}, e) \models D$. □

We therefore have two characterisation theorems

Theorem 5.1.16. *For every signature $\tau$ and every formula $\varphi(x) \in \mathrm{FO}(\tau)$ which is preserved under simulations and has a minimal model w.r.t. $\mathcal{EL}$, there is a logically equivalent concept $C_\varphi \in \mathcal{EL}(\tau)$.*

Proof. Theorem 5.1.9 shows that $\varphi$ is equivalent to a disjunction $\bigsqcup\{X^n_{\mathfrak{I},d} \mid (\mathfrak{I}, d) \models \varphi(x)\}$ of characteristic $\mathcal{EL}$-concepts over $\tau$. Since $\varphi$ has a minimal model there is some $(\mathfrak{I}, d) \models \varphi(x)$ such that $\varphi$ is logically equivalent to $\bigsqcup\{X^n_{\mathfrak{I},d} \mid (\mathfrak{I}, d) \models \varphi(x)\}$ which is as discussed on page 178 logically equivalent to $X^n_{\mathfrak{I},d}$. □

Theorem 5.1.17. *For every signature $\tau$ and every formula $\varphi(x) \in \mathrm{FO}(\tau)$ which is preserved under simulations and preserved under direct products, there is a logically equivalent concept $C_\varphi \in \mathcal{EL}(\tau)$.*

Note that we do not distinguish between being preserved under arbitrary products or being preserved under finite direct products, i.e. being preserved under all direct products where the index set $I$ is finite: [34, Theorem 6.3.14] tells us that every FO-sentence which is preserved under finite direct products, is also preserved under arbitrary products.

Proof of Theorem 5.1.17.

We may w.l.o.g. assume that $\tau$ comprises only the symbols in $\varphi$, i.e. that $\tau$ is finite.

Theorem 5.1.9 yields that $\varphi$ must be equivalent to a disjunction $\bigsqcup\{X^n_{\mathfrak{H},e} \mid (\mathfrak{H}, e) \models \varphi\}$ of characteristic $\mathcal{EL}$-concepts $X^n_{\mathfrak{H},e}$. We show that $\varphi$ has a minimal model w.r.t. $\mathcal{EL}^{\sqcup}$: Let $I := \{X^n_{\mathfrak{H},e} \mid (\mathfrak{H}, e) \models \varphi\}$ and set $(\mathfrak{I}_i, d_i) := (\mathfrak{M}(i), d)$ for all $i \in I$.

Since we have $(\mathfrak{H}, e) \Longrightarrow (\mathfrak{I}_i, d_i)$ for all models $(\mathfrak{H}, e)$ of $\varphi$ with $i = X^n_{\mathfrak{H},e}$, we have $(\mathfrak{I}_i, d_i) \models \varphi$ for all $i \in I$ and hence $(\bigtimes_{i\in I} \mathfrak{I}_i, \bar{d}) \models \varphi$.

We show that $\varphi$ has a minimal model w.r.t. $\mathcal{EL}^{\sqcup}$: If $(\bigtimes_{i\in I} \mathfrak{I}_i, \bar{d}) \models D$ then Lemma 5.1.14 yields for all models $(\mathfrak{H}, e)$ of $\varphi$ that $(\bigtimes_{i\in I} \mathfrak{I}_i, \bar{d}) \Longrightarrow (\mathfrak{H}, e)$ and therefore that $(\mathfrak{H}, e) \models D$. But our reflexion on page 178 upon minimal models shows that there is some $(\mathfrak{I}, d) \models \varphi$ such that $\varphi$ is logically equivalent to $\bigsqcup\{X^n_{\mathfrak{I},d} \mid (\mathfrak{I}, d) \models \varphi\}$ which, in turn, is logically equivalent to $X^n_{\mathfrak{I},d}$. □

### 5.1.5 Characterisation of $\mathcal{EL}^{\sqcup}$-TBoxes and $\mathcal{EL}$-TBoxes

In comparison to more expressive languages of the $\mathcal{ALC}$-family, TBoxes considerably add expressiveness to $\mathcal{EL}$: Following the scheme with which we treated all other families it should be enough to define a global notion of simulation, which, regarding the assymetric nature observed in local simulation looks somehow like

$$\mathfrak{I} \overset{\exists}{\Longrightarrow} \mathfrak{H} \iff \forall d \in \Delta^{\mathfrak{I}} : \exists e \in \Delta^{\mathfrak{H}} : (\mathfrak{I}, d) \Longrightarrow (\mathfrak{H}, e)$$

This notion, however, can only preserve global existential concepts, i.e. conjunctions and disjunctions of the form $\bigsqcup_{i\leq n} \bigsqcap_{j\leq m} \exists u.C_{ij}$ where $C_{ij}$ is some $\mathcal{EL}^{\sqcup}$-concept. Invariance under disjoint unions would get rid of the disjunctions on the global level, but there is no hope to preserve universal global concepts of the like $\forall u.C$.

Even the symmetrisation $\mathfrak{I} \overset{\exists}{\Longleftrightarrow} \mathfrak{H}$ defined as $\mathfrak{I} \overset{\exists}{\Longrightarrow} \mathfrak{H}$ and $\mathfrak{H} \overset{\exists}{\Longrightarrow} \mathfrak{I}$, i.e. for all $d \in \Delta^{\mathfrak{I}}$ there is some $e \in \Delta^{\mathfrak{H}}$ with $(\mathfrak{I}, d) \Longrightarrow (\mathfrak{H}, e)$ and vice versa, and is not enough to preserve $\mathcal{EL}$-TBoxes as the following example shows.

Example 5.1.18. Let $\mathfrak{I}$ be defined as $\Delta^{\mathfrak{I}} := \{a, b, c\}$, $r^{\mathfrak{I}} := \{(c, a), (c, b)\}$ with $A^{\mathfrak{I}} := \{a\}$ and $B^{\mathfrak{I}} := \{b\}$ and let $\Delta^{\mathfrak{H}} := \{d, e, f, g, h\}$ with $r^{\mathfrak{H}}\{(d, e), (d, f), (g, h)\}$ and $A^{\mathfrak{H}} := \{e, h\}$ and $B^{\mathfrak{H}} := \{f\}$. Then $\mathfrak{I} \overset{\exists}{\Longleftrightarrow} \mathfrak{H}$ but $\mathfrak{I} \models \{\exists r.B \sqsubseteq \exists r.A\}$ yet $\mathfrak{H} \not\models \{\exists r.B \sqsubseteq \exists r.A\}$.

The reason lies in the hidden negation of the concept-inclusion $\exists r.B \sqsubseteq \exists r.A$.

Rewritten in $\mathcal{ALC}u$ it would read $\forall u.\neg\exists r.B \sqcup \exists r.A$. Another model-theoretic notion is needed on the local level and this is equi-simulation.

### The Equi-Similar Fragment of FO

DEFINITION 5.1.19. $(\mathfrak{I}, d)$ and $(\mathfrak{H}, e)$ are *equisimilar*, $(\mathfrak{I}, d) \Longleftrightarrow (\mathfrak{H}, e)$ if $(\mathfrak{I}, d) \Longrightarrow (\mathfrak{H}, e)$ and $(\mathfrak{H}, e) \Longrightarrow (\mathfrak{I}, d)$. Analogously we set $(\mathfrak{I}, d) \Longleftrightarrow_n (\mathfrak{H}, e)$ if $(\mathfrak{I}, d) \Longrightarrow_n (\mathfrak{H}, e)$ and $(\mathfrak{H}, e) \Longrightarrow_n (\mathfrak{I}, d)$ ◇

We define $\mathcal{EL}^{\neg}$ to be the boolean closure of $\mathcal{EL}$, i.e. for every signature $\tau$ the concepts of $\mathcal{EL}^{\neg}$ are recursively defined by

$$C ::= F \mid D \sqcap E \mid \neg D$$

where $F \in \mathcal{EL}(\tau)$ and $D, E$ are $\mathcal{EL}^{\neg}$ concepts. We denote with $\mathcal{EL}^{\neg}(\tau)$ the set of all $\mathcal{EL}^{\neg}$-concepts over $\tau$. We use the usual abbreviations $\sqcup$, $\rightarrow$ etc. We may allow $\neg\exists r.C$ to be denoted as $\forall r.\neg C$. However, it is not possible to obtain alternating quantifications like $\exists r.\forall r.(\neg)C$

It should be noted that it is not necessary to define $\mathcal{EL}^{\neg}$ as boolean closure over $\mathcal{EL}^{\sqcup}$: The following lemma shows that every $\mathcal{EL}^{\sqcup}$-concept can be expressed as disjunctions of $\mathcal{EL}$-concepts which, in turn, can be expressed in $\mathcal{EL}^{\neg}$.

LEMMA 5.1.20. *Every $C \in \mathcal{EL}^{\sqcup}(\tau)$ is equivalent to $\bigsqcup_{i<n} C_i$ where $C_i \in \mathcal{EL}(\tau)$.*

PROOF. The proof is carried out via induction up then structure of $C \in \mathcal{EL}(\tau)$. If $C$ is atomic it is already in the desired form. With the induction hypothesis, we obtain for each disjunct in $C \sqcup D$ a concept of the desired form so that $C \sqcup D$ is logically equivalent to $\bigsqcup_{i<n} C_i \sqcup \bigsqcup_{i<m} D_i$ which is in desired form.

For $C \sqcap D$ we use the induction hypothesis and distributivity laws obtaining

$$C \sqcap D \equiv (\bigsqcup_{i<n} C_i) \sqcap (\bigsqcup_{j<m} D_j) \equiv \bigsqcup_{i<n} (C_i \sqcap \bigsqcup_{j<m} D_j) \equiv \bigsqcup_{i<n} \bigsqcup_{j<m} (C_i \sqcap D_j)$$

where each $C_i \sqcap D_j \in \mathcal{EL}(\tau)$, so that the concept has the desired form. And finally for $\exists r.D$ the induction hypothesis yields $\exists r. \bigsqcup_{i<n} D_i$ where $\exists$ commutes with disjunctions so that we obtain $\bigsqcup_{i<n} \exists r.D_i$. Every $\exists r.D_i$ is an $\mathcal{EL}(\tau)$-concept and so the concept is in the desired form. □

For all finite $\tau$ and all $\tau$-interpretation $(\mathfrak{I}, d)$ we define characteristic $\mathcal{EL}^{\neg}$-concepts for each $n < \omega$ as follows: $Z^n_{\mathfrak{I},d} := X^n_{\mathfrak{I},d} \sqcap \bigsqcap\{\neg X^n_{\mathfrak{H},e} \mid (\mathfrak{H}, e) \not\Longrightarrow_n (\mathfrak{I}, d)\}$ where $X^n_{\mathfrak{H},e}$ are the characteristic $\mathcal{EL}$-concepts over $\tau$. $Z^n_{\mathfrak{I},d}$ is well defined as there are only finitely many different characteristic $\mathcal{EL}$-concepts over $\tau$ on each level $n < \omega$. For the following proposition we use the rank-function for $\mathcal{ALC}$.

Proposition 5.1.21 . *Let $\tau$ be a finite signature and let $(\mathfrak{I}, d)$ and $(\mathfrak{H}, e)$ be $\tau$-interpretations. Then for all $n < \omega$ the following items are equivalent:*

1. $(\mathfrak{H}, e) \models Z^n_{\mathfrak{I},d}$
2. $(\mathfrak{I}, d) \Longleftrightarrow_n (\mathfrak{H}, e)$
3. $Th_n(\mathfrak{I}, d) = Th_n(\mathfrak{H}, e)$

*where* $Th_n(\mathfrak{I}, d) := \{C \in \mathcal{EL}^\neg(\tau) \mid (\mathfrak{I}, d) \models C$ *and* $\operatorname{rank} C \le n\}$.

Proof.
'1. $\Longrightarrow$ 2.': If $(\mathfrak{H}, e) \models Z^n_{\mathfrak{I},d}$ then $(\mathfrak{H}, e) \models X^n_{\mathfrak{I},d}$ and so $(\mathfrak{I}, d) \Longrightarrow_n (\mathfrak{H}, e)$. From $(\mathfrak{H}, e) \models X^n_{\mathfrak{H},e}$ follows with Proposition 5.1.5 and the assumption that $(\mathfrak{H}, e) \models Z^n_{\mathfrak{I},d}$ that $\neg X^n_{\mathfrak{H},e}$ cannot be a conjunct of $Z^n_{\mathfrak{I},d}$ and so $(\mathfrak{H}, e) \Longrightarrow_n (\mathfrak{I}, d)$.
'2. $\Longrightarrow$ 3.': We use induction upon the structure of $\mathcal{EL}^\neg$-concepts to prove the claim. If $C \in \mathcal{EL}(\tau)$ then $C$ is preserved by $n$-simulation. Since $(\mathfrak{I}, d) \Longrightarrow_n (\mathfrak{H}, e)$ and $(\mathfrak{H}, e) \Longrightarrow_n (\mathfrak{I}, d)$, $(\mathfrak{I}, d) \models C$ iff $(\mathfrak{H}, e) \models C$. All other cases are trivially obtained by the induction hypothesis.
'3. $\Longrightarrow$ 1.': Every characteristic concept $Z^n_{\mathfrak{H},e}$ has rank $\le n$. We show $(\mathfrak{I}, d) \models Z^n_{\mathfrak{I},d}$: Clearly $(\mathfrak{I}, d) \models X^n_{\mathfrak{I},d}$; but $(\mathfrak{I}, d)$ also satisfies all the negated conjuncts. Assume $\neg X^n_{\mathfrak{K},c}$ is amongst the negated conjuncts and $(\mathfrak{I}, d) \models X^n_{\mathfrak{K},c}$. Proposition 5.1.5 yields that $(\mathfrak{K}, c) \Longrightarrow_n (\mathfrak{I}, d)$ in contrast to the definition $(\mathfrak{K}, c) \not\Longrightarrow_n (\mathfrak{I}, d)$. So $Z^n_{\mathfrak{I},d} \in \mathrm{Th}_n(\mathfrak{I}, d)$ and since $\mathrm{Th}_n(\mathfrak{I}, d) = \mathrm{Th}_n(\mathfrak{H}, e)$, we have $Z^n_{\mathfrak{I},d} \in \mathrm{Th}(\mathfrak{H}, e)$ whence 1. follows. □

We lift the restriction of $\tau$ being finite.

Definition 5.1.22 . Let $\tau$ be arbitrary and $\mathfrak{I}$ a $\tau$-interpretation.

1. A set $\Gamma \subseteq \mathcal{EL}^\neg(\tau)$ is called an $\exists r$-type of $d$ in $\mathfrak{I}$ if for every finite subset $\Gamma_0 \subseteq \Gamma$ we have $(\mathfrak{I}, d) \models \exists r. \bigsqcap \Gamma_0$.
2. An $\exists r$-type is $\Gamma$ *realised* at $d$ in $\mathfrak{I}$ if there is some $r$-successor $d'$ of $d$ in $\mathfrak{I}$ such that $(\mathfrak{I}, d') \models \Gamma$.
3. A $\tau$-interpretation is $\mathcal{EL}^\neg$-saturated if for all $r \in \mathsf{N_R}$ and every $d \in \Delta^{\mathfrak{I}}$ every $\exists r$-type of $d$ is realised at $d$.

◇

Every $\mathcal{ALC}$-saturated interpretation and in particular every $\omega$-saturated interpretation is also $\mathcal{EL}^\neg$-saturated and every $\mathcal{EL}^\neg$ saturated interpretation is $\mathcal{EL}$-saturated. The inverse direction of the claims is in general not true.

Proposition 5.1.23 . *$\mathcal{EL}^\neg$ has the Hennessy-Milner-Property, i.e. for every signature $\tau$ and $\mathcal{EL}^\neg$-saturated $\tau$-interpretations $(\mathfrak{I}, d)$ and $(\mathfrak{H}, e)$ we have*

$$Th(\mathfrak{I}, d) = Th(\mathfrak{H}, e) \iff (\mathfrak{I}, d) \Longleftrightarrow (\mathfrak{H}, e)$$

*where $Th(\mathfrak{I}, d) := \{C \in \mathcal{EL}^\neg(\tau) \mid (\mathfrak{I}, d) \models C\}$ and analogously for $(\mathfrak{H}, e)$.*

Proof. Assume $(\mathfrak{I}, d)$ and $(\mathfrak{H}, e)$ are both $\mathcal{EL}^\neg$-saturated with $\text{Th}(\mathfrak{I}, d) = \text{Th}(\mathfrak{H}, e)$. Then $(\mathfrak{I}, d)$ and $(\mathfrak{H}, e)$ are both $\mathcal{EL}$-saturated and in particular $(\mathfrak{I}, d) \models C \implies (\mathfrak{H}, e) \models C$ for every $C \in \mathcal{EL}(\tau)$. The Henness-Milner-Property of $\mathcal{EL}$ yields $(\mathfrak{I}, d) \Longrightarrow (\mathfrak{H}, e)$. We can obtain the opposite direction in a similar fashion. □

Lemma 5.1.24 . *Every formula $\varphi(x) \in \text{FO}(\tau)$ which is invariant under equi-simulation is already invariant under n-equi-simulation for some $n < \omega$.*

Proof. Assume $\varphi$ is invariant under equi-simulation. Then $\varphi$ is invariant under bisimulation and hence is logically equivalent to some $\mathcal{ALC}$-concept $C_\varphi$. This concept is invariant under $n$-bisimulation for some $n < \omega$ and we have

$$(\mathfrak{I}, d) \Longleftrightarrow (\mathfrak{I}^T_d, d) \Longleftrightarrow_n (\mathfrak{I}^T_d{\restriction}n, d) \Longleftrightarrow (\mathfrak{H}^T_e{\restriction}n, e) \Longleftrightarrow_n (\mathfrak{H}^T_e, e) \Longleftrightarrow (\mathfrak{H}, e)$$

where $\mathfrak{I}^T_d$ is the tree-unravelling of $\mathfrak{I}$ in $d$ and $\mathfrak{I}^T_d{\restriction}n$ are all elements reachable within $n$-many steps from the root $d$; similarly for $\mathfrak{H}^T_e$ and $\mathfrak{H}^T_e{\restriction}n$. With this chain we obtain $(\mathfrak{I}, d) \models \varphi$ iff $(\mathfrak{H}, e) \models \varphi$. □

Theorem 5.1.25 . *Let $\tau$ be an arbitrary signature. If $\varphi(x) \in \text{FO}(\tau)$ is invariant under equi-simulation then there is a concept $C \in \mathcal{EL}^\neg(\tau)$ which is logically equivalent to $\varphi$.*

Proof. We assume w.l.o.g. that $\tau$ contains exactly the signature symbols occurring in $\varphi$ and therefore that $\tau$ is finite. Since $\varphi$ is invariant under equi-simulation, there is some $n < \omega$ such that $\varphi$ is invariant under $n$-equi-simulation. For this $n$ let $C_\varphi := \bigsqcup\{Z^n_{\mathfrak{I},d} \mid (\mathfrak{I}, d) \models \varphi\}$ where $Z^n_{\mathfrak{I},d}$ is the characteristic $\mathcal{EL}^\neg$-concept for $(\mathfrak{I}, d)$ on level $n$. The disjunction in $C_\varphi$ is finite and thus $C_\varphi$ is a concept in $\mathcal{EL}^\neg$.

$\varphi \models C_\varphi$ but $C_\varphi \models \varphi$ as well: Assume $(\mathfrak{H}, e) \models C_\varphi$. Then $(\mathfrak{H}, e) \models Z^n_{\mathfrak{I},d}$ where $Z^n_{\mathfrak{I},d}$ is a conjunct in $C_\varphi$. Proposition 5.1.21 shows $(\mathfrak{I}, d) \Longleftrightarrow_n (\mathfrak{H}, e)$ and since $(\mathfrak{I}, d) \models \varphi(x)$ and $\varphi$ is invariant under $n$-equi-simulation we have $(\mathfrak{H}, e) \models \varphi(x)$. □

## The Globally Equi-Similar Fragment of FO

Let $\tau$ be an arbitrary signature. Then $\mathcal{EL}u^\neg$-concepts are recursively defined as follows:

$$C ::= B \mid D \sqcap E \mid \neg D \mid \exists u.D$$

where $B \in \mathcal{EL}(\tau)$. The appropriate notion of global equi-simulation is defined in the following way:

Definition 5.1.26 . $\mathfrak{I}$ is globally equisimilar to $\mathfrak{H}$, $\mathfrak{I} \overset{\forall}{\Longleftrightarrow} \mathfrak{H}$, if both of the following holds

1. for all $d \in \Delta^{\mathfrak{I}}$ there is $e \in \Delta^{\mathfrak{H}}$ with $(\mathfrak{I}, d) \Longleftrightarrow (\mathfrak{H}, e)$
2. for all $e \in \Delta^{\mathfrak{H}}$ there is $d \in \Delta^{\mathfrak{I}}$ with $(\mathfrak{I}, d) \Longleftrightarrow (\mathfrak{H}, e)$

For each $n < \omega$ we define $\mathfrak{I} \overset{\forall}{\Longleftrightarrow}_n \mathfrak{H}$ analogously, requiring $(\mathfrak{I}, d) \Longleftrightarrow_n (\mathfrak{H}, e)$. ◇

We extend the notion by writing $(\mathfrak{I}, d) \overset{\forall}{\Longleftrightarrow} (\mathfrak{H}, e)$ if $\mathfrak{I} \overset{\forall}{\Longleftrightarrow} \mathfrak{H}$ and $(\mathfrak{I}, d) \Longleftrightarrow (\mathfrak{H}, e)$ and analogously $(\mathfrak{I}, d) \overset{\forall}{\Longleftrightarrow}_n (\mathfrak{H}, e)$. A simple proof by induction shows that $\mathcal{EL}u^\neg$-concepts are preserved under equi-simulation.

The set of global $\mathcal{EL}u^\neg$-concepts is recursively defined as by

$$C ::= \exists u.F \mid D \sqcap E \mid \neg D$$

where $F$ is an $\mathcal{EL}^\neg$-concept and $D, E$ are global $\mathcal{EL}u^\neg$-concepts. Note that $\exists u.\top$ is logically equivalent to $\top$. For each $n < \omega$ the global characteristic $\mathcal{EL}u^\neg$-concept over a finite signature $\tau$ is defined, using characteristic $\mathcal{EL}^\neg$-concepts $Z^n_{\mathfrak{I},d}$ over $\tau$, as

$$Z^n_{\mathfrak{I}} := \bigsqcap\{\exists u.Z^n_{\mathfrak{I},d} \mid d \in \Delta^{\mathfrak{I}}\} \sqcap \forall u.\bigsqcup\{Z^n_{\mathfrak{I},d} \mid d \in \Delta^{\mathfrak{I}}\}.$$

Every global characteristic $\mathcal{EL}u^\neg$-concept is a global $\mathcal{EL}u^\neg$-concept.

Proposition 5.1.27 . *For every finite signature $\tau$ and $\tau$-interpretations $\mathfrak{I}$ and $\mathfrak{H}$ and all $n < \omega$ the following statements are equivalent:*

1. $\mathfrak{H} \models Z^n_{\mathfrak{I}}$

2. $\mathfrak{I} \overset{\forall}{\Longleftrightarrow}_n \mathfrak{H}$

3. *$Th_n(\mathfrak{I}) = Th_n(\mathfrak{H})$*

*with $Th_n(\mathfrak{I}) := \{C \in \mathcal{ELu}^{\neg}(\tau) \mid \mathfrak{I} \models C,\ C$ is a global concept and* $\operatorname{rank} C \leq n\}$*. The rank function applied is the one for $\mathcal{ALCu}$.*

Proof.

'1.$\Longrightarrow$ 2.:' Since $\mathfrak{H} \models \bigsqcap\{\exists u.Z^n_{\mathfrak{I},d} \mid d \in \Delta^{\mathfrak{I}}\}$, for every element $d$ in $\mathfrak{I}$ there is an element $e$ in $\mathfrak{H}$ such that $(\mathfrak{H}, e) \models Z^n_{\mathfrak{I},d}$ and Proposition 5.1.21 yields $(\mathfrak{I}, d) \Longleftrightarrow_n (\mathfrak{H}, e)$. On the other hand, $\mathfrak{H} \models \forall u. \bigsqcup\{Z^n_{\mathfrak{I},d} \mid d \in \Delta^{\mathfrak{I}}\}$ and so for every $e$ in $\mathfrak{H}$ there is $d$ in $\mathfrak{I}$ such that $(\mathfrak{H}, e) \models Z^n_{\mathfrak{I},d}$. From Proposition 5.1.21 we obtain $(\mathfrak{I}, d) \Longleftrightarrow_n (\mathfrak{H}, e)$. This shows $\mathfrak{I} \overset{\forall}{\Longleftrightarrow}_n \mathfrak{H}$.

'2.$\Longrightarrow$ 3.:' If $\exists u.F \in \mathrm{Th}_n(\mathfrak{I})$ then there is $d$ in $\mathfrak{I}$ such that $(\mathfrak{I}, d) \models F$. Since $\mathfrak{I} \overset{\forall}{\Longleftrightarrow}_n \mathfrak{H}$ there is $e$ in $\mathfrak{H}$ with $(\mathfrak{I}, d) \Longleftrightarrow_n (\mathfrak{H}, e)$ and Proposition 5.1.21 yields $(\mathfrak{H}, e) \models F$. Therefore $\exists u.F \in \mathrm{Th}_n(\mathfrak{H})$. The induction hypothesis yields the step cases for conjunction and negation. $\mathrm{Th}_n(\mathfrak{H}) \subseteq \mathrm{Th}_n(\mathfrak{I})$ is shown in the same way.

'3.$\Longrightarrow$ 1.:' As mentioned above, $Z^n_{\mathfrak{I}}$ is a global concept and its rank is $\leq n$. Since $\mathfrak{I} \models Z^n_{\mathfrak{I}}$, it is contained in $\mathrm{Th}_n(\mathfrak{I})$ which by assumption entails $\mathfrak{H} \models Z^n_{\mathfrak{I}}$. □

Corollary 5.1.28 . *The following statements are equivalent:*

1. $(\mathfrak{H}, e) \models Z^n_{\mathfrak{I}} \sqcap Z^n_{\mathfrak{I},d}$

2. $(\mathfrak{I}, d) \overset{\forall}{\Longleftrightarrow}_n (\mathfrak{H}, e)$

3. *$Th_n(\mathfrak{I}, d) = Th_n(\mathfrak{H}, e)$*

*with $Th_n(\mathfrak{I}, d) := \{C \in \mathcal{ELu}^{\neg}(\tau) \mid (\mathfrak{I}, d) \models C$ and* $\operatorname{rank} C \leq n\}$.

Proof. We settle on '2.$\Longrightarrow$ 3.' Assume $B \in \mathcal{EL}(\tau)$. Then $(\mathfrak{I}, d) \models B$ and since $\mathcal{EL}(\tau) \subseteq \mathcal{EL}^{\neg}(\tau)$ for all signatures $\tau$, Proposition 5.1.21 yields $(\mathfrak{H}, e) \models B$.

The step cases for conjunction and negation follow immediately from the induction hypothesis. In case $C = \exists u.D$ there is $d' \in \Delta^{\mathfrak{I}}$ such that $(\mathfrak{I}, d') \models D$. Since $\mathfrak{I} \overset{\forall}{\Longleftrightarrow}_n \mathfrak{H}$ there is $e' \in \Delta^{\mathfrak{H}}$ with $(\mathfrak{I}, d') \Longleftrightarrow_n (\mathfrak{H}, e')$. With $\mathfrak{I} \overset{\forall}{\Longleftrightarrow}_n \mathfrak{H}$ follows $(\mathfrak{I}, d') \overset{\forall}{\Longleftrightarrow}_n (\mathfrak{H}, e')$ and the induction hypothesis yields $(\mathfrak{H}, e') \models D$, which entails $\mathfrak{H} \models \exists u.D$; in particular we obtain $\exists u.D \in \mathrm{Th}_n(\mathfrak{H}, e)$. The inclusion $\mathrm{Th}_n(\mathfrak{H}, e) \subseteq \mathrm{Th}_n(\mathfrak{I}, d)$ is shown by the same rationale. □

We lift the restriction for $\tau$ to be finite.

Definition 5.1.29. Let $\tau$ be arbitrary. Then

1. $\Gamma \subseteq \mathcal{EL}u^{\neg}(\tau)$ is called an *$\exists u$-type* of $\mathfrak{I}$ if for all finite subsets $\Gamma_0 \subseteq \Gamma$ we have $\mathfrak{I} \models \exists u. \bigsqcap \Gamma_0$.
2. An $\exists u$-type $\Gamma$ is said to be *realised* in $\mathfrak{I}$ if there is $d \in \Delta^{\mathfrak{I}}$ such that $(\mathfrak{I}, d) \models \Gamma$.
3. A $\tau$-interpretation $\mathfrak{I}$ is *$\mathcal{EL}u^{\neg}$-saturated* if it is $\mathcal{EL}^{\neg}$-saturated and every $\exists u$-type of $\mathfrak{I}$ is realised in $\mathfrak{I}$.

◇

Proposition 5.1.30. *$\mathcal{EL}u^{\neg}$ has the Hennessy-Milner-Property, i.e. for arbitrary $\tau$ and $\mathcal{EL}u^{\neg}$-saturated $\tau$-interpretations $\mathfrak{I}$ and $\mathfrak{H}$ we have*

$$Th(\mathfrak{I}, d) = Th(\mathfrak{H}, e) \iff (\mathfrak{I}, d) \overset{\forall}{\longleftrightarrow} (\mathfrak{H}, e)$$

*where $Th(\mathfrak{I}, d) := \{C \in \mathcal{EL}u^{\neg}(\tau) \mid (\mathfrak{I}, d) \models C\}$ and analogously for $Th(\mathfrak{H}, e)$.*

Lemma 5.1.31. *For $\mathcal{EL}u^{\neg}$-saturated $\tau$-interpretations $\mathfrak{I}$ and $\mathfrak{H}$ we have*

$$Th(\mathfrak{I}) = Th(\mathfrak{H}) \iff \mathfrak{I} \overset{\forall}{\longleftrightarrow} \mathfrak{H}$$

*where $Th(\mathfrak{I}) := \{C \in \mathcal{EL}u^{\neg}(\tau) \mid C$ is a global concept and $\mathfrak{I} \models C\}$*

Proof. Let $d \in \Delta^{\mathfrak{I}}$ be arbitrary. Then the set $\Gamma := \{C \in \mathcal{EL}^{\neg} \mid (\mathfrak{I}, d) \models C\}$ is an $\exists u$-type realised at $d$. In particular for every finite subset $\Gamma_0 \subseteq \Gamma$ the global concept $\exists u. \bigsqcap \Gamma_0$ is in $\mathrm{Th}(\mathfrak{I})$ and by assumption $\exists u. \bigsqcap \Gamma_0$ is also in $\mathrm{Th}(\mathfrak{H})$. Hence $\Gamma$ is an $\exists u$-type in $\mathfrak{H}$ and, since $\mathfrak{H}$ is $\mathcal{EL}u^{\neg}$-saturated, realised at some $e \in \Delta^{\mathfrak{H}}$. Hence $(\mathfrak{I}, d)$ and $(\mathfrak{H}, e)$ share the same $\mathcal{EL}^{\neg}$ theory and are $\mathcal{EL}^{\neg}$-saturated. Thus the Hennessy-Milner-Property for $\mathcal{EL}^{\neg}$ yields $(\mathfrak{I}, d) \longleftrightarrow (\mathfrak{H}, e)$. With the same argument one shows that for every $e \in \Delta^{\mathfrak{H}}$ there is $d \in \Delta^{\mathfrak{I}}$ such that $(\mathfrak{I}, d) \longleftrightarrow (\mathfrak{H}, e)$. The only-if direction is obtained immediately from the fact that $\mathcal{EL}u^{\neg}$-concepts are invariant under global equi-simulation. □

Proof of Proposition 5.1.30. $(\mathfrak{I}, d)$ and $(\mathfrak{H}, e)$ share the same $\mathcal{EL}^{\neg}$-theory and the Hennessy-Milner-Property for $\mathcal{EL}^{\neg}$ yields $(\mathfrak{I}, d) \longleftrightarrow (\mathfrak{H}, e)$. We obtain global equi-simulation by from Lemma 5.1.31. The only-if direction is again yielded by the invariance of $\mathcal{EL}u^{\neg}$ under global equi-simulation. □

Proposition 5.1.32. *Every FO-formula $\varphi(x)$ that is invariant under global equi-simulation is invariant under global $n$-equi-simulation for some $n < \omega$.*

Due to the symmetric nature of equi-simulation, the proof applies the exact same rationale as Lemma 2.1.25 which states the analogue statement for $\mathcal{ALC}$ and bisimulation.

Proof. Assume this would be wrong. Then for every $n < \omega$ there are interpretations $(\mathfrak{J}, d), (\mathfrak{H}, e)$ such that $(\mathfrak{J}, d) \overset{\forall}{\longleftrightarrow}_n (\mathfrak{H}, e)$, but $(\mathfrak{J}, d) \models \varphi(x)$, yet $(\mathfrak{H}, e) \models \neg\varphi(x)$. Let $\tau_\varphi$ be the set of signature symbols in $\varphi$, which is thus finite. We set

$$\mathfrak{Z}^n := \{Z^n_{\mathfrak{J}} \sqcap Z^n_{\mathfrak{J},d} \mid \exists(\mathfrak{J}, d), (\mathfrak{H}, e) : (\mathfrak{J}, d) \overset{\forall}{\longleftrightarrow}_n (\mathfrak{H}, e)\}$$

where $Z^n_{\mathfrak{J}} \sqcap Z^n_{\mathfrak{J},d}$ is the characteristic $\mathcal{ELU}^\neg$-concept over $\tau_\varphi$. $\mathfrak{Z}^n$ is finite, as the number of characteristic $\mathcal{ELU}^\neg$-concepts is finite. No $\mathfrak{Z}^n$ is empty.

Let $\mathfrak{Z} := \{\bigsqcup \mathfrak{Z}^n \mid n < \omega\}$. Then every finite subset of $\mathfrak{Z}$ is satisfiable with $\varphi(x)$: Let $\mathfrak{Z}_0 \subseteq \mathfrak{Z}$ be finite and let $m := \max\{n \mid \bigsqcup \mathfrak{Z}^n \in \mathfrak{Z}_0\}$. Then, there is some $(\mathfrak{J}, d) \models \varphi(x)$ such that $Z^m_{\mathfrak{J}} \sqcap Z^m_{\mathfrak{J},d} \in \mathfrak{Z}^m$.

Also by definition of $\mathfrak{Z}^m$, there is $(\mathfrak{H}, e)$ such that $(\mathfrak{J}, d) \overset{\forall}{\longleftrightarrow}_m (\mathfrak{H}, e)$ and $(\mathfrak{H}, e) \models \neg\varphi(x)$. From $(\mathfrak{J}, d) \overset{\forall}{\longleftrightarrow}_m (\mathfrak{H}, e)$ follows $(\mathfrak{J}, d) \overset{\forall}{\longleftrightarrow}_n (\mathfrak{H}, e)$ for all $n \leq m$ and so $Z^n_{\mathfrak{J}} \sqcap Z^n_{\mathfrak{J},d} \in \mathfrak{Z}^n$ for all $n \leq m$. Hence $(\mathfrak{J}, d) \models \bigsqcup \mathfrak{Z}^n$ for all $n \leq m$ and so $\mathfrak{J} \models \mathfrak{Z}_0$. Since $(\mathfrak{J}, d) \models \varphi(x)$ this shows that $\mathfrak{Z}_0 \cup \{\varphi(x)\}$ is satisfiable. Compactness of FO shows that $\mathfrak{Z} \cup \{\varphi(x)\}$ is satisfiable by some interpretation $(\mathfrak{K}, g)$.

For this $(\mathfrak{K}, g)$ we show that $\mathfrak{Z}_{\mathfrak{K},g} := \{Z^n_{\mathfrak{K}} \sqcap Z^n_{\mathfrak{K},g} \mid n < \omega\}$ is satisfiable with $\neg\varphi(x)$. Let therefore $\mathfrak{Z}_0 \subseteq \mathfrak{Z}_{\mathfrak{K},g}$ be finite and let $m := \max\{n \mid Z^n_{\mathfrak{K}} \sqcap Z^n_{\mathfrak{K},g} \in \mathfrak{Z}_0\}$. Since $(\mathfrak{K}, g) \models \mathfrak{Z}$, there is $(\mathfrak{J}, d)$ and $Z^m_{\mathfrak{J}} \sqcap Z^m_{\mathfrak{J},d} \in \mathfrak{Z}^m$ such that $(\mathfrak{K}, g) \models Z^m_{\mathfrak{J}} \sqcap Z^m_{\mathfrak{J},d}$ and hence $(\mathfrak{J}, d) \overset{\forall}{\longleftrightarrow}_m (\mathfrak{K}, g)$. By definition of $\mathfrak{Z}^m$, there is $(\mathfrak{H}, e)$ such that $(\mathfrak{J}, d) \overset{\forall}{\longleftrightarrow}_m (\mathfrak{H}, e)$ and $(\mathfrak{H}, e) \models \neg\varphi(x)$. Thus $(\mathfrak{H}, e) \models \mathfrak{Z}_0$ and so $\mathfrak{Z}_0$ and $\neg\varphi(x)$ are satisfiable. Compactness of FO shows $\mathfrak{Z}_{\mathfrak{K},g} \cup \{\neg\varphi(x)\}$ is satisfiable by some interpretation $(\mathfrak{H}, e)$. We have $\varphi(x) \dashv (\mathfrak{K}, g) \equiv_{\mathcal{ELU}^\neg} (\mathfrak{H}, e) \models \neg\varphi(x)$.

Let $(\mathfrak{K}^*, g)$ and $(\mathfrak{H}^*, e)$ be the $\omega$-saturated extensions of $(\mathfrak{K}, g)$ and $(\mathfrak{H}, e)$ respectively. Since both of them satisfy the same FO-theory as their originals, they still share the same $\mathcal{ELU}^\neg$-theory and $(\mathfrak{K}^*, g) \models \varphi(x)$ whilst $(\mathfrak{H}^*, e) \models \neg\varphi(x)$. The Hennessy-Milner-Property shows for both saturated interpretations $(\mathfrak{K}^*, g) \overset{\forall}{\longleftrightarrow} (\mathfrak{H}^*, e)$. But this is a contradiction as $\varphi$ is assumed to be invariant under global equi-simulation. Hence, there must be some $n < \omega$ such that $\varphi$ is invariant under global $n$-equi-simulation. □

Theorem 5.1.33. *Let $\tau$ be an arbitrary signature. For every formula $\varphi(x) \in \mathrm{FO}(\tau)$ which is invariant under global equi-simulation, there is a logically equivalent concept*

$C \in \mathcal{ELu}^{\neg}(\tau)$.

Proof. Let $\varphi(x)$ be such a formula. We may w.l.o.g. assume that $\tau$ comprises exactly the signature symbols in $\varphi$. In particular $\varphi(x)$ is invariant under global $n$-equi-simulation. We show that $\varphi(x)$ is logically equivalent to $C_\varphi$ with $C_\varphi :=$ $\bigsqcup\{Z^n_{\mathfrak{I}} \sqcap Z^n_{\mathfrak{I},d} \mid (\mathfrak{I}, d) \models \varphi(x)\}$ where $C_\varphi$ is well defined as there are just finitely many characteristic $\mathcal{ELu}^{\neg}$-concepts on each level $n < \omega$.

$\varphi \models C_\varphi$ follows immediately. For the opposite direction assume $(\mathfrak{H}, e) \models C_\varphi$. There is $(\mathfrak{I}, d)$ such that $Z^n_{\mathfrak{I}} \sqcap Z^n_{\mathfrak{I},d}$ is a disjunct in $C_\varphi$ and $(\mathfrak{H}, e) \models Z^n_{\mathfrak{I}} \sqcap Z^n_{\mathfrak{I},d}$. Hence $(\mathfrak{I}, d) \overset{\forall}{\rightleftarrows}_n (\mathfrak{H}, e)$ and since $(\mathfrak{I}, d) \models \varphi(x)$ and $\varphi$ is invariant under $n$-equi-simulation we obtain $(\mathfrak{H}, e) \models \varphi$. □

### Characterisation of $\mathcal{EL}^{\neg}$-TBoxes and $\mathcal{EL}^{\sqcup}$-TBoxes Respectively

A $\mathcal{EL}^{\neg}$-concept inclusion is an expression $C \sqsubseteq D$, where $C, D \in \mathcal{EL}^{\neg}(\tau)$ for some $\tau$. $\mathcal{EL}^{\neg}$-TBoxes are finite sets of $\mathcal{EL}^{\neg}$ concept-inclusions. Similarly $\mathcal{EL}^{\sqcup}$-concept inclusions recruit their components $C, D$ from $\mathcal{EL}^{\sqcup}(\tau)$ and likewise $\mathcal{EL}$-concept inclusions are constructed from $\mathcal{EL}$-concepts. It is thus clear what $\mathcal{EL}^{\sqcup}$-TBoxes and $\mathcal{EL}$-TBoxes are.

Every such $\mathcal{EL}$-, $\mathcal{EL}^{\sqcup}$- and $\mathcal{EL}^{\neg}$-TBox is an $\mathcal{ALC}$-TBox and interpreted as such. Every $\mathcal{EL}$-, $\mathcal{EL}^{\sqcup}$- and $\mathcal{EL}^{\neg}$-concept inclusion $C \sqsubseteq D$ can be expressed as $\mathcal{ELu}^{\neg}$-concept $\forall u.C \rightarrow D$. Hence every $\mathcal{EL}$-, $\mathcal{EL}^{\sqcup}$- and $\mathcal{EL}^{\neg}$-TBox is invariant under global equi-simulation.

Proposition 5.1.34. *Every $\mathcal{EL}^{\neg}$-TBox $\mathcal{T}$ is invariant under disjoint unions, i.e.*

$$(\forall i \in I : \mathfrak{I}_i \models \mathcal{T}) \Longrightarrow \biguplus_{i \in I} \mathfrak{I}_i \models \mathcal{T}$$

We already know that $\mathcal{ALC}$-TBoxes are invariant under disjoint unions and $\mathcal{EL}^{\neg}$-TBoxes form a subset of $\mathcal{ALC}$-TBoxes.

Theorem 5.1.35. *For all signatures $\tau$ and every sentence $\varphi \in \mathrm{FO}(\tau)$ we have: if $\varphi$ is invariant under equi-simulation and invariant under disjoint union then $\varphi$ is logically equivalent to some $\mathcal{EL}^{\sqcup}$-TBox over $\tau$.*

The proof applies the same rationale as all the characterisation proofs for TBoxes presented so far. For the sake of contradiction, the opposite of the claim is assumed, which leads to two interpretations which realise same $\mathcal{EL}^{\sqcup}$-types but are distinguished by the sentence $\varphi$. Indeed, these interpretations need to satisfy the same $\mathcal{EL}^{\neg}$-types! To obtain these interpretations, the disjoint union is used: we show that every $\mathcal{EL}^{\neg}$-type of the model which violates $\varphi$ is satisfiable with $\varphi$. Now, however, we have to be careful, as we have to derive our claim w.r.t. the $\mathcal{EL}^{\sqcup}$-consequences of $\varphi$. We therefore have to show that we can transform $\mathcal{EL}^{\neg}$-concept subsumptions accordingly to obtain $\mathcal{EL}^{\sqcup}$-concept subsumptions.

Proof.

$$\operatorname{cons}\varphi := \{C \sqsubseteq D \mid C, D \in \mathcal{EL}^{\sqcup}(\tau) \text{ and } \varphi \models C \sqsubseteq D\}$$

By compactness of FO it suffices to show $\operatorname{cons}\varphi \models \varphi$. Assume this is not the case. Then there must be some model $\mathfrak{H}$ of $\operatorname{cons}\varphi$ such that $\mathfrak{H} \models \neg\varphi$. Let $T := \{p \subseteq \mathcal{EL}^{\sqcup}(\tau) \mid p \cup \{\varphi\} \text{ satisfiable }\}$. For every $p \in T$ there is a model $(\mathfrak{I}_p, d_p)$ of $p \cup \{\varphi\}$. Since $\varphi$ is invariant under disjoint union, $\mathfrak{I} := \biguplus \mathfrak{I}_p$ is a model of $\varphi$. Let $\mathfrak{K} := \mathfrak{I} \uplus \mathfrak{H}$. Then $\mathfrak{K} \models \operatorname{cons}\varphi$ because every $\mathcal{EL}^{\sqcup}$-concept-subsumption is invariant under disjoint unions. But we also have $\mathfrak{K} \models \neg\varphi$ because $\mathfrak{H} \models \neg\varphi$ and $\varphi$ is invariant under disjoint unions, too.

We set $\mathfrak{K}^*$ and $\mathfrak{I}^*$ to be the $\omega$-saturated extensions of $\mathfrak{K}$ and $\mathfrak{I}$ respectively and show $\mathfrak{I}^* \xleftrightarrow{\forall} \mathfrak{K}^*$. Since $\mathcal{EL}u^{\neg}$ has the Hennessy-Milner-Property, it suffices to show that for all $d \in \Delta^{\mathfrak{I}^*}$ there is $e \in \Delta^{\mathfrak{K}^*}$ such that $(\mathfrak{K}^*, e) \models \operatorname{Th}(\mathfrak{I}^*, d)$ where $\operatorname{Th}(\mathfrak{I}^*, d) := \{C \in \mathcal{EL}^{\neg}(\tau) \mid (\mathfrak{I}^*, d) \models C\}$ and vice versa.

For all $d \in \Delta^{\mathfrak{I}^*}$ we have that every finite subset $\Gamma_0 \subseteq \operatorname{Th}(\mathfrak{I}^*, d)$ is satisfied[1] at some $d_{\Gamma_0} \in \Delta^{\mathfrak{I}}$. Hence $\mathfrak{I} \models \exists u. \bigsqcap \Gamma_0$ and so $\mathfrak{K} \models \exists u. \bigsqcap \Gamma_0$ for every finite $\Gamma_0 \subseteq \operatorname{Th}(\mathfrak{I}^*, d)$. Thus $\operatorname{Th}(\mathfrak{I}^*, d)$ is an $\exists u$-type in $\mathfrak{K}$ and since $\mathfrak{K}^*$ is in particular $\mathcal{EL}u^{\neg}$-saturated, there is some $e \in \Delta^{\mathfrak{K}^*}$ such that $(\mathfrak{K}^*, e) \models \operatorname{Th}(\mathfrak{I}^*, d)$.

If $e \in \Delta^{\mathfrak{K}^*}$ then every finite subset $\Gamma_0 \subseteq \operatorname{Th}(\mathfrak{K}^*, e)$ is satisfiable with $\varphi$: For the sake of contradiction assume the opposite. Then for every $\mathfrak{H} \models \varphi$ we have $(\mathfrak{H}, g) \models \neg \bigsqcap \Gamma_0$ for all $g \in \Delta^{\mathfrak{H}}$.

Let $\tau_{\Gamma_0}$ be the signature symbols occurring in $\Gamma_0$, thus $\tau_{\Gamma_0}$ is finite, and let $\mathfrak{Z}_{\Gamma_0} := \{Z^n_{\mathfrak{H},g} \mid (\mathfrak{H}, g) \models \bigsqcap \Gamma_0\}$ where $n = \operatorname{rank} \bigsqcap \Gamma_0$ and $Z^n_{\mathfrak{H},g}$ be the characteristic $\mathcal{EL}^{\neg}$-concept over $\tau_{\Gamma_0}$. Then $\bigsqcup \mathfrak{Z}_{\Gamma_0}$ is logically equivalent to $\bigsqcap \Gamma_0$.

By assumption, for every model $\mathfrak{H}$ of $\varphi$ every $g \in \Delta^{\mathfrak{H}}$ satisfies $\neg \bigsqcup \mathfrak{Z}_{\Gamma_0}$ and so we have $\varphi \models \{\top \sqsubseteq \neg Z^n_{\mathfrak{H},g} \mid Z^n_{\mathfrak{H},g} \in \mathfrak{Z}_{\Gamma_0}\}$. Every $Z^n_{\mathfrak{H},g}$ is of the form $X^n_{\mathfrak{H},g} \sqcap \bigsqcap\{\neg X^n_{\mathfrak{H}',g'} | (\mathfrak{H}', g') \not\Rightarrow_n (\mathfrak{H}, g)\}$, where $X^n_{\mathfrak{H},g}$ and each $X^n_{\mathfrak{H}',g'}$, respectively, is the characteristic $\mathcal{EL}$-concept over $\tau_{\Gamma_0}$. Therefore each $\top \sqsubseteq \neg Z^n_{\mathfrak{H},g}$ can be equival-ently rewritten into an $\mathcal{EL}^{\sqcup}$-concept subsumption $X^n_{\mathfrak{H},g} \sqsubseteq \bigsqcup\{X^n_{\mathfrak{H}',g'} \mid (\mathfrak{H}',g') \not\Rightarrow_n (\mathfrak{H},g)\}$.

[1] For otherwise $\mathfrak{I}$ would have been a model of the *FO*-sentence expressing $\forall u. \neg \bigsqcap \Gamma_0$. Since $\mathfrak{I}$ and $\mathfrak{I}^*$ share the same FO-theory also $\mathfrak{I}^* \models \forall u. \neg \bigsqcap \Gamma_0$, contradicting that $d$ satisfies $\bigsqcap \Gamma_0$.

This means that each of these concept subsumptions is present in cons $\varphi$; But then $(\mathfrak{K}, e) \not\models \Gamma_0$ since $(\mathfrak{K}, e) \models$ cons $\varphi$; a contradiction. Hence $\Gamma_0$ is satisfiable with $\varphi$. Therefore, for every finite subset $\Gamma_0 \subseteq \mathrm{Th}(\mathfrak{K}, e)$ there is some $d \in \Delta^{\mathfrak{I}}$ with $(\mathfrak{I}, d) \models \Gamma_0$. Since $\mathfrak{I}^*$ is in particular $\mathcal{EL}u^{\neg}$-saturated, there is $d' \in \Delta^{\mathfrak{I}^*}$ such that $(\mathfrak{I}^*, d') \models \mathrm{Th}(\mathfrak{K}^*, e)$.

This shows that $\varphi \dashv \mathfrak{I}^* \overset{\forall}{\Longleftrightarrow} \mathfrak{K}^* \models \neg\varphi$. But this is a contradiction to $\varphi$ being invariant under global equi-simulation. Hence cons $\varphi \models \varphi$. □

Since every $\mathcal{EL}^{\neg}$-TBox is equivalent to some FO-sentence which is invariant under global equi-simulation and invariant under disjoint union (see Proposition 5.1.34), this theorem shows that every $\mathcal{EL}^{\neg}$-TBox is logically equivalent some $\mathcal{EL}^{\sqcup}$-TBox.

Indeed, $\mathcal{EL}^{\sqcup}$-TBoxes are merely a syntactic variant of $\mathcal{EL}^{\neg}$-TBoxes:

Observation 5.1.36 . *Every $\mathcal{EL}^{\neg}$-TBox can logically equivalent be rewritten as $\mathcal{EL}^{\sqcup}$-TBox.*

We first define the functions $\cdot^{+\mathrm{NNF}}$ and $\cdot^{-\mathrm{NNF}}$ which will be used to obtain the negation normal form for $\mathcal{EL}^{\neg}$-concepts, where $F$ denotes an $\mathcal{EL}$-concept and $D$ and $E$ denote $\mathcal{EL}^{\neg}$-concepts:

$$
\begin{aligned}
F^{+\mathrm{NNF}} &:= F & F^{-\mathrm{NNF}} &:= \neg F\\
(D \sqcap E)^{+\mathrm{NNF}} &:= D^{+\mathrm{NNF}} \sqcap E^{+\mathrm{NNF}} & (D \sqcap E)^{-\mathrm{NNF}} &:= D^{-\mathrm{NNF}} \sqcup E^{-\mathrm{NNF}}\\
(\neg D)^{+\mathrm{NNF}} &:= D^{-\mathrm{NNF}} & (\neg D)^{-\mathrm{NNF}} &:= D^{+\mathrm{NNF}}
\end{aligned}
$$

In order to avoid a non-determinism, we require that it is first checked whether the argument of $\cdot^{+\mathrm{NNF}}$ and $\cdot^{-\mathrm{NNF}}$ is an $\mathcal{EL}$-concept. A $\mathcal{EL}^{\neg}$-concept $C$ is then in *negation normal form* (NNF), if $C = C^{+\mathrm{NNF}}$. It is clear that an $\mathcal{EL}^{\neg}$-concept $C$ is logically equivalent to $C^{+\mathrm{NNF}}$.

Proof of Observation 5.1.36. We assume w.l.o.g. that the $\mathcal{EL}^{\neg}$-TBox consists of only one concept subsumption of the form $\top \sqsubseteq C$. Let $E := C^{+\mathrm{NNF}}$. We can transfer $E$ into a conjunctive normal form $\bigsqcap_{i<n} \bigsqcup_{j<m_i} E_{ij}$ where for each $i < n$ and $j < m_i$ we have that either $E_{ij} = D_{ij}$ or $E_{ij} = \neg D_{ij}$ and $D_{ij}$ is an $\mathcal{EL}$-concept. Each disjunction $\bigsqcup_{j<m_i} E_{ij}$ can in turn be split into two sets of disjuncts $N_i$ and $P_i$, where $N_i = \{D_{ij} \mid E_{ij} = \neg D_{ij}\}$ contains all negated $\mathcal{EL}$-concepts and $P_i = \{D_{ij} \mid E_{ij} = D_{ij}\}$ contains all positive $\mathcal{EL}$-concepts of $\bigsqcup_{j<m_i} E_{ij}$.

Then $\bigsqcup_{j<m_i} E_{ij}$ is logically equivalent to $(\neg \bigsqcap N_i) \sqcup (\bigsqcup P_i)$ and hence $C$ is logically equivalent to $\bigsqcap_{i<n} N_i \to P_i$. We set $\mathcal{T} := \{N_i \sqsubseteq P_i \mid i < n\}$ and observe that $\mathcal{T}$ is an $\mathcal{EL}^{\sqcup}$-TBox which is logically equivalent to $\{\top \sqsubseteq C\}$. □

With the forestalling remarks about the hidden negation and Observation 5.1.36 it may not be surprising anymore that TBoxes lend so much expressive power to $\mathcal{EL}^{\sqcup}$ concepts that their TBoxes even capture the expressiveness of $\mathcal{EL}^{\neg}$-TBoxes.

This observation also shows that global equi-simulation was just enough to characterise $\mathcal{EL}^{\sqcup}$-TBoxes and did not 'exceed' the necessary. The question arises whether direct products will, similar to the characterisation of $\mathcal{EL}$, yield a characterisation for $\mathcal{EL}$-TBox fragments.

### Characterisation of $\mathcal{EL}$-TBoxes as FO-Fragment

In what follows we shall prove that the FO-Fragment which is invariant under global equi-simulation, disjoint unions and is preserved under direct products, coincides with the fragment of the $\mathcal{EL}$-TBoxes.

Since e.g. $\exists u.A$ is preserved under direct products and global equi-simulation, direct products alone are not sufficient to characterise $\mathcal{EL}$-TBoxes. Hence we need the invariance under disjoint unions. The fact that the property of being preserved under direct products singles out the $\mathcal{EL}$-TBoxes amongst the $\mathcal{EL}^{\sqcup}$-TBoxes analogously to the $\mathcal{EL}$-concepts amongst the $\mathcal{EL}^{\sqcup}$-concepts, suggests that being preserved direct product is indeed a natural notion. As already stated under Theorem 5.1.17, sentences which are invariant under finite direct products are already invariant under arbitrary products so that we do not distinguish between them.

Theorem 5.1.37 . *For all signatures $\tau$ and every sentence $\varphi \in \mathrm{FO}(\tau)$ we have: if $\varphi$ is invariant under equi-simulation, invariant under disjoint union and preserved under direct products then $\varphi$ is logically equivalent to some $\mathcal{EL}$-TBox over $\tau$.*

Proof. Let $\operatorname{cons}\varphi := \{C \sqsubseteq D \mid C, D \in \mathcal{EL}(\tau) \text{ and } \varphi \models C \sqsubseteq D\}$. We shall show that there is a finite subset $\mathcal{T} \subseteq \operatorname{cons}\varphi$ such that $\mathcal{T}$ is logically equivalent to $\varphi$. Since $\varphi \models \operatorname{cons}\varphi$ we merely have to show $\mathcal{T} \models \varphi$ and because FO is compact, it suffices to prove $\operatorname{cons}\varphi \models \varphi$.

Assume for the sake of contradiction that $\operatorname{cons}\varphi \not\models \varphi$. Then there is a $\tau$-interpretation $\mathfrak{I} \models \operatorname{cons}\varphi$ yet $\mathfrak{I} \not\models \varphi$. We shall show that for every $d \in \Delta^{\mathfrak{I}}$ there is a model of $\mathrm{Th}_{\mathcal{EL}^{\neg}}(\mathfrak{I}, d) \cup \{\varphi\}$:

Let $d \in \Delta^{\mathfrak{I}}$ be arbitrary, and $\mathrm{Th}_{\mathcal{EL}}(\mathfrak{I}, d) := \{C \in \mathcal{EL}(\tau) \mid d \in C^{\mathfrak{I}}\}$. Assume

there is $C \in \mathcal{EL}$ such that $d \notin C^{\mathfrak{I}}$ and $\mathrm{Th}_{\mathcal{EL}}(\mathfrak{I}, d) \cup \{\neg C, \varphi\}$ would not be satisfiable. By compactness, there is a finite subset $T \subseteq \mathrm{Th}_{\mathcal{EL}}(\mathfrak{I}, d)$ such that $T \cup \{\neg C, \varphi\}$ is not satisfiable. In particular $(\mathfrak{H}, e) \models \neg(\bigsqcap T \sqcap \neg C)$ for every model $\mathfrak{H}$ of $\varphi$ and every $e \in \Delta^{\mathfrak{H}}$. Hence every model $\mathfrak{H}$ of $\varphi$ satisfies $\bigsqcap T \sqsubseteq C$ and so $\bigsqcap T \sqsubseteq C \in \operatorname{cons} \varphi$, contradicting the fact that $\mathfrak{I} \models \operatorname{cons} \varphi$. Hence there is $(\mathfrak{H}_C, e_C) \models T \cup \{\varphi, \neg C\}$.

We set now $(\mathfrak{K}_d, g) := \bigtimes\{(\mathfrak{H}_C, e_C) \mid C \in \mathcal{EL}(\tau) \setminus \mathrm{Th}_{\mathcal{EL}}(\mathfrak{I}, d)\}$. Since $\varphi$ is preserved under direct products we obtain $\mathfrak{K}_d \models \varphi$. However, $\mathcal{EL}$ is even invariant under direct products, i.e. $(\mathfrak{K}_d, g) \models \mathrm{Th}_{\mathcal{EL}}(\mathfrak{I}, d)$ but also if $(\mathfrak{K}_d, g) \models D$ then all $(\mathfrak{H}_C, e_C) \models D$. Hence for no $C \in \mathcal{EL}(\tau) \setminus \mathrm{Th}_{\mathcal{EL}}(\mathfrak{I}, d)$ we have $(\mathfrak{K}_d, g) \models C$ and so $\mathrm{Th}_{\mathcal{EL}}(\mathfrak{K}_d, g) = \mathrm{Th}_{\mathcal{EL}}(\mathfrak{I}, d)$. This shows $(\mathfrak{K}_d, g) \models \mathrm{Th}_{\mathcal{EL}^{\neg}}(\mathfrak{I}, d) \cup \{\varphi\}$.

Let $P := \{p \subseteq \mathcal{EL}^{\neg}(\tau) \mid p \cup \{\varphi\} \text{ is satisfiable}\}$ and define $\mathfrak{K}^*$ to be the $\omega$-saturated extension of $\biguplus_{p \in P}(\mathfrak{K}_p, g_p)$ where $\mathfrak{K}_p$ is a model of $\varphi$ and $(\mathfrak{K}_p, g_p) \models p$. Let $\mathfrak{H}^* := \mathfrak{K}^* \uplus \mathfrak{I}^*$ where $\mathfrak{I}^*$ is the $\omega$-saturated extension of $\mathfrak{I}$. As $\varphi$ is invariant under disjoint union we have $\mathfrak{H}^* \models \neg\varphi$. Both interpretations $\mathfrak{H}^*$ and $\mathfrak{K}^*$ are $\mathcal{EL}^{\neg}$-saturated.

We shall show that $\mathfrak{K}^* \overset{\forall}{\rightleftarrows} \mathfrak{H}^*$. Since $\mathfrak{H}^*$ is the disjoint union of $\mathfrak{K}^*$ and $\mathfrak{I}^*$, we have $(\mathfrak{K}^*, g) \rightleftarrows (\mathfrak{H}^*, g)$ for all $g \in \Delta^{\mathfrak{K}^*}$. Let $d \in \Delta^{\mathfrak{I}^*}$. Then every finite subset $T \subseteq \mathrm{Th}_{\mathcal{EL}^{\neg}}(\mathfrak{I}^*, d)$ is satisfied at some $d' \in \Delta^{\mathfrak{I}}$, which is, as shown above, satisfiable with $\varphi$. By compactness $\mathrm{Th}_{\mathcal{EL}^{\neg}}(\mathfrak{I}^*, d) \cup \{\varphi\}$ is satisfiable and so $p_d := \mathrm{Th}_{\mathcal{EL}^{\neg}}(\mathfrak{I}^*, d) \in P$. Hence $\mathrm{Th}_{\mathcal{EL}^{\neg}}(\mathfrak{K}_{p_d}, g_{p_d})$ is an $\exists u$-type in $\mathfrak{K}^*$ and thus realised at some $g$. The Hennessy-Milner-Property of $\mathcal{EL}^{\neg}$ shows that $(\mathfrak{K}^*, g) \rightleftarrows (\mathfrak{H}^*, d)$ which entails $\mathfrak{K}^* \overset{\forall}{\rightleftarrows} \mathfrak{H}^*$.

But $\varphi$ is invariant under global equi-simulation, implying that $\mathfrak{H}^* \models \varphi$ which derives the contradiction. Hence $\operatorname{cons} \varphi \models \varphi$ showing that $\varphi$ must be logically equivalent to some $\mathcal{EL}$-TBox over $\tau$. □

In this chapter, a characterisation for $\mathcal{EL}$-concepts and $\mathcal{EL}$-TBoxes were given. But mere preservation under simulation was not enough to characterise the $\mathcal{EL}$-concepts as the FO-fragment which is preserved under simulation is indeed $\mathcal{EL}^{\sqcup}$. The missing property was preservation under direct products; $\mathcal{EL}$ is thus the FO-fragment which is preserved under simulation and direct products.

The preservation under direct products also implies another property, namely having the minimal model property. A minimal model of some $\mathcal{EL}$-concept $C$ is an interpretation which satisfies exactly the consequences of $C$. It turns out that $\mathcal{EL}$ corresponds exactly to the FO-fragment which is preserved under simulation and where each formula has the minimal model property.

A similar picture could be witnessed for $\mathcal{EL}$-TBoxes. But it also becomes clear that TBoxes lift the expressive power considerably compared to e.g. $\mathcal{ALC}$. The reason could be detected in the hidden negations in concept inclusions. Hence a global version of simulation is not enough to characterise $\mathcal{EL}$-TBoxes as FO-fragment. The answer was a symmetricised notion of simulation, namely equi-simulation.

The problem of concept subsumption is to decide whether or not $\mathcal{T} \models C \sqsubseteq D$ holds for two given $\mathcal{EL}$-concepts $C$ and $D$. The hidden negation in TBoxes is also the reason why in $\mathcal{EL}$ concept subsumption is not interreducible with deciding whether a concept is satisfiable. This is, however, possible in description logics which possess negation, like $\mathcal{ALC}$, as there a concept $C$ is unsatisfiable iff $C \sqsubseteq \bot$ and similarly, a concept subsumption $C \sqsubseteq D$ holds iff $\neg C \sqcap D$ is unsatisfiable. Hence deciding concept satisfiability and concept subsumption is from a complexity theoretical point of view equally hard.

But $\mathcal{EL}$ does not have negation. Indeed in the variant of $\mathcal{EL}$, where $\bot$ is excluded from the logical symbols, *every* concept is satisfiable. In our definition of $\mathcal{EL}$ which includes $\bot$, exactly those $\mathcal{EL}$-concepts which do not contain $\bot$ are satisfiable. Hence concept satisfiability is decidable in linear time: we have to parse the concept in its full length and check whether or not $\bot$ occurs. $\mathcal{EL}$-concept subsumption however cannot be reduced to concept satisfiability in $\mathcal{EL}$, as $\mathcal{EL}$ has no means of expressing $\neg C \sqcap D$. Instead, concept subsumption for $\mathcal{EL}$ (w.r.t. arbitrary $\mathcal{EL}$-TBoxes) is already PTIME-complete [66] and thus harder than satisfiability.

The FO-fragment which is invariant under equi-simulation are those $\mathcal{EL}^{\neg}$-concepts, which form the boolean closure over $\mathcal{EL}$-concepts. It turned out that $\mathcal{EL}^{\sqcup}$-TBoxes are equivalent to $\mathcal{EL}^{\neg}$-TBoxes and that the global notion of equi-simulation characterises $\mathcal{EL}^{\sqcup}$-TBoxes as FO-Fragment. The complexity of concept subsumption w.r.t. to arbitrary TBoxes rises for $\mathcal{EL}^{\sqcup}$ to EXPTIME-completeness [67] and is thus equally hard as deciding $\mathcal{ALC}$-concept subsumption [121, 122].

# 6. Application

Of course, the question arises in which concrete way the model-theoretic characterisation of the expressiveness of the DLs treated in previous chapters are suitable for application. It turns out that the characterisations give properties at hand with which one can recognise whether or not a concept or TBox of a logic is expressible in a weaker logic, i.e. a fragment of this logic. E.g. if we could decide for an $\mathcal{ALCI}$-TBox that it is invariant under global $\mathcal{ALC}$-bisimulation then this $\mathcal{ALCI}$-TBox must be equivalent to some $\mathcal{ALC}$-TBox. The following definition will formalise this as a problem:

Definition 6.0.38 . Let $\mathcal{L}_1$ and $\mathcal{L}_2$ be description logics. A TBox $\mathcal{T}$ is $\mathcal{L}_1$-rewritable, if it is equivalent to some $\mathcal{L}_1$-TBox over the same signature. $\mathcal{L}_1$-to-$\mathcal{L}_2$-rewritability is the problem to decide whether a given $\mathcal{L}_1$-TBox is $\mathcal{L}_2$-rewritable. ◇

## 6.1 The $\mathcal{ALCI}$-to-$\mathcal{ALC}$ Rewritability Problem

In what follows we shall first show that we can decide $\mathcal{ALC}$-rewritability for $\mathcal{ALCI}$-TBoxes and that we can decide $\mathcal{EL}$-rewritability for $\mathcal{ALC}$-TBoxes. The results of this section have been established in co-authorship with Carsten Lutz and Frank Wolter and have been published in [92].

The algorithm used to decide $\mathcal{ALC}$-rewritability for $\mathcal{ALCI}$-TBoxes is based on the type elimination method [110, 106] and determines whether or not for a given $\mathcal{ALCI}$-TBox $\mathcal{T}$ there are two globally $\mathcal{ALC}$-bisimilar interpretations $\mathfrak{I}$ and $\mathfrak{H}$, such that $\mathfrak{I} \not\models \mathcal{T}$ and $\mathfrak{H} \models \mathcal{T}$. In this case, obviously, $\mathcal{T}$ is not invariant under global $\mathcal{ALC}$-bisimulation and hence cannot be equivalent to an $\mathcal{ALC}$-TBox.

Otherwise, if no such two interpretations can be found, for every two globally $\mathcal{ALC}$-bisimilar interpretation $\mathfrak{I}$, $\mathfrak{H}$ we have $\mathfrak{I} \models \mathcal{T}$ iff $\mathfrak{H} \models \mathcal{T}$, showing that $\mathcal{T}$ is invariant under global $\mathcal{ALC}$-bisimulation.

In this case, $\mathcal{T}$ is equivalent to some first order sentence which is invariant under global $\mathcal{ALC}$-bisimulation and since $\mathcal{ALCI}$-TBoxes are invariant under disjoint unions (Proposition 3.1.21) the characterisation theorem (Theorem 2.3.7) for $\mathcal{ALC}$-TBoxes shows that $\mathcal{T}$ must be equivalent to some $\mathcal{ALC}$-TBox.

Indeed, in order to show that $\mathcal{T}$ is not invariant under global bisimulation, it is not necessary to determine whether there are two fully globally $\mathcal{ALC}$-bisimilar interpretations, but merely whether there are two interpretations $\mathfrak{I}$ and $\mathfrak{H}$ such that $\mathfrak{I} \not\models \mathcal{T}$ and $\mathfrak{H} \models \mathcal{T}$ and for every $d \in \Delta^{\mathfrak{I}}$ there is some $e \in \Delta^{\mathfrak{H}}$ such that $(\mathfrak{I}, d) \Longleftrightarrow (\mathfrak{H}, e)$: With $\mathfrak{K} := \mathfrak{I} \uplus \mathfrak{H}$ we obtain that $\mathfrak{K}$ is globally bisimilar to $\mathfrak{H}$ and since $\mathcal{ALCI}$-TBoxes are invariant under disjoint unions, we have $\mathfrak{K} \models \mathcal{T}$ if and only if $\mathfrak{I} \models \mathcal{T}$, showing that $\mathfrak{K} \not\models \mathcal{T}$.

In what follows we shall give the algorithm that decides non-$\mathcal{ALCI}$-to-$\mathcal{ALC}$-rewritability for $\mathcal{ALCI}$-TBoxes, keeping in mind that this algorithm is meant to construct $\mathfrak{I}$ and $\mathfrak{H}$ as just mentioned. We shall now define a special sort of type, which is not a complete type as known from model theory [109, 34]. A *complete* FO-*type* over a signature $\tau$ is a satisfiable set of formulae $\Gamma \subseteq \mathrm{FO}(\tau)$, such that for all formulae $\varphi \in \mathrm{FO}(\tau) \setminus \Gamma$ we have $\Gamma \cup \{\varphi\}$ is unsatisfiable. $\Gamma$ is maximal or complete in the sense that it cannot be extended. The types we are about to define are 1-types as mentioned in Section 'Remark on Saturation and Types' (on page 44) but they are not complete in $\mathcal{ALCI}(\tau)$. Yet they are complete w.r.t. to all subconcepts that appear in the *closure* $\operatorname{clos}\mathcal{T}$ of a given TBox $\mathcal{T}$. We therefore recursively define for every $\mathcal{ALCI}$-concept $C$

$$\operatorname{clos} C := \begin{cases} \{A, \neg A\} & \text{if } C = A \\ \{C, \neg C\} \cup \operatorname{clos} D \cup \operatorname{clos} E & \text{if } C = D \sqcap E \\ \operatorname{clos} D & \text{if } C = \neg D \\ \{C, \neg C\} \cup \operatorname{clos} D & \text{if } C = \exists r.D \\ \{C, \neg C\} \cup \operatorname{clos} D & \text{if } C = \exists r^{-}.D \end{cases}$$

and for every $\mathcal{ALCI}$-TBox $\mathcal{T}$ we set $\operatorname{clos}\mathcal{T} := \bigcup\{\operatorname{clos} C \cup \operatorname{clos} D \mid C \sqsubseteq D \in \mathcal{T}\}$.

We set $t^{\mathfrak{I}}(d) := \{C \in \operatorname{clos}\mathcal{T} \mid d \in C^{\mathfrak{I}}\}$. Let tp be a subset of the power-set $\mathfrak{P}(\operatorname{clos}\mathcal{T})$ of $\operatorname{clos}\mathcal{T}$ which contains all those $t \subseteq \operatorname{clos}\mathcal{T}$ which are satisfiable and maximal w.r.t. $\operatorname{clos}\mathcal{T}$, i.e. $t \cup \{C\}$ is unsatisfiable for all $C \in \operatorname{clos}\mathcal{T} \setminus t$. Let furthermore $\mathrm{tp}\,\mathcal{T} \subseteq \mathrm{tp}$ contain all those $t \in \mathrm{tp}$ such that $\mathcal{T} \cup t$ is satisfiable.

Starting out with the set $Y_0$ that will be defined below, the algorithm will successively remove all elements until we are left with a set $Z$ whose properties shall be set out as follows: Let $Z$ be defined as the set of all pairs $(s, S)$ in

$$Y_0 := \{(s, S) \in \mathrm{tp} \times \mathfrak{P}(\mathrm{tp}\,\mathcal{T}) \mid \forall A \in \mathsf{N_C}.\forall t \in S : A \in s \iff A \in t\}$$

such that for each $(s, S) \in Z$ there are interpretations $\mathfrak{I}, \mathfrak{H}$ and a bisimulation $B$ between $\mathfrak{I}$ and $\mathfrak{H}$ with

1. $\mathfrak{H} \models \mathcal{T}$
2. for all $d_0 \in \Delta^{\mathfrak{I}}$ there is $e_0 \in \Delta^{\mathfrak{H}}$ such that $(d_0, e_0) \in B$, i.e. $(\mathfrak{I}, d_0) \iff (\mathfrak{H}, e_0)$
3. there is $d \in \Delta^{\mathfrak{I}}$ with $(\mathfrak{I}, d) \models s$ and $S = \{t^{\mathfrak{H}}(e) \mid (d, e) \in B\}$

After Algorithm 6.1.2 we shall give an intuition for those pairs $(s, S)$ whilst discussing the 'asymmetry' of the algorithm.

Observation 6.1.1 . *$\mathcal{T}$ is not $\mathcal{ALC}$-rewritable iff there is $s \in \mathrm{tp} \setminus \mathrm{tp}\mathcal{T}$ and $S \subseteq \mathrm{tp}\mathcal{T}$ such that $(s, S) \in Z$.*

Proof. If $\mathcal{T}$ is not $\mathcal{ALC}$-rewritable then $\mathrm{cons}_{\mathcal{ALC}}(\mathcal{T}) \not\models \mathcal{T}$ where

$$\mathrm{cons}_{\mathcal{ALC}}\mathcal{T} := \{C \sqsubseteq D \mid C, D \in \mathcal{ALC}(\tau), \mathcal{T} \models C \sqsubseteq D\}$$

for otherwise compactness of $\mathcal{ALCI}u$ yields that there is a finite subset $\mathcal{T}_0 \subseteq \mathrm{cons}_{\mathcal{ALC}}\mathcal{T}$ such that $\mathcal{T}_0 \models \mathcal{T}$ which contradicts the non-$\mathcal{ALC}$-rewritability of $\mathcal{T}$. Now let $\mathfrak{I}$ be a model of $\mathrm{cons}_{\mathcal{ALC}}(\mathcal{T})$ such that $\mathfrak{I} \not\models \mathcal{T}$ and set $\mathfrak{H} := \biguplus_{p \in T} \mathfrak{H}_p$ where $(\mathfrak{H}_p, d_p) \models p \cup \mathcal{T}$ and $T := \{p \subseteq \mathcal{ALC}(\tau) \mid p \cup \mathcal{T}$ is satisfiable $\}$. The invariance of $\mathcal{ALCI}$-TBoxes under disjoint unions, yields that $\mathfrak{H} \models \mathcal{T}$.

Take the $\omega$-saturated extensions $\mathfrak{I}^*$ of $\mathfrak{I}$ and $\mathfrak{H}^*$ of $\mathfrak{H}$ respectively and assume for the sake of contradiction that there is $d \in \Delta^{\mathfrak{I}^*}$ such that for all $e \in \Delta^{\mathfrak{H}^*}$ we have $(\mathfrak{H}^*, e) \not\models \mathrm{Th}_{\mathcal{ALC}}(\mathfrak{I}^*, d)$. Then $\mathrm{Th}_{\mathcal{ALC}}(\mathfrak{I}^*, d) \notin T$ and from the compactness of $\mathcal{ALCI}u$ we infer there is a finite subset $p \subseteq \mathrm{Th}_{\mathcal{ALC}}(\mathfrak{I}^*, d)$ such that $\mathcal{T} \models p \sqsubseteq \bot$. Hence $p \sqsubseteq \bot \in \mathrm{cons}_{\mathcal{ALC}}\mathcal{T}$ and so $\mathfrak{I} \models p \sqsubseteq \bot$. Since $\omega$-saturation preserves first order theories, $\mathfrak{I}^* \models p \sqsubseteq \bot$ which is a contradiction to $(\mathfrak{I}, d) \models p$.

So for all $d \in \Delta^{\mathfrak{I}^*}$ there is $e \in \Delta^{\mathfrak{H}^*}$ such that $\mathrm{Th}_{\mathcal{ALC}}(\mathfrak{I}^*, d) = \mathrm{Th}_{\mathcal{ALC}}(\mathfrak{H}^*, e)$ for which the Hennessy-Milner-Property of $\mathcal{ALC}$ yields $(\mathfrak{I}^*, d) \iff (\mathfrak{H}^*, e)$.

We now construct $(s, S) \in Z$. Since $\mathfrak{I}^* \not\models \mathcal{T}$, there is $C \sqsubseteq D \in \mathcal{T}$ and some $d \in \Delta^{\mathfrak{I}^*}$ such that $(\mathfrak{I}^*, d) \models C \sqcap \neg D$. Set $s := t^{\mathfrak{I}}(d)$ and set $S := \{e \in \Delta^{\mathfrak{H}^*} \mid (\mathfrak{I}^*, d) \iff (\mathfrak{H}^*, e)\}$. Clearly $s \in \mathrm{tp} \setminus \mathrm{tp}\,\mathcal{T}$ and $A \in s$ iff $A \in t$ for all $A \in \mathsf{N_C}$ and all $t \in S$. It follows that $(s, S) \in Z$ which proves the if-direction.

If on the other hand $(s, S) \in Z$ and $s \in \text{tp} \setminus \text{tp}\,\mathcal{T}$, there is a pointed interpretation $(\mathfrak{I}, d)$ such that $(\mathfrak{I}, d) \models s$ which entails that $\mathfrak{I} \not\models \mathcal{T}$. Item 2 entails that if $\mathcal{T}$ is equivalent to some $\mathcal{ALC}$-TBox then $\mathfrak{H} \models \mathcal{T}$ implies $\mathfrak{I} \models \mathcal{T}$. Since this is false, $\mathcal{T}$ cannot be $\mathcal{ALC}$-rewritable. □

Before we give the algorithm, we need the following notions: For elements $s_0, s_1 \in \text{tp}$ we define $s_0 \rightsquigarrow_r s_1$ and read $s_1$ is a possible $r$-successor of $s_0$ if both of the following holds

1. if $\neg\exists r.C \in s_0$ then $C \notin s_1$
2. if $\neg\exists r^{-}.C \in s_1$ then $C \notin s_0$

We lift this definition to pairs $(s_0, S_0), (s_1, S_1) \in Y_0$ by setting $(s_0, S_0) \rightsquigarrow_r (s_1, S_1)$ if

1. $s_0 \rightsquigarrow_r s_1$
2. for all $t_0 \in S_0$ there is $t_1 \in S_1$ such that $t_0 \rightsquigarrow_r t_1$.

Algorithm 6.1.2 . The algorithm is given by the following rules which are applied to each $Y_n$, starting with $Y_0$ as defined above and leading after each application to some new set $Y_{n+1}$:

1. If $(s_0, S_0) \in Y_n$ and $\exists r.C \in s_0$ and there is no possible $r$-successor $(s_1, S_1) \in Y_n$ for $(s_0, S_0)$ with $C \in s_1$ then $Y_{n+1} := Y_n \setminus \{(s_0, S_0)\}$.
2. If $(s_1, S_1) \in Y_n$ and $\exists r^{-}.C \in s_1$ and there is no $(s_0, S_0) \in Y_n$ with $C \in s_0$ for which $(s_1, S_1)$ is a possible $r$-successor then $Y_{n+1} := Y_n \setminus \{(s_1, S_1)\}$.
3. If $(s_0, S_0) \in Y_n$ with $t_0 \in S_0$ and $\exists r.C \in t_0$ and there is no possible $r$-successor $(s_1, S_1)$ of $(s_0, S_0)$ with $t_1 \in S_1$ such that $C \in t_1$ and $t_0 \rightsquigarrow_r t_1$ then $Y_{n+1} := Y_n \setminus \{(s_0, S_0)\}$.

Since $Y_0$ is finite, the rules can only be applied finitely many times, until sequence stabilises in $Y_m$, where none of the above rules is applicable.

The asymmetry of the rules where a symmetric counterpart of ($r$3) is missing needs explaining. As noted earlier, it is enough to synthesise two interpretations $\mathfrak{I}$ which potentially violates $\mathcal{T}$ and $\mathfrak{H} \models \mathcal{T}$ such that for every $d \in \Delta^{\mathfrak{I}}$ there is some $e \in \Delta^{\mathfrak{H}}$ with $(\mathfrak{I}, d) \Longleftrightarrow (\mathfrak{H}, e)$. It is important to understand that $s$ represents some element $d \in \Delta^{\mathfrak{I}}$ with $(\mathfrak{I}, d) \models s$ whilst every $t \in S$ represents a an element $e \in \Delta^{\mathfrak{H}}$ such that $(\mathfrak{H}, e) \models t$.

For $\mathfrak{I}$, we are interested in realising the type $s$, i.e. after the algorithm has stopped, $(s, S)$ has either been deleted or its element $d \in \Delta^{\mathfrak{I}}$ satisfies $s$. Therefore the rules $(r1)$ and $(r2)$ delete $s$ as soon as no corresponding successors or predecessors for $d$ can be found amongst the types.

For $\mathfrak{H}$, we can be more generous: we simply need to ensure the realisation of $t$ by an element $e$ in 'forward direction', which is done by rule $(r3)$: if some $t \in S$ has no corresponding successor, $(s, S)$ and therefore $t$ is deleted. The reason for this asymmetric view is that bisimulation works only in 'forward direction', too. Whatever predecessors $e$ might have, they never become effective for the bisimulation. Hence, in order to realise the 'backward direction' of $t$ we can simply realise the type $t$ in a disjoint connected component $(\mathfrak{K}_t, e_t)$ and connect the predecessors of $e_t$ with $e$, as done in the proof of Lemma 6.1.5.

Proposition 6.1.3. $Y_m = Z$

This proposition essentially shows that the algorithm is sound and complete: $Y_m \subseteq Z$ shows that only pairs $(t, s)$ are identified which meet requirements of $Z$, hence the algorithm is sound. The algorithm is also complete because whenever some element is present in $Z$, it will also be contained in the final set of the algorithm.

We shall show first $Y_m \subseteq Z$ and then $Z \subseteq Y_m$. Showing $Y_m \subseteq Z$ will be divided into several lemmas, where always $Y_m \neq \emptyset$ is assumed. These lemmas will show the following intermediate results:

1. There is some interpretation $\mathfrak{I}$ with carrier-set $Y_m$ such that for all $(s, S) \in Y_m$ we have $C \in s$ iff $(\mathfrak{I}, (s, S)) \models C$.
2. There is some interpretation $\mathfrak{H} \models \mathcal{T}$ where triples $(s, S, t)$ with $(s, S) \in Y_m$ and $t \in S$ form part of the carrier set, such that for every such triple we have $(\mathfrak{H}, (s, S, t)) \models C$ iff $C \in t$.
3. $(\mathfrak{I}, (s, S)) \Longleftrightarrow (\mathfrak{H}, (s, S, t))$ for all $((s, S), (s, S, t)) \in B$.

Lemma 6.1.4. *There is some interpretation $\mathfrak{I}$ with carrier-set $Y_m$ such that for all $(s, S) \in Y_m$ we have $C \in s$ iff $(\mathfrak{I}, (s, S)) \models C$.*

Proof. We show for every $(s, S) \in Y_m$ that there is a pointed $\tau$-interpretation $(\mathfrak{I}, d) \models s$ and an interpretation $\mathfrak{H} \models \mathcal{T}$ such that for all $d_0 \in \Delta^{\mathfrak{I}}$ there is $e_0 \in \Delta^{\mathfrak{H}}$ with $(\mathfrak{I}, d_0) \Longleftrightarrow_{\mathcal{ALC}} (\mathfrak{H}, e_0)$. We begin with constructing $\mathfrak{I}$: Set

$$\begin{aligned}
\Delta^{\mathfrak{I}} &:= Y_m \\
A^{\mathfrak{I}} &:= \{(s, S) \in Y_m \mid A \in s\} \quad \text{for all} \quad A \in \mathsf{N_C} \\
r^{\mathfrak{I}} &:= \{((s_0, t_0), (s_1, S_1)) \in \Delta^{\mathfrak{I}} \times \Delta^{\mathfrak{I}} \mid (s_0, S_0) \rightsquigarrow_r (s_1, S_1)\} \quad \text{for all} \quad r \in \mathsf{N_R}.
\end{aligned}$$

We show that $(\mathfrak{I}, (s, S)) \models C$ iff $C \in s$ for all $(s, S) \in \Delta^{\mathfrak{I}}$. The proof is carried via induction upon the structure of $C$:

Since $\mathsf{N_C} \subseteq \operatorname{sig} \mathcal{T}$ and $s$ comprises every subconcept of $\mathcal{T}$ itself or its negation, $s$ is complete in the sense that $A \in s$ or $\neg A \in s$ for all $A \in \mathsf{N_C}$. The induction hypothesis trivially yields the cases for conjunctions and negations.

Assume $\exists r.C \in s$. Then there must be $(s', S') \in Y_m$ with $(s, S) \rightsquigarrow_r (s', S')$ and $C \in s'$, for otherwise, (r1) could be applied to delete $(s, S)$ in contradiction to the assumption that $Y_m$ is stable under rule application. The definition of $r^{\mathfrak{I}}$ shows that this $(s', S')$ is an $r$-successor of $(s, S)$ which, according to the induction hypothesis, satisfies $C$. Hence $(\mathfrak{I}, (s, S)) \models \exists r.C$.

Conversely, if $\exists r.C \in \operatorname{clos} \mathcal{T}$ and $(\mathfrak{I}, (s, S)) \models \exists r.C$ then there is $(s', S')$ such that $(s, S) \rightsquigarrow_r (s', S')$ and $(\mathfrak{I}, (s', S')) \models C$. The induction hypothesis yields $C \in s'$. But since $(s, S) \rightsquigarrow_r (s', S')$, we have that $\neg\exists r.C \notin s$ and since $s$ is complete w.r.t. $\operatorname{clos} \mathcal{T}$, we have $\exists r.C \in s$.

Analogously, since $Y_m$ is stable under rule ($r2$), one can show that $\exists r^-.C \in s$ then there is $(s', S') \in Y_m$ with $(s', S') \rightsquigarrow_r (s, S)$ and $C \in s'$. By the definition of $r^{\mathfrak{I}}$, $(s', S')$ is an $r$-predecessor of $(s, S)$ and the induction hypothesis yields $(\mathfrak{I}, (s', S')) \models C$. Hence $(\mathfrak{I}, (s, S)) \models \exists r^-.C$.

On the other hand, assume $\exists r^-.C \notin s$. Since $s$ is complete w.r.t. to $\operatorname{clos} \mathcal{T}$, we find $\neg\exists r^-.C \in s$ and by the definition of $\rightsquigarrow_r$, we have for all predecessors $(s', S')$ of $(s, S)$ that $C \notin s'$. The induction hypothesis yields that $C \notin s'$ iff $(\mathfrak{I}, (s', S')) \models \neg C$ and so for all $r$-predecessors of $(s', S')$ of $(s, S)$ in $\mathfrak{I}$ we have $(\mathfrak{I}, (s', S')) \models \neg C$. Hence $(\mathfrak{I}, (s, S)) \models \neg\exists r^-.C$ □

Lemma 6.1.5 . *There is some interpretation $\mathfrak{H} \models \mathcal{T}$ where triples $(s, S, t)$ with $(s, S) \in Y_m$ and $t \in S$ form part of the carrier set, such that for every such triple we have $(\mathfrak{H}, (s, S, t)) \models C$ iff $C \in t$.*

Proof. We construct $\mathfrak{H}$, where $\mathfrak{H}$ is a composition of several interpretations. At first we set $\mathfrak{K}_0$ in an analogous way as we did above:

$$\Delta^{\mathfrak{K}_0} := \{(s, S, t) \mid (s, S) \in Y_m, t \in S\}$$
$$A^{\mathfrak{K}_0} := \{(s, S, t) \in \Delta^{\mathfrak{H}} \mid A \in t\} \quad \text{for all} \quad A \in \mathsf{N_C}$$
$$r^{\mathfrak{K}_0} := \{((s_0, S_0, t_0), (s_1, S_1, t_1)) \mid (s_0, S_0) \rightsquigarrow_r (s_1, S_1) \text{ and } t_0 \rightsquigarrow_r t_1\} \text{ for all } r \in \mathsf{N_R}.$$

For every $(s, S) \in Y_m$ and $t \in S$ we define $(\mathfrak{K}_t, e)$ to be a model of $t \cup \mathcal{T}$ and set $\mathfrak{H}_0 := \mathfrak{K}_0 \uplus \biguplus\{\mathfrak{K}_t \mid t \in S \text{ and } (s, S) \in Y_m\}$. We now define $\mathfrak{H}$ to be $\mathfrak{H}_0$ but add for each $r \in \mathsf{N_R}$ the following $r$-edges:

For every $r \in \mathsf{N_R}$ we set $r^+$ to be the set containing those $(e', (s, S, t)) \in \Delta^{\mathfrak{K}_t} \times \Delta^{\mathfrak{K}_0}$ where $t$ contains $\exists r^-.C$ and $e'$ is the $r$-predecessor of $e$ in $\mathfrak{K}_t$ with $(\mathfrak{K}_t, e) \models t$ and $(\mathfrak{K}_t, e') \models C$.

We then set $\Delta^{\mathfrak{H}} := \Delta^{\mathfrak{H}_0}$, $A^{\mathfrak{H}} := A^{\mathfrak{H}_0}$ for every $A \in \mathsf{N_C}$ and $r^{\mathfrak{H}} := r^{\mathfrak{H}_0} \cup r^+$ for all $r \in \mathsf{N_R}$.

We shall proceed in two further steps.

1. for every $(s, S, t) \in \Delta^{\mathfrak{H}}$ we have $(\mathfrak{H}, (s, S, t)) \models t$
2. for every $t_0 \in \operatorname{tp} \mathcal{T}$ and every $e \in \Delta^{\mathfrak{K}_t}$ we have $(\mathfrak{H}, e) \models t_0$ iff $(\mathfrak{K}_t, e) \models t_0$.

We start with the first step and show for all $C \in \operatorname{clos} \mathcal{T}$ by induction upon the structure of $C$ that $(\mathfrak{H}, (s, S, t)) \models C$ iff $C \in t$ for all $(s, S, t) \in \Delta^{\mathfrak{K}_0}$.

Since $t$ is complete w.r.t. $\operatorname{clos} \mathcal{T}$, the definition of $A^{\mathfrak{K}_0}$ yields that $(\mathfrak{H}, (s, S, t)) \models A$ iff $A \in t$. Negation and conjunction are yield by the induction hypothesis.

Let $(s, S, t) \in \Delta^{\mathfrak{H}}$ be arbitrary and assume $\exists r.C \in t$ then, since $t \in S$ by definition of $\Delta^{\mathfrak{K}_0}$ and ($r2$) is not applicable to $Y_m$, there must be some $(s', S') \in Y_m$ with $(S, s) \rightsquigarrow_r (s', S')$ and some $t' \in S'$ containing $C$ such that $t \rightsquigarrow_r t'$. Hence $(s', S', t') \in \Delta^{\mathfrak{K}_0}$ and $(s', S', t')$ is an $r$-successor of $(s, S, t)$. The induction hypothesis yields that $(\mathfrak{H}, (s', S', t')) \models C$ and so $(\mathfrak{H}, (s, S, t)) \models \exists r.C$.

Assume conversely that $\exists r.C \in \operatorname{clos} \mathcal{T}$ and $(\mathfrak{H}, (s, S, t)) \models \exists r.C$. By the definition of $r^{\mathfrak{H}}$, $(s, S, t)$ can only be an $r$-predecessor to elements from $\Delta^{\mathfrak{K}_0}$. Hence $(\mathfrak{H}, (s, S, t)) \models \exists r.C$ iff there is some $r$-successor $(s', S', t') \in \Delta^{\mathfrak{K}_0}$ such that $(\mathfrak{H}, (s', S', t')) \models C$. From the induction hypothesis follows that $C \in t'$. Since $t \rightsquigarrow_r t'$ by definition of $r^{\mathfrak{K}_0}$, we have $\neg\exists r.C \notin t$ and since $t$ is complete w.r.t. $\operatorname{clos} \mathcal{T}$ it follows that $\exists r.C \in t$.

Let $\exists r^-.C \in t$. Since we do not have an analogue rule to ($r2$) for the range of $Y_0$, we can for no $(s_0, S_0, t_0) \in \Delta^{\mathfrak{K}_0}$ with $t = t_0$ rely on the existence of an $r$-predecessor $(s_1, S_1, t_1) \in \Delta^{\mathfrak{K}_0}$. But by the construction of $\mathfrak{H}$, for every $(s, S, t) \in \Delta^{\mathfrak{H}}$ with $\exists r^-.C$ there is $(\mathfrak{K}_t, e) \models t \cup \mathcal{T}$ such that for some $r$-predecessor $e'$ of $e$ in $\mathfrak{K}_t$ we have $(\mathfrak{K}_t, e') \models C$ and $(e', (s, S, t)) \in r^{\mathfrak{H}}$. Hence $(\mathfrak{H}, (s, S, t)) \models \exists r^-.C$.

Assume now $\exists r^-.C \in \operatorname{clos} \mathcal{T}$ and $(\mathfrak{H}, (s, S, t)) \models \exists r^-.C$. Either the $r$-predecessor which satisfies $C$ is from $(\mathfrak{K}_t, e)$. Since this predecessor is by definition of $r^{\mathfrak{H}}$ also a predecessor of $e$ in $\mathfrak{K}_t$, we have $(\mathfrak{K}_t, e) \models \exists r^-.C$. But $(\mathfrak{K}, e) \models t$ and since $t$ is complete w.r.t. $\operatorname{clos} \mathcal{T}$, it follows that $\exists r^-.C \in t$.

In case the $r$-predecessor satisfying $C$ is some element $(s', S', t') \in \Delta^{\mathfrak{K}_0}$ we have $t' \rightsquigarrow_r t$ by definition of $r_0^{\mathfrak{K}}$ and $C \in t'$ according to the induction hypothesis. The definition of $\rightsquigarrow_r$ forbids $t$ to contain $\neg\exists r^-.C$ and since $t$ is complete w.r.t. $\operatorname{clos} \mathcal{T}$ we obtain $\exists r^-.C \in t$.

We continue with the second step and show that for every $t \in \operatorname{tp} \mathcal{T}$ and every $e \in \Delta^{\mathfrak{K}_t}$ we have $(\mathfrak{H}, e) \models C$ iff $C \in \operatorname{tp} \mathcal{T}$ for all $C \in \operatorname{clos} \mathcal{T}$:

Let $t \in \operatorname{tp} \mathcal{T}$ be arbitrary. We show by induction upon the structure of concepts for all $C \in t_0$ and all $t_0 \in \operatorname{tp} \mathcal{T}$ and all $e_0 \in \Delta_t^{\mathfrak{K}}$ that $(\mathfrak{K}_t, e_0) \models t_0$ implies $(\mathfrak{H}, e_0) \models C$ iff $C \in t_0$.

The base case is clear and conjunction and negation are trivially obtained from the induction hypothesis. Assume we have $\exists r.C \in t_0$ and $(\mathfrak{K}_t, e_0) \models t_0$. Then there is $e_1 \in \Delta^{\mathfrak{K}_t}$ such that $(\mathfrak{K}_t, e_1) \models C$. Let $t_1 \in \operatorname{tp} \mathcal{T}$ such that $(\mathfrak{K}_t, e_1) \models t_1$. Since $C \in \operatorname{clos} \mathcal{T}$ it follows that $C \in t_1$ and thus by induction hypothesis that $(\mathfrak{H}, e_1) \models C$. In $\mathfrak{H}$, $e_1$ is still an $r$-successor of $e_0$ and so $(\mathfrak{H}, e_0) \models \exists r.C$.

In case $\exists r.C \in \operatorname{clos} \mathcal{T}$ we have $(\mathfrak{H}, e_0) \models \exists r.C$ then some $r$-successor which satisfies $C$ which is either from $\mathfrak{K}_t$ or from $\mathfrak{K}_0$. In the former case there is some $r$-successor $e_1 \in \Delta^{\mathfrak{K}_t}$ of $e_0$ which satisfies $C$ and for which the induction hypothesis yields that $C \in t_1$ where $t_1 \in \operatorname{tp} \mathcal{T}$ with $(\mathfrak{K}_t, e_1) \models t_1$. Hence $(\mathfrak{K}_t, e_0) \models \exists r.C$ and since $t$ is complete w.r.t. $\operatorname{clos} \mathcal{T}$, $\exists r.C \in t$ follows.

Assume the former case were wrong, i.e. there is no $r$-successor of $e_0$ from $\mathfrak{K}_t$ which satisfies $C$. Hence, by the construction of $\mathfrak{H}$, the only $r$-successor satisfying $C$ must be $(s, S, t) \in \Delta^{\mathfrak{K}_0}$. With the result $(\mathfrak{H}, (s, S, t)) \models C$ iff $C \in t$ for all $C \in \operatorname{clos} \mathcal{T}$ from above and the fact that $C \in \operatorname{clos} \mathcal{T}$ we obtain $C \in t$. But since $e_0$ is then also a $r$-predecessor of $e \in \Delta^{\mathfrak{K}_t}$ with $(\mathfrak{K}_t, e) \models t$ we derive a contradiction to the assumption that no $r$-successor of $e_0$ exists in $\mathfrak{K}_t$ which satisfies $C$.

Let $\exists r^-.C \in t_0$ with $(\mathfrak{K}_t, e_0) \models t_0$. There is some $r$-predecessor $e_1$ of $e_0$ in $\mathfrak{K}_t$ such that $(\mathfrak{K}_t, e_1) \models C$. Since $C \in \operatorname{clos} \mathcal{T}$ we have $C \in t_1$ where $t_1 \in \operatorname{tp} \mathcal{T}$ and $(\mathfrak{K}_t, e_1) \models t_1$. The induction hypothesis yields $(\mathfrak{H}, e_1) \models C$ and since $e_1$ is still an $r$-predecessor of $e_0$ in $\mathfrak{H}$ we obtain $(\mathfrak{H}, e_0) \models \exists r^-.C$.

For the last case let $\exists r^-.C \in \operatorname{clos} \mathcal{T}$ and $(\mathfrak{H}, e_0) \models \exists^- r.C$. Then there is some predecessor $e_1$, which, by the construction of $\mathfrak{H}$, must be in $\mathfrak{K}_t$: elements from $\mathfrak{K}_0$ can at most be $r$-successors of $e_0$. Hence the induction hypothesis yields that $C \in t_1$, where $t_1 \in \text{tp}\,\mathcal{T}$ and $(\mathfrak{K}_t, e_1) \models t_1$. Therefore $(\mathfrak{K}_t, e_1) \models C$ and so $(\mathfrak{K}_t, e_0) \models \exists r^-.C$. Since $t_0$ is complete with respect to clos $\mathcal{T}$ we obtain $\exists r^-.C \in t_0$.

It remains to show that $\mathfrak{H} \models \mathcal{T}$. We just have shown that every element in $\mathfrak{H}$ satisfies some type $t \in \text{tp}\,\mathcal{T}$. Let $e \in \Delta^{\mathfrak{H}}$ be arbitrary and let $(\mathfrak{H}, e) \models t$ where $t \in \text{tp}\,\mathcal{T}$. For the sake of contradiction, assume there is $C \sqsubseteq D \in \mathcal{T}$ such that $(\mathfrak{H}, e) \models C \sqcap \neg D$. Then $C, \neg D \in \text{clos}\,\mathcal{T}$ and since $t$ is complete w.r.t. to clos $\mathcal{T}$ we have $C, \neg D \in t$. A contradiction to the fact that $t \in \text{tp}\,\mathcal{T}$. □

Lemma 6.1.6 . *We define a relation* $B := \{((s, S), (s, S, t)) \mid (s, S) \in Y_m \text{ and } t \in S\}$ *and show that* $B$ *is a bisimulation.*

Proof. We show that **II** has a winning strategy in the bisimulation game, if she maintains configurations $(\mathfrak{I}, (s_0, S_0); \mathfrak{H}, (s_0, S_0, t_0))$ s.t. $((s_0, S_0), (s_0, S_0, t_0)) \in B$: For all $(s_0, S_0) \in Y_m$ we have $S_0 \neq \emptyset$ and so $B \neq \emptyset$. By the definition of $(s_0, S_0)$, it follows that $s_0$ contains exactly the same atomic concepts as $t_0$; hence $(s_0, S_0)$ and $(s_0, S_0, t_0)$ are atomically equivalent. The start-configuration, therefore, is valid, as will be all other configurations of this kind.

Assume that **I** challenges **II** by moving in $\mathfrak{I}$ from $(s_0, S_0)$ via an $r$-labelled edge where $r \in \mathsf{N_R}$ to $(s_1, S_1)$. By definition of $r^{\mathfrak{I}}$, we have $(s_0, S_0) \rightsquigarrow_r (s_1, S_1)$ and in particular $S_0 \rightsquigarrow_r S_1$, which entails that there is $t_1 \in S_1$ such that $t_0 \rightsquigarrow_r t_1$. Hence $(s_0, S_0, t_0) \rightsquigarrow_r (s_1, S_1, t_1)$ and so $(s_1, S_1, t_1)$ is an $r$-successor of $(s_0, S_0, t_0)$ in $\mathfrak{H}$ which satisfies the requirement for the configurations.

Assume **I** challenges **II** by moving in $\mathfrak{H}$ from $(s_0, S_0, t_0)$ via an $r$-labelled edge to $(s_1, S_1, t_1)$. The definition of $r^{\mathfrak{H}}$ yields that $(s_0, S_0, t_0) \rightsquigarrow_r (s_1, S_1, t_1)$ whence $(s_0, S_0) \rightsquigarrow_r (s_1, S_1)$ follows, by the definition of $\rightsquigarrow_r$. The definition of $r^{\mathfrak{I}}$ entails that $(s_1, S_1)$ is an $r$-successor of $(s_0, S_0)$ in $\mathfrak{I}$ which satisfies the requirement for the configuration. □

Lemma 6.1.7 . *If* $d_1$ *is an* $r$*-successor of* $d$*, and we set* $s_1 := t^{\mathfrak{I}}(d_1)$ *and* $S_1 := \{t^{\mathfrak{H}}(e_1) \mid (d_1, e_1) \in B\}$ *we have that* $(s_1, S_1) \in Z$ *and* $(s_0, S_0) \rightsquigarrow_r (s_1, S_1)$.

Proof. For all $t \in S_1$ we have for all $A \in \mathsf{N_C}$ that $A \in t$ where $t = t^{\mathfrak{H}}(e_1)$ iff $(\mathfrak{H}, e_1) \models A$ iff $(\mathfrak{I}, d_1) \models A$, since $(d_1, e_1) \in B$, iff $A \in s_1$. Hence $(s_1, S_1) \in Y_0$. $(s_1, S_1)$ also satisfies all three requirements and so $(s_1, S_1) \in Z$.

We show $(s_0, S_0) \rightsquigarrow_r (s_1, S_1)$: $(\mathfrak{I}, d) \models s_0$, so for every $\neg\exists r.C \in s_0$ we have $(\mathfrak{I}, d) \models \forall r.\neg C$ and hence $C \notin s_1$. Since $d_1$ is an $r$-successor of $d$ we have if $C \in s_0$ then $\neg\exists r.C \notin s_1$. So $s_0 \rightsquigarrow_r s_1$.

For every $t \in S_0$ we have $t = t^{\mathfrak{H}}(e)$ with $(d, e) \in B$ for some $e \in \Delta^{\mathfrak{H}}$. Since $(d, e) \in B$, there exists an $r$-successor $e_1$ of $e$ such that $(d_1, e_1) \in B$. By definition of $S_1$ we have $t^{\mathfrak{H}}(e_1) \in S_1$ which, as just shown, entails $t \rightsquigarrow_r t_1$ since $e_1$ is an $r$-successor of $e$. It follows that $(s_0, S_0) \rightsquigarrow_r (s_1, S_1)$. □

Analogously we can show for every $r$-predecessor $d_1$ of $d$ that $(s_1, S_1) \rightsquigarrow_r (s_0, S_0)$ where $s_1 = t^{\mathfrak{I}}(d_1)$ and $S_1 := \{t^{\mathfrak{H}}(e_1) \mid (d_1, e_1) \in B\}$.

Proof of Proposition 6.1.3. We show $Y_m \subseteq Z$: Let $(s, S) \in Y_m$ be arbitrary. We check all three points of the definition of $Z$:

1. Lemma 6.1.5 shows that $\mathfrak{H} \models \mathcal{T}$.

2. Lemma 6.1.6 shows the existence of an element $(s, S, t) \in \Delta^{\mathfrak{H}}$ such that $(\mathfrak{I}, (s, S)) \Longleftrightarrow (\mathfrak{H}, (s, S, t))$.

3. Lemma 6.1.4 shows $(\mathfrak{I}, (s, S)) \models s$ and Lemma 6.1.6 shows the existence of $B$. We have $S \subseteq \{t^{\mathfrak{H}}(e) \mid (d, e) \in B\}$ because every $((s, S), (s, S, t)) \in B$ for all $t \in S$ according to Lemma 6.1.6. Since $B$ is defined in a way such that $((s, S), (s, S, t)) \in B$ for all $t \in S$, we have we also have $\{t^{\mathfrak{H}}(e) \mid (d, e) \in B\} \subseteq S$.

We have to show that $Z \subseteq Y_m$. We shall show by induction upon $n < \omega$ that $(s, S) \in Y_n$ for all $(s, S) \in Z$. Since $Z \subseteq Y_0$ the claim is true for $n = 0$.

We have to show that $Z \subseteq Y_{n+1}$. By the induction hypothesis we know that $Z \subseteq Y_n$. Assume for the sake of contradiction $Z$ would not be subset of $Y_{n+1}$. Then there is $(s_0, S_0) \in Z$ which has been deleted by some rule application from $Y_{n+1}$.

Since $(s_0, S_0) \in Z$ there are interpretations $\mathfrak{I}, \mathfrak{H}$ such that

1. $\mathfrak{H} \models \mathcal{T}$

2. There is a bisimulation $B$ such that for all $d_0 \in \Delta^{\mathfrak{I}}$ there is $e_0 \in \Delta^{\mathfrak{H}}$ with $(d_0, e_0) \in B$

3. there is $d \in \Delta^{\mathfrak{I}}$ with $(\mathfrak{I}, d) \models s_0$ and $S_0 = \{t^{\mathfrak{H}}(e) \mid (d, e) \in B\}$

We now distinguish the following cases

(r1) Assume, $(s_0, S_0)$ would have been deleted by (r1) because $\exists r.C \in s_0$ and there is no $(s_1, S_1) \in Y_n$ with $(s_0, S_0) \rightsquigarrow_r (s_1, S_1)$ and $C \in s_1$.

Since $(\mathfrak{I}, d) \models s_0$ there is some $r$-successor $d_1$ of $d$ in $\mathfrak{I}$ such that $C \in d_1$.

Set $s_1 = t^{\mathfrak{I}}(d_1)$ and $S_1 := \{t^{\mathfrak{H}}(e_1) \mid (d_1, e_1) \in B\}$. Lemma 6.1.7 shows that $(s_1, S_1) \in Z$ and $(s_0, S_0) \rightsquigarrow_r (s_1, S_1)$. The induction hypothesis $Z \subseteq Y_n$ yields that $(s_1, S_1) \in Y_n$ and so (r1) is not applicable for $(s_0, S_0)$. A contradiction to the assumption (r1) deleted $(s_0, S_0)$.

(r2) Assume $(s_0, S_0)$ would have been deleted by (r2) because $\exists r^-.C \in s_0$ and there is no $(s_1, S_1)$ with $(s_1, S_1) \rightsquigarrow_r (s_0, S_0)$ and $C \in s_1$.

Since $(\mathfrak{I}, d) \models s_0$ there is $d_1 \in \Delta^{\mathfrak{I}}$ with $(d_1, d) \in r^{\mathfrak{I}}$ and $C \in d_1$. Let $s_1 := t^{\mathfrak{I}}(d_1)$ and $S_1 := \{t^{\mathfrak{H}}(e_1) \mid (d_1, e_1) \in B\}$. As mentioned after the proof of Lemma 6.1.7, we have $(s_1, S_1) \rightsquigarrow_r (s_0, S_0)$ and $(s_1, S_1) \in Z$.

The induction hypothesis yields that $(s_1, S_1) \in Y_n$ and therefore (r2) is not applicable to $(s_0, S_0)$ in contradiction to the assumption.

(r3) Assume $(s_0, S_0)$ would have been deleted by (r3) because there is $t_0 \in S_0$ with $\exists r.C \in t_0$ and there is no $(s_1, S_1)$ with $(s_0, S_0) \rightsquigarrow_r (s_1, S_1)$ and $C \in t_1$.

Since $t_0 = t^{\mathfrak{H}}(e)$ for some $e \in \Delta^{\mathfrak{H}}$ with $(d, e) \in B$, there is an $r$-successor $e_1$ of $e$ in $\mathfrak{H}$ such that $(\mathfrak{H}, e_1) \models C$. Since $(d, e) \in B$ there is an $r$-successor $d_1$ of $d$ in $\mathfrak{I}$ such that $(d_1, e_1) \in B$. Set $s_1 := t^{\mathfrak{I}}(d_1)$ and $S_1 := \{t^{\mathfrak{H}}(e') \mid (d_1, e') \in B\}$.

Clearly $t^{\mathfrak{H}}(e_1) \in S_1$ and $C \in t^{\mathfrak{H}}(e_1)$ since $C \in \operatorname{clos} \mathcal{T}$ and $t^{\mathfrak{H}}(e_1)$ is complete w.r.t. $\operatorname{clos} \mathcal{T}$. Lemma 6.1.7 shows that $(s_1, S_1) \in Z$ and $(s_0, S_0) \rightsquigarrow_r (s_1, S_1)$. The induction hypothesis entails that $(s_1, S_1) \in Y_n$ and therefore (r3) was not applicable to $(s_0, S_0)$.

This finally shows that $Z \subseteq Y_n$ for all $n < \omega$ and in particular that $Z \subseteq Y_m$. □

As we recruit the elements in $Y_0$ from $\mathfrak{P}(\operatorname{tp} \mathcal{T})$, the set $Y_0$ is in the worst case in the order of $2^{2^{|\mathcal{T}|}}$ and hence the $\mathcal{ALCI}$-to-$\mathcal{ALC}$-rewritability problem is decidable in 2-EXPTIME w.r.t. to size of $\mathcal{T}$. As of yet, a lower bound has not been found.

Clearly polynomial rewritings have been devised to transcribe $\mathcal{ALCI}$-TBoxes into $\mathcal{ALC}$-TBoxes [44, 33] but these rewritings introduce new signature symbols and do not fit in our setting where the retention of signatures is central.

The question nevertheless arises, how big such a rewritten TBox theoretically can be. Assume, a given $\mathcal{ALCI}$-TBox $\mathcal{T}'$ over the signature $\tau$ is found to be rewritable into an $\mathcal{ALC}$-TBox $\mathcal{T}$. We can w.l.o.g. assume that $\mathcal{T} = \{\top \sqsubseteq C\}$ with $C \in \mathcal{ALC}(\tau)$.

At first, we look at rank $C$. Let therefore $k := \operatorname{rank} \mathcal{T}'$. Every type $t \in \mathrm{tp}$ has rank $t = k$. These types can only influence the elements in their neighbourhood, i.e. successors and predecessors, up to depth $k$. Hence $t$ influences elements in a neighbourhood of diameter $2k$ and so the concept $C$, as well, only needs to specify properties of elements up to (forward) depth $2k$; we shall give a proof for this intuition:

Proposition 6.1.8. *Every global $\mathcal{ALCI}u$-concept $C$ with* rank $C \leq k$ *which is preserved under $\overset{g}{\Longleftrightarrow}$ is already preserved under $\overset{g}{\Longleftrightarrow}_{2k}$.*

Lemma 6.1.9. *Let $\mathfrak{I}$ and $\mathfrak{H}$ be two $\tau$-interpretation and $\mathfrak{I}^F$ and $\mathfrak{H}^F$ their forest-unravellings. If $\mathfrak{I} \overset{g}{\Longleftrightarrow}_{2k} \mathfrak{H}$ then $\mathfrak{I}^F \overset{gi}{\Longleftrightarrow}_k \mathfrak{H}^F$.*

Proof. It is important to keep in mind, that predecessors are unique in forest-unravellings. We prove the claim by induction upon $k < \omega$. The claim is clear for $k = 0$. Let $\mathfrak{I} \overset{g}{\Longleftrightarrow}_{2(k+1)} \mathfrak{H}$ and assume at first $\bar{d}' \in \Delta^{\mathfrak{I}^F}$ has a predecessor $\bar{d} \in \Delta^{\mathfrak{I}^F}$ in distance $k+1$, i.e. $\bar{d}' = \bar{d} \cdot r_1 \cdot d_1 \cdots r_{k+1} \cdot d_{k+1}$ is in depth $k+1$ of the subtree of $\bar{d}$. Then there is some $\bar{e} \in \Delta^{\mathfrak{H}^F}$ such that $(\mathfrak{I}^F, \bar{d}) \Longleftrightarrow_{2(k+1)} (\mathfrak{H}^F, \bar{e})$ and some $\bar{e}' = \bar{e} \cdot r_1 \cdot e_1 \cdots r_{k+1} \cdot e_{k+1}$ such that $(\mathfrak{I}^F, \bar{d} \cdots d_i) \Longleftrightarrow_{2(k+1)-i} (\mathfrak{H}^F, \bar{e} \cdots e_i)$ for all $i \in \{1, \ldots, k+1\}$. Essentially, we require that $\bar{e}'$ is the answer of **II** after being challenged for $k$ rounds by **I** moving in $\mathfrak{I}^F$ from $\bar{d}$ to $\bar{d}'$.

We have to show that $(\mathfrak{I}^F, \bar{d}') \overset{i}{\Longleftrightarrow}_{k+1} (\mathfrak{H}^F, \bar{e}')$. Let **I** challenge **II** in the $\mathcal{ALCI}$-bisimulation game by moving to the $r_{k+1}$-predecessor $\bar{d}'' := \bar{d} \cdot r_1 \cdot d_1 \cdots r_k \cdot d_k$ of $\bar{d}'$, say. Then **II** can answer with $\bar{e}'' := \bar{e} \cdot r_1 \cdot e_1 \cdots r_k \cdot e_k$: We have that $\bar{d}$ is in distance $k$ to $\bar{d}''$ and equally for $\bar{e}$ and $\bar{e}''$. Furthermore $(\mathfrak{I}^F, \bar{d}) \Longleftrightarrow_{2k} (\mathfrak{H}^F, \bar{e})$ and $(\mathfrak{I}^F, \bar{d} \cdots d_i) \Longleftrightarrow_{2k-i} (\mathfrak{H}^F, \bar{e} \cdots e_i)$ for all $i \in \{1, \ldots, k\}$. Hence the induction hypothesis yields $(\mathfrak{I}^F, \bar{d}'') \overset{i}{\Longleftrightarrow}_k (\mathfrak{H}^F, \bar{e}'')$. The same arguments hold if **I** challenges **II** by moving to the predecessor in $\mathfrak{H}^F$.

Assume **I** challenges **II** by moving in $\mathfrak{I}^F$ from $\bar{d}'$ to $\bar{d}'' := \bar{d}' \cdot r_{k+2} \cdot d_{k+2}$. Because we have $(\mathfrak{I}^F, \bar{d}') \Longleftrightarrow_{k+1} (\mathfrak{H}^F, \bar{e}')$, there is an $r_{k+2}$-successor $\bar{e}'' := \bar{e}' \cdot r_{k+2} \cdot e_{k+2}$ of $\bar{e}'$ such that $(\mathfrak{I}^F, \bar{d} \cdots d_i) \Longleftrightarrow_{2k-i} (\mathfrak{H}^F, \bar{e} \cdots e_i)$ for all $i \in \{1, \ldots, k+2\}$. In particular is $\bar{d} \cdot r_1 \cdot d_1 \cdot r_2 \cdot d_2$ in distance $k$ of $\bar{d}''$ and $\bar{e} \cdot r_1 \cdot e_1 \cdot r_2 \cdot e_2$ in distance $k$ of $\bar{e}''$ which entails $(\mathfrak{I}^F, \bar{d} \cdots d_{i+2}) \Longleftrightarrow_{2k-i} (\mathfrak{H}^F, \bar{e} \cdots e_{i+2})$ for all $i \in \{1, \ldots, k\}$. Again, the induction hypothesis yields $(\mathfrak{I}^F, \bar{d}'') \overset{gi}{\Longleftrightarrow}_k (\mathfrak{H}^F, \bar{e}'')$. This shows $(\mathfrak{I}, \bar{d}') \overset{i}{\Longleftrightarrow}_{k+1} (\mathfrak{H}^F, \bar{e}')$.

We cover the case where $\bar{d}'$ does not have a predecessor in distance $k+1$. Take its predecessor $d$ with maximal distance $\ell$ to $\bar{d}' := d \cdot r_1 \cdot d_1 \cdots r_\ell \cdot d_\ell$. The element $d$ is the root of a tree in $\mathfrak{I}^F$. Since $\mathfrak{I}^F \Longleftrightarrow_{2k} \mathfrak{H}^F$ there is a root $e \in \Delta^{\mathfrak{H}^F}$ with $(\mathfrak{I}^F, d) \Longleftrightarrow_{2(k+1)} (\mathfrak{H}^F, e)$, and so there is $\bar{e}' := e \cdot r_1 \cdot e_1 \cdots r_\ell \cdot e_\ell$ such that $(\mathfrak{I}^F, d \cdots d_i) \Longleftrightarrow_{2(k+1)-i} (\mathfrak{H}^F, e \cdots e_i)$ for all $i \in \{1, \ldots, \ell\}$.

In case **I** challenges **II** in the $\mathcal{ALCI}$-bisimulation game by moving to a predecessor, we still have that $d$ and $e$ are in maximal distance $< k$ to the (unique) predecessors of $\bar{e}'$ and $\bar{d}'$ and $(\mathfrak{I}^F, d \cdots d_i) \Longleftrightarrow_{2k-i} (\mathfrak{H}^F, e \cdots e_i)$ for all $i \in \{1, \ldots, \ell - 1\}$. This allows to use the induction hypothesis which yields $\mathcal{ALCI}$-k-bisimilarity. If **I** challenges **II** in the $\mathcal{ALCI}$-bisimulation game by moving to a successor, then either the successor. Since $(\mathfrak{I}^F, \bar{d}') \Longleftrightarrow_{k+1} (\mathfrak{H}^F, \bar{e}')$, **II** can answer this challenge yielding $(\mathfrak{I}^F, \bar{d}'') \Longleftrightarrow_k (\mathfrak{H}^F, \bar{e}'')$ for successors $\bar{d}''$ of $\bar{d}'$ and $\bar{e}''$ of $\bar{e}'$. Either $\bar{d}''$ has now the root $d$ in distance $k$, for which the proof from above shows $(\mathfrak{I}^F, \bar{d}'') \overset{i}{\Longleftrightarrow}_k (\mathfrak{H}^F, \bar{e}'')$, or we are still in the situation where the successor has no root in distance $k$. But then $(\mathfrak{I}^F, d \cdots d_i) \Longleftrightarrow_{2(k+1)-i} (\mathfrak{H}^F, e \cdots e_i)$ for all $i \in \{1, \ldots, \ell + 1\}$ and the induction hypothesis applies again. This shows $(\mathfrak{I}, \bar{d}') \overset{i}{\Longleftrightarrow}_{k+1} (\mathfrak{H}^F, \bar{e}')$ and hence proves together with the result from above that $\mathfrak{I}^F \overset{gi}{\Longleftrightarrow}_{k+1} \mathfrak{H}^F$ □

Proof of Proposition 6.1.8. Let $\mathfrak{I} \models C$ and $\mathfrak{I} \overset{g}{\Longleftrightarrow}_{2k} \mathfrak{H}$. We show $\mathfrak{H} \models C$. Since $\mathfrak{I} \overset{g}{\Longleftrightarrow} \mathfrak{I}^F$ we have $\mathfrak{I}^F \models C$. The rank of $C$ implies that $C$ is preserved under global $\mathcal{ALCI}$-$k$-bisimulation and with Lemma 6.1.9 we obtain $\mathfrak{I}^F \overset{gi}{\Longleftrightarrow}_k \mathfrak{H}^F$, so that $\mathfrak{H}^F \models C$. Since $\mathfrak{H}^F \overset{g}{\Longleftrightarrow} \mathfrak{H}$ we obtain $\mathfrak{H} \models C$. □

Hence an $\mathcal{ALC}$-rewritable $\mathcal{ALCI}$-TBox of depth $k$ can be rewritten as $\mathcal{ALC}$-TBox of at most rank $2k$. This estimation is sharp: Let $\mathcal{T}' := \{\exists r^-.\top \sqsubseteq \exists r.\top\}$ then $\mathcal{T} := \{\top \sqsubseteq \forall r.\exists r.\top\}$ is a logically equivalent $\mathcal{ALC}$-TBox with minimal rank. Indeed, this result can be obtained by checking that all TBoxes formed from concepts

$$\{\top, \bot, \exists r.\top, \exists r.\bot, \forall r.\top, \forall r.\bot\}$$

with rank equal to 1 are not equivalent to $\mathcal{T}'$ by creating a counterexample. It is now not difficult to extend this example to $\mathcal{T}' := \{(\exists r^-)^n.\top \sqsubseteq (\exists r)^n.\top\}$ and $\mathcal{T} := \{\top \sqsubseteq (\forall r)^n.(\exists r)^n.\top\}$ for every $n < \omega$. It follows that rank $\mathcal{T} = 2 \cdot$ rank $\mathcal{T}'$.

Now $|C|$ and therefore $|\mathcal{T}|$ can be estimated by looking at characteristic concepts. We can estimate a lower bound of the length of a characteristic concept $X^{n+1}_{\mathfrak{I},d}$ by estimating how many characteristic concepts there are on level $n$: e.g. $X^1_{\mathfrak{I},d}$ can contain $\exists r.X^0_{\mathfrak{I},d'}$ for every possible $r \in \mathsf{N_R}$ and every possible characteristic concept $X^0_{\mathfrak{I},d'}$ as subconcepts. There are $|\mathsf{N_R}| \cdot 2^{|\mathsf{N_C}|}$-many concepts of the form $\exists r.X^0_{\mathfrak{I},d'}$ which could be contained in $X^1_{\mathfrak{I},d}$. $X^2_{\mathfrak{I},d}$ could therefore contain $|\mathsf{N_R}| \cdot 2^{|\mathsf{N_R}| \cdot 2^{|\mathsf{N_C}|}}$ many. It is not difficult to see that we get a tower of powers.

Hence the length grows very fast very large and is for level $n$ in the order of a tower of powers of height $n$. So the length of a characteristic concept on level $n$ can be estimated by a tower of powers at whose top we have $|\tau|$ as exponent.

As discussed earlier, the rank of the translation of the TBox $\mathcal{T}'$ is $n \leq 2 \cdot \operatorname{rank} \mathcal{T}'$ and so the translation could be in the order of a tower of powers of height $2 \cdot \operatorname{rank} \mathcal{T}'$ with $|\tau|$ as top exponent. Apart from this considerations tighter bounds for rewritings in the same signature have not been determined.

## 6.2 The $\mathcal{ALC}$-to-$\mathcal{EL}$ Rewritability Problem

We shall now give two algorithms which together decide the $\mathcal{ALC}$-to-$\mathcal{EL}$ Rewritability problem: The first one which will check whether or not a given TBox $\mathcal{T}$ is invariant under global equi-simulation and the second one which will check whether or not a given TBox is preserved under products. The characterisation theorem for $\mathcal{EL}$-TBoxes (Theorem 5.1.37) yields, if both algorithms return a positive result, that the given $\mathcal{ALC}$-TBox can be expressed as $\mathcal{EL}$-TBox.

We start with the algorithm which is meant to determine whether or not $Z \subseteq \operatorname{tp} \mathcal{T} \times \operatorname{tp} \setminus \operatorname{tp} \mathcal{T}$ is empty, where Z contains all pairs $(s, t)$ such that there are pointed interpretations $(\mathfrak{I}, d)$ and $(\mathfrak{H}, e)$

1. $(\mathfrak{H}, e) \overset{\forall}{\Longleftrightarrow} (\mathfrak{I}, d)$
2. $(\mathfrak{H}, e) \models s$ and $(\mathfrak{I}, d) \models t$
3. $\mathfrak{H} \models \mathcal{T}$.

Observation 6.2.1 . *$\mathcal{T}$ is not $\mathcal{EL}$-rewritable iff there is $s \in tp \setminus \operatorname{tp} \mathcal{T}$ and $t \in \operatorname{tp} \mathcal{T}$ such that $(s, t) \in Z$.*

The algorithm is inspired by the one used to determine $\mathcal{ALCI}$-to-$\mathcal{ALC}$ rewritability: In order to determine the pairs in $Z$, we shall consider the components of pairs $(s, t) \in \operatorname{tp} \mathcal{T} \times (\operatorname{tp} \setminus \operatorname{tp} \mathcal{T})$ as possible distinguished elements in pointed interpretations $(\mathfrak{I}, t)$ and $(\mathfrak{H}, s)$ with $(\mathfrak{I}, t) \models t$ and $(\mathfrak{H}, s) \models s$. But how can we determine the entire of $\mathfrak{I}$ and $\mathfrak{H}$?

Let us decouple $\mathfrak{H}$ from $\mathfrak{I}$ and neglect their interplay for a moment. How could we build a model of $\mathcal{T}$ such that $(\mathfrak{H}, s) \models s$? Clearly $s \in \operatorname{tp} \mathcal{T}$ and therefore it is complete w.r.t. to concept names i.e. for all $A \in \mathsf{N_C}$ either $A \in s$ or $\neg A \in s$. We simply label $s \in \Delta^{\mathfrak{H}}$ accordingly.

However, if $\exists r.C \in s$ then we must have an $r$-successor $\tilde{s}' \in \Delta^{\mathfrak{H}}$ that satisfies $C$. Since $\mathfrak{H}$ is supposed to satisfy $\mathcal{T}$, this $\tilde{s}'$ must satisfy an element from $\mathrm{tp}\,\mathcal{T}$. Again, we can consider this element $\tilde{s}'$ being identical to the $s' \in \mathrm{tp}\,\mathcal{T}$ containing all concepts from $\mathrm{clos}\,\mathcal{T}$ satisfied by $\tilde{s}'$. Or conversely possible $r$-successors of $s$ are at first all $s' \in \mathrm{tp}\,\mathcal{T}$ which are not 'prohibited' by $s$ because, e.g. $\exists r.\neg D \in s$ and $D \in s'$.

Connecting $s$ and $s'$ and inductively carrying on from $s'$ will eventually lead to an interpretation $(\mathfrak{H}, s) \models s$ and even $\mathfrak{H} \models \mathcal{T}$. If for some element $s_0$ with $\exists r.C \in s_0$, occurring along the way, no possible $r$-successor $s_0'$ exists, such that $C \in s_0'$, we must remove $s_0$ as possible candidate: This could lead to several backtracking steps, deleting more elements which now turn out to lack $s_0$ as only possible $r'$-successor.

Though, we not only need to construct $\mathfrak{H}$ but also $\mathfrak{I}$ such that $(\mathfrak{H}, s) \Longleftrightarrow (\mathfrak{I}, t)$. So we have to coordinate both procedures which create $(\mathfrak{H}, s)$ and $(\mathfrak{I}, t)$: Since every pair $(s, t) \in Z$ is contained in

$$Y_0 := \{(s, t) \in \mathrm{tp}\,\mathcal{T} \times \mathrm{tp} \mid \forall A \in \mathsf{N_C} : A \in s \iff A \in t\}$$

we could simply delete pairs from $Y_0$ for which we cannot find appropriate successors such that $(\mathfrak{H}, s) \Longrightarrow (\mathfrak{I}, t)$ and $(\mathfrak{I}, t) \Longrightarrow (\mathfrak{H}, s)$. This must eventually transform $Y_0$ into $Z$. Hence we need to determine whether $\mathfrak{H}$ and $\mathfrak{I}$ exist, such that $(\mathfrak{H}, s) \Longrightarrow (\mathfrak{I}, t)$ and $(\mathfrak{I}, t) \Longrightarrow (\mathfrak{H}, s)$ and, at the same time, $\mathfrak{I} \overset{\forall}{\Longleftrightarrow} \mathfrak{H}$.

Unfortunately, there is a technical difficulty concerning the determination of $Z$: Assume we were about to test whether $(s, t) \in Y_0$ is also in $Z$ by constructing interpretations $(\mathfrak{I}, t)$ and $(\mathfrak{H}, s)$. Assume furthermore that $\exists r.C \in t$.

In order obtain $(\mathfrak{I}, d) \Longleftrightarrow (\mathfrak{H}, e)$ we have to assure that $(\mathfrak{I}, d) \Longrightarrow (\mathfrak{H}, e)$ and $(\mathfrak{H}, e) \Longrightarrow (\mathfrak{I}, d)$. Hence for the $r$-successor $t_0$ of $t$ which we have to introduce as witness of $\exists r.C$ there must be some $r$-successor $s_0$ of $s$ satisfying $(\mathfrak{I}, t_0) \Longrightarrow (\mathfrak{H}, s_0)$. But in particular we also have to assure $(\mathfrak{H}, s) \Longrightarrow (\mathfrak{I}, t)$ and so there must also be some $r$-successor $t_1'$ in $\mathfrak{I}$ such that $(\mathfrak{H}, s_1') \Longrightarrow (\mathfrak{I}, t_1')$ and so on. Figure 6.1 illustrates this seemingly infinite ascending chain.
Lemma 6.2.2 further down however states that this chain stabilises in elements $d^*$ and $e^*$ as the Figure 6.1 suggests. But the subconcept $C$ in $\exists r.C \in t$ might be an $\mathcal{ALC}$-concept which is not preserved under $\Longrightarrow$ ; we cannot guarantee that $(\mathfrak{I}, t_0) \models C$ and $(\mathfrak{I}, t_0) \Longrightarrow (\mathfrak{I}, d^*)$ implies $(\mathfrak{I}, d^*) \models C$. Hence it is not sufficient to take $d^*$ instead of $t_0$.

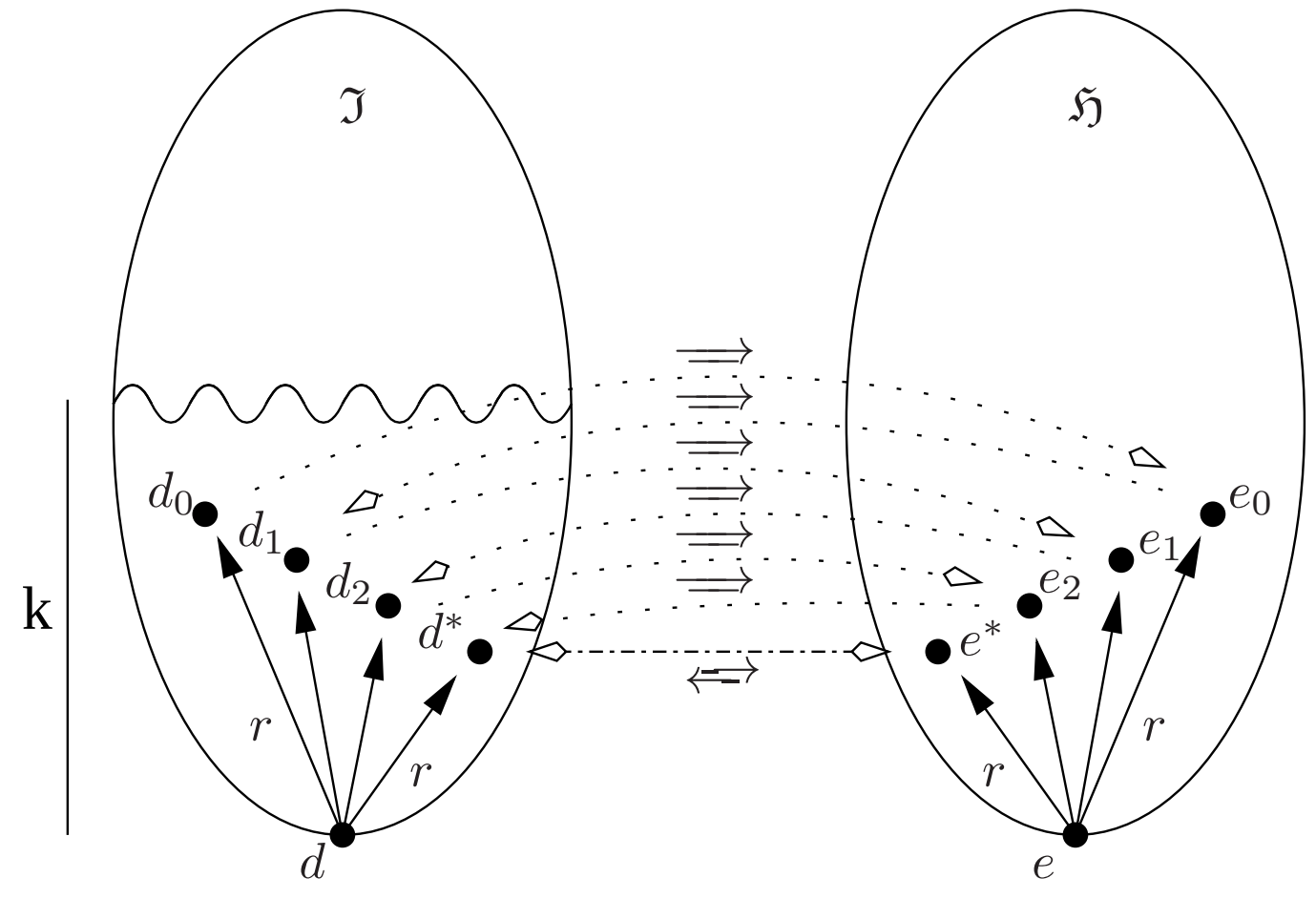

Figure 6.1: The ascending chain of simulating $r$-successors, oscillating between $\mathfrak{I}$ and $\mathfrak{H}$, stabilises in two equisimilar elements $d^*$ and $e^*$.

We therefore have to make the constraint that we can find a quadruple $(s_0, t_0, s_1, t_1)$, where (Cf. Rule 1 in Algorithm 6.2.5)

1. $t_0$ is the witness for $\exists r.C \in t$
2. $s_0$ ensures $\mathfrak{I} \overset{\forall}{\Longleftrightarrow} \mathfrak{H}$ by providing $(\mathfrak{I}, t_0) \Longleftrightarrow (\mathfrak{H}, s_0)$
3. $t_1 := t^{\mathfrak{I}}(d^*)$ and $s_1 := s^{\mathfrak{H}}(e^*)$ are the types of the end points of the reciprocal simulation chain from $t_0$ and $t_1$

If now $\exists r.C \in t_0$ we shall need an $r$-successor $t'_0$ of $t_0$ with $(\mathfrak{I}, t'_0) \vDash t'_0$ and $C \in t'_0$ with $t'_1$ and $s'_1$ being the elements provided by Lemma 6.2.2. Additionally we have to keep book about the development of $t_1$ and $s_1$: they need to have $r$-successors $t'_2$ and $s'_2$, respectively, such that again $(\mathfrak{I}, t'_2) \Longleftrightarrow (\mathfrak{H}, s'_2)$ but also that $(\mathfrak{I}, t'_0) \Longrightarrow (\mathfrak{I}, t'_2)$. If the latter is not satisfied, we cannot guarantee $(\mathfrak{I}, t_0) \Longrightarrow (\mathfrak{I}, t_1)$. If the former is not satisfied $\mathfrak{H} \overset{\forall}{\Longleftrightarrow} \mathfrak{I}$ is endangered. As consequence we obtain an even bigger tuple $(s'_0, t'_0, s'_1, t'_1, s'_2, t'_2)$. Thus, the tuples in $Z$ grow.

But the tuples do not grow indefinitely: In contrast to $\mathfrak{H}$, where every element needs to satisfy some element of tp $\mathcal{T}$ to ensure $\mathfrak{H} \vDash \mathcal{T}$, it suffices that *some* element in $\mathfrak{I}$ satisfies some $t \in \text{tp} \setminus \text{tp}\,\mathcal{T}$ to create a counterexample for $\mathfrak{I} \vDash \mathcal{T}$. Since every $t \in \text{tp} \setminus \text{tp}\,\mathcal{T}$ is finite, $\bigsqcap t$ is $\ell$-local, where $\ell = \max\{\text{rank}\, C \mid C \in t\}$. I.e. it is enough if we ensure that possible $r$-successors of the distinguished element $t$ satisfy some element $t' \in \text{tp}$ up to rank $t - 1$ where $t'$ is not forbidden by $t$. This will ensure $(\mathfrak{I}, t) \models t$. As soon as we reach after at most rank $\mathcal{T}$-many steps $t'$ with rank $t' = 0$, any continuation is acceptable.

Similarly we have to ensure that all concepts in newly introduced $t_i$ components are satisfied which leads to Rule 4 in Algorithm 6.2.5. In contrast to those $t \in \text{tp} \setminus \text{tp}\,\mathcal{T}$, we cannot shrink the rank of the $s \in \text{tp}\,\mathcal{T}$, as also successor elements must satisfy $\mathcal{T}$ and therefore must satisfy some $s \in \text{tp}\,\mathcal{T}$. Hence an increase in length of the tuples caused by elements from tp $\mathcal{T}$ would not stop.

But as mentioned before: it suffices to have one counterexample for $\mathcal{T}$ in $\mathfrak{I}$ to defy $\mathfrak{I} \models \mathcal{T}$. Hence a disjoint union $\mathfrak{I} \uplus \mathfrak{H}$ is not a model of $\mathcal{T}$. So, for an $r$-successor $s_0'$ of $s_0$, say, being a witness for $\exists r.C \in s_0$, we need not provide those end-points of the reciprocal simulation chain: we can assume that $s_0$ is contained in $\mathfrak{I}$, which preserves $\mathfrak{I} \overset{\forall}{\Longleftrightarrow} \mathfrak{H}$. We can therefore make due with an auxiliary component, later simply represented by an $s$ without index, which takes care of realising witnesses for $\exists r.C \in s$ (Cf. Rule 2, Rule 3 and Rule 5 in Algorithm 6.2.5).

After stating and proving Lemma 6.2.2 which was announced above, we shall give the algorithm and its necessary definitions.

Lemma 6.2.2. *Let $(\mathfrak{I}, d) \Longleftrightarrow (\mathfrak{H}, e)$ and let $\mathfrak{I}$, $\mathfrak{H}$ be $\omega$-saturated. For every $r$-successor $d'$ of $d$ in $\mathfrak{I}$ there is an $r$-successor $d^*$ of $d$ in $\mathfrak{I}$ and an $r$-successor $e^*$ of $e$ in $\mathfrak{H}$ such that*

$$(\mathfrak{I}, d') \Longrightarrow (\mathfrak{I}, d^*) \Longleftrightarrow (\mathfrak{H}, e^*)$$

To this end, we prove the following

Lemma 6.2.3. *Let $X := \{d_0 \in \Delta^{\mathfrak{I}} \mid (d, d_0) \in r^{\mathfrak{I}}$ and $(\mathfrak{I}, d') \Longrightarrow (\mathfrak{I}, d_0)\}$. We show that $X$ has a maximal element w.r.t. $\Longrightarrow$*

Proof. We have to show that there is some element $d^* \in X$ such that for all $d_0 \in X$ we have $(\mathfrak{I}, d_0) \Longrightarrow (\mathfrak{I}, d^*)$ implies $(\mathfrak{I}, d^*) \Longrightarrow (\mathfrak{I}, d_0)$.

In order to apply Zorn's Lemma we have to show that every $\Longrightarrow$-chain $K \subseteq X$ has an upper $\Longrightarrow$-bound $d^+$ in $X$. We shall apply the following steps: We show that $\Gamma := \biguplus_{d_0 \in K} \text{Th}_{\mathcal{EL}}(\mathfrak{I}, d_0)$ is satisfied by some $r$-successor $d^+$ of $d$ and then, using the $\omega$-saturatedness of $\mathfrak{I}$, that $d^+$ simulates every element in $K$. From Zorn's Lemma we then obtain the maximal element $d^*$.

Let $\Theta := \bigcup_{d_0 \in K} \{d_0\} \times \text{Th}_{\mathcal{EL}}(\mathfrak{I}, d_0)$ and let $\Theta_0 \subseteq \Theta$ be finite. Set

$$K' := \{d_0 \in K \mid \exists C \in \text{Th}_{\mathcal{EL}}(\mathfrak{I}, d_0) : (d_0, C) \in \Theta_0\},$$

then $K'$ is a finite $\Longrightarrow$-chain. Take the $\Longrightarrow$-greatest element $d_1$ in $K'$. We have $(\mathfrak{I}, d_0) \Longrightarrow (\mathfrak{I}, d_1)$ for all $d_0 \in K'$ and, since all $\mathcal{EL}$ concepts are preserved by $\Longrightarrow$, we also have $\mathrm{Th}_{\mathcal{EL}}(\mathfrak{I}, d_0) \subseteq \mathrm{Th}_{\mathcal{EL}}(\mathfrak{I}, d_1)$. Hence $(\mathfrak{I}, d_1) \models \bigcup_{d_0 \in K} \mathrm{Th}_{\mathcal{EL}}(\mathfrak{I}, d_0)$. This shows that every finite subset of $\Gamma$ is satisfied at some $r$-successor of $d$

Hence $\Gamma$ is an $r$-type of $d$. Since $\mathfrak{I}$ is $\omega$-saturated, $\Gamma$ must be realised by some $r$-successor $d^+$. But also, since $\mathfrak{I}$ is $\omega$-saturated, the Hennessy-Milner-Property for $\mathcal{EL}$ (Proposition 5.1.30) yields for every $d_1 \in \Delta^{\mathfrak{I}}$ with $\mathrm{Th}_{\mathcal{EL}}(\mathfrak{I}, d_1) \subseteq \mathrm{Th}_{\mathcal{EL}}(\mathfrak{I}, d^+)$ that $(\mathfrak{I}, d_1) \Longrightarrow (\mathfrak{I}, d^+)$. By definition of $X$, $d^+$ must be in $X$

This shows that every $\Longrightarrow$-chain $K \subseteq X$ has an upper bound in $X$ and Zorn's Lemma yields that there must be a $\Longrightarrow$-maximal element $d^* \in X$. □

Proof of Lemma 6.2.2. Let $X$ be defined as in Lemma 6.2.3 and let $d^*$ be a maximal element of $X$. Since $(\mathfrak{I}, d) \Longleftrightarrow (\mathfrak{H}, e)$ there is some $r$-successor $e^*$ of $e$ in $\mathfrak{H}$ such that $(\mathfrak{I}, d^*) \Longrightarrow (\mathfrak{H}, e^*)$. But we also have $(\mathfrak{H}, e^*) \Longrightarrow (\mathfrak{I}, d^*)$:

$(\mathfrak{I}, d) \Longleftrightarrow (\mathfrak{H}, e)$ provides $d'$ such that $(\mathfrak{H}, e^*) \Longrightarrow (\mathfrak{I}, d')$. Transitivity entails for $(\mathfrak{I}, d_1) \Longrightarrow (\mathfrak{I}, d^*) \Longrightarrow (\mathfrak{H}, e^*) \Longrightarrow (\mathfrak{I}, d')$ that $d' \in X$ and $(\mathfrak{I}, d^*) \Longrightarrow (\mathfrak{I}, d')$. Since $d^*$ is $\Longrightarrow$-maximal in $X$, we have $(\mathfrak{I}, d') \Longrightarrow (\mathfrak{I}, d^*)$ and therefore that $(\mathfrak{H}, e^*) \Longrightarrow (\mathfrak{I}, d^*)$. □

### 6.2.1 Algorithm to Determine Invariance under Equi-Simulation

Let $\mathrm{clos}\,\mathcal{T}$ be defined as for $\mathcal{ALC}$-TBoxes. tp is the set of all satisfiable subsets of $\mathrm{clos}\,\mathcal{T}$ and and $\mathrm{tp}\,\mathcal{T}$ are all those $s \in \mathrm{tp}$ such that $\mathcal{T} \cup s$ is satisfiable. We denote with $\mathrm{tp}^k := \{t \in \mathrm{tp} \mid \forall C \in t : \mathrm{rank}\, C \leq k\}$.

For $t, t' \in \mathrm{tp}$ we define $t \rightsquigarrow_r t'$ if $\neg\exists r.C \in t$ implies $C \notin t'$. Indeed this relation is the same as the one used for the $\mathcal{ALCI}$-to-$\mathcal{ALC}$-rewritability problem, as $\neg\exists r^-.C \in t' \implies C \notin t$ is trivially true for all $t, t' \subseteq \mathcal{ALC}(\tau)$.

For all $k, \ell \geq 0$ with $k + \ell \leq \mathrm{rank}\,\mathcal{T}$ we define

$$X^k_\ell := (s, s_0, t_0, \dots s_\ell, t_\ell) \mid \forall i \in \{0, \dots \ell\} : s, s_i \in \mathrm{tp}\,\mathcal{T} \text{ and } t_i \in \mathrm{tp}^k\}$$

Then $Z^k_\ell$ is the subset of $X^k_\ell$ such that for every $(s, s_0, t_0, \dots s_\ell, t_\ell) \in Z^k_\ell$ there are pointed interpretations $(\mathfrak{I}_0, d_0), \dots, (\mathfrak{I}_\ell, d_\ell)$ and $(\mathfrak{H}, e), (\mathfrak{H}_0, e_0), \dots, (\mathfrak{H}_\ell, e_\ell)$ such that

1. $(\mathfrak{I}_i, d_i) \models t_i$ for all $i \in \{0, \dots, \ell\}$
2. $(\mathfrak{H}, e) \models s$ and $(\mathfrak{H}_i, e_i) \models s_i$ for all $i \in \{0, \dots, \ell\}$

3. $\mathfrak{H} \models \mathcal{T}$ and $\mathfrak{H}_i \models \mathcal{T}$ for all $i \in \{0, \dots, \ell\}$

4. $(\mathfrak{I}_i, d_i) \overset{\forall}{\Longleftrightarrow} (\mathfrak{H}_i, e_i)$ for all $i \in \{0, \dots, \ell\}$

5. $(\mathfrak{I}_i, d_i) \Longrightarrow (\mathfrak{I}_{i+1}, d_{i+1})$ for all $i \in \{1, \dots, \ell - 1\}$

6. $(\mathfrak{H}, e) \Longrightarrow (\mathfrak{H}_0, e_0)$

These definitions give rise to the following observation:

Observation 6.2.4. *$\mathcal{T}$ is not invariant under $\overset{\forall}{\Longleftrightarrow}$ iff there is $k \leq \operatorname{rank} \mathcal{T}$ as well as $t \in \operatorname{tp} \setminus \operatorname{tp} \mathcal{T}$ and $s \in \operatorname{tp} \mathcal{T}$ such that $(s, s, t) \in Z_0^k$*

In this sense, for all $k, \ell \geq 0$ with $k + \ell \leq \operatorname{rank} \mathcal{T}$, the set $Z_\ell^k$ is a target set to which our algorithm will have to reduce the corresponding $X_\ell^k$ or rather its subset $Y_\ell^k$ from which Algorithm 6.2.5 will set out. If we can show that the set $Final_\ell^k \subseteq Y_\ell^k$ which the algorithm leaves behind equals to $Z_\ell^k$ then Observation 6.2.4 tells us that we simply need to check $\mathrm{Final}_0^k$ for tuples $(s, s, t)$ with $t \in \operatorname{tp} \setminus \operatorname{tp} \mathcal{T}$ and $s \in \operatorname{tp} \mathcal{T}$ to decide whether or not $\mathcal{T}$ is preserved by global equi-simulation. We shall thus use $Z_\ell^k$ to show that the algorithm is correct.

Algorithm 6.2.5. As initial sets for the algorithm, we use $Y_\ell^m$, with $0 \leq \ell, m$ and $\ell + m \leq \operatorname{rank} \mathcal{T}$, where $Y_\ell^m$ consists of all tuples $(s, s_0, t_0, \dots s_\ell, t_\ell) \in X_\ell^k$ such that for all $A \in \mathsf{N_C}$ we have

1. $A \in s \Longrightarrow A \in s_0$

2. $A \in t_i \Longrightarrow A \in t_{i+1}$ for all $i \in \{0, \dots, \ell\}$

3. $A \in t_i \iff A \in s_i$ for all $i \in \{0, \dots, \ell\}$

Starting with $Y_\ell^m$ the algorithm applies the following rules to each set $Y_\ell^m$ obtaining a new set after each application: Let $(s, s_0, t_0, \dots s_\ell, t_\ell) \in Y_\ell^m$ and for the purpose of this algorithm, set $m - 1 := 0$ if $m = 0$. We remove this tuple from $Y_\ell^m$, i.e.

We set $Y_\ell^m := Y_\ell^m \setminus \{(s, s_0, t_0, \dots t_\ell, s_\ell)\}$ whenever

Rule 1

$m > 0$ and $\exists r.C \in t_0$ and there is no $(s', s'_0, t'_0, \dots, s_\ell, t_\ell, t'_{\ell+1}, s'_{\ell+1}) \in Y^{m-1}_{\ell+1}$ with

1. $C \in t'_0$
2. $t_0 \rightsquigarrow_r t'_0$ and $t_0 \rightsquigarrow_r t'_1$, $s_0 \rightsquigarrow_r s'_1$
3. $t_i \rightsquigarrow_r t'_{i+1}$ and $s_i \rightsquigarrow_r s'_{i+1}$ for all $i \in \{1, \dots \ell\}$

Rule 2
$\exists r.C \in s_0$ and there is no $(s', s'_0, t'_0, \dots s'_\ell, t'_\ell) \in Y^{m-1}_\ell$ with

1. $C \in s'$
2. $t_0 \rightsquigarrow_r t'_0$, $s_0 \rightsquigarrow_r s'_0$ and $s_0 \rightsquigarrow_r s'$
3. $t_i \rightsquigarrow_r t'_i$ and $s_i \rightsquigarrow_r s'_i$ for all $i \in \{1, \dots \ell\}$

Rule 3
$\exists r.C \in s$ and there is no $(s', s'_0, t'_0, \dots, s'_\ell, t'_\ell) \in Y^{m-1}_\ell$ with

1. $C \in s'$
2. $t_0 \rightsquigarrow_r t'_0$, $s_0 \rightsquigarrow_r s'_0$ and $s \rightsquigarrow_r s'$,
3. $t_i \rightsquigarrow_r t'_i$ and $s_i \rightsquigarrow_r s'_i$ for all $i \in \{1, \dots \ell\}$

Rule 4
$m > 0$ and for some $i \in \{1, \dots, \ell\}$ there is $\exists r.C \in t_i$ and there is no $(s', s'_0, t'_0, s'_1, t'_1, \dots, s_{\ell-i+1}, t_{\ell-i+1}) \in Y^{m-1}_{\ell-i+1}$ with

1. $C \in t'_0$
2. $t_i \rightsquigarrow_r t'_0$ and $t_i \rightsquigarrow_r t'_1$, $s_i \rightsquigarrow_r s'_1$
3. $t_h \rightsquigarrow_r t'_{h-i+1}$, $s_h \rightsquigarrow_r s'_{h-i+1}$ for $i < h \leq \ell$

Rule 5
for some $i \in \{1, \dots, \ell\}$ there is $\exists r.C \in s_i$ and there is no $(s', s'_0, t'_0, \dots, s_{\ell-i}, t_{\ell-i}) \in Y^{m-1}_{\ell-i}$ with

1. $C \in s'$
2. $s_i \rightsquigarrow_r s'$ and $t_i \rightsquigarrow_r t'_0$, $s_i \rightsquigarrow_r s'_0$
3. $t_h \rightsquigarrow_r t'_{h-i}$, $s_h \rightsquigarrow_r s'_{h-i}$ for $i < h \leq \ell$

Please note that the requirement $m > 0$ in RULE 1 and RULE 4 is redundant: We assume $\exists r.C \in t$, which implies $m > 0$. But it is made explicit in order to show that $\ell + 1$ is not 'out of bounds'.

Since for all $\ell, m \geq 0$ with $\ell + m \leq \operatorname{rank} \mathcal{T}$ all sets $Y_\ell^m$ are finite, this algorithm terminates as soon as no rule is applicable. We denote with $Final_\ell^m$ the subset of $Y_\ell^m$ which is left after the algorithm terminates.

### 6.2.2 Proof of Correctness for the Algorithm

Algorithm 6.2.5 starts out with sets $Y_\ell^m$ whose shape and form has been discussed in quite some length: In essence the tuples keep track of several parallel simulations which originally emanated from two equisimilar points. The algorithm deletes all those tuples from $Y_\ell^m$ for which either the simulation cannot be maintained (mainly condition 2. and 3. in the rules) or for which the types $s$, $t$, and $s_i$, $t_i$ with $i \leq \ell$ cannot be realised (condition 1.). Thus only those tuples are left, whose types can be realised and appropriate simulations and equi-simulations exists; confer to properties 1–6 of $Z_\ell^k$ on page 213 or to Table 6.1 for a diagram.

In order to prove correctness for Algorithm 6.2.5, we have to show that $Z_\ell^m = \mathrm{Final}_\ell^m$. We split this task into Proposition 6.2.6, showing $Z_\ell^m \subseteq \mathrm{Final}_\ell^m$ , i.e. for all $m \in \{0, \dots, k\}$ no tuple is deleted by any rule r(1)–r(5) from $Y_\ell^m$ which is in $Z_\ell^m$ and Proposition 6.2.8 showing $\mathrm{Final}_\ell^m \subseteq Z_\ell^m$.

PROPOSITION 6.2.6 . *For all $m, \ell < \omega$ with $m + \ell \leq \operatorname{rank} \mathcal{T}$ we have $Z_\ell^m \subseteq Final_\ell^m$.*

Since this claim is trivial if $Z_\ell^m = \emptyset$, we shall assume that $Z_\ell^m \neq \emptyset$. The proof will be carried out by induction upon $m$, where we shall infer from the induction hypothesis that $Z_\ell^m \subseteq \mathrm{Final}_\ell^m$ for all $\ell$ with $m + \ell \leq \operatorname{rank} \mathcal{T}$. The induction hypothesis is proven separately. We set $t^{\mathfrak{I}}(d) := \{C \in \operatorname{clos} \mathcal{T} \mid (\mathfrak{I}, d) \models C\}$ and analogously

$$t_m^{\mathfrak{I}}(d) := \{C \in \operatorname{clos} \mathcal{T} \mid (\mathfrak{I}, d) \models C \text{ and } \operatorname{rank} C \leq m\}.$$

LEMMA 6.2.7 . $Z_\ell^0 \subseteq Y_\ell^0$.

For the proof of Lemma 6.2.7 we need to hark back to the constructions we shall be developing in the step cases. We shall therefore, without circular reasoning, present the step-cases first and then prove the base case.

PROOF OF PROPOSITION 6.2.6. Let $(m+1)+\ell \leq \operatorname{rank} \mathcal{T}$. Assume $(s, s_0, t_0, \dots, s_\ell, t_\ell) \in Z_\ell^{m+1}$. The following setup can be inferred, where w.l.o.g. $\mathfrak{H}$ and for all $i \in$

$\{0, \ldots, \ell\}$ $\mathfrak{I}_i, \mathfrak{H}_i$ can be assumed to be $\omega$-saturated:

$$
\begin{array}{ccccccc}
 & & t_0 & & t_i & & t_{i+1} \\
 & & \perp\!\!\perp & & \perp\!\!\perp & & \perp\!\!\perp \\
 & & (\mathfrak{I}_0, d_0) & \Longrightarrow & (\mathfrak{I}_i, d_i) & \Longrightarrow & (\mathfrak{I}_{i+1}, d_{i+1}) \\
 & & {>}\!\Updownarrow & & {>}\!\Updownarrow & & {>}\!\Updownarrow \\
(\mathfrak{H}, e) & \Longrightarrow & (\mathfrak{H}, e_0) & & (\mathfrak{H}_i, e_i) & & (\mathfrak{H}_{i+1}, e_{i+1}) \\
\top\!\!\top & & \top\!\!\top & & \top\!\!\top & & \top\!\!\top \\
s & & s_0 & & s_i & & s_{i+1}
\end{array}
$$

Table 6.1: Tuples in $Z^m_\ell$ represent the depicted situation, where $1 \leq i < \ell$

For the sake of contradiction assume $(s, s_0, t_0, \ldots, s_\ell, t_\ell)$ has been deleted from $Y^{m+1}_\ell$ by one of the rules (r1)–(r5). For each rule a contradiction is derived by constructing a tuple, required by the rule to not exist, which will be contained, depending on the rule, in $Z^m_{\ell+1}$, $Z^m_\ell$, $Z^m_1$ or $Z^m_0$ respectively. The induction hypothesis then yields the contradiction.

### Rule 1

We have $m + 1 > 0$; assume $\exists r.C \in t$. Using the diagram above, we can infer that there is some $r$-successor $d'_0$ of $d_0$ in $\mathfrak{I}_0$ such that $(\mathfrak{I}_0, d'_0) \models C$. Since $\mathfrak{I}_0 \overset{\forall}{\Longleftrightarrow} \mathfrak{H}_0$ there is some $e'_0$ in $\mathfrak{H}_0$ such that $(\mathfrak{I}_0, d'_0) \Longleftrightarrow (\mathfrak{H}_0, e'_0)$. Note that $e'_0$ is not necessarily an $r$-successor of $e_0$. We set $t'_0 := t^{\mathfrak{I}_0}_m(d'_0)$, $s'_0 := t^{\mathfrak{H}_0}(e'_0)$ and $s := s'_0$. Since $d'_0$ is an $r$-successor of $d_0$, we have $t_0 \rightsquigarrow_r t'_0$ and $C \in t'_0$ follows from the definition of $t'_0$.

Since $(\mathfrak{I}_0, d_0) \Longleftrightarrow (\mathfrak{H}_0, e_0)$ we find with Lemma 6.2.2 an $r$-successor $d^*_0$ of $d_0$ and $e^*_0$ of $e_0$ such that $(\mathfrak{I}_0, d'_0) \Longrightarrow (\mathfrak{I}_0, d^*_0) \Longleftrightarrow (\mathfrak{H}_0, e^*_0)$. We set $t'_1 := t^{\mathfrak{I}_0}_m(d^*_0)$ and $s'_1 := t^{\mathfrak{H}_0}(e^*_0)$ and obtain, since $d^*_0$ is an $r$-successor of $d_0$ and $e^*_0$ is an $r$-successor of $e_0$, that $t_0 \rightsquigarrow_r t'_1$ and $s_0 \rightsquigarrow_r s'_1$.

Incrementing $i$ stepwise by 1 from 0 up to $\ell - 1$, we find the appropriate candidates for $d'_i$ and $e'_i$ as follows: Assume we have already found $d'_i$ in the previous step. Then, since $(\mathfrak{I}_i, d_i) \Longrightarrow (\mathfrak{I}_{i+1}, d_{i+1})$, there is some $r$-successor of $d'_{i+1}$ of $d_{i+1}$ such that $(\mathfrak{I}_i, d'_i) \Longrightarrow (\mathfrak{I}_{i+1}, d'_{i+1})$. Since $(\mathfrak{I}_{i+1}, d_{i+1}) \Longleftrightarrow (\mathfrak{H}_{i+1}, e_{i+1})$ we find $r$-successors $d^*_{i+1}$ of $d_{i+1}$ and $e^*_{i+1}$ of $e_{i+1}$ such that $(\mathfrak{I}_{i+1}, d'_{i+1}) \Longrightarrow (\mathfrak{I}_{i+1}, d^*_{i+1}) \Longleftrightarrow (\mathfrak{H}_{i+1}, e^*_{i+1})$.

Transitivity now yields $(\mathfrak{I}_i, d'_i) \Longrightarrow (\mathfrak{I}_{i+1}, d^*_{i+1}) \Longleftrightarrow (\mathfrak{H}_{i+1}, e^*_{i+1})$ and we set $t'_{i+2} := t^{\mathfrak{I}_{i+1}}_m(d^*_{i+1})$ and $s'_{i+2} := s^{\mathfrak{H}_{i+1}}(e^*_{i+1})$. We obtain $t_i \rightsquigarrow_r t'_{i+1}$ and $s_i \rightsquigarrow_r s'_{i+1}$ for $i \in \{0, \ldots, \ell\}$. Note that we have an index-shift; both, the assignment and the sub-sequent statement are correct in their indices:

$$
\begin{array}{ccccccccc}
 & (\mathfrak{I}_0, d_0) & & \Longrightarrow & (\mathfrak{I}_i, d_i) & \Longrightarrow & (\mathfrak{I}_{i+1}, d_{i+1}) & & \\
 & \top\!\!\top & & & \top\!\!\top & & \top\!\!\top & & \\
 & t_0 & & & t_i & & t_{i+1} & & \\
\swarrow_r & & \searrow_r & & & \searrow_r & & \searrow_r & \\
t'_0 & & & t'_1 & & t'_{i+1} & & & t'_{i+2} \\
\perp\!\!\perp & & & \perp\!\!\perp & & \perp\!\!\perp & & & \perp\!\!\perp \\
(\mathfrak{I}_0, d'_0) & \Longrightarrow & & (\mathfrak{I}_0, d^*_0) & \Longrightarrow & (\mathfrak{I}_i, d^*_i) & & \Longrightarrow & (\mathfrak{I}_{i+1}, d^*_{i+1})
\end{array}
$$

Therefore $(s', s'_0, t'_0, \dots s'_{\ell+1}, t'_{\ell+1}) \in Z^m_{\ell+1}$. The induction hypothesis $Z^m_{\ell+1} \subseteq \mathit{Final}^m_{\ell+1}$ together with $\mathit{Final}^m_{\ell+1} \subseteq Y^m_{\ell+1}$, yields $Z^m_{\ell+1} \subseteq Y^m_{\ell+1}$ and so Rule 1 is not applicable to $(s, s_0, t_0, \dots, s_\ell, t_\ell)$ in $Y^{m+1}_\ell$.

Rule 2

As $(\mathfrak{H}_0, e_0) \models s_0$ and $\exists r.C \in s_0$ there there is $e'_0 \in \Delta^{\mathfrak{H}}$ with $(e_0, e'_0) \in r^{\mathfrak{H}}$ and $e'_0 \in C^{\mathfrak{H}}$. By Lemma 6.2.2 we can find an $r$-successor $e^*_0$ of $e_0$ in $\mathfrak{H}_0$ and an $r$-successor $d^*_0$ of $d_0$ in $\mathfrak{I}_0$ such that $(\mathfrak{H}_0, e'_0) \Longrightarrow (\mathfrak{H}_0, e^*_0) \Longleftrightarrow (\mathfrak{I}_0, d^*_0)$. Of course, still $\mathfrak{I}_0 \overset{\forall}{\Longleftrightarrow} \mathfrak{H}_0$. We set $s' := t^{\mathfrak{H}_0}(e'_0)$, $s'_0 := t^{\mathfrak{H}_0}(e^*_0)$ and $t'_0 := t^{\mathfrak{I}_0}_m(d^*_0)$, which implies $C \in s'$. Since $e'_0$ and $e^*_0$ are $r$-successors of $e$ we have $s_0 \rightsquigarrow_r s'$ and $s_0 \rightsquigarrow_r s'_0$. The same argument shows $t_0 \rightsquigarrow_r t'_0$.

With the same step-wise construction from the case Rule 1 we obtain for each $i \in \{0, \dots, \ell - 1\}$ $r$-successors $d^*_{i+1}$ of $d_{i+1}$ and $e^*_{i+1}$ of $e_{i+1}$ such that $(\mathfrak{I}_i, d^*_i) \Longrightarrow (\mathfrak{I}_{i+1}, d^*_{i+1}) \Longleftrightarrow (\mathfrak{H}_{i+1}, e^*_{i+1})$. We set $t_{i+1} := t^{\mathfrak{I}_{i+1}}_m(d^*_{i+1})$ and $s := t^{\mathfrak{H}_{i+1}}(e^*_{i+1})$ for each $i \in \{0, \dots, \ell - 1\}$ and thus obtain $t_i \rightsquigarrow_r t'_i$ and $s_i \rightsquigarrow_r s'_i$ for all $i \in \{0, \dots, \ell\}$.

It shows $(t', s', s'_0, t'_1, s'_1, \dots, t'_\ell, s'_\ell) \in Z^{m-1}_\ell$ and with the induction hypothesis $(t', s', s'_0, t'_1, s'_1, \dots, t'_\ell, s'_\ell) \in Y^{m-1}_\ell$ is inferred. Hence Rule 2 was not applicable to $(s, s_0, t_0, \dots, s_\ell, t_\ell)$ in $Y^{m+1}_\ell$, which derives the contradiction.

Rule 3

The case is similar to the case Rule 2: Assume $\exists r.C \in s$. The setup displayed in the first diagram shows that we can choose an $r$-successor $e'$ of $e$ with $(\mathfrak{H}, e') \models C$. Accordingly there is some $r$-successor $e'_0$ of $e_0$ such that $(\mathfrak{H}, e') \Longrightarrow (\mathfrak{H}_0, e'_0)$.

Lemma 6.2.2 yields $r$-successors $d^*_0$ of $d_0$ and $e^*_0$ of $e_0$ such that $(\mathfrak{H}_0, e'_0) \Longrightarrow (\mathfrak{H}, e^*_0) \Longleftrightarrow (\mathfrak{I}, d^*_0)$ and therefore that $(\mathfrak{H}, e) \Longrightarrow (\mathfrak{H}, e^*_0) \Longleftrightarrow (\mathfrak{I}, d^*_0)$ We set $t_0 := t^{\mathfrak{I}_0}_m(d^*_0)$, $s_0 := t^{\mathfrak{H}_0}(e^*_0)$ and $s := t^{\mathfrak{H}}(e')$. The $r$-successorship for each of them yields

$s \rightsquigarrow_r s'$, $s_0 \rightsquigarrow_r s_0'$ as well as $t_0 \rightsquigarrow_r t_0'$.

Analogously to Rule 1 we can find $t_i'$ and $s_i'$ such that $(s', s_0', t_0', \dots, t_\ell', s_\ell') \in Z_\ell^m$. Using the induction hypothesis one can argue in the same fashion of Rule 2 that Rule 3 is not applicable to $(s, s_0, t_0, \dots, t_\ell, s_\ell)$ in $Y_\ell^{m+1}$.

Rule 4 uses the rationale of Rule 1 and Rule 5 use the same rationale as Rule 2. □

Proof of Lemma 6.2.7. Let $Y_\ell^0(n)$ denote the set $Y_\ell^0$ after the $n$-th rule application. We shall show by induction upon $n$ that

$$\text{for all } n < \omega \text{ and all } \ell \leq \operatorname{rank} \mathcal{T} \text{ we have } Z_\ell^0 \subseteq Y_\ell^0(n).$$

Note that we do need the definition $m - 1 := 0$ if $m = 0$ used in Algorithm 6.2.5!

The base case is clear from the fact that $Y_\ell^0(0)$ is the initial set which contains $Z_\ell^0$: Let $(s, s_0, t_0, \dots, t_\ell, s_\ell) \in Z_\ell^0$ then from the definition of $Z_\ell^0$ we obtain all requirements 1–4 so that $(t, s, s_0, t_1, s_1, \dots, t_\ell, s_\ell)$ is a member of the initial $Y_\ell^0(0)$.

Assume now $Z_\ell^0 \subseteq Y_\ell^0(n)$. Clearly rule Rule 1 and Rule 4 are not applicable to $Y_\ell^0(n+1)$ as $m > 0$ is required. We show that none of the rules Rule 2, Rule 3 and Rule 5 is applicable to this tuple:

Let $(s, s_0, t_0, \dots, t_\ell, s_\ell) \in Z_\ell^0$. Assume $\exists r.C \in s_0 / \exists r.C \in s$. The construction method used in the step case for Rule 2/Rule 3 created a tuple $(s', s_0', t_0', \dots, t_\ell', s_\ell')$ which is in $Z_\ell^0$ and satisfies all requirements imposed on $(s', s_0', t_0', \dots, t_\ell', s_\ell')$ in Rule 2/Rule 3. By induction hypothesis $Z_\ell^0 \subseteq Y_\ell^0(n)$ and so Rule 2/Rule 3 is not applicable to $(t, s, s_0, t_1, s_1, \dots, t_\ell, s_\ell)$ in $Y_\ell^0(n)$.

For Rule 5, assume for some $i \in 1, \dots, \ell$ we have $\exists r.C \in s_i$. Again we find $(s', s_0', t_0', \dots s_{\ell-i}', t_{\ell-i}') \in Z_1^0$ such that all requirements imposed on it by Rule 5 are met. By induction hypothesis, $Z_1^0 \subseteq Y_1^0(n)$ and so Rule 5 cannot be applied to $(s, s_0, t_0 \dots, t_\ell, s_\ell)$ in the $n+1$-st loop pass of the algorithm.

Since none of the rules is applicable, it shows that $(s, s_0, t_0, \dots, t_\ell, s_\ell) \in Y_\ell^0(n+1)$ and therefore that $Z_\ell^0 \subseteq Y_\ell^0$ for all stages of $Y_\ell^0$. □

Proposition 6.2.8 . *For all $m, \ell < \omega$ with $m + \ell \leq \operatorname{rank} \mathcal{T}$ we have $Final_\ell^m \subseteq Z_\ell^m$*

Again, the claim is trivial if $Final_\ell^m = \emptyset$ and we hence assume $Final_\ell^m \neq \emptyset$. We show that we can construct for each element in $Final_\ell^m$ interpretations as required by $Z_\ell^m$, which proves that this element must be an element of $Z_\ell^m$.

To this end, we consider the components of the tuples $\bar{d} := (s, s_0, t_0, \dots, s_\ell, t_\ell)$ of $Z_\ell^m$ as being indexed from 0 to $2\ell + 3$. For a tuple $\bar{d}$ we write $\bar{d}_i$ for the $i$-th component. We define the interpretation $\mathfrak{I}$ by setting

$$\Delta^{\mathfrak{I}} := \bigcup_{m+\ell \le \operatorname{rank} \mathcal{T}} \{(\bar{d}, i) \mid \bar{d} \in \mathit{Final}^{m}_{\ell} \text{ and } 0 \le i \le 2\ell + 3\}$$

and $A^{\mathfrak{I}} := \{(\bar{d}_i, i) \mid A \in \bar{d}_i\}$. For each $r \in \mathsf{N_R}$ we define $((\bar{d}, i), (\bar{d}', j)) \in r^{\mathfrak{I}}$, where $\bar{d} := (s, s_0, t_0, \dots, t_\ell, s_\ell)$, whenever one of the following condition holds

(r1) $\bar{d}' = (s', s'_0, t'_0, \dots, t'_{\ell+1}, s'_{\ell+1}) \in \mathit{Final}^{m-1}_{\ell+1}$ with (cf. rule (r1))
$t_0 \rightsquigarrow_r t'_0$ and $t_k \rightsquigarrow_r t'_{k+1}$, $s_k \rightsquigarrow_r s'_{k+1}$ for $0 \le k \le \ell$
and $(\bar{d}_i, \bar{d}'_j) \in \{(t_0, t'_0)\} \cup \{(t_k, t'_{k+1}), (s_k, s'_{k+1}) \mid 0 \le k \le \ell\}$

(r2) $\bar{d}' := (s', s'_0, t'_0, \dots, s'_\ell, t'_\ell) \in \mathit{Final}^{m-1}_{\ell}$ with (cf. rule (r2))
$s_0 \rightsquigarrow_r s'$ and $t_k \rightsquigarrow_r t'_k$, $s_k \rightsquigarrow_r s'_k$ for $0 \le k \le \ell$
and $(\bar{d}_i, \bar{d}'_j) \in \{(s_0, s')\} \cup \{(t_k, t'_k), (s_k, s'_k) \mid 0 \le k \le \ell\}$,

(r3) $\bar{d}' := (s', s'_0, t'_0, \dots, s'_\ell, t'_\ell) \in \mathit{Final}^{m-1}_{\ell}$ with (cf. rule (r3))
$s \rightsquigarrow_r s'$ and $t_k \rightsquigarrow_r t'_k$, $s_k \rightsquigarrow_r s'_k$ for $0 \le k \le \ell$
and $(\bar{d}_i, \bar{d}'_j) \in \{(s, s')\} \cup \{(t_k, t'_k), (s_k, s'_k) \mid 0 \le k \le \ell\}$

(r4) there is $i \in \{1, \dots, \ell\}$ and $\bar{d}' := (s', s'_0, t'_0, \dots, s'_{\ell-i+1}, t'_{\ell-i+1}) \in \mathit{Final}^{m-1}_{\ell-i+1}$
such that (cf. rule (r4)) $t_i \rightsquigarrow_r t'_0$ and $t_i \rightsquigarrow_r t'_1$, $s_i \rightsquigarrow_r s'_1$
and for all $h \in \{i, \dots, \ell\}$ we have $t_h \rightsquigarrow_r t'_{h-i+1}$ and $s_h \rightsquigarrow_r s'_{h-i+1}$
and $(\bar{d}_i, \bar{d}'_j) \in \{(t_i, t'_0)\} \cup \{(t_h, t'_{h-i+1}), (s_h, s'_{h-i+1}) \mid i \le h \le \ell\}$.

(r5) there is $i \in \{1, \dots, \ell\}$ and $\bar{d}' := (s', s'_0, t'_0, \dots, s'_{\ell-i}, t'_{\ell-i}) \in \mathit{Final}^{m-1}_{\ell-i}$
such that (cf. rule (r5)) $s_i \rightsquigarrow_r s'$ and $s_i \rightsquigarrow_r s'_0$, $t_i \rightsquigarrow_r t'_0$
and for all $h \in \{i, \dots, \ell\}$: $t_h \rightsquigarrow_r t'_{h-i}$ and $s_h \rightsquigarrow_r s'_{h-i}$
and $(\bar{d}_i, \bar{d}'_j) \in \{(s_i, s')\} \cup \{(t_h, t'_{h-i}), (s_h, s'_{h-i}) \mid i \le h \le \ell\}$.

Lemma 6.2.9 . *Let* $(\bar{d}, i) \in \Delta^{\mathfrak{I}}$ *with* $\bar{d}_i \in \operatorname{tp}^m$. *For all* $C \in \mathit{clos}\,\mathcal{T}$ *with* rank $C \le m$ *we have* $(\mathfrak{I}, (\bar{d}, i)) \models C \iff C \in \bar{d}_i$.

Proof. The proof is carried out for fixed $m$ by induction upon rank $C$. If rank $C = 0$ then $C$ is atomic and the definition of $A^{\mathfrak{I}}$ readily yields $(\mathfrak{I}, (\bar{d}, i)) \models C \iff C \in \bar{d}_i$.

Assume rank $C = n + 1$ and $n + 1 \le m$. If $C = \neg D$ we may show $(\mathfrak{I}, (\bar{d}, i)) \models D \iff D \in \bar{d}_i$ instead since $\bar{d}_i$ is complete with respect to $\operatorname{clos}\mathcal{T}$. Similarly, since $D_0 \sqcap D_1 \in d_i$ implies $D_0, D_1 \in d_i$ by the definition of $\operatorname{clos}\mathcal{T}$, we can show $(\mathfrak{I}, (\bar{d}, i)) \models D_i \iff D_i \in \bar{d}_i$ for $i \in \{0, 1\}$ instead. Hence the step case reduces to show $(\mathfrak{I}, (\bar{d}, i)) \models \exists r.D \iff \exists r.D \in \bar{d}_i$.

Assume $C = \exists r.D \in \operatorname{clos} \mathcal{T}$ with $0 < \operatorname{rank} C \leq m$ and $(\mathfrak{I}, (\bar{d}, i)) \models \exists r.D$. Then there is some $(\bar{d}', j) \in \Delta^{\mathfrak{I}}$ such that $((\bar{d}, i), (\bar{d}', j)) \in r^{\mathfrak{I}}$ with $(\mathfrak{I}, (\bar{d}', j)) \models D$. Hence, by definition of $r^{\mathfrak{I}}$, $(\bar{d}_i, \bar{d}_j)$ is some pair, say $(t_k, t'_k) \in \operatorname{tp}^m \times \operatorname{tp}^{m-1}$.

The induction hypothesis yields $D \in t'$. From the definition of $r$ we also know that then $t \rightsquigarrow_r t'$ (check all 5 items) and the definition of $\rightsquigarrow_r$ shows that $D \in t'$ implies $\neg\exists r.D \notin t$. Since $\exists r.D \in \operatorname{clos} \mathcal{T}$ and $t$ is complete w.r.t. $\operatorname{clos} \mathcal{T}$ up to rank $m$ we infer that $\exists r.D \in t$.

This argument can be used for all possible combinations $(s, s')$, $(s_0, s')$, $(s_k, s'_k)$, $(t_k, t'_k)$ etc, where $s, s_k \in \operatorname{tp} \mathcal{T} \subseteq \operatorname{tp}^{\operatorname{rank} \mathcal{T}}$ and $(t_k, t'_k) \in \operatorname{tp}^m \times \operatorname{tp}^{m-1}$, showing for all $(\bar{d}, i) \in \Delta^{\mathfrak{I}}$ that $(\mathfrak{I}, (\bar{d}, i)) \models \exists r.D$ implies $\exists r.D \in d_i$.

For the only-if direction, assume $\exists r.D \in \bar{d}_i$. In case $\bar{d}_i = t_0$ we use the non-applicability of Rule 1 to infer the existence of $\bar{d}' \in Final^{m-1}_{\ell+1}$ with $D \in \bar{d}'_i$ (where $\bar{d}'_i = t'_0$). The definition of $r^{\mathfrak{I}}$ yields $((\bar{d}, i), (\bar{d}', i)) \in r^{\mathfrak{I}}$. With the induction hypothesis we conclude that $(\mathfrak{I}, (\bar{d}', i)) \models D$ and hence that $(\mathfrak{I}, (\bar{d}, i)) \models \exists r.D$.

Analogously we can use non-applicability of Rule 2 if $\bar{d}_i = s_0$, Rule 3 if $\bar{d}_i = s$, the one of Rule 4 if $\bar{d}_i = t_k$ and the non-applicability for Rule 5 if $\bar{d}_i = s_k$ for any $k \in \{1, \ldots, \ell\}$ where $\ell$ is the length of the tuple $\bar{d}$. This shows for all $(\bar{d}, i) \in \Delta^{\mathfrak{I}}$ that $(\mathfrak{I}, (\bar{d}', i)) \models D$ implies $(\mathfrak{I}, (\bar{d}, i)) \models \exists r.D$ □

Lemma 6.2.10. *Let* $\bar{d} := (s, s_0, t_0, \ldots, t_\ell, s_\ell)$, *and* $(\bar{d}, i), (\bar{d}, j) \in \Delta^{\mathfrak{I}}$.

$$(\bar{d}_i, \bar{d}_j) \in \{(t_k, t_{k+1}) \mid 0 \leq k < \ell\} \implies (\mathfrak{I}, (\bar{d}, i)) \Longrightarrow (\mathfrak{I}, (\bar{d}, j))$$

$$(\bar{d}_i, \bar{d}_j) \in \{(t_k, s_k) \mid 1 \leq k \leq \ell\} \implies (\mathfrak{I}, (\bar{d}, i)) \Longleftrightarrow (\mathfrak{I}, (\bar{d}, j))$$

Proof. For the following relation, we shall prove that it is a simulation

$$S := \bigcup_{m+\ell \leq \operatorname{rank} \mathcal{T}} \{((\bar{d}, i), (\bar{d}, j)) \mid \bar{d} \in Z^m_\ell \text{ with } \bar{d}_i = t_k \text{ and } \bar{d}_j = t_{k'}, 0 \leq k \leq k' < \ell\}$$

Assume $((\bar{d}, i), (\bar{d}, j)) \in S$. All tuples in $Y^m_\ell$ satisfy $A \in d_i \implies A \in d_j$ for all $A \in \mathsf{N_C}$ and in particular those which are in $Final^m_\ell$. With Lemma 6.2.9 we infer $(\mathfrak{I}, (\bar{d}, i)) \models A \implies (\mathfrak{I}, (\bar{d}, j)) \models A$. Hence it only remains to show that $S$ is closed under the FORTH condition:

Let $\bar{d} \in Z^m_\ell$ and $((\bar{d}, i), (\bar{d}, j)) \in S$. Assume $((\bar{d}, i), (\bar{d}', i')) \in r^{\mathfrak{I}}$.

Case 1

If $\bar{d}'$ is longer than $\bar{d}$ then the elements are connected due to $(r1)$.

1. Either $d_i = t_0$, then $\bar{d}'_{i'} = t'_0$ or $\bar{d}'_{i'} = \bar{d}'_{i+2} = t'_1$
2. or there is $k \in \{1, \dots, \ell\}$ (shift!) with $\bar{d}_i = t_k$ and $\bar{d}'_{i'} = t'_{k+1}$.

In either case, $(\bar{d}, j) = t_{k'}$ is connected via an $r$-edge to $(\bar{d}', j+2)$ with $\bar{d}'_{j+2} = t_{k'+1}$ for each $k' \in \{0, \dots, \ell\}$ . Since $k \le k'$, by definition of $S$, we have $k < k+1 \le k'+1$ and so $((\bar{d}', i'), (\bar{d}', j')) \in S$.

Case 2

If $\bar{d}'$ has the same length as $\bar{d}$ then the $r$-edge is due to either (r2) or (r3) or (r4). We treat the latter in Case 3. Is the $r$-edge due to (r2) and (r3) we know that $(\bar{d}', i') = (\bar{d}', i)$ and also that $((\bar{d}, j), (\bar{d}', j)) \in r^{\mathfrak{I}}$. Since we have $k \le k'$ we know by definition of $S$ that $((\bar{d}', i), (\bar{d}', j)) \in S$.

Case 3

If the $r$-edge is due to (r4) then there is some $h \in \{1, \dots, \ell\}$ with $\bar{d}' \in Z^{m-1}_{\ell-h+1}$. In this case, $k \le h \le \ell$ and $d_i = t_k$ where $(\bar{d}, i)$ is via some $r$-edge connected to $(\bar{d}', i')$ with $\bar{d}'_{i'} = t_{k-h+1}$. Similarly $(\bar{d}, j)$ is connected to $(\bar{d}', j')$ with $\bar{d}'_{j'} = t_{k'-h+1}$. Again we obtain that $((\bar{d}', i'), (\bar{d}, j')) \in S$.

If the $r$-edge is due to (r5) the argument is analogously to where the edge is caused by (r4). □

Lemma 6.2.11 . *Let* $\bar{d} := (s, s_0, t_0, \dots, t_\ell, s_\ell)$, *and* $(\bar{d}, i), (\bar{d}, j) \in \Delta^{\mathfrak{I}}$.

$$(\bar{d}_i, \bar{d}_j) = \{(s, s_0)\} \quad \Longrightarrow \quad (\mathfrak{I}, (\bar{d}, i)) \Longrightarrow (\mathfrak{I}, (\bar{d}, j))$$

Proof. For the following relation we shall prove that it is a simulation.

$$S := \bigcup_{m+\ell \le \operatorname{rank} \mathcal{T}} \{((\bar{d}, i), (\bar{d}, j)) \mid \bar{d}_i = s \text{ and } \bar{d}_j = s_0\}$$

Again, $Y^m_\ell$ only contains tuples that satisfy $A \in s \Longrightarrow A \in s_0$ for all $A \in \mathsf{N_C}$. Lemma 6.2.9 yields that then $(\mathfrak{I}, (\bar{d}, i)) \models A \Longrightarrow (\mathfrak{I}, (\bar{d}, j)) \models A$ for all $A \in \mathsf{N_C}$. It remains to show the FORTH property:

Let $((\bar{d}, i), (\bar{d}, j)) \in S$. If $((\bar{d}, i)(\bar{d}', i')) \in r^{\mathfrak{I}}$ then only due to (r3). But this section of the definition of $r^{\mathfrak{I}}$ also stipulates that $((\bar{d}, j), (\bar{d}', j'))$ with $(\bar{d}', j') = s'_0$. Hence $((\bar{d}', i'), (\bar{d}', j')) \in S$. □

Lemma 6.2.12. *Let* $\bar{d} := (s, s_0, t_0, \dots, t_\ell, s_\ell)$*, and* $(\bar{d}, i), (\bar{d}, j) \in \Delta^{\mathfrak{I}}$.

$$(\bar{d}_i, \bar{d}_j) \in \{(t_k, s_k) \mid 1 \le k \le \ell\} \quad \Longrightarrow \quad (\mathfrak{I}, (\bar{d}, i)) \Longleftrightarrow (\mathfrak{I}, (\bar{d}, j))$$

Proof. We show that for each $k \in \{0, \dots, \ell\}$: **II** has a winning strategy in the simulation game $G(\mathfrak{I}, (\bar{d}, i); \mathfrak{I}, (\bar{d}, j))$ with $\bar{d}_i = t_k$ and $\bar{d}_j = s_k$, by maintaining configurations

$$(\mathfrak{I}, (\bar{d}', i'); \mathfrak{I}, (\bar{d}', j')) \text{ such that } \bar{d}'_{i'} = t_h \text{ and } \bar{d}'_{j'} = s_{h'} \text{ where } 0 \le h \le h' \le \ell \quad (*)$$

where $\bar{d}' \in Y^m_\ell$. By definition of $Y^m_\ell$, we know for each such configuration that $A \in d'_{i'}$ implies $A \in d'_{j'}$ is satisfied for all $A \in \mathsf{N_C}$ because $h \le h'$. Lemma 6.2.9 furnishes $(\mathfrak{I}, (\bar{d}', i')) \models A \Longrightarrow (\mathfrak{I}, (\bar{d}', j')) \models A$ for all $A \in \mathsf{N_C}$, and so all such configurations are admissible. In particular, **II** does not lose the 0-th round, as the start-configuration $(\mathfrak{I}, (\bar{d}, i); \mathfrak{I}, (\bar{d}, j))$ satisfies requirement $(*)$.

Assume the game has reached configuration $(\mathfrak{I}, (\bar{d}, i); \mathfrak{I}, (\bar{d}, j))$, which satisfies requirement $(*)$ and where $\bar{d} \in Y^m_\ell$. Let **I** challenge **II** by moving for some $r \in \mathsf{N_R}$ to some $r$-successor $(\bar{d}', i')$ of $(\bar{d}, i)$. Except for (r1) we have in all parts (r2)–(r4) of the definition of $r^{\mathfrak{I}}$ that if $(\bar{d}', i')$ is an $r$-successor of $(\bar{d}, i)$ where $\bar{d}_i = t_k$ then $\bar{d}'_{i'} = t'_k$ with the same index $k$. In (r2)–(r4) we then also find that $(\bar{d}, j)$ has an $r$-successor $(\bar{d}', j')$ where $\bar{d}_j = s_{k'}$ and $\bar{d}'_{j'} = s'_{k'}$. So, **II** can achieve a new configuration $(\mathfrak{I}, (\bar{d}', i'); \mathfrak{I}, (\bar{d}', j'))$ that meets requirement $(*)$.

The only exception is (r1). Let **I** challenge **II** by moving choosing $r \in \mathsf{N_R}$ and moving from $(\bar{d}, i)$ to some $r$-successor $(\bar{d}', i')$ where $\bar{d}' \in Z^{m-1}_{\ell+1}$. If $\bar{d}_i = t_0$ then we either have $\bar{d}'_{i'} = t'_0$ or $\bar{d}'_{i'} = t'_1$. In all other cases where $1 \le k \le \ell$ we have $\bar{d}'_{i'} = t'_{k+1}$. But (r1) also endows $(\bar{d}, j)$ where $\bar{d}_j = s_{k'}$ with an $r$-successor $(\bar{d}', j')$ where $\bar{d}'_j = s_{k'+1}$ for each $k' \in \{0, \dots, \ell\}$ So in any case of $k \in \{0, \dots, \ell\}$, **II** can provide a configuration $(\mathfrak{I}, (\bar{d}', i'); \mathfrak{I}, (\bar{d}', j'))$ with $k < k+1 \le k' + 1$, which therefore satisfies $(*)$.

We now show for each $k \in \{0, \dots, \ell\}$: **II** has a winning strategy in the simulation game $G(\mathfrak{I}, (\bar{d}, j); (\bar{d}, i))$ with $d_j = s_k$ and $d_i = t_{k'}$ if she maintains configurations

$$(\mathfrak{I}, (\bar{d}', j'); \mathfrak{I}, (\bar{d}', i')) \text{ so that either } d'_{j'} = s \text{ and } d'_{i'} = t_0 \text{ or } d'_{j'} = s_k \text{ and } d'_{i'} = t_k \quad (*)$$

where $0 \le k \le \ell$ and $\bar{d}' \in Y^m_\ell$. For each such configuration we know by definition of $Y^m_\ell$ that $A \in s$ implies $A \in s_0$ and also that $A \in s_k$ iff $A \in t_k$ for all $k \in \{0, \dots, \ell\}$ and all $A \in \mathsf{N_C}$. Lemma 6.2.9 then shows that $(\mathfrak{I}, (\bar{d}', j')) \models A \Longrightarrow (\mathfrak{I}, (\bar{d}', i')) \models A$ for all $A \in \mathsf{N_C}$ and all configurations satisfying $(*)$. Since the start configuration is of this kind, **II** does not lose the 0-th round.

Assume the game has reached configuration $(\mathfrak{I}, (\bar{d}, j); \mathfrak{I}, (\bar{d}, i))$, which satisfies requirement $(*)$ and where $\bar{d} \in Y^m_\ell$. Let **I** challenge **II** by moving for some $r \in \mathsf{N_R}$ to some $r$-successor $(\bar{d}', j')$ of $(\bar{d}, j)$.

If $\bar{d}' \in Z^{m-1}_{\ell+1}$, this $r$-successor is due to (r1) and hence $(\bar{d}, j) = s_k$ and $(\bar{d}', j') = s'_{k+1}$ for some $k \in \{0, \dots, \ell\}$. In this case **II** can move from $(\bar{d}, i)$ with $\bar{d}_i = t_k$ via an $r$-edge, due to (r1), to $(\bar{d}', i')$ with $\bar{d}'_{i'} = t'_{k+1}$. The resulting configuration $(\mathfrak{I}, (\bar{d}', j'); \mathfrak{I}, (\bar{d}', i'))$ meets $(*)$.

If $\bar{d}' \in Z^{m-1}_\ell$, this $r$-successor is either due to (r4), which will be treated further down, or (r2) and (r3) respectively. If $(\bar{d}, j) = s_k$ with $k \in \{1, \dots \ell\}$ then $(\bar{d}', j') = s'_k$ with the same index $k$. Both, (r2) and (r3) then provide an $r$-successor $(\bar{d}', i')$ for $(\bar{d}, i)$ such that $d'_{i'} = t'_k$. If **II** moves via $r$ to $(\bar{d}', i')$, the resulting configuration meets $(*)$.

If however $(\bar{d}, j) = s_0$ or $(\bar{d}, j) = s$, the $r$-successor could be $(\bar{d}', j') = s'_0$ or $(\bar{d}', j') = s'$ depending on (r2) and (r3). In either case, since the configuration met $(*)$, we know $\bar{d}_i = t_0$ and both (r2) and (r3) furnish an $r$-successor $(\bar{d}', i')$ of $(\bar{d}, i)$ with $(\bar{d}', i') = t'_0$. **II** can play this $r$-successor $(\bar{d}', i')$ and obtains a new configuration which meets $(*)$.

If $\bar{d}' \in Z^{m-1}_{\ell-h}$ or $\bar{d}' \in Z^{m-1}_{\ell-h+1}$, where $1 \leq h \leq \ell$, then the $r$-successor is due to (r4) or (r5). Since the current configuration meets $(*)$, $(\bar{d}, j) = s_k$ implies $(\bar{d}, i) = t_k$ for all $k \in \{h, \dots, \ell\}$. Assume, in the case of (r4), **I** challenges **II** by moving to $(\bar{d}', j')$ with $d'_{j'} = s'_{k-h}$. (r4) and (r5) provide an $r$-successor $(\bar{d}', i')$ of $(\bar{d}, i)$ with $\bar{d}'_{i'} = t'_{k-h}$. By choosing $(\bar{d}', i')$, **II** obtains a configuration which satisfies the desired properties of $(*)$. The same argument applies for (r5).

In case of (r5), **I** could challenge **II** by moving to $(\bar{d}', j')$ with $d'_{j'} = s'$, then (r5) also connects $(\bar{d}, i)$ with $(\bar{d}', i')$ where $\bar{d}'_{i'} = t'_0$. Moving to this $r$-successor, **II** obtains an admissible configuration in the sense of $(*)$.

We showed at first that $(\mathfrak{I}, (\bar{d}, i)) \Longrightarrow (\mathfrak{I}, (\bar{d}, j))$ and we have just shown now that $(\mathfrak{I}, (\bar{d}, j)) \Longrightarrow (\mathfrak{I}, (\bar{d}, i))$. This proves Lemma 6.2.12 □

Proof of Proposition 6.2.8. Let $\mathfrak{I}$ be the interpretation we have just defined. We can now extract $\mathfrak{H}$ from $\mathfrak{I}$: Let

$$\Delta^{\mathfrak{H}} := \{(\bar{d}, i) \in \Delta^{\mathfrak{I}} \mid \bar{d} = (s, s_0, t_0 \dots, s_\ell, t_\ell) \text{ and } \bar{d}_i \in \{s, s_0, \dots s_\ell\}\}$$

For all $r \in \mathsf{N_R}$ we define $A^{\mathfrak{H}} := A^{\mathfrak{I}} \cap \Delta^{\mathfrak{H}}$ and $r^{\mathfrak{H}} := r^{\mathfrak{I}} \cap (\Delta^{\mathfrak{H}} \times \Delta^{\mathfrak{H}})$.

We have $\mathfrak{I} \overset{\vee}{\Longleftrightarrow} \mathfrak{H}$: On the one hand $\mathfrak{H}$ is a sub-interpretation of $\mathfrak{I}$ on the other hand for every element $(\bar{d}, i) \in \Delta^{\mathfrak{I}}$ with $\bar{d} = (s, s_0, t_0, \dots, s_\ell, d_\ell)$ and $d_i \in \{t_0, \dots, t_\ell\}$ we know from Lemma 6.2.12 $(\mathfrak{I}, (\bar{d}, i)) \Longleftrightarrow (\mathfrak{H}, (\bar{d}, i+1))$.

Furthermore $\mathfrak{H} \models \mathcal{T}$: Let $(\bar{d}, j) \in \Delta^{\mathfrak{H}}$ arbitrary, where $\bar{d} = (s, s_0, t_0, \dots s_\ell, t_\ell)$ and $d_j = s$ or $d_j = s_k$ where $0 \leq k \leq \ell$. Let w.l.o.g. $d_j = s$. By definition of $\bar{d}$, we know that $s \in \operatorname{tp} \mathcal{T}$. In particular, we know $\neg C \in s$ or $D \in s$, for every $C \sqsubseteq D \in \mathcal{T}$. Hence every $(\mathfrak{H}, s) \models C \to D$ for all $C \sqsubseteq D \in \mathcal{T}$. Since this is true for all elements in $\mathfrak{H}$ we have $\mathfrak{H} \models \mathcal{T}$.

For each $\bar{d} \in \mathit{Final}^m_\ell$ with $\bar{d} = (s, s_0, t_0, \dots, s_\ell, t_\ell)$ we show that it satisfies the six conditions for being in $Z^m_\ell$:

1. $(\mathfrak{I}, (\bar{d}, i)) \models t_k$ where $i = 2 + 2k$ and $0 \leq k \leq \ell$
2. $(\mathfrak{H}, (\bar{d}, 0)) \models s$ and $(\mathfrak{H}_i, (\bar{d}, j)) \models s_k$ where $j = 2k + 1$ and $0 \leq k \leq \ell$
3. $\mathfrak{H} \models \mathcal{T}$
4. $(\mathfrak{H}, (\bar{d}, i)) \overset{\vee}{\Longleftrightarrow} (\mathfrak{I}, (\bar{d}, i+1)$ where $i = 2k + 1$ and $0 \leq k < \ell$
5. $(\mathfrak{I}, (\bar{d}, i)) \Longrightarrow (\mathfrak{I}, (\bar{d}, i+2))$ where $i = 2 + 2k$ and $0 \leq k < \ell$
6. $(\mathfrak{H}, (\bar{d}, 0)) \Longrightarrow (\mathfrak{H}, (\bar{d}, 1))$

1. and 2. are shown by Lemma 6.2.9. 3. Has just been argued at the beginning of this proof. 4. is yielded by $\mathfrak{I} \overset{\vee}{\Longleftrightarrow} \mathfrak{H}$, as argued in the beginning of this proof, and by Lemma 6.2.12. 5. is entailed by Lemma 6.2.10 and finally 6. follows from Lemma 6.2.11. This concludes the proof for $\mathit{Final}^m_\ell \subseteq Z^m_\ell$. □

### 6.2.3 Algorithm to Determine Preservation under Direct Products

In the characterisation of $\mathcal{EL}$-TBoxes we did not distinguish between being preserved under arbitrary direct products or being preserved merely under finite direct products. Indeed, for FO-sentences and in particular for $\mathcal{ALC}$-TBoxes this property coincides (cf. [34, Theorem 6.3.14]).

But for direct products $\mathfrak{I}_0 \times \dots \times \mathfrak{I}_n$ over a family $(\mathfrak{I}_i)_{i \leq n}$ of finitely many interpretations, we can determine the direct product by inductively forming pairwise products: $\mathfrak{H}_0 := \mathfrak{I}_0$ and $\mathfrak{H}_{k+1} := \mathfrak{H}_k \times \mathfrak{I}_{k+1}$ with $k < n$. If $\mathfrak{I}_k$ is a model $\mathcal{T}$ for each $k \in \{0, \dots, n\}$ then $\mathfrak{H}_k \models \mathcal{T}$ for each $k \in \{0, \dots n\}$ or $\mathcal{T}$ is not preserved under direct products. Hence the problem reduces to check for pairs of models of $\mathcal{T}$ whether $\mathcal{T}$ is preserved under direct products.

We shall show that it suffices to check for certain tree-interpretations satisfying elements from tp $\mathcal{T}$ whether or not their direct product amongst them lead to new tree-interpretations satisfying elements from tp $\mathcal{T}$. If this is not the case then $\mathcal{T}$ is not preserved under direct products.

Proposition 6.2.13. $\iff$ *respects* $\times$.

Proof. We have to show that $(\mathfrak{I}_0 \times \mathfrak{I}_1, (d_0, d_1)) \iff (\mathfrak{H}_0 \times \mathfrak{H}_1, (e_0, e_1))$ whenever $(\mathfrak{I}_0, d_0) \iff (\mathfrak{H}_0, e_0)$ and $(\mathfrak{I}_1, d_1) \iff (\mathfrak{H}_1, e_1)$: Assume the premise is true.

**II** has a winning-strategy by maintaining configurations $(\mathfrak{I}_0 \times \mathfrak{I}_1, (d_0', d_1')) \iff (\mathfrak{H}_0 \times \mathfrak{H}_1, (e_0', e_1'))$ such that $(\mathfrak{I}_0, d_0') \iff (\mathfrak{H}_0, e_0')$ and $(\mathfrak{I}_1, d_1') \iff (\mathfrak{H}_1, e_1')$.

We first show that for each such configuration $(d_0', d_1')$ and $(e_0', e_1')$ are atomically equivalent: Assume $(d_0', d_1') \in A^{\mathfrak{I}_0 \times \mathfrak{I}_1}$. Then $d_0' \in A^{\mathfrak{I}_0}$ and $d_1' \in A^{\mathfrak{I}_1}$. By assumption, $d_0'$ and $e_0'$ are atomically equivalent and so are $d_1'$ and $e_1'$. Hence $e_0' \in A^{\mathfrak{H}_0}$ and $e_1' \in A^{\mathfrak{H}_1}$ and by the definition of the direct product $(e_0', e_1') \in A^{\mathfrak{H}_0 \times \mathfrak{H}_1}$. With the same arguments we obtain the only-if direction.

In particular the start-configuration is of the desired kind, and so **II** does not lose the 0-th round. Assume now the game has reached the configuration $(\mathfrak{I}_0 \times \mathfrak{I}_1, (d_0', d_1')) \iff (\mathfrak{H}_0 \times \mathfrak{H}_1, (e_0', e_1'))$ with the desired property, and **I** challenges **II** by moving in $\mathfrak{I}_0 \times \mathfrak{I}_1$ from $(d_0', d_1')$ for some $r \in \mathsf{N_R}$ via an $r$-edge $(d_0'', d_1'')$. By the definition of the direct product we have $(d_0', d_0'') \in r^{\mathfrak{I}_0}$ and $(d_1', d_1'') \in r^{\mathfrak{I}_0}$.

Since the configuration is of the desired kind, there is an $r$-successor $e_0''$ of $e_0'$ in $\mathfrak{H}_0$ and some $r$-successor of $e_1''$ of $e_1'$ in $\mathfrak{H}_1$ such that $(\mathfrak{I}_0, d_0'') \iff (\mathfrak{H}_0, e_0'')$ and $(\mathfrak{I}_1, d_1'') \iff (\mathfrak{H}_1, e_1'')$. By the definition of the direct product we have that $(e_0'', e_1'')$ is an $r$-successor of $(e_0', e_1')$ in $\mathfrak{H}_0 \times \mathfrak{H}_1$. **II** can move via the $r$-edge to $(e_0'', e_1'')$ and obtains a new configuration which is of the desired type.

With the same arguments we can show that **II** can find an appropriate answer to any challenge from **I** in $\mathfrak{H}_0 \times \mathfrak{H}_1$ and thus that **II** has a winning-strategy in this game. □

Corollary 6.2.14. $\overset{g}{\iff}$ *respects* $\times$.

Using the same arguments one can show that $\iff_n$ and $\overset{g}{\iff}_n$ respect $\times$, too.

Recall that the forest unravelling (page 56) $\mathfrak{I}^F = \bigcup_{d \in \Delta^{\mathfrak{I}}} \mathfrak{I}_d$ is the union of all tree-unravellings $\mathfrak{I}_d$ in every element $d \in \Delta^{\mathfrak{I}}$. It is necessarily a disjoint union, as tree-unravellings consist of path-elements, which then all start with the root element $d$ in which they have been unravelled. (cf. page 33).

Since $\overset{g}{\iff}$ respects $\times$ we can for all interpretations $\mathfrak{I}, \mathfrak{H}$ change over to their forest unravellings $\mathfrak{I}^F$ and $\mathfrak{H}^F$. Then $\mathfrak{I}\times\mathfrak{H}$ is, since $\mathfrak{I}^F \overset{g}{\longleftrightarrow} \mathfrak{I}$ and $\mathfrak{H}^F \overset{g}{\longleftrightarrow} \mathfrak{H}$, globally bisimilar to $\mathfrak{I}^F \times \mathfrak{H}^F$. In particular $\mathfrak{I} \times \mathfrak{H} \models \mathcal{T}$ for some given $\mathcal{T}$ iff $\mathfrak{I}^F \times \mathfrak{H}^F \models \mathcal{T}$.

Lemma 6.2.15 . $|clos\,\mathcal{T}| \leq 2|\mathcal{T}|$

Proof. We first show the $|\text{clos}\,C| \leq 2|C|$ for $\mathcal{ALC}$-concepts $C$. If $A \in \mathsf{N_C}$ we have $\text{clos}\,A = \{A, \neg A\}$ and hence the claim is true in the atomic case. In case $C = \neg D$ then $|\text{clos}\,C| = |\text{clos}\,D|$ by definition of clos , where $|\text{clos}\,D|$ is by the induction hypothesis estimated as $|\text{clos}\,D| \leq 2|D| \leq 2|C|$.

For $C = D \sqcap E$ we have $|\text{clos}\,C| = |\text{clos}\,D \cup \text{clos}\,E \cup \{C, \neg C\}|$ by definition of clos which is bound by $|\text{clos}\,D| + |\text{clos}\,E| + 2$. The induction hypothesis yields $|\text{clos}\,D| + |\text{clos}\,E| + 2 \leq 2|D| + 2|E| + 2$ where $|D| + |E| = |C| - 1$, hence $2|D| + 2|E| + 2 = 2|C|$ which yields all in all $|\text{clos}\,C| \leq 2|C|$.

Finally for $C = \exists r.D$ we have $|\text{clos}\,C| = |\text{clos}\,D| \cup \{C, \neg C\}$ and so the induction hypothesis yields $|\text{clos}\,C| \leq |\text{clos}\,D| + 2 \leq 2|D| + 2$ where $|D| \leq |C| - 1$. Hence we obtain $|\text{clos}\,C| \leq 2|C|$.

We then have $|\text{clos}\,\mathcal{T}| = |\bigcup\{\text{clos}\,C \cup \text{clos}\,D \mid C \sqsubseteq D \in \mathcal{T}\}| \leq \sum\{2|C| + 2|D| \mid C \sqsubseteq D \in \mathcal{T}\}| \leq 2\sum\{|C \sqsubseteq D| \mid C \sqsubseteq D \in \mathcal{T}\}| \leq 2|\mathcal{T}|$. □

Proposition 6.2.16 . *An $\mathcal{ALC}$-TBox $\mathcal{T}$ is not preserved under direct products iff there are two tree-models $\mathfrak{I}$ and $\mathfrak{H}$ of $\mathcal{T}$ with out-degree at most* $2^{(|\mathcal{T}|^2)}$ *such that $\mathfrak{I} \times \mathfrak{H} \not\models \mathcal{T}$.*

In the proof, we shall be working with pruned tree-interpretations $\mathfrak{H}_0, \mathfrak{H}_1$, where unnecessary branches of $\mathfrak{I}$ and $\mathfrak{H}$ have been removed. The root $(d_0, d_1)$ of the forest-interpretation $\mathfrak{H}_0 \times \mathfrak{H}_1$ locally violates $\mathcal{T}$, i.e. if $\mathcal{T} = \{\top \sqsubseteq C\}$ then $(\mathfrak{H}_0 \times \mathfrak{H}_1, (d, e)) \models \neg C$.

Nevertheless, the estimation for the out-degree $2^{(|\mathcal{T}|^2)}$ seems to be high. Indeed, in the proof not the out-degree is estimated but the number of elements connected to the root of $\mathfrak{H}_0 \times \mathfrak{H}_1$ in depth rank $\mathcal{T}$ is estimated. If $2^{(|\mathcal{T}|^2)}$ is high, at which *possible* out-degree are we actually looking? Considered that there are at most $2^{|\mathcal{T}|}$ different types $t \subseteq \text{clos}\,\mathcal{T}$ each element in a tree-interpretation can have any subset of types as successors. Hence a tree of depth 2 could possibly branch into $2^{|\mathcal{T}|} \cdot |\mathfrak{P}(2^{|\mathcal{T}|})| = 2^{|\mathcal{T}| \cdot 2^{|\mathcal{T}|}}$ many subtrees without being redundant.

Lemma 6.2.17 . *An $\mathcal{ALC}$-TBox $\mathcal{T}$ is not preserved under direct products iff there are two tree-models $\mathfrak{I}$ and $\mathfrak{H}$ of $\mathcal{T}$ such that $\mathfrak{I} \times \mathfrak{H} \not\models \mathcal{T}$.*

Proof. The only-if direction is immediate. If $\mathcal{T}$ is not invariant under direct products, there are two models $\mathfrak{I}, \mathfrak{H}$ of $\mathcal{T}$ such that for some concept inclusion $C \sqsubseteq D \in \mathcal{T}$ there is some $(d, e) \in (C \sqcap \neg D)^{\Delta^{\mathfrak{I} \times \mathfrak{H}}}$, hence $\mathcal{T}$ is violated.

Let $\mathfrak{I}_d$ and $\mathfrak{H}_e$ be the tree-unravelling for $\mathfrak{I}$ in $d$ and, respectively, the tree-unravelling of $\mathfrak{H}$ in $e$. $\mathfrak{I}_d$ and $\mathfrak{H}_e$ are both models of $\mathcal{T}$: since $\mathcal{ALC}$-TBoxes are preserved under global bisimulation we have $\mathfrak{I}^F$ and $\mathfrak{H}^F$ are models of $\mathcal{T}$ and the invariance of $\mathcal{T}$ under disjoint unions yields that $\mathfrak{I}_d$ and $\mathfrak{H}_e$ are models of $\mathcal{T}$.

We have $(\mathfrak{I}, d) \iff (\mathfrak{I}_d, d)$ and analogously $(\mathfrak{H}, e) \iff (\mathfrak{H}_e, e)$ and Proposition 6.2.13 yields $(\mathfrak{I} \times \mathfrak{H}, (d, e)) \iff (\mathfrak{I}_d \times \mathfrak{H}_e, (d, e))$ which shows that $(\mathfrak{I}_d \times \mathfrak{H}_e, (d, e) \models C \sqcap \neg D$, hence $\mathcal{T}$ is violated in $\mathfrak{I}_d \times \mathfrak{H}_e$. □

Proof of Proposition of 6.2.16. The only-if direction is again immediate. To prove the if-direction let $\mathcal{T} = \{\top \sqsubseteq C\}$ which is not preserved under direct products and let $\mathfrak{I}_0, \mathfrak{I}_1$ be the tree-models provided by Lemma 6.2.17 with root-elements $d_0$ and $d_1$ such that $(\mathfrak{I}_0 \times \mathfrak{I}_1, (d_0, d_1)) \models \neg C$.

In the following we shall prune subtrees from the tree-interpretations $\mathfrak{I}_0$ and $\mathfrak{I}_1$. Let $\Gamma_i \subseteq \Delta^{\mathfrak{I}_i}$, $i \in \{0, 1\}$ be minimal sets, such that

1. $d_i \in \Gamma_i$
2. If $d_i' \in \Gamma_i$ and $(\mathfrak{I}_i, d_i') \models \exists r.C$ for some $\exists r.C \in \operatorname{clos} \mathcal{T}$ then there is $d_i'' \in \Gamma_i$ such that $(d_i', d_i'') \in r^{\mathfrak{I}_i}$ and $(\mathfrak{I}_i, d_i'') \models C$.
3. If $(d_0', d_1') \in \Gamma_0 \times \Gamma_1$ such that there is a path in $\mathfrak{I}_0 \times \mathfrak{I}_1$ from $(d_0, d_1)$ to $(d_0', d_1')$ of length $\ell \leq \operatorname{rank} \mathcal{T}$ and $(\mathfrak{I}_0 \times \mathfrak{I}_1, (d_0', d_1')) \models \exists r.C$ for some $\exists r.C \in \operatorname{clos} \mathcal{T}$ with $\operatorname{rank} \exists r.C \leq \operatorname{rank} - \ell$ then there is $(d_0'', d_1'') \in \Gamma_0 \times \Gamma_1$ such that $((d_0', d_1'), (d_0'', d_1'')) \in r^{\mathfrak{I}_0 \times \mathfrak{I}_1}$ and $(\mathfrak{I}_0 \times \mathfrak{I}_1, (d_0'', d_1'')) \models C$.

We define $\mathfrak{H}_i := \mathfrak{I}_i \restriction \Gamma_i$ for $i \in \{0, 1\}$. Note that $\mathfrak{H}_0$ and $\mathfrak{H}_1$ are not necessarily cut off in some depth: We will show that $\mathfrak{H}_0$ and $\mathfrak{H}_1$ satisfy $\mathcal{T}$; since they are tree-interpretations, this can make them infinite (assume $\mathcal{T} \models \{\top \sqsubseteq A, A \sqsubseteq \exists r.\top\}$). We show that $\mathfrak{H}_0 \times \mathfrak{H}_1, (d_0, d_1) \models \neg C$.

We show for each $i \in \{0, 1\}$ and all $C \in \operatorname{clos} \mathcal{T}$

1. for all $e \in \Delta^{\mathfrak{H}_i} : (\mathfrak{H}_i, e) \models C \iff (\mathfrak{I}_i, e) \models C$
2. for the root element $(d_0, d_1) \in \Delta^{\mathfrak{H}_0 \times \mathfrak{H}_1}$ we have $(\mathfrak{H}_0 \times \mathfrak{H}_1, (d_0, d_1)) \models C$ iff $(\mathfrak{I}_0 \times \mathfrak{I}_1, (d_0, d_1)) \models C$

We show the first claim by induction upon the structure of the concepts $C \in \operatorname{clos} \mathcal{T}$. Let $e \in \Delta^{\mathfrak{H}_i}$. Then $(\mathfrak{H}_i, e)$ and $(\mathfrak{I}_i, e)$ are atomically equivalent. Since $\operatorname{clos} \mathcal{T}$ is closed under negation and contains the separate conjuncts of conjunctions, the induction hypothesis readily applies in both cases.

In case $(\mathfrak{H}_i, e) \models \exists r.C$, there is an $r$-successor $e'$ of $e$ such that $(\mathfrak{H}_i, e') \models \exists r.C$. The induction hypothesis yields $(\mathfrak{H}_i, e') \models C$. Since $\mathfrak{H}_i$ is a restriction of $\mathfrak{I}_i$, we have that $(e, e') \in r^{\mathfrak{I}_i}$ iff $(e, e') \in r^{\mathfrak{H}_i}$ and so that $(\mathfrak{I}_i, d) \models \exists r.C$. If $(\mathfrak{I}_i, e) \models \exists r.C$, requirement 2. ensures that there is $e' \in \Delta^{\mathfrak{H}_i}$ such that $(e, e') \in r^{\mathfrak{H}_i}$ and $(\mathfrak{H}_i, e') \models C$. Hence $(\mathfrak{H}_i, e) \models \exists r.C$.

We show the following claim for all elements $(d_0', d_1')$ reachable from the root $(d_0, d_1)$ via a path of length $\ell < \operatorname{rank} \mathcal{T}$: For all $C \in \operatorname{clos} \mathcal{T}$ with $\operatorname{rank} C \leq \operatorname{rank} \mathcal{T} - \ell$ we have $(\mathfrak{H}_0 \times \mathfrak{H}_1, (d_0', d_1')) \models C$ iff $(\mathfrak{I}_0 \times \mathfrak{I}_1, (d_0', d_1')) \models C$.

The claim is clear for atomic concepts. Since $\operatorname{clos} \mathcal{T}$ is closed under negation and contains the conjuncts of conjunctions, the induction hypothesis trivially applies. Let $\exists r.C \in \operatorname{clos} \mathcal{T}$ and let $(d_0', d_1')$ be reachable from the root $(d_0, d_1)$ via a path of length $\ell$ and assume $\operatorname{rank} \exists r.C \leq \operatorname{rank} \mathcal{T} - \ell$. Then $(\mathfrak{H}_0 \times \mathfrak{H}_1, (d_0', d_1')) \models \exists r.C$ iff $((d_0'', d_1''), (d_0'', d_1'')) \in r^{\mathfrak{H}_0 \times \mathfrak{H}_1}$ and $(\mathfrak{H}_0 \times \mathfrak{H}_1, (d_0'', d_1'')) \models C$. If $((d_0'', d_1''), (d_0'', d_1'')) \in r^{\mathfrak{H}_0 \times \mathfrak{H}_1}$ then $((d_0'', d_1''), (d_0'', d_1'')) \in r^{\mathfrak{I}_0 \times \mathfrak{I}_1}$ and the induction hypothesis yields $(\mathfrak{I}_0 \times \mathfrak{I}_1, (d_0'', d_1'')) \models C$, whence $(\mathfrak{I}_0 \times \mathfrak{I}_1, (d_0', d_1')) \models \exists r.C$ is concluded. The only-if direction is argued by $\Gamma_0 \times \Gamma_1$ is closed under requirement 3.

We now estimate the out-degree:

Each element $e$ in $\mathfrak{H}_i$ obtains at most $|\mathcal{T}|$ many successors from requirement 2: Each successor of $e$ is due to some concept $\exists r.C \in \operatorname{clos} \mathcal{T}$ with $e \in (\exists r.C)^{\mathfrak{I}_i}$ where the cardinality of $\operatorname{clos} \mathcal{T}$ is estimated by $|\operatorname{clos} \mathcal{T}| \leq 2|\mathcal{T}|$. Clearly, since for every $C \in \operatorname{clos} \mathcal{T}$ we also have $\neg C \in \operatorname{clos} \mathcal{T}$, $e$ satisfies at most half of the concepts in $\operatorname{clos} \mathcal{T}$. Hence 2. introduces at most $|\mathcal{T}|$ successors. Since condition 3. is only required up to depth $\operatorname{rank} \mathcal{T}$ all elements in greater depth have only up to $|\mathcal{T}|$ successors.

If $(e_0, e_1)$ is a pair in the tree $\mathfrak{H}_0 \times \mathfrak{H}_1$ with root $(d_0, d_1)$ then both elements $e_0$ and $e_1$ appeared in the same depth of each tree-interpretation $\mathfrak{H}_0$ and $\mathfrak{H}_1$ respectively. Hence the number of pairs $(e_0, e_1) \in \Delta^{\Gamma_0 \times \Gamma_1}$ is for fixed $e_0$ bounded by the number of elements $e_1$ in the same depth of $\mathfrak{H}_1$ as $e_0$ in $\mathfrak{H}_0$. For each element $(e_0, e_1) \in \Delta^{\Gamma_0 \times \Gamma_1}$, connected by a path of length $\leq \operatorname{rank} \mathcal{T}$ from the root, requirement 3. adds $n$-successor pairs to $\Gamma_0 \times \Gamma_1$. Hence we can estimate the number of elements in depth $n < \operatorname{rank} \mathcal{T}$ by counting the successors of depth $n+1$ which are bounded by the number of elements in depth $n \cdot |\mathcal{T}|$. This bound on elements in $(e_0, e_1) \in \Delta^{\Gamma_0 \times \Gamma_1}$ for fixed depth from the root $(d_0, d_1)$ then also serves as bound for the maximal out-degree that an element can have.

We can use these two considerations to synthesise a function $f$ which recurs-ively estimates the maximal number of successors for each element and each depth in $\mathfrak{H}_0$:

$$f(n) := \begin{cases} |\mathcal{T}| + |\mathcal{T}| \cdot 1 & n = 0 \\ |\mathcal{T}| + |\mathcal{T}| \cdot f(n-1) & \text{otherwise} \end{cases}$$

Clearly, the root $d_i$ has $\frac{|\mathrm{clos}\,\mathcal{T}|}{2} = |\mathcal{T}|$-many successors plus $|\mathcal{T}|$-many additional successors for the root element $(d_0, d_1)$. The recursion exploits that it is an estimation for both interpretations $\mathfrak{H}_0$ and $\mathfrak{H}_1$: The number of elements $(e_0, e_1)$ for fixed $e_0$ depends on the number of $r$-successors of the predecessor of $e_1$. Hence there are $f(n-1)$ many. Each $(e_0, e_1)$ has $|\mathcal{T}|$-many successors so requirement 3 adds at most $|\mathcal{T}| \cdot f(n-1)$ to $e_0$ plus at most $|\mathcal{T}|$-many that are introduced by requirement 2.

After rewriting $f$, we shall give an explicit form of $f$: $f(n) = 2 \cdot (\sum_{k=0}^{n} |\mathcal{T}|^{k+1})$. The proof of this equation is conducted by induction upon $n < \omega$ for which the base case is clear. In the step-case

$$\begin{aligned} f(n+1) &= |\mathcal{T}| + |\mathcal{T}| \cdot f(n) \overset{\mathrm{IH}}{=} |\mathcal{T}| + |\mathcal{T}| \cdot 2 \cdot (\textstyle\sum_{k=0}^{n} |\mathcal{T}|^{k+1}) \\ &= |\mathcal{T}| + 2 \cdot (\textstyle\sum_{k=1}^{n+1} |\mathcal{T}|^{k+1}) = 2 \cdot (\textstyle\sum_{k=0}^{n+1} |\mathcal{T}|^{k+1}) \end{aligned}$$

The partial sum is then explicitly given for $|\mathcal{T}| > 1$ by

$$f(n) = 2 \cdot (\sum_{k=0}^{n} |\mathcal{T}|^{k+1}) = 2 \cdot \frac{|\mathcal{T}|^{n+1} - 1}{|\mathcal{T}| - 1} \leq |\mathcal{T}|^{n+2}$$

Since Item 3 only requires successors for elements up to depth $\mathrm{rank}\,\mathcal{T}$, all successors of $\mathfrak{H}_0$ and $\mathfrak{H}_1$ in depth $> \mathrm{rank}\,\mathcal{T}$ have only up to $|\mathcal{T}|$ successors. We will hence reach the greatest number of successors per element in depth $\mathrm{rank}\,\mathcal{T}$: $f(\mathrm{rank}\,\mathcal{T}) = |\mathcal{T}|^{\mathrm{rank}\,\mathcal{T}+2} \leq (2^{|\mathcal{T}|})^{|\mathcal{T}|} \leq 2^{(|\mathcal{T}|^2)}$, where the first inequality is true for all $|\mathcal{T}| > 1$ as $\mathrm{rank}\,\mathcal{T} < |\mathcal{T}|$: The proof for $|\mathcal{T}|^{\mathrm{rank}\,\mathcal{T}+2} \leq 2^{(|\mathcal{T}|^2)}$ can be carried out by hand for $|\mathcal{T}| \in \{2, 3\}$ and then by induction, where

$$(n+1)^{n+3} \leq 2^{(n+1)^2} \text{ if (IH) } \frac{(n+1)^{n+3}}{n^{n+2}} \leq \frac{2^{(n+1)^2}}{2^{n^2}} \text{ if } (1 + \frac{1}{n})^{n+2}(n+1) \leq 2^{2n+1}$$

$$\text{if } 2^{n+2}(n+1) \leq 2^n \cdot 2^{n+1} \text{ if } 2(n+1) \leq 2^n$$

whereat the latter is true for all $n \geq 3$. □

Definition 6.2.18. We define the class of *potential tree-models for $\mathcal{T}$* to exactly comprise all tree-models $\mathfrak{J}$ of depth rank $\mathcal{T}$ such that there is a mapping $t : \Delta^{\mathfrak{J}} \longrightarrow$ tp $\mathcal{T}$ such that

1. for all $d_1 \in \Delta^{\mathfrak{J}}$ which is reachable from the root $d_0 \in \Delta^{\mathfrak{J}}$ via a path of length $\ell \leq \operatorname{rank} \mathcal{T}$, we have

$$(\mathfrak{J}, d_1) \models A \iff A \in t(d_1) \text{ for all } A \in \mathsf{N}_\mathsf{C}$$

2. if $d_1 \in \Delta^{\mathfrak{J}}$ which is reachable from the root $d_0 \in \Delta^{\mathfrak{J}}$ via a path of length $\ell < \operatorname{rank} \mathcal{T}$ and $\exists r.C \in t(d_1)$ then there is $t_2 \in \Delta^{\mathfrak{J}}$ with $(t_1, t_2) \in r^{\mathfrak{J}}$ such that $C \in t(d_2)$.

3. for all $r \in \mathsf{N}_\mathsf{R}$ and all $(d_1, d_2) \in r^{\mathfrak{J}}$ we have $t(d_1) \rightsquigarrow_r t(d_2)$.

◇

We justify the name 'potential tree-models for $\mathcal{T}$' by proving the following observation:

Observation 6.2.19. *Every potential tree-model of $\mathcal{T}$ can be continued to a model of $\mathcal{T}$*

Proof. Let $\mathfrak{J}$ be a potential tree-model of $\mathcal{T}$ with root $d_0$ and let

$$\operatorname{leaf} \mathfrak{J} := \{d \in \Delta^{\mathfrak{J}} \mid d \text{ is a leaf in depth of rank } \mathcal{T} \text{ in } \mathfrak{J}\}.$$

Recall that every $t \in \operatorname{tp} \mathcal{T}$ is satisfiable with $\mathcal{T}$ (page 6.1). For each $d \in \operatorname{leaf} \mathfrak{J}$ let $\mathfrak{K}_d$ such that $(\mathfrak{K}_d, c_d) \models \mathcal{T} \cup t(d)$ where $t : \Delta^{\mathfrak{J}} \longrightarrow \operatorname{tp} \mathcal{T}$ is as in Definition 6.2.18. We create a new interpretation $\mathfrak{K}$ being the disjoint union of $\mathfrak{J}$ with all $\mathfrak{K}_d$ where $d \in \operatorname{leaf} \mathfrak{J}$ and additionally add edges $(d, c')$ to $r^{\mathfrak{K}}$ if $(c_d, c') \in r^{\mathfrak{K}_d}$ for all $d \in \operatorname{leaf} \mathfrak{J}$. Hence we have connected a leaf $d$ with all successors of a node that satisfies $t(d)$ and so $d \in C^{\mathfrak{K}}$ for all $C \in t(d)$.

To prove that $\mathfrak{K} \models \mathcal{T}$ we show that every element in $\mathfrak{K}$ satisfies some $t \in \operatorname{tp} \mathcal{T}$. The claim is clear for every element in $\mathfrak{K}_d$ for every $d \in \operatorname{leaf} \mathfrak{J}$. We concentrate on elements $d \in \Delta^{\mathfrak{J}}$ and show via induction upon $n < \omega$ that $(\mathfrak{K}, d) \models t(d)$ for all elements in distance $\operatorname{rank} \mathcal{T} - n \geq 0$.

The claim is true for $n = 0$, as any element $d$ in distance rank $\mathcal{T}$ from $d_0$ was a leaf in $\mathfrak{J}$ and is hence connected to all successors of $c_d$ from $\mathfrak{K}_d$ which proves the claim. For $\operatorname{rank} \mathcal{T} - (n + 1) \geq 0$ let $d \in \Delta^{\mathfrak{K}}$. If $d$ is a leaf in $\mathfrak{K}$, it was a leaf in $\mathfrak{J}$ and so $t(d)$ cannot have contained any concept of form $\exists r.C$ for any $r \in \mathsf{N_R}$ and $C \in \operatorname{clos} \mathcal{T}$, so $(\mathfrak{I}, d) \models t(d)$. As $(\mathfrak{I}, d) \Longleftrightarrow (\mathfrak{H}, d)$ we obtain $(\mathfrak{H}, d) \models t(d)$.

If $d$ is not a leaf, we prove by induction upon the structure of $C \in \operatorname{clos} \mathcal{T}$ that $(\mathfrak{I}, d) \models C$ iff $C \in t(d)$. The claim is true by Item 1 of Definition 6.2.18 for atomic concepts. Since $\operatorname{clos} \mathcal{T}$ contains conjuncts of conjunctions and is closed under negation, the induction hypothesis applies. Assume $\exists r.C \in t(d)$. Then Item 2 of 6.2.18 guarantees the existence of an $r$-successor $d' \in \Delta^{\mathfrak{I}}$ such that $C \in t(d')$. By induction hypothesis $(\mathfrak{K}, d') \models t(d')$ which proves $(\mathfrak{K}, d) \models \exists r.C$.

In case $\exists r.C \notin t(d)$ then, since $t$ is complete with respect to $\operatorname{clos} \mathcal{T}$, we have $\neg\exists r.C \in t(d)$. Hence for all $t \in \operatorname{tp} \mathcal{T}$ with $C \in t$ we have $t(d) \not\rightsquigarrow_r t$ and so Item 3 of 6.2.18 guarantees that $C \notin t(d')$ for all $r$-successors $d'$ of $d$. Again completeness of $t$ w.r.t. $\operatorname{clos} \mathcal{T}$ entails $\neg C \in t(d')$ for all $r$-successors $d'$ when with the induction hypothesis $(\mathfrak{I}, d) \models \neg\exists r.C$ is inferred. □

Clearly, if $\mathfrak{I}$ is a tree-model of $\mathcal{T}$ then $(\mathfrak{I} \restriction \operatorname{rank} \mathcal{T})$ is a potential tree-models of $\mathcal{T}$, where $(\mathfrak{I} \restriction \operatorname{rank} \mathcal{T})$ contains only those elements reachable via a path of length up to $\operatorname{rank} \mathcal{T}$ from the root.

Proposition 6.2.20 . *An $\mathcal{ALC}$-TBox $\mathcal{T}$ is not preserved under direct products, iff there are two potential tree-models $\mathfrak{I}, \mathfrak{H}$ with maximal out-degree $2^{|\mathcal{T}|^2}$ such that $(d, e) \in (C \sqcap \neg D)^{\mathfrak{I} \times \mathfrak{H}}$ for the roots $d$ of $\mathfrak{I}$ and $e$ of $\mathfrak{H}$, respectively, for some $C \sqsubseteq D \in \mathcal{T}$.*

Proof. The if-direction is proved using the tree-models $\mathfrak{I}, \mathfrak{H}$ yielded by the Proof of Proposition 6.2.16, for whose roots $d \in \Delta^{\mathfrak{H}}$ and $e \in \Delta^{\mathfrak{H}}$ we have $(\mathfrak{I} \times \mathfrak{H}, (d, e)) \models C \sqcap \neg D$ for some $C \sqsubseteq D \in \mathcal{T}$. Their restriction $\mathfrak{I} \restriction \operatorname{rank} \mathcal{T}$ and $\mathfrak{H} \restriction \operatorname{rank} \mathcal{T}$ are both potential tree-models of $\mathcal{T}$ with maximal out-degree $2^{|\mathcal{T}|^2}$. For the roots $d \in \Delta^{\mathfrak{H}}$ and $e \in \Delta^{\mathfrak{H}}$ we then have $(\mathfrak{I}, d) \Longleftrightarrow_{\operatorname{rank} \mathcal{T}} (\mathfrak{I} \restriction \operatorname{rank} \mathcal{T}, d)$ and $(\mathfrak{H}, e) \Longleftrightarrow_{\operatorname{rank} \mathcal{T}} (\mathfrak{H} \restriction \operatorname{rank} \mathcal{T}, e)$ and so Proposition 6.2.13 yields $(\mathfrak{I} \restriction \operatorname{rank} \mathcal{T} \times \mathfrak{H} \restriction \operatorname{rank} \mathcal{T}, (d, e)) \Longleftrightarrow_{\operatorname{rank} \mathcal{T}} (\mathfrak{I} \times \mathfrak{H}, (d, e))$ which in turn entails $(\mathfrak{I} \restriction \operatorname{rank} \mathcal{T} \times \mathfrak{H} \restriction \operatorname{rank} \mathcal{T}, (e, d)) \models C \sqcap \neg D$.

Assume there are two potential tree-models $\mathfrak{I}_0, \mathfrak{H}_0$ of $\mathcal{T}$ such that at their roots $d$ of $\mathfrak{I}$ and $e$ of $\mathfrak{H}$ we have $(\mathfrak{I} \times \mathfrak{H}, (d, e)) \models C \sqcap \neg D$ for some $C \sqsubseteq D \in \mathcal{T}$. Then Observation 6.2.19 yields continuations $\mathfrak{I}$ and $\mathfrak{H}$, respectively which are both models of $\mathcal{T}$. Since $\mathfrak{I} \restriction \operatorname{rank} \mathcal{T} = \mathfrak{I}_0$ and similarly for $\mathfrak{H}$, we have $(\mathfrak{I}, d) \Longleftrightarrow_{\operatorname{rank} \mathcal{T}} (\mathfrak{I}_0, d)$ and $(\mathfrak{H}, d) \Longleftrightarrow_{\operatorname{rank} \mathcal{T}} (\mathfrak{H}_0, d)$ and Proposition 6.2.13 yields $(\mathfrak{I} \times \mathfrak{H}, (d, e)) \Longleftrightarrow_{\operatorname{rank} \mathcal{T}} (\mathfrak{I}_0 \times \mathfrak{H}_0, (d, e))$, which shows $(\mathfrak{I} \times \mathfrak{H}, (d, e)) \models C \sqcap \neg D$. Hence there are two models $\mathfrak{I}, \mathfrak{H}$ of $\mathcal{T}$ but $\mathfrak{I} \times \mathfrak{H} \not\models \mathcal{T}$ which shows the only-if direction. □

Corollary 6.2.21 . *Whether or not $\mathcal{T}$ is preserved under direct products can be de-*

*cided in* NEXPTIME.

Proof. Every $\mathcal{ALC}$-TBox $\mathcal{T}$ can be rendered into the form $\{\top \sqsubseteq C\}$. So assume w.l.o.g. that $\mathcal{T} = \{\top \sqsubseteq C\}$. The steps of the algorithm are

1. Create two tree-interpretations $\mathfrak{I}$, $\mathfrak{H}$ with out-degree at most $2^{n^2}$.
2. Test whether $\mathfrak{I}\restriction\mathrm{rank}\,\mathcal{T}$ and $\mathfrak{H}\restriction\mathrm{rank}\,\mathcal{T}$ are potential tree-models of $\mathcal{T}$
3. If not, yield true. Otherwise, form the direct product and yield whether the root node satisfies $C$.

We use a non-deterministic Turing-machine to execute the algorithm. If this non-deterministic Turing-machine yields false for one computation then $\mathcal{T}$ is not preserved under direct products.

Step 1. Let $n := |\mathcal{T}|$. There are at most $2^n$ different labellings with symbols from $\mathsf{N_C}$ for each node. Hence for every node the non-deterministic Turing-machine has to choose whether or not to realise an $r$-successor. This amounts to $|\mathsf{N_R}| \cdot 2^n \leq 2^{2n}$ many steps in which each the Turing-machine splits up. We then follow each copy of the Turing-machine separately.

The Turing-machine builds the tree depth-layer by depth-layer. For each node in the same depth the Turing-machine creates a successor set as just explained where the number of nodes in each successor set is bound by $2^{n^2}$. Hence a layer in depth $k$ has at most $2^{k \cdot n^2}$ many nodes.

Let the Turing-machine create tree-interpretations of depth 2rank $\mathcal{T}$, which will be explained later. Every tree, then, has at most

$$\sum_{k=0}^{2\cdot\mathrm{rank}\,\mathcal{T}} 2^{k\cdot n^2} = \sum_{k=0}^{2\cdot\mathrm{rank}\,\mathcal{T}} \frac{2^{2\cdot\mathrm{rank}\,\mathcal{T}\cdot n^2}}{2^k} = 2^{2\cdot\mathrm{rank}\,\mathcal{T}\cdot n^2} \cdot \sum_{k=1}^{2\cdot\mathrm{rank}\,\mathcal{T}} \frac{1}{2^k} \leq 2^{2\cdot\mathrm{rank}\,\mathcal{T}\cdot n^2+1}$$

many nodes, for which each at most $2^{2n}$ many steps are needed to determine the successor set. Hence each copy of the machine needs, to create both interpretations at most $2 \cdot 2^{2\cdot n+2\cdot\mathrm{rank}\,\mathcal{T}\cdot n^2+1} \leq 2^{2\cdot n+2\cdot(n-2)\cdot n^2+2} \leq 2^{n^3}$ many steps whenever $n \geq 3$ i.e. all nonempty TBoxes.

Step 2. For each interpretation we test whether $C$ is satisfied at every node up to depth rank $\mathcal{T}$. If so, then for every node up to depth rank $\mathcal{T}$ there is a mapping $t : \Delta \longrightarrow \mathrm{tp}\,\mathcal{T}$ and requirements 1–3 in Definition 6.2.18 are trivially satisfied. Hence we determine whether or not the two tree-interpretations in Step 1 are potential tree-models of $\mathcal{T}$.

We hence have to consider $2^{n^3}$ subtrees of size $2^{n^3}$ for which we have to check whether each subtree is a model of the $\mathcal{ALC}$-concept $C$ whose length is estimated as $|C| < n$. Each check can be performed in at most $|C| \cdot 2^{n^3} < n \cdot 2^{n^3}$ steps an so for both tree-interpretations together we need at most $2 \cdot 2^{n^3} \cdot n \cdot 2^{n^3} \leq 2^{3n^3}$ many steps.

Step 3. Assume that both interpretations $\mathfrak{I}$ and $\mathfrak{H}$ computed in Step 1 are potential tree-models of $\mathcal{T}$ with roots $d \in \Delta^{\mathfrak{I}}$ and $e \in \Delta^{\mathfrak{H}}$. We then create their direct product for their elements up to depth rank $\mathcal{T}$ and end up with a set of size $2^{n^3} \cdot 2^{n^3} = 2^{2 \cdot n^3}$ which we can compute together with the extensions for all role names (2) and all concept names (3) in at most $2^{2 \cdot n^3} \overset{(2)}{+} n \cdot 2^{4 \cdot n^3} \overset{(3)}{+} n \cdot 2^{2 \cdot n^3}$ many steps. The model checking then takes another $n \cdot 2^{2n^3}$ steps.

Combining the results of all steps together we get at most

$$2^{2n+n^3} + 2^{3 \cdot n^3} + 2^{2 \cdot n^3} + n \cdot 2^{4 \cdot n^3} + n \cdot 2^{2 \cdot n^3} + n \cdot 2^{2 \cdot n^3} \leq 4n \cdot 2^{4n^3} \leq 2^{5n^3}$$

steps to termination for a non-deterministic Turing machine, which shows that determining whether $\mathcal{ALC}$-TBoxes are not preserved under direct products is in NEXPTIME. □

In [92] it is shown that the algorithm is NEXPTIME-complete.

Both, Algorithm 6.1.2 and Algorithm 6.2.5 together with its decision procedure for whether or not an $\mathcal{ALC}$-TBox is preserved under direct products, do not yield TBox rewritings, but simply decide whether or not such an rewriting exists. They hence decide whether two theories of different logics are logically equivalent. Such an approach, which exploits characterisation results in this way is novel.

# 7. Conclusion

The model-theoretic properties of description logics were, so far, mainly investigated on concept level [85]. Because of the correspondence of description logic on concept-level to modal logics [121, 88], results of the well investigated model theory of modal logics could be transferred to description logic—but only on concept level.

The special model theoretic nature of TBoxes, an original part of description logics, has not been investigated so far. Thus, this thesis closes a gap in the understanding of description logics: We have imported notions and techniques from the model-theoretic investigation of modal-logics and classical logic in order to characterise the expressivity of some very important description logics like $\mathcal{ALC}$, $\mathcal{ALCQIO}$ and $\mathcal{EL}$ where we have put a special focus on TBoxes.

The characterisation results for the $\mathcal{ALC}$-family on concept level, which are mostly known from modal logic, are all based on some adaption of bisimulation. A characterisation result for the extension of the description logic in hand with the universal role can be obtained by defining its adaption of bisimulation on the global level. Characterisations for TBoxes make use of this global bisimulation but also need some form of invariance under disjoint union.

The initial characterisation results for $\mathcal{ALC}$, $\mathcal{ALCI}$ and $\mathcal{ALCQ}$ on all levels, i.e. on concept level, on concept level with universal role and on TBox level, show that the notions we have chosen to characterise these description logics are in no way ad hoc but are natural notions in connexion with the description logic family $\mathcal{ALC}$.

Invariance under disjoint union implies not only being preserved under disjoint union but also being preserved under generated subinterpretations. This is not problematic for $\mathcal{ALC}$, $\mathcal{ALCI}$ and $\mathcal{ALCQ}$ but it gets more involved for description logics with individuals.

Since individuals are interpreted as unique elements, a disjoint union of inter-pretations can no longer performed as naïve set-theoretic union of disjoint parts. Instead, conditions had to be given when a disjoint union is admissible and techniques were developed in order to unify possible multiplicity of individuals.

Nominal disjoint unions, the disjoint union used for the characterisation of $\mathcal{ALCQIO}$-TBoxes, simply required that each of the individuals was present in exactly one of the disjoint parts which were meant to be unified. Hence no multiplicity of individuals could occur. Disjoint unions for $\mathcal{ALCO}$-TBoxes required that all individuals were present in some of the disjoint parts but a allowed individuals to occur multiple times. It additionally required that if a individual occurred multiple times then these elements needed to be bisimilar. The same was requirement was made for disjoint unions in the case of $\mathcal{ALCQO}$.

In order to unify these multiple individuals, $\mathcal{ALCO}$ allowed to simply factorise the disjoint union by the largest bisimulation relation. This turned out to be for $\mathcal{ALCQO}$ more complicated. Instead of factorising, a forest-unravelling was performed for every disjoint part. For the subtrees that had individuals in their root, new subtrees were synthesised such that individuals with the same name were roots of isomorphic subtrees. These trees and hence the individuals could then be unified by a factorisation.

Looking at the different techniques needed to characterise $\mathcal{ALCQIO}$, $\mathcal{ALCO}$ and $\mathcal{ALCQO}$, the difficulty level seems to rise which makes $\mathcal{ALCQIO}$ appear more balanced in its expressivity: individuals which are the cause for the break down of the invariance under normal disjoint unions are met by inverse roles which reduce the notion of generated substructure to simply connected component. It remains to be seen whether this interaction between individuals and inverse roles is coincidental or whether a fundamental reason can be found.

$\mathcal{EL}$ gave a slightly different picture as the model-theoretic notion is not bisimulation but simulation. Simulation is asymmetric in nature but the model theoretic results are analogue to $\mathcal{ALC}$ on concept level. The characterisation for $\mathcal{EL}$-concepts needed, since $\mathcal{EL}$ does not incorporate disjunctions $\sqcup$, an extra condition: either the minimal model property or preservation under direct products.

For $\mathcal{EL}$-TBoxes a global version of simulation is not sufficient but rather the global version of a symmetricised notion of simulation: global equi-simulation. $\mathcal{EL}$-TBoxes are then characterised under this global equi-simulation, disjoint union and being preserved under direct products.

We have learnt that TBoxes lend to $\mathcal{EL}$ a considerable amount of additional expressivity, witnessed by the fact that global equi-simulation is necessary to char-acterise $\mathcal{EL}$-TBoxes. Another interesting fact is, that $\mathcal{EL}$-concepts and $\mathcal{EL}$-TBoxes are preserved under direct products. From a model-theoretic point of view, it would be interesting to know how $\mathcal{EL}$ and $\mathcal{EL}$-TBoxes relate to $\mathcal{ALC}$ and $\mathcal{ALC}$-TBoxes w.r.t. to the formation of direct products: Is $\mathcal{EL}$ the fragment of $\mathcal{ALC}$ which is invariant under direct products? Are $\mathcal{EL}$-TBoxes the fragment of $\mathcal{ALC}$-TBoxes which are preserved under direct products? The questions are open and we shall meet them in the section on Future Work again.

Finally the TBox-rewritability introduced a new class of problems, namely deciding whether or not a TBox can be equivalently expressed in a weaker language using the same vocabulary. Hence we give a decision procedure for whether or not two theories of different logics are logically equivalent. This application of model-theoretic characterisation is interesting in its own right.

The characterisations can have potential applications in areas where the expressive power of TBoxes and description logics per se are central. Such areas could be TBox approximations [120, 82]: As discussed in the beginning, the reasoning complexity rises with the expressivity and hence reasoning does not scale for very expressive description logics (see e.g. the example ontology FMA in [99]) As a solution, TBoxes for these expressive description logics are approximated by TBoxes of less expressive languages, e.g. $\mathcal{EL}$ [111]. The reasoning is then performed on the approximant rather than on the original TBox. As price for the speed up, reasoning becomes incomplete w.r.t. to the original TBox.

Similarly, the results can also be interesting for TBox modularisation [83]: In modularisation parts of a given TBox $\mathcal{T}$ are extracted into a TBox $\mathcal{T}_0 \subseteq \mathcal{T}$, the module, such that $\mathcal{T}_0$ is contains only the part which is relevant to the reasoning task. 'Relevant' is different depending on the application in hand.

In both cases the expressive power of the description logic is of central importance and thus our results may contribute for future work.

The thesis gives a model-theoretical basis to major DLs apart from the thorough investigations of DLs in terms of complexity theory and reasoning entertained for so many years before. The expressivity of the languages treated is now understood in model-theoretic terms and can be used as a theoretical foundation to build on.

Apart from that, the author hopes that this thesis can serve as introductory text to the model-theoretic foundations of the treated logics, without claiming to be complete nor extensive. We give a brief summary of the chapters before setting out future work.

## 7.1 Summary

### Chapter 2

The chapter recapitulates in a rather patient manner the model-theoretic notions and results which can be found condensed in [61]. It gives different views on bisimulation and explains in detail the interconnexions between the notions like type, saturation, the Hennessy-Milner-Property and characteristic concepts. The well known characterisation of van Benthem is recalled together with its proof. The following section extrapolates the results to a characterisation of $\mathcal{ALC}u$-concepts, i.e. $\mathcal{ALC}$ extended by a universal role, which is also known as $\forall\mathcal{ML}$ in modal logics. This paves the way to the $\mathcal{ALC}$-TBox characterisation, where every $\mathcal{ALC}$-TBox is equivalent to some $\mathcal{ALC}u$-concept, so that the $\mathcal{ALC}$-TBoxes effectively form a fragment of $\mathcal{ALC}u$-concepts.

For the characterisation of $\mathcal{ALC}$-TBoxes, the notion of global bisimulation, introduced in the $\mathcal{ALC}u$-section, is used and invariance under disjoint unions. The characterisation gives us an insight in the expressiveness of $\mathcal{ALC}$-TBoxes: nothing can be expressed which is sensitive to disjoint unions, generated substructures or which is not invariant under global bisimulation. In particular this means that it must be invariant under tree-unravellings.

### Chapter 3

Chapter 3 treats the model-theoretic characterisation of the expressiveness of two extensions of $\mathcal{ALC}$, namely $\mathcal{ALCI}$ and $\mathcal{ALCQ}$. Syntax and semantics and all important notions introduced for $\mathcal{ALC}$ are adapted and explicitly stated. One can observe that $\mathcal{ALCI}$, but in particular $\mathcal{ALCQ}$ behave analogously to $\mathcal{ALC}$, if only the notions of saturation, bisimulation and the logic itself fit together, witnessed by the Hennessy-Milner-Property.

In particular, the stratification of the bisimulation relation together with the characteristic concepts for $\mathcal{ALCQ}$ is developed in detail. Along with this, not only the extension with a simple universal role for $\mathcal{ALCQ}$ is investigated but also the extension with counting universal role. For both logics, a characterisation on concept level for $\mathcal{ALCI}$ and $\mathcal{ALCQ}$ as well as for their extensions with universal quantifiers and finally for TBoxes is given.

Here, we focus on the extension with nominals namely $\mathcal{ALCO}$, $\mathcal{ALCQO}$, and $\mathcal{ALCQIO}$. Nominals are sometimes treated as constants, but we felt that treating them as predicates which are interpreted as singleton sets is more appropriate: The local nature of bisimulation is preserved and thus the stratified bisimulation relation can be captured by characteristic $\mathcal{ALCO}$-concepts.

On concept level, also for concepts with universal role, $\mathcal{ALCO}$, $\mathcal{ALCQO}$ and $\mathcal{ALCQIO}$ behave very much like their counterparts without individuals. Many propositions and theorems can be, with caution, transferred and thus the characterisations can be swiftly derived. Since $\mathcal{ALCQIO}$ might be, apart from $\mathcal{EL}$ the most interesting DL treated in this thesis, so the chapter is kept fairly independent.

But $\mathcal{ALCO}$ is more challenging, when it comes to the characterisation of $\mathcal{ALCO}$-TBoxes. The invariance of disjoint unions interferes with the special nature of nominals: a naive disjoint union destroys the uniqueness of elements, a naive decomposition into substructures may leave nominals uninterpreted, i.e. empty.

As adaption, both, the definition of subinterpretation and disjoint union are altered. In particular, the formation of disjoint unions is only allowed over interpretations in which nominals with the same name are bisimilar. For then the disjoint union can be factorised by the bisimulation relation, merging all nominals with the same name.

This approach gets in particular technical for $\mathcal{ALCQO}$ as simple factorisation does not preserve the exact number of successors. An involved construction first performs tree-unravelling on the interpretations, then renders trees into uniform trees, forms the disjoint union and factorises the result, obtaining an interpretation which is bisimilar to a naive disjoint-union, but ensures that every nominal is interpreted by exactly one element.

As recompense for all the trouble the reader is treated with a construction for minimal globally bisimilar companions which has been, to the best of our knowledge, open so far. The construction can be extended to find $\mathcal{ALCQIO}$-bisimilar companions.

It turns out that $\mathcal{ALCQIO}$ is much simpler than $\mathcal{ALCQO}$, as a simple pick-and-mix procedure of connected components already yields an adequate adaption for the notion of disjoint unions which allows a characterisation of $\mathcal{ALCQIO}$-TBoxes.

Chapter 5

This Chapter treats the characterisation of $\mathcal{EL}$. The model-theoretic notions used to characterise $\mathcal{ALC}$ transfer relatively smoothly to its weak fragment. Though, the common model-theoretic notion for $\mathcal{EL}$, simulation, does not only preserve $\mathcal{EL}$ itself but $\mathcal{EL}^{\sqcup}$, i.e. $\mathcal{EL}$ extended with 'OR'. To obtain $\mathcal{EL}$ itself, another property is needed. Exchangeably the minimal model property or being preserved under direct products can be used to complete the characterisation.

But the peculiarities carry on, when TBoxes have to be characterised. It turns out that the concept inclusions introduce a hidden negation, which lifts the expressiveness of $\mathcal{EL}$-TBoxes so much so that a symmetric notion of simulation, so called equi-simulation, is needed to characterise $\mathcal{EL}$-TBoxes.

Hence Chapter 5 gave rise to the investigation of several fragments on concept level: The one which is preserved by simulation ($\mathcal{EL}^{\sqcup}$), the one which is preserved by simulation and direct products ($\mathcal{EL}$) and finally $\mathcal{EL}u^{\neg}$, the logic which is invariant under global equi-simulation.

We finally identify $\mathcal{EL}$-TBoxes as $\mathcal{EL}u^{\neg}$-TBoxes which are preserved by direct products. This also justifies the characterisation of $\mathcal{EL}$ by simulation and direct products on the concept level.

Chapter 6

Chapter 6, finally, gives an application of the characterisation results: They can be used to decide the rewritability problem for TBoxes, i.e. whether a given TBox can be equivalently expressed in a weaker logic using the same signature symbols. To this end, two algorithms are presented. One, which decides whether an $\mathcal{ALCI}$-TBox can be expressed as $\mathcal{ALC}$-TBox and one which decides whether an $\mathcal{ALC}$-TBox can be expressed as an $\mathcal{EL}$-TBox.

Whilst the former merely has to check whether the TBox is invariant under global bisimulation, the latter has to check whether the TBox is invariant under global equi-simulation and whether the TBox is preserved under direct products.

## 7.2 Future Work

The open questions and areas for future work are plenty. An immediate question is whether the characterisation results can be extended to DLs which incorporate RBoxes like $\mathcal{SROIQ}$ [77] and $\mathcal{SHOIQ}$ [81]. One would have to give a notion of bisimulation that preserves role hierarchies, e.g. where edge labels are considered to be subsets of $\mathsf{N_R}$ rather than elements from $\mathsf{N_R}$ which must be respected by this bisimulation. The idea of treating statements like trans $r$ similar to individuals by restricting the view to some class $\mathbb{K}$ of interpretation which interpret $r$ as transitive relation, might prove to be useful.

Another, very interesting question is, how the expressivity of a language can be increased if fresh signature symbols are allowed to be introduced. More concrete, is it e.g. possible to express an $\mathcal{ALC}$-TBox $\mathcal{T}$ over $\tau$ by an $\mathcal{EL}$-TBox $\mathcal{T}'$ over $\tau'$ where $\tau \subseteq \tau'$ such that $\mathcal{T}' \models \mathcal{T}$ and for all $\tau$-interpretations $\mathfrak{I} \models \mathcal{T}$ there is a $\tau'$-interpretation $\mathfrak{H} \models \mathcal{T}'$ such that $\mathfrak{H} \cong_\tau \mathfrak{I}$. In this case $\mathfrak{H}$ is called a conservative extension [96, 56, 94, 8] of $\mathfrak{I}$.

Further instances of rewriting problems come to mind like $\mathcal{ALCQ}$-to-$\mathcal{ALC}$-rewritability and many more. They all have not been investigated as of yet. Connected to this is the question 'is it decidable for a given TBox in $\mathcal{ALC}$ whether or not it is rewritable under signature extension in $\mathcal{EL}$'? Again, $\mathcal{ALC}$ and $\mathcal{EL}$ are in this problem interchangeable with other description logics.

A bit aside but not less important are questions that arose during the consideration of direct products. During their investigation, we became aware that even the precise FO-fragment which is preserved under the formation of direct products is not known[1]. On the other hand, the FO-fragment of all sentences which are preserved under reduced products is precisely known: It is the Horn-fragment of FO. In the context of DLs and in particular TBoxes a lot of fragments are called Horn-fragment, e.g. DL-Lite$_{Horn}$ [5, 6](model theoretic investigation in [92]). But is it really *the* Horn-fragment, i.e. does it comprise exactly all DL-Lite concepts which are preserved under the formation of reduced products? And how do $\mathcal{EL}$ and $\mathcal{EL}$-TBoxes relate to $\mathcal{ALC}$ and *ALC*-TBoxes, respectively, considering direct products?

[1][34] referred to [133], who questions whether or not the discovered subset of the FO-fragment is actually the whole fragment.

# Biography

Robert Piro was born on 17 March 1978 in Celle, Germany and raised in Mörfelden-Walldorf since he was five years old. There, he attended Bertha-von-Suttner-Schule a comprehensive high school until the age of 19.

After completing his national service in the German army, he started an apprenticeship with BHF-BANK in Frankfurt/Main and completed in 2000 the apprenticeship as certified bank clerk.

He then began his studies in Computer Science at Technical University (TU) of Darmstadt and transferred after two years to Mathematics. He studied Model Theory and Modal Logic under Professor Martin Otto at the Faculty of Mathematics of the TU-Darmstadt and graduated with 'very good' in 2008 with his master thesis 'Lindströmsche Sätze für Modale Logiken' (Lindström Theorems for Modal Logics).

In the same year, he won the PhD-scholarship from the University of Liverpool and studied under Professor Frank Wolter at the Computer Science Department of the University of Liverpool until 2011. He submitted his PhD-Thesis 'Model-theoretic Characterisations of Description Logics' in 2012.

Since 2011 he is employed as research assistant the University of Oxford in the Information Systems Group under Professor Ian Horrock and conducts research in parallel reasoning.

Throughout his time at the University of Darmstadt, the University of Liverpool and the University of Oxford he worked as student demonstrator.

Robert Piro is interested in the model theoretic foundations of description logic (DL) including the characterisation of the expressiveness of DLs and the parallelisation of automated reasoning with its complexity theoretic foundations.